\documentclass[rmp,reprint,amsmath,superscriptaddress,amssymb,floatfix]{revtex4-1}

\usepackage{bm}
\usepackage[retainorgcmds]{IEEEtrantools}
\usepackage{graphicx}
\usepackage{mathrsfs}
\usepackage{amsmath}
\usepackage{amssymb}
\usepackage{xcolor}
\usepackage{amsfonts}
\usepackage{times,txfonts} 
\usepackage{nicefrac}
\usepackage[colorlinks=true,linkcolor=blue,urlcolor=blue,citecolor=blue,pdfusetitle]{hyperref} 
\hypersetup{pdfauthor={Name}} 
\usepackage[normalem]{ulem}
\usepackage{bbm} 
\usepackage{bbold}  
\usepackage{soul}

\usepackage{makeidx}

\sethlcolor{red}

\makeindex

\DeclareMathOperator{\tr}{tr}
\newcommand{\vvec}{\text{vec}}
\newcommand{\trans}{^{\text{T}}}

\newcommand{\im}{{\rm i}}
\newcommand{\ii}{{\rm i}} 

\newcommand{\cur}[1]{I_{\rm{#1}}} 
 
\newcommand{\Jpart}{\cur{N}} 
\newcommand{\Jmag}{\cur{M}} 
\newcommand{\curOp}[1]{\mathcal{I}_{#1}} 

\newcommand{\Hs}{H_{\rm S}} 
\newcommand{\Hseff}{H_{\rm eff}} 
\newcommand{\Ns}{N_{\rm S}}
\newcommand{\Hint}{H_{\rm I}}
\newcommand{\Hi}{H_{\rm I}} 
\newcommand{\Hbath}{H_{\rm B}}
\newcommand{\Hb}{H_{\rm B}} 
\newcommand{\Nb}{N_{\rm B}} 
\newcommand{\Htot}{H_{\rm tot}} 
\newcommand{\HtotTF}{H_{\rm tot}^{\rm TF}} 
\newcommand{\Htarget}{H_{\rm T}} 
\newcommand{\Liouv}{\mathcal{L}}
\newcommand{\Contr}{\mathcal{C}}

\newcommand{\id}{\mathbb{1}} 
\newcommand{\aop}{a}
\newcommand{\bop}{b}
\newcommand{\cop}{c}
\newcommand{\aopd}{{a^{\dagger}}}
\newcommand{\bopd}{{b^{\dagger}}}
\newcommand{\copd}{{c^{\dagger}}} 
\newcommand{\szero}{\sigma^0} 
\newcommand{\sx}{\sigma^x} 
\newcommand{\sy}{\sigma^y} 
\newcommand{\sz}{\sigma^z} 
\newcommand{\splus}{\sigma^+} 
\newcommand{\sm}{\sigma^-} 
\newcommand{\su}{\sigma^u} 
\newcommand{\sd}{\sigma^d} 

\newcommand{\ups}{1} 
\newcommand{\dws}{-1} 

\newcommand{\ip}[1]{{\mbox{\boldmath$#1$}}} 
\newcommand{\ind}[1]{{\bf #1}\index{#1}} 
\newcommand{\indTwo}[2]{{\bf #1}\index{#2}} 

\newcommand{\npop}{\bar{n}}

\newcommand{\rhos}{\rho_{\rm S}} 
\newcommand{\rhob}{{\bar\rho_{\rm B}}}
\newcommand{\rhobI}{{\rho_{\rm B}}}
\newcommand{\rhosb}{{\rho_{\rm SB}}}
\newcommand{\rhoss}{\rho_{\rm ss}}
\newcommand{\rhossvec}{\vec{\rho}_\text{ss}}

\newcommand{\SO}[1]{\mathcal{ #1 }} 
\newcommand{\Diss}{\SO{D}}
   
\newcommand{\Dim}{{\rm }D_{dim}}

\newcommand{\Lind}[1]{D[#1]} 

\newcommand{\expval}[1]{\left< #1 \right>}

\newcommand{\ket}[1]{\left|#1\right>}
\newcommand{\bra}[1]{\left<#1\right|}
\newcommand{\braket}[2]{\left<#1|#2\right>}

\newcommand{\f}[1]{\mbox{\boldmath$#1$}}
\newcommand{\ord}[1]{{\cal O}{\left\{#1\right\}}}
\newcommand{\trace}[1]{{\rm tr}\left\{ #1 \right\}}
\newcommand{\ptrace}[2]{{\rm tr}_{#1}\left\{ #2 \right\}}
\newcommand{\traceB}[1]{{\rm tr_B}\left\{ #1 \right\}}

\newcommand{\abs}[1]{{\left| #1 \right|}}
\newcommand{\nn}{\nonumber\\}

\newcommand{\Rect}{{\mathcal{R}}}

\newcommand{\Drude}{\mathcal{D}_{W}}   
\newcommand{\cond}{\sigma}
\newcommand{\condreg}{\sigma_{\rm reg}}

\newcommand{\uniexpo}{\phi}

\definecolor{asparagus}{rgb}{0.53, 0.66, 0.42} 
\newcommand{\adp}[1]{\textcolor{black}{#1}} \definecolor{dgreen}{rgb}{0.0, 0.5, 0.0}
\newcommand{\gtl}[1]{\textcolor{black}{#1}} 
 
\newcommand{\gsc}[1]{\textcolor{black}{#1}}
 
\newcommand{\alphaq}{{\alpha_Q}}    
\newcommand{\betaq}{{\beta_Q}}    

\begin{document}

\title{Non-equilibrium boundary-driven quantum systems: models, methods and properties}

\author{Gabriel T. Landi} 
\email{gtlandi@gmail.com}
\affiliation{Instituto de F\'isica da Universidade de S\~ao Paulo,  05314-970 S\~ao Paulo (Brazil)}

\author{Dario Poletti} 
\email{dario\_poletti@sutd.edu.sg} 
\affiliation{Science, Mathematics and Technology Cluster and Engineering Product Development Pillar, Singapore University of Technology and Design, 8 Somapah Road in 487372 Singapore (Singapore)} 

\author{Gernot Schaller} 
\email{g.schaller@hzdr.de}
\affiliation{Helmholtz-Zentrum Dresden-Rossendorf, 
Bautzner Landstra{\ss}e 400 in 01328 Dresden (Germany)}
\affiliation{
Institut f\"ur Theoretische Physik, Technische Universit\"at Berlin, D-10623 Berlin (Germany)}



\begin{abstract}
Recent years have seen tremendous progress in the theoretical understanding of quantum systems driven dissipatively 
by coupling them to different baths at their edges. 
This was possible because of the concurrent advances in the {\it models} used to represent these systems, the {\it methods} employed, and the analysis of the emerging {\it phenomenology}.  
Here we aim to give a comprehensive review of these three integrated research directions. 
We first provide an overarching view of the models of boundary-driven open quantum systems, both in the weak and strong coupling regimes. 
This is followed by a review of state-of-the-art analytical and numerical methods, both exact, perturbative and approximate.
Finally, we discuss the transport properties of some paradigmatic one-dimensional chains, with an emphasis on disordered and quasiperiodic systems, 
the emergence of rectification and negative differential conductance, and the role of phase transitions, 
\gsc{and we give an outlook on further research options}.       
\end{abstract}

\maketitle{}

\def\thefootnote{}\footnotetext{The authors are listed in alphabetical order.} \def\thefootnote{\arabic{footnote}}\setcounter{footnote}{0}

\tableofcontents


\section{\label{sec:intro}Introduction}

A piece of material placed in contact with two baths at different temperatures can reach a \ind{non-equilibrium steady-state} (NESS) characterized by a current of heat from one bath to the other. 
This corresponds to the simplest scenario of non-equilibrium systems most of us have an intuition of.   
The currents are generated by differences in the baths, and the type of current that emerges depends on the properties of the bath, of the system and their coupling. 
The ability to control the transport properties of a system can result in devices such as diodes, transistors, and energy converters, which are at the core of a broad range of applications (c.f.~\cite{LiLi2012} for a review on ``phononics'' and~\cite{BenentiWhitney2017, DubiDiVentra2011} for reviews on thermoelectricity). 
A deeper understanding of transport at the quantum scale can lead the way towards significant progress in nanotechnologies.         

\begin{figure}[!b]
    \centering
    \includegraphics[width=\columnwidth]{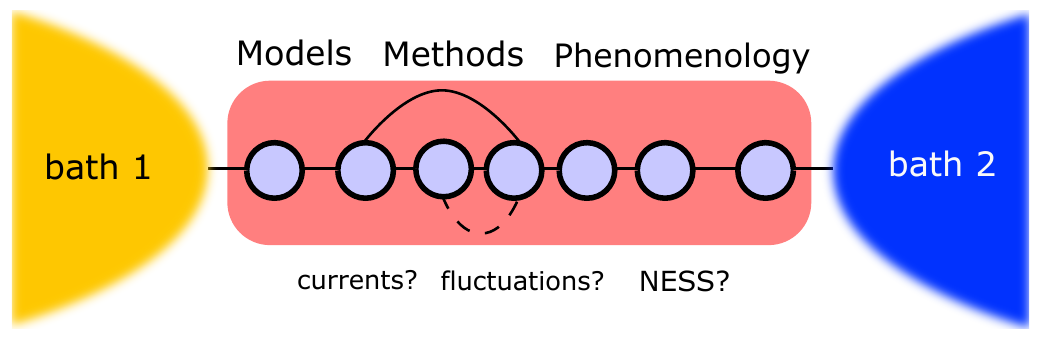}
    \caption{Boundary-driven systems are composite open systems that are locally coupled to external baths (yellow left and blue right) and internally via a Hamiltonian. This \gsc{enables NESSs (non-equilibrium steady states) characterized by stationary currents.}
    }
    \label{FIG:1dtransportsketch}
\end{figure}

Recent years have witnessed significant advances in the study of quantum systems coupled, at their edges, to different baths, as  depicted in Fig.~\ref{FIG:1dtransportsketch}.
We refer to these as \ind{boundary-driven systems}.
The 
possibly different temperatures or chemical potentials of the baths can cause a current, and the basic question to be addressed is what are the transport properties of the NESS, e.g. the properties of the currents generated.  
This review focuses on this type of scenario in the quantum regime. 
This is particularly relevant in light of experimental advances, which make setups like the one in Fig.~\ref{FIG:1dtransportsketch} directly accessible in the laboratory. 
The review is split into three fundamental questions:  finding the equations of motion, 
the development of analytical and numerical tools to study them, and the classification of the phenomenology that emerges, especially for strongly correlated quantum systems. 

To give an idea of the problems \adp{which will be reviewed,} we briefly consider heat transport between two baths at different temperatures through a bulk material. 
It was found by Fourier~\cite{Fourier1822}, that the heat current $\cur{}$ often reads
\begin{equation}\label{history_diffusive_flux}
    \cur{} = -\kappa_c\frac{\Delta T}{L}\,,
\end{equation}
where $L$ is the length of the material \gsc{in the direction of the current}, $\Delta T$ is the temperature difference and $\kappa_c$ is the thermal conductivity (often considered as constant).
One notices that in Eq.~\eqref{history_diffusive_flux},  known as \ind{Fourier's law}, the current decays as $\cur{}\sim 1/L$, which is referred to as \ind{diffusive transport}.

However, digging deeper one finds that transport can be a much richer field. 
First of all, the conductivity is actually a function of both the temperature $T$ and system size $L$. 
The temperature dependence of the conductivity causes the  current to change nonlinearly with temperature. 
Microscopically, this may e.g. result from reservoir modes controlling occupations at the boundaries. 
Importantly, a nonlinear response may cause the current to be very different in magnitude (and not just in sign) if one flips the temperature gradient $\Delta T$ to $-\Delta T$~\cite{BenentiPeyrard2016}. 
This means that the system can be used as a diode/rectifier~\cite{Terraneo2002, Li2004a}.   
\adp{For the dependence on the system size, often one finds that the current follows an algebraic dependence 
\begin{equation}
    \cur{} \sim 1/L^\alpha.\label{eq:generic_cur_scaling} 
\end{equation}
For $\alpha=0$, the current is independent of the system size,  known as \emph{ballistic} transport. 
As observed for Fourier's law, the case $\alpha=1$ is called diffusive. 
For $0<\alpha<1$ the transport is called superdiffusive, and for $\alpha>1$, subdiffusive. 
In the limit $\alpha\rightarrow\infty$ the current goes to zero, which implies that the system is an insulator. 
In this case the current decays faster than algebraically with the system size, e.g. $\cur{} \sim \exp\left(-L/L_0\right)$ where $L_0$ is referred to as the localization length~\footnote{In other contexts, the transport scenarios are sometimes defined by focusing on the thermodynamic limit $L\to \infty$ and looking at the conductivity $\kappa_c := - \cur{} L/\Delta T$ [Eq.~\eqref{history_diffusive_flux}]. Then, subdiffusiveness would lead to $\kappa_c\to 0$ (when $L\to \infty$) and hence would be referred to as insulating. Similarly,  superdiffusiveness would lead to $\kappa_c \to \infty$ and thus be referred to as ballistic. In this review, we focus on finite $L$, as this allows for deeper insights on the scaling behavior of transport mechanisms~\cite{Tritt2004}.}.
For a summary of transport exponents, see Table~\ref{tab:scaling} in Sec.~\ref{sec:properties}.
}

The techniques to derive the dependence of the conductivity on the system size are also relevant. 
If one considers transport due to phonons, it is natural to model the system as a chain of coupled harmonic oscillators, which often works well to describe the heat capacity of bulk materials. 
However, this turns out to result in \gtl{a ballistic current}~\cite{Rieder1967,Aoki2001}.
%
In classical systems it is found that nonlinearities in the Hamiltonian describing the system, are a key ingredient to recover a size-independent conductivity in  Fourier's law in Eq.~(\ref{history_diffusive_flux})~\cite{Casati1984} (see also~\cite{Dhar2008} for a recent review). 
It is thus crucial to use the correct models to study the transport properties of a system, as models with different characteristics can result in extremely different \gtl{dependencies of the currents} on the system and bath parameters. 
\emph{This is the essence of boundary-driven systems}.       

Given the complexity of the systems studied, \gtl{researchers} typically 
face three important questions:  
\begin{itemize}
    \item How can one model the dynamics of \gsc{system and baths?}
    \item Which analytical/numerical methods could one use\gsc{?}
    \item What different 
    \gsc{transport properties can one expect}? 
\end{itemize}
Each of these questions could be the focus of a review. However, we aim to give enough breadth and depth of information, so that someone starting in the field, or focusing in only some of these questions, can have a more comprehensive perspective. 
We thus hope to foster more research and advances in the field of boundary-driven systems.      
\gsc{We provide answers to the above questions in Secs.~\ref{sec:models},~\ref{sec:methods}, and~\ref{sec:properties}, respectively.}


%
%

Given the structure of the review, a reader who is already familiar with some of the questions, should be able to dive deep in the other issues. 
The theme of transport is extremely vast and has been studied for more than 200 years. 
It is thus important not only that we specify what we aim to review, but also what we chose not to. 
Our focus on quantum systems coupled at the their extremities to baths (i.e., dissipative boundary-driven ones) implies that we do not address classical systems (for a review, see~\cite{Dhar2008}). 
Regarding the models and methods, we chose not to review linear response theory, Landauer-B\"uttiker theory and Green's functions, as they have already been reviewed, e.g. in~\cite{wang2014a, DharHanggi2012,NikolicThygesen2012, ZimbovskayaPederson2011, Aeberhard2011, ProciukDunietz2010, MeirWingreen1992, CaroliSaint-James1971} or in textbooks such as~\cite{economou2006,HaugJauho2008,Ryndyk2016}.    
\gtl{Similarly, although some of our methods apply to non-Markovian systems, most of our focus will be on Markovian (time-local) quantum dynamics. For reviews specific on non-Markovianity, see~\cite{Breuer2015,Rivas2014,devega2017a}.}
As for recent experimental works, prominent results on strongly interacting systems have been reviewed in~\cite{BertiniZnidaric2020}, and hence we will only mention those experiments closer to the phenomenology discussed in Sec.~\ref{sec:properties}. 
We also focus exclusively on \gsc{systems without explicit external time-dependence}, and as such we will not touch on studies of periodically driven systems and their thermodynamic properties~\cite{BenentiWhitney2017, Kosloff2013, VinjanampathyAnders2016, MillenXuereb2016, VANDENBROECK2015,KohlerHanggi2005,GrifoniHanggi1998}.

\section{\label{sec:models}Models for boundary-driven open systems}



 
\subsection{Paradigms of open quantum system dynamics}\label{ssec:paradigm} 

We begin by discussing mathematical models of quantum transport. 
The natural framework for this is the theory of open quantum systems. 
In standard treatments it is assumed that the global density matrix of the system and its surrounding  baths evolves unitarily, according to the \ind{Liouville$-$von Neumann equation} \gtl{(we set $\hbar = 1$ throughout)}, 
\begin{equation}\label{preamble_vonNeumann_eq}
    \frac{d\rhosb}{d t} = -\im [\Htot, \rhosb]\,,
\end{equation}
where 
$\Htot = \Hs + \Hb + \Hint$
is the total Hamiltonian, encompassing system $\Hs$, bath $\Hb$ and their interaction $\Hint$.
This naturally includes the case of multiple  baths.

The initial conditions are usually taken such that $S$ and $B$ are uncorrelated; that is, in a product state $\rhosb(0) = \rhos(0)\otimes \rhobI(0)$. 
The global state at time $t$ is then 
\begin{equation}\label{preamble_global_unitary}
    \rhosb(t) = U(t) \rhosb(0) U^\dagger(t)\,, 
    \qquad U(t) = e^{-\im \Htot t}\,.
\end{equation}
The main goal in open quantum systems is to obtain the evolution of the systems reduced density matrix $\rhos(t) = \traceB{\rhosb(t)}$, where $\traceB{\ldots}$ stands for the partial trace over the bath.
This can be done at the level of the map~\eqref{preamble_global_unitary}, or the differential equation~\eqref{preamble_vonNeumann_eq}.
The former leads to the so-called Kraus maps and the latter to master equations~\footnote{
The name ``master equation'' was first coined in a paper about cosmic ray showers~\cite{Nordsieck1940}, where it played the role of a central equation from which all other results could be derived, i.e. a ``master key''.    
}.

Considering first the map~\eqref{preamble_global_unitary}, 
it has been shown that the most general open dynamics is given by~\cite{kraus1971a}
\begin{equation}\label{preamble_kraus}
    \rhos(t) = \sum\limits_\alpha K_{\alpha,t} \rhos(0) K_{\alpha,t}^\dagger\,,
\end{equation}
where $K_{\alpha,t}$ are (time-dependent) operators satisfying $\sum_\alpha K_{\alpha,t}^\dagger K_{\alpha,t} = \id$.
For any choice of such operators $K_\alpha$, the map~\eqref{preamble_kraus} is said to be completely positive and trace preserving (CPTP). 
This means that it maps density matrices to density matrices
(preserves hermiticity, normalization, and positivity -- even for extensions of the map over ancillas~\cite{NielsenChuang2000})~\footnote{
The specific form of the operators $K_\alpha$ can be obtained from the global map~\eqref{preamble_global_unitary}~\cite{Stinespring1955}. 
We mention two ways to do this. 
First, if we decompose the bath's initial state as $\rhobI(0) = \sum_n q_n \ket{n}\bra{n}$, then one may verify that $K_{nm} = \sqrt{q_n} \langle m | U |n \rangle$ (which is still an operator acting on the system). 
The index $\alpha$ in $K_\alpha$ is then a collective one $\alpha = (n,m)$. 
Alternatively, one may decompose the unitary as $U = \sum_j A_j \otimes B_j$, where $A_j$ and $B_j$ are operators acting only on $S$ and $B$, respectively. 
Taking the partial trace of~\eqref{preamble_global_unitary} then leads to  $\rhos(t) = \sum_{j, k} C_{jk} \; A_j \rhos(0) A_k^\dagger$, where 
$C_{jk} = \traceB{B_k^\dagger B_j \rhobI(0)}$ is positive semidefinite.
Normalization now implies $\sum_{jk} C_{jk} A_k^\dagger A_j = \id$.
This is not yet in Kraus' form~\eqref{preamble_kraus}, but it can be made so by first diagonalizing $C_{jk}$ as $C_{jk} = \sum_{\alpha} u_{j\alpha} \xi_{\alpha} u_{k\alpha}^*$ with eigenvalues $\xi_\alpha$ and eigenvectors $u_{j\alpha}$. 
Defining new operators $K_\alpha = \sqrt{\xi_\alpha} \sum_j u_{j\alpha}^* A_j$ then yields the form~\eqref{preamble_kraus}.
}.

In practice, the Kraus form can be inconvenient, and it is preferable to write the evolution of $\rhos(t)$ as a differential equation, in general of the form~\cite{nakajima1958a,zwanzig1960a}
\begin{equation}\label{preamble_time_non_local_MEq}
    \frac{d\rhos}{dt} = \int\limits_0^t dt' \mathcal{K}_{t-t'}\Big[\rhos(t')\Big]\,,
\end{equation}
where $\mathcal{K}_{t-t'}$ is a linear superoperator, called the \ind{memory kernel}.  
This equation must be linear in $\rhos$, since~\eqref{preamble_kraus} is also linear. 
But it will in general be non-local in time, i.e. the evolution at time $t$ will depend on the history of the system, a notion of non-Markovianity.

\gtl{Eq.}~\eqref{preamble_time_non_local_MEq} is usually complicated to derive and to solve~\cite{devega2017a}.
Instead, we usually search for simpler time-local equations, of the form~\cite{weiss1993,mandel1995,Breuer2002,schlosshauer2007, RivasHuelgaBook}
\begin{equation}\label{preamble_time_local_MEq}
    \frac{d\rhos}{dt} = -\im[\Hs, \rhos]+ \Diss(\rhos)\,,
\end{equation}
which we will refer to as \ind{quantum master equations (QMEs)}. 
The last term, usually called the \ind{dissipator}, is a linear superoperator that takes into account the effects of the reservoir. 
\gsc{Additionally, one may encounter corrections to the Hamiltonian part which are discussed in Sec.~\ref{SEC:weak_coupling}.}
The precise form of $\Diss(\rhos)$ depends on the situation at hand. 
\gsc{This emphasizes the importance}
of a \ind{microscopic derivation}, where $\Diss(\rhos)$ is derived from a physical model of system-bath interactions. 
This will be the main topic of this section.

To give an example, consider a single spin 1/2 (qubit) with Hamiltonian $\Hs = -h \sz/2$ and Pauli matrix $\sz$.
A very popular QME for this system, describing the contact with a bath at inverse temperature $\beta = 1/T$ \gtl{(we also set $k_{\rm B} = 1$ throughout)}, is (see Sec.~\ref{SEC:weak_coupling})
\begin{equation}\label{preamble_qubit_example}
    \frac{d\rhos}{d t} = -\im [\Hs,\rhos]
    + \gamma (1-f) \Lind{\sm}(\rhos) + \gamma f \Lind{\splus}(\rhos)\,,
\end{equation}
where the last two terms form the dissipator $\Diss(\rhos)$. 
Here $\gamma>0$ is the coupling strength to the bath, $f = (e^{\beta h}+1)^{-1}$ is the Fermi-Dirac distribution and 
\begin{equation}\label{preamble_Lindblad_dissipator}
    \Lind{L}(\rhos) = L \rhos L^\dagger - \frac{1}{2} \{L^\dagger L, \rhos\},
\end{equation} 
is called a \ind{Lindblad dissipator} with jump operators $L$.
%
%
\gtl{The dynamics reflects the interplay between the term $\Lind{\sm}$, which annihilates an excitation with rate $\gsc{\gamma}(1-f)$, and the term $\Lind{\splus}$, which creates an excitation with rate $\gsc{\gamma}f$.}
Eq.~\eqref{preamble_qubit_example} accurately describes many setups, from quantum optics to condensed matter (often with different parametrizations for $\gamma$ and $f$).
It also has many simple and nice properties:
e.g. it relaxes with rate $\gamma$ to a unique fixed point, which is the thermal state 
\begin{equation}\label{preamble_qubit_thermal_state}
\rhos^\text{eq} = \frac{e^{-\beta \Hs}}{\tr\left\{e^{-\beta \Hs}\right\}} = f \ket{\ups}\bra{\ups} + (1-f)\ket{\dws}\bra{\dws}\,, 
\end{equation}
with average occupation $\expval{\splus \sm}_\text{th} = f$.

One of the most powerful features of the Kraus representation~\eqref{preamble_kraus} is that it establishes the basic structure that any map should satisfy in order to be CPTP. 
%
%
Similarlry one could ask, given a time-local QME of the form~\eqref{preamble_time_local_MEq}, what is the most general structure that the dissipator $\Diss(\rhos)$ may have to ensure that the dynamics is CPTP? 
For such a QME and any valid density matrix $\rhos(0)$, the evolved state $\rhos(t)$ will continue to be a valid physical state for all times $t$.

The answer was given independently by Lindblad~\cite{Lindblad1976} 
and Gorini, Kossakowski and Sudarshan~\cite{Gorini1976}. 
Namely, if the master equation has the form 
\begin{equation}\label{EQ:GKSL}
    \frac{d\rhos}{dt} = -\im [\Hs, \rhos]+\sum\limits_k D[L_k](\rhos)\,, 
\end{equation}
for \emph{any} set of operators $\{L_k\}$ and $D[L_k]$ defined in~\eqref{preamble_Lindblad_dissipator}, then it is guaranteed that the dynamics will be CPTP. 
Equations of this form -- \gtl{e.g.~Eq.~\eqref{preamble_qubit_example} --} are called Gorini-Kossakowski-Sudarshan-Lindblad (GKSL) QMEs.

The GKSL form only ensures that physical states are mapped to physical states. 
But it says nothing about which kinds of jump operators should be used to model actual thermal baths. 
To do that, one must \emph{derive} GKSL equations from models of system-bath reservoirs, for which there is no unique optimal route. 
Different methods require different approximations and are only accurate in different regimes. 
A thorough appreciation of the limitations and advantages of each method is therefore crucial in order to properly tackle boundary-driven systems.
This will be the overarching theme of this section.


\subsection{\label{sec:lme_stage}Setting the stage through the lens of local master equations}

We begin by discussing some prototypical Hamiltonians we deal with, e.g. lattices with spins or fermions or bosons. 
We will often consider 1D lattices containing $L$ sites. 
This could be a spin chain, where each site $i$ is associated with Pauli operators $\sigma^\alpha_i$,  $\alpha \in \{x,y,z,+,-\}$. 
A typical Hamiltonian is the \ind{XXZ model} in the presence of a magnetic field, described by
\begin{equation}
\label{XXZ}
    \Hs = -J \sum\limits_{i=1}^{L-1} \Big(
    \sx_i \sx_{i+1} + \sy_i \sy_{i+1} + \Delta \sz_i\sz_{i+1}
    \Big) + \sum\limits_{i=1}^L h_i \sz_i\,,
\end{equation}
where the first term describes local nearest-neighbor interactions, $h_i$ are local magnetic fields, and $\Delta$ is the anisotropy: When $\Delta = 1$  Eq.~(\ref{XXZ}) is called the \ind{Heisenberg model} and when $\Delta=0$ it is called the \ind{XX model} \begin{equation}\label{XX}
    \Hs = -J \sum\limits_{i=1}^{L-1} \left(
    \sx_i \sx_{i+1} + \sy_i \sy_{i+1} 
    \right) + \sum\limits_{i=1}^L h_i \sz_i\,.
\end{equation}
The XXZ chain is said to be an interacting model, while the XX is said to be non-interacting. 
Although in both cases 
the spins clearly interact, this terminology can be understood through the \ind{Jordan-Wigner transformation}~\cite{JordanWigner1928, Lieb1961}, which maps spin operators into a set fermionic operators $\{c_i\}$, e.g. 
via the mapping 
\begin{equation}\label{Jordan_Wigner}
    c_i = (-\sz_1) (-\sz_2)\ldots (-\sz_{i-1}) \sm_i\,.
\end{equation}
This is designed to satisfy the canonical algebra $\{c_i,c_j^\dagger\} = \delta_{ij}$, and satisfies $\splus_i \sm_i = c_i^\dagger c_i$, so that $\sz_i = 2 c_i^\dagger c_i - \id$. 
To apply this to the XX chain~\eqref{XX}, we note that $\sx_i \sx_{i+1} + \sy_i \sy_{i+1} = 2(\splus_i \sm_{i+1} + \sm_i\splus_{i+1})$.
Up to a constant, this yields the \ind{tight-binding model}~\footnote{
\gsc{Instead of fermions, we often} work with the bosonic version of the tight-binding model~\eqref{tight_binding}. In this case the Hamiltonian is identical, but the operators satisfy $[c_i,c_j] = 0$ and $[c_i,c_j^\dagger] = \delta_{ij}$.}
\begin{equation}\label{tight_binding}
    H = - 2J \sum\limits_{i = 1}^{L-1} \left(c_i^\dagger c_{i+1} + c_{i+1}^\dagger c_i \right) + \sum\limits_{i=1}^L 2h_i~c_i^\dagger c_i\,.
\end{equation}
The tight-binding model describes free fermions hopping through a lattice, which is why the XX chain is regarded as non-interacting.
Conversely, for the XXZ chain we get an additional term  $\Delta (2c_i^\dagger c_i-\id)(2c_{i+1}^\dagger c_{i+1}-\id)$, which is  quartic in the $c$'s.
%

The typical  boundary-driven scenario is to couple this 1D chain to a bath  at sites $i =1$ and $i=L$. 
The most naive way to do this is to  use dissipators like in Eq.~\eqref{preamble_qubit_example}, which leads to 
a \ind{Local Master Equation (LME)}
\begin{equation}\label{preamble_LME_basic}
    \frac{d\rhos}{d t} = -\im [\Hs,\rhos] + \Diss_1(\rhos) + \Diss_L(\rhos)\,,
\end{equation}
where 
\begin{IEEEeqnarray}{rCl}
\label{preamble_Lindblad_dissipator_sites_general}
\IEEEnosubnumber*
\IEEEyessubnumber*
\label{preamble_Lindblad_dissipator_sites}
    \Diss_i(\rhos) 
    &=& \gamma_i (1-f_i) \Lind{\sm_i}(\rhos) + \gamma_i f_i \Lind{\splus_i}(\rhos)\\[0.4cm]
    \IEEEyessubnumber*
    &=& \gamma_i \left(\frac{1-\eta_i}{2}\right) \Lind{\sm_i}(\rhos) + \gamma_i \left(\frac{1+\eta_i}{2}\right) \Lind{\splus_i}(\rhos)\,,
    \IEEEeqnarraynumspace
    \label{preamble_Lindblad_dissipator_sites2}
\end{IEEEeqnarray}
with  $\gamma_i$ describing the coupling strength to each bath $i \in\{1,L\}$. 
Here, $\eta_i$ and $f_i = (1+\eta_i)/2$ are just two equivalent parametrizations.
If $f_i = \expval{\splus_i \sm_i}_\text{th}$, then $\eta_i = \expval{\sz_i}_\text{th}$. 
For instance, $f_i=0$ ($\eta_i=-1$) corresponds to a bath that tries to impose the contact site to a spin down state, $f_i=1$ ($\eta_i=1$) imposes spin up and $f_i=1/2$ ($\eta_i=0$) imposes an infinite temperature (maximally mixed) state. 
This would only happen, however, if that site were isolated. 
Due to the \gtl{internal coupling between all sites in the chain}, there will be a competition between $f_i$ ($\eta_i$) and the Hamiltonian couplings.
\gtl{As a consequence, the steady-state will generally differ from what the baths are trying to impose.}


The interest is usually in the NESS obtained as the long-time solution of Eq.~\eqref{preamble_LME_basic}.
For different $f_1$ and $f_L$, this NESS is characterized by a magnetization current (see Sec.~\ref{sec:transport_and_currents} for a detailed discussion)
\begin{equation}
    \label{XXZ_current}
    \Jmag = -2J \expval{\sx_i \sy_{i+1} - \sy_i \sx_{i+1}}\,. 
\end{equation}
In the NESS, $\Jmag$ is actually independent of the site, since the current entering one site is the same as that leaving to the other. 
The question, therefore, is how $\Jmag$ depends on the system parameters and what kinds of transport regimes emerge. 

For instance, the XX model~\eqref{XX} with $h_i=0$ can be solved analytically~\cite{Karevski2009,AsadianBriegel2013,Znidaric2010b}, as will be reviewed in Sec.~\ref{sec:non_interacting}. 
\gtl{Assuming $\gamma_1 = \gamma_L = \gamma$,}
one finds that 
    $\Jmag = \big[16\gamma J^2/(16J^2+\gamma^2)\big] (f_1 - f_L)$.
There will thus be a flow whenever there is a bias between $f_1$ and $f_L$.
Moreover, the fact that $\Jmag$ is independent of $L$ shows that transport in the XX chain is ballistic. 
In addition, the average magnetization in each site will be 
\begin{equation}\label{preamble_XX_magnetization}
    \langle \sz_{1,L} \rangle =m^* \pm \frac{\gamma}{\gtl{16}J^2}\Jmag\,, 
\qquad 
    \langle \sz_{2,\ldots,L-1} \rangle = m^*\,, 
\end{equation}
where \gtl{$m^* = (f_1 + f_L - 1)$}.
This constant magnetization profile  also indicates that transport is ballistic.

These results illustrate well some features of LMEs. 
In a nutshell, they correspond to  
\emph{``using local dissipators in spatially extended \gsc{systems}''}
and are thus very simple to construct. 
They also have some nice properties. 
First, they are GKSL by construction, so the dynamics is guaranteed to always be physical, in the sense that it always yields valid density matrices. 
Second, the dissipators $\mathcal{D}_i$ act only locally on the first and last sites, greatly simplifying numerical and analytical calculations.
And third, they provide a straightforward recipe for building other configurations, such as different chain geometries or multiple  baths acting on multiple sites. 
For these reasons, LMEs are often used as the starting point in transport studies with quantum chains. 

Even though LMEs produce a CPTP dynamics, they need not reproduce the behavior of standard thermal baths (in the literature, they have been referred to as non-equilibrium or ``magnetization'' baths~\cite{ZnidaricGoold2017,VarmaZnidaric2019,Schuab2016a}). 
\gtl{For instance, Eq.~\eqref{preamble_LME_basic} does not properly thermalize the system when the two baths have $f_1 = f_L=f$.
Indeed, 
Eq.~\eqref{preamble_XX_magnetization} predicts $\langle \sz_i \rangle = m^* = 2f-1$, while for a thermal state $e^{-\beta H}$ we would have$\langle \sz_i\rangle = 0$, since we set $h_i = 0$.
Such a discrepancy also appears at the level of the full density matrix.
For example, Eq.~\eqref{preamble_LME_basic} with $f_1 = f_L$ predicts that all sites should be uncorrelated~\cite{AsadianBriegel2013}, which is again not the case for the thermal state $e^{-\beta H}$.}
%
These limitations motivate the search for better models for describing the system-bath dynamics, which will be the focus of Secs.~\ref{SEC:weak_coupling} and~\ref{sec:strong_coupling}. 

Notwithstanding these deficiencies, LMEs are still extremely useful in transport studies, since they provide a convenient tool for describing the injection/ extraction of excitations. 
Moreover, it is often found that the \ind{transport regime} (ballistic, diffusive etc.) is independent of the choice of the boundary driving.
This cannot be proven in general, but rigorous results for diffusive systems have been derived in~\cite{Znidaric2019}. 
The reason why this is important is because, very often, one is interested in knowing how different ingredients in a Hamiltonian affect the ensuing transport. 
A nice example are quasi-periodic and disordered systems, which will be reviewed in Sec.~\ref{ssec:dis_quasi}. 
In this case, different types of disorder lead to dramatically different transport regimes. 
And while LMEs may not faithfully describe actual heat baths, they may suffice to determine the latter.


\subsection{Weak system-bath coupling} \label{SEC:weak_coupling}


\subsubsection{\label{sssec:redfield}Redfield master equation}

The usual starting point in the derivation of a master equation for the system density matrix is the partition of the Hamiltonian $H=\Hs+\Hi+\Hb$ into a system part $\Hs$, a bath (reservoir) part $\Hb$, and an interaction part $\Hi$.
The first two act on different Hilbert spaces (and thus commute), whereas $\Hi$ can generally be decomposed as
\begin{align}\label{EQ:HI_ip}
    \Hi = \sum_\alpha A_\alpha \otimes B_\alpha,
\end{align}
with system and bath coupling operators $A_\alpha$ and $B_\alpha$, respectively.
Such a tensor product decomposition can always be obtained.
Additionally, it may prove convenient to choose the coupling operators $A_\alpha$, $B_\alpha$  individually Hermitian, but we will proceed without this assumption.
Altogether, system and reservoir constitute the full universe, which evolves unitarily according to the von-Neumann Eq.~\eqref{preamble_vonNeumann_eq}.
Since we aim at weak-coupling representations, it is useful to switch to an interaction picture with respect to $\Hs + \Hb$,
by defining $\ip{\rhosb}(t)\equiv e^{+\ii (\Hs+\Hb) t} \rhosb e^{-\ii (\Hs+\Hb)t}$ (in the following, we use boldface symbols to represent operators in the interaction picture).
The von Neumann equation is then changed to 
\begin{align}\label{EQ:vonneumann_ip}
    \dot{\ip{\rho}}_{\rm SB} = -\ii \left[\ip{\Hi}(t), \ip{\rhosb}(t)\right]\,,
\end{align}
where 
$\ip{\Hi}(t) = \sum_\alpha \ip{A_\alpha}(t) \otimes \ip{B_\alpha}(t)$ with $\ip{A_\alpha}(t) = e^{+\ii \Hs t} A_\alpha e^{-\ii \Hs t}$ and
$\ip{B_\alpha}(t) = e^{+\ii \Hb t} B_\alpha e^{-\ii \Hb t}$.
This equation is still exact but involves all the reservoir degrees of freedom. 
Integrating both sides in time and solving formally for $\ip{\rhosb}(t)$, we get
$\ip{\rhosb}(t) = \rhosb(0) -\ii \int_0^t \left[\ip{\Hi}(t'), \ip{\rhosb}(t')\right] dt'$,
which we can re-insert into Eq.~(\ref{EQ:vonneumann_ip}). 
Tracing out the reservoir degrees of freedom, we then obtain for the system density matrix in the interaction picture $\ip{\rhos} \equiv \traceB{\ip{\rhosb}}$
\begin{align}
\label{temporary1251231}
    \dot{\ip{\rho}_{\rm S}} &= -\ii~ \traceB{\left[\ip{\Hi}(t), \rhosb(0)\right]}\nn
    &\qquad- \int_0^t dt' \traceB{\left[\ip{\Hi}(t), \left[\ip{\Hi}(t'), \ip{\rhosb}(t')\right]\right]}\,.
\end{align}
This equation is still exact, but not yet closed, as the r.h.s.~still depends on the density matrix $\ip{\rhosb}(t)$ of the full universe.
At the initial time, system and reservoir are \gsc{assumed} in a product state $\rho_0 = \rhos^0 \otimes \rhob$.
Here, $\rhob$ is typically an equilibrium reservoir state (e.g. a Gibbs state), and we assume $[\Hb, \rhob]=0$.
For many reasonable interaction Hamiltonians and reservoir density matrices, we also have that $\traceB{B_\alpha\rhob}=0$, such that only the second term on the r.h.s. of~\eqref{temporary1251231} remains~\footnote{
\gsc{This condition can always be met} by the transformation $\Hs \to \Hs+\sum_\alpha g_\alpha A_\alpha$ and $B_\alpha\to B_\alpha-g_\alpha$, with suitably chosen $g_\alpha=\trace{B_\alpha \rhob}\in\mathbb{C}$, that leaves the total Hamiltonian invariant but redefines the system \gsc{and interaction Hamiltonians}.}.

To proceed, we assume the system-reservoir interaction is small, $\Hi=\ord{\lambda}$, where $\lambda$ is a dimensionless bookkeeping parameter that is later set to unity. 
We can thus close Eq.~\eqref{temporary1251231} by inserting the \ind{Born approximation} for all times,
$\ip{\rhosb}(t) = \ip{\rhos}(t) \otimes \rhob + \ord{\lambda}$.
Due to the double commutator in~\eqref{temporary1251231}, the error in doing this is of $\ord{\lambda^3}$.
%
%
The partial trace on the r.h.s. of~\eqref{temporary1251231} can be expressed in terms of the \indTwo{reservoir correlation functions}{correlation function}
\begin{align}\label{EQ:res_corr_func}
    C_{\alpha\beta}(t_1,t_2) &= \traceB{\ip{B_\alpha}(t_1) \ip{B_\beta}(t_2) \rhob}\nn
    &= \traceB{\ip{B_\alpha}(t_1-t_2) B_\beta \rhob} \equiv C_{\alpha\beta}(t_1-t_2)\,, 
\end{align}
which depend only on the time difference since $[\Hb,\rhob]=0$. 
For Gibbs states of the reservoir, at inverse temperature $\beta$, the correlation functions also obey the \ind{Kubo-Martin-Schwinger (KMS) relations}
\begin{align}\label{EQ:kms}
  C_{\alpha\bar\alpha}(\tau)=C_{\bar\alpha\alpha}(-\tau-\ii\beta)\,,
\end{align}
that eventually imprint the thermal properties of the reservoir onto the system~\cite{Kubo1957,martin1959a}.

Making the correlation functions explicit results in a \ind{non-Markovian master equation} in integro-differential form:
\begin{align}\label{EQ:nmme_ip}
    \ip{\dot{\rho}_\text{s}}&= - \sum_{\alpha\beta} \int_0^t dt' \Big\{C_{\alpha\beta}(t-t') \left[\ip{A_\alpha}(t), \ip{A_\beta}(t') \ip{\rhos}(t')\right]\nn
    &\quad\qquad\qquad + C_{\beta\alpha}(t'-t) \left[\ip{\rhos}(t') \ip{A_\beta}(t'), \ip{A_\alpha}(t)\right]\Big\}\,.
\end{align}
This equation preserves both trace and hermiticity of the system density matrix but can only be solved efficiently for sufficiently simple reservoir correlation functions (e.g. exponentially decaying ones~\cite{kleinekathoefer2004a}).
Positivity of $\rhos$ is also no longer guaranteed, except in special cases~\cite{maniscalco2007a}.

The integrand on the r.h.s.~of~\eqref{EQ:nmme_ip} is a product of (typically) rapidly decaying correlation functions with the (slowly varying) system density matrix.
This allows one to perform the Markov approximation in two steps: 
The first step renders the equation time-local by replacing $\ip{\rhos}(t')\to \ip{\rhos}(t)$, yielding the \indTwo{Redfield-I master equation}{Redfield master equation}~\cite{Redfield1965}
\begin{align}\label{EQ:redfieldI_ip}
    \ip{\dot{\rho}_\text{s}} &= - \sum_{\alpha\beta} \int_0^t dt' \Big\{C_{\alpha\beta}(t-t') \left[\ip{A_\alpha}(t), \ip{A_\beta}(t') \ip{\rhos}(t)\right]\nn
    &\quad \qquad \qquad + C_{\beta\alpha}(t'-t) \left[\ip{\rhos}(t) \ip{A_\beta}(t'), \ip{A_\alpha}(t)\right]\Big\}\,.
\end{align}
The coefficients still depend on time (even after transforming back to the Schr\"odinger picture).    
Therefore, by the same reasoning (fast decay of the correlation functions) the upper integration bounds can, after the substitution $\tau=t-t'$, be sent to infinity. 
This yields the
\indTwo{Redfield-II master equation}{Redfield master equation}
\begin{align}\label{EQ:redfieldII_ip}
    \ip{\dot{\rho}_\text{s}} &=-\sum_{\alpha\beta} \int_0^\infty d\tau~\Big\{ C_{\alpha\beta}(\tau) \left[\ip{A_\alpha}(t), \ip{A_\beta}(t-\tau) \ip{\rhos}(t)\right]\nn
    &\quad \qquad \qquad+ C_{\beta\alpha}(-\tau) \left[\ip{\rhos}(t) \ip{A_\beta}(t-\tau),  \ip{A_\alpha}(t)\right]\Big\}\,.
\end{align}
%
Back in the Schr\"odinger picture, this has the advantage that it is not only time-local, but also has constant coefficients:
\begin{align}\label{EQ:redfieldII_sp}
    \dot\rho_\text{S} &= -\ii [\Hs, \rhos(t)]\\
    &\qquad-\sum_{\alpha\beta} \int_0^\infty C_{\alpha\beta}(+\tau) \left[A_\alpha, e^{-\ii \Hs \tau} A_\beta e^{+\ii \Hs \tau} \rhos(t)\right]d\tau\nn
    &\qquad-\sum_{\alpha\beta} \int_0^\infty C_{\beta\alpha}(-\tau) \left[\rhos(t) e^{-\ii \Hs \tau} A_\beta e^{+\ii \Hs \tau},  A_\alpha\right]d\tau\,.\nonumber
\end{align}
\gsc{The remaining integrals can be explicitly computed by inserting Fourier transforms of the reservoir correlation functions, as \gtl{will be done below} in Eq.~\eqref{EQ:res_corr_func_ft}, and invoking the \ind{Sokhotski-Plemelj theorem}
\begin{align}\label{EQ:sokhotski_plemelj}
\frac{1}{2\pi} \int_0^\infty e^{+\ii\omega\tau} d\tau = \frac{1}{2} \delta(\omega) + \frac{\ii}{2\pi} {\cal P} \frac{1}{\omega}\,,
\end{align}
where ${\cal P}$ denotes the Cauchy principal value.}
It is straightforward to show that both Redfield versions I and II unconditionally preserve trace and hermiticity of the system density matrix.
But they do  not necessarily preserve positivity, which has led to extensive efforts to correct for this shortcoming~\cite{gaspard1999a,kirsanskas2018a,farina2019a,ptaszynski2019a}.
Additionally, they may not exactly thermalize the system with the reservoir temperature, but, when the perturbative assumptions employed in the derivation are valid, one may show that violations of both positivity and thermodynamic consistency, will also be small, of $\ord{\lambda^3}$~\cite{thingna2012a}.
One also finds that the \gtl{so-called} fluctuation relations~\cite{esposito2006a} are not necessarily obeyed by the Redfield master equation~\cite{hussein2014a}.

\subsubsection{\label{sssec:GKSL}Global GKSL master equation}

The Redfield-II QME is not in GKSL form [Eq.~\eqref{EQ:GKSL}] and, in general, does not relax the system to its local equilibrium state.
To arrive at a GKSL generator, an additional approximation is necessary.
We return to Eq.~(\ref{EQ:redfieldII_ip}) and make the interaction-picture time-dependence of the coupling operators explicit
\begin{align}
\label{GME_secular_part}
    \ip{A_\alpha}(t) = \sum_{ab} \bra{a} A_\alpha \ket{b} e^{+\ii (E_a-E_b) t} \ket{a}\bra{b},
\end{align}
with the \ind{system energy eigenbasis} defined by
\begin{align}\label{EQ:energy_eigenbasis}
    \Hs \ket{a} = E_a \ket{a}\,.
\end{align}
Neglecting all terms that oscillate in $t$ amounts to applying the \ind{secular approximation}, which generally yields a dissipator of GKSL form (even in presence of degeneracies).
\gsc{This approximation in general differs from the similar \ind{rotating wave approximation} that is performed on the level of the initial Hamiltonian instead~\cite{maekelae2013a}.}
After some algebra, the result is, in terms of  $L_{ab} \equiv \ket{a}\bra{b}$  
\begin{align}\label{EQ:me_bms}
    \ip{\dot{\rho}_\text{S}} &= -\ii \sum_{ab} \sigma_{ab} \left[L_{ab}, \ip{\rhos}\right]\\
    &\qquad+ \sum_{abcd} \gamma_{ab,cd} \left[L_{ab} \ip{\rhos} L_{cd}^\dagger - \frac{1}{2} \left\{L_{cd}^\dagger L_{ab}, \ip{\rhos}\right\}\right] \equiv {\cal L} (\ip{\rhos})\,,\nonumber
\end{align}
which is usually referred to as a \ind{global master equation, (GME)} \gsc{-- as it leads to dissipators that act globally on the system. An alternative term is \ind{Born-Markov-secular master equation}.}
\gsc{We have also used} the calligraphic notation for the \ind{superoperator} ${\cal L}$, and defined the coefficients~\footnote{Note that when demanding the  coupling operators $A_\alpha$ in Eq.~\eqref{EQ:HI_ip} to be individually Hermitian~\cite{Breuer2002} -- which can always be achieved --  Eq.~\eqref{EQ:me_bms} falls back to the known results.
}
\begin{IEEEeqnarray}{rCl}
\gamma_{ab,cd} &=& \delta_{E_b-E_a,E_d-E_c} \sum_{\alpha\beta} \gamma_{\alpha\beta}(E_b-E_a) \bra{a}A_\beta\ket{b} \bra{c} A_\alpha^\dagger \ket{d}^*,  \nonumber \\[-0.1cm]
\label{EQ:sigma_bms} \\[-0.1cm]
\sigma_{ab} &=& \delta_{E_a,E_b} \sum_{\alpha\beta} \sum_{c}   \frac{\sigma_{\alpha\beta}(E_a-E_c)}{2\ii} \bra{c} A_\beta\ket{b} \bra{c}A_{\alpha}^\dagger \ket{a}^* \,. \nonumber
\end{IEEEeqnarray}
Here, $\gamma_{ab,cd}$ and the \ind{Lamb-shift} Hamiltonian elements $\sigma_{ab}$ are defined in terms of the even and odd Fourier transforms of the reservoir \indTwo{correlation functions}{correlation function} in Eq.~(\ref{EQ:res_corr_func})
\begin{IEEEeqnarray}{rCl}
\gamma_{\alpha\beta}(\omega) &=& \int C_{\alpha\beta}(\tau) e^{+\ii\omega\tau}d\tau\,,\nonumber\\[-0.1cm]
    \label{EQ:res_corr_func_ft}\\[-0.1cm]
\sigma_{\alpha\beta}(\omega) &=& \int C_{\alpha\beta}(\tau)\text{sgn}(\tau) e^{+\ii\omega\tau}d\tau = \frac{\ii}{\pi} {\cal P} \int \frac{\gamma_{\alpha\beta}(\omega')}{\omega-\omega'} d\omega'\,,\nonumber    
\end{IEEEeqnarray}
\gsc{where the relation between the two follows from Eq.~\eqref{EQ:sokhotski_plemelj}.} 
The secular approximation leads to the Kronecker-$\delta$ functions in Eqs.~\eqref{EQ:sigma_bms}.
Transforming back to the Schr\"odinger picture simply amounts to the system Hamiltonian $\Hs$ re-appearing in the commutator term.
The proof that Eq.~\eqref{EQ:me_bms} is in GKSL form relies only on proving positive definiteness of the dampening matrix $\gamma_{ab,cd}$ and Hermiticity of the Lamb-shift Hamiltonian.

Since $\sigma_{ab}\propto \delta_{E_a,E_b}$, the Lamb-shift in Eq.~\eqref{EQ:me_bms} commutes with the system Hamiltonian $[\Hs, \sum_{ab} \sigma_{ab} L_{ab}]=0$.
For  non-degenerate $\Hs$, this term will  merely affect the dynamics of the coherences in the energy eigenbasis.
Beyond the secular approximation, however, it may become important for near or exact degeneracies of $\Hs$, where it can select a preferred pointer basis~\cite{schultz2009a,trushechkin2021a}.
For thermal reservoirs, the KMS relations ~\eqref{EQ:kms} can be used to show that the the thermal state
$\rhoss\propto e^{-\beta\Hs}$ is a stationary state of the GME Eq.~\eqref{EQ:me_bms}.

This becomes even more apparent in the case when not only the spectrum of $\Hs$ is non-degenerate, $\delta_{E_a,E_b}=\delta_{ab}$, but also the set of Bohr (transition) frequencies $\{\omega_{ab} = E_a - E_b\}$.
Then, Eq.~\eqref{EQ:me_bms} completely decouples populations and coherences in the system energy eigenbasis.
The coherences $\rho_{ij}\equiv\bra{i}\rhos\ket{j}$ with $i\neq j$ simply  decay ($\gamma_{ai,ai}\ge 0$) according to 
\begin{align}
    \dot{\rho}_{ij} &= \left[
    - \frac{1}{2}\left(\sum_a \gamma_{ai,ai}+\sum_a \gamma_{aj,aj}\right) - 
    \ii (\sigma_{ii}-\sigma_{jj})\right] \rho_{ij}\,,
\end{align}
while the populations obey a classical \ind{Pauli master equation}
\begin{align}\label{EQ:GLOBAL_PAULI}
    \dot\rho_{aa} = \sum_b \bigg\{W_{ab} \rho_{bb} - W_{ba} \rho_{aa}\bigg\}\,,
\end{align}
with transition rates from eigenstate $\ket{b}$ to eigenstate $\ket{a}$ given by $W_{ab} = \gamma_{ab,ab}\ge 0$. 
Thermalization then follows from the fact that the transition rates obey \ind{detailed balance}, 
$W_{ab}/W_{ba} = e^{-\beta (E_a - E_b)}$,
which is a consequence of the KMS condition~\eqref{EQ:kms}.
%
%
Thermalization is also found for grand-canonical bath states~\cite{bulnes_cuetara2016a}.

Despite this, we stress that a GKSL form only ensures positivity, not that the dynamics is in fact physically accurate~\cite{strasberg2022}.
Moreover, having the thermal state as a fixed point only grants compliance 
with thermodynamics in the parameter regime that allows to perform the necessary approximations~\cite{spohn1978b,duemcke1979a}.
It may fail e.g. beyond weak-coupling or when the system energy splittings are small in comparison to the system-reservoir coupling, which can happen when the system consists of many components (see the discussion around Eq.~\eqref{EQ:lme_weak_int_coup}).
Beyond GKSL generators, exact thermalization is a feature expected only for vanishing coupling strength~\cite{mori2008a,fleming2011a, XuWang2017, thingna2012a}.
Higher-order perturbative expansions~\cite{laird1991a,jang2002a,schroeder2007a, ThingnaWang2014} are required to see the effects of the system-reservoir coupling strength in the steady-state solution.

\subsubsection{Additivity for multiple reservoirs}\label{SEC:additivity}

The previous derivation considered only one reservoir at equilibrium. 
When multiple reservoirs \gsc{-- labeled by the index $\nu$ --} are present, the  bath Hamiltonian  may be decomposed as
$\Hb = \sum_\nu \Hb^{(\nu)}$
with the individual contributions acting on different Hilbert spaces.
\gsc{Usually, it is assumed that} each bath is held in a local equilibrium state
    $\rhob = \underset{\nu}{\otimes} \rhob^{(\nu)}$ with
    $\rhob^{(\nu)} =  e^{-\beta_\nu (\Hb^{(\nu)}-\mu_\nu \Nb^{(\nu)})}/Z_\text{B}^{(\nu)}$
characterized by local inverse temperatures $\beta_\nu$ and chemical potentials $\mu_\nu$.
The system bath-coupling in Eq.~\eqref{EQ:HI_ip} \gsc{can now be} written as 
    $\Hi = \sum_\alpha A_\alpha \otimes \sum_\nu B_\alpha^{(\nu)}$,
\gsc{where the $A_\alpha$ may be a complete set of operators for the system and the possibility $B_\alpha^{(\nu)}\to 0$ allows to consider reservoir-specific couplings. 
With this, one can follow the very same steps in the derivation of a master equation for weak couplings.}

Since we are working in a frame where the first order correlation function vanishes,
$\trace{B_\alpha^{(\nu)} \rhob^{(\nu)}}=0$, this also applies to product terms between different reservoirs, such that 
$\traceB{\ip{B_\alpha^{(\nu)}}(t_1) \ip{B_\beta^{(\nu')}}(t_2) \rhob} = \delta_{\nu\nu'} C_{\alpha\beta}^{(\nu)}(t_1,t_2)$.
Consequently, a weak-coupling treatment yields an additive decomposition of the dissipators
$\dot{\rhos} = -\ii [\Hs, \rhos] + \sum_\nu \mathcal{D}_\nu \rhos$,
with one dissipator for each reservoir $\nu$.  
Notice that to define \ind{additivity} strictly, each dissipator $\mathcal{D}_\nu$ should also only depend on the parameters of the reservoir $\nu$, as is the case here.
For instance, in the case of  thermal reservoirs, each GME dissipator will  thermalize the system with its associated reservoir $\mathcal{D}_\nu e^{-\beta_\nu (\Hs-\mu_\nu \Ns)}=0$.

\gsc{The additivity property has important consequences:
First, from a practical point of view, it enables one to split the problem into smaller pieces, i.e., to derive the dissipator by treating the system as if it was coupled only to one reservoir.
Second, it allows one to properly define the currents entering or leaving the system (Sec.~\ref{sssec:globalcontinuityequation}).
Unfortunately, additivity is not preserved for stronger system-reservoir coupling, see Sec.~\ref{sec:strong_coupling}, and the definition of currents may require microscopic approaches, as exposed in Sec.~\ref{SEC:fcs_mic}.
}

\subsubsection{Local GKSL master equation}\label{SEC:local_gksl_master_equation}

For multiple reservoirs,  
the interaction Hamiltonian, typically, only couples the reservoirs to finite portions of the system.
In regimes where the system-internal coupling strengths are smaller than 
the system-reservoir coupling strengths, the LME approach is often more applicable \gsc{than the GME, see also Sec.~\ref{ssec:connection_exact}.}
We therefore now turn to a more detailed look on how to derive LMEs, introduced in Sec.~\ref{sec:lme_stage}.
We first show how they can be derived from \ind{collisional models}~\cite{Rau1963,Englert2002,Scarani2002} (also called \ind{repeated interactions}), where the open dynamics is replaced by a series of sequential collisions with small bath units. 
\gtl{Then we proceed to show how the methods put forth in Sec.~\ref{sssec:redfield} can be adapted to yield LMEs instead.}

\paragraph*{Derivation based on collisional models}

The basic idea of a collisional model is to describe the open dynamics as a series of (unitary) collisions, which involve only the system and a small piece of the bath (often called \indTwo{ancillas}{ancilla}). 
The ancillas are independent and identically prepared, usually in thermal states.
Each interaction lasts for a short period of time, after which the ancilla is discarded and never participates again in the dynamics. 
In the next step a fresh new ancilla is introduced and the process restarts anew. 
This therefore generates a stroboscopic dynamics, whose continuous time limit can be shown to be a LME.
The refreshing of the ancillas is consistent with the idea of a Markovian bath. 
And the sequential nature of the collisions fit with Boltzmann’s molecular chaos hypothesis (\emph{Stosszahlansatz}), where at any given time the system would only be interacting with a fraction of the environment (which is subsequently reset due to the bath's inherent complexity).

This connection, within the context of boundary-driven systems, was first established in~\cite{Karevski2009}.
Since then, collisional models have experienced a revival of interest. 
This is partially because they allow to construct a thermodynamically consistent models~\cite{Barra2015,Pereira2018,DeChiara2018,strasberg2017a,strasberg2019b}.

For the purpose of illustration, following~\cite{LandiKarevski2014}, we consider the derivation of Eq.~\eqref{preamble_LME_basic} with a system spin Hamiltonian $\Hs$ containing $L$ sites, as in Eq.~\eqref{XXZ}.  
We start by focusing on a single collision. 
In addition to the $L$ sites of the chain, we use two extra spins as ancillas, labeled $0$ and $L+1$, such that the total Hamiltonian becomes
\begin{equation}\label{LME_collisional_HT}
H_T = \Hs + h_0 \sz_0 + h_{L+1} \sz_{L+1} + \sqrt{\gamma}(V_{0,1} + V_{L,L+1})\,,
\end{equation}
where $V_{i,j} = \sm_i \splus_j + \splus_i \sm_j$.
The system starts in an arbitrary state $\rhos$, whereas the two new spins are prepared in thermal states $\rho_0$ and $\rho_{L+1}$ like Eq.~\eqref{preamble_qubit_thermal_state}, but with Fermi-Dirac occupations $f_0$ and $f_{L+1}$ given by $f_i = (e^{\beta_i h_i}+1)^{-1}$, $i = 0,L+1$.
The joint system then evolves for a time $\tau$, resulting in $e^{-\im H_T \tau} (\rho_0 \otimes \rhos \otimes \rho_{L+1}) e^{\im H_T \tau}$.
We are only interested in the reduced state of the system, which is found by tracing out $0$ and $L+1$; that is, 
\begin{equation}\label{LME_collisional_map}
    \rhos' = \tr_{0,L+1}\left\{ e^{-\im H_T \tau} \big(\rho_0 \otimes \rhos \otimes \rho_{L+1}\big) e^{\im H_T \tau} \right\}\,.
\end{equation}
This can be viewed as the basic building block of a collisional model, since 
after this interaction, the ancillas are discarded and fresh new ones are introduced.
One then simply reapplies the same map of Eq.~\eqref{LME_collisional_map}, but with $\rhos'$ as input. 
Repeating the procedure leads to a set of states $\rhos^0$, $\rhos^1$, $\rhos^2$, $\ldots$ that describe the stroboscopic open evolution of the system, in steps of $\tau$. 

The above construction applies for collisions with arbitrary duration. 
To obtain the LME, we now consider the continuous-time limit $\tau\to 0$, so that the unitaries $e^{\gsc{\pm}\im H_T \tau}$ can be expanded in a Taylor series.
Due to the partial trace in Eq.~\eqref{LME_collisional_map}, many terms vanish and, to leading order in $\tau$, we are left with
\begin{equation}\label{LME_collisional_expansion}
\rhos' = \rhos - \im \tau [\Hs, \rhos] +  \tau^2 \left( \mathcal{D}_1(\rhos) + \mathcal{D}_L(\rhos)\right)\,, 
\end{equation}
where $\mathcal{D}_1(\rhos) = -\frac{\gamma}{2}\ptrace{0,L+1}{[V_{01},[V_{01}, \rho_0 \otimes \rhos \otimes \rho_{L+1}]]}$ and similarly for $\mathcal{D}_L$. 
Using the specific form of $V_{i,j}$, together with the fact that the initial states of sites $0$ and $L+1$ are thermal, one finds \gsc{for $i\in\{1,L\}$ the dissipators}
\begin{equation}\label{eq:local_diss}
    \mathcal{D}_i = \gamma(1-f_i) D[\sigma^-_i] + \gamma f_i D[\sigma^+_i]\,,
\end{equation}
where $f_1 := f_0$ and $f_L := f_{L+1}$. 
These are exactly the GKSL dissipators appearing in the LME~\eqref{preamble_Lindblad_dissipator_sites}. 

As the last step, we now \gsc{rewrite Eq.~\eqref{LME_collisional_expansion} to form the time-derivative} $d\rhos/dt \simeq (\rhos' - \rhos)/\tau$. 
%
\gsc{On taking the limit $\tau\to 0$, }
however, we encounter a problem. 
The dissipative term is of the order $\tau^2$ and would thus vanish if this limit was taken naively. 
The continuous-time limit corresponds to small $\tau$, but the limit $\tau \equiv 0$ would mean no interaction at all. 
One should therefore interpret the continuous-time limit as a physical limit, where $(\rhos'- \rhos)/\tau$ is sufficiently smooth to be interpreted as a derivative, but the last term in Eq.~\eqref{LME_collisional_expansion} is nonetheless finite. 
One way to implement this mathematically is to rescale the interaction strengths $\sqrt{\gamma}$ in Eq.~\eqref{LME_collisional_HT} to  $\sqrt{\gamma/\tau}$. 
That is, the interaction is taken to be very short, but also very strong.
This idea is also used in classical Brownian motion, to introduce the  $\delta$-correlated noise in the Langevin equation~\cite{Coffey2004}.
With this proviso, we \gsc{precisely obtain} 
Eq.~\eqref{preamble_LME_basic}. 
Thus, LMEs indeed can be viewed as the continuous-time limit of collisional models. 

In the literature, LMEs of the form Eq.~\eqref{preamble_LME_basic} are often said to represent ``\ind{magnetization baths}''~\cite{ZnidaricGoold2017,VarmaZnidaric2019,Schuab2016a}. 
This is used  to emphasize that they are not describing actual thermal baths, but instead act so as to force the magnetization of the boundary sites to point in certain directions.
The collisional model derivation above provides a clear interpretation of this idea.
At each collision, the spins $0$ and $L+1$ tend to polarize the magnetization of the boundary sites ($1$ and $L$).
But since the interaction time $\tau$ is very short, 
this effect cannot propagate deeply within the chain \gsc{after just one collision}, and thus remains confined to the first and last sites.
This is ultimately the reason why collisional models with local ancillas yield local jump operators. 
In principle, the same idea can be used to derive GMEs, but then the ancillas have to interact with multiple parts of the chain.
Recently, it was shown~\cite{Cattaneo2021} that one may algorithmically reproduce any Markovian evolution through a suitably chosen collisional model.

\paragraph*{Derivation based on weak internal couplings}

It is tempting  to ask whether LMEs can somehow be viewed as a special limit of GMEs.
Indeed, Eq.~(\ref{EQ:me_bms}) is formulated in terms of the global system energy eigenbasis Eq.~(\ref{EQ:energy_eigenbasis}).
Let us consider what happens if the system is composed  of different subsystems (for now we consider just two but this will be straightforwardly generalized)
\begin{align}\label{EQ:lme_weak_int_coup}
    \Hs = H_\text{S}^1 + H_\text{S}^2 + \xi H_\text{S}^{12},
\end{align}
with local Hamiltonians $H_\text{S}^\mu$ and  intra-system interaction $H_\text{S}^{\mu\nu}$,  quantified by the dimensionless bookkeeping parameter $\xi$.
In the limit $\xi\to 0$, the system energy eigenbasis Eq.~(\ref{EQ:energy_eigenbasis}) factorizes as
$\ket{a} \to \ket{a_1} \otimes \ket{a_2}$ while 
$E_a \to E_{a_1} + E_{a_2}$,
where $H_\text{S}^\mu \ket{a_\mu} = E_{a_\mu} \ket{a_\mu}$.
With this limit, one may often retrieve an LME from the GME (provided the coupling operators $A_\alpha$ in the Hamiltonian are local).

However, this does not always work. 
In particular when the subsystems are identical, $\Hs$  becomes degenerate when $\xi\to 0$.
%
%
Whereas superpositions of global eigenstates with the same eigenvalue are still global eigenstates, they need not be given by tensor products of local eigenstates, if global symmetries are conserved.
An example for this would be the master equation for superradiant decay~\cite{gross1982a}: Although here one has a cloud of non-interacting atoms (i.e., $\xi=0$ from the beginning), the jump operators remain global due to a globally conserved symmetry (the collective coupling induces angular momentum conservation), and hence do not reduce to local jump operators~\footnote{
For this model, one has for $N$ two-level atoms in total $2^N$ energy eigenstates, but due to the globally conserved angular momentum only at most $N+1$ of them couple to each other. 
If one performs the secular approximation in the local energy eigenbasis, couplings to coherences between degenerate energy eigenstates must be kept in the global master equation.
By contrast, choosing the angular momentum eigenstates as eigenbasis (which also diagonalizes the Hamiltonian), one obtains a simple rate equation in each angular momentum subspace.
}.
It is therefore advisable  to retrieve the LME from the Redfield master equation~(\ref{EQ:redfieldII_ip}) by performing the limit $\xi\to 0$ on it. 

\begin{figure}
    \centering
    \includegraphics[width=\columnwidth]{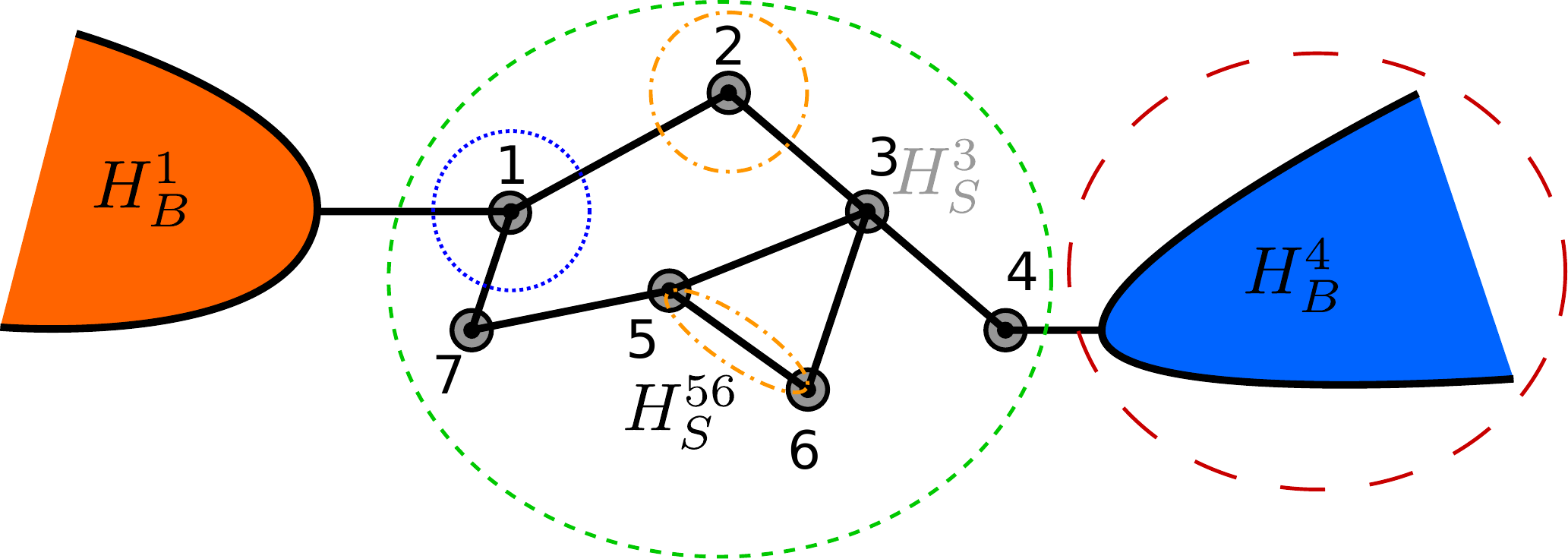}
    \caption{
Sketch of the considered setup. Small grey spheres correspond to sites with Hamiltonians $\Hs^\nu$, and connecting black lines correspond to couplings (either bonds $\Hs^{\mu\nu}$ or system-reservoir couplings $\Hint^\nu$). 
GMEs are obtained by treating only system-reservoir couplings (traversing the green short-dashed sphere) perturbatively. 
Conversely, LMEs result from treating both system-reservoir couplings and system-internal couplings (e.g.~traversing the blue-dotted sphere) perturbatively.
Currents through various interfaces (dashed and/or dotted curves) can be meaningfully defined by balance equations of locally conserved observables (Sec.~\ref{sec:transport_and_currents}). 
For instance, we discuss the global energy balance of the system (green short-dashed) in Eq.~\eqref{Currents_H_evo} and the local energy current entering a boundary site (blue dotted) from the reservoir in Eq.~\eqref{EQ:Ecurrent_local}. 
Local  balances of internal sites, as well as internal bonds (orange dash-dotted), are discussed in Eqs.~\eqref{Currents_site_resolved_Ok} and~\eqref{Currents_unitary_bond_continuity}, respectively. 
The energy balances of the reservoir (red long-dashed) are treated using Full Counting Statistics in Sec.~\ref{sec:full_counting}. 
}
\label{fig:f_sketch_localsetup}
\end{figure}

Alternatively, one may also derive the LME by treating the system-bath interactions $\Hint$, and the internal system interactions $\Hs^{\mu \nu}$, on equal footing, as we now outline. 
We start from the usual decomposition of the universe Hamiltonian 
$\Htot=\Hs+\Hb+\Hint$. 
The system is assumed to have an internal structure, such that $\Hs$ can be written as 
\begin{align}
    \Hs = \Hs^0 + \Hs^1 = \sum_\nu \Hs^\nu + \sum_{\mu<\nu} \Hs^{\mu\nu}\,,
\end{align}
where the free parts of the system $\Hs^\nu$ act on different Hilbert spaces, such that $\left[\Hs^\nu,\Hs^\mu\right]=0$, and $\Hs^{\mu \nu}$ represents pair interactions between these parts. 
In addition, we assume \gsc{multiple reservoirs} $\Hb = \sum_\nu \Hb^{\nu}$, with each \gsc{bath} $\Hb^\nu$ coupled only \emph{locally} to subsystem $\nu$ according to an interaction (c.f. Eq.~\eqref{EQ:HI_ip})
\begin{align}
    \Hint = \sum_\nu \Hint^\nu 
    = \sum_\nu \sum_\alpha A_\alpha^\nu \otimes B_\alpha^\nu\,.
\end{align}
We can naturally choose some of the $\Hint^\nu = 0$, allowing for  boundary-driven setups with internal parts not directly coupled to any reservoir as in Fig.~\ref{fig:f_sketch_localsetup}.
By construction, the system coupling operators $A_\alpha^\nu$ only act on their respective sites, such that $\left[\Hs^\nu, A_\alpha^{\mu\neq \nu}\right]=0$, and similarly for the reservoir couplings.

The problem now involves the interplay between two small couplings,  $\Hint=\ord{\lambda}$ and $\Hs^{\mu\nu}=\ord{\xi}$, with small dimensionless bookkeeping parameters $\lambda$ and $\xi$.
These terms may be of similar strength, but they are both small in comparison to the other parts of the Hamiltonian.
We therefore use an interaction picture with respect to the free part of the
system Hamiltonian and the reservoirs
\begin{align}
    H_0 = \sum_\nu \Hs^\nu + \sum_\nu \Hb^\nu\,.
\end{align}
In this picture (marked by bold symbols below), the exact von-Neumann equation for the full universe reads
\begin{align}
    \ip{\dot\rho} = -\ii \left[\ip{\Hs^1}(t),\ip{\rho}(t)\right] 
    -\ii \left[\ip{\Hint}(t),\ip{\rho}(t)\right]\,,
\end{align}
analogous to Eq.~\eqref{temporary1251231}.
The r.h.s.~is small ($\ord{\xi}+\ord{\lambda}$) and we can use an equivalent perturbative treatment as for the derivation of the global master equation in Secs.~\ref{sssec:redfield} and~\ref{sssec:GKSL}, with only small modifications.
We formally integrate the above equation, but in contrast to the treatment there, we insert the solution only in the second commutator term.
Then performing the partial trace over the reservoir degrees of freedom yields
the exact equation
\begin{align}\label{local_deriv_step2}
    \ip{\dot\rho_\text{S}} &= -\ii \left[\ip{\Hs^1}(t), \traceB{\ip{\rho}(t)}\right]
    -\ii~\traceB{\left[\ip{\Hint}(t'), \rho_0\right]}\nn
    &\qquad-\int_0^t \traceB{\left[\ip{\Hint}(t), \left[\ip{\Hs^1}(t') + \f{\Hint}(t'), \ip{\rho}(t')\right]\right]} dt'\,.
\end{align}
First, with using a factorizing initial condition 
$\rho(0) = \rhos^0 \underset{\nu}{\otimes} \rhob^\nu$ and the assumption that
$\traceB{\ip{\Hint}(t) \rhob}=0$, the second commutator term on the first line above vanishes identically.
Second, we extend this assumption to all times with the \ind{Born approximation} $\ip{\rho}(t) = \ip{\rhos}(t) \underset{\nu}{\otimes} \rhob^\nu + \ord{\lambda}+\ord{\xi}$.
Considering that $\ip{\rhos}(t)$ is only an approximation to the exact reduced density matrix $\traceB{\ip{\rho}(t)}$, neglecting the remainder terms in the first commutator and in the nested double commutator leads to 
\begin{align}\label{local_deriv_step3}
    \ip{\dot\rho_\text{S}} &= -\ii \left[\ip{\Hs^1}(t), \ip{\rhos}(t)\right]\nn
    &\qquad-\sum_\nu \int_0^t \ptrace{\nu}{\left[\ip{\Hint^\nu}(t), \left[\ip{\Hint^\nu}(t'), \ip{\rhos}(t') \otimes \rhob^\nu\right]\right]} dt'\nn
    &\qquad+\ord{\lambda\xi}+\ord{\xi^2}
    + \ord{\lambda^2\xi} + \ord{\lambda \xi^2} + \ord{\lambda^3}\,,
\end{align}
where we have neglected the mixed terms between system and reservoir and also between different reservoirs.
Furthermore, we have already performed all the trivial partial traces.
The step in going from Eq.~\eqref{local_deriv_step2} to Eq.~\eqref{local_deriv_step3} highlights the terms being discarded in a local approach. 
The first line contains the intra-system interactions: Upon transforming back to the Schr\"odinger picture, it restores together with $\Hs^0$ the full system Hamiltonian.
More importantly, the double commutator term involves only system operators acting on the specific subsystem $\nu$ since, due to the different interaction picture, we have
$\ip{\Hint^\nu}(t) = \sum_\alpha e^{+\ii \Hs^\nu t} A_\alpha^\nu e^{-\ii \Hs^\nu t} \otimes e^{+\ii \Hb^\nu t} B_\alpha^\nu e^{-\ii \Hb^\nu t}$.    
It yields in fact the same \gsc{(LME)} dissipator that one would obtain if subsystem $\nu$ was exclusively coupled to its adjacent reservoir (see e.g.~\cite{schaller2022a} for an application in the Fermi-Hubbard model). 
Hence, the above equation generates a local non-Markovian master equation, and following the standard Markov approximations discussed in Sec.~\ref{sssec:redfield} one can also derive local versions of Redfield I/II equations~\eqref{EQ:redfieldI_ip} and~\eqref{EQ:redfieldII_ip}.
Accordingly, the final secular approximation only requires the splittings of the local system parts $\Hs^\nu$ to be large, 
and it is not invalidated by possibly many degeneracies of $\sum_\nu \Hs^\nu$.
Eventually, one obtains the sum of local dissipators, but the \gsc{commutator from the first line of Eq.~\eqref{local_deriv_step3}}
will mediate an interaction between distant system sites (and reservoirs).
Upon transforming back to the Schr\"odinger picture, the full system Hamiltonian (\gsc{anything inside the green short-dashed circle} in Fig.~\ref{fig:f_sketch_localsetup}) is restored, but dissipative contributions (also Lamb shift terms) are the same as if one would derive the \gsc{master equation from Sec.~\ref{sssec:GKSL}} for the coupled parts only.
It hence always generates LMEs, see e.g.~\cite{Wichterich2007} for specific applications.
Here, the term ``local'' refers to the chosen partition of the system, for which a suitable choice is dictated by the relation between coupling strengths.
Often $\Hs^\nu$ are thought of as individual bosonic or fermionic modes, but they can also be complicated interacting systems themselves and in this sense, there may be varying degrees of locality depending on the context.
\gtl{We also point out approaches that effectively provide interpolations between LMEs and GMEs, such as~\cite{Seah2018} and~\cite{lidar2001a}.}

\paragraph*{Derivation for negligible system Hamiltonian}

Yet another derivation -- the \ind{singular coupling limit} -- that leads to LMEs, arises in the limit where the system Hamiltonian is small in comparison to the system-reservoir coupling.
The Born- and Markov approximations are applied as before, the only difference is that due to the small system Hamiltonian one can neglect the time-dependence of the system coupling operators in the  interaction picture -- rendering a secular approximation unnecessary. 
The simple outcome of this procedure is an equation of GKSL form that can technically also be obtained from Eq.~\eqref{EQ:me_bms} by setting all system energies zero~\cite{palmer1977a}.
In the Schr\"odinger picture it reads (assuming hermitian coupling operators $A_\alpha=A_\alpha^\dagger$)
\begin{align}
    \dot\rhos &= -\ii \left[\Hs, \rhos\right]-\ii \sum_{\alpha\beta} \frac{\sigma_{\alpha\beta}(0)}{2\ii}\left[A_\alpha A_\beta, \rhos\right]\nn
    &\qquad
    + \sum_{\alpha\beta} \gamma_{\alpha\beta}(0)
    \left[A_\beta \rhos A_\alpha - \frac{1}{2} \left\{A_\alpha A_\beta, \rhos\right\}\right]\,,
\end{align}
where the Fourier transforms of the reservoir correlation functions of Eq.~\eqref{EQ:res_corr_func_ft} are evaluated at vanishing arguments. 
Since it follows from the KMS relation Eq.~\eqref{EQ:kms} that $\gamma_{\alpha\bar{\alpha}}(0)=\gamma_{\bar{\alpha}\alpha}(0)$, one finds that the completely mixed \gsc{(infinite temperature) state $\rhos\propto\id$ is one (not necessarily unique)} stationary solution of the above master equation.
\gsc{
As the Lindblad operators coincide with those in the coupling Hamiltonian Eq.~\eqref{EQ:HI_ip}, phenomenologic approaches to boundary-driven systems like Eq.~\eqref{preamble_Lindblad_dissipator_sites} may actually be microscopically motivated in the singular coupling limit. 
The interaction with further (non-singular) reservoirs may nevertheless introduce interesting stationary non-equilibrium properties as may also be obtained by combining the singular coupling limit with conventional GME treatments~\cite{schultz2009a}.} 

\subsection{\label{sec:transport_and_currents}Transport properties and currents}

The usual boundary-driven scenario consists of a system coupled to multiple baths, which is left to evolve until it reaches a steady-state. 
The latter, being out-of-equilibrium, is characterized by a stationary  currents from one bath to another~\cite{DharHanggi2012}. 
It is therefore essential to correctly determine these currents for each model.
This includes not only the currents of energy but also other observables like  particle number or magnetization/spin.

The crucial aspect, when dealing with transport, is to correctly account for all possible sources and sinks of the quantity in question. 
Take energy, for instance. 
It only makes sense to talk about energy \emph{transport}, if the amount of energy leaving one region is the same as the amount entering the other. 
If, in between, energy is spontaneously created (e.g. by an external work agent), or if some energy is trapped in the interaction potential, this must be properly taken into account. 
This introduces the idea of \ind{continuity equations}. 
That is, equations which identify how the changes in a given quantity are linked to fluxes of that quantity to other regions of space. 
There are two types of continuity equations one may look at: \indTwo{system-resolved}{system-resolved continuity equations}, which look only at  net currents from the bath to the system, and \indTwo{site-resolved}{site-resolved continuity equation}, which analyzes the flows \emph{within} the system.

\subsubsection{System-resolved continuity equations}\label{sssec:globalcontinuityequation}

Consider a master equation of the form
\begin{equation}\label{Currents_M}
    \frac{d\rhos}{dt} = -\im[\Hs, \rhos] + \sum\limits_\nu \Diss_\nu(\rhos)\,, 
\end{equation}
where $\Hs$ describes all internal energies of the system and  $\Diss_\nu$ summarizes the net effect of bath $\nu$ (which may also contain Lamb-shift contributions).
In the case of Redfield or GMEs, $\Diss_\nu$ has support over the entire system, while for LMEs it acts only on specific sites.  

A system-resolved continuity equation for $\expval{\Hs}$ is readily found using Eq.~\eqref{Currents_M}:
\begin{equation}\label{Currents_H_evo}
    \frac{d\langle \Hs \rangle}{d t} = \sum\limits_\nu \tr\Big\{ \Hs \Diss_\nu(\rhos)\Big\} := \sum\limits_\nu \cur{E}^\nu.
\end{equation}
Each term in the r.h.s.~can be identified with the flow of energy entering the system (\gsc{inside the short-dashed green circle in Fig.~\ref{fig:f_sketch_localsetup}}) from reservoir $\nu$.
Changes in energy of the system are thus entirely due to the fluxes arriving from each bath.
At steady-state, $d\langle \Hs\rangle/dt = 0$ and all fluxes balance out: $\sum_\nu \cur{E,ss}^\nu = 0$.
For just two baths, this becomes $\cur{E,ss}^1 = - \cur{E,ss}^2$, meaning that all energy entering from bath 1 leaves towards bath 2.

Next we consider some other operator $\mathcal{O}$ of the system. 
From Eq.~\eqref{Currents_M}, its rate of change will be given by
\begin{equation}\label{Currents_O_evo_gen}
    \frac{d\langle \mathcal{O} \rangle}{d t} = \im \langle [\Hs, \mathcal{O}]\rangle + \sum\limits_\nu \trace{\mathcal{O} \Diss_\nu(\rhos)}. 
\end{equation}
The last term is naturally associated with the ``flow of $\mathcal{O}$'' to each bath, but now we also have the term $[\Hs,\mathcal{O}]$. 
When this is non-zero, $\mathcal{O}$ can be spontaneously created or destroyed, even if the system is isolated.
We therefore typically speak of transport of $\mathcal{O}$ only when $[\Hs,\mathcal{O}] = 0$, in which case we get
\begin{equation}\label{Currents_O_evo_conserved}
    \frac{d\langle \mathcal{O} \rangle}{d t} =  \sum\limits_\nu \tr\Big\{ \mathcal{O} \Diss_\nu(\rhos)\Big\} = \sum\limits_\nu \cur{\mathcal{O}}^\nu\,.
\end{equation}
For example, if the Hamiltonian $\Hs$ conserves the total particle number, \gtl{$[\Hs,\Ns] = 0$, }
we can naturally define the particle current  via
\begin{align}\label{Currents_N_evo}
    \frac{d\langle \Ns \rangle}{d t} =  \sum\limits_\nu \tr\Big\{ \Ns \Diss_\nu(\rho)\Big\} = \sum\limits_\nu \cur{N}^\nu\,.
\end{align}
Similarly, we can consider transport of magnetization, $\mathcal{M} = \sum_i \sigma_z^i$, in spin chains.
However, some Hamiltonians may not conserve $\mathcal{M}$. One such example is the $XYZ$ chain:
\begin{equation}
    \label{Currents_XYZ}
    \Hs = -\sum\limits_{i=1}^{L-1} \Big( 
    J_x \sx_i \sx_{i+1} + 
    J_y \sy_i \sy_{i+1} +
    J_z \sz_i \sz_{i+1} \Big) + \sum\limits_{i=1}^L h_i \sz_i\,,
\end{equation}
One may verify that $\mathcal{M}$ is only conserved in the XXZ limit [$J_x = J_y$; Eq.~\eqref{XXZ}].
Thus, one may study magnetization transport in XXZ or XX chains, but not in the full XYZ model.

\subsubsection{\label{sssec:unitary_continuity_equations}Site-resolved continuity equations: unitary components}
 
Eq.~\eqref{Currents_O_evo_gen} can also be employed in the case where $\mathcal{O}$ is a local operator, associated either to a site or a bond. 
This will give rise to \indTwo{site-resolved continuity equations}{site-resolved continuity equation}.
In this case, the unitary term $\im [\Hs,\mathcal{O}]$ is crucial.
And since it is independent of the type of QME being used, we treat it separately in this section.
For concreteness we assume that the system has a 1D nearest-neighbor Hamiltonian 
\begin{equation}\label{Currents_H_nn}
    \Hs = \sum\limits_{k=1}^L \Hs^k + \sum\limits_{k=1}^{L-1} \Hs^{k,k+1}\,. 
\end{equation}
The indices in each term clarify in which Hilbert spaces it has support. 
For instance, $\Hs^{2,3}$ acts only on sites 2 and 3, and thus commutes with any operator which does not pertain to these two sites.
Let $\mathcal{O}^k$ denote a generic local operator with support only over site $k$. 
Then
$[\Hs,\mathcal{O}^k] = [\Hs^k, \mathcal{O}^k] + [\Hs^{k-1,k}, \mathcal{O}^k] + [\Hs^{k,k+1}, \mathcal{O}^k]$. 
This helps identify what is required for a proper site-resolved continuity equation for $\mathcal{O}^k$. 
First, one must assume that $\mathcal{O}^k$ is ``locally conserved'', in the sense that $[\Hs^k, \mathcal{O}^k] = 0$. 
Eq.~\eqref{Currents_O_evo_gen} then gives (without the dissipative part)
\begin{equation}\label{Currents_site_resolved_Ok}
    \frac{d\langle \mathcal{O}^k \rangle}{dt} = \ii \langle [\Hs^{k-1,k}, \mathcal{O}^k] \rangle + \ii \langle [\Hs^{k,k+1}, \mathcal{O}^k]\rangle\,.
\end{equation}
It is tempting to associate $\ii \langle [\Hs^{k-1,k}, \mathcal{O}^k] \rangle$ with the flow from site $k-1$ to $k$ and 
$-\ii \langle [\Hs^{k,k+1}, \mathcal{O}^k]\rangle$ with the flow leaving $k$ to $k+1$.
But this only makes sense if it agrees with the flow \emph{entering} $k+1$ from $k$, obtained by looking at 
%
$\mathcal{O}^{k+1}$:
\begin{equation*}
    \frac{d\langle \mathcal{O}^{k+1} \rangle}{dt} = \ii \langle [\Hs^{k,k+1}, \mathcal{O}^{k+1}] \rangle + \ii \langle [\Hs^{k+1,k+2}, \mathcal{O}^{k+1}]\rangle\,.
\end{equation*}
The term we are looking for is  $\im \langle [\Hs^{k+1,k}, \mathcal{O}^{k+1}]\rangle$. 
However, in general 
$-\ii \langle [\Hs^{k,k+1}, \mathcal{O}^k]\rangle \neq \im \langle [\Hs^{k,k+1}, \mathcal{O}^{k+1}]\rangle$.
That is, the quantity leaving $k$ to $k+1$ is not necessarily the same as that entering $k+1$ from $k$. 
\gtl{From this, we conclude that a} 
\ind{site-resolved continuity equation} for a set of local operators $\mathcal{O}^k$ is  only possible when
\begin{equation}\label{Currents_Ok_condition}
    [\Hs^{k,k+1},~\mathcal{O}^k + \mathcal{O}^{k+1}] = 0\,.
\end{equation}
Together with $[\Hs^k, \mathcal{O}^k] = 0$, this implies that 
\begin{equation}\label{Currents_local_continuity_condition}
    [\Hs, \sum_k \mathcal{O}^k] = 0\,.
\end{equation}
Hence, the condition~\eqref{Currents_local_continuity_condition} is seen to be equivalent to the one we used in going from Eq.~\eqref{Currents_O_evo_gen} to Eq.~\eqref{Currents_O_evo_conserved}, namely that $[\Hs, \mathcal{O}] = 0$, with $\mathcal{O} = \sum_k \mathcal{O}^k$.
Consequently, both the system- and the site-resolved approaches are founded on similar assumptions. 
The latter, however, generally provide an additional level of detail.


When Eq.~\eqref{Currents_Ok_condition} is satisfied, we may unambiguously define the current of \gtl{a given observable} $\mathcal{O}^k$ leaving $k$ towards $k+1$ as 
\begin{equation}\label{Currents_local_current_def}
    \cur{\mathcal{O}}^{k,k+1} = - \im \langle [\Hs^{k,k+1}, \mathcal{O}^k]\rangle\,,
\end{equation}
so that Eq.~\eqref{Currents_site_resolved_Ok} becomes
\begin{equation}\label{Currents_site_resolved_unitary_continuity}
    \frac{d\langle \mathcal{O}^k \rangle}{dt} = \cur{\mathcal{O}}^{k-1,k} - \cur{\mathcal{O}}^{k,k+1}\,.
\end{equation}
These definitions hold for all sites, provided we define $\cur{\mathcal{O}}^{0,1} = \cur{\mathcal{O}}^{L,L+1} = 0$. 
%
%
%
%
For instance, in the XXZ chain   Eq.~\eqref{Currents_local_current_def} yields
$\cur{M}^{k,k+1} = -2 J \langle \sx_k \sy_{k+1} - \sy_{k} \sx_{k+1} \rangle$, 
which is Eq.~\eqref{XXZ_current}.   
The structure of Eq.~\eqref{Currents_local_current_def} also naturally invites one to define a corresponding \ind{current operator}
\begin{equation}\label{current_operator}
    \mathcal{I}_{\mathcal{O}}^{k,k+1} = - \im [\Hs^{k,k+1}, \mathcal{O}^k],
\end{equation}
such that $\langle \mathcal{I}_{\mathcal{O}}^{k,k+1} \rangle = \cur{\mathcal{O}}^{k,k+1}$.
Current operators find many uses, especially when dealing with unitary dynamics, such as in Kubo's formula. 

For energy, the situation is a bit more delicate since $\Hs$ has contributions from both sites and bonds [Eq.~\eqref{Currents_H_nn}]. 
In some cases, such as when $\Hs^{k,k+1}$ are very small, it may be reasonable to analyze a site-resolved continuity equation for the local energies $\Hs^k$. 
However, in order for this to exist, one must satisfy Eq.~\eqref{Currents_Ok_condition}; that is, $[\Hs^{k,k+1}, \Hs^k + \Hs^{k+1}]= 0$.
As an example, take $\Hs^{k,k+1} = a_k^\dagger a_{k+1} + a_{k+1}^\dagger a_k$ and suppose $\Hs^k = \epsilon_k a_k^\dagger a_k$, with  different energies $\epsilon_k$ for each site. 
This interaction preserves the number of quanta, so Eq.~\eqref{Currents_Ok_condition} is satisfied for $\mathcal{O}^k = a_k^\dagger a_k$. 
However, it will not preserve the local energies when the $\epsilon_k$ are different. 

Instead, we can construct a local energy current via a \ind{bond-resolved continuity equation}~\cite{wu2009a}.
We first rewrite Eq.~\eqref{Currents_H_nn} as a sum over bonds, $\Hs = \sum_{i=1}^{L-1} \tilde{H}_{\rm S}^{i,i+1}$, for some new operators $\tilde{H}_{\rm S}^{i,i+1}$. 
%
%
A continuity equation for $\tilde{H}_{\rm S}^{i,i+1}$ will then read 
\begin{equation}\label{Currents_unitary_bond_continuity}
    \frac{d \langle \tilde{H}_{\rm S}^{i,i+1} \rangle}{d t} = 
      \im \langle [\tilde{H}_{\rm S}^{i-1,i}, \tilde{H}_{\rm S}^{i,i+1}] \rangle 
    + \im \langle [\tilde{H}_{\rm S}^{i+1,i+2},\tilde{H}_{\rm S}^{i,i+1}] \rangle\,.
\end{equation}
The first term is the current entering bond $(i,i+1)$  from bond $(i-1,i)$, and the second is the current going from bond $(i,i+1)$ to $(i+1,i+2)$. 
The latter will always coincide with  the current entering $(i+1,i+2)$ from $(i,i+1)$, as one may explicitly verify by constructing an analogous equation for $d \langle \tilde{H}_{\rm S}^{i+1,i+2} \rangle/d t$.
With the proper boundary conditions, one also finds that the total Hamiltonian is a constant of motion.

\subsubsection{Site-resolved continuity equations}\label{sssec:localcontinuityequation}

We now include the effects of dissipation through Eq.~\eqref{Currents_M}.
We first consider the LME case, where each dissipator $\Diss_\nu$ acts only on a specific site (so that $\nu$ can be linked to $k$). 
For a set of local observables $\mathcal{O}^k$ satisfying~\eqref{Currents_Ok_condition}, 
Eq.~\eqref{Currents_site_resolved_unitary_continuity} becomes 
\begin{equation}\label{Currents_Ok_LME}
\frac{d\langle \mathcal{O}^k \rangle}{dt} = \cur{\mathcal{O}}^{k-1,k} - \cur{\mathcal{O}}^{k,k+1} + \cur{\mathcal{O},\text{diss}}^k\,,
\end{equation}
where $\cur{\mathcal{O},\text{diss}}^k = \tr\big\{\mathcal{O}^k \Diss_k(\rhos)\big\}$.
In the important case where the baths couple to the first and last sites, we get
\begin{IEEEeqnarray}{rCl}
    \frac{d\langle \mathcal{O}^1 \rangle}{dt} &=& -\cur{\mathcal{O}}^{1,2} + \cur{\mathcal{O},\text{diss}}^1\,,
    \\[0.2cm]
    \frac{d\langle \mathcal{O}^k \rangle}{dt} &=&
    \cur{\mathcal{O}}^{k-1,k} - \cur{\mathcal{O}}^{k,k+1}\,, \qquad  k = 2,\ldots,L-1\,,
    \\[0.2cm]
    \frac{d\langle \mathcal{O}^1 \rangle}{dt} &=& \cur{\mathcal{O}}^{L-1,L} + \cur{\mathcal{O},\text{diss}}^L\,.
\end{IEEEeqnarray}
In the steady state, $d\langle \mathcal{O}^k \rangle/dt = 0$ and all currents coincide
\begin{equation}
    \cur{\mathcal{O},\text{diss}}^1 = \cur{\mathcal{O}}^{1,2} = \cur{\mathcal{O}}^{2,3} = \ldots = \cur{\mathcal{O}}^{L-1,L} = - \cur{\mathcal{O},\text{diss}}^L \equiv \cur{\mathcal{O}}\,.
\end{equation}
yielding a unique current $\cur{\mathcal{O}}$ through the system. 
For instance, with $\mathcal{O}^k= \sigma_z^i$ and LME baths~\eqref{preamble_Lindblad_dissipator_sites2},
we get 
$\cur{M,{\text{diss}}}^i = \gamma_i (\eta_i - \langle \sz_i \rangle)$.


A similar analysis can also be done for the energy.
Once again, one must distinguish between a bond-, or site-resolved description.
For bonds Eq.~\eqref{Currents_unitary_bond_continuity} is appended with dissipative currents $\tr\big\{ \tilde{H}_{\rm S}^{k,k+1} \big( \mathcal{D}_k + \mathcal{D}_{k+1}\big) \rhos \big\}$. 
For LMEs, however, site-resolved equations for the local energies $\Hs^k$ are in a sense more natural.
The reason is that, as reviewed in Sec.~\ref{SEC:local_gksl_master_equation}, LMEs can be microscopically derived assuming the system-system interactions $\Hs^{k,k+1}$ are small. 
In this case, provided Eq.~\eqref{Currents_Ok_condition} is satisfied, the site-resolved continuity equation will have the form
    $\frac{d\langle \Hs^k\rangle}{dt} = \cur{E}^{k-1,k} - \cur{E}^{k,k+1} + \cur{E,\text{diss}}^k\,$, 
where $\cur{E}^{k,k+1} = - \im \langle [\Hs^{k,k+1}, \Hs^k]\rangle$.
The last term is thus interpreted as the \ind{local energy current} entering site $k$ via the adjacent reservoir
\begin{align}\label{EQ:Ecurrent_local}
    \cur{E,\text{diss}}^k = \tr\big\{ \Hs^k \Diss_k(\rhos)\big\}\,,
\end{align}
and analogous considerations for the particle current lead to
\begin{align}\label{EQ:Mcurrent_local}
    \cur{N,\text{diss}}^k = \tr\big\{ \Ns^k \Diss_k(\rhos)\big\}\,,
\end{align}
with particle number operator $\Ns^k$ of site $k$.
\gtl{Explicit formulas for $\cur{E,{\text{diss}}}^k$, for the XXZ chain in the presece of magnetic fields, are provided in~\cite{Mendoza-Arenas2013}.}
The local energy currents $\cur{E,\text{diss}}^k$ from bath $k$ to site $k$, is generally different than
$\tr\big\{\Hs \Diss_k(\rhos)\big\}$, 
%
%
since the latter has additional contributions $\tr\big\{\Hs^{k,k\pm1} \Diss_k(\rhos)\big\}$  [c.f. Eq.~\eqref{Currents_H_nn}].
However, using the same bookkeeping notation of Sec.~\ref{SEC:local_gksl_master_equation}, where the system-bath couplings are of order $\lambda$, and the system-system couplings of order $\xi$, we see that these extra contributions will be of order $\lambda^2\xi$, so
\begin{equation}\label{Currents_local_global_LME_approx}
    \tr\big\{\Hs \Diss_k(\rhos)\big\} = \cur{E,\text{diss}}^k + O(\lambda^2\xi)\,. 
\end{equation}

For Redfield or GMEs, each dissipator $\Diss_\nu$ in Eq.~\eqref{Currents_M} will act on entire chain, even if the coupling is to a specific site.
Thus, instead of Eq.~\eqref{Currents_Ok_LME} one will have 
\begin{equation}\label{Currents_Ok_LME2}
\frac{d\langle \mathcal{O}^k \rangle}{dt} = \cur{\mathcal{O}}^{k-1,k} - \cur{\mathcal{O}}^{k,k+1} +\sum\limits_\nu \tr\Big\{\mathcal{O}^k \Diss_\nu(\rhos)\Big\}\,,
\end{equation}
where each term in the sum represents the action of bath $\nu$ on site $k$. 
This reflects the non-local character of the dissipators. 
For this reason, one often does not assess site-resolved equations for global dissipators, focusing instead only on system-resolved equations such as Eqs.~\eqref{Currents_H_evo} or~\eqref{Currents_O_evo_conserved}. 
In fact, as pointed out in~\cite{Wichterich2007}, the local currents can even be unphysical in GMEs, due to the secular approximation (a wokraround was put forth in~\cite{Kamiya2015}).

\subsection{Thermodynamics of LMEs and GMEs\label{sec:td_lme_gme}}

\gtl{Sec.~\ref{sec:transport_and_currents} illustrated how to write down  consistent expressions for the first law of thermodynamics in the form of continuity equations for the energy. 
Here we show that a similar argument can also be made concerning the second law.
The key quantity of interest, in this case, is the irreversible entropy production~\cite{LandiPaternostro2021}.
Unlike energy, entropy does not satisfy a continuity equation:
Part of the change in the von Neumann entropy of the system $S[\rhos] = - \tr\big\{ \rhos \ln \rhos\big\}$ can be associated to a flow of entropy to the reservoirs. 
But there is also another part -- the entropy production rate $\dot{\Sigma}$ -- associated to the irreversible nature of the process.}
%
%
\gtl{From usual thermodynamic arguments, the entropy flux to a thermal reservoir at temperature $T_\nu$ is  $\cur{Q}^\nu/T_\nu$, where $\cur{Q}^\nu$ is the heat current {\gsc entering the reservoir}. 
The entropy balance equation for the system should therefore read $dS/dt = \dot{\Sigma} + \sum_\nu \cur{Q}^\nu/T_\nu$.
While $dS/dt$ may in general have any sign, the entropy production rate is by construction always non-negative, $\dot{\Sigma} \geqslant 0$, which is a mathematical statement of the second law. 
}

\gtl{Determining $\dot{\Sigma}$ is thus a relevant task for any open quantum process. 
The difficulty is that 
since the baths are traced out in a microscopic derivation, the value of $\cur{Q}^\nu$ (which is a bath quantity) can be wrongly assessed from the reduced description of the master equation. 
In fact, as will be discussed below, this might even lead to apparent violations of the second law. 
In~\cite{LandiPaternostro2021} it was shown that this is only fully resolved if one has access to the full system-bath dynamics.
Here, we discuss instead how the latter might be approximated in different types of master equations of the form~\eqref{Currents_M}, to arrive at thermodynamically consistent expressions. 
}

We start with GMEs. From  the global energy and particle currents [Eqs.~\eqref{Currents_H_evo} and \eqref{Currents_N_evo}],  we define the  \ind{heat current} to each bath  $\cur{Q,\rm GME}^\nu = \cur{E}^\nu - \mu_\nu \cur{N}^\nu$.
Additionally, we assume, as in Sec.~\ref{sssec:GKSL}, that each dissipator targets a thermal state at temperature $\beta_\nu$ and chemical potential $\mu_\nu$; that is, $\Diss_\nu(\rho_{\text{ss, GME}}^\nu) = 0$, where $\rho_{\text{ss, GME}}^\nu = e^{-\beta_\nu(\Hs - \mu_\nu \Ns)}/Z_\nu$.
From classical thermodynamic considerations, the change in entropy of reservoir $\nu$  is related to its energy and particle change via $dU_\nu =T_\nu dS_\nu + \mu_\nu dN_\nu$,
so $\dot{S}_\nu=\beta_\nu\left(\frac{dU_\nu}{dt}-\mu_\nu \frac{dN_\nu}{dt}\right)$. %
Assuming energy and particle conservation between system and reservoir, 
\gtl{which is generally true under weak coupling conditions, this can be approximated to $\dot{S}_\nu=-\beta_\nu \cur{Q,\rm GME} = - \beta_\nu (\cur{E}^\nu - \mu_\nu \cur{N}^\nu)$.}
%
The entropy production, computed as the rate of change of the entropy of the universe  is thus~\cite{Alicki_1979}
\begin{align}\label{EQ:sigma_dot_global}
    \dot\Sigma_\text{GME} 
    &\approx -\frac{d}{dt} \trace{\rhos\ln\rhos} -\sum_\nu \beta_\nu \left[\cur{E}^\nu - \mu_\nu \cur{N}^\nu\right]\nn
    &= \gsc{-\trace{\dot{\rhos} \ln \rhos} 
    - \sum_\nu\beta_\nu\trace{\left(\Hs - \mu_\nu \Ns\right) \left({\cal D}_\nu \rhos\right)}}\nn
    &= -\sum_\nu \trace{\left({\cal D}_\nu \rhos\right) \ln \rhos}
    + \sum_\nu \trace{\left({\cal D}_\nu \rhos\right) \ln \rho^\nu_{ss,\rm GME}}\nn
    &= -\sum_\nu \trace{\left({\cal D}_\nu \rhos\right) \left[\ln \rhos - \ln\rho^\nu_{\text{ss},\rm GME}\right]} \ge 0\,.
\end{align}
%
In the third line, we substituted Eq.~\eqref{Currents_M} for $\dot{\rhos}$ in the 1st term, and represented the 2nd by a logarithm  (the partition function does not contribute since ${\cal D}_\nu$ is traceless).
%
The inequality in the last line follows from Spohn's inequality~\cite{spohn1978b} applied to each term in the sum. 

Similar considerations can be made for LMEs.
Crucially, in this case the second law must be formulated in terms of the local energy and particle currents $\cur{E,\text{diss}}^k$ and $\cur{N,\text{diss}}^k$ [Eqs.~\eqref{EQ:Ecurrent_local} and~\eqref{EQ:Mcurrent_local}], 
defined in terms of the local Hamiltonians $\Hs^k$ for each site $k$ [c.f. Eq.~\eqref{Currents_H_nn}]. 
The local heat currents are then  $\cur{Q,\text{LME}}^k = \cur{E,\text{diss}}^k-\mu_k \cur{N, \text{diss}}^k$.
With this proviso, we find a similar inequality:
\begin{align}\label{EQ:sigma_dot_local}
    \dot{\Sigma}_\text{LME} &= -\frac{d}{dt} \trace{\rhos \ln \rhos} 
    -\sum_k \beta_k \cur{Q,\rm LME}^k \ge 0\,,    
\end{align}
but now defined in terms of local currents. 
The derivation is similar to Eq.~\eqref{EQ:sigma_dot_global}, but exploiting the fact that the local dissipators $\Diss_k^\text{LME}$ thermalize only the local states 
$\rho^k_{ss,\rm LME}=e^{-\beta_k(\Hs^k-\mu_k \Ns^k)}/Z_k$.
This allows us to write 
$\dot{\Sigma}_\text{LME} = \sum_k \tr\Big\{ \Diss_k \rhos \Big[ \ln \rhos - \ln \rho^k_{ss,\rm LME}\Big]\Big\} \geq 0$.~\footnote{This does not require that all sites are connected to a bath: Since the $\Diss_k$ act locally, the sum is only over those sites which are actually connected to a bath.}

%


There are (at least) two physical arguments why for LMEs, Eq.~\eqref{EQ:sigma_dot_local} must be formulated in terms of local heat currents $\cur{Q,\text{LME}}^k$.
First, in a microscopic derivation, the LME is a good approximation only when $\Hs^{k,k+1}$ are small, in which case the local and global currents are approximately the same [Eq.~\eqref{Currents_local_global_LME_approx}]. 
Second, if the LME stems from a collisional model (Sec.~\ref{SEC:local_gksl_master_equation}), 
it is shown in~\cite{DeChiara2018,Barra2015}
that $\cur{Q,\text{LME}}^k$ is the actual heat current flowing to the ancillas, while the terms associated with $\Hs^{k,k+1}$ are related to the work cost of turning the bath interactions on and off. 
This is consistent with the fact that the last term in Eq.~\eqref{EQ:sigma_dot_local} should be the change in entropy of the environments. 

To summarize, in formulating the second law, one should use global currents for GMEs and local currents for LMEs. 
Otherwise, one may arrive at apparent violations of the second law. 
An example was discussed in~\cite{Levy2014}, which considered two bosonic modes with frequencies $\omega_{1(2)}$ coupled to baths at  $\beta_{1(2)}$.
They used LMEs, but employed global currents to compute the entropy production rate, leading to  
$\dot{\Sigma}  \propto  (\omega_1+\omega_2)(\beta_2 - \beta_1)(n_1-n_2)$, where $n_k = (e^{\beta_k\omega_k} -1)^{-1}$.
As one may verify, it is possible to tune $\omega_{1(2)}$ so as to always induce $\dot{\Sigma} <0$, which suggested an inadequacy of LMEs in describing the second law. 
This was reconciled in Refs.~\cite{DeChiara2018,Barra2015,Pereira2018} with the help of a collision model. The resulting  entropy production rate, derived in terms of local currents, now has the form 
$\dot{\Sigma} \propto (\beta_2 \omega_2- \beta_1 \omega_1)(n_1-n_2)$, which depends only on the products $\beta_i\omega_i$, being  strictly non-negative and zero iff $\beta_1\omega_1 = \beta_2\omega_2$.

\subsection{\label{sec:conundrum} Local~vs.~global~vs.~Redfield master equations}

As illustrated in Sec~\ref{SEC:weak_coupling}, even within the weak-coupling paradigm, there are still many QMEs one can derive, each based on different (and sometimes opaque) approximations.
This may lead one to ask which is the ``\emph{best}'' QME to use in a given situation.
%
In this section, we provide a rough guideline. 
LMEs (Secs.~\ref{sec:lme_stage} and~\ref{SEC:local_gksl_master_equation})  work well if both internal system-system coupling and system-reservoir couplings are weak.
They do not {\gsc exactly} thermalize the system when the baths are at the same temperature but they are very easy to use. 
GMEs (Sec.~\ref{sssec:GKSL}) rely heavily on the secular approximation, and so 
tend to work better if the internal system-system interactions are strong, since large energy splittings better justify the secular approximation.
They may also produce unphysical internal currents~\cite{Wichterich2007},
but global currents are reliable.
Finally, Redfield equations (Sec.~\ref{sssec:redfield}) often capture the best of both worlds, in that they are compare well with exact solutions \gsc{(Sec.~\ref{ssec:connection_exact})}.
But they can produce unphysical states, since they are not CPTP.


QMEs are benchmarked by comparing them with models that allow for exact solutions.  
Sec.~\ref{ssec:connection_exact} provides a detailed example.
Ref.~\cite{Rivas2010b} studied LMEs and GMEs in bosonic systems. 
They first considered a single harmonic oscillator coupled to a bath (where LMEs and GMEs coincide), and found that QMEs work well, except at very short times. 
Afterwards they considered two bosonic modes, with operators $a_1$ and $a_2$, coupled to two baths with operators $\{b_k\}$ and $\{c_k\}$. 
The dynamics was modeled with a Hamiltonian of the form
\begin{IEEEeqnarray}{rCl}
\label{conundrum_Rivas_H}
    H &=& \Omega(a_1^\dagger a_1 + a_2^\dagger a_2) + \sum\limits_k \Big( \omega_{1k} b_k^\dagger b_k + \omega_{2k} c_k^\dagger c_k\Big)
    \\[0.2cm]
    \nonumber
    &&+ \xi~\vartheta(a_1,a_2) + \lambda \sum\limits_k \Big[ g_{1k}~\vartheta(a_1,b_k) + g_{2k}~\vartheta(a_2,c_k)\Big]\,,
\end{IEEEeqnarray}
where $\vartheta(a,b) = a^\dagger b + b^\dagger a$ is  a shorthand notation 
and -- as in our discussion around Eq.~\eqref{EQ:lme_weak_int_coup} --  $\xi$ gauges the internal system coupling while $\lambda$ the overall system-environment interaction (with $g_{ik}$ being an additional, dimensionless parameter). 

Comparing the LME and GME resulting from this Hamiltonian with the exact solution~\cite{karrlein1997a, SerafiniBook}, they found that for small  $\xi$, the LME works better than the GME. 
Conversely, for large $\xi$ the GME tends to be the better choice. 
Qualitatively similar results are found for fermions~\cite{Ribeiro2015,Mitchison2018}; see also Sec.~\ref{ssec:connection_exact}.
Ref.~\cite{Purkayastha2016}  extended these results to Redfield  equations and found that, over a large parameter regime, it yielded dramatically better results than the LME and GME.
The Hamiltonian~\eqref{conundrum_Rivas_H} assumes a hopping interaction, which conserves the number of quasi-particles.
In contrast, a ``position-position'' interaction $(a+a^\dagger)(b+b^\dagger)$ was studied in~\cite{Gonzalez2017}.
The conclusions were similar, but spanned a much wider range of parameter space. 
\gtl{Recently,~\cite{Potts2021} compared the LME and GME approaches with a specific emphasis on the first law of thermodynamics; i.e., on the proper identification of thermal currents.}
Ref.~\cite{Mitchison2018} also addressed  the additivity of multiple reservoirs: as discussed in Sec.~\ref{SEC:additivity},  LMEs and GMEs enjoy the convenient property that reservoirs combine additively.
Comparing with exact solutions, the authors found that, in general, this is not a good assumption, as also revealed by  strong-coupling methods (Sec.-\ref{sec:strong_coupling}). 
\gtl{The studies above have focused on the NESS. Transient behavior was studied in~\cite{Scali2021}, which showed that LMEs generally perform better. This is due to the fact that the secular approximation tends to destroy key dynamical features of GMEs.}

All these analyzes address what is the ``best QME'' by comparing with exact solutions.
However, these are only available for special classes of systems (usually non-interacting). 
Little is known for interacting systems.
Ref.~\cite{hartmann2020a} studied spin-boson models and benchmarked the Redfield master equation with a pseudomode approach~\cite{garraway1997a,Imamoglu1996, Garraway1997,TamascelliPlenio2018}, these were found to agree better than GKSL equations.
These findings can be contrasted with the results in~\cite{mccauley2020a}, which show that for systems weakly coupled via a constant spectral density, one may find GKSL master equations that are superior to the Redfield approach.
\cite{XuPoletti2019a} compared LMEs and Redfield master equations with exact solutions, for large spin chains of up to 21 sites (simulated using tensor networks).
They found Redfield equations performed significantly better than LMEs in the presence of strong interactions within the system. 
A comparison with GMEs was not possible for being too computationally costly.
%



\subsection{\label{ssec:connection_exact}Connection with exactly solvable systems}

Despite the complexity of open quantum systems there exist some cases where the dynamics is exactly solvable. 
This includes the pure-dephasing limit of the spin-boson model~\cite{LeggettZwerger1987,lidar2001a}, where the dynamics is simple since the interaction commutes with the system Hamiltonian and thus cannot change the system energy.
Additionally, quadratic models also allow for an exact solution, which can be obtained by various methods.
In the bosonic case for example, \gsc{exact non-Markovian master equations for} systems of coupled oscillators have been studied~\cite{karrlein1997a}.
Non-interacting models can also be  treated with  non-equilibrium Green's functions~\cite{economou2006,wang2014a, DharHanggi2012,NikolicThygesen2012, ZimbovskayaPederson2011, Aeberhard2011, ProciukDunietz2010, MeirWingreen1992, CaroliSaint-James1971, HaugJauho2008},
including slowly driven systems~\cite{Bhandari2021}.
Complications arise, though, when interactions are considered~\cite{MeirWingreen1992,meir1993a}.

In this section we expose  a simpler approach, that fully suffices to treat models with quadratic Hamiltonian.
Although we exemplify the model for fermions, a similar derivation works for bosons.
We focus on reservoirs modeled by 1D tight-binding chains, but  
this is not a restriction since any chain can be mapped into a set of non-interacting modes and vice-versa (see Sec.~\ref{sec:strong_coupling}).
The advantage of the homogeneous tight-binding chain is that it can be analytically diagonalized even for finite chain lengths.

We consider two chain reservoirs $\alpha\in\{L,R\}$, described by $N_\alpha$ fermionic operators $d_{i,\alpha}$ each,  and modeled by a tight-binding Hamiltonian [Eq.~\eqref{tight_binding}]
$\Hb^{(\alpha)} = \epsilon \sum_{i=1}^{L_\alpha} d_{i,\alpha}^\dagger d_{i,\alpha} + \tau \sum_{i=1}^{L_\alpha-1}\left[d_{i,\alpha}^\dagger d_{i+1,\alpha} + d_{i+1,\alpha}^\dagger d_{i,\alpha}\right]$,
where $\epsilon\in\mathbb{R}$ and $\tau>0$.
The system, in turn,  is described by $N$ non-interacting sites, with operators $d_i$, and Hamiltonian 
$\Hs = \sum_{i,j=1}^N h_{ij} d_i^\dagger d_j$. 
The coupling to the baths occurs at sites 1 and $N$, and has the form
$\Hi = \tau_L (d_{1,L}^\dagger d_1+d_1^\dagger d_{1,L})
+ \tau_R (d_{1,R}^\dagger d_N + d_N^\dagger d_{1,R})$,
with tunnel amplitudes $\tau_\alpha>0$.
The reservoirs are diagonalized with the transformation 
$d_{i,\alpha} = \sqrt{\frac{2}{L_\alpha+1}} \sum_{k=1}^{L_\alpha} \sin\left(\frac{\pi i k}{L_\alpha+1}\right) c_{k,\alpha}$, 
to a new set of operators $c_{k,\alpha}$, leading to:
\begin{align}\label{EQ:diagonal_form}
    \Hb^{(\alpha)} = \sum_{k=1}^{L_\alpha} \epsilon_{k\alpha}c_{k,\alpha}^\dagger c_{k,\alpha}\,,\qquad
    \epsilon_{k\alpha}=\epsilon-2\tau \cos\left(\frac{\pi k}{L_\alpha+1}\right) \,.
\end{align}
In turn, the system-bath interactions change to
\begin{align}
    \Hi &= \sum_{k=1}^{N_L} t_{kL} \left[c_{kL}^\dagger d_1 + \text{H.c.}\right]+\sum_{k=1}^{N_R} t_{kR}  \left[c_{kR}^\dagger d_N + \text{H.c.}\right]\,,
\end{align}
with  $t_{k\alpha} = \tau_\alpha \sqrt{\frac{2}{L_\alpha+1}} \sin\left(\frac{\pi k}{L_\alpha+1}\right)$.
Diagonalizing $\Hb^{(\alpha)}$ has the advantage that we can explicitly eliminate the reservoir operators from the Heisenberg equations of motion: 
\gsc{$\f{\dot{d}_1} = -\ii \sum_j h_{1j} \f{d_j} -\ii \sum_k t_{kL} \f{c_{kL}}$ for the first site, 
$\f{\dot{d}_a} = -\ii \sum_j h_{aj} \f{d_j}$ for  $2 \le a \le N-1$, 
$\f{\dot{d}_N} = -\ii \sum_j h_{Nj} \f{d_j} -\ii \sum_k t_{kR} \f{c_{kR}}$ 
for the last site, and
$\f{\dot{c}_{kL}} = -\ii \epsilon_{kL} \f{c_{kL}} -\ii t_{kL} \f{d_1}$ and
$\f{\dot{c}_{kR}} = -\ii \epsilon_{kR} \f{c_{kR}} -\ii t_{kR} \f{d_N}$ for the left and right reservoir operators, respectively.}

An exact solution is  possible  using non-equilibrium Green functions~\cite{HaugJauho2008}.
In App.~\ref{APP:exact_solution} we provide a more direct exact approach based on Laplace transforms.
It applies to reservoirs in the star representation~\cite{schaller2014}, and explicitly incorporates the initial conditions, which complies with generalized formulations of non-equilibrium thermodynamics~\cite{esposito2010b}.
The baths are characterized by the \ind{spectral coupling density} $\Gamma_\alpha(\omega) = 2\pi \sum_k \abs{t_{k\alpha}}^2 \delta(\omega-\epsilon_{k\alpha})$ (also termed \ind{bare tunneling rate} in this context), which in our case becomes
\begin{align}\label{EQ:specdens_chain}
    \Gamma_\alpha(\omega) 
    &= \frac{2\tau_\alpha^2}{\tau}\sqrt{1-\frac{(\omega-\epsilon)^2}{4\tau^2}} \Theta(4\tau^2-(\omega-\epsilon)^2)\,,
\end{align}
which has strict finite support.
This is  known as the semi-elliptical, or Newns, spectral density~\cite{Newns1969,Mitchison2018}.
%
Non-constant spectral densities~\cite{schaller2009a,topp2015a} can be used to model non-Markovian effects~\cite{zedler2009a}, 
and finite support can even lead to phenomena like \ind{bound states}~\cite{longhi2007a,jussiau2019a}.


\begin{figure}
    \centering
    \includegraphics[width=\columnwidth,clip=true]{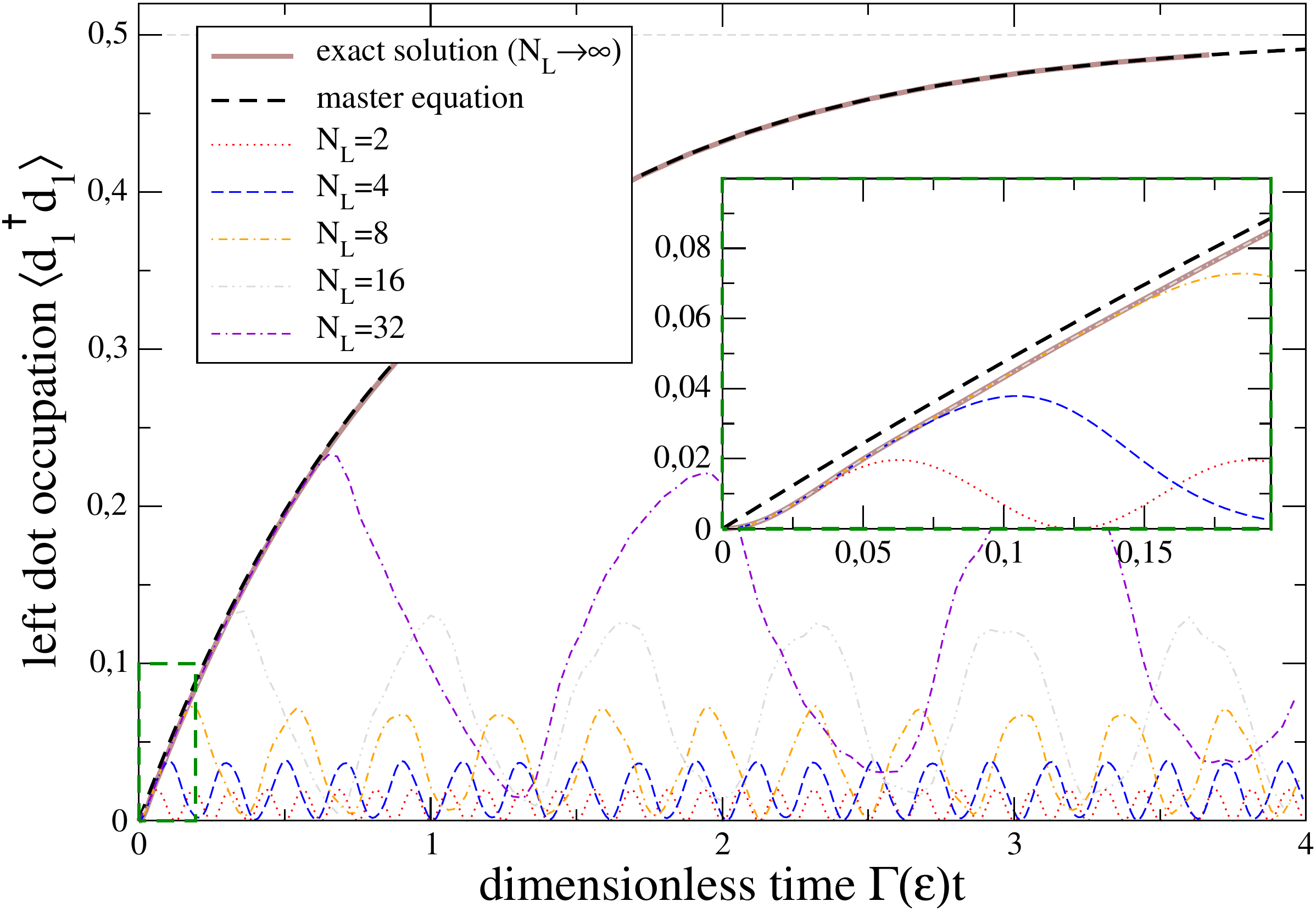}
    \caption{Time-dependent occupation of a single dot of energy $\varepsilon$ coupled to a 
    tight-binding chain of different lengths $N_L$. 
    As $N_L\to\infty$, the reservoir  drags the system towards a stationary occupation, but for finite chain lengths, recurrences occur.
    For the chosen parameters (weak system-reservoir coupling), the QME (dashed) captures well the dynamics,
    except at small times (inset).
    The horizontal dashed line at $0.5$ is the long-time steady-state occupation.
    Other parameters $h_{11}=h_{22}=\epsilon\equiv\varepsilon$, $h_{12}=0$, $\mu_\nu=\varepsilon$, $\tau=\varepsilon$, $\tau_\nu=0.1 \varepsilon$, $\Gamma_\nu(\varepsilon)\beta_\nu = 2\tau_\nu^2 \beta_\nu /\tau = 0.02$.}
    \label{fig:leftdotoccupation}
\end{figure}

Considering first the simplified case of a single dot coupled to only one reservoir, we can compare the exact solution for the  dot occupation $\expval{d_1^\dagger d_1}$, for various bath sizes,  with the solution arising from a GKSL treatment, as shown in  Fig.~\ref{fig:leftdotoccupation}.
One can see that  finite-size baths (system and reservoir) lead to back-flows of information (here, a fermionic particle tunneling back to the system), which one can use as a non-Markovianity measure~\cite{breuer2012a}.
If the dot energy were significantly outside the support of $\Gamma_\alpha(\omega)$, a bound state may emerge~\cite{longhi2007a,jussiau2019a}, which is not captured by a Markovian master equation (not shown).

Next we go back to the original 2-bath configuration, and consider the steady-state. We assume for simplicity the  \ind{wideband limit}, where $\Gamma_\alpha(\omega)\to\Gamma_\alpha$. 
Formally, this can be achieved by taking
    $\tau,\tau_\alpha \to \infty$, such that 
    $\frac{2 \tau_\alpha^2}{\tau} \equiv \Gamma_\alpha = \text{const}$.
%
In this limit, the Green's function $G_{ij}(z)$ (defined in App.~\ref{APP:exact_solution} below Eq.~\eqref{eq:appendix_temporary141029831092}) has only isolated poles with negative real part.
We may thus drop the initial-state dependence and all poles with a negative real part in the inverse Laplace transform, such that eventually, the stationary limit simplifies to
\begin{align}\label{EQ:sysocc_exactss}
    \expval{d_i^\dagger d_j}_t &\stackrel{t\to\infty}{\to} 
    \int \Gamma_L(\omega)  G_{i1}^*(0^+-\ii\omega) G_{j1}(0^+-\ii\omega) f_L(\omega)  \frac{d\omega}{2\pi}\\
    &\qquad+ \int \Gamma_R(\omega) G_{i2}^*(0^+-\ii\omega) G_{j2}(0^+-\ii\omega) f_R(\omega) \frac{d\omega}{2\pi}\,.\nonumber
\end{align}
which coincides with Green's functions approaches~\cite{economou2006,wang2014a, DharHanggi2012,NikolicThygesen2012, ZimbovskayaPederson2011, Aeberhard2011, ProciukDunietz2010, MeirWingreen1992, CaroliSaint-James1971, HaugJauho2008}.

We compare this with the LME, GME and Redfield equations~\cite{Levy2014,hofer2017a,Gonzalez2017,farina2020a} by considering a two-site
system (double quantum dot), with 
$d_{1(2)} \equiv d_{L(R)}$, 
$h_{11}=h_L$, $h_{22}=h_R$ and $h_{12} = h_{21} = h$.
The LME from Sec.~\ref{SEC:local_gksl_master_equation} becomes, in this case,
\begin{align}\label{EQ:dqd_local}
    \dot{\rhos} &= -\ii \left[\Hs, \rhos\right] + \sum_{\alpha = L,R} 
    \Gamma_\alpha \Big\{(1-f_\alpha(h_\alpha)] \Lind{d_\alpha}(\rhos) 
    +  f_\alpha(h_\alpha) \Lind{d_\alpha^\dagger}(\rhos)\Big\}.
    %
\end{align}
For the GME (\ref{EQ:me_bms}), we must first diagonalize $\Hs$:
\begin{align}
    \Hs = \left(d_L^\dagger, d_R^\dagger\right)\left(\begin{array}{cc}
        h_{L} & h \\
        h & h_{R}
    \end{array}
    \right)
    \left(\begin{array}{c}
      d_L\\
      d_R
    \end{array}\right)
    = \epsilon_- c_-^\dagger c_- + \epsilon_+ c_+^\dagger c_+\,.
\end{align} 
Finally, the Redfield equation~\eqref{EQ:redfieldII_sp}, assuming $h_L=h_R=\epsilon$, and dropping the principal values in Eq.~\eqref{EQ:sokhotski_plemelj}, becomes
\begin{align}\label{EQ:redfield_dqd}
    \dot{\rhos} &= -\ii \left[\Hs, \rhos\right]\nn
    &\quad-\frac{\Gamma_L}{4} f_L(\epsilon+h) \left\{\left[d_L,(d_L^\dagger+d_R^\dagger)\rhos\right]+\left[\rhos(d_L+d_R),d_L^\dagger\right]\right\}\nn
    &\quad-\frac{\Gamma_L}{4} f_L(\epsilon-h) \left\{\left[d_L,(d_L^\dagger-d_R^\dagger)\rhos\right]+\left[\rhos(d_L-d_R),d_L^\dagger\right]\right\}\nn
    &\quad-\frac{\Gamma_L}{4} f_L^-(\epsilon+h) \left\{\left[d_L^\dagger,(d_L+d_R)\rhos\right]+\left[\rhos(d_L^\dagger+d_R^\dagger),d_L\right]\right\}\nn
    &\quad-\frac{\Gamma_L}{4} f_L^-(\epsilon-h) \left\{\left[d_L^\dagger,(d_L-d_R)\rhos\right]+\left[\rhos(d_L^\dagger-d_R^\dagger),d_L\right]\right\}\nn
    &\quad+ (L \leftrightarrow R)\,,
\end{align}
where $f_L^-(\omega)\equiv1-f_L(\omega)$.
For small internal couplings $h\to 0$, this falls back to the LME~\eqref{EQ:dqd_local}.

\begin{figure*}
    \begin{tabular}{cccc}
    \includegraphics[height=5.5cm]{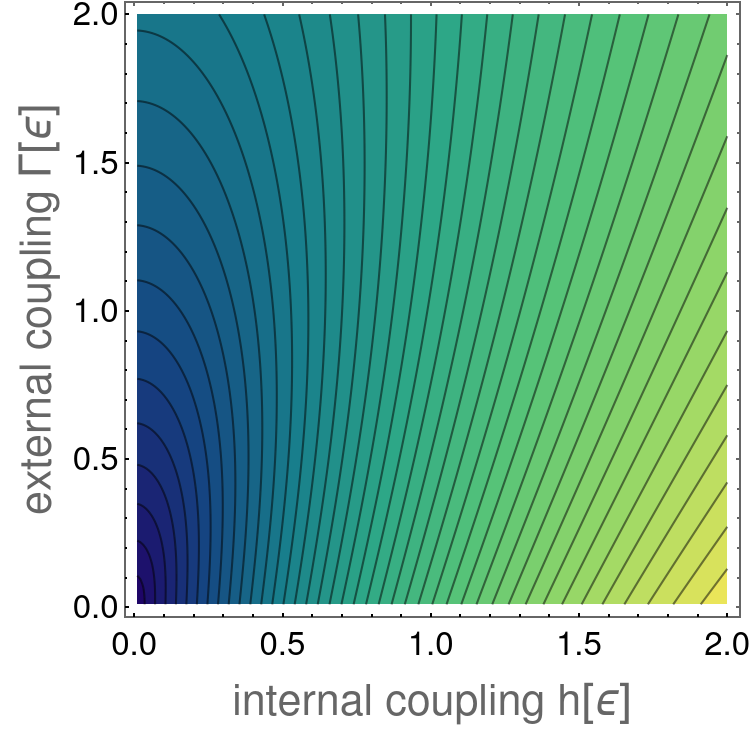} & 
    \includegraphics[height=5.5cm]{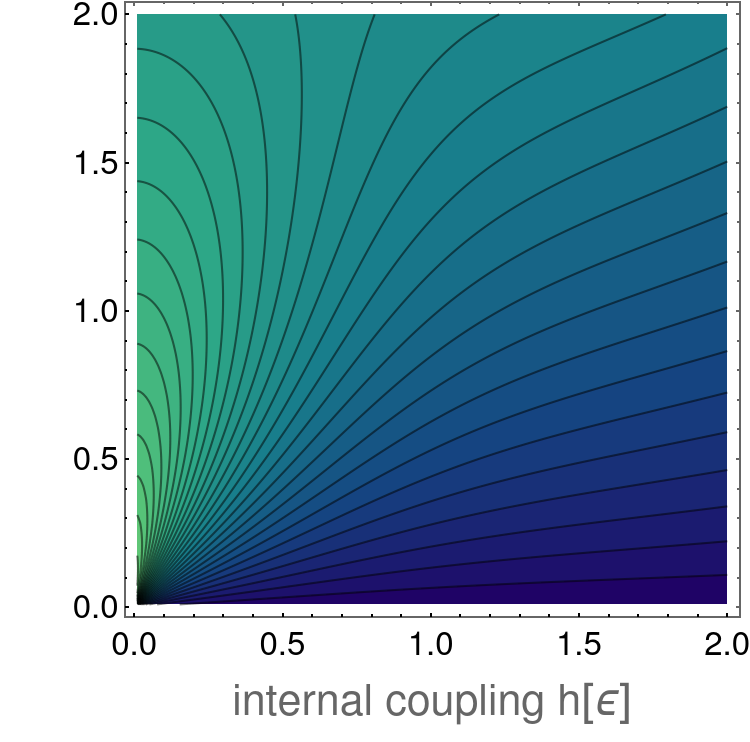} &
    \includegraphics[height=5.5cm]{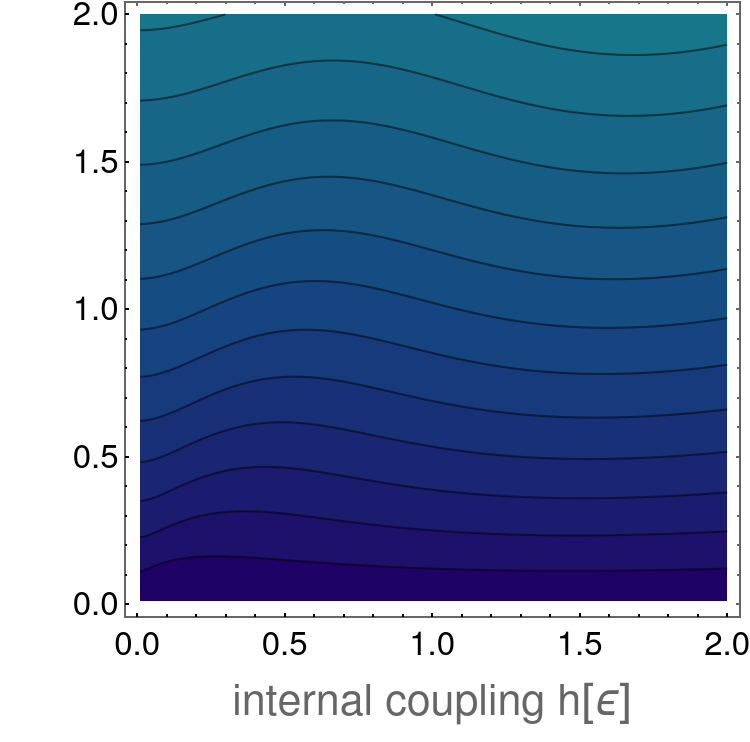} &
    \includegraphics[height=5.5cm]{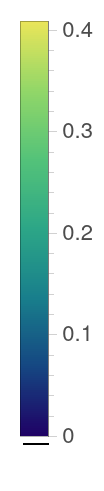}
    \end{tabular}
    \caption{
    \gsc{
    Contour plot of the trace distance between the NESS single-particle density matrix  [Eq.~(\ref{EQ:spdensmat})] of the exact solution and the perturbative solutions for LME $D(\rho_1^\text{EX}, \rho_1^\text{LME})$ (left panel),  GME $D(\rho_1^\text{EX}, \rho_1^\text{GME})$ (middle panel), and  Redfield equation $D(\rho_1^\text{EX}, \rho_1^\text{RED})$ (right panel); dark/blue means agreement of solutions.
    The plots are done as a function of the dimensionless internal coupling $h_{12}/\epsilon=h_{21}/\epsilon\equiv h/\varepsilon$ (horizontal axes) and the dimensionless external coupling $\Gamma_L/\epsilon = \Gamma_R/\epsilon = \Gamma/\epsilon$ (vertical axes) with contours
    in steps of 0.01, assuming different chemical potentials and a  wideband limit.
    Parameters: $h_L=h_{11}=h_{22}=h_R\equiv\epsilon$, $\beta_\nu \epsilon=1$, $\mu_L=\epsilon=-\mu_R$.}}
    \label{fig:loglerror}
\end{figure*}

In Fig.~\ref{fig:loglerror} we compare the
\ind{trace distance}
\begin{equation}\label{trace_distance}
    D(\rho,\sigma) = \trace{|\rho-\sigma|}/2,
\end{equation}
between the \indTwo{single-particle density matrix}{single-particle density matrix}
\begin{align}\label{EQ:spdensmat}
    \rho_1 = \left(\begin{array}{cc}
    \langle{d_1^\dagger d_1}\rangle & \langle{d_1^\dagger d_2}\rangle\\
    \langle{d_2^\dagger d_1}\rangle & \langle{d_2^\dagger d_2}\rangle
    \end{array}\right)\frac{1}{\langle{d_1^\dagger d_1}\rangle + \langle{d_2^\dagger d_2}\rangle},
\end{align}
of the stationary exact solution (EX), versus those from the LME, GME and Redfield (RED) equations (left, middle and right panels respectively).
The horizontal and vertical axes are the  internal system coupling $h$ and the system bath coupling $\Gamma$ (in units of $\epsilon =1$).
We see that the QMEs  complement each other and have their own region of validity~\cite{DeChiara2018, hofer2017a, Gonzalez2017}: For large $h$ and small $\Gamma$, the GME performs well, while for small $h$ the LME is better.
One can also see that the Redfield equation yields good results throughout, despite not being in GKSL form.
In fact, the superiority of the Redfield equation has been observed for a number of models where an exact solution exists. 
The development of GKSL equations reaching  Redfield accuracy at steady state is therefore currently an interesting route of research~\cite{kirsanskas2018a,kleinherbers2020a,mccauley2020a,trushechkin2021a}.

A qualitatively similar picture arises for the currents:
For non-interacting electronic transport, the exact stationary particle current can be computed using \ind{Landauer's formula}~\cite{landauer1957a,buettiker1986a,nazarov2009} 
\begin{align}\label{EQ:landauer}
    \cur{N}^\text{EX} = \frac{1}{2\pi} \int T(\omega)\left[f_L(\omega)-f_R(\omega)\right] d\omega\,,
\end{align}
where $0 \le T(\omega) \le 1$ is the transmission function~\cite{HaugJauho2008}.
%
%
%
The comparison with the LME, GME and RED (not shown) is qualitatively similar to Fig.~\ref{fig:loglerror}. 
There are regimes, however, where  although the steady state is poorly approximated, both LME and GME by chance give acceptable results for the currents.

%

\subsection{\label{sec:strong_coupling}Strong system-bath coupling} 

The physics of strong system-bath coupling has been an intense research topic, both in classical and quantum systems~\cite{campisi2011a,TalknerHanggi2020}.    
In strong coupling, the QMEs of Sec.~\ref{SEC:weak_coupling} are not applicable, the influence of multiple reservoirs no longer simply adds up (compare with Sec.~\ref{SEC:additivity}),  the currents may need to be computed using Full Counting Statistics (Sec.~\ref{sec:full_counting}), since in multiple particles may be emitted/absorbed at once~\cite{schaller2013a}, and  reservoirs may cross-talk~\cite{talarico2020a}.
However, strong coupling  also offers possibilities.
For example, with non-additive reservoirs 
one may build an autonomous refrigerator with just two levels~\cite{mu2017a}, whereas for additive reservoirs this requires at least three~\cite{linden2010a}.

In this section we review techniques for describing  boundary-driven systems in the strong coupling regime.
There are two main strategies: 
The first is to extend the system with a portion of the bath which, after careful transformations, can be weakly coupled to a \ind{residual bath}. 
This includes the \ind{reaction coordinate} (Sec.~\ref{sec:reaction_coordinates}) and  \ind{polaron} approaches (Sec~\ref{sec:polaron}).
The other strategy is to unitarily evolve system and baths together, 
which can be done using the \ind{star-to-chain} and \ind{thermofield} transformations (Sec.~\ref{sec:star_to_chain_thermofield}), as well as other methods reviewed in Sec.~\ref{sec:extended}.  
The problem can also be approached with Green's function (Sec.~\ref{ssec:connection_exact}),
\gsc{higher-order calculations~\cite{schroeder2007a,kast2013a}, generalized thermodynamic arguments~\cite{aurell2018a,perarnau_llobet2018a}, surrogate Hamiltonians~\cite{GatzKosloff2016}, phenomenological collective modes~\cite{cabot2017a}, Feynman-Vernon influence functionals~\cite{jin2010a,yang2020a}, among others.}

\subsubsection{Reaction coordinates}\label{sec:reaction_coordinates}

The approximations used in the derivation of QMEs can be easily violated when the system-reservoir coupling strength is not small, if very short timescales are considered or when the system has  near-degenerate levels.
Some of these restrictions can be overcome if the framework is applied in a different frame, where the boundaries between system and reservoir are shifted.
Transitions between different frames can be  realized by Bogoliubov transforms~\cite{WoodsPlenio2014}.
To properly address the strong-coupling limit, one should make sure to start from a Hamiltonian that maintains a lower spectral bound.
For example, considering a single bosonic reservoir with dimensionless system coupling operator $S$, this would be guaranteed when
\begin{align}\label{EQ:hamrc1}
    \Htot = \Hs + \sum_k \omega_k \left(b_k^\dagger + \frac{h_k}{\omega_k} S^\dagger\right) 
    \left(b_k + \frac{h_k^*}{\omega_k} S\right)\,,
\end{align} 
where $\omega_k, h_k$ are constants.
Upon expanding the second term, we get the usual system and reservoir Hamiltonian, together with a renormalization 
\begin{align}
    \Delta \Hs = \sum_k \frac{\abs{h_k}^2}{\omega_k} S^2 
    = \frac{1}{2\pi} \int_0^\infty \frac{\Gamma^{(0)}(\omega)}{\omega}  S^2d\omega\,,
\end{align}
where
\begin{align}\label{EQ:spectral_density}
    \Gamma^{(0)}(\omega) = 2\pi \sum_k \abs{h_k}^2 \delta(\omega-\omega_k)\,.
\end{align}
is the \indTwo{spectral (coupling) density}{spectral coupling density}~\cite{LeggettZwerger1987,xu2019a,mascherpa2020a}.

Beyond weak coupling, a perturbative treatment of the system reservoir interaction $h_k$ is not applicable.
Instead, we apply a \ind{Bogoliubov transform}s
$b_k = \sum_q \left(u_{kq} B_q + v_{kq} B_q^\dagger\right)$,
to new bosonic annihilation operators $B_q$, where $u_{kq}\in\mathbb{C}$ and $v_{kq}\in\mathbb{C}$ are coefficients that have to ensure proper commutation relations for the $B_q$. 
The idea is to choose them so as to recast $\Htot$ in the form
\begin{align}\label{EQ:hamrc2}
    \Htot &= \Hs + \Omega_1 \left(B_1^\dagger + \frac{\lambda_1}{\Omega_1} S\right)\left(B_1 + \frac{\lambda_1}{\Omega_1} S\right)\\
    &\qquad+\sum_{q>1} \Omega_q \left(B_q^\dagger + \frac{H_q}{\Omega_q} \left(B_1+B_1^\dagger\right)\right)\left(B_q + \frac{H_q^*}{\Omega_q} \left(B_1+B_1^\dagger\right)\right)\,,\nonumber
\end{align}
i.e., such that $S$ couples only to a single mode $B_1$ -- called the \ind{reaction coordinate} -- with coupling strength $\lambda_1$ and energy $\Omega_1$.
In turn, $B_1$ couples to the residual reservoir modes $B_{q>1}$, via new coupling constants $H_q$, which parametrize a residual spectral density
\begin{align}
    \Gamma^{(1)}(\omega) = 2\pi \sum_{k>1} \abs{H_k}^2 \delta(\omega-\Omega_k)\,.
\end{align}
Finding the Bogoliubov coefficients is generally tedious, but for an infinitely large reservoir with dense level spacing, explicit formulas for $\lambda_1$, $\Omega_1$, and $\Gamma^{(1)}(\omega)$ can be derived~\cite{StrasbergBrandes2016,nazir2019a}:
\begin{align}\label{EQ:energy_rc}
    \Omega_1^2 &= \frac{\int_0^\infty \omega \Gamma^{(0)}(\omega) d\omega}{\int_0^\infty \frac{\Gamma^{(0)}(\omega)}{\omega} d\omega}\,,
    \\[0.2cm]
    \lambda_1^2 &= \frac{1}{2\pi \Omega_1} \int_0^\infty \omega \Gamma^{(0)}(\omega) d\omega\,.
    \\[0.2cm]
    \Gamma^{(1)}(\omega) &= \frac{4 \lambda_1^2 \Gamma^{(0)}(\omega)}{\left[\frac{1}{\pi} {\cal P} \int_{-\infty}^{+\infty} \frac{\Gamma^{(0)}(\omega')}{\omega'-\omega} d\omega'\right]^2 + \left[\Gamma^{(0)}(\omega)\right]^2}\,,
\end{align}
where, in the last line, a analytic continuation as an odd function, $\Gamma^{(n)}(\omega') = -\Gamma^{(n)}(-\omega')$, is understood under the integral.
Multiple variants of such mapping exist, c.f.~\cite{WoodsPlenio2014,nazir2019a}.

The usefulness of this approach lies in the fact that, even though the original couplings $h_k$ are strong, the new couplings $H_k$ to the residual bath can be weak.
To see this, notice that  Eq.~\eqref{EQ:spectral_density} is $\mathcal{O}(h_k^2)$, 
while the new spectral density $\Gamma^{(1)}$ is 
$\mathcal{O}(1)$.
Hence, although $S$ and $B_1$ couple strongly, $B_1$ and its residual bath do not~\cite{MartinazzoBurghardt2011,StrasbergBrandes2016},
which was benchmarked in~\cite{IlesSmithNazir2014,ilessmith2016a}. 
This has even been observed to hold for even stronger residual couplings~\cite{correa2019a}.
The effects of system-reservoir correlations~\cite{IlesSmithNazir2014} or non-Markovian dynamics~\cite{ilessmith2016a} within the original system also come within reach.
The price one has to pay is that the reaction coordinate needs to be treated
explicitly, which already for two bosonic reservoirs can be challenging~\cite{anto_sztrikacs2021a}.

The reaction-coordinate mappings can -- with slight modifications -- be applied to fermionic systems as well,
where the explicit treatment of a fermionic reaction coordinate is computationally not as costly as for bosonic ones.
This was used in~\cite{restrepo2019a,SchallerStrasberg2018,StrasbergEsposito2018} to discuss quantum non-equilibrium thermodynamics of fermionic systems.

\begin{figure}
\includegraphics[width=\columnwidth,clip=true]{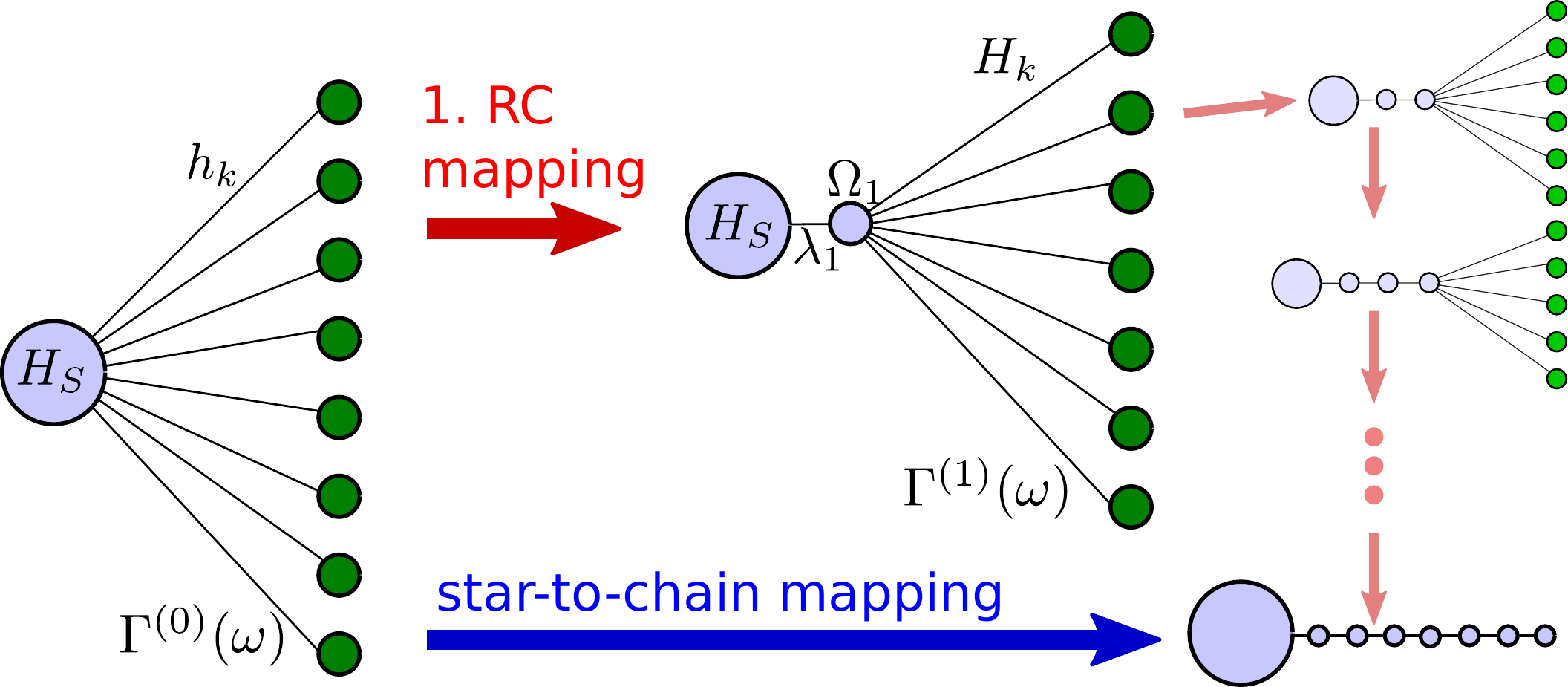}
\caption{\label{FIG:mapping_sketch}
Sketch of a single reaction-coordinate (RC) mapping (top left arrow) that effectively transfers one degree of freedom from the reservoir (dark green circles) to the system (light blue circles) generating an enlarged supersystem that requires an explicit treatment. 
Recursive applications (on the right in faint colors) generate a chain configuration, effectively implementing the star-to-chain mapping (bottom long arrow) described in Sec.~\ref{sec:star_to_chain_thermofield}.}
\end{figure}

The mapping above is independent of $\Hs$, such that one can use these approaches to study interacting systems~\cite{BrenesGoold2020,TamascelliPlenio2019,strasberg2019b}. 
In addition, the transformed Hamiltonian is equivalent to the original  upon identifying $S \to (B_1 + B_1^\dagger)$ and 
$\Hs\to \Hs + \Omega_1 \left(B_1^\dagger + \frac{\lambda_1}{\Omega_1} S^\dagger\right)\left(B_1 + \frac{\lambda_1}{\Omega_1} S\right)$. 
The procedure can thus be applied recursively, which would eventually map the ``star-like'' system-reservoir interaction, where the system is coupled to each bath mode; into a chain, where the system and each bath site are only coupled to nearest neighbors (Fig.~\ref{FIG:mapping_sketch}).
The typical intention of the reaction-coordinate formalism is to stop after a few iterations:
As a rule of thumb, the transformed spectral densities tend to become more and more structureless with each iteration, which at some point enables a Markovian description~\cite{MartinazzoBurghardt2011,WoodsPlenio2014}.
For methods requiring a chain representation however, it is  possible to perform the star-to-chain mapping directly, see  
 Sec.~\ref{sec:star_to_chain_thermofield}.
Alternatively, since Bogoliubov transforms are invertible, one may also exploit reverse reaction coordinate mappings to map long chains with structureless reservoirs into a single quantum dot with highly structured reservoirs, which can be treated with non-equilibrium Green's functions~\cite{martensen2019a,ehrlich2021a}. 

If the system is coupled to multiple reservoirs, the reaction coordinate mapping is applied individually to each bath. 
Hence, the resulting QME will still add up.
However, a dissipator for reservoir $\nu$ now depends on the system plus reaction coordinates Hamiltonian, and hence on the parameters of the original reservoirs.
As such, the original reservoirs no longer enter additively, in the sense of Sec.~\ref{SEC:additivity}.

We can also generalize the formalism to multiple reaction coordinates.
%
For example, at each step of the mapping one may split the support of $\Gamma^{(n)}(\omega)$ into intervals, and introduce a reaction coordinates for each interval. 
This will lead to tree-type network topologies that can be more efficient  than a chain mapping~\cite{huh2014a,mascherpa2020a,pleasance2020a}.
Alternatively, one may represent $\Gamma^{(0)}(\omega)$ by a sum of simpler (e.g. Lorentzian) functions, which also introduces multiple collective coordinates at once. 
This is often called the pseudo-mode approach~\cite{garraway1997a,Imamoglu1996,pleasance2020a, Garraway1997}, 
and has been used to describe  strong-coupling and non-Markovianity.
In~\cite{BrenesGoold2020}, this was  combined with multi-site GKSL baths (Sec.~\ref{sec:znidaricprosenbath}) and tensor networks (Sec.~\ref{sec:tensor_networks}),  with a specific focus on applications in many-body thermal machines.

\subsubsection{Star-to-chain and Thermofield}\label{sec:star_to_chain_thermofield}

Recursively applying  the reaction-coordinate method takes the \ind{star configuration}~\eqref{EQ:hamrc1} into a \ind{chain configuration} for the bath (Fig.~\ref{FIG:mapping_sketch}).
Since they can be interpreted as successive Bogoliubov transforms, their joint application is also a Bogoliubov transform.
The \ind{star-to-chain mapping} can thus be performed on a single step.
In this section, we describe this idea in more detail. 
In addition, we also show how to combine it with 
the \ind{thermofield transformation}, which ensures that the resulting chain reservoirs are empty, thus providing simple initial conditions for simulations.

\paragraph*{Star-to-chain transformation.}
Consider, for simplicity, a continuous bath  with 
Hamiltonian 
$\Hbath = \int \omega(k) b^\dagger(k)b(k) dk \label{eq:Hbath_continuum}$ 
and  coupling 
\begin{align}
    \Hi = S\int h(k) [b(k)+b^\dagger(k)] dk\,, \label{eq:Hint_continuum} 
\end{align}
where $S$ acts only on the system and $h(k)$ is the coupling density.
The \indTwo{spectral function}{spectral coupling density}  -- analogous to Eq.~\eqref{EQ:spectral_density} -- is thus $J(\omega)=\pi h^2[k(\omega)]\frac{d k(\omega)}{d\omega}$~\cite{BullaVojta2005,LeggettZwerger1987}. 

To apply the star-to-chain mapping directly, one may either discretize the spectrum, (as in numerical renormalization group~\cite{Wilson1975, BullaPruschke2008}), use orthogonal polynomials~\cite{PriorPlenio2010, ChinPlenio2010}, or perform numerical optimization via a cost function~\cite{CaffarelKrauth1994, DordaArrigoni2014}. 
%
%
In \ind{direct discretization}, we split the range of frequencies of the bath in different intervals, and  use a set of basis functions to describe the bath operators within each range, thus recovering a discrete bath. 
A strategy which properly weights  different energy scales is the logarithmic discretization, 
which considers intervals at frequencies $\omega_n = \omega_{max} \Lambda^{-n}$ where $\Lambda>1$~\cite{BullaPruschke2008}. 
In each interval, one can choose basis functions which are plain waves in that interval and $0$ elsewhere, so that $\int_0^{\omega_{max}}d\omega = \sum_n \int^{\omega_n}_{\omega_{n+1}}d\omega$.

Conversely, with orthogonal polynomials one introduces a new set of bosonic operators $B_n=\int U_n(k) b(k) dk$, with $n=0,1,\dots$  
and $U_n(k)=h(k)\rho_n^{-1}\pi_n(k)$.
Here $\rho_n$ is a normalization factor
and $\pi_n(k)$ are orthogonal polynomials with inner product $\int h^2(k) \pi_n(k)\pi_m(k) dk = \rho_n \delta_{n,m}$;
Since all information about  the environment is contained in  $J(\omega)$~\cite{LeggettZwerger1987}, 
one  may choose $\omega(k)$ and $h^2(k)$ freely, as long as $J(\omega)$ is preserved.
A useful choice is  $\omega(k)=\omega_c k$ and $h^2(k) = \omega_c J(\omega(k))/\pi$. 
This transforms the bath and the interaction into 
\begin{align}
    \Hbath &= \omega_c\left[\sum_n \alpha_n B_n^\dagger B_n +  (\sqrt{\beta_{n+1}} B^\dagger_{n+1} B_n  +\text{H.c.})\right]\,, \label{eq:Hbath_continuum_converted}
    \\[0.2cm]
    \Hi &= S \sqrt{\frac{\eta^{}_J}{\pi}} (B_0 + B_0^\dagger)\,,  
    \qquad 
    \eta^{}_J = \int J(\omega) d\omega,
    \label{eq:Hint_continuum_converted}
\end{align}
which follows from the recursion relation $k \pi_n(k) = \alpha_n \pi_n(k) + \beta_n \pi_{n-1}(k) + \pi_{n+1}(k)$ (that holds when $\omega(k)=\omega_c k$). 
Choosing $\pi_{-1}(k)=0$, $\pi_1(k)=1$,
the coefficients $\alpha_n,\beta_n$ in $\Hb$ are then given by 
$\beta_0=0$, $\beta_n=\gamma_n/\gamma_{n-1}$, $\gamma_n=\int h^2(k) \pi_n(k)^2 dk$, and  
$\alpha_n = \gamma_n^{-1}\int k h^2(k) \pi_n(k)^2 dk$,
which are amenable to numerical evaluation~\cite{Gautschi2005}.      
These coefficients are bounded if the spectrum is finite.
And if the measure $h^2(k)dk$ belongs to the Szeg\"o class~\cite{Szego1939}, they converge to a value independent of $n$.
Hence, deep  within the chain, the couplings will be uniform, as is also the case in reaction-coordinates~\cite{WoodsPlenio2014}. %
%
Additional details can be found in~\cite{ChinPlenio2010}.

If the bath is already discrete, a Lanczos or  Householder transform will turn tridiagonal the originally diagonal bath Hamiltonian, i.e. a chain. 
The important step is to choose as the first vector the sum of all modes weighted by their coupling with the system~\cite{DeVegaWolf2015}.      

Converting the original problem to a chain configuration, allows it to be more readily tackled with tensor networks (Sec.~\ref{sec:tensor_networks}).
However, this is not always advantageous. 
For instance, for fermionic baths~\cite{WolfSchollwock2014, RamsZwolak2020, LuHaverkort2019} the star configuration may lead to a slower growth of entanglement, limited by the large number of fully occupied states in the Fermi sea, which do not participate in the evolution. 

\paragraph*{Thermofield transformation} 
The star-to-chain mapping allows one to study the evolution of the system plus baths together, as a single  system.
A difficulty  of bosonic chains, however, is that they may have large local occupations at high temperatures. 
Using the \ind{thermofield transformation}~\cite{BlasoneVitiello2011, ArakiWoods1963, TakahashiUmezawa1996, Bargmann1961, DeVegaBanuls2015}, we can map the chain, at least at initial times, to two empty chains.
This can be particularly useful in tensor network algorithms~(Sec.~\ref{sec:tensor_networks})
The thermofield transformation is employed in bosonic baths modeled with bilinear couplings to the system.
It applies to arbitrary coupling strengths and could also be extended to consider  initial correlations between system and bath.
Extensions to fermionic systems are discussed in~\cite{DeVegaBanuls2015, BlasoneVitiello2011, SchwarzWeischselbaum2018, NuselerPlenio2019}. 

Given a bath, defined by modes $b_k$, the idea is to introduce another bath, with modes $\cop_k$,  identical to the physical one, such that for thermal density matrices $\rho_\text{B}$, the  expectation value can be written as $\tr( \; \cdot \; \rho_\text{B}) = \langle \Omega | \cdot | \Omega \rangle$, with 
\begin{align}
    |\Omega\rangle = \bigotimes\limits_k\left(   \sum_{n_k=0}^\infty \frac{e^{-\beta \omega_k n_k/2}}{\sqrt{1+\npop_k}} |n_k\rangle_{b_k}|n_k\rangle_{c_k}\right),
\end{align}
where $\npop_{k} = 1/(e^{\beta \omega_k}-1)$ and 
$|n\rangle_{b_k}, |n\rangle_{c_k}$ are Fock states of $b_k,c_k$.
The state $|\Omega\rangle$ is called the \ind{thermal vacuum}. 
It satisfies $\exp(-\beta \Hb)/ Z_{\rm B} = \ptrace{C}{\ket{\Omega}\bra{\Omega}}$,
and can be obtained from the global vacuum $|\bm{0}\rangle$ of $\bop_k,\cop_k$ via a \ind{thermal Bogoliubov transformation} 
$|\Omega\rangle = e^{-\ii G}|\bm{0}\rangle$,
with $G = \im\sum \theta_k\left( \bopd_k\copd_k - \bop_k\cop_k \right)$ and $\tanh(\theta_k)=\exp(-\beta\omega_k/2)$. 
This implies that $|\Omega\rangle$ is the vacuum of a new set of bosonic modes $\aop_{1,k}= e^{-\ii G} \bop_k e^{\ii G}$ and $\aop_{2,k}= e^{-\ii G} \cop_k e^{\ii G}$.

We can now simulate the evolution of a system, initially prepared in the pure state $|\psi\rangle$ and coupled to a thermal bath of harmonic oscillators, as the evolution of the pure state $|\psi\rangle\otimes_k |0\rangle_{\aop{1,k}}|0\rangle_{\aop{2,k}}$.
However, to do so we must transform the Hamiltonian from the $\bop_k$, $\cop_k$  to the $\aop_{1,k}$, $\aop_{2,k}$ representation. 
It is much more convenient to use a slightly different Hamiltonian     
$\HtotTF = \Htot - \sum_k \omega_k c_k^\dagger \cop_k$.
Adding such a decoupled term, that only depends on the $\cop_k$ modes, has no physical consequences. 
Using $\HtotTF$, uncouples the modes $\aop_{1,k},\aop_{2,k}$ in the equations of motion,  resulting in a much simpler treatment. 
A similar idea also leads to better numerical performance of tensor network methods when studying finite temperatures~\cite{KarraschMoore2012}.

As an example,  consider the  Hamiltonian 
\begin{align}
    \Htot 
    =&\Hs + \sum_k \omega_k b_k^\dagger\bop_k + \sum_k h_k(a_{\rm S}^\dagger \bop_{k} + a_{\rm S} b_{k}^\dagger)\,,   
\end{align}
where $a_{\rm S}$, $a_{\rm S}^\dagger$ act on the system only.
The thermofield Hamiltonian $\HtotTF$, after a thermal Bogoliubov transformation, becomes 
\begin{align}
    \HtotTF =& \Hs + \sum_k \omega_k \left(a_{1,k}^\dagger\aop_{1,k} - a_{2,k}^\dagger\aop_{2,k}\right) \nonumber\\
    & + \sum_{i\in\{1,2\}} \sum_k h_{i,k}(a_{\rm S}^\dagger \aop_{i,k} + a_{\rm S} a_{i,k}^\dagger), 
\end{align}
with new couplings $h_{1,k}=h_k\cosh{(\theta_k)}$, $h_{2,k}=h_k\sinh{(\theta_k)}$. 
This can then be used together with the star-to-chain mapping, transforming a single thermal bath into two empty semi-infinite chains~\cite{GuoPoletti2018, SchwarzWeischselbaum2018, ChenPoletti2020}. 
%

\subsubsection{Polaron treatments}\label{sec:polaron} 

A convenient tool to study the strong-coupling regime with bosonic environments is the \ind{polaron master equation}.
The starting point is a system-environment Hamiltonian parametrized as in Eq.~\eqref{EQ:hamrc1}.
The \indTwo{polaron}{polaron transform} or \indTwo{Lang-Firsov}{Lang-Firsov transform} transform~\cite{lang1963a} amounts to a global unitary transformation of system and reservoir
\begin{align}
U_\text{p} = \exp\left\{S \sum_k \left(\frac{h_k^*}{\omega_k}b_k^\dagger - \frac{h_k}{\omega_k}b_k\right)\right\}\,,
\end{align}
where the $h_k$ ad $\omega_k$ are defined in Eq.~\eqref{EQ:hamrc1}.
As one may verify, this yields
$U_\text{p} S U_\text{p}^\dagger = S$ and 
$U_\text{p} b_k U_\text{p}^\dagger = b_k - \frac{h_k^*}{\omega_k} S$.
Hence, under this transform Eq.~\eqref{EQ:hamrc1} becomes
\begin{align}
H_\text{tot}' 
&= \underbrace{\traceB{U_\text{p} \Hs U_\text{p}^\dagger \frac{e^{-\beta\sum_k \omega_k b_k^\dagger b_k}}{Z}}}_{\Hs'} \otimes \id_\text{B}+\sum_k \omega_k b_k^\dagger b_k\nn
&\quad+\underbrace{\left[U_\text{p} \Hs U_\text{p}^\dagger - \traceB{U_\text{p} \Hs U_\text{p}^\dagger \frac{e^{-\beta\sum_k \omega_k b_k^\dagger b_k}}{Z}} \otimes \id_\text{B}\right]}_{\Hint'}\,,\nonumber
\end{align} 
%
where we  added and subtracted the same term to make the renormalized interaction $\Hint'$  obey
$\tr_B\big\{\Hint' \rhob \big\}=0$, such that the standard derivation of a GME or LME from Sec~\ref{SEC:weak_coupling} can be followed.
To apply a perturbative treatment, it is  not necessary that the $h_k$ are small, but rather that $\Hint'$ is.
In fact, since the polaron transform is unitary, none of these terms diverge when $h_k \to \infty$.
In the weak-coupling limit, the standard QME is reproduced, and for pure-dephasing models ($[\Hs, S]=0$), the system  is  untouched $\Hs'=\Hs$.
More involved statements on the thermodynamic consistency of the polaron approach are possible, including fluctuation theorems~\cite{schaller2013a,krause2015a}.

Such favorable properties have been used to advocate the polaron approach as capable of interpolating between the weak and strong coupling regimes~\cite{wang2015a,wang2015b}.
Furthermore, it has been found to yield consistent thermodynamic results, even for  driven systems~\cite{gelbwaser_klimovsky2015a}.
Various improvements have been suggested, 
including variational polaron transforms~\cite{mccutcheon2011a} and mixtures of polaron transforms and reaction-coordinate mappings~\cite{waechtler2020a}.
Finally, we remark that for multiple reservoirs, a polaron transform designed to modify the coupling to one bath will dress the couplings to others~\cite{brandes2005a,schaller2013a}, making the non-additivity very explicit.

\subsubsection{Evolution of system and large finite baths}\label{sec:extended}

\begin{figure}
    \centering
    \includegraphics[width=\columnwidth]{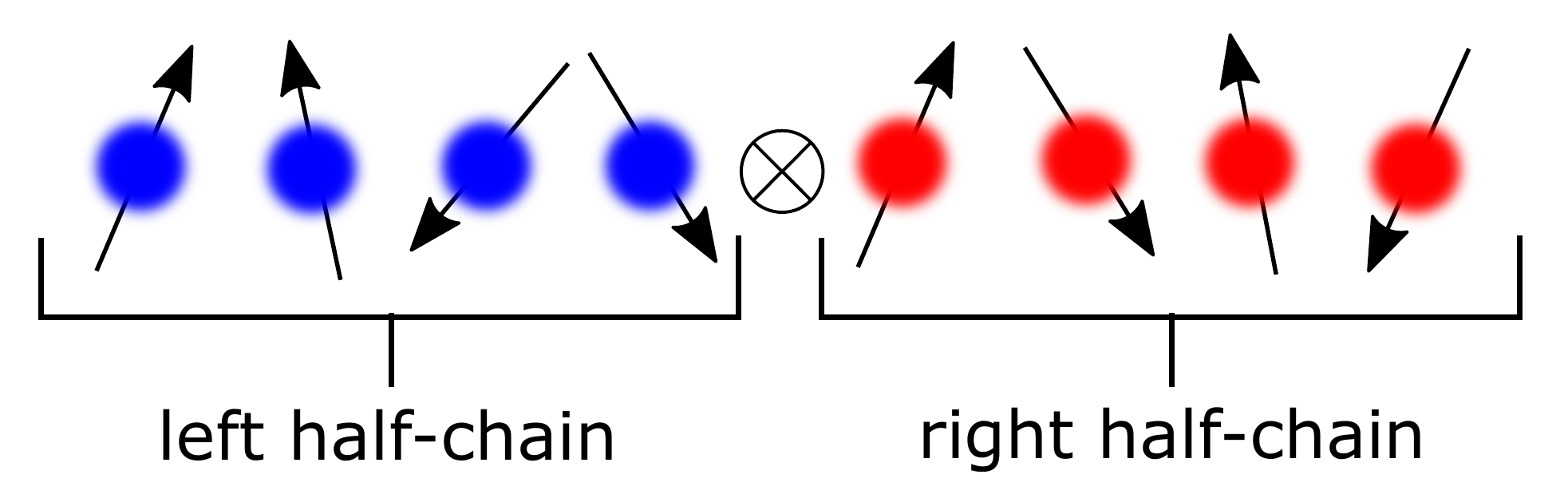}
    \caption{Unitary (Hamiltonian) dynamics can be simulated for two long but finite spin chains prepared in an initial product state, with each chain in a different pure or mixed state (different colors).}
    \label{fig:segmented_preparation}
\end{figure}

The methods above rely on quadratic interactions and the harmonicity of the bath Hamiltonian.
However, baths composed of interacting particles are often more realistic.
Hence, depending on the model, it may not be possible to use the above mentioned methods, requiring one to treat system and bath almost on the same footing.
The bath does not have to be infinite in size, but as long as the system does not feel any finite-size effects, and the dynamics has approximately reached a steady state, one can still learn about its transport properties.

For instance, in Ref.~\cite{KarraschMoore2013} the authors studied the dynamics of an XXZ chain with different types of perturbations, and within a quantum Ising chain, when each half was prepared at a different temperature, as depicted in Fig.~\ref{fig:segmented_preparation}. 
They showed that the system relaxes quickly to a steady current value in the presence of a nonzero Drude weight (Sec.~\ref{ssec:integrability}). 
In~\cite{MascarenhasSavona2017}, the authors considered a spin chain in the center, coupled at the edges to two chains  prepared in different \adp{(non-thermal)} states.
Ref.~\cite{LjubotinaProsen2017,ZnidaricLjubotina2018} considered a similar setup, and studied the evolution of the time-integrated current, 
$\int_0^t \cur{} dt' \propto t^{\phi}$, with  $\phi$ characterizing the transport regime according to  Table~\ref{tab:scaling}.
One can also consider the unitary evolution of two {\it different} integrable chains, each prepared in a different thermal state~\cite{BiellaFazio2016, BiellaMazza2019}. 
The contact interaction between the two chains was chosen to break the integrability, and induce the formation of a growing region supporting steady transport, until finite-size effects come into play. 
This idea was actually preceded in other works, such as.~\cite{PonomarevHanggi2011}, which the authors studied thermalization in bosonic systems. 

In a recent study~\cite{XuPoletti2021} on a setup analogous to Fig.~\ref{fig:segmented_preparation}, the authors considered pure state preparations of each half chain.
They showed that in the regime of weak coupling between the two halves, the emergence of a steady current is typical. 
This means that for any initial condition formed of energy eigenstates within  an energy window, the resulting long lasting current converges, in the thermodynamic limit, to the expected current from microcanonical preparations of each half.

A slightly different approach was taken in~\cite{MascharenasDeVega2019}, where additional dissipation at the edges of the baths was used, to extend the time for which the system could be studied. 
More recently, a different way of periodically refreshing the baths was put forward~\cite{PurkayasthaGoold2020}, and a (hierarchy of) master equations for the description of finite size baths was introduced in~\cite{RieraStrasberg2021}.  
For non-interacting systems, works along these lines with fermionic-mesoscopic baths were introduced in~\cite{AjisakaProsen2012, AjisakaBarra2013}. 


\subsection{\label{sec:full_counting}Full counting statistics and generalized master equations}

Full Counting Statistics (FCS) asks for the number of jump processes during a given timespan.
If e.g. the jumps are net particle transfers, the time-derivative of that number yields the particle current. 
And similarly, if they come with net energy changes, one can construct energy currents. 
FCS goes beyond the average, however, and also accounts for fluctuations.
We begin with a phenomenological introduction  in Sec.~\ref{SEC:fcs_me}, followed by a rigorous microscopic derivation in Sec.~\ref{SEC:fcs_mic}.
Tools on how to actually extract the results  are discussed in Sec.~\ref{SEC:fcs_sol}.

\subsubsection{Phenomenological introduction}\label{SEC:fcs_me}

When studying the dynamics of GKSL-QMEs~\eqref{EQ:GKSL},
one may wonder about the statistics of the \ind{quantum jumps} described by the
terms $L_\alpha \rhos L_\alpha^\dagger$~\cite{wiseman2010}. 
To track it, we can decompose the full Liouvillian into a no-jump evolution $\SO{L}_0$ and a jump-term of interest~\cite{brandes2008a}, i.e., 
\begin{align}
    \SO{L} = \SO{L}_0+\SO{L}_1,\qquad\qquad \SO{L}_1\rhos \equiv \gamma_{\bar\alpha} L_{\bar\alpha}\rhos L_{\bar\alpha}^\dagger\,,
\end{align}
for a single certain chosen process $\bar{\alpha}$~\footnote{One may generalize to many different jumps.}
Using a \ind{Dyson series}, the full propagator $\SO{P}(t,t_0) = e^{\SO{L} (t-t_0)}$ can be expressed as sequences of jump-free evolutions,
$\SO{P}_0(t) = e^{\SO{L}_0 t}$,
interrupted by a variable number of quantum jumps:
\begin{IEEEeqnarray}{rCl}\label{eq:dyson_series}
    &&\SO{P}(t,t_0) = \SO{P}_0(t-t_0)
    + \int\limits_0^t dt_1 \SO{P}_0(t-t_1) \SO{L}_1 \SO{P}_0(t_1-t_0)\\
    &&\qquad +\int\limits_0^t dt_2 \int\limits_0^{t_2} dt_1 \SO{P}_0(t-t_2) \SO{L}_1 \SO{P}_0(t_2-t_1) \SO{L}_1 \SO{P}_0(t_1-t_0) 
    +\ldots. \nonumber
\end{IEEEeqnarray}
The first term describes a trajectory without jumps, the second term one with a single quantum jump at $0<t_1<t$, the third one with two quantum jumps at $0<t_1<t_2<t$, and so on.

To select trajectories with a specific number of jumps one introduces a
\ind{counting field} $\chi$~\cite{levitov1996a,esposito2007b}, tilting the Liouvillian according to 
$\SO{L}(\chi) = \SO{L}_0 + \SO{L}_1 e^{+\ii\chi}$.
In the Dyson series of $\SO{P}(\chi,t) = e^{\SO{L}(\chi) t}$,
terms containing $n$ jumps then go with $e^{\ii n\chi}$. 
Using the orthogonality relation 
$\int_{-\pi}^\pi \frac{d\chi}{2\pi} e^{\ii (n-m)\chi} = \delta_{n,m}$
we can then select the propagator associated to a dynamics with specifically $n$ jumps up to time $t$:
\begin{align}
    \SO{P}^{(n)}(t) = \frac{1}{2\pi} \int_{-\pi}^{+\pi} \SO{P}(\chi,t) e^{-\ii n \chi} d\chi\,.
\end{align}
The associated probability $p_n(t)$ of obtaining $n$ quantum jumps during the time interval $[t_0,t]$
is then simply the trace of the conditional dynamics $\SO{P}^{(n)}(t) \rhos(t_0)$:
\begin{align}\label{EQ:pnoft}
    p_n(t) = \frac{1}{2\pi} \int_{-\pi}^{+\pi} \trace{e^{\SO{L}(\chi) (t-t_0)} \rhos(t_0)} e^{-\ii n \chi} d\chi\,.
\end{align}
The explicit evaluation is numerically tedious, and analytically  possible only in very special cases~\cite{schaller2010b}.
Instead, it is more convenient to look at the \ind{moment-generating function} 
\begin{equation}\label{EQ:FCS_MGF}
    M(\chi,t) 
    = \tr\big\{ \rho(\chi,t)\},
    \qquad 
    \rho(\chi,t) := e^{\SO{L}(\chi) (t-t_0)} \rhos(t_0)\,.
\end{equation}
where $\rho(\chi,t)$ is the solution of the \ind{generalized master equation},
$\dot{\rho}(\chi,t) = \Liouv(\chi) \rho(\chi,t)$.
From $M(\chi,t)$, we obtain the \ind{cumulant-generating function}  $C(\chi,t) = \ln M(\chi,t)$, which is the QME equivalent of the Levitov-Lesovik formula for non-interacting electrons~\cite{levitov1993a,klich2003a,schoenhammer2007a}.
Specific techniques to extract the lowest cumulants will be discussed in Sec.~\ref{SEC:fcs_sol}.

The questions raised in \indTwo{FCS}{Full Counting Statistics} are universal: 
They have been rigorously treated in both GMEs~\cite{schaller2014} and LMEs~\cite{garrahan2010a,znidaric2014b,znidaric2014c}, where approximate methods have been developed for large and interacting 1D systems~\cite{carollo2017a,carollo2018a}.
Some GKSL extensions include feedback interventions~\cite{brandes2010a,wiseman2010}, or the analysis of factorial cumulants~\cite{Stegmann2015,Stegmann2020}.
Beyond the GKSL framework, FCS was used in Redfield~\cite{hussein2014a,jin2020a} and non-Markovian QMEs~\cite{flindt2007a,flindt2008a,braggio2009a}.
And electronic transport setups allow to test these concepts also experimentally~\cite{gustavsson2006a,fujisawa2006a,flindt2009a,utsumi2010a,wagner2017a,Kurzmann2019}.
Alternatively, one may also ask for the \ind{waiting-time distribution} $\Omega(t)$~\cite{cohen_tannoudji1986a} between jumps, which is related to the probability of no jump by $\Omega(t)=-\frac{d}{dt}p_0(t)$. 
This can also be extended to the waiting times between different jumps~\cite{brandes2008a}.

As an example,  consider the QME~\eqref{EQ:me_bms}.
If we only want the total number of jumps (termed \ind{dynamical activity}),
we can tilt all jump operators identically,
$L_{ab} \rhos L_{cd}^\dagger \to L_{ab} \rhos L_{cd}^\dagger e^{+\ii\chi}$.
Conversely, to count the particle current we use 
$L_{ab} \rhos L_{cd}^\dagger \to L_{ab} \rhos L_{cd}^\dagger e^{+\ii\chi(N_a-N_b)}$, (the coefficients $\gamma_{ab,cd}$ impose that $N_d-N_c=N_b-N_a$).
Similarly, the energy current can be counted with 
$L_{ab} \rhos L_{cd}^\dagger \to L_{ab} \rhos L_{cd}^\dagger e^{+\ii\chi(E_a-E_b)}$.
For multiple reservoirs, specific counting fields can track the exchanges with each individual bath.
Sometimes, however, a phenomenological identification of jump terms is not  obvious.
In these cases, one must resort instead to a microscopic derivation, see below.

\subsubsection{Microscopic derivation}\label{SEC:fcs_mic}



Ref.~\cite{Esposito2009} provides a microscopic approach to connect changes in the bath with quantum sumps in the system.
The assumption is that the reservoir observable of interest $\hat{O}$ (e.g. particle number or energy) commutes with the reservoir Hamiltonian $[\hat{O}, \Hb]=0$.
For a typical reservoir, the absolute value of such observables may assume infinite values.
However, one only needs to track their changes during the time-interval $[0,t]$.
These can be obtained from a \ind{two-point measurement} scheme, where we measure $\hat O$ at time $0$ and again at time $t$:
Letting
$\hat{O} = \sum_\ell O_\ell \ket{\ell}\bra{\ell}$, in the first measurement, the outcome $O_\ell$ occurs with probability $p_\ell = \trace{\ket{\ell}\bra{\ell} \rhob}$, and projects $\rhob$ to 
$\rhob^{(\ell)}/p_\ell$, where 
$\rhob^{(\ell)} = \ket{\ell}\bra{\ell} \rhob \ket{\ell}\bra{\ell}$.
Since we only measure the reservoir, 
this does not affect the system.
Averaging over all outcomes of the first measurement, the moment-generating function is 
\begin{align}\label{EQ:FCS_MGF_unitary}
M(\chi,t) = \sum_\ell \tr_{\rm SB}\Big\{e^{\ii \chi (\hat{O}-O_\ell)} \f{U}(t) \rhos^0 \otimes \rhob^{(\ell)} \f{U^\dagger}(t)\Big\}\,,
\end{align}
where $\f{U}(t)$ is the full evolution operator in the interaction picture.
This is similar to Eq.~\eqref{EQ:FCS_MGF} but from the perspective of the system plus bath unitary dynamics.
Analogously, derivatives with respect to $\chi$ generate the moments of the distribution of changes in $\hat{O}$.

We can also rewrite Eq.~\eqref{EQ:FCS_MGF_unitary} as
$M(\chi,t)=\tr_{\rm SB} \rho_{\rm SB}(\chi,t)$, where 
$\rho_{\rm SB}(\chi,t) = \f{U_{+\chi/2}}(t) (\rho_0 \otimes \bar{\rho}_{\rm B}) \f{U_{-\chi/2}^\dagger}(t)$
and 
$\f{U_{+\chi/2}}(t) = e^{+\ii \hat{O} \frac{\chi}{2}} \f{U}(t) e^{-\ii \hat{O} \frac{\chi}{2}}$.
As a consequence,  $\rho_{\rm SB}(\chi,t)$ will evolve according to
\begin{equation}
    \frac{d\f{\rho}_{\rm SB}(\chi,t)}{dt} = -\ii \left(
    \f{H_{\chi/2}}(t) \f{\rho}_{\rm SB}(\chi,t)
    - 
    \f{\rho}_{\rm SB}(\chi,t)
    \f{H_{-\chi/2}}(t)
    \right)\,,
\end{equation}
with the tilted Hamiltonian
\begin{align}\label{EQ:hamint_fcs}
\f{H_{\chi/2}}(t) = e^{+\ii\hat{O} \frac{\chi}{2}} \f{\Hint}(t) e^{-\ii\hat{O} \frac{\chi}{2}}
= \sum_\alpha \f{A_\alpha}(t) \otimes e^{+\ii\hat{O} \frac{\chi}{2}} \f{B_\alpha}(t) e^{-\ii\hat{O} \frac{\chi}{2}}
\,.
\end{align}
The tools of Sec.~\ref{SEC:weak_coupling} can now be employed to trace out the bath and derive a generalized QME for $\rhos = \traceB{\rho_{\rm SB}}$, which will have the form $\dot\rhos = {\cal L}(\chi) \rhos$.
The moment-generating function will then be given by Eq.~\eqref{EQ:FCS_MGF}.
Formally, $\mathcal{L}(\chi)$ 
looks similar to the GKSL generator derived of Sec.~\ref{SEC:weak_coupling}, with the exception that the jump terms will now be proportional to
\begin{align}\label{correlation_function_counting_field}
C_{\alpha\beta}^\chi(t_1,t_2) &\equiv \trace{e^{-\ii\hat{O} \frac{\chi}{2}} \f{B_\alpha}(t_1) e^{+\ii\hat{O} \frac{\chi}{2}} 
e^{+\ii\hat{O} \frac{\chi}{2}} \f{B_\beta}(t_2) e^{-\ii\hat{O} \frac{\chi}{2}} \bar{\bar{\rho}}_{\rm B}}\,.
\end{align}
In the other terms of Eq.~\eqref{EQ:FCS_MGF} (where $\rhos$ appears either on the left or right, cf. Eq.~\eqref{EQ:me_bms}), the standard correlation function~(\ref{EQ:res_corr_func}) applies.
For energy in particular ($\hat{O}=\Hb$), Eq.~\eqref{correlation_function_counting_field} simplifies to $C_{\alpha\beta}^\chi(\tau) = C_{\alpha\beta}(\tau-\chi)$.

The FCS formalism counts net transfers from the reservoir, i.e., those crossing the red long-dashed circle in Fig.~\ref{fig:f_sketch_localsetup}.
For the particle current, this typically coincides with the phenomenologic replacements above since typical system-reservoir-couplings do not create particles: Those leaving the reservoir must enter the system, and are thus tracked using either approach. 
The secular approximation (which neglects the energy in the system-reservoir interaction Hamiltonian) enforces an analogous property for the energy. 
However, for non-secular QMEs, the phenomenologic introduction of energy counting fields is usually not obvious. 
In the Redfield Eq.~\eqref{EQ:redfield_dqd}, for instance, a microscopic derivation would yield tilted jump terms of the form
$f_L(\epsilon\pm h) \hat{A} \rhos \hat{B} \to f_L(\epsilon\pm h) \hat{A} \rhos \hat{B} e^{-\ii\chi_L(\epsilon\pm h)}$
and 
$[1-f_L(\epsilon\pm h)] \hat{A} \rhos \hat{B} \to [1-f_L(\epsilon\pm h)] \hat{A} \rhos \hat{B} e^{+\ii\chi_L(\epsilon\pm h)}$, 
counting energies positively when they enter the left reservoir.

Finally, although we focused on QMEs, the FCS is defined via Eq.~\eqref{EQ:hamint_fcs}, and can thus be extracted from other (perturbative and non-perturbative) methods, see e.g. Refs.~\cite{schoenhammer2007a,saito2008a,simine2012a,friedman2018a,kilgour2019a}. 

\subsubsection{Extracting Full Counting Statistics}\label{SEC:fcs_sol}

The computation of \ind{moments} or \ind{cumulants} is generally simpler than obtaining the full distribution~\eqref{EQ:pnoft}:
The $k$-th moment $\langle n^k\rangle$ or the $k$-th cumulant $\langle\langle n^k \rangle\rangle$
are found by differentiating the moment (cumulant) generating function $M(\chi)$ (or $C(\chi)$), defined in~\eqref{EQ:FCS_MGF}; that is, 
$\expval{n^k} \equiv (-\ii \partial_\chi)^k M(\chi,t)|_{\chi=0}$ and
$\langle\langle n^k \rangle\rangle
\equiv (-\ii \partial_\chi)^k C(\chi,t)|_{\chi=0}$\,.
If the steady-state is unique, one may show that 
the long-time cumulant-generating function is given by the \ind{dominant eigenvalue} of $\Liouv(\chi)$~\cite{touchette2009a}
\begin{align}
    C(\chi,t) = \ln M(\chi,t) \stackrel{t\longrightarrow\infty}{\to} \lambda_\text{dom}(\chi) t\,,
\end{align}
which is the eigenvalue with largest real part. 
This eigenvalue obeys $\lambda_\text{dom}(0) = 0$ and corresponds to the stationary state as $\chi\to 0$ (see Sec.~\ref{sec:vectorization} for more on the spectrum of $\Liouv$).
For simple systems, the eigenvalues of ${\cal L}(\chi)$ may be directly accessible.
If not, one may perform a series expansion on the leading coefficients of the characteristic polynomial~\cite{bruderer2014a,friedman2019a}, which allows the lowest cumulants to be extracted analytically.

Often, one is only interested in the first two cumulants, i.e., the average current and its fluctuations. 
These can be extracted without computing $C(\chi,t)$.
Differentiating $M(\chi,t)$ with respect to $\chi$ and using the fact that ${\cal L}(0)$ is traceless, we find that the average current is 
\begin{equation}
    I(t) \equiv \frac{d}{dt} \langle \langle n \rangle\rangle_t = -\ii\partial_\chi \trace{{\cal L}(\chi) \rho(\chi,t)}|_{\chi=0}
    = -\ii \trace{{\cal L}'(0) \rho(t)},
    \nn
\end{equation}
which depends only on $\rho(t)$.
Similarly, the current fluctuations can be written as 
\begin{equation}
     \frac{d}{dt} \langle\langle n^2 \rangle\rangle_t
     = -\trace{{\cal L}''(0)\rho(t)} -2\ii \trace{{\cal L}'(0) \sigma(t)}\,,
\end{equation}
where $\sigma(t) \equiv-\ii \partial_\chi \frac{\rho(\chi,t)}{\trace{\rho(\chi,t)}}\Big|_{\chi=0}$ is an auxiliary quantity.
Importantly,it need not be computed via $\rho(\chi,t)$, but can  be obtained by solving the differential equation
\begin{align}\label{FCS_sigma_differential_equation}
    \dot{\sigma} = -\ii {\cal L}'(0) \rho(t) - I(t) \rho(t) + {\cal L}(0) \sigma(t),
\end{align}
with $\sigma(0)=0$ (which implies that $\trace{\sigma(t)}=0$). 
These results can be generalized to driven systems~\cite{benito2016b,restrepo2019a}.
In the steady-state, Eq.~\eqref{FCS_sigma_differential_equation} reduces to the algebraic equation $\mathcal{L}\sigma_\text{ss} = \ii \mathcal{L}'(0) \rho_\text{ss} + \cur{ss} \rho_\text{ss}$, which can be solved using standard linear algebra routines.\footnote{Since $\mathcal{L}$ is not invertible,  this  has an infinite number of solutions, and one must pick the one satisfying $\trace{\sigma(t)}=0$.}

\subsection{Dephasing/Bulk noises}\label{sec:bulk_noise}

\begin{figure}
    \centering
    \includegraphics[width=\columnwidth]{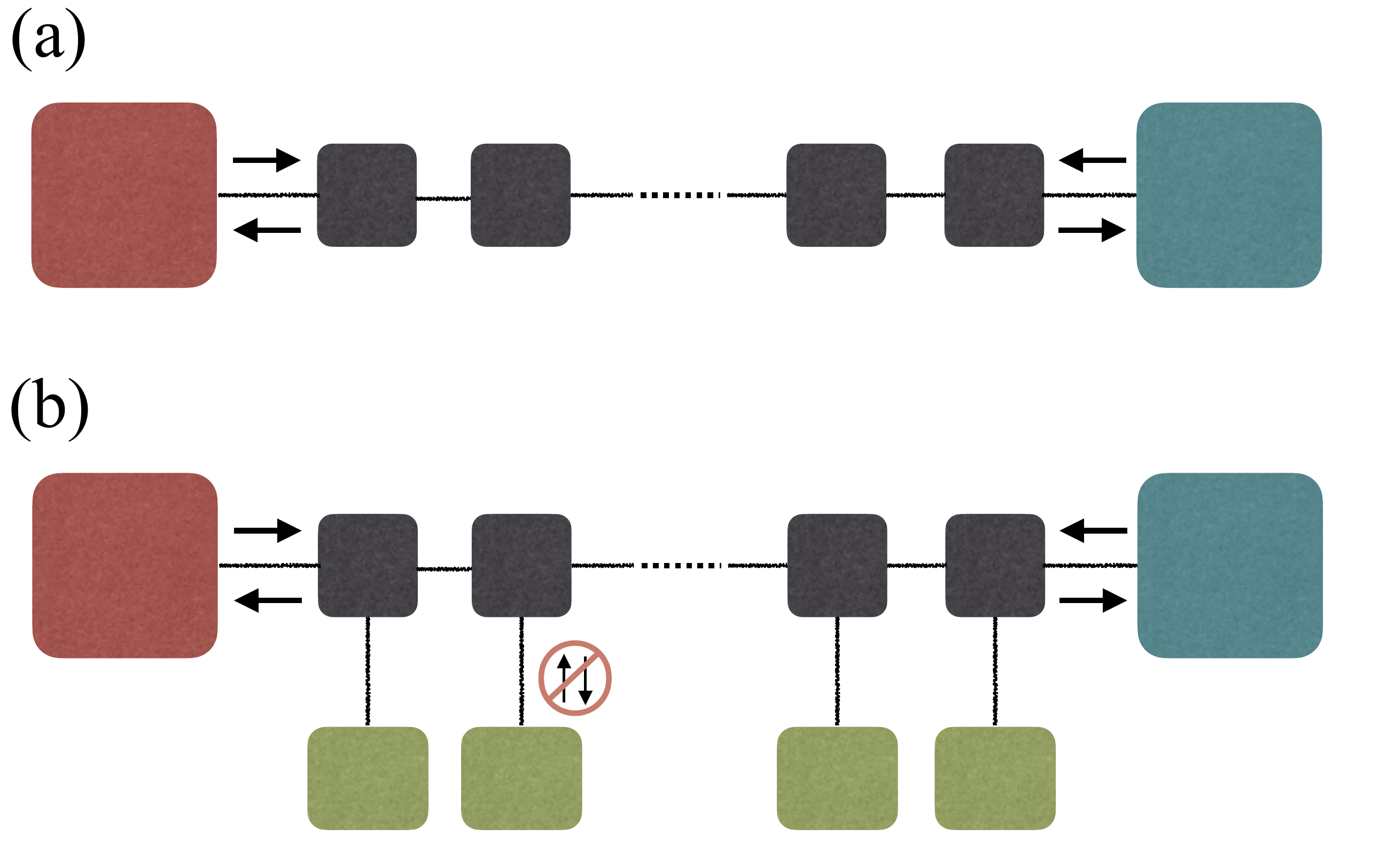}
    \caption{Comparison between (a) a standard boundary-driven model and (b) a model involving bulk noises. In the latter, additional baths act on all sites, adjusted so that no current flows between the bulk baths and the system.}
    \label{fig:bulk_noises}
\end{figure}

%
%
\ind{Bulk noises} refer to a series of methods designed to induce diffusive transport, even in non-interacting chains. 
In the classical literature they are usually termed \ind{self-consistent baths}~\cite{Bolsterli1970}, whereas 
in mesoscopics they are called \ind{B\"uttiker probes}~\cite{buettiker1986a}. 
The boundary-driven community usually employs the quantum information name \ind{dephasing}. 
The idea is illustrated in Fig.~\ref{fig:bulk_noises}: one introduces virtual reservoirs acting on all sites, designed to inject noise without currents.
The interesting physics emerging from this is reviewed in  Sec.~\ref{ssec:dephasing_and_transport}.

For concreteness, consider a 1D tight-binding Hamiltonian~\eqref{tight_binding} with $L$ sites.
Dephasing can be described by a dissipator of the form\footnote{More generally, the term dephasing stands for a GKSL dissipator $\Diss[L]$ with Hermitian jump operator $L$:
Any state satisfying $[\rhos,L]$ will be a fixed point;
they tend to destroy coherences in the eigenbasis of $L$, while leaving the diagonal entries unchanged, increasing the system's entropy~\cite{polkovnikov2011a}.
Eq.~\eqref{bulk_dephasing}, for example, tends to destroy coherences between different positions. 
Using the logarithmic sum inequality, one can show that the entropy increase is monotonic.
}
\begin{equation}\label{bulk_dephasing}
    D_i^\text{deph}(\rhos) = \Gamma \left( c_i^\dagger c_i \rhos c_i^\dagger c_i - \frac{1}{2} \{ (c_i^\dagger c_i)^2, \rhos\}\right),    
\end{equation}
with rate $\Gamma$. 
In spin systems, the dephasing might instead be written as 
\begin{equation}\label{eq:spin_dephasing}
D_i^\text{deph}(\rhos) = \Gamma \left( \sz_i \rho \sz_i - \rho\right)\,, 
\end{equation}
where we used $(\sz_i)^2 = \id$. 
The dissipator~\eqref{bulk_dephasing} induces no particle current at the operator level, although there may still be an energy current~\cite{Mendoza-Arenas2013,Werlang2015}.

The full QME for a system subject to dephasing will have the form
\begin{equation}\label{bulk_M}
    \frac{d \rhos}{d t} = -\im [\Hs,\rhos] + \mathcal{D}_\text{b}(\rhos) + \sum\limits_{i=1}^L D_i^\text{deph}(\rhos)\,, 
\end{equation}
where $\mathcal{D}_\text{b}(\rhos)$ refers generically to the boundary dissipators.
As discussed in Sec.~\ref{sec:non_interacting}, the effects of dephasing on the steady-state are  dramatic~\cite{AsadianBriegel2013,Znidaric2010, Karevski2009}: For any $\Gamma>0$, it will \emph{always} lead to diffusive transport for sufficiently large chains.

An alternative to dephasing are self-consistent reservoirs~\cite{Bolsterli1970}. 
Instead of~\eqref{bulk_dephasing}, one adds local dissipators
$D_i^\text{sc}(\rhos) = \Gamma (1\pm N_i^\text{sc}) D[c_i] + \Gamma  N_i^\text{sc} D[c_i^\dagger]$,
where $+$ ($-$) refers to bosons (fermions).
%
The parameters $N_i^\text{sc}$ are then chosen to match the local occupation of the system, 
$N_i^\text{sc} = \langle c_i^\dagger c_i \rangle$
(hence the name self-consistent).
This enforces zero particle current, since 
$\cur{i}^\text{sc} = tr \big\{c_i^\dagger c_i D_i^{sc}(\rho)\big\} = \Gamma (N_i^\text{sc} - \langle c_i^\dagger c_i \rangle) = 0$.
Although similar in spirit, the two methods are different: self-consistent baths 
only enforce zero current on average,
while dephasing does it at the operator level. 
Thus, while they may predict the same steady-state currents, the steady-state density matrices will be different.
For instance, in non-interacting Hamiltonians, self-consistent baths lead to Gaussian steady-states, while dephasing does not~\cite{Malouf2018}.

\subsection{Other heuristic methods}\label{ssec:heuristics}

We discuss here two additional approaches that are also employed in the literature to deal with strong coupling,
quasistatic equilibrium reservoirs (Sec.~\ref{sec:quasistatic}) and  multi-site baths (Sec.~\ref{sec:znidaricprosenbath}).
Other approaches, such as surrogate Hamiltonians~\cite{BaerKosloff1997, TorronteguiKosloff2016}, are not discussed as they have not yet been applied, to the best of our knowledge, in boundary-driven systems.

\subsubsection{Quasistatic equilibrium reservoirs}\label{sec:quasistatic} 

In most QMEs, the reservoirs are typically kept at constant equilibrium states $\rhob\propto e^{-\beta (\Hb - \mu \Nb)}$.
We can also model \ind{meso-reservoirs} --  i.e., reservoirs of finite size, but which always remain very close to local equilibrium -- 
by fixing its dynamics to be the form~\cite{schaller2014b,amato2020a}
\begin{align}
    \rhob(t) = \frac{e^{-\beta(t)(\Hb-\mu(t) \Nb)}}{\trace{e^{-\beta(t)(\Hb-\mu(t) \Nb)}}}\,,
\end{align}
with time-dependent parameters $\beta(t)$ and $\mu(t)$ determined from  energy and particle currents.
To find these, we write the currents to the bath as
    $\cur{E}^\text{res} \equiv \frac{d}{dt} \int \rho_\text{d}(\omega) \omega n_\pm(\omega,t) d\omega$ and
    $\cur{N}^\text{res} \equiv \frac{d}{dt} \int \rho_\text{d}(\omega) n_\pm(\omega,t) d\omega$, 
where $\rho_\text{d}(\omega)$ denotes the \ind{density of states} of the (fermionic or bosonic) meso-reservoir and $n_\pm(\omega,t)\equiv [e^{\beta(t)[\omega-\mu(t)]}\pm 1]^{-1}$ are Fermi-Dirac or Bose-Einstein distributions, respectively.
Following the chain rule, the time derivative will generate terms proportional to $\dot\mu$ and $\dot\beta$, with the prefactors interpreted as a
\ind{heat capacitance} and \ind{charge capacitance}.
Imposing energy and matter conservation, we can then equate 
$\cur{E(N)}^\text{res}$ with minus the currents entering the system [e.g. Eq.~\eqref{Currents_H_evo}], resulting in nonlinear 1st order differential equations for $\beta(t)$ and $\mu(t)$.
For $N$ meso-reservoirs this will yield $2N$ coupled equations.
Solving these equations numerically thus allows one to track the evolution of quasistatic equilibrium reservoirs.


\subsubsection{Multi-site GKSL baths for non-integrable spin chains}
\label{sec:znidaricprosenbath}

\begin{figure}
    \centering
    \includegraphics[width=\columnwidth,clip=true]{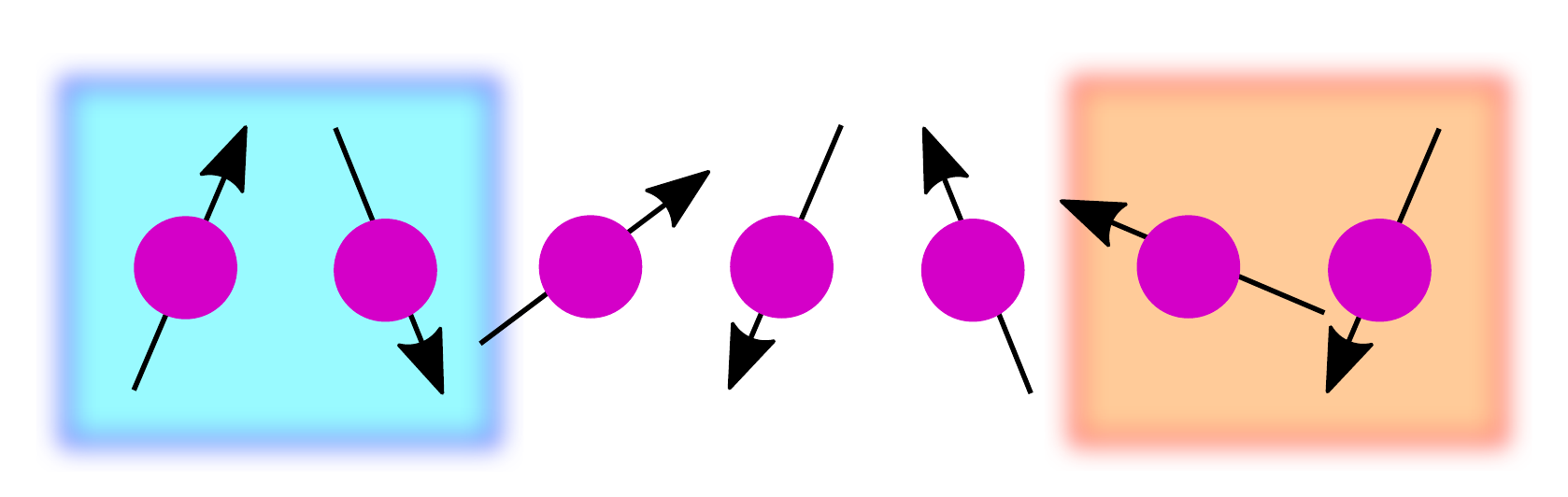}
    \caption{Depiction of the multi-site GKSL thermal baths model described in Sec.~\ref{sec:znidaricprosenbath}. The last two sites at each edge are in contact with a bath which drive them towards \gtl{different target states.}
    } 
    \label{fig:ZnidaricProsenBath}
\end{figure}

LME dissipators lack information on interactions within the system. 
Motivated by this,~\cite{Prosen2009} considered non-integrable spin chains where entire chunks (instead of single sites) were coupled to the baths (Fig.~\ref{fig:ZnidaricProsenBath}). 
The jump operators were chosen so as to push these chunks to specific thermal states. 
This cooperates with the non-integrability of the chain, yielding an accurate model of heat transport.         

To better understand the idea, we focus on two-site dissipators. 
They can be parametrized as
$\Diss (\rho) = \sum_{i,j} \gamma_{ij}\big( \Gamma_i\rho \Gamma_j^\dagger -\frac{1}{2}\left\{ \Gamma_j^\dagger \Gamma_i, \rho \right\}\big)$,
with jump operators $\Gamma_j$ acting only on the two sites involved.
For spin 1/2 systems, the indices $i,j$  take $16$  values since, in each site, we can act with $\sigma^\pm=(\sx\pm\ii\sy)/2$, 
$\su =(\id+ \sz)/2$, $\sd =(\id- \sz)/2$.
The matrix $\gamma_{ij}$ thus has $256$ entries, which can be adjusted to impose that the fixed point of $D(\rho)$ is a thermal state $\rho_T$ on the two sites (due to the other Hamiltonian terms, the steady-state of the whole chain will generally differ).
Since $\rho_T$ only has $16$ entries, there is significant freedom in choosing the $\gamma_{ij}$.
Ref.~\cite{Prosen2009} chose them so that all excited modes decayed at the same rate.
In~\cite{PalmeroPoletti2019} the authors chose rates satisfying detailed-balance, which improves the performance by producing states closer to the thermal one at lower temperatures. 
An extension to baths coupling to more than 2 sites was studied in~\cite{Guimar2016}.
 
The target state $\rho_T\propto e^{-\beta_T \Htarget}$ is determined by $\beta_T$ and the target Hamiltonian $\Htarget$, for which two choices have been considered. 
Ref.~\cite{Mendoza-Arenas2014a} took $\Htarget$ as the reduced Hamiltonian of the two sites, including their interactions, but neglecting any couplings to other sites. 
Alternatively,~\cite{ZnidaricRossini2010,Znidaric2011b, MendozaArenasScardicchio2019} took $\rho_T$ as the reduced density matrix of the two sites, if the entire chain was in equilibrium, $\rho_T\propto \ptrace{3\dots L}{\exp[-\beta (\Hs-\mu M)]}$ where $\mu$ is the chemical potential and $M$ is the total magnetization.

To evaluate the consistency of the method one may couple the chain to only one dissipator, or to two at the same temperature, and check whether
the steady-state behaves like a thermal state
%
The first thing to point out is that $\rhoss$ does not typically approach a thermal state at $\beta_T$, but at some other $\beta_{\rm S} < \beta_T$. 
If $\beta_T$ is small enough, the system will thus approach the thermal state $\propto \exp(-\beta_{\rm S} \Hs)$.
One can then compute expectation values $\expval{O}_{S}$ of different observables, and check whether they are consistent with those at the corresponding thermal state $\expval{O}_{\beta_{\rm S}}$.
This defines an effective temperature $\beta_O$ for each $O$.
If $\rhos$ indeed approaches a thermal state, then $\beta_O$ will be the same for all observables.
%
%
Using this criterion,~\cite{Znidaric2011b,PalmeroPoletti2019},  found that the dispersion in  $\beta_O$'s was smaller at high temperatures, and when they had smaller support (e.g. one-site observables rather than two-sites).    

One may also consider the trace distance [Eq.~\eqref{trace_distance}]
between the reduced density matrix of a portion of $\rhoss$, and that of the same portion in  a thermal state~\cite{Mendoza-Arenas2014a, ZanociSwingle2021}.
This measure is  meaningful
since it bounds the differences between expectation values of any operator with finite eigenvalues~\cite{Mendoza-Arenas2014a}.

The above methodology can be implemented with tensor networks~\cite{Prosen2009, ZnidaricRossini2010, Znidaric2011b, Mendoza-Arenas2014a, PalmeroPoletti2019, MendozaArenasScardicchio2019} [Sec.~\ref{sec:tensor_networks}], resulting in a powerful approach to study heat transport in strongly interacting quantum system.
This has been used to study transport in chains,  with and without disorder, which we review in Sec.~\ref{sec:properties}. 
For instance, Fig.~\ref{fig:MBL_Merged}(b) was obtained with this tool.  
Unfortunately, the method is only accurate at relatively large temperatures. 
%
However, in this regime, and for large strongly correlated systems, it is more efficient than other methods like GMEs or LMEs:
The former requires diagonalizing $\Hs$ and the latter does not faithfully represent thermal baths at strong coupling. 
Multi-site baths also require a smaller number of sites than a thermo-field plus star-to-chain mapping (Sec.~\ref{sec:star_to_chain_thermofield}).
%

\section{\label{sec:methods}Methods for boundary-driven open systems}


This section provides methods specifically designed for boundary-driven problems.
We will not review methods designed for open quantum systems, but which have not been applied to boundary driven problems. 
These include
variational and cluster mean-field methods~\cite{Weimer2015, JinRossini2016}, quantum Monte Carlo~\cite{NagySavona2018}, corner-space renormalization~\cite{FinazziCiuti2015},
and neural-network ansatzes~\cite{NagySavona2019, HartmannCarleo2019, VicentiniCiuti2019, YoshiokaHamazaki2019, LuoClark2020}. 
We will also not discuss microscopically exact approaches, like non-equilibrium Green's functions~\cite{economou2006,wang2014a, DharHanggi2012,NikolicThygesen2012, ZimbovskayaPederson2011, Aeberhard2011, ProciukDunietz2010, MeirWingreen1992, CaroliSaint-James1971, HaugJauho2008}, hierarchies of equations of motion~\cite{tanimura1989a,tanimura1990a} or path integrals~\cite{AleinerGlazman2002, MuhlbacherRabani2008, SegalReichman2010, HutzenThorwart2012}.

\subsection{Vectorization}  \label{sec:vectorization}

Irrespective of the type of system-bath interaction, many boundary-driven problems are described by linear time-local evolution equations, of the form 
\begin{equation}\label{vec_M}
    \frac{d\rho}{dt} = \Liouv (\rho)\,, 
\end{equation}
where $\Liouv$, called the \ind{Liouvillian}, is a linear superoperator.
Since $\rho$ is an operator, $\Liouv(\rho)$ will involve left and right multiplication. 
Notwithstanding, this equation is in the exact same spirit as a linear matrix-vector equation like $dx/dt = A x$.

Since matrix-vector equations are standard in scientific computing, it is often  convenient to recast Eq.~\eqref{vec_M} in this form. 
This can be done introducing a \ind{vectorization} operation, which maps a matrix $M$ into a vector $\vvec(M) \equiv \vec{M}$ (both notations will be used interchangeably) by stacking the columns~\cite{Vectorization}; e.g.,
\begin{equation}\label{vec_example}
\vvec \begin{pmatrix}
a & b \\ c & d \end{pmatrix} 
=
\begin{pmatrix} a \\ c \\ b \\ d \end{pmatrix}.
\end{equation}
The most important property of vectorization is the following:
Given arbitrary matrices $A$, $B$, $C$~\footnote{If one defines vectorization by stacking rows, instead of columns, then one should use instead $\vvec(ABC)=(A \otimes C\trans) \vvec(B)$.} 
\begin{equation}\label{vec_property}
    \vvec(ABC) = (C\trans \otimes A) \vvec(B)\,.
\end{equation} 
This will allow us to write the superoperator $\Liouv$ as a matrix $\hat{\Liouv}$, and hence recast~\eqref{vec_M} as 
\begin{equation}\label{vec_M_vec}
    \frac{d\vec{\rho}}{d t} = \hat{\Liouv} \vec{\rho}.
\end{equation}
For example, take  $ \Liouv(\rho) = -\im [H,\rho] + \sum_k D[L_k](\rho)$.
%
Terms such as $H\rho$ can be written as $H\rho \id$, so that 
$\text{vec}(H\rho) = (\id\otimes H) \vec{\rho}$.
Proceeding similarly with the other terms yields the vectorized Liouvillian
\begin{IEEEeqnarray}{rCl}
\label{vec_vectorized_L}
\hat{\mathcal{L}} &=& -\im (\id\otimes H - H\trans \otimes \id) 
\\[0.2cm]&&
+ \sum\limits_k  \Bigg[L_k^* \otimes L_k - \frac{1}{2} \id \otimes L_k^\dagger L_k - \frac{1}{2} (L_k^\dagger L_k)\trans \otimes \id\Bigg]\,,
\nonumber
\end{IEEEeqnarray}
which is now a matrix of dimension $d^2\times d^2$, instead of a superoperator.

Vectorization can also be viewed as the operation taking
\begin{equation}\label{vec_def_Hilbert_space}
    |i\rangle\langle j | \to |j\rangle\otimes |i\rangle,
\end{equation}
known as the \ind{Choi-Jamiolkowski isomorphism}~\cite{Choi,Jamiolkowski}.
As a consequence, a  $d$-dimensional density matrix 
$ \rho = \sum_{ij} \rho_{ij} |i\rangle\langle j |$,
is vectorized to 
$\vvec(\rho) = \sum_{ij} \rho_{ij} |j\rangle\otimes |i\rangle$, of length $d^2$.
The inner product between two vectorized operators is related to the \ind{Hilbert-Schmidt inner product}  
\begin{equation}\label{vec_Hilbert_Schmidt}
    \vvec(A)^\dagger \vvec(B) =\tr(A^\dagger B).
\end{equation}
Accordingly, the normalization condition $\tr(\rho)=1$ is mapped into 
$\vvec(\id)^\dagger \;\vvec(\rho) = 1$,
where 
$\vvec(\id) = \sum_{i} |i\rangle\otimes |i\rangle$
is the vectorized identity.

\subsubsection{Spectral properties of the Liouvillian}\label{sec:spec_prop_Liouv}

Vectorization converts the master Eq.~\eqref{vec_M} into the standard matrix-vector equation~\eqref{vec_M_vec} for $\vec{\rho}$. 
The formal solution is then simply 
\begin{equation}\label{vec_M_vec_sol}
    \vec{\rho}(t) = e^{\hat{\Liouv} t} \vec{\rho}(0)\,,
\end{equation}
which naturally leads to the problem of computing the exponential of the
non-Hermitian matrix $\hat{\Liouv}$. 
To gain some intuition, let us first assume that $\hat{\Liouv}$ is diagonalizable.
Being non-Hermitian, however, it will have different \ind{right and left eigenvectors} $\vec{x}_\alpha$ and $\vec{y}_\alpha$:
\begin{equation}\label{vec_eigenvectors}
    \hat{\Liouv} \vec{x}_\alpha = \lambda_\alpha \vec{x}_\alpha, \qquad \qquad 
    \vec{y}_\alpha^{\;\dagger} \hat{\Liouv} = \lambda_\alpha \vec{y}_\alpha^{\;\dagger}\,,
\end{equation}
associated with the eigenvalue $\lambda_\alpha$ (sometimes called \ind{rapidities}~\cite{Kimura2017}). 
If $S$ is a matrix with columns $\vec{x}_\alpha$, then $\vec{y}_\alpha^\dagger$ are the rows of $S^{-1}$. 
Hence 
$\vec{y}_\alpha^\dagger \vec{x}_{\alpha'} = \delta_{\alpha,\alpha'}$,
and 
\begin{equation}\label{vec_L_decomp}
    \hat{\Liouv} = \sum\limits_\alpha \lambda_\alpha \vec{x}_\alpha \vec{y}_\alpha^{\; \dagger} = S \Lambda S^{-1}, 
\end{equation}
where $\Lambda = \text{diag}(\lambda_1, \lambda_2, \ldots)$.
The rapidities $\lambda_\alpha$ are in general complex.
In particular for GKSL generators $\hat{\Liouv}$ one finds that they have 
non-positive real parts (indicating a decay towards the steady-state) and come in complex conjugate pairs (if $\lambda_\alpha$ is an eigenvalue, then so is $\lambda_\alpha^*$)~\cite{albert2014a}.
In fact, the pairing is a necessary condition for $\vec{\rho}(t)$ to be Hermitian at all $t$.

The steady-state $\rhoss$ is the \ind{fixed point} of Eq.~\eqref{vec_M_vec}:
\begin{equation}\label{vec_steady_state_vectorized_equation}
    \hat{\Liouv}\rhossvec = 0\,. 
\end{equation}
As shown in~\cite{Evans1977,Evans1979,Frigerio1978,Baumgartner2008}, GKSL equations always have \emph{at least} one fixed point $\rhoss$.
However, if some rapidity is purely imaginary, the system may never relax towards it, but instead oscillate indefinitely in a dark subspace~\cite{DAbbruzzo2021,Buca2021}. 
%
Moreover, $\rhoss$ may not be  unique~\cite{Buca2012,BucaJaksch2019,ManzanoHurtado2014,Thingna2021,schaller2010b},
although this is often the case~\cite{nigro2019a,Evans1977}.

For concreteness, we  henceforth assume that $\rhoss$ is unique. 
Eq.~\eqref{vec_steady_state_vectorized_equation} then shows  that \emph{$\rhossvec =\vec{x}_0$ is the  right eigenvector of $\hat{\Liouv}$ with eigenvalue $\lambda_0 = 0$.}
In addition, since the dynamics is trace preserving, 
it must also follow that 
$\vvec(\id)^\dagger \hat{\Liouv} = 0$.
Thus, $\vvec(\id) = \vec{y}_0$ is the left eigenvector of $\hat{\Liouv}$ with eigenvalue $\lambda_0 = 0$.
%
Assuming Eq.~\eqref{vec_L_decomp} holds,
one may write
\footnote{
\gtl{If $\hat{\mathcal{L}}$ is not diagonalizable,
we get instead 
$e^{\hat{\Liouv} t} = \sum_\alpha e^{\lambda_\alpha t} \left(\sum_{k=0}^{N_\alpha-1} t^k {\cal A}_{\alpha k}\right)$,
where $N_\alpha$ denotes the multiplicity of eigenvalue $\lambda_\alpha$, and
${\cal A}_{\alpha k}$ are unknown matrices which follow from the Jordan-block decomposition of $\hat{\Liouv}$.
They can be determined by differentiating both sides of this equation, which generates a set of equations for jointly determining  $\lambda_\alpha$ and ${\cal A}_{\alpha k}$.}
}
$e^{\hat{\Liouv} t} = \sum_\alpha e^{\lambda_\alpha t} \vec{x}_\alpha \vec{y}_\alpha^{\;\dagger}$,
so that Eq.~\eqref{vec_M_vec_sol} becomes
\begin{equation}\label{vec_sol_12341231}
\vec{\rho}_t = \sum_\alpha c_\alpha e^{\lambda_\alpha t} \vec{x}_\alpha\,,    
\end{equation}
where $c_\alpha = \vec{y}_\alpha^{\; \dagger} \vec{\rho}_0$ are determined by the initial conditions.
The evolution will thus be a linear combination of the right-eigenvectors $\vec{x}_\alpha$, each with weight $c_\alpha$ and evolving in time with an exponential dependence $e^{\lambda_\alpha t}$.
The spectral properties thus allow us to classify the decay modes $\alpha$ into oscillatory ($\text{Re}(\lambda_\alpha) = 0$), purely decaying ($\text{Im}(\lambda_\alpha) = 0$), decaying spirals ($\text{Re}(\lambda_\alpha) < 0$, $\text{Im}(\lambda_\alpha) \neq 0$) and steady states ($\lambda_\alpha=0$)~\cite{albert2014a}.

We can single out $\lambda_0 = 0$ in Eq.~\eqref{vec_sol_12341231}. 
Due to normalization, $c_0 =\trace{\rho_0} = 1$, so that
\begin{equation}\label{vec_M_sol_spectral}
    \vec{\rho_t} = \rhossvec + \sum\limits_{\alpha\neq 0} e^{\lambda_\alpha t} c_\alpha \vec{x}_\alpha\,.
\end{equation}
If $\text{Re}(\lambda_\alpha) < 0$ for all $\alpha \neq 0$, 
the system will relax always to $\rhoss$ as $t\to \infty$, for any initial condition.

The full Liouvillian spectrum can be computed in the case of pure loss models~\cite{Torres2014,Nakagawa2021}, which was used  in Ref.~\cite{Buca2020}  to study the Liouvilian spectrum of the XXZ Hamiltonian under pure loss processes.
Similar methods were used in~\cite{MedvedyevaProsen2016,Ziolkowska2020,Essler2020} to study models under the presence of bulk noises (Sec.~\ref{sec:bulk_noise}).

The computation of the steady-state [Eq.~\eqref{vec_steady_state_vectorized_equation}] is equivalent to finding the eigenvector of  $\hat{\mathcal{L}}$ with zero eigenvalue. 
As this is much simpler for Hermitian problems, in~\cite{CiracBanuls2015}, the authors introduced a  workaround by solving instead 
\begin{equation}\label{vec_LdL}
    \hat{\mathcal{L}}^\dagger \hat{\mathcal{L}} \rhossvec = 0\,. 
\end{equation}
Clearly, the eigenvectors of $\hat{\mathcal{L}}^\dagger \hat{\mathcal{L}}$ are different from those of $\hat{\mathcal{L}}$, but the one with $\lambda_\alpha=0$ is the same, since $\hat{\mathcal{L}}\rhossvec = 0$  implies $\hat{\mathcal{L}}^\dagger \hat{\mathcal{L}}\rhossvec = 0$.
The advantage is that $\hat{\mathcal{L}}^\dagger \hat{\mathcal{L}}$ is 
also positive semi-definite.
This is thus a variational problem, where one searches for the eigenvector of the ``effective Hamiltonian''   $\hat{\mathcal{L}}^\dagger \hat{\mathcal{L}}$ with smallest eigenvalue. 
Unlike in the standard variational principle, here the smallest eigenvalue is explicitly known $\lambda_\alpha=0$, such that the distance from zero can be used as a measure of convergence. 

This method is particularly useful when implemented with tensor network
(Sec.~\ref{sec:tensor_networks}).
There are however two main disadvantages: First, even if $\hat{\mathcal{L}}$ is  local (e.g. nearest-neighbor interactions), $\hat{\mathcal{L}}^\dagger \hat{\mathcal{L}}$ will in general be non-local~\cite{Gangat2017}.
Second, $\hat{\mathcal{L}}^\dagger \hat{\mathcal{L}}$ is often ill-conditioned. 
In fact, if the  relaxation gap of $\hat{\mathcal{L}}$ is $\lambda_1$, the first excited state of  $\hat{\mathcal{L}}^\dagger\hat{\mathcal{L}}$ is $\sim|\lambda_1|^2$, which can be very small.

Inverting a matrix is typically a lot simpler than exponentiating it. For this reason, \ind{Laplace-transform} methods can often provide a convenient advantage.
Laplace transforming Eq.~\eqref{vec_M_vec} leads to
$\vec{\rho}(z) = (z - \hat{\Liouv})^{-1} \vec{\rho}(0)$, which has poles precisely at $z_\alpha = \lambda_\alpha$.
Fortunately, steady-state expectation values can be obtained via the final value theorem, 
\begin{equation*}
    \expval{A}_{\rm ss} \equiv \lim\limits_{t\to\infty} \trace{A \rho(t)} =
    \lim\limits_{z\to 0}~z~\vvec(\id)^\dagger \Big[(A\trans \otimes \id) (z- \hat{\Liouv})^{-1} \Big] \vec{\rho}(0)\,.   
\end{equation*}
In addition,  if one can decompose the Liouvillian $\hat{\Liouv}=\hat{\Liouv}_0+\hat{\Liouv}_1$, into a simple part $\hat{\Liouv}_0$ and a perturbation $\hat{\Liouv}_1$, it is possible to set up a series expansion of the solution 
as
$(z - \hat{\Liouv})^{-1} = \sum_{n=0}^\infty \left[{\cal P}_0(z) \hat{\Liouv}_1\right]^n {\cal P}_0(z)$ where 
${\cal P}_0(z) = (z- \hat{\Liouv}_0)^{-1}$.
This is the analogue of Eq.~(\ref{eq:dyson_series}) in the $z$-domain.
When the structure of $\hat{\Liouv}_0$ is simple, it is often possible to transform back the solutions to the time domain.
Other perturbative methods are reviewed in Sec.~\ref{sec:perturbative}.


The vectorization recipe of Eqs.~\eqref{vec_example} or~\eqref{vec_def_Hilbert_space} chooses a specific basis to represent $\hat{\Liouv}$ and other operators.
For instance, in the case of $2\times 2$ matrices this basis is
\begin{IEEEeqnarray*}{rCl}
\nonumber
    q_1 = \begin{pmatrix} 1 & 0 \\ 0 & 0 \end{pmatrix},
    &\quad& 
    q_2 = \begin{pmatrix} 0 & 0 \\ 1 & 0 \end{pmatrix}
    \quad
    q_3 = \begin{pmatrix} 0 & 1 \\ 0 & 0 \end{pmatrix},
    \quad
    q_4 = \begin{pmatrix} 0 & 0 \\ 0 & 1 \end{pmatrix},
\nonumber    
\end{IEEEeqnarray*}
which, when vectorized, leads to the four basis vectors $\vec{q}_1 = (1,0,0,0)$, $\vec{q}_2 = (0,1,0,0)$, etc.
With the inner product~\eqref{vec_Hilbert_Schmidt}, however, one may generalize this for arbitrary bases of operators. 
For instance, another natural choice are the Pauli matrices (supplemented with $\id$), 
which are orthogonal with respect to~\eqref{vec_Hilbert_Schmidt} (but not ortho\emph{normal}).
It can also be useful to split populations (diagonals) and coherences (off-diagonals) in some chosen basis, e.g. in models where their evolutions are decoupled.
Given a set of orthogonal matrices $\{q_i\}$,
the Liouvillian can always be written as 
\begin{equation}
    \hat{\Liouv} = \sum_{ij} \Liouv_{ij} \vec{q}_i\; \vec{q}_j^\dagger,
    \qquad 
    \Liouv_{ij} 
    = \frac{\tr\big\{ q_i^\dagger \Liouv(q_j)\big\}}{\tr(q_i^\dagger q_i) \tr(q_j^\dagger q_j)}\,.
\end{equation}
%
For spin chains, a natural choice of basis is the set of tensor products  $\sigma^{i_1}\otimes \sigma^{i_2}\otimes \ldots \otimes \sigma^{i_N}$ with $i_n \in \{0,x,y,z\}$.
In this case, one also has the freedom of choosing how to order the corresponding vectorized space. 
In fact, the standard vectorization recipe in Eq.~\eqref{vec_def_Hilbert_space} is not good, as it leads to long-range interactions. 
Alternative orderings were discussed  in~\cite{PeresCasagrandeLandi2020}, and may lead to significant numerical advantages in tensor network simulations (Sec.~\ref{sec:tensor_networks}).


\subsubsection{Symmetries}\label{ssec:symmetries} 

Symmetries can significantly reduce  the vector space of a computation. 
For instance, in the unitary (Hamiltonian) dynamics of a spin chain with number conservation, such as the XXZ model with $L$ spins [Eq.~(\ref{XXZ})], one does not need to store $2^L$ entries, 
but only $\binom{L}{N_u} = L!/ [N_u! (L-N_u)!]$ where $N_u$ is the number of spins up. 
%
As discussed in~\cite{Buca2012} for GKSL QMEs, there are two main types of symmetries: A ``weak'' symmetry is a unitary $S$ obeying 
\begin{align}
    S\Liouv(\rho)S^\dagger = \Liouv(S\rho S^\dagger).
\end{align} 
Conversely, a ``strong'' symmetry is one in which
\begin{align}
    [S, H] = 0, \; \text{and} [S,L_k]= 0 ,
\end{align}
where $H$ is the Hamiltonian and $L_k$ are the jump operators.
Strong symmetries imply the weak one. 
As shown in~\cite{Buca2012}, for the strong symmetry there is at least one steady state for each symmetry sector.
The evolution of elements within a symmetry sector always remain in that sector, allowing simulations to be performed with a smaller vector space. 
For instance, if both the Hamiltonian and  jump operators preserve the total magnetization (e.g. an XXZ chain with dephasing), there is a strong symmetry and hence multiple steady states. 
The dynamics within a subspace with $N_u$ spins up is then represented by a density matrix with $\{(L!)/[(N_u!(L-N_u)!)]\}^2$ elements, as in the unitary case. 
 

Typically, however, the total magnetization or the total number of particles is not conserved due to the dissipators at the boundaries [c.f. Eq.~\eqref{preamble_Lindblad_dissipator_sites}].
In this case only a weak symmetry applies, and the steady-state is unique.
If one is interested solely in $\rhoss$, it may suffice to study much smaller matrices.
To see why, consider the XXZ chain under the LME~\eqref{preamble_Lindblad_dissipator_sites}.
Instead of using the Pauli basis $\ket{\sigma_1,\ldots,\sigma_L}$, we bundle the states in the form $\ket{\vec{m}_M, M}$, where $M$ is the total magnetization and $\vec{m}_M$ labels all states with magnetization $M$. 
We can then decompose
\begin{equation}\label{vec_symmetries_rho_general}
    \rho = \sum\limits_{M,M'} \sum\limits_{\vec{m}_M, \vec{m}_{M'}'} \rho_{M,M'}^{\vec{m}_M, \vec{m}_{M'}'} \ket{\vec{m}_M, M} \bra{\vec{m}_{M'}', M'}\,.
\end{equation}
Terms such as $\sm_i \rho \splus_i$ will generate transitions between different $M,M'$.
But they do so by the same amount, i.e.,  $M \to M-1$ and $M' \to M'-1$. 
The difference $M-M'$ is hence preserved, which defines the different symmetry sectors. 
Furthermore, if $\rhoss$ is unique, it must belong to a sector with  $M-M'=0$, since these are the only ones with unit trace. 
Hence, $\rhoss$ lives in a space with dimension $(2L)!/[(L)!(L)!]$.
Examples of works which rely heavily o symmetries are~\cite{SaProsen2020, GuoPoletti2019, GuoPoletti2017b, Znidaric2013, Znidaric2013b, LeePoletti2021b}. 
Some  properties of the emerging currents are discussed in Sec.~\ref{sec:properties}. 
Another example, which can help to build some intuition, is superradiant decay and other Dicke-like collective processes~\cite{gross1982a,Garraway2011}, where it is much more convenient to use the large-spin representation instead of local Pauli matrices.

\subsection{\label{sec:non_interacting}Non-interacting systems and Lyapunov equation }

We consider here a generic fermionic or bosonic system with $L$ sites,  and annihilation operators $a_i$. 
The enormous complexity of boundary-driven systems lies in the exponentially large Hilbert space dimension. 
In some cases, however, the full dynamics is captured 
entirely by pairwise correlations, such as $\langle a_j^\dagger a_i \rangle$.
This will be the case for QMEs such as~\eqref{EQ:GKSL}, where the Hamiltonian is at most quadratic in $a_i$ and $a_i^\dagger$, and the jump operators are at most linear.
Systems of this form are said to be \indTwo{Gaussian}{Gaussian systems}, or \indTwo{non-interacting}{non-interacting systems}. 
It is convenient to organize the correlations into an $L\times L$ \ind{covariance matrix (CM)}\footnote{
Gaussian states (assuming $\langle a_i \rangle = 0$) have the form 
$\rho = e^{-\sum_{i,j} M_{ij} a_i^\dagger a_j}/Z$ where $M$ is a matrix related to $C$ by $C = (e^{M} \pm 1)^{-1}$ for fermions/bosons respectively. Moreover, $Z = \det(1 \pm e^{-M})^{\pm 1}$.
}
\begin{equation}\label{nonInt_CM_simple}
     C_{ij} = \langle a_j^\dagger a_i \rangle - \langle a_j^\dagger \rangle \langle a_i \rangle\,,
\end{equation}
where the choice of ordering $a_j^\dagger a_i$, instead of $a_i^\dagger a_j$, is to simplify subsequent calculations.
In some cases, a more general definition is necessary, which includes terms such as $\langle a_i a_j \rangle$. 
This will be discussed further below.
When the Hamiltonian is quadratic and the jump operators are linear, $C$ will satisfy 
the \gtl{\ind{Lyapunov equation}}
\begin{equation}\label{nonInt_lyap}
    \frac{dC}{dt} = -(W C + C W^\dagger) + D\,,
\end{equation}
where $W$ and $D$ are matrices that depend on the parameters of the QME (see below for an example).
The  solution is 
\begin{equation}
    C(t) = e^{-W t} C(0) e^{- W^\dagger t} + \int\limits_0^t dt'~e^{-W(t-t')} D e^{- W^\dagger (t-t')}\,.
\end{equation}
Convergence to a steady-state will thus depend on whether the eigenvalues of $W$ have positive real parts. 
When this is the case, the steady-state will satisfy 
\begin{equation}
    W C + C W^\dagger = D\,.
\end{equation}
Routines for solving this equation can be found in most scientific computing libraries, allowing one to study sizes as large as $L = 10000$~\cite{BaterlsStewart1972, GolubVanLoan1979}.    

The Lyapunov equation is not restricted to GKSL dynamics, also appearing in other non-interacting open systems~\cite{Purkayastha2022}.
To provide a concrete example, consider a tight-binding Hamiltonian $\Hs = \sum_{ij} h_{ij} a_i^\dagger a_j$ evolving according to the GKSL master equation 
\begin{IEEEeqnarray}{rCl}
    \frac{d\rhos}{dt} = -\im [\Hs,\rhos] + \sum\limits_{i,j} 
     \gamma_{ji}^- \bigg[a_i \rhos a_j^\dagger - \frac{1}{2} \{a_j^\dagger a_i, \rhos\} \bigg]\nonumber \\[0.2cm]
+ \gamma_{ij}^+ \bigg[ a_i^\dagger \rhos a_j - \frac{1}{2}\{a_j a_i^\dagger, \rhos\}\bigg],
\label{nonInt_M_simp}
\end{IEEEeqnarray}
for Hermitian positive semi-definite matrices $\gamma^+$ and $\gamma^-$, with entries $\gamma_{ij}^+$ and $\gamma_{ij}^-$ (mind the reverse ordering of the indices in $\gamma^-$).
To derive the Lyapunov equation it suffices to assume
$\langle a_i \rangle = 0$, since the 
form of the final result will not be affected by this term.
We thus write  $dC_{ij}/dt = \tr\{ a_j^\dagger a_i \;d\rhos/dt\}$ and use Eq.~\eqref{nonInt_M_simp}, together with the fermionic or bosonic algebras.
As a result, we find precisely Eq.~\eqref{nonInt_lyap}, with 
\begin{equation}
W = \im h + (\gamma^- \pm \gamma^+)/2\,, \qquad 
D = \gamma^+,
\end{equation}
where the upper sign is for fermions and the lower for bosons.
In LMEs,  $\gamma^\pm$ are diagonal (Sec.~\ref{SEC:local_gksl_master_equation}): 
For fermions,  $\gamma_{ii}^+ = \gamma_i f_i$ and $\gamma_{ii}^- = \gamma_i(1-f_i)$, with $f_i$ being the Fermi-Dirac distributions.
For bosons, one has instead $\gamma_{ii}^+ = \gamma_i N_i$ and $\gamma_{ii}^- = \gamma_i(1+N_i)$, where
$N_i$ are now the Bose-Einstein distributions. 
The Lyapunov matrices therefore become identical for fermions and bosons: $W = \im h + \gamma/2$ and $D_{ii} = \gamma_i n_i$,  with $n_i$ being either $f_i$ or $N_i$, and $\gamma = \text{diag}(\gamma_i)$.

Next suppose that one has only nearest-neighbor hopping in $\Hs$, so that $h_{ii} = \epsilon$ and $h_{i,i\pm 1} = -J$ (c.f. Eq.~\eqref{tight_binding}). 
Moreover, suppose that the system is coupled to two LME baths in the first and last sites, with equal coupling strengths $\gamma$, but different $n_1$, $n_L$.
As shown in~\cite{Karevski2009,Znidaric2010b,AsadianBriegel2013}, the solution of Eq.~\eqref{nonInt_lyap} will be a 
\ind{Toeplitz matrix}, with diagonals 
$\langle a_j^\dagger a_j \rangle = (n_1+n_L)/2 + (\gtl{\gamma}/2J) x(\delta_{j,L}-\delta_{j,1})$
and a constant first off-diagonal 
$\langle a_{j+1}^\dagger a_j \rangle = \im x$, where 
$x=\gtl{\gamma} J (n_L - n_1)/(\gtl{\gamma}^2 + 4 J^2)$ (all other entries being zero).
The current between sites $i$ and $i+1$ is $\Jpart = \im J \langle a_{i+1}^\dagger a_i - a_i^\dagger a_{i+1} \rangle = 2x$.
This is the solution discussed in Sec.~\ref{sec:lme_stage}, and provides the prototypical example of ballistic transport, where 
$\Jpart$ is independent of $L$, and the occupations are uniform along the chain, except in the first and last sites.

The example above led to a closed set of equations for $C_{ij}$.
This is no longer the case when the Hamiltonian contains pairing/squeezing terms $a_i^\dagger a_j^\dagger$, or when the dissipators contain terms such as $a_i^\dagger \rhos a_j^\dagger$ and $a_i \rhos a_j$.
Examples include the XY spin chain, or squeezed baths, respectively.
In these cases one may use instead Majorana/quadrature operators 
$q_i = a_i + a_i^\dagger$ and $p_i = \ii(a_i^\dagger - a_i)$, 
and define a new $2L\times 2L$ covariance matrix
$\Theta_{ij} = \frac{1}{2} \langle \{ X_i,X_j\} \rangle - \langle X_i \rangle \langle X_j \rangle$, where $\bm{X} = (q_1,\ldots,q_L,p_1,\ldots,p_L)$. 
This will follow a Lyapunov equation, of the same form as~\eqref{nonInt_lyap}, but with new matrices $W$ and $D$ of size $2L$~\cite{Barthel2021}.

Next we turn to the somewhat peculiar case of  dephasing [Eq.~\eqref{bulk_dephasing}], defined by jump operators $L_i = a_i^\dagger a_i$. 
Since they are not linear, the dynamics will not be Gaussian, and one would expect the equation for $C$ (or $\Theta$) to no longer be closed (i.e., depend on higher order moments).
For dephasing however, it turns out that the equations close.
In fact, one may readily verify that 
$\sum_\ell \tr \big\{a_j^\dagger a_i D_\ell^\text{deph}(\rhos)\big\}  = -C_{ij} (1-\delta_{ij})$, which depends only on $C$.
Dephasing thus causes the off-diagonal elements to  decay, while keeping the diagonal ones intact.
Next, suppose we append to Eq.~\eqref{nonInt_M_simp} a set of dephasing baths on all sites, $\sum_i \Gamma ~D[a_i^\dagger a_i]$, of strength $\Gamma$.
The resulting equation for $C$ will then be given by~\cite{AsadianBriegel2013,Znidaric2010,Znidaric2011c,Malouf2018} 
\begin{equation}\label{nonInt_deph_eq}
    \frac{dC}{dt} = - (WC + C W^\dagger) + D - 2\Gamma \Delta(C)\,, 
\end{equation}
where $\Delta(C) = C - \text{diag}(C_{11}, \ldots,C_{LL})$ is the off-diagonal part of $C$. 

Due to this new element, Eq.~\eqref{nonInt_deph_eq} is no longer a Lyapunov equation, although it is still a closed set of linear equations for $C$ (and hence still tractable). 
A proposal to simulate this dynamics in trapped ions was put forth in~\cite{BermudezPlenio2013}.

The solution of a uniform tight-binding chain can also be obtained analytically when dephasing is present~\cite{AsadianBriegel2013,Znidaric2010}. 
The resulting current is 
\begin{equation}\label{dephasing_current_tight_binding}
    \cur{N} = \frac{\gtl{2\gamma} J}{4J^2 + \gtl{\gamma}^2 + 2 \gtl{\gamma} \Gamma (L-1)} (n_L - n_1)\,, 
\end{equation}
which now depends on $L$. 
In fact, for any $\Gamma \neq 0$, there is always a sufficiently large size for which $\cur{N} \sim 1/L$.
Dephasing therefore renders the transport diffusive for any $\Gamma > 0$.
The occupations $\langle a_i^\dagger a_i \rangle$ are also modified, and vary linearly between $n_1$ and $n_L$.
  
\subsection{Third quantization}\label{sec:third_quantization} 

This section combines concepts from Secs.~\ref{sec:vectorization} and~\ref{sec:non_interacting}.
Eq.~\eqref{vec_property} illustrates how  vectorized operators, such as $\hat{\mathcal{L}}$, live in a doubled Hilbert space. 
This is also clear from the Choi-Jamiolkowski isomorphism in Eq.~\eqref{vec_def_Hilbert_space}. 
Consider, for concreteness, a system composed of  bosons, with operators $a_i$. 
Eq.~\eqref{vec_property} then implies that 
$\vvec(a_i \rho) = (\id\otimes a_i) \vvec(\rho)$ and $\vvec(\rho a_i) = (a_i\trans \otimes \id) \vvec(\rho)$. 
In the vectorized space, we therefore end up with two species of particles, the ``left-multiplying kind'' and the ``right-multiplying kind''.
That is, we can define $b_i = \id \otimes a_i$ and $c_i = a_i \otimes \id$, and therefore establish the mappings $(a_i~\bullet) \to b_i$ and $(\bullet~a_i)\to c_i^\dagger$. 
The problem has thus been recast into a doubled Hilbert space, inhabited by two bosonic species.
For fermions this recipe also works, but
some subtleties arise, see~\cite{Prosen2008} for details.

This is the idea behind ``third quantization'', first introduced in~\cite{Prosen2008} and built upon in~\cite{Prosen2010b, ProsenSeligman2010, GuoPoletti2017, GuoPoletti2018b, BanchiZanardi2014, YamanakaSasamoto2021,ProsenZunkovic2010}.
It is equivalent to the superfermion approach of~\cite{Dzhioev2011}, although the latter requires only matrices of size $2L$, instead of $4L$.

The situation simplifies when when $\Hs$ is at most quadratic in the $a_i$, and the jump operators $L_k$ are at most linear [Sec.~\ref{sec:non_interacting}].
In this case, Eq.~\eqref{vec_vectorized_L}  becomes a quadratic form in the new set of $4L$ operators $\bm{d} = (b_1,b_1^\dagger,c_1,c_1^\dagger, ,\ldots,b_L,b_L^\dagger,c_L,c_L^\dagger)$:
\begin{equation}\label{3rd_liouv_gen}
    \hat{\Liouv} = \sum\limits_{\mu, \nu=1}^{4L} \Lambda_{\mu\nu} d_\mu^\dagger d_\nu\,,
\end{equation}
with a $4L\times 4L$ matrix of coefficients $\Lambda_{\mu\nu}$.
For instance, the master equation~\eqref{nonInt_M_simp} can be rewritten as 
$\hat{\Liouv} = - \sum\_{ij} (W_{ij} b_i^\dagger b_j + W_{ij}^\dagger c_i^\dagger c_j) 
    + \sum\_{ij} (\gamma_{ij}^- c_i b_j + \gamma_{ij}^+ c_j^\dagger b_i^\dagger)$\,,
%
where $W$ is defined below Eq.~\eqref{nonInt_lyap}.


Third quantization generalizes the Lyapunov equation~\eqref{nonInt_lyap}. 
If one is interested only in the covariance matrix and currents, then Eq.~\eqref{nonInt_lyap} suffices.
However, with Eq.~\eqref{3rd_liouv_gen} one now has access to the full quadratic Liouvillian, and therefore, to its entire spectrum. 
In fact, as shown in~\cite{Prosen2008}, Eq.~\eqref{3rd_liouv_gen} can always be put in diagonal form, as 
$\hat{\Liouv} = -2\sum_{k=1}^{2L} \lambda_k f'_k f_k$,
where $f_k$ and $f'_k$ are normal modes of the Liouvillian, which respectively annihilate and create a particle in mode $k$ (since $\hat{\mathcal{L}}$ is not Hermitian, $f_k'  \neq f_k^\dagger$). 
In turn, the $c$-number quantities $\lambda_k$ are the rapidities in Eq.~\eqref{vec_L_decomp}. 
For tight-binding Hamiltonians or local boundary drives,  their computation can simplify considerably, and may even allow for solutions in closed form~\cite{Yueh2005, Kouachi2006, Willms2008, FonsecaKowalenko2020}.

\subsection{Perturbative approaches} \label{sec:perturbative} 

It is often impossible to find the steady-state analytically [Eq.~\eqref{vec_M}], but it may  be possible to gain important insights using \ind{Liouvillian perturbation theory}. 
Consider a  Liouvillian of the form
$\Liouv = \Liouv_0 + \mu \Liouv_1$,
where $\mu$ is a small parameter and $\Liouv_0$ and $\Liouv_1$ are in GKSL form.
We decompose $\rhos$ in a power series in $\mu$, as  $\rhos = \sum_{i} \mu^i \rho_i$.
\gtl{The steady-state equation $\mathcal{L}(\rhos) = 0$ then results in a set of recurrence relations, $\mathcal{L}_0(\rho_0) = 0$ and $\mathcal{L}_0(\rho_n) = - \mathcal{L}_1(\rho_{n-1})$, for $n \geqslant 1$, }, which can be solved sequentially
[see e.g.~\cite{MichelMahler2004, Znidaric2019, LenarcicProsen2015}].       
This approach is particularly relevant for boundary-driven problems where the bath couplings are the perturbation. 

In some cases the steady-state $\rho_0$ of $\mathcal{L}_0$ may be degenerate, forcing one to resort to degenerate perturbation theory instead.
For example, 
if one chooses $\Liouv_0=-\im [\Hs, \bullet]$,
the steady-state can be any density matrix diagonal in the eigenbasis of $\Hs$.
This degeneracy can be lifted when $\Liouv_0$ also includes local dephasing terms like $\Lind{a_\ell^\dagger a_\ell}$ or $\Lind{\sz_l}$ [Sec.~\ref{sec:bulk_noise}], leading to a unique $\rho_0$, which is usually the identity matrix (infinite temperature state).
%
If the Hamiltonian is number conserving (e.g. conserves the total number of particles or the total magnetization), then $\rho_0$ would still be degenerate, and formed by mixed states proportional to the identity in each number sector.

Perturbation theory for the evaluation of the steady state, and the relaxation gap, has been used in various works on boundary-driven problems, see e.g. Refs.~\cite{FlindtAnttiPekka2010, Znidaric2015, Znidaric2011, Znidaric2019, ZnidaricZunkovicProsen2011, ZnidaricHorvat2013, BucaProsen2014, GuoPoletti2016, ZnidaricLjubotina2018, LenarcicProsen2015} and also more general problems of open quantum systems [see~\cite{GarciaRipollCirac2009, PolettiKollath2012, PolettiKollath2013, SciollaKollath2015, MedvedyevaZnidaric2016, CaiBarthel2013}]. 


\subsection{Tensor network methods} \label{sec:tensor_networks}

\indTwo{Tensor networks}{tensor networks} form an extremely effective class of numerical methods for boundary-driven strongly interacting quantum systems. 
In the following we review the general ideas and strategies, with a focus on open quantum systems and boundary-driven transport problems.
For general reviews see~\cite{Schollwock2005, Schollwock2011, Orus2014, VerstraeteCirac2008, SilviMontangero2019} and for reviews touching open quantum systems see~\cite{Daley2014, WeimerOrus2019,BertiniZnidaric2020}. 
We  also call attention to~\cite{BrenesGoold2020}, which developed tensor network methods for  handling dissipative, boundary-driven, thermal machines, and we also highlight various publicly available tensor network libraries~\cite{itensor, JaschkeCarr2018, DolfiTroyer2014, AlAssamJaksch2017, KaoChen2015, JaschkeDeVega2019}.

\subsubsection{Introduction to tensor networks and matrix product states} 

We consider a state $\ket{\psi}=\sum c_{\sigma_1,\dots, \sigma_L}\ket{\sigma_1,\dots,\sigma_L}$ representing a chain with $L$ sites, each with $d$ local levels.
One can always write the coefficients $c_{\sigma_1,\ldots, \sigma_L}$ as a product of tensors, each with a number of indices  depending on the geometry of the system considered. 
A  particularly effective case is that of 1D systems, in which one can write
\begin{align}
    c_{\sigma_1,\dots, \sigma_L}=\sum_{a_0,\dots,a_L} M^{\sigma_1}_{a_0,a_1}M^{\sigma_2}_{a_1,a_2}\dots M^{\sigma_L}_{a_{L-1},a_L}\,, \label{eq:MPS_ansatz}
\end{align}
where, for each $\sigma_l$, $M^{\sigma_l}$ is a matrix with auxiliary indices $a_{l-1}$ and $a_l$. 
This is referred to as a \ind{matrix product state} (MPS).
MPSs are not the only type of tensor network. 
But since most boundary-driven systems studied are 1D, where MPS work particularly well, we will henceforth concentrate on this structure.
The MPS ansatz is exact if we allow the size of the matrices $M^{\sigma_l}$ to grow unboundedly. 
However, this would be of little interest. 
The advantage, instead, is that accurate representations can often be obtained even with fixed maximum sizes $D_{\rm B}$, called the \ind{bond dimension}.
This is because \ind{area laws} predict that the entanglement entropy of the ground state of gapped short-range Hamiltonians grows as $L^{\Dim-1}$, where $\Dim$ is the dimensionality~\cite{BombelliSorkin1986,  AudenaertWerner2002, EisertPlenio2010, Srednicki1993, PlenioCramer2005,Wolf2006}.
In 1D critical systems the entropy grows at most logarithmically~\cite{VidalKitaev2003, LaTorreVidal2004}.

For a fixed  $D_{\rm B}$, the memory requirements go as $Ld^2{D_{\rm B}}^2$, i.e., linearly with $L$, a significant advantage compared to the exponential scaling of the coefficients $c_{\sigma_1,\dots, \sigma_L}$. 
Given the MPS ansatz in Eq.~(\ref{eq:MPS_ansatz}), tensor network  algorithms aim to find the matrices $M^{\sigma_l}$ which most accurately approximate a particular state, for fixed $D_{\rm B}$. 
From a mathematical point of view, MPSs are thus simply an algorithm to approximate a vector in a high dimensional space, by a product of matrices. 
It can be implemented beyond the evaluation of ground states, to study  time evolutions of unitary and dissipative systems, and in the computation of steady states. 
In most of these scenarios however, there is no guarantee that the ansatz will be accurate.
It has been shown in~\cite{BrandaoPerezGarcia2015, CubittPerezGarcia2015} that area laws also exist for rapidly mixing dissipative systems, and in~\cite{GullansHuse2019b} the authors showed the existence of NESSs with local equilibrium, which can be represented effectively with tensor networks.  
In any case, one can systematically study the equilibrium and non-equilibrium properties of a system as a function of the maximum bond dimension $D_{\rm B}$, and estimate the accuracy of these results. 
%

\subsubsection{Tensor networks and boundary-driven systems}\label{ssec:tensor_networks_open} 
 
Any normalized vector can represent a physical state, but not every matrix, even with unit trace, represents a physical density matrix, since the latter must be  positive semidefinite. 
Various approaches have been suggested to express density matrices as tensor networks in a way that preserves positivity, either using the structure of Eq.~\eqref{AnaSol_rho_decomp} below~\cite{minganti2018a,WernerMontangero2016,WeimerOrus2019}, or via purification techniques~\cite{NielsenChuang2000}. 
However, these approaches have not been largely pursued. 
One possible reason is that QMEs guarantee that any accurate representation of steady-states will be physical.
In other words, if one approximates  the steady state well enough, relevant physical properties can be accurately evaluated, even if the density matrix is not exactly positive semidefinite.    

One can thus choose to write the density matrix as a \ind{Matrix Product Operator}
\begin{align}
    \rho &= \sum_{ {\sigma_1,\dots,\sigma_L \atop \sigma'_1,\dots,\sigma'_L}} \rho_{\sigma_1,\dots,\sigma_L}^{\sigma'_1,\dots,\sigma'_L} \;\; | \;\sigma_1, \dots,\sigma_L \rangle \langle \sigma'_1,\dots, \sigma'_L |\\ 
    &= \sum_{ {\sigma_1,\dots,\sigma_L, \sigma'_1,\dots,\sigma'_L \atop a_0,\dots,a_L}} M^{\sigma_1,\sigma'_1}_{a_0,a_1}\dots M^{\sigma_L,\sigma'_L}_{a_{L-1},a_L} \;\; | \;\sigma_1, \dots,\sigma_L \rangle \langle \sigma'_1,\dots, \sigma'_L |\,,\nonumber
\end{align}   
where each $M^{\sigma_l,\sigma'_l}$ is a matrix with indices $a_{l-1}$ and $a_l$, of dimension smaller than the bond dimension $D_{\rm B}$. 
Together with vectorization [Sec.~\ref{sec:vectorization}], this ansatz can then be used to directly study the time-evolution and the steady-state. 
For a review of methods for time evolving tensor networks, see~\cite{GarciaRipoll2006, PaeckelHubig2019, ZnidaricGoold2017}. 
Another approach is to compute the steady state as the ground state of the superoperator $\hat{\mathcal{L}}^\dagger\hat{\mathcal{L}}$,
as discussed in Eq.~\eqref{vec_LdL} and Ref.~\cite{CiracBanuls2015}. 
Similarly, one may use ground-state-like search algorithms, but looking for the null vector of the non-Hermitian superoperator $\hat{\mathcal{L}}$~\cite{MascarenhasSavona2015, BaireyArad2020}. 
Finally, one may also evolve the density matrix in the Heisenberg picture~\cite{HartmannPlenio2009}: 
Ref.~\cite{ClarkPlenio2010} showed that depending on the observables studied,  this can lead to significant improvements in efficiency, and even to exact solutions.   

An important question is when to stop the simulation. 
This depends on the quantities of interest, which in our case is usually the current.
In the steady state, the current is the same in all bonds of the system, but in the transient, this may not be the case.
It is thus common to compute the standard deviation of the current over different bonds, and stop when this falls below a chosen threshold.

\subsubsection{Approaches to improve the performance} 

Although tensor networks allow one to study  large, strongly interacting, boundary-driven systems, the numerical calculations can still be very demanding. 
It is thus worth considering different ways to improve the numerical performance of the algorithm.
Tensor networks can be used to describe any type of quantum systems, whether bosonic, fermionic, spins, or mixtures. 
Bosonic systems can be particularly demanding, because the local Hilbert space dimension $d$ may be large.
At large temperatures, the relevant terms of $M^{\sigma_l,\sigma'_l}$ are usually those for which $\sigma_l \approx \sigma'_l$.
Only considering such terms can result in a significant reduction of memory requirements~\cite{GuoPoletti2015}.             

Another improvement is to account for  symmetries (Sec.~\ref{ssec:symmetries}), more notably the conservation of a quantum number such as total particle number or total magnetization. 
For instance, in Eq.~\eqref{eq:MPS_ansatz} one can also write $M^{\sigma_l}_{(n_{l-1},a_{l-1}),(n_{l},a_{l})}$ where $a_{l-1}$ and $a_{l}$ still represent the auxiliary indices, and $n_{l}$ represents the sum of the physical indices $\sigma_l$ until site $l$, i.e., $n_l=\sum_{m=1}^l \sigma_m$, which implies that $n_l = \sigma_l + n_{l-1}$.
This more detailed data structure allows to save considerable memory, as one needs to deal with smaller matrices (although many more of them)~\cite{McCulloch2007, HubigSchollwock2017, SilviMontangero2019}. 
This can be generalized to open quantum systems, but one needs to account for the quantum numbers in both  the bra and the ket of the density matrix. 
Moreover, while the Hamiltonian does not change the total quantum number, the dissipators generally do.
A scheme that automatically adapts to such changes in quantum number was introduced in~\cite{GuoPoletti2019}.
Taking this into account, however, works best  when the fluctuations in the particle number are small, e.g. far from the infinite temperature state. 
In the latter, a simpler non-number conserving code would typically be more effective.
Similar ideas in a system with non-Abelian symmetry were studied in~\cite{Moca2022}.

Parallelization of the time evolution also improves the performance.
Ref.~\cite{Schollwock2011} discusses (pages 84-85) how to apply the evolution operator to the MPS in parallel, alternating between even and odd bonds.
This can be done while preserving the canonical structure of the MPS,  allowing to reduce the local bond dimensions in parallel. 
Another alternative is the algorithm developed in~\cite{StoudenmireWhite2013}, and recently adapted for time evolution in~\cite{SecularJaksch2020}. 
In this case one divides the sweep over the full system into sweeps over different portions of the system, and distributes these internal sweeps to different computational nodes, letting the nodes communicate once they reach their shared bond. 

Recently,~\cite{RamsZwolak2020, WojtowiczZwolak2020} presented a way to significantly reduce the bond dimension, based on the realization that, for fermionic systems, the star configuration (Sec.~\ref{sec:star_to_chain_thermofield}) requires lower bond dimension than the chain configuration (see also~\cite{BrenesGoold2020}).
The authors also grouped the eigenmodes of the two baths in increasing order of energy, allowing the correlations between baths modes to be concentrated within a narrow energy band, close to their Fermi levels. 
This is particularly effective in the common scenarios that an energy mode from one bath is scattered to a mode of similar energy in a different bath. 
Still on the topic of the star-to-chain mapping,~\cite{TamascelliPlenio2019,NuselerPlenio2019} obtained improvements specifically designing a thermofield transformation approach (Sec.~\ref{sec:star_to_chain_thermofield}) for 
system-bath couplings with Hermitian bath operators.
Tensor networks algorithms have also been used to simulate Redfield QMEs [Eqs.~\eqref{EQ:redfieldI_ip} and \eqref{EQ:redfieldII_ip}], reaching system sizes much larger than what can be obtained by exact diagonalization~\cite{XuPoletti2019a}.  
%

\subsection{Analytical solutions for interacting  chains} \label{sec:analytical_MPA} 

Analytical solutions of interacting boundary-driven models are seldomly available, due to the enormous mathematical complexities involved. 
When they are, it usually concerns restricted scenarios or extremes of parameter space. 
Notwithstanding, these solutions offer invaluable insight, precisely because they are \emph{exact}.
Here we review the most famous example: an XXZ chain~\eqref{XXZ} under maximally biased local dissipators, which can be solved in terms of MPSs.
This was first discovered in~\cite{Prosen2011,Prosen2011b} and  generalized in~\cite{Karevski2013}. 
Similar ideas were subsequently used in~\cite{Popkov2013a,Popkov2016}, to show that there exist special configurations, associated to twisted Lindblad operators of the form $\sigma_x \cos\theta -\ii \sigma_y+ \sigma_z \sin\theta$, where the NESS is found to be in a product state, despite carrying a finite current (which is highly unusual).
A matrix-product solution, in similar spirit, was also found for the open Fermi-Hubbard chain in~\cite{Prosen2014,Popkov2015}. 
Very recently, these developments were generalized to encompass general XYZ chains and twisted jump operators~\cite{PopkovZadnik2020, PopkovZadnik2020b}.

The system is taken to be an XXZ chain [Eq.~\eqref{XXZ}], with zero magnetic field and LME dissipators acting on the  first and last sites,  as in Eq.~\eqref{preamble_LME_basic}. 
The drives are taken as $f_1 = 1$ and $f_L = 0$, meaning the dissipators try to fully polarize the first and last spins in opposite directions.
The master equation~\eqref{preamble_LME_basic} is thus described by only two jump operators, $\sqrt{\gamma} \splus_1$ and $\sqrt{\gamma} \sm_L$, where $\gamma$ is the coupling strength.
We mention upfront that the solution obtained in this limit is \emph{not} representative of what is found for other values of $f_i$. 
For instance, in this case transport is found to be subdiffusive ($\cur{} \sim 1/L^2$) when $\Delta = 1$.
Conversely, at high temperatures ($f_1 = 0.5 + \delta$ and $f_L = 0.5 - \delta$), it is superdiffusive~\cite{Znidaric2011}, see Sec.~\ref{sec:properties} for more details.

The solution of~\cite{Prosen2011,Prosen2011b} starts by decomposing the steady-state  as 
\begin{equation}\label{AnaSol_rho_decomp}
    \rho = \frac{S S^\dagger}{\tr(SS^\dagger)},
\end{equation}
for some operator $S$. 
This is always possible, for any $\rho$, since $SS^\dagger$ is always a positive semi-definite operator.
However, such a decomposition is often not useful since, when plugged into the master equation~\eqref{preamble_LME_basic}, it  leads to a nonlinear equation for $S$. 
The main innovation of~\cite{Prosen2011,Prosen2011b} was to prove that $S$, in this particular problem, is given by a \ind{matrix product ansatz}. 
One can approach the problem as follows.
Let $\mathcal{H}_i$ denote the Hilbert space of each spin degree of freedom (so that both $\rho$ and $S$ live in $\mathcal{H}_1 \otimes \ldots \otimes \mathcal{H}_L$). 
We then introduce an auxiliary space $\mathcal{A}$ and write $S$ as
\begin{equation}\label{AnaSol_MPO_ansatz}
    S = \bra{\phi}\Omega_1 \otimes \ldots \otimes \Omega_L \ket{\psi}\,, 
\end{equation}
where $\Omega_i \in \mathcal{H}_i \otimes \mathcal{A}$ are operators that live in the joint Hilbert space of site $i$ and the auxiliary space. 
Moreover, $\ket{\phi}$ and $\ket{\psi}$ are states which act only on $\mathcal{A}$. 
As a consequence, the sandwich in Eq.~\eqref{AnaSol_MPO_ansatz} effectively traces out $\mathcal{A}$, leaving $S$ as a state only in $\mathcal{H}_1 \otimes \ldots \otimes \mathcal{H}_L$, as it should.

The goal then is to adjust the parameters in Eq.~\eqref{AnaSol_MPO_ansatz} so that $\rho$ is ultimately a steady-state solution of~\eqref{preamble_LME_basic}. 
This involves, first of all, deciding what properties the auxiliary space $\mathcal{A}$ should have. 
The methods used for doing so are non-trivial, so we only state the result.
Since $\Omega_i \in \mathcal{H}_i \otimes \mathcal{A}$, we can parametrize
\begin{equation}
\Omega_i = A_0 \szero_i + A_+ \splus_i + A_- \sm_i + A_z \sz_i\,,
\end{equation}
for some operators $A_\alpha \in \mathcal{A}$.
This allows us to write~\eqref{AnaSol_MPO_ansatz} as 
\begin{equation}\label{AnaSol_MPO_ansatz_2}
    S = \sum\limits_{\alpha_1, \ldots, \alpha_L \in \{0,+,-,z\}} \langle \phi | A_{\alpha_1} \ldots A_{\alpha_L} |\psi\rangle~\sigma^{\alpha_1} \otimes \ldots \otimes \sigma^{\alpha_L}\,. 
\end{equation}
Hence, $S$ is written as a linear combination of all possible basis elements $\sigma^{\alpha_1} \otimes \ldots \otimes \sigma^{\alpha_L}$.

Refs.~\cite{Prosen2011,Prosen2011b} showed that all coefficients involving $\sz_i$ can be chosen to vanish.
This leaves only three operators to be specified: $A_0$, $A_+$ and $A_-$. 
Next, they showed that $\mathcal{A}$ can be chosen as an infinite dimensional Hilbert space with basis elements $|r\rangle$, $r = 0,1,2,3,\ldots$ and  
\begin{align}
    A_0 &= |0\rangle\langle 0| + \sum\limits_{r=1}^\infty a_r^0 \ket{r}\bra{r}\,,\\
    A_+ &= (\ii \gamma/2) |0\rangle\langle 1| + \sum\limits_{r=1}^\infty a_r^+ \ket{r}\bra{r+1}\,,\\
    A_- &= |1\rangle\langle 0| + \sum\limits_{r=1}^\infty a_r^- \ket{r+1}\bra{r}\,, 
\end{align}
where $a_r^\alpha$ are complicated coefficients depending on  $\gamma$ and  $\Delta$  (see~\cite{Prosen2011b} for the explicit definitions).
For $\Delta \neq 1$, these operators are found to satisfy a $q$-deformed SU(2) algebra,  which reduces to the standard SU(2) algebra for $\Delta=1$~\cite{Karevski2013}.
Finally, it suffices to take both $|\phi\rangle$ and $|\psi\rangle$ to be $\ket{0}$. 
A systematic calculation of arbitrary correlation functions from this solution can be found in~\cite{Buca2016}.

\begin{figure}
    \centering
    \includegraphics[width=\columnwidth]{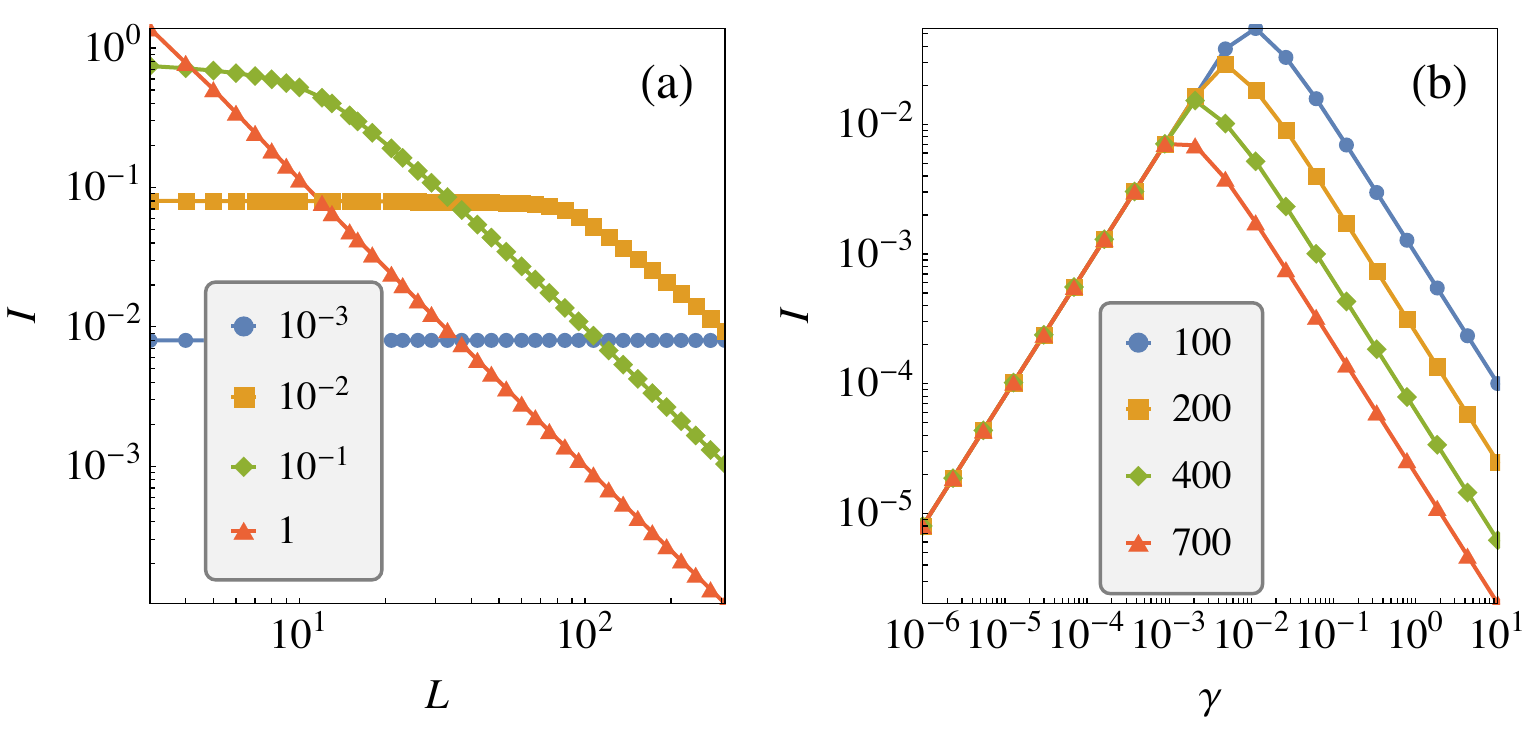}
    \caption{Example of the steady-state currents in the Heisenberg chain (XXZ with $\Delta = 1$) obtained from the exact solution in Eq.~\eqref{AnaSol_current_example}. 
    (a) Dependence on the size, for different coupling strengths $\gamma$.
    (b) As a function of $\gamma$, for different sizes $L$.
    }
    \label{fig:AnaSol_currents}
\end{figure}

Extracting physical observables from this formal solution is non-trivial for $\Delta \neq 1$.
But when $\Delta = 1$, it can be shown that the magnetization current~\eqref{XXZ_current} reduces to~\cite{Landi2015a}
\begin{equation}\label{AnaSol_current_example}
    \cur{} = \frac{2}{\gamma} \frac{(B^{L-1})_{00}}{(B^L)_{00}}\,, 
    \quad 
    B_{k\ell} = 2 \Big|k - \frac{\ii}{(2\gamma)}\Big|^2 \delta_{k\ell} + \ell^2 \delta_{k,\ell-1} + \Big|\ell - \frac{\ii}{\gamma}\Big|^2 \delta_{k,\ell+1}
\end{equation}
with $k, \ell = 0,1,2,\ldots,L$.
Eq.~\eqref{AnaSol_current_example} therefore involves the entries of the powers $B^{L-1}$ and $B^L$ of a $L+1$-dimensional matrix, allowing one to compute the current for arbitrarily large sizes. 
Plots of the behavior of $\cur{}$~vs.~$L$ and $\gamma$ are shown in Fig.~\ref{fig:AnaSol_currents}. 
As it can be seen, the presence of interactions dramatically alters the current profile, as compared e.g. to the ballistic scenario of  Sec.~\ref{sec:non_interacting}. 
The dependence of $\cur{}$ on $L$ is found to depend sensibly on $\gamma$. 
For small $\gamma$, transport is roughly ballistic, while for large $\gamma$ it becomes subdiffusive, scaling as $\cur{} \propto 1/L^2$. 
For sufficiently large $L$, it always eventually becomes subdiffusive for any $\gamma>0$.

\subsection{Quantum trajectories approach} \label{sec:quantum_trajectories} 

A method  commonly used to describe open quantum systems, including boundary-driven problems (e.g.~\cite{Wichterich2007}), are \ind{quantum trajectories}, also known as quantum jumps or Monte-Carlo wave functions~\cite{Carmichael1993,wiseman2010, DalibardMolmer1992, DumRitsch1992, MolmerDalibard1993, DumGardiner1992, Gardiner2004, Breuer2002, PlenioKnight1998,Jacobs2006}. 
The main idea is not to solve the master equation, but to stochastically evolve wavefunctions, such that the average of these wavefunctions reproduces the dynamics of the QME.

A GKSL QME of the form in Eq.~\eqref{EQ:GKSL}  can be rewritten as
\begin{equation}\label{eq:GKSL_jump}
    \frac{d\rhos}{dt} = -\ii \left(\Hseff \; \rhos - \rhos \; \Hseff^\dagger \right)+\sum\limits_k L_k^{} \rhos L_k^\dagger\,, 
\end{equation} 
where $\Hseff$ is a non-Hermitian Hamiltonian given by
\begin{equation}\label{eq:Hjump_eff}
    \Hseff = \Hs -\frac{\ii}{2} \sum_k L_k^\dagger L_k\,. 
\end{equation}  
The last term in~(\ref{eq:GKSL_jump}) represents discrete jumps, while the first represents a no-jump evolution.
As a method for quantum trajectories, one can thus use the following algorithm:  evolve a wavefunction $\ket{\psi(t)}$ under $\Hseff$, by a small time $\delta t$:
\begin{equation}\label{eq:Hjump_psitilde}
    \ket{\tilde{\psi}(t+\delta t)} = \left[\id-\ii \Hseff \delta t\right] \ket{\psi(t)}\,.  
\end{equation} 
Since $\Hseff$ is not Hermitian, the norm of the state reduces to  $\langle \tilde{\psi}(t+\delta t) | \tilde{\psi}(t+\delta t) \rangle \simeq 1 - \delta p$ where 
$\delta p = \delta t \sum_k  \langle \psi(t) | L_k^\dagger L_k | \psi(t) \rangle := \sum_k\delta p_k$.
Here, $\delta p_k$ is interpreted as the probability of performing the \indTwo{quantum jump}{quantum jumps} $L_k$, and $1-\delta p$ as the probability of performing no jump during time interval $\delta t$.
Thus, with probability $\delta p_k$, the state is updated to
\begin{align}
    \gsc{\ket{\psi(t+\delta t)} \to \ket{\psi_\text{jump}^k}} = \frac{L_k \ket{\psi(t)}}{\sqrt{\bra{\psi(t)} L_k^\dagger L_k \ket{\psi(t)}}}\,,
\end{align}
while with probability $1-\delta p$, it goes to
\begin{align}
    \gsc{\ket{\psi(t+\delta t)} \to \ket{\psi_\text{stay}}} = \frac{\ket{\tilde{\psi}(t+\delta t)}}{\sqrt{\braket{\tilde{\psi}(t+\delta t)}{\tilde{\psi}(t+\delta t)}}}\,.
\end{align}
One may then verify that the ensemble-averaged evolution,
$\overline{\rho(t +\delta t)} = 
(1-\delta p) |\psi_\text{stay}\rangle\langle\psi_\text{stay}|
+\sum_k \delta p_k |\psi_\text{jump}^k\rangle\langle \psi_\text{jump}^k|$,
can be written as  to order $\delta t^2$ as
\begin{equation*}
   \overline{\rho(t +\delta t)} \simeq \rho(t) - \im \delta t \Big[\Hseff \rho(t) - \rho(t) \Hseff^\dagger\Big]+ \delta t \sum_k L_k \rho(t) L_k^\dagger\,,
\end{equation*}
which is equivalent to Eq.~\eqref{eq:GKSL_jump}.

This also implies that the expectation value of any observable $O$ can also be evaluated by averaging over the trajectories 
$\langle O \rangle = \trace{\rho(t) O} = \overline{\langle \psi(t) | O | \psi(t) \rangle }$.
Computing a single trajectory this way requires significantly less effort than solving Eq.~\eqref{eq:GKSL_jump}.
One has to make sure that this numerical advantage is not scotched by the required sampling (a process which is easily parallelizable though).

It is natural to try to bring together quantum trajectories with tensor networks (Sec.~\ref{sec:tensor_networks}). 
This has been reviewed in~\cite{Daley2014}, which also  reviews the quantum trajectories approach in
many-body settings. 
The benefits of such an approach where studied  in~\cite{BonnesLauchli2014}, which analyzed the growth of the bond dimension required for an accurate description of the system.
An important element identified by the authors is the size of the local Hilbert space $d$, which must be squared for density matrices. This has important consequences on the memory requirements of the algorithms. 
Quantum trajectories can also be preferred when the relaxation time to the NESS is exponentially long~\cite{BenentiZnidaric2009}.
Recently, Ref.~\cite{WolffKollath2020} considered different ways of computing two-time correlations with quantum trajectories, showing that also for this problem the preferred approach depends on the problem at hand. 


\section{\label{sec:properties}Properties of boundary-driven open systems}

\subsection{Overview of different types of steady transport regimes}\label{ssec:overviewtransport}

\begin{figure}[ht]
\includegraphics[width=\columnwidth]{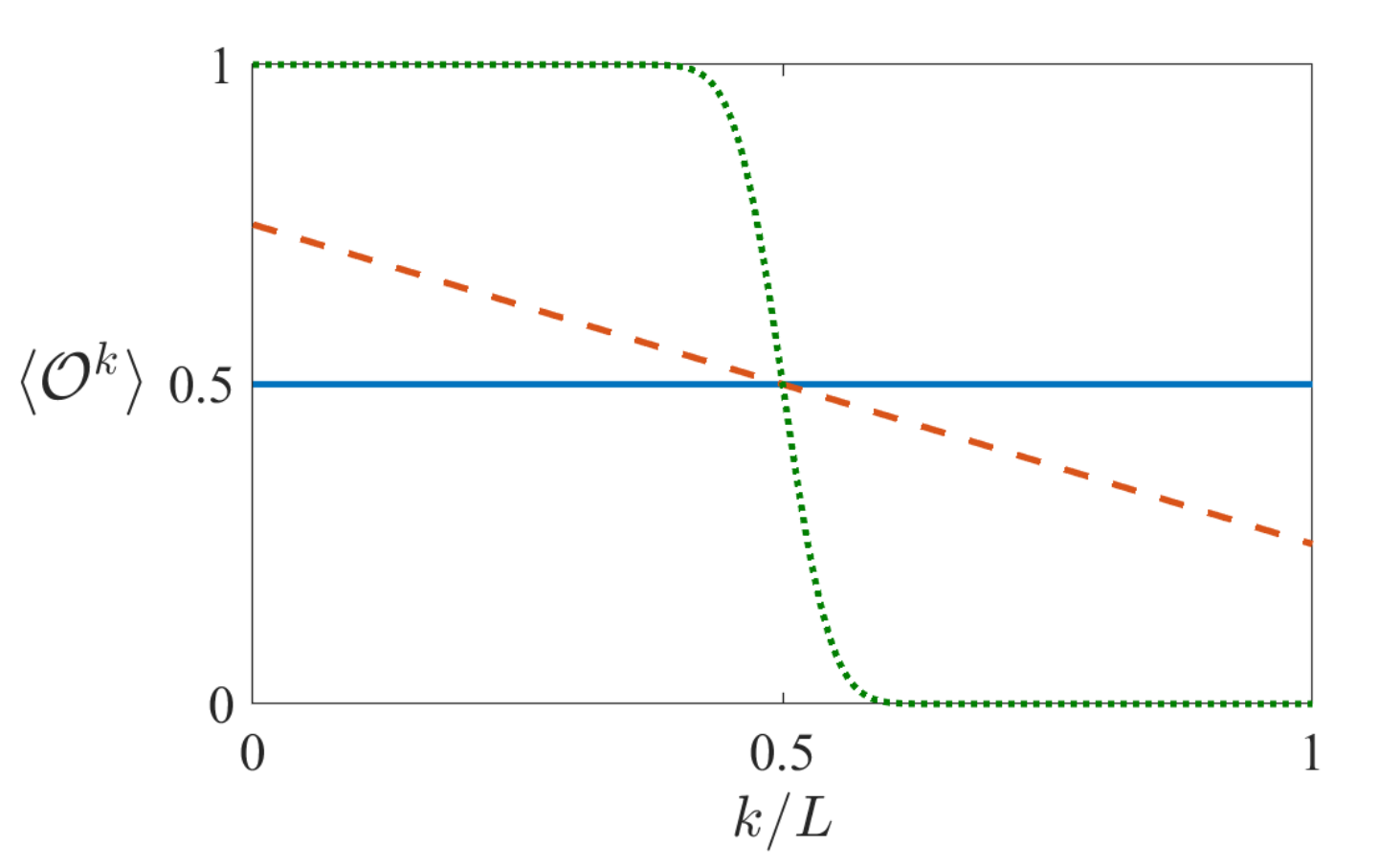} 
\caption{Qualitative profile of a local quantity $\langle \mathcal{O}^k \rangle$ versus position $k$ for three transport regimes: ballistic (blue continuous line), diffusive (red dashed line) and insulating (green dotted line). In all cases the dissipative boundary driving tries to impose the value $1$ at the left edge, and the value $0$ at the right edge as e.g. discussed in Sec.~\ref{sssec:localcontinuityequation}.}
\label{fig:profiles} 
\end{figure}

We begin with an introduction on the phenomenology of transport in boundary-driven systems.
This is usually characterized by how the current (of e.g. energy or matter) $\cur{}$ scales with the system size $L$.
For sufficiently large $L$ this usually has the form $\cur{} \sim 1/L^\alpha$ [Eq.~(\ref{eq:generic_cur_scaling})].
The different transport regimes where discussed in Sec.~\ref{sec:intro}, and are summarized in Table~\ref{tab:scaling}.
They also manifest as signatures in the expectation values of the quantity considered (e.g. spin or energy).
For instance, ballistic transport yields roughly constant values through the system that are independent of $L$. 
This is illustrated by the blue-solid line in Fig.~\ref{fig:profiles}, which depicts a generic
quantity $\langle \mathcal{O}^k\rangle$ as a function of the position $k$ in a chain of length $L$, when the left (right) bath tries to impose the value $1$ ($0$). 
Conversely, for diffusive transport one would observe a linear profile (red-dashed lines).
At the boundaries, however, there is usually a mismatch with respect to the value the baths are trying to impose.
Finally, in an insulating regime, since there is no current the profile will be constant over large parts of the chain. 
But since the quantity must also be adjusted to the one imposed by the baths, this will result in a step-like behavior, as depicted by the green-dotted line in Fig.~\ref{fig:profiles}.

The classification of transport exponents according to the scaling $\cur{} \sim 1/L^\alpha$ [Eq.~\eqref{eq:generic_cur_scaling}] can alternatively be performed in isolated (unitary) systems, by analyzing  the spreading of wavepackets.
An initially localized packet will spread along the system with a standard deviation that scales as 
\begin{equation}\label{spreading_exponent}
    \sqrt{\langle \Delta x^2 \rangle} \sim t^{~\uniexpo},
\end{equation}
with some exponent $\uniexpo$.
Diffusive transport corresponds to $\uniexpo = 1/2$. 
Other transport regimes are shown in Table~\ref{tab:scaling}.

\begin{table}
    \centering
    \caption{Scaling exponents for the different transport regimes, when analyzed from the perspective of the current, or from the perspective of the spreading of a localized wavepacket. The two exponents are connected by $\uniexpo = 1/(\alpha+1)$. }
    \begin{tabular}{c|c|c}
    Transport regime & $\cur{} \sim 1/L^\alpha$ &  $\sqrt{\langle \Delta x^2 \rangle} \sim t^{\uniexpo}$ \\[0.2cm] \hline 
    Ballistic & $\alpha = 0$ & $\uniexpo = 1$\\ 
    Superdiffusive & $0<\alpha<1$ & $1/2<\uniexpo < 1$\\ 
    Diffusive & $\alpha = 1$ & $\uniexpo = 1/2$\\
    Subdiffusive & $\alpha > 1$ & $\uniexpo < 1/2$ \\ 
    Localized & $\alpha = \infty$ & $\uniexpo = 0$ 
    \end{tabular}
    \label{tab:scaling}
\end{table}

\gtl{To relate the exponents $\alpha$ and $\uniexpo$, we notice from Eq.~\eqref{spreading_exponent} that the characteristic time it takes for a particle to travel through the chain will be $\tau \sim L^{1/\uniexpo}$. 
In the NESS, the current should be proportional to the rate of particles flowing through, $\cur{} \sim L/\tau \sim L/L^{1/\uniexpo}$. 
Comparing this with $\cur{} \sim 1/L^\alpha$ then yields
\begin{equation}\label{scaling_exponents_relation}
    \uniexpo = \frac{1}{\alpha+1}\,,
\end{equation}
(c.f.~Table~\ref{tab:scaling}).
The above identification, however, crucially relies on the somewhat arbitrary  hypothesis of a universal transport exponent. 
While true in many cases, this has been reported to  breakdown in certain models~\cite{Varma2017,PurkayasthaKulkarni2016}.}

\subsection{Integrability and transport}\label{ssec:integrability} 
One might think that integrable systems are ballistic and non-integrable systems are diffusive, but reality is much more interesting~\footnote{By integrable, we refer to Bethe-ansatz integrable models that have an infinite number of local conservation laws, as discussed in~\cite{BertiniZnidaric2020}. For a more exhaustive discussion on integrability in quantum systems, see~\cite{CauxMossel2011}.}.
For example, integrable systems can present ballistic, diffusive, anomalous and insulating transport properties~\cite{Prosen2011, Znidaric2011}.
The mechanisms behind such richness are still a topic of research. 
Similarly, non-integrable models can also exhibit ballistic transport. 
In particular at $T=0$, systems with a gapless phase always support ballistic transport -- independent of whether they are integrable or not~\cite{Kohn1964, ScalapinoZhang1993, ShastrySutherland1990}. 
See~\cite{Mastropietro2013} for a  recent study of an XXZ chain equipped with next-nearest neighbor interactions to break integrability.
Away from $T=0$, transport can be qualitatively different. 
Ballistic transport has been observed in finite-temperature non-integrable systems, both  classical~\cite{LebowitzScaramazza2018} and 
quantum~\cite{BrenesGoold2018}. 
In the latter, the authors considered an XXZ chain with a single, local, impurity in the center, rendering the system non-integrable.
However, since the system is composed of two ballistic halves, ballistic transport was still preserved.
A small density of impurities, however, renders the transport diffusive~\cite{Znidaric2020}.
And when the clean system behaves subdiffusively, impurities can also increase transport~\cite{Znidaric2022} (c.f. Sec~\ref{ssec:dephasing_and_transport}).

We now provide an overview of some guiding principles, starting with the concept of Drude weight in linear response. 
Considering the conductivity of a system as a function of frequency $\sigma(\omega)$, which can be obtained e.g. from Kubo's formula~\cite{Green1952, Green1954, Kubo1957}, the AC response is obtained when $\omega\rightarrow 0$, with ballistic transport  signaled by an infinite conductivity. 
It is thus useful to write the real part of the conductivity, $\cond' = \text{Re}(\cond)$ as 
$\cond'(\omega) = \Drude \delta(\omega) + \condreg(\omega)$,
where $\Drude$ is the \ind{Drude weight}, and  $\condreg(\omega)$ is the regular portion of the conductivity. 
A non-zero Drude weight thus implies ballistic transport.
The Drude weight of a 1D system in equilibrium with inverse temperature $\beta$ is given by 
\begin{align} 
\Drude = \lim_{t\rightarrow\infty}\lim_{L\rightarrow\infty} \frac{\beta}{4 L t} \int_{-t}^t \langle \curOp{L}(0)\curOp{L}(t') \rangle_{\beta} dt'\,,
\end{align} 
where $\langle\dots\rangle_\beta$ denotes the average over a thermal state at temperature $\beta$, and $\curOp{L} = \sum_l \curOp{l}$ is the sum of all the local current operators (defined in Eq.~\eqref{current_operator}).

If the $\curOp{l}$ are conserved at system-internal junctions, transport is guaranteed to be ballistic~\cite{ZotosPrelovsek1997} because  $\langle \curOp{L}(0)\curOp{L}(t') \rangle_{\beta}$ becomes a constant in $t$, and hence the Drude weight is finite.  
For instance, in the XXZ model [Eq.~(\ref{XXZ})], the energy current operator $\mathcal{I}_E^{l}$ is given, using Eq.~(\ref{Currents_unitary_bond_continuity}), by 
\begin{align}
\mathcal{I}_E^{l} = \Delta( \mathcal{I}_M^{l-1,l}\sz_{l+1} +  \sz_{l-1}\mathcal{I}_M^{l,l+1} ) + J \mathcal{I}_M^{l-1,l+1}\sz_l.  \label{eq:nrg_cur} 
\end{align} 
where $\mathcal{I}_M^{l,l+1} = - 2 J (\sigma_l^x \sigma_{l+1}^y - \sigma_l^y \sigma_{l+1}^x )$ [c.f.~Eq.~\eqref{XXZ_current}]. 
It follows that energy transport is ballistic for any parameter value~\cite{ZotosPrelovsek1997}. 
On the other hand, $[\sum_l\mathcal{I}_M^{l,l+1},H_{XXZ}]\ne 0$ and hence spin transport is not guaranteed to be ballistic.

%
A lower bound to the Drude weight is provided by the 
\ind{Mazur bound}~\cite{Mazur1969, Suzuki1971, ZotosPrelovsek1997} 
\begin{align}
\Drude \ge \lim_{L\rightarrow\infty} \frac{\beta}{2L} \sum_k \frac{ | \langle \curOp{L} Q_k \rangle |^2 }{ \langle Q_k^\dagger Q_k \rangle }\,,
\end{align}
where $Q_k$ are orthogonal conserved quantities which, if they are (quasi-)local, have an overlap with $\curOp{l}$ which is not exponentially small as $L$ increases.
If the system has an extensive number of such quantities, then the Drude weight $\Drude$ can be non-zero, and transport will be guaranteed to be ballistic. 
Finding the $Q_k$ may be non-trivial, however.
For the XXZ chain they were computed in~\cite{Prosen2011,ProsenIlievski2013}.
Using these results one may show that spin transport in XXZ chains, in the zero magnetization sector, is ballistic for $|\Delta|<1$~\cite{Prosen2011}, and for any $\Delta$, when away from the zero magnetization sector~\cite{ZotosPrelovsek1997}; this can also be shown using the Bethe-ansatz at $T=0$~\cite{ShastrySutherland1990} or with numerical methods~\cite{HeidrichMeisnerBrenig2005}. 
In a similar manner one can study the Hubbard model, where the energy current is not conserved, but the current operator has a finite overlap with conserved quantities, resulting in a nonzero Mazur bound~\cite{ZotosPrelovsek1997} (see also~\cite{KarraschHeidrichMeisner2016, Karrasch2017, IlievskiDeNardis2017}). 
For charge and spin currents, the current operators have finite overlap with conserved quantities outside of the half-filling sector, and thus transport is ballistic in these regimes~\cite{ZotosPrelovsek1997, GarstRosch2001, Karrasch2017, IlievskiDeNardis2017}.

\subsection{Transport in the XXZ chain}\label{ssec:XXZchainexample}

\begin{figure}
\includegraphics[width=\columnwidth]{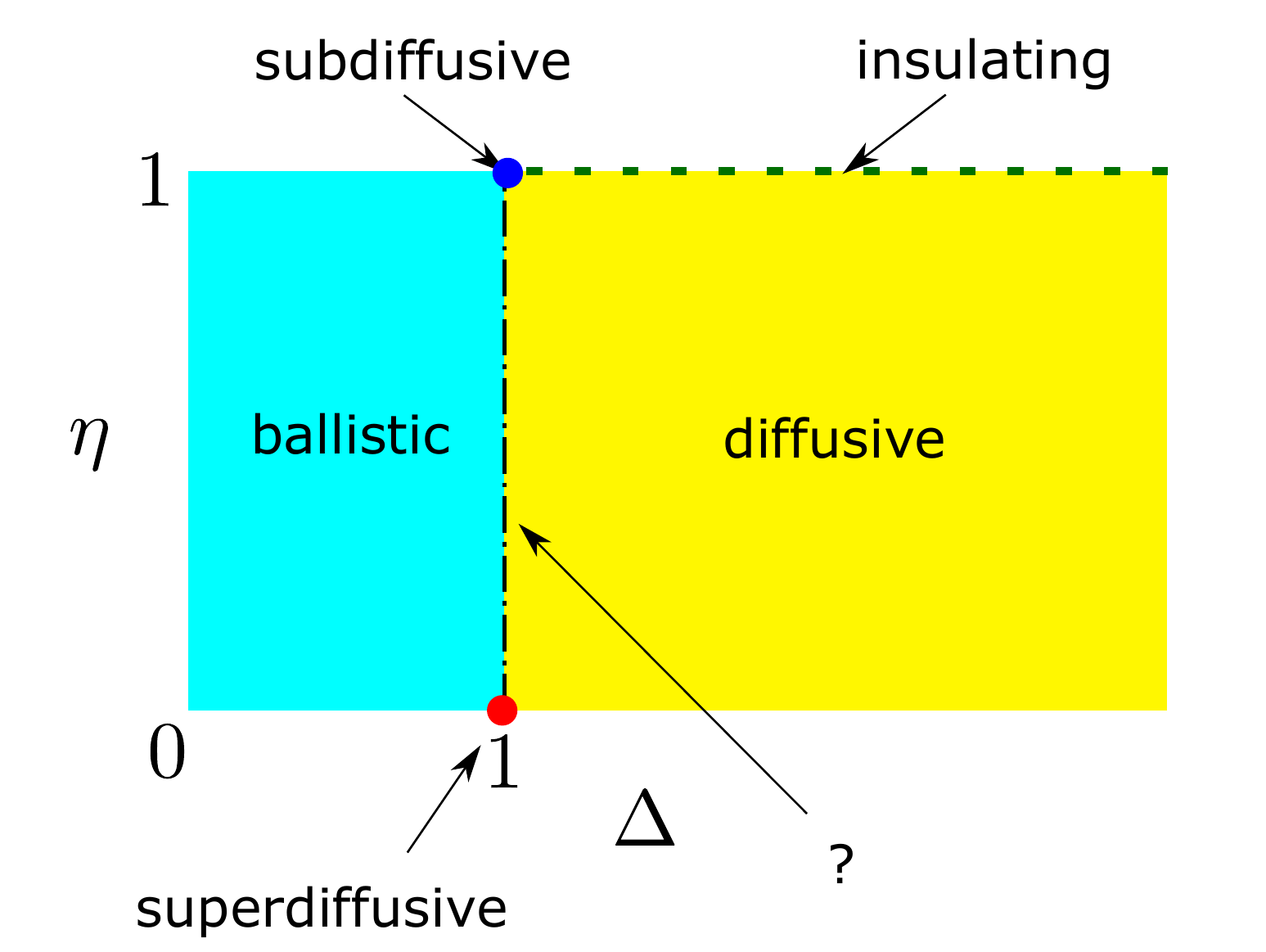} 
\caption{Phase diagram of magnetization transport regimes in the boundary-driven XXZ chain with driving (\ref{preamble_Lindblad_dissipator_sites2}) for different anisotropies $\Delta$ and driving bias $\eta$.}
\label{fig:XXZ_Map} 
\end{figure}

To illustrate the richness of transport behaviors in interacting systems, we review  some of the main results for boundary-driven XXZ spin chains [Eq.~\eqref{XXZ}] (see also~\cite{BertiniZnidaric2020} and  below for additional details).
We assume LME boundary driving given by Eq.~\eqref{preamble_Lindblad_dissipator_sites2}.  
The spin current transport diagram as a function of the anisotropy $\Delta$ and the bias $\eta_1=-\eta_L=\eta$ is shown in Fig.~\ref{fig:XXZ_Map}. 
For $\Delta < 1$,  transport is ballistic for any value of $\eta$.
For $\Delta=1$ it is superdiffusive for  $\eta\simeq 0$, with exponent $\alpha = -1/2$~\cite{Znidaric2011}, but subdiffusive for $\eta=1$, with exponent $\alpha=-2$~\cite{Prosen2011, Landi2015a} (c.f. Table~\ref{tab:scaling}). 
A superdiffusive regime was also predicted from hydrodynamic theories~\cite{DeNardisIlievski2019, GopalakrishnanVasseur2019}, which have recently been verified experimentally in ultracold atoms~\cite{JepsenKetterle2020} and  antiferromagnetic materials~\cite{ScheieTennat2021}~\footnote{The scenarios reproduced in the experiments are not boundary-driven.}. 
Superdiffusive transport was also observed for energy transport~\cite{BrenesGoold2020}.  
For $\Delta>1$ and $\eta<1$, numerical evidence suggests spin transport is diffusive~\cite{Znidaric2011,JesenkoZnidaric2011}, but at $\eta=1$ one finds insulating behaviour~\cite{Prosen2011}. 
The region $\Delta = 1$ for $0<\eta<1$ still requires further exploration.   
Similarly rich behavior can be found in non-interacting systems with long-range tunneling~\cite{PurkayasthaAgarwalla2021}.

%
As mentioned earlier, the energy current through an XXZ chain is ballistic because the corresponding current operator is a conserved quantity. 
%
However, for $\eta=1$ and $\Delta>1$, the spin current decreases exponentially with $L$, and the magnetization profile acquires the sigmoid form in Fig.~\ref{fig:profiles}. 
Such insulating spin current behavior may seem at odds with ballistic (without scattering) energy transport. 
But there is no contradiction because the energy current in this case is exactly $0$ due to the symmetries of the system~\cite{Schuab2016a}.

\subsection{Disordered and quasiperiodic systems}\label{ssec:dis_quasi} 


Disordered system have been intensely studied for decades, particularly since the discovery of Anderson localization~\cite{Anderson1958}: 
For 1D non-interacting systems, any small amount of disorder makes the system an insulator.
In interacting systems, Anderson localization  was studied  in~\cite{GiamarchiSchulz1987, GiamarchiSchulz1988}, a line of research that was invigorated after the discovery of many-body localization~\cite{GornyiPolyakov2005, BaskoAltshuler2006}. 
Comprehensive reviews can be found in~\cite{NandkishoreHuse2015, ParameswaranVasseur2018, AbaninPapic2017, LuitzLev2017, AbaninSerbyn2019}. 
Here, we focus on boundary-driven systems of the type~(\ref{preamble_Lindblad_dissipator_sites2}).
The transport properties depend significantly on whether disorder is uncorrelated, as in Anderson localization, or correlated, as in quasi-periodic potentials, such as in the Aubry-Andr\'e-Harper (AAH) model~\cite{Harper1955,AubryAndre1980, Hofstadter1976}. 
We thus review them separately in Secs.~\ref{sssec:dis} and~\ref{ssssec:quasi}.
%

\subsubsection{Uncorrelated disorder}\label{sssec:dis} 

The standard model for the study of transport in disordered system is an XXZ chain [Eq.~(\ref{XXZ})] with  local potentials $h_l$. 
For uncorrelated disorder, $h_l$ is taken from a uniform distribution on $[-h,h]$, where $h$  quantifies the disorder strength.
We have seen in Fig.~\ref{fig:XXZ_Map} that when $h=0$ and $\eta$ is small 
[c.f. Eq.~(\ref{preamble_Lindblad_dissipator_sites2})], the system is ballistic for $\Delta<1$ and diffusive for $\Delta>1$;
for $\Delta=1$ it is superdiffusive. 
We also know from Anderson's work~\cite{Anderson1958}  that, in the absence of interactions, the system becomes an insulator for any value of $h$.
Analytical results for an XX chain with large disorder can be found in~\cite{Monthus2017}.  

\begin{figure}[ht]
\includegraphics[width=\columnwidth]{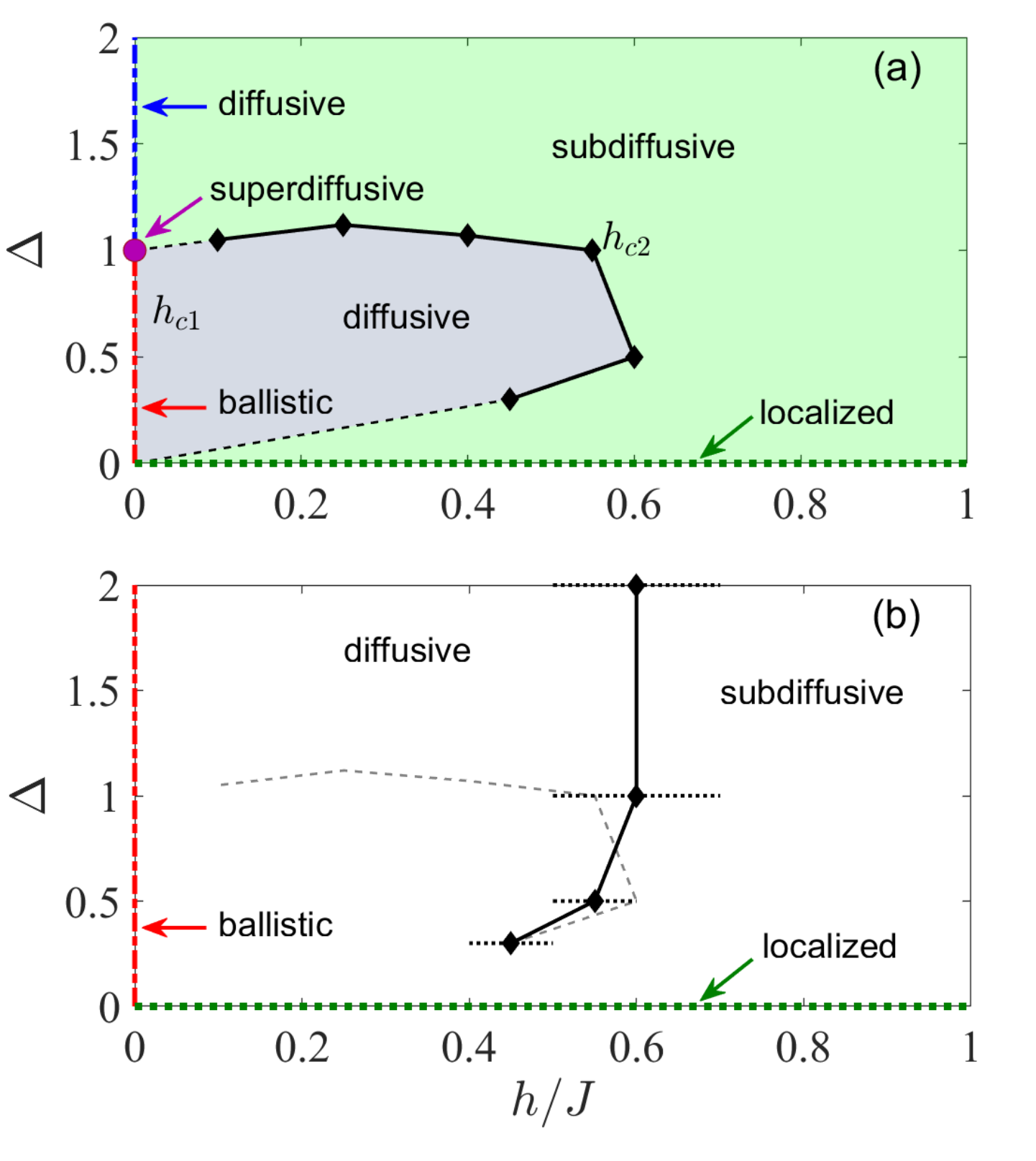}
\caption{Maps of spin (a) and energy (b) transport characteristics for the model~\eqref{XXZ} with local uncorrelated disordered potential $h_l$. (a) Boundary driving given by Eq.~\eqref{preamble_Lindblad_dissipator_sites}. The transition lines at $h_{c1}$ and $h_{c2}$ represent transitions from ballistic to diffusive and diffusive to subdiffusive, respectively. 
The transition region, represented by black dots, is evaluated from simulations with $L$ up to $400$ spins. Adapted from Ref.~\cite{ZnidaricVarma2016}. 
(b) Boundary driving as described in Sec.~\ref{sec:znidaricprosenbath}. The black continuous line shows the transition line between diffusive and subdiffusive transport, with the dotted lines indicating the confidence level. The grey dashed line represent the transition from diffusive to subdiffusive for particle transport as shown in panel (a). Adapted from~\cite{MendozaArenasScardicchio2019}. We thank J.J. Mendoza-Arenas and M. \v{Z}nidari\v{c} for providing the data.} 
\label{fig:MBL_Merged} 
\end{figure}

The high temperature transport diagram in the XXZ chain is depicted in Fig.~\ref{fig:MBL_Merged}(a), taken from~\cite{ZnidaricVarma2016}. 
These were simulated for small $\eta$ using tensor networks (Sec.~\ref{sec:tensor_networks}), with $L$ up to $400$.
For $\Delta>1$, the transport is subdiffusive for any $h>0$. 
Conversely, for $\Delta<1$, transport is diffusive for $h>h_{c1}=0$, and subdiffusive for $h>h_{c2}$.
For even larger disorder strengths $h>h_{c3}$, not shown in Fig.~\ref{fig:MBL_Merged}(a), the system is localized.  
The emergence of a subdiffusive region for spin current is aligned with other works~\cite{AgarwalDemler2015, PotterParameswaran2015, VoskAltman2015, BarLevReichman2015, AgarwalKnap2017, BeraEvers2017}, including experiments~\cite{LuschenBloch2017, BordiaBloch2017}. 
However, the origin of the subdiffusive behavior is still under investigation. 
In particular, \cite{AgarwalDemler2015, GopalakrishnanKnap2016, PotterParameswaran2015} point to the presence of rare regions with very large disorder, that can significantly slow down the dynamics, known as Griffiths phases~\cite{Griffiths1969}. 
Interestingly though, subdiffusive transport can also occur in systems with correlated disorder (Sec.~\ref{ssssec:quasi}), in which the above-mentioned regions do not occur~\cite{WeinerBera2019}.
Numerical simulations initially did not find any evidence of such regions~\cite{SchulzZnidaric2020}. 
\adp{However, more recent studies involving much larger systems together with} local dephasing [c.f.~Eq.~(\ref{bulk_dephasing})] of random magnitude, and acting on all sites except the boundaries, found clear signatures of the emergence of a Griffiths phase, signaled by power-law tails in the distribution of the resistance~\cite{TaylorScardicchio2020}.     

We next turn to energy transport. 
In ordered systems we previously showed it differs significantly from spin transport, since energy is a conserved quantity, while the latter is not. 
In disordered systems the local energy current is no longer a conserved quantity but, as we will see later, transport is strongly affected by the limiting case of disorder going to $0$, in which case the current is conserved and transport becomes ballistic.  
A diagram of the energy transport properties at high temperature and zero magnetization, in the disordered XXZ chain [Eq.~(\ref{XXZ})] is depicted in Fig.~\ref{fig:MBL_Merged}(b), adapted from~\cite{MendozaArenasScardicchio2019}. 
The boundary driving is modeled by the two-site reservoirs reviewed in Sec~\ref{sec:znidaricprosenbath}, i.e., baths which act on two sites and which are designed to locally thermalize these sites.
In particular, the authors used the target state, acting at the edges, such that $\rho_{T,l}\propto \exp\left[-\beta_l (H_l - \mu_l M_l)\right]$ where $H_l$ is the Hamiltonian acting on the two edge sites, and $M_l$  the total magnetization of $l$ and $l+1$.
Earlier studies of energy transport in the disordered XXZ model can also be found  in~\cite{VarmaScardicchio2017}, where two halves of the system were prepared in different states and then allowed to  evolve unitarily (c.f. Sec.~\ref{sec:extended});  diffusive transport was found for small disorder.  
In Fig.~\ref{fig:MBL_Merged}(b) we see that for any value of $\Delta$, as soon as $h>0$, energy transport goes from ballistic to diffusive, and only at a non-zero critical field it can become subdiffusive.
The dashed gray line in Fig.~\ref{fig:MBL_Merged}(b) shows the boundary between diffusive and subdiffusive transport for spin currents (Fig.~\ref{fig:MBL_Merged}(a)). 
\adp{For finite magnetization with $h=0$, both the particle and the energy current operators are ballistic. In this case,} the energy current mimics more closely the particle current~\cite{MendozaArenasScardicchio2019}.     

A qualitatively similar description is also found for a non $U(1)$ symmetric model, such as an XYZ chain with local disorder [Eq.~(\ref{Currents_XYZ})]. 
This was studied in~\cite{SchulzScardicchio2018}, where it was shown that the location of the transition between diffusive and subdiffusive occurs at disorder strengths which depend on the amount of the XY anisotropy ($J_x-J_y$).       

Energy transport was also studied in the localized phase, where it is  strongly suppressed. 
An interesting advantage of the localized phase is that one can exploit the $l$-bit representation~\cite{HuseOganesyan2014, RosScardicchio2015} based on extensive in number, quasi-local integrals of motion, which can be computed at polynomial cost as a function of the size~\cite{KhemaniSondhi2016, KulshreshthaSimon2019, WahlSimon2017}. 
This was used in~\cite{WuEckardt2019}, allowing them to go up to $100$ spins. 
Both in the non-interacting and weakly interacting regimes, the authors found that the temperature $T$ dependence of the heat conductivity $\sigma$ follows Mott's law~\cite{Mott1969} $\sigma \propto \exp[(-T_0/T)^{1/(d+1)}]$,
where $d$ is the dimensionality of the system and $T_0$ is an interaction-dependent temperature above which the conductivity becomes more sizeable. 

We  also  comment on systems beyond the XXZ chain, but not necessarily studied in the context of boundary-driven open quantum systems. 
One interesting setup is the Fermi-Hubbard model: It does not fully localize when the disorder does not differentiate between spin up and spin down~\cite{PrelovsekZnidaric2016}. 
Instead, the charge degrees of freedom can be localized and non-ergodic, while the spin degrees of freedom are delocalized. 
Full localization can be obtained  using  spin-dependent tunneling~\cite{SrodaMierzejewski2019}.
In Ref.~\cite{IadecolaZnidaric2019} the authors considered a spin ladder and showed that the presence of symmetries, independent of the strength of disorder, can be used to construct exponentially large subspaces, which can be localized or ballistic.    
Another interesting setup is a single particle in a lattice, with a disordered potential, and  coupled to a bosonic chain~\cite{PrelovsekMierzejewski2018, MierzejewskiBonca2019, MierzejewskiPrelovsek2020}~\footnote{Note that only Ref.~\cite{MierzejewskiPrelovsek2020} considers a boundary-driven setup.}. 
The authors modeled the effect of the latter using Fermi's golden rule, and also included a rate equation to describe the injection/removal of particles at the edges.
This  allowed them to study up to $10^4$ sites, and  also systems in two dimensions. 
In 1D,~\cite{MierzejewskiBonca2019} found that strongly interacting bosons (hard-core bosons), made the transport subdiffusive, while weak interactions  resulted in diffusive transport at long times. 
Conversely,~\cite{MierzejewskiPrelovsek2020}  showed that, while transport can be subdiffusive in 1D, in 2D it is always diffusive, even for large disorder.







\subsubsection{Correlated disorder}\label{ssssec:quasi} 


As discussed in the previous subsection, in 1D non-interacting  systems any small amount of uncorrelated disorder turns the system into an insulator. 
In other words, there is no \ind{mobility edge}, i.e., an energy threshold which differentiates localized from delocalized energy states.  
Mobility edges are  possible, however, with correlated disorder, such as quasiperiodic potentials. 
We consider a  non-interacting lattice of free fermions [Eq.~\eqref{tight_binding}], with local potential $h_{\alphaq,l}\equiv h_l$ given by~\cite{GaneshanDasSarma2015}    
\begin{align} 
h_{\alphaq,l} = \lambda \frac{2 \cos(2\pi \betaq l + \phi)}{1-\alphaq \cos(2\pi\betaq l + \phi)}\,, \label{eq:V_AAH}   
\end{align}   
where $\betaq = (1+\sqrt{5})/2$ is the Golden mean (other Diophantine numbers lead to similar results). 
This is a generalization of the Aubry-Andr\'{e}-Harper (AAH)
model, which is obtained setting $\alphaq=0$~\cite{Harper1955, AubryAndre1980, Hofstadter1976}.   
In the AAH model there exist a critical value of $\lambda$ below which all eigenstates are delocalized and above which all eigenstates are exponentially localized. 
This is already a remarkable difference from uncorrelated noise for which any amount of disorder localizes all eigenstates in 1D. 
The transport is found to be ballistic for $\lambda<1$ and insulating for $\lambda>1$. 
At  $\lambda=1$ all eigenstates are critical~\cite{Ostlund1983} and particle transport is subdiffusive, with  $\cur{} \propto L^{-1.4}$~\cite{PurkayasthaKulkarni2018, PurkayasthaKulkarni2017b}. 

Here, we add a note on the system sizes $L$ considered to compute $\cur{}$. 
It was shown in~\cite{PurkayasthaKulkarni2018, VarmaZnidaric2017, SutradharBanerjee2019} 
that \adp{particle} transport differs quantitatively  when $L$ is a Fibonacci number or not. 
For instance in~\cite{VarmaZnidaric2017} it was  shown that in the subdiffusive region ($\lambda=1$) the current scaled as $\cur{} \propto L^{-1.38}$ when $L$ are powers of $2$, while $\cur{} \propto L^{-1.27}$ when they are Fibonacci numbers. 
A related study, based on entanglement entropy, was put forth in~\cite{RoySharma2019}.
%
Another subtlety of the model is that, at $\lambda=1$,  one finds different transport coefficients for boundary-driven system and a unitary Green-Kubo analysis~\cite{PurkayasthaKulkarni2018}. 
In particular, the latter can predict superdiffusion (due to the fast spreading of the tails of a wavepacket). 
This has been shown to be due to the fact that the Green-Kubo formalism takes $L\to\infty$ first, and then $t\rightarrow \infty$; conversely, boundary drives can only take $L\to\infty$ since the infinite time limit is implicit~\cite{Purkayastha2019}. 
Hence, for systems which are sensitive to boundary conditions (as is the case here), this results in different predictions, depending on the method used. 
If one is  interested in finite systems, then boundary drives are the most accurate approach. 
Such subtleties are not present for diffusive transport~\cite{Znidaric2019}. 

\begin{figure}[ht]
\includegraphics[width=\columnwidth]{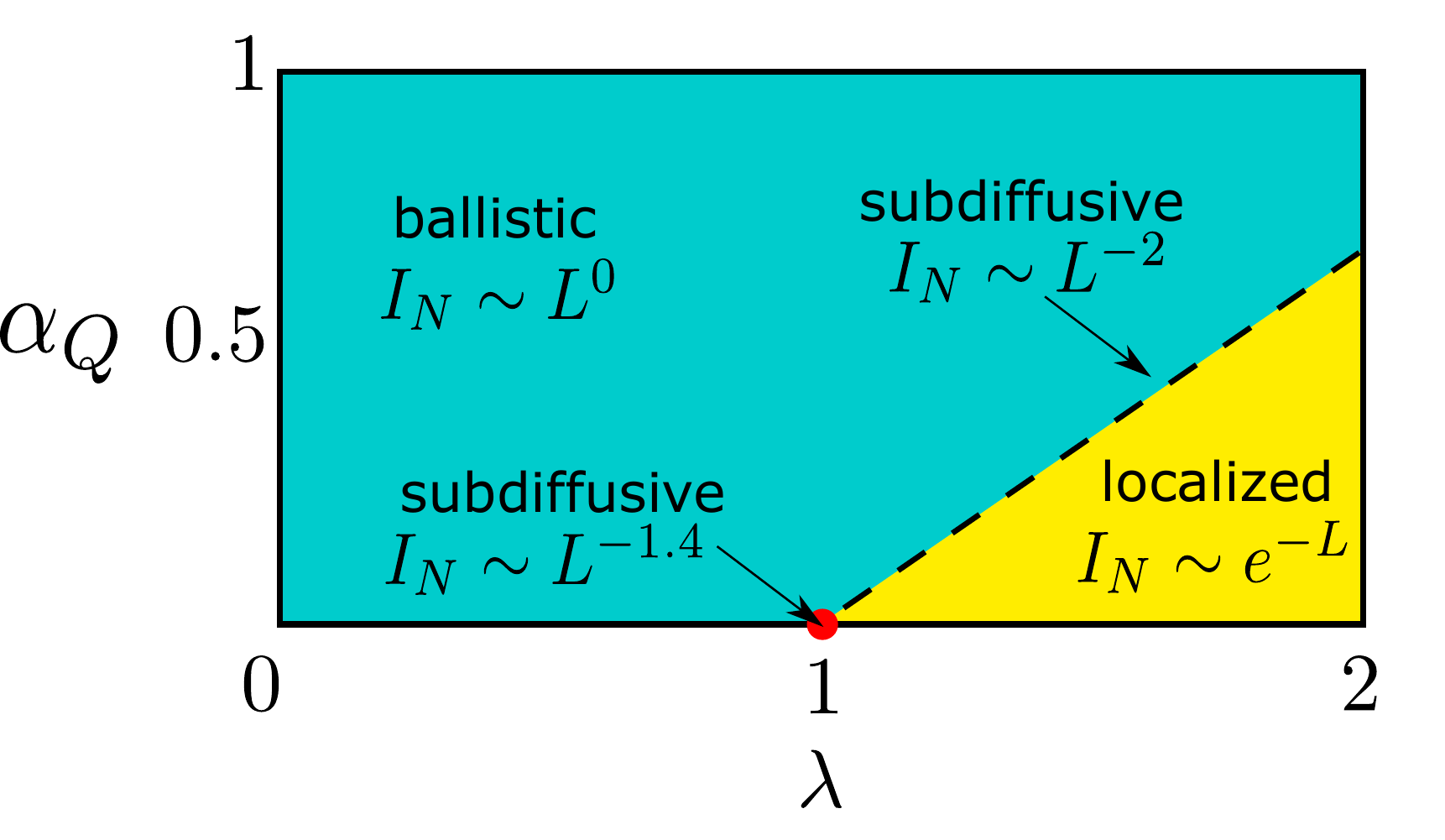}
\caption{Phase diagram of the \adp{high temperature} particle transport properties of the modified AAH model with potential (\ref{eq:V_AAH}). The ballistic region (top left, turquoise), is separated from the localized region (bottom right, yellow), by a subdiffusive line with $\cur{N}\sim L^{-2}$ and a subdiffusive point at $\alphaq=0$ and $\lambda=1$ with $\cur{N}\sim L^{-1.4}$. The symbol $\cur{N}$ stands for the NESS particle current and $L$ represents the system size. 
Adapted from Ref.~\cite{PurkayasthaKulkarni2017b}.}
\label{fig:AAH_map_Purkashtaya} 
\end{figure}

When $\alphaq > 0$ in Eq.~\eqref{eq:V_AAH}, the system presents a \ind{mobility edge}, i.e., below a certain energy all  eigenstates are delocalized, and above it they are localized (or vice versa). 
The presence of a mobility edge shifts the critical value of $\lambda$ at which transport goes from ballistic to insulating.
Furthermore, while the system is still subdiffusive at the critical value 
the exponent for $\alphaq>0$ is given by $\cur{N} \propto L^{-2}$, and thus differs from the case $\alphaq=0$. 
This is illustrated in Fig.~\ref{fig:AAH_map_Purkashtaya}, based on~\cite{PurkayasthaKulkarni2017b}, which presents the parameter regions with ballistic and diffusive \adp{particle} transport, and the critical line with the two  scalings  for $\alphaq=0$ and $\alphaq\ne 0$.  

A mobility edge can also be obtained by modifying the AAH model to include beyond nearest-neighbor tunneling~\cite{DasSarmaXie1988, BoersHolthaus2007, LiDasSarma2017}, which was studied experimentally by~\cite{LuschenBloch2018}. 
Power-law hopping  can result in algebraically localized modes, instead of exponential, and this can lead to different forms of zero-temperature particle transport, from ballistic to superdiffusive to insulating, as well as the presence of mobility edges in 1D~\cite{SahaPurkayastha2019}. 
Furthermore, when transport is modeled by B\"uttiker probes, see Sec.~\ref{sec:bulk_noise}, the current can depend nontrivially on the magnitude of the coupling to the probes~\cite{SahaAgarwalla2022}.

Equilibrium localization properties for an AAH potential ($\alpha_Q =0$) have also been studied  in the presence of interactions, both theoretically~\cite{IyerHuse2013, CookmeyerMoore2020} and experimentally~\cite{SchreiberBloch2015,  LuschenBloch2017}, also in higher dimensions~\cite{BordiaBloch2017}. 
Interactions with $\alphaq\ne 0$ were studied in Refs.~\cite{LiDasSarma2015, ModakMukerjee2015, ModakMukerjee2018}. 
Additionally, the possible existence of a non-ergodic phase violating the eigenstate thermalization hypothesis, while having volume law entanglement entropy, has been studied in~\cite{Srednicki1994}. 

Focusing on $\alphaq=0$,~\cite{ZnidaricLjubotina2018} showed that for the region of $\lambda\lesssim 1.5$, any small $\Delta$ makes the \adp{particle/magnetization transport} diffusive. 
However, this requires much larger system sizes, even up to sizes of $L=1000$ sites~\cite{Znidaric2020b}. 
Since for $\Delta = 0$, the system is insulating when $\lambda>1$, the diffusion constant diverges as  $\Delta\rightarrow 0$ when $\lambda<1$, and tends to zero otherwise.
These results indicate that even for strongly localized integrable systems, small integrability-breaking perturbations can result in diffusive dynamics. 
Ref.~\cite{ZnidaricLjubotina2018} showed that altering the potential at chosen sites can significantly affect the transport, thus opening another  door to engineer transport properties in quantum systems.      
    
\begin{figure}
    \centering
    \includegraphics[width=\columnwidth]{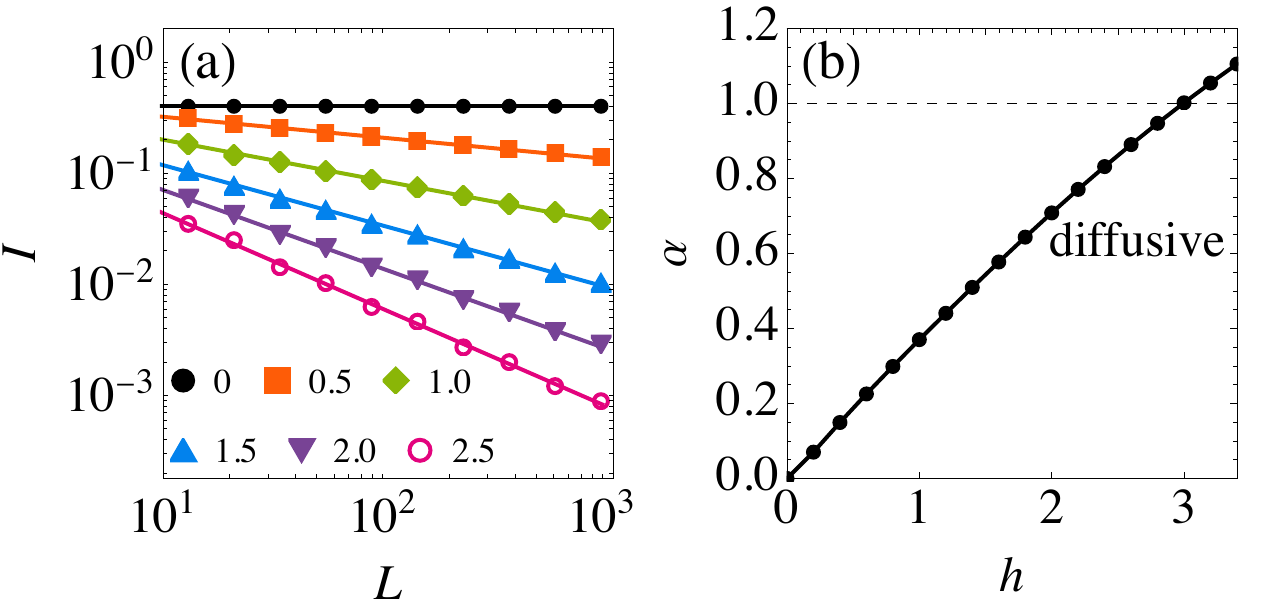}
    \caption{(a) Magnetization current as a function of $L$ for the non-interacting Fibonacci model~\eqref{eq:Fibonacci_V}, for different values of $h$, from $h=0$ (top) to $h=2.5$ (bottom) in steps of 0.5.
    For sufficiently large $L$, all curves behave like $\cur{} \sim 1/L^\alpha$ [Eq.~\eqref{eq:generic_cur_scaling}], with an exponent $\alpha$ that depends on $h$. 
    (b) $\alpha$~vs.~$h$, extracted from the simulations in figure (a).
    Adapted from~\cite{Lacerda2021}.}
    \label{fig:fibonacci}
\end{figure}

Another particularly interesting  quasi-periodic potential is the \ind{Fibonacci chain}, defined by the local potential~\cite{Ostlund1983, KohmotoTang1987, HiramotoAbe1988, HiramotoKohmoto1992, KohmotoTang1983, KaluginLevitov1986, SutherlandKohmoto1987}
\begin{align} 
h_l = \frac{h}{2} \left(2 V(l g) -1\right),
\label{eq:Fibonacci_V}
\end{align}
where $h$ is the magnitude, $g=(1+\sqrt{5})/2$  and $V(x) = \lfloor x + g\rfloor - \lfloor x \rfloor$, with 
$ \lfloor x \rfloor$ denoting the integer part of $x$.
Unlike the AAH model, the non-interacting ($\Delta=0$) Fibonacci model is critical for any value of $h$, showing eigensystem fractality, and anomalous transport~\cite{Ostlund1983, KohmotoTang1987, HiramotoAbe1988, HiramotoKohmoto1992, KohmotoTang1983, KaluginLevitov1986, MacePiechon2016, SutherlandKohmoto1987, VarmaZnidaric2017}. 
In fact, the particle transport varies continuously from ballistic to insulating as $h$ increases.
This is illustrated in Fig.~\ref{fig:fibonacci}, which shows simulations computed using the Lyapunov equation [Sec.~\ref{sec:non_interacting}].

Concerning interacting Fibonacci chains, in the low-temperature regime, bosonization techniques showed that repulsive interactions can lead to a metal-insulator transition, and a power-law dependence of the conductivity with $L$~\cite{VidalGiamarchi1999, VidalGiamarchi2001}. 
The high temperature case was studied in~\cite{MaceAlet2019, VarmaZnidaric2019}, where it was shown that for large enough  $h$ and $\Delta$, the system becomes subdiffusive and can show many-body localization. 
However, it is important to point out that, independent of the magnitude of $h$, for small $\Delta$ the current is diffusive. 
Further studies are required for larger values of $\Delta$, especially close to $\Delta=1$. 
In Ref.~\cite{ChiaracaneGoold2021}, although not a boundary-driven setup, the authors found signatures of superdiffusive behavior for $h< \Delta$, and  subdiffusive to insulating behavior for $h>\Delta$. 
Similar results were also found for the energy current in~\cite{ChiaracaneGoold2021b}.

In the AAH model [Eq.~(\ref{eq:V_AAH}) with $\alphaq=0$], the eigenstates for $\lambda=1$ are neither localized nor delocalized.
Instead, they are critical and display fractal properties. 
Conversely, in the Fibonacci model [Eq.~(\ref{eq:Fibonacci_V})], they are critical for any value of $h$. 
The fractal nature of the steady-state magnetization profile was shown in Ref.~\cite{VarmaZnidaric2017}.
To characterize said fractal structure, the authors had go to almost $40000$ sites, which was made possible by recasting the problem as a  Lyapunov equation [Sec.~\ref{sec:non_interacting}].        
To conclude, we mention that contrasting the AAH potential~\eqref{eq:V_AAH} with a uniform distribution may not provide a fair comparison between uncorrelated and correlated disorder, since the potential magnitudes are significantly different. 
An interesting workaround was pursued in~\cite{SetiawanPixley2017}, which used the AAH potential~(\ref{eq:V_AAH}) but with randomly chosen local phases $\phi$ on each site.
With this they showed that for small $\lambda$ and high temperatures, the correlated disordered model led to a higher current when compared with the uncorrelated one. 
Conversely, for large $\lambda$ the situation was inverted (although for both setups the currents decayed exponentially with $L$, since both are insulators).

\subsection{Dephasing and transport}
\label{ssec:dephasing_and_transport}

\gtl{In this section we discuss how transport is affected by the presence of dephasing baths (Sec.~\ref{sec:bulk_noise}) acting locally on all sites of a chain.
In short, dephasing is expected to always render the transport diffusive for a sufficiently large chain size $L$.
This is based on evidence from various models: First, in non-interacting chains, with homogeneous~\cite{AsadianBriegel2013,Malouf2018,Znidaric2010,Znidaric2011c, BermudezPlenio2013}, disordered~\cite{ZnidaricHorvat2013,TaylorScardicchio2020} and quasiperiodic~\cite{Dwiputra2021,Lacerda2021} potentials.
And second, in interacting systems, both homogeneous~\cite{MendozaArenaClark2014} and disordered~\cite{ZnidaricGoold2017}.}

To illustrate the idea, consider the nearest-neighbor homogeneous tight-binding chain studied in Sec.~\ref{sec:non_interacting}. 
The current, in this case, is given by
Eq.~\eqref{dephasing_current_tight_binding} which scales as 
\begin{equation}\label{NAT_dephasing_current_scaling}
    \cur{} = \frac{a}{b + \Gamma L}\,,
\end{equation}
where $a$ and $b$ are constants. 
In the absence of dephasing, $\Gamma \equiv 0$, the transport is ballistic. 
But for any non-zero $\Gamma$, there will always be a sufficiently large $L$ for which the current starts to behave diffusively, as $\cur{} \sim 1/L$.

\gtl{For other models, involving either interactions, disorder, etc., the functional form of the current $I(\Gamma,L)$ will no longer be given by Eq.~\eqref{NAT_dephasing_current_scaling}.
Notwithstanding, one can still predict that if $\Gamma$ is much larger than any other scale in the problem, any Hamiltonian contribution should be washed away, so the current is expected to behave similarly to  Eq.~\eqref{NAT_dephasing_current_scaling}:
\begin{equation}\label{NAT_current_scaling_large_Gamma}
    I(\Gamma,L) = \frac{c_\text{deph}}{\Gamma L}\,, 
    \qquad 
    (\Gamma \text{ large})\,,
\end{equation}
where $c_\text{deph}$ is a constant.
The interesting case, therefore, is when $\Gamma$ is moderate, as this should lead to an intricate competition between the Hamiltonian, which is responsible for the system's natural transport $\cur{} \sim 1/L^\alpha$, and the dephasing, which tends to make the transport diffusive ($I\sim 1/L$). 
In fact, in this case there should be a characteristic length scale $L_\Gamma$ at which the system transitions from one regime to the other~\cite{ZnidaricGoold2017}. 
That is
\begin{equation}\label{NAT_current_scaling_general}
    I(\Gamma,L) = \begin{cases}
    c_0/L^\alpha & L \ll L_\Gamma, \\[0.2cm]
    c_\Gamma/L & L \gg L_\Gamma,
    \end{cases}
\end{equation}
where $c_0$ and $c_\Gamma$ are constants, with the former being independent of $\Gamma$. 
The length scale $L_\Gamma$, where the transition occurs, can be estimated if one assumes the single-transport exponent hypothesis Eqs.~\eqref{spreading_exponent} and~\eqref{scaling_exponents_relation}, from which the typical length scale for the spreading of wavepackets is found to behave as $\sqrt{\langle \Delta x^2 \rangle} \sim \tau^{1/(\alpha+1)}$. 
Since dephasing introduces a characteristic time scale $\tau \sim 1/\Gamma$,  $L_\Gamma$ should scale as 
\begin{equation}\label{NAT_dephasing_scale}
    L_\Gamma \sim \Gamma^{-1/(\alpha+1)}.
\end{equation}
Since $\alpha>0$, $L_\Gamma$ is always decreasing in $\Gamma$. 
Hence, for large $\Gamma$ the diffusive behavior should become visible for small chain sizes, while for low $\Gamma$, very large $L$'s might be necessary.
}

Finally, the functional dependence of  $c_\Gamma$ in Eq.~\eqref{NAT_current_scaling_general} can also be estimated by requiring continuity when $L = L_\Gamma$. 
That is, $c_0/L_\Gamma^\alpha = c_\Gamma/L$, which yields 
\begin{equation}\label{coefficient_dephasing_gamma}
    c_\Gamma = c_0~ \Gamma^{(\alpha-1)/(\alpha+1)}\,.
\end{equation}
One can  use this predict the following:
Suppose $L$ is sufficiently large, so that $L \gg L_\Gamma$ for a wide range of $\Gamma$'s. 
Both~\eqref{NAT_current_scaling_large_Gamma} and~\eqref{NAT_current_scaling_general} will predict diffusive transport, but with a non-trivial dependence on $\Gamma$: 
\begin{equation}\label{dephasing_current_two_gamma_behaviors}
    I(\Gamma,L) L = \begin{cases}
    c_0 \Gamma^{(\alpha-1)/(\alpha+1)} & \Gamma \text{ small},\\[0.2cm]
    c_\text{deph}/\Gamma & \Gamma \text{ large}\,,
    \end{cases}
\end{equation}
where ``small'' and ``large'' are defined with respect to the energy scales of the system Hamiltonian. 
Particularly noteworthy, even though the transport is diffusive due to  dephasing, the exponent $\alpha$ of the dephasing-free system still plays a very clear role.

\gtl{The presence of noise is often expected to to reduce the current and thus deleterious for transport.
For instance, in the ballistic example of Eq.~\eqref{NAT_dephasing_current_scaling}, the current with dephasing is always smaller than that without it ($\Gamma=0$).
There are situations, however, where noise can actually be beneficial. 
The term \ind{dephasing-assisted transport} (or noise-assisted) is used to describe those situations in which  additional noise source increases the current. 
}

This effect was first investigated in~\cite{Plenio2008,OlayaCastro2008}, and a series of follow up papers~\cite{Caruso2009,RebentrostAspuruGuzik2009,Chin2010,Chin2012,DanielManzano2013,Contreras-Pulido2014,peaks1}.
The interest  was on light harvesting in biological molecules, modeled by  spin 1/2 (or tight-binding) networks.
The noise was introduced by means of a dephasing dissipator (Sec.~\ref{sec:bulk_noise}) acting on all sites of the lattice.
Experimental simulations  were carried out in optical setups~\cite{Viciani2015,Biggerstaff2016} and trapped ions~\cite{Cormick2016,Maier2019}. 
Related studies were also carried out using B\"uttiker probes (Sec.~\ref{sec:bulk_noise}) in~\cite{Kilgour2015,Kilgour2016}.

Dephasing-assisted transport always occurs in systems exhibiting \emph{subdiffusive} transport~\cite{ZnidaricGoold2017} (with insulators being a limiting case).
This can be seen from Eq.~\eqref{dephasing_current_two_gamma_behaviors}:
Subdiffusivity implies $\alpha > 1$ and hence $(\alpha-1)/(\alpha+1) > 0$. 
For small $\Gamma$ the function $\cur{} \sim \Gamma^{(\alpha-1)/(\alpha+1)}$ will thus be  monotonically increasing in $\Gamma$, which is \emph{exactly} the dephasing-assisted transport. 
This was explored in Ref.~\cite{ZnidaricHorvat2013,Lacerda2021}, who studied the XX model in the presence of disorder and Fibonacci potentials, respectively.
Mobility edges were studied in~\cite{Dwiputra2021}, and  applications to thermal machines  in~\cite{ChiaracaneGoold2021b}. 
In the context of many-body systems, this was studied in Refs.~\cite{Mendoza-Arenas2013,MendozaArenaClark2014}, who considered  interacting fermionic chains with LME dissipators~\eqref{preamble_Lindblad_dissipator_sites2} at the boundaries, and dephasing dissipators on all sites.
The system was simulated using tensor network methods (Sec.~\ref{sec:tensor_networks}) up to $120$ sites. 
Finally, in \cite{Mendoza_ArenasPlenio2014} the authors considered a system made by multiple lattices, and found an even larger transport enhancement by a type of noise that induces incoherent coupling to neighboring lattices.

\subsection{Negative differential conductance}\label{ssec:ndc}

In linear response, the greater the bias at the boundaries of a driven system, the larger is the response. 
For example, the larger the temperature difference, the larger the heat current, as visible in Fourier's law, Eq.~(\ref{history_diffusive_flux}). 
From the early works of Esaki~\cite{Esaki1958, EsakiStiles1966, EsakiTsu1970}, it was shown that it is possible to have \ind{negative differential conductance} (NDC) or resistance, which means that an increase of the bias would result in a decrease of the response. 
NDC is a key ingredient in the realization of transistors, including thermal ones~\cite{LiCasati2006}.      

Ref.~\cite{EsakiTsu1970} showed that superlattices, and the consequent presence of a structured density of states, can result in a resonant response which naturally leads to NDC. 
Many other mechanisms resulting in resonant behavior do as well, e.g. Coulomb blockade~\cite{HeijMooij1999}, spin-charge separation~\cite{CavaliereKramer2004}, ferromagnetic interactions~\cite{RolfPissarczykLoth2017}, conformational changes~\cite{MujicaKubiak2003}, or electron interaction with vibrational modes~\cite{GaudiosoHo2000, GalperinNitzan2005} in molecules.      

NDC has been observed in semiconductor quantum dots~\cite{WeisPloog1993}, carbon nanotubes~\cite{ZhouDai2000, PopDai2005} and small molecular systems~\cite{ChenTour1999, HalbritterWihaly2008}. 
It has been reviewed, with an emphasis on molecular electronics, in Refs.~\cite{ZimbovskayaPederson2011, XuDubi2015}, or in more general reviews on nano-electronics~\cite{Kastner1992, AnantramLeonard2006}. 
We thus focus on reviewing negative differential conductance in boundary-driven strongly interacting spin systems, which were not discussed in those earlier reviews. 

Due to interactions, spin chains are an ideal set-up for the emergence of NDC.
Recently, it was shown that it can also appear in classical Heisenberg chain models with varying magnetic field~\cite{Bagchi2015}. 
In~\cite{BenentiZnidaric2009, BenentiRossini2009}, the authors showed that strongly interacting XXZ chains [Eq.~(\ref{XXZ})] can present strong NDC under LME boundary driving.
In the notation of Eq.~\eqref{preamble_Lindblad_dissipator_sites}, strong bias ($f_1=1-f_L=0$) forces the edge spins to be either up or down.
If $\Delta>1$, the bulk Hamiltonian, together with the large bias, will favor the presence of two large ferromagnetic domains at the edges.
Because of their insulating nature this lowers the current, and for large enough interactions and biases, the system can turn into an insulator. 
The stability of such an insulating phase is due to the properties of the Hamiltonian, and for large enough interactions the steady state is given by two large domains, which can only be destabilized by magnons -- excitations at the boundary between the  domains. 
However, for large enough $\Delta$, such excitations are gapped and thus exponentially localized. 
As a consequence they cannot propagate and reach the baths, thereby prohibiting the decay of the system to a different state.
\adp{A different explanation of the phenomenon can be found in~\cite{Mendoza-Arenas2013}, where it is stressed that for $\Delta\gg 1$ the baths preferentially couple to high energy states with low conductivity. 
Specifically, one configuration becomes particularly important, which is the one with two, half-chain-long, ferromagnetic domains, as this is a dark state of the baths alone, for which the domain boundary is farthest from the baths. 
All other states are exponentially suppressed. }         

One can also ask on the robustness of NDC.
First, it is robust against various types of integrability-breaking terms, thus it is not related to the integrable nature of the XXZ chain. 
For instance, in Ref.~\cite{BenentiRossini2009} the authors also added an integrability-breaking staggered magnetic field and still observed it. 
In Refs.~\cite{MendozaArenaClark2014, DroennerCarmele2017} the authors considered instead a dissipative perturbation, as an additional dephasing in the local $\sz$ basis [Eq.~(\ref{eq:spin_dephasing})].  
Since the mechanism for such strong rectification is related to the properties of the bulk Hamiltonian,  dephasing is found to significantly reduce, or even  remove, NDC.
This is another manifestation of dephasing-assisted transport  [Sec.~\ref{ssec:dephasing_and_transport}]. 
Another way to remove NDC is to increase the range of the spin-spin interactions, or add strong disorder, as studied in Ref.~\cite{DroennerCarmele2017}. 
Long range interactions opposes the formation of two large and opposite  domains at the edges, thus resulting in a more gradual spin magnetization profile.
Particularly chosen local potentials can also significantly reduce the NDC for one bias and reinforce it for the opposite one, thus resulting in strong spin current rectification~\cite{LeePoletti2019, LenarcicProsen2015} [Sec.~\ref{ssec:rect}].


\subsection{Rectification}\label{ssec:rect} 




\indTwo{Rectification}{rectification} is an effect that has widespread application in electronics. %
It describes, in the widest sense, an asymmetric response of a two-terminal system to reflection.
The \ind{electric diode}, for instance, makes the flow of electric current in one direction significantly different from the other, if the driving bias is inverted. 
While electric diodes are widespread, the implementation of a \ind{heat diode} appears much more challenging.   
A key requirement is the presence of reflection symmetry breaking in the system. 
%
%
Given our focus on quantum boundary-driven systems, we are going to discuss rectification of different quantities, e.g. spin, particle and heat currents. 
We also mention key developments in classical rectification, which form the basis for the developments in quantum systems (for comprehensive reviews, see~\cite{LiLi2012,BenentiPeyrard2016}). 
There has also been significant interest in experimental realization of thermal rectifiers in various  setups, such as oxides~\cite{Starr1936, Brattain1951}, nanotubes~\cite{Chang2006}, nanoribbons~\cite{Hu2009}, quantum dots~\cite{ScheibnerMolenkamp2008} and systems with superconducting components~\cite{FornieriGiazotto2015, MartinezPerezGiazotto2015, BoursGiazotto2019, IorioGiazotto2021}. 
A review of  the experimental efforts can be found in~\cite{Roberts2011}.  

The rectification performance is measured by the \ind{rectification coefficient} $\Rect$ defined as 
$\Rect = -\cur{f}/\cur{b}$,
where $\cur{f}$, $\cur{b}$ are the current with the bias in the {\it ``forward''} and 
{\it ``backward''} (swapped) directions.
The minus sign is used because the current, after inverting the bias driving the current, is in the opposite direction.
An alternative figure of merit is 
$\Contr = \left| \frac{\cur{f}+\cur{b}}{\cur{f}-\cur{b}}\right|$,
which is sometimes also denoted the rectification coefficient, but which we will refer to as {\it contrast}.
A system which does not rectify has $\Rect = 1$ and $\Contr \rightarrow 0$, while a perfect rectifier (where the system is an insulator in one direction) has $\Rect \rightarrow 0,\infty$ and $\Contr\rightarrow 1$.
Since $\Contr$ is confined between $0$ and $1$, it helps to observe sharp transitions between non-rectifying and strongly rectifying behavior~\cite{Balachandran2018}. 


To better understand the requirements for rectification, we  consider the heat current through a system coupled to two baths ($L$ and $R$), at temperatures $T_L$ and $T_R$.
We can thus write $\cur{f}=\cur{}(\bar{T}, \Delta T)$ and $\cur{b}=\cur{}(\bar{T}, -\Delta T)$ where  $\bar{T}=(T_L+T_R)/2$ and $\Delta T = T_L - T_R$. 
Each current  can be written as a series expansion in powers of $\Delta T$:
\begin{align}
    \cur{}\left(\bar{T}, \Delta T\right) = \sum_{n>0} \alpha_n \cur{n}\left(\bar{T}\right) \Delta T^n\,. \label{eq:exp_cur_deltaT}    
\end{align}
It is thus clear that in order to have rectification,  $\cur{}(\bar{T}, \Delta T)$ should be nonlinear in $\Delta T$, and the expansion in Eq.~(\ref{eq:exp_cur_deltaT}) should include even powers of $n$. 
For a study on the role of quadratic and quartic conductivities as a function of temperature, see~\cite{YangZhang2018}. 

For rectification to occur, it is necessary to break reflection symmetry. 
This can be done by considering either (i) a non-reflection symmetric system~\cite{ChioquettaDrumond2020, WangLifa2019}, (ii) equivalent baths with asymmetric couplings~\cite{PurkayasthaKulkarni2016} or (iii) different baths (e.g. baths with different magnetic fields~\cite{ArracheaAligia2009} or different statistics~\cite{WuSegal2009, WuSegal2009b}). 
These conditions, however, are necessary but not sufficient.

%

Interactions tend to play a key role in rectification. 
For instance, considering the Landauer formula~\eqref{EQ:landauer}, for non-interacting transport, one finds that  when the transmission function is reflection-symmetric, 
there will be no rectification:
Reversing the bias amounts to exchanging  $f_L(\omega)\leftrightarrow f_R(\omega)$, and the current just reverses sign. 
This therefore excludes rectification effects for a broad class of models.
In contrast, Ref.~\cite{MeirWingreen1992} provided a formally exact current formula for transport through an interacting region, which has also recently been done for  boundary-driven LMEs~\cite{jin2020a}.
Said formula is not antisymmetric under reservoir exchanges, and hence
rectification is in principle possible. 
%
%
In the remainder of this section we review general ways of obtaining nonlinear responses and thus rectification. 
Sec.~\ref{sssec:non_linearities} discusses the role of nonlinearities and interactions within the system, while Sec.~\ref{sssec:asym_baths} considers asymmetric baths.      



\subsubsection{Nonlinearities and interactions within the system}\label{sssec:non_linearities}   

The authors of~\cite{Terraneo2002, Li2004a}  greatly advanced our understanding of heat rectification in classical systems. 
They showed that to obtain a nonlinear response, one can build  rectifiers from a concatenation of chains of nonlinear  oscillators.
%
In this case, rectification occurs because each portion of the chain has a very different, temperature-dependent, power spectrum. 
It results that for a certain pair of temperatures $T_L, T_R$, there is a large overlap between the spectra of the different portions of the chain (resulting in large currents), but if the temperatures are swapped, this overlap is significantly suppressed (resulting in small currents). 

Nonlinear responses have  been used to rectify the propagation of waves~\cite{LepriCasati2011, MascharenasGerace2016}, and they have also been employed in quantum many-body systems.
A prototypical example was presented in~\cite{WerlangValente2014}, which considered two spins coupled via a $\sigma^z_1\sigma^z_2$ coupling. 
Each spin feels a different local magnetic field, and is coupled to a different heat bath. 
The interplay between the structure of the system Hamiltonian, and the coupling of the system to the baths, can result in perfect rectification (zero current in one direction). 
This approach was later pursued for more than two spins in~\cite{Pereira2019}.

Considering extended boundary-driven systems, one of the first works to give strong rectification was~\cite{Balachandran2018}, which considered an XXZ chain [Eq.~(\ref{XXZ})] segmented in two halves,  one with anisotropy $\Delta=0$, and the other with  $\Delta\ne 0$~\footnote{With a Jordan-Wigner transformation [Eq.~\eqref{Jordan_Wigner}], the XXZ chain can be seen as a fermionic lattice, where the anisotropy is plays the role of the interactions.}.
The authors considered spin transport with LME dissipators acting on the 1st and last sites, as in Eq.~\eqref{preamble_Lindblad_dissipator_sites}.
One bath had $f = 0.5$ (infinite temperature) and the other $f=0$.
It was shown that for anisotropy $\Delta > 1+\sqrt{2}$, the system had very large rectification ($\Rect\approx 10^4$) even for small chains. 
And in the thermodynamic limit, the system became a perfect heat diode. 
This occurs because, in reverse bias, the baths drive the system towards a state with a large excitation gap, which thus is stable towards spin excitations arising from the Hamiltonian. In the forward bias the state is not gapped, so excitations can propagate. 

An analysis of the stability of this rectifier, as well as further properties, were provided in Refs.~\cite{Balachandran2018, LeePoletti2019, LeePoletti2021b}.  
For instance, if  $\Delta$ is non-zero in both sides of the chain, then the value of $\Delta$ at which the energy gap occurs in reverse bias increases with the system size~\cite{LeePoletti2021}. 
As a result, there is no perfect insulator in the thermodynamic limit, although the system can still become a very strong rectifier. 
Interestingly, this setup can also be used to rectify heat, as studied in Ref.~\cite{Balachandran2019}, with possibly large values of heat current rectification ($\Rect\approx 400$). 
Adding a ring structure in the middle of the chain can increase  rectification, which occurs together with the formation of entanglement between the spins in the ring~\cite{PoulsenZinner2021}.

An XXZ spin chain which also shows large spin  rectification, is one with homogeneous anisotropy, but local magnetic fields  pointing in different directions in each half of the chain~\cite{LenarcicProsen2015,LeePoletti2019}.
%
%
Gradual changes of the local magnetic field also result in rectification~\cite{LifaLi2009,Pereira2017,LandiKarevski2014}. 
In particular, \cite{LandiKarevski2014} showed that when $\Delta \equiv 0$, the rectification is identically zero, underlining once again the importance of interactions.      
Moreover, in Ref.~\cite{Pereira2017}, the author showed that in spin chains described by LMEs, rectification can also reach an extreme situation where even the direction of the current does not change sign, as the baths are inverted, i.e. the energy current flow in the same direction even after inverting the baths. 
Even though this seems thermodynamically inconsistent at first sight,  it can actually be explained by the existence of a work term in LMEs~\cite{DeChiara2018,Pereira2018} (c.f. Eq.~\eqref{EQ:sigma_dot_local}).

The role of interactions in rectification is also prominent in bosonic chains~\cite{PurkayasthaKulkarni2016} and quantum dots~\cite{Stopa2002, ScheibnerMolenkamp2008, ChenLei2008, PoltlBrandes2009, KuoChang2010, MichaelisBeenakker2006, SvenssonLinke2013,SierraSanchez2014, MarcosViciosoSanchez2018, TangWang2018, Zimbovskaya2020}. 
In Ref.~\cite{SchallerCelardo2016}, the authors relied on collective effects such as \ind{configurational blockade} to significantly enhance the rectification. 
Alternatively, rectification can also be enhanced using long-range interactions~\cite{Pereira2013,Chen2015}.

\subsubsection{Nonlinearities in the baths and role of quantum statistics}\label{sssec:asym_baths}  

Important insights into the origin of nonlinear terms in Eq.~(\ref{eq:exp_cur_deltaT}) were developed in~\cite{Segal2008, WuSegal2009, WuSegal2009b}. 
The authors considered systems coupled to two reservoirs, and found two sufficient conditions for the emergence of thermal rectification:
Baths with different energy-dependent densities of states; and different statistical properties of system and  baths, plus asymmetric system-bath couplings. 
They studied the Hamiltonian $\Hs=\sum_n E_n \ket{n} \bra{n}$, weakly coupled to the baths via $\Hi = \sum_\nu S \lambda_{\nu}B_{\nu}$, with $B_{\nu}$ an operator acting on bath $\nu$, $\lambda_{\nu}$ a coupling constant and $S=\sum_{nm} S_{n,m} \ket{n} \bra{m}$, an operator acting only in the system. 
They considered  Markovian baths and weak coupling, such that the evolution could be described by a Pauli master equation for the populations $p_n$ of the system energy levels [see Eq.~(\ref{EQ:GLOBAL_PAULI})]  
\begin{align*}
    \dot{p}_n = \sum_{\nu,m} |S_{n,m}|^2 p_m(t) k^{\nu}_{m\rightarrow n}(T_{\nu}) - p_n(t)\sum_{\nu,m}|S_{n,m}|^2 k^{\nu}_{n\rightarrow m}(T_{\nu})\,,
\end{align*}
with transition rates 
\begin{align}
    k^{\nu}_{m\rightarrow n}(T_{\nu}) = \lambda_{\nu}^2 \int_{-\infty}^{\infty} d\tau e^{\im (E_{m}-E_n)\tau} \langle B_{\nu}(\tau) B_{\nu}(0) \rangle_{T_{\nu}}  ,    
\end{align}
where $\langle B_{\nu}(\tau) B_{\nu}(0) \rangle_{T_{\nu}}$ are bath \ind{correlation functions} [Eq.~(\ref{EQ:res_corr_func})].   
%
The authors then observed that if the system is a harmonic oscillator with $\Hs = \omega a^\dagger a$
and $S \propto a+a^\dagger$, then
\begin{equation*}
    \cur{HO}(T_L, T_R) = - \frac{\omega\left[ n_{\rm B}(\omega,T_L) - n_{\rm B}(\omega,T_R) \right]}{n_{\rm B}(-\omega,T_L)/k^L(T_L) + n_{\rm B}(-\omega,T_{R})/k^R(T_R)}, 
\end{equation*}
where $k^{\nu}(T_{\nu})=k^{\nu}(T_\nu)_{n\rightarrow n-1}$ is independent of $n$, and where $n_{\rm B}(\omega,T_{\nu})= ( e^{\omega/T_{\nu}} - 1)^{-1}$.
%
Conversely, if one has a two-level system $\Hs=\frac{\omega} 2 \sigma^z$, with  $S=\sigma^x$, then
\begin{align*}
    \cur{S}(T_L, T_R) = - \frac{\omega\left[ n_{\rm S}(\omega,T_L) - n_{\rm S}(\omega,T_R) \right]}{n_{\rm S}(-\omega,T_L)/k^L(T_L) + n_{\rm S}(-\omega,T_R)/k^R(T_R) }, \label{eq:cur_spin}  
\end{align*}
with $n_{\rm S}(\omega,T_{\nu})= ( e^{\omega/T_{\nu}} + 1)^{-1}$. 
In both cases the numerator is antisymmetric with respect to a swap  $T_L \leftrightarrow T_R$, but the denominator may not be. 
It is thus sufficient to study the denominator to find criteria for the occurrence of rectification.         

The decay rate $k^{\nu}(T)$ can be written as~\cite{WuSegal2009b}
\begin{align}
    k^{\nu}(T) = 2\pi\lambda_{\nu}^2 \frac{ \int d\epsilon e^{-\epsilon/T}\rho_{\text{d},\nu}(\epsilon) g_{\nu}(\epsilon,\omega)}{\int d\epsilon e^{-\beta\epsilon}\rho_{\text{d},\nu}(\epsilon) },
\end{align}
where $\rho_{\text{d},\nu}(\epsilon)$ is the density of states (see Sec.~\ref{sec:quasistatic}) of bath $\nu$, while $\lambda_\nu^2 g_{\nu}(\epsilon,\omega)$ characterizes the system-bath coupling [c.f.~Eq.~(\ref{EQ:spectral_density})]. Assuming $\lambda_L=\lambda_R$ and $g_L=g_R$, it follows that if  
\begin{align}
    \frac{ \int d\epsilon e^{-\epsilon/T}\rho_{\text{d},L}(\epsilon) g_L(\epsilon,\omega)}{\int d\epsilon e^{-\beta\epsilon}\rho_{\text{d},L}(\epsilon) } \ne \frac{ \int d\epsilon e^{-\epsilon/T}\rho_{\text{d},R}(\epsilon) g_R(\epsilon,\omega)}{\int d\epsilon e^{-\beta\epsilon}\rho_{\text{d},R}(\epsilon) }, \label{eq:rect_cond}
\end{align}
then rectification is possible. 
For example, as long as $g_{\nu}(\epsilon,\omega)$ depends on $\epsilon$, it is sufficient that $\rho_{\text{d},L}(\epsilon)\ne \rho_{\text{d},R}(\epsilon)$, and at least one of them depends on $\epsilon$ as well.
%
Thus, for instance, if both baths are harmonic with constant density of states, there is no rectification.

An example of this behavior, in the  context of quantum spin chains, can be found in Ref.~\cite{ArracheaAligia2009}. 
The authors considered an XX  chain coupled on one edge to a semi-infinite XY chain, and on the other to a semi-infinite XX chain, which assumed the role of the reservoirs. 
Each bath was subject to different magnetic field biases, resulting in rectification.
%
Ref.~\cite{Bijay2021} showed that, while it is not possible to obtain rectification in a completely harmonic system plus bath setup, it is possible to obtain rectification when \adp{different portions of each edge of the chain are coupled to baths at different temperature}.
%
%
Similarly, Ref.~\cite{Pereira2017b} showed that classical harmonic chains subject to temperature-dependent potentials could also lead to rectification.



Another sufficient condition for the emergence of rectification, first discussed in~\cite{WuSegal2009, WuSegal2009b}, is when system and baths are composed of particles with different statistics. 
An intuitive way to think about it is that, if the bath is harmonic and the system is a spin,  the latter can be thought of as a strongly anharmonic system (as is exemplified by a Holstein-Primakoff transform), and thus there is an effective nonlinearity in the overall system plus baths setup.   
Another way to look at this is to note that 
different quantum statistics make it  impossible to describe transport with linear scattering theory (see e.g. Ref.~\cite{BenentiWhitney2017}).
A difference in statistics is in fact the mechanism which allows the linear models described in~\cite{YanLi2009, SauloPereira2020, Balachandran2019a} to show rectification.   
    
%

The fact that rectification is  possible does not imply that it will be strong.
Ref.~\cite{Balachandran2019a}  considered a bosonic chain with a quasi-periodic potential (first introduced in~\cite{GaneshanDasSarma2015}) and two identical spins baths at different temperatures. 
%
%
This set-up 
allows rectification because of the different statistics in the system and in the baths, and because the quasi-periodic potential breaks reflection symmetry.
The quasi-periodic potential also induces mobility edges [c.f. Eq.~(\ref{eq:V_AAH})] which significantly enhances the rectification. 
%
%
%
In fact,  the authors showed that this could be used to tune the magnitude of the rectification over three orders of magnitude.  
This is because localized modes can be moved into different positions by tuning the potential parameters, and in the scenarios in which one localized mode is at one of the edges of the system, but no other localized mode is present at the other edge, then strong rectification emerges.

%

\subsection{Beyond 1D systems}\label{ssec:beyond1d}

Until now we have discussed transport in  boundary-driven systems which are mainly 1D. 
Extending beyond 1D geometries opens various possibilities. 
First, systems which are integrable in 1D may not be integrable in 2D, e.g. the 1D XXZ chain is integrable, but an XXZ ladder, or the 2D XXZ model, are not.
Second, there is more flexibility on what parameters to tune, and one can also introduce qualitatively different new terms, such as gauge fields, in the Hamiltonian.
%
%
%
Here we focus on ladders dissipatively driven far from equilibrium. 
Results obtained via the Kubo formula, for closed quantum systems, were reviewed in~\cite{BertiniZnidaric2020}.  



Spin ladders are setups which clearly show that integrability is not strictly associated with ballistic transport, nor that non-integrability corresponds to diffusive transport. 
The following refers to results  for local dissipators of the form~(\ref{preamble_Lindblad_dissipator_sites2}).  
In Ref.~\cite{Znidaric2013b}, the author considered an integrable spin ladder and showed that while for finite magnetization the transport is ballistic, in the zero magnetization sector transport can be anomalous. 
For non-integrable ladders the author found numerical evidence of diffusive transport.
However, in Ref.~\cite{Znidaric2013} it was shown that a non-integrable spin ladder, with XX coupling for bonds in the legs and XXZ coupling within the rungs (i.e., $\Delta\ne 0$), has invariant subspaces which permit ballistic spin transport within them.  
The existence of such subspaces, with different transport properties, has been investigated as a tool to control the transport properties of the system~\cite{ManzanoHurtado2014, ManzanoCao2016, ThingnaCao2016, ManzanoHurtado2018, ThingnaCao2020}. 
In the Fermi-Hubbard model, analyzed in Ref.~\cite{ProsenZnidaric2012} for small driving $\eta_i$ [see Eq.~(\ref{preamble_Lindblad_dissipator_sites2}) after conversion from fermions to spins], the authors showed the emergence of diffusive transport except for the non-interacting limit, where it was ballistic, and for infinite interaction strengths, where it was anomalous. 
Moreover, for extreme boundary drivings $\eta_i=\pm 1$, the system  became an insulator.


The addition of an extra dimension also allows to explore the effect of different Hamiltonian terms and the resulting phases of matter. 
For instance, one can consider a non-interacting bosonic ladder with gauge fields in each leg:
\begin{align}\label{eqn:Hs}
      H_\text{VM} &= -\left(  J^{\|} \sum_{l,p}  e^{i\left( -1 \right)^{p+1}\phi/2} \; \aopd_{l,p}\aop_{l+1,p}
 + J^{\perp}\sum_{l} \aopd_{l,1}\aop_{l,2} + \text{H.c} \right)\nn
 &\qquad+ V\sum_{l,p}\aopd_{l,p}\aop_{l,p}\,. 
\end{align}
Here $\phi$ is the gauge field phase, $\aop_{l,p}$ ($\aopd_{l,p}$) is the bosonic annihilation (creation) operator at the $l$-th rung and $p$-th leg of the ladder, and $J^{\|}$, $J^{\perp}$ are the tunnelling amplitudes along the legs and rungs respectively. 
Such a system presents a phase transition between a Meissner phase, in which the ground state has a current circulating on the edges,  but not within the rungs (known as chiral current), and a vortex phase for which the current in the ground state presents a different number of vortices~\cite{Kardar1986, Granato1990}. 
This system has been studied both theoretically (including interactions) and experimentally, with different techniques and experimental setups~\cite{Kardar1986, Granato1990, DennistonTang1995, Nishiyama2000, OrignacGiamarchi2001, DonohueGiamarchi2001, vanOudenaardenMooij1996, AtalaBloch2014}.
However, only recently the effects of the phase transition on the boundary-driven transport properties have been explored~\cite{GuoPoletti2016, RivasMartinDelgado2017, GuoPoletti2017b, XingPoletti2020}. 

Given the non-interacting nature of the problem, transport is always ballistic, independently of the strength of the gauge field $\phi$. 
However, tuning $\phi$ can significantly alter the current across the ladder, and within the system. 
For instance, in Ref.~\cite{GuoPoletti2016} the authors considered LME baths of the type~\eqref{preamble_Lindblad_dissipator_sites2} (adapted for bosons), and showed that chiral currents can  emerge, but only if the baths are coupled to certain sites.
They also showed that it is possible to tune the system into a perfect insulator. 
Importantly, the non-equilibrium properties of this system depend significantly on whether the two bands of the spectrum have an energy gap between them or not (see further discussion in Sec.~\ref{ssec:phasetransitions} on phase transitions). 
As the gap opens between the bands, e.g. by varying  $\phi$ or $J^{\perp}$, the current is significantly reduced.
In Ref.~\cite{RivasMartinDelgado2017} the authors considered a similar setup, under the effect of GME thermal baths [as in Eq.~(\ref{EQ:me_bms})], and showed the robustness of the chiral, heat and particle currents versus the presence of disorder.
In Ref.~\cite{XingPoletti2020} the authors used non-equilibrium Green's functions, which unlike GME/LME, is exact, to study chiral, heat and particle currents for a broad range of system parameters. 
They showed how, at low temperatures, these three currents were significantly different depending on whether the underlying ground state is in the Meissner or vortex phase, and on the presence/absence of a gap between the two bands of the energy spectrum of the system. 
For instance, if one of the temperatures is large enough to allow transport via the upper band, then even a small amount of particle current in this band would result in a significant change to the heat current, especially if the gap between the bands is large. 
If one compares the current pattern in the ground state with the one in a low-temperature NESS, it is also found that the pattern is more robust to the coupling with the baths for parameters such that the ground state is in the Meissner phase, while it is less robust for an underlying vortex phase. 
The boundary-driven bosonic system with a gauge field has also been studied for interacting particles, in particular hardcore bosons in Ref.~\cite{GuoPoletti2017b}, where it was found that the effect of the gauge field on the current becomes less prominent as the filling increases.

Similarly to ladders, in ring configurations the magnetic field can significantly affect the transport properties. 
To be more specific,  we focus on rings with two external baths coupled to different sites. 
A ring can behave as an Aharonov-Bohm interferometer~\cite{AharonovBohm1959} which is sensitive to an external magnetic field, as long as the ring is smaller than the phase-coherence length. 
Ring set-ups have been studied intensely, in particular when they contain quantum dots; see Refs.~\cite{YacobyShtrikman1995, SchusterShtrikman1997, HolleitnerBlick2001, SigristBichler2004} and Refs.~\cite{Akera1993, YeyatiButtiker1995, HackenbroichWeidenmuller1996, BruderSchoeller1996} for  pioneering experimental and theoretical works, respectively. 
Various phenomena have been observed, including resonant tunneling~\cite{ShahbazyanRaikh1994,MourokhSmirnov2002} and cotunneling~\cite{Akera1993, LossSukhorukov2000}, Fano physics~\cite{KubalaKonig2002, SilvaGefen2002, UedaEto2003, KobayashiIye2003}, Kondo correlations~\cite{GerlandOreg2000, HofstetterSchoeller2001,BoeseSchoeller2002, KimHershfield2002}, the influence of Coulomb interaction on transport, on quantum coherence and current statistics~\cite{BruderSchoeller1996, HackenbroichWeidenmuller1996,  KonigGefen2001, KonigGefen2002, Weidenmuller2003, hiltscher2010a, UrbanFazio2008, urban2009a}, and also entanglement~\cite{LossSukhorukov2000}. 
We have provided here some of the key references, but we acknowledge that a fair discussion of transport in rings and quantum dots set-ups would require its own review.  
%
%
One interesting aspect of rings is that currents can follow  different paths, depending on the parameters. 
For instance, while overall the current enters the system at one point and leaves in another, within the ring it could follow an overall clock-wise or counter-clock-wise path, or even move on both sides of the ring in the same direction~\cite{XuPoletti2019b}, while spin and energy currents can follow different patterns. 

Last but not least, we also point out that molecules are a prominent example of beyond 1D system, in which it is very important to study transport properties. 
Given the vastness of the research done on molecular transport, we direct the reader to Refs.~\cite{DubiDiVentra2011, NitzanRatner2003, Tao2009, ZimbovskayaPederson2011, SegalAgarwalla2016}.

\subsection{Phase transitions }\label{ssec:phasetransitions}


%
Phase transitions represent abrupt transitions that take place when one changes an external parameter, such as temperature.
In classical phase transitions, such as the melting of ice, the driving of the transition are thermal fluctuations, while quantum phase transitions (QPTs)~\cite{Sachdev2011}, such as the superfluid-to-Mott-insulator transition~\cite{FisherFisher1989, JakschZoller1998, GreinerBloch2002}, are driven by quantum fluctuations\footnote{This includes not only the more common ground-state transitions, but also 
\ind{excited state} phase transitions~\cite{cejnar2021a}, in which the level density changes abruptly, and \ind{topological phase transitions}~\cite{hasan2010a}, where some eigenstates undergo topological changes.}.
%
%
%
Conversely, \ind{dissipative phase transitions}, are driven by a competition between non-equilibrium fluctuations produced by the coupling to the environment, and internal system parameters.

To provide an example of a dissipative phase transition, consider the electron shuttle, which -- analogous to a ball-pendulum between two capacitor plates~\cite{kim2015a} -- can be realized by a single electron transistor mounted on a harmonic oscillator system~\cite{gorelik1998a}.
The single electron transistor can be modeled by a LME (e.g. of Redfield type such as Eq.~\eqref{EQ:redfieldII_sp}) to describe electronic tunneling to the leads, and the oscillator by a local Fokker-Planck equation (equivalent to an oscillator-local master equation of Redfield-type in the classical limit)~\cite{novotny2003a,strasberg2021a}.
The tunneling depends non-linearly on the oscillator position, and the phase transition manifests as a sharp onset of autonomous self oscillations.
This transition has been studied in various works, highlighting e.g. the FCS of electron transfers~\cite{flindt2004a} (Sec.~\ref{sec:full_counting}), or its thermodynamic properties~\cite{waechtler2019a} (Sec.~\ref{sec:transport_and_currents}).
Further experimental~\cite{park2000a,scheible2004a,moskalenko2009a,koenig2012a} and theoretical~\cite{joachim2000a,shekhter2003a,galperin2007a,galperin2008a} studies demonstrate the use of LMEs to explain dissipative non-equilibrium phenomena.

There are many natural questions that one can then ask regarding phase transitions and boundary-driven systems, for example (i) how does a phase transition affect the transport properties of a system? (ii) can transport be used as an indicator of the occurrence of a phase transition? (iii) are the character and position of the critical point, and nature of the phases, altered or induced by the presence of the reservoirs?

%

%
In Markovian systems described by a GKSL master equation~(\ref{EQ:GKSL}), the Liouvillian $\Liouv$, instead of the Hamiltonian, becomes the central object of interest~\cite{morrison2008a,Kessler2012, minganti2018a}. 
Thus, while a QPT can be associated with the ground-state of the Hamiltonian, a dissipative transition is associated with the NESS, which is nothing but the (right) eigenstate of $\Liouv$ with vanishing eigenvalue [c.f.~Eq.~\eqref{vec_eigenvectors}]. 
Similarly, the Hamiltonian energy gap, which closes at the critical point for QPTs is replaced by the gap with the next eigenvalue of $\Liouv$~\cite{minganti2018a}.
In some systems, dissipation only affects a transition that already occurs in the isolated system, e.g. shifting the critical point~\cite{morrison2008a,bhaseen2012a,DalidovichKennett2009}. In other models, it may induce entirely new behavior.
%

In previous sections we have seen that in systems driven by LMEs, transport properties can be significantly altered at the transition point. 
For instance, in the XXZ chain, transport changes from ballistic to diffusive to insulating as $\Delta$ crosses the critical value $\Delta_c=1$  (Fig.~\ref{fig:XXZ_Map}). 
At $\Delta_c=1$, transport can be superdiffusive for small boundary driving, and subdiffusive for large boundary driving. 
As mentioned in Sec.~\ref{ssec:dis_quasi}, it is also possible to have phase transitions without interaction.
For instance, in the AAH model, which uses the potential of Eq.~(\ref{eq:V_AAH}), the system is insulating in one phase and ballistic in the other, and shows anomalous tranport at the transition  (Fig.~\ref{fig:AAH_map_Purkashtaya}). 
Another well-studied example, see also Sec.~\ref{ssec:beyond1d}, is a ladder in the presence of a magnetic field (c.f.~Eq.~(\ref{eqn:Hs})). 
The QPT takes the ground state  from a Meissner phase, characterized by currents only at the edges, to a vortex phase, in which the ground state has vortices of currents. 
The transport properties of this setup have been discussed in Sec.~\ref{ssec:beyond1d}.
%
%
%
%
Here we add that there is also a significant change in the relaxation time scale towards the steady state~\cite{GuoPoletti2017}. 
More precisely, the rapidity of the slowest decaying state scales  as $L^{-3}$, except at the phase transition where it goes as $L^{-5}$. 
As shown in~\cite{Prosen2008, Znidaric2015}, this is due to the change in the momentum $k$ dependence of the energy dispersion relation from $\omega(k)\propto k^2$ to $k^4$. 
Interestingly, the model in Eq.~(\ref{eqn:Hs}) shows also an excited state QPT in which a gap opens in the density of states. 
%
%

Continuing the review of non-interacting systems, in~\cite{ProsenPizorn2008} the authors considered the $XY$ model [Eq.~(\ref{Currents_XYZ}) with $J_z=0$]. 
In particular, they parametrized the Hamiltonian with $J_x=J(1+\delta)/2$, $J_y=J(1-\delta)/2$, $h_i=h$ and a coupling to baths described by an LME of the type of Eq.~(\ref{preamble_Lindblad_dissipator}). 
They found the existence of a critical magnetic field magnitude $h_c/J=1-\delta^2$: 
Below this critical value, the single particle correlations present long-range order and the operator space entanglement entropy (OSEE), which is a measure of the complexity of the operator~\cite{ProsenPizorn2007, Prosen2009}, grows linearly with $L$.
Above the critical value the correlations decay and the OSEE saturates to a constant value. 
This transition is not related to a phase transition in the ground state of the system, but to dynamical properties of the system~\cite{ProsenPizorn2009}. 
The out-of-equilibrium transition can also be observed when using a Redfield-II master equation (\ref{EQ:redfieldII_ip}), as shown in~\cite{ProsenZunkovic2010}. 

In bosonic systems one can also consider the effects of Bose-Einstein condensation. 
In~\cite{VorbergEckardt2013} the authors showed that the presence of multiple macroscopically occupied states can significantly enhance transport in  non-interacting bosonic chains.

As already mentioned above, the fact that a phase transition affects transport also implies that
with GMEs one can use the current to identify signatures of phase transitions occurring already in isolated systems in~\cite{VoglBrandes2011, SchallerBrandes2014}. 
An example is the critical nature of the eigenstates of the AAH or Fibonacci models (Sec.~\ref{ssec:dis_quasi}).
%
In such scenarios, the authors of~\cite{VarmaZnidaric2017} found that the magnetization profile shows a fractal behavior.

Phase transitions can turn systems into 
a strong rectifier (see Sec.~\ref{ssec:rect} for a more in-depth discussion), and this phenomenology can, in itself, be a probe of the phase transition, \adp{see e.g.~\cite{Balachandran2018} and Sec.~\ref{ssec:rect}}.   
%
%
In a completely different set-up, discussed in~\cite{SchallerCelardo2016}, the authors considered a system  coupled differently to each of the two baths, and observed that when the system undergoes a phase transition the rectification becomes significantly stronger. 
It is a challenging task to develop  accurate models to describe the out-of-equilibrium dynamics  close to criticality.
Recent advances in this direction were obtained in~\cite{waechtler2020a}, which used polaron transforms and reaction coordinates (Secs.~\ref{sec:reaction_coordinates} and~\ref{sec:polaron}) to obtain thermodynamically consistent equations.
%

\section{Summary and Outlook} 

In this manuscript we  reviewed theoretical aspects of boundary-driven systems, focusing on models to describe them, Sec.~\ref{sec:models}, techniques to analyze them, Sec.~\ref{sec:methods}, and  the emerging phenomenology, Sec.~\ref{sec:properties}.
The models studied in Sec.~\ref{sec:models} have revealed that the local and global master equations are complementary tools capable to address different parameter regimes of open non-equilibrium systems.
Transport theory shows in particular that for a proper thermodynamic interpretation the computed currents must properly reflect the corresponding energy balances between system and reservoirs.
We have also exposed methods to explore the strong-coupling regime.
However, \gsc{apart from exact solutions}, these methods are only capable of exploring small regions of the parameter space.
This means that none of these methods should be used naively beyond their range of validity.
In particular GKSL approaches can be deceptive in the sense that a formally correct solution can be very far from being physically correct. 

Many important questions remain largely unexplored. 
For example, already within the exposed theoretical framework, it would be interesting to consider varying degrees of locality by treating only certain internal junctions of a multipartite system perturbatively.
This would yield dissipators acting nontrivially only on a few sites of a larger system, and it would be interesting to learn how the spectral changes of random Liouvillians~\cite{timm2009a,can2019a,lange2021a}, as a function of locality~\cite{denisov2019a,wang2020a}, are linked to physical phenomena.
Furthermore, while symmetries in the full counting statistics such as the fluctuation theorem~\cite{crooks1999a} are naturally reflected in GMEs~\cite{Esposito2009,campisi2011a}, they are less well explored in LMEs.
For example, the standard thermodynamic uncertainty relation for Pauli-type rate equations~\cite{barato2015a,gingrich2016a} need not be respected by LMEs or Redfield equations~\cite{liu2021a}. 
The fluctuation theorem alone would enforce weaker uncertainty relations~\cite{timpanaro2019a,hasegawa2019a}. 
To explore the long-time limit of strongly interacting systems, or simply very large systems with small energy gaps, would also require to advance the corresponding theoretical description: 
Here, both GMEs and LMEs have issues, since approximations like the secular one may fail.
With such issues resolved, it would be interesting to consider the dynamics and transport characteristics of larger and strongly interacting systems~\cite{navez2010a,queisser2014a}.
Systems for which a rich dynamics can be expected are e.g. Fermi-Hubbard models~\cite{wu2019a,kleinherbers2020a,wu2020a,kolovsky2020a} or higher-dimensional spin systems (for example  higher-dimensional lattices or irregular spin networks~\cite{farhi2001a} with quantum information applications).
Then, one could go beyond mere heat transport applications, and also revisit the development of  correlations inside the system from the perspective of information processing.
Still, with their contractive properties, GKSL master equations can only treat a fraction of all possible dynamics.
Self-oscillatory systems~\cite{gorelik1998a,novotny2003a} e.g. require some nonlinear ingredient and it is questionable whether this can be achieved with GKSL approaches.

These arguments encourage the search for alternative approaches for treating open systems.
The few exactly solvable models can assist as benchmarks in this endeavor.
Small steps include discarding or avoiding certain approximations, such as the secular one~\cite{hartmann2020a,mccauley2020a}.
More ambitiously, one may also consider methods to find approximations to the full Kraus map dynamics, aiming at a picture consistent with thermodynamic laws.

As for the methods discussed in Sec.~\ref{sec:methods}, we would like to add a few comments.   
Boundary-driven problems are usually remarkably difficult to solve, owing to the large Hilbert space dimension, together with the need for working with density matrices, instead of pure states. %
In this aspect, the use of tensor network methods is currently revolutionizing the field, with new approaches being invented all the time, see Sec.~\ref{sec:tensor_networks}.
Tools such as entanglement entropy greatly assisted in the development of tensor networks for closed systems, but are not readily extensible to open dynamics. 
This is part of the challenge that must be overcome. 
The other challenge is to gain a better understanding of the typical tensor structures of boundary-driven systems. 
Most open system models use MPSs. 
But other structures, such as tree tensor networks or neural network states could offer advantages in certain cases. 

Once again, analytical solutions offer valuable insight. 
The results discussed in Sec.~\ref{sec:analytical_MPA} clearly highlight the underlying tensor structure of the NESS but, unfortunately, hold only for very special setups.
In parallel to the development of novel tensor network algorithms, the discovery of new models amenable to analytical treatment is therefore highly desirable.

Another very related challenge is in the description of boundary-driven systems with different geometries. 
This does not necessarily mean higher dimensionality; it could very well mean more complex bond structures, such as the network illustrated in Fig.~\ref{fig:f_sketch_localsetup}. 
These systems appear frequently in biological processes, molecular transport, nuclear magnetic resonance, optomechanics and quantum-dot devices. 
Most solution methods can handle this kind of geometry, but few are actually optimized for it.
Bridging this gap is another important open challenge in the field. 

In Sec.~\ref{sec:properties} we have reviewed the different transport properties and phenomena that can emerge in boundary-driven quantum systems.   
We have described systems with very different transport properties, i.e., insulating, subdiffusive, diffusive, superdiffusive and ballistic; and showed phase diagrams for prototypical models in clean conditions,
with disorder and with a quasi-periodic potential (Figs.~\ref{fig:XXZ_Map}-\ref{fig:AAH_map_Purkashtaya}).
%
Many of these diagrams, however, are still incomplete.
%
For instance, in Fig.~\ref{fig:XXZ_Map}, it is still unknown what the transport properties in the vertical line for anisotropy $\Delta=1$ are.
%
The models presented are also often studied only close to infinite temperature, as only in this regime can one investigate large enough systems to be able to  extract the transport exponent.
This is made worse by the very small values that the currents tend to reach.
%
%
%
%
The same type of problem occurs when studying negative differential conductance (Sec~\ref{ssec:ndc}) or rectification (Sec~\ref{ssec:rect}), which  occurs primarily when the bath imposes  large biases, and not small ones near the infinite temperature state. 
%
%
%
%
%

For interacting systems, most studies have relied on tensor network methods to describe the steady states (Sec.~\ref{sec:tensor_networks}).
However, for large system sizes the bond dimensions needed to describe the steady state accurately can be too large to make computations practical. 
This is particularly relevant in systems beyond the nearest neighbor 1D chains. 
In Sec.~\ref{ssec:beyond1d} we  showed that geometries such as ladders, or 2D systems, can have significantly different properties compared to their 1D counterpart. 
For instance an XXZ chain could be integrable and ballistic, while an XXZ ladder with the same magnitude of anisotropy would be non-integrable and diffusive~\cite{Znidaric2013b}. 
Even more interesting, because of the presence of ulterior symmetries in the ladders, it is even possible that the same system is ballistic in some symmetry sectors but diffusive in others~\cite{Znidaric2013}. 

Going beyond 1D nearest neighbor chains also allows one to introduce other terms in the Hamiltonian, such as gauge fields on plaquettes, which can induce quantum phase transitions in the ground state or and abrupt changes in the density of states (excited state QPTs),  significantly affecting transport across the system. 
Interestingly, in ladders such QPTs can occur already for non-interacting systems~\cite{Kardar1986, Granato1990}, and hence can be studied  up to very large system sizes. 
However, interactions can affect their transport properties, and this is still a vastly unexplored area. 
%
%
%

Phase transitions can significantly affect the transport properties of the system, as reviewed in Sec.~\ref{ssec:phasetransitions}.
%
%
We stress here that this can be seen from two parallel and equally interesting points of view. 
From one side, one can vary a parameter to induce a phase transition, as a way to control transport properties in a device. 
Conversely, since a phase transition can drastically change the transport properties, one can design detectors which function by measuring currents in systems which can go through a phase transition, if some external parameters are varied.
Additionally, the impact of topological and excited state phase transition on transport properties requires more research~\cite{benito2016a}.

%
Another topic currently under intense study concerns the complexity of boundary-driven steady-states. 
This has been analyzed through the operator space entanglement entropy (OSEE; Sec.~\ref{ssec:phasetransitions}),  the levels statistics of the steady state~\cite{ProsenZnidaric2013}, or  the effective Hamiltonian~\cite{SaProsen2020c}. 
%
The OSEE in 1D steady states may follow an area or a volume law. 
The level spacing of the steady-state, or the effective Hamiltonian, can follow a Poisson distribution for integrable or localized systems, or a Wigner-Dyson distribution for non-integrable ones. 
This mirrors the behavior of unitary chaotic systems, in which the change from Poisson to Wigner-Dyson of the energy level spacing  is associated with a passage from regular to chaotic dynamics~\cite{Casati1980, BGS1984, AlessioRigol2016}. 
For the overall open system setup, one needs to take into account that rapidities, unlike eigenvalues of a Hamiltonian, can be complex numbers and thus one needs to consider the distance in a 2D space~\cite{SaProsen2020, LiChan2021}.  
Recently, it was shown that Liovillians of GKSL form with randomly chosen jump operators~\cite{can2019a, CanGopalakrishnan2019} tend, in the thermodynamic limit, to a universal lemon-like shaped spectrum of rapidities in the complex plane~\cite{denisov2019a,timm2009a,wang2020a}. 
Further analyses on random Liouvillians were undertaken e.g. in Refs.~\cite{lange2021a,CanGopalakrishnan2019, SaProsen2020b, SaProsen2020c}. 
Depending on the symmetries in the system, other universality classes in the level-spacing distribution of the rapidities have been uncovered~\cite{HamazakiUeda2020}, and it has been shown in an increasing number of works that the level spacing statistics of non-integrable open systems tend to one of these classes~\cite{HamazakiUeda2020, SaProsen2020, RubioGarcia2021, LiChan2021, AkemannProsen2019}.    
There has also been recent developments in  understanding the relation between  the scaling of the rapidities with system size, and the scaling of the current~\cite{MoriShirai2020}.
The relaxation gap, in particular does not necessarily scale in the same way as the current~\cite{Znidaric2015}. 

A more recent field of research, which was not covered in this review, is that of periodically driven systems. 
For example, Ref.~\cite{ProsenIlievski2011} found that the driving could result in the emergence of long-range spin-spin correlations.
In Ref.~\cite{PurkayasthaDubi2017} the authors realized that the instantaneous current can be orders of magnitude larger than the average current,  making it experimentally measurable. 
Periodic drivings bring a whole new dimension to the study of transport because they can be used to model engines, refrigerators and heat pumps, c.f. Refs.~\cite{BenentiWhitney2017,Lacerda2022, VinjanampathyAnders2016, Kosloff2013, MillenXuereb2016, GooldSkrzypczyk2016, VANDENBROECK2015,SwitkesGossard1999, KaestnerKashcheyevs2015, Thouless1983, XiaoNiu2010, PekolaAverin2013}.  
The modeling of open quantum systems subject to time-dependent driving must be considered carefully, however. 
For periodic driving and weak coupling, one may rely on Floquet master equations~\cite{GrifoniHanggi1998,Breuer2002}. 
In the strong coupling regime, one may instead employ methods simulating both  system and  bath
(Secs.~\ref{sec:star_to_chain_thermofield} and~\ref{sec:extended}), which are extendable to driven systems.

Finally, in this review we have not covered experimental papers in details, although we have cited a number of experimental results in various sections. 
An important future research direction is the  proposal of novel experiments in new setups, such as ultracold atoms~\cite{DamanetDaley2019b} or trapped ions~\cite{BermudezPlenio2013},  both of which allow a high degree of control~\cite{BlochZwerger2008, BlattRoos2012} and can lead to applications in atomtronics~\cite{SeamanHolland2007, AmicoKwek2017, AmicoYakimenko2020}. 
We highlight, for instance, experiments which use ultracold atoms
to create two reservoirs, connected by a tunable channel through which both particles and energy can flow~\cite{BrantutEsslinger2012, BrantutGeorges2013, HusmannBrantut2015, LebratEsslinger2018, HusmannEsslinger2018, KrinnerEsslinger2016, StadlerEsslinger2012, KrinnerEsslinger2015}.
%
Even more importantly, experiments would be the first step towards the design and production of new materials and devices, such as tunable spin and heat current rectifiers, transistors and sensors. 

\begin{acknowledgments}
The authors would like to thank the referees involved in the peer-review process, for the detailed and constructive reports, which helped us significantly improve the quality of this manuscript.
We also thank B. K. Agarwalla, V. Balachandran, G. Benenti, B. Bu\v{c}a, A. Eckardt, J.P. Garrahan, J. Goold, C. Guo, Z. L. Lim, E. Pereira, J. J. Mendoza-Arenas, T. Prosen, A. Purkayastha, L. S\'{a}, A. Scardicchio,  D. Segal, P. Stegmann, S. R. Taylor, J. Thingna, X. Xu, and M. \v{Z}nidari\v{c} for feedback on early draft of this work. We thank in particular J. J. Mendoza-Arenas and M. \v{Z}nidari\v{c} for sharing their data for Figs.~\ref{fig:MBL_Merged}.      
D.P. acknowledges support from the Ministry of Education of Singapore AcRF MOE Tier-II (projects No. MOE2018-T2-2-142 and MOE-T2EP50120-0019). 
G.S. acknowledges support by the Helmholtz high-potential program and support by the DFG (project ID 278162697 - CRC 1242).
G.T.L. and D.P. acknowledge the financial support and hospitality of the International Centre for Theoretical Physics (ICTP) in Trieste, Italy, where this project began.
\end{acknowledgments}

\appendix

\section{Exact solution via Laplace transforms}\label{APP:exact_solution}

Here, we provide details for the exact time-dependent solution discussed in Sec.~\ref{ssec:connection_exact}.
The star representation has the advantage that, by computing the \ind{Laplace-transform} $C_{k\alpha}(z) = \int_0^\infty \f{c_{k\alpha}}(t) e^{-z t} dt$ we can eliminate the reservoir modes
\begin{align}
    C_{kL}(z) &= \frac{c_{kL}}{z+\ii\epsilon_{kL}} - \ii \frac{t_{kL} D_1(z)}{z+\ii\epsilon_{kL}}\,,\nn
    C_{kR}(z) &= \frac{c_{kR}}{z+\ii\epsilon_{kR}} - \ii \frac{t_{kR} D_N(z)}{z+\ii\epsilon_{kR}}\,,
\end{align}
leaving only the Laplace-transformed equations for the system operators to solve
\begin{align}
    z D_a(z)-d_a &= -\ii \sum_j h_{aj} D_j(z)\qquad:\qquad 2 \le a \le N-1\,,\nn
    z D_1(z)-d_1 &= -\ii \sum_j h_{1j} D_j(z) -\ii \sum_k \frac{t_{kL} c_{kL}}{z+\ii\epsilon_{kL}}\nn
    &\qquad- \sum_k \frac{\abs{t_{kL}}^2}{z+\ii\epsilon_{kL}} D_1(z)\,,\nn
    z D_N(z)-d_N &= -\ii \sum_j h_{Nj} D_j(z) -\ii \sum_k \frac{t_{kR} c_{kR}}{z+\ii\epsilon_{kR}}\nn &\qquad- \sum_k \frac{\abs{t_{kR}}^2}{z+\ii\epsilon_{kR}} D_N(z)\,.
\end{align}
Thereby, one can express the Laplace-transformed system operators $D_j(z)$ in terms of the initial system ($d_j$) and reservoir ($c_{k\alpha}$) operators, and by performing inverse Laplace transforms one can compute in principle all observables exactly.

To be more specific, we now consider the case $N=2$.
Then, we can recast the equations for the system annihilation operators as
\begin{align}
    G^{-1}(z) 
    \left(\begin{array}{c}
    D_1(z)\\
    D_2(z)
    \end{array}\right)
    &=  \left(\begin{array}{c}
    d_1\\
    d_2
    \end{array}\right)
    -\ii \sum_k    \left(\begin{array}{c}
    \frac{t_{kL}}{z+\ii\epsilon_{kL}} c_{kL}\\
    \frac{t_{kR}}{z+\ii\epsilon_{kR}} c_{kR}
    \end{array}\right)\,,\\
    G^{-1}(z) &= \left[z \cdot \id + \ii \left(\begin{array}{cc}
    h_{11} & h_{12}\\
    h_{21} & h_{22}
    \end{array}\right) + \sum_k \left(\begin{array}{cc}
    \frac{\abs{t_{kL}}^2}{z+\ii\epsilon_{kL}} & 0\\
    0 & \frac{\abs{t_{kR}}^2}{z+\ii\epsilon_{kR}}
    \end{array}\right)\right]\,,\nonumber
\label{eq:appendix_temporary141029831092}
\end{align}
where $G_{ij}(z)$ is the Green's function. By inversion, we can write this in the form
\begin{align}
    D_i(z) &= G_{i1}(z) d_1 + G_{i2}(z) d_2\nn
    &\qquad-\ii \sum_k \frac{t_{kL} G_{i1}(z)}{z+\ii\epsilon_{kL}} c_{kL} 
    -\ii \sum_k \frac{t_{kR} G_{i2}(z)}{z+\ii\epsilon_{kR}} c_{kR}\,.
\end{align}
Employing the initial product assumption
\begin{align}
    \rho_0 = \rhos^0 \otimes \frac{e^{-\beta_L (\Hb^{(L)} - \mu_L \Nb^{(L)})}}{Z_\text{B}^{(L)}} \otimes \frac{e^{-\beta_R (\Hb^{(R)} - \mu_R \Nb^{(R)})}}{Z_\text{B}^{(R)}},
\end{align}
one then finds via the inverse Laplace transform (Bromwich integral),
that system observables can be expressed in terms of initial expectation values as
\begin{align}\label{EQ:sysocc_exact}
    \expval{d_i^\dagger d_j}_t &= g_{i1}^*(t) g_{j1}(t) \expval{d_1^\dagger d_1}_0 + g_{i2}^*(t) g_{j2}(t) \expval{d_2^\dagger d_2}_0\nn
    &\qquad+ g_{i1}^*(t) g_{j2}(t) \expval{d_1^\dagger d_2}_0 + g_{i2}^*(t) g_{j1}(t) \expval{d_2^\dagger d_1}_0\nn
    &\qquad+ \sum_k \abs{t_{kL}}^2 g_{i1kL}^*(t) g_{j1kL}(t) f_L(\epsilon_{kL})\nn
    &\qquad+ \sum_k \abs{t_{kR}}^2 g_{i2kR}^*(t) g_{j2kR}(t) f_R(\epsilon_{kR})\,,
\end{align}
where the initial reservoir properties are encoded in the Fermi functions $f_\alpha(\epsilon_{k\alpha})$ and $g_{ij}(t)$ and $g_{ijk\alpha}(t)$ are the inverse Laplace transforms of $G_{ij}(z)$ and $-\ii \frac{G_{ij}(z)}{z+\ii\epsilon_{k\alpha}}$, respectively.

As a first observation, \gsc{although for finite $L_\alpha$ the evolution is quasi-periodic,} a stationary long-term limit may arise from the exact dynamics:
When we make the reservoirs larger $L_\alpha\to\infty$, their single-particle spectra from Eq.~\eqref{EQ:diagonal_form} continuously cover the interval $[\epsilon-2\tau,\epsilon+2\tau]$ \gsc{and we can use the spectral coupling density of Eq.~\eqref{EQ:specdens_chain}}.
To invert the Laplace transform, we can then analytically calculate expressions like
\begin{align}\label{EQ:tunnel_integral}
    \sum_k \frac{\abs{t_{k\alpha}}^2}{z+\ii\epsilon_{k\alpha}} &= \frac{1}{2\pi} \int \frac{\Gamma_\alpha(\omega)}{z+\ii\omega} d\omega
    \nn&= \frac{\tau_\alpha^2}{2\tau^2} \left(\sqrt{1+\frac{4\tau^2}{(z+\ii\epsilon)^2}}-1\right)(z+\ii\epsilon)\,,
\end{align}
which has a branch cut discontinuity along the imaginary axis in the interval $\ii[-\epsilon-2\tau,-\epsilon+2\tau]$, which has to be considered when inverting the Laplace transform.

\bibliography{BoundaryBib}

\begin{thebibliography}{729}%
\makeatletter
\providecommand \@ifxundefined [1]{%
 \@ifx{#1\undefined}
}%
\providecommand \@ifnum [1]{%
 \ifnum #1\expandafter \@firstoftwo
 \else \expandafter \@secondoftwo
 \fi
}%
\providecommand \@ifx [1]{%
 \ifx #1\expandafter \@firstoftwo
 \else \expandafter \@secondoftwo
 \fi
}%
\providecommand \natexlab [1]{#1}%
\providecommand \enquote  [1]{``#1''}%
\providecommand \bibnamefont  [1]{#1}%
\providecommand \bibfnamefont [1]{#1}%
\providecommand \citenamefont [1]{#1}%
\providecommand \href@noop [0]{\@secondoftwo}%
\providecommand \href [0]{\begingroup \@sanitize@url \@href}%
\providecommand \@href[1]{\@@startlink{#1}\@@href}%
\providecommand \@@href[1]{\endgroup#1\@@endlink}%
\providecommand \@sanitize@url [0]{\catcode `\\12\catcode `\$12\catcode
  `\&12\catcode `\#12\catcode `\^12\catcode `\_12\catcode `\%12\relax}%
\providecommand \@@startlink[1]{}%
\providecommand \@@endlink[0]{}%
\providecommand \url  [0]{\begingroup\@sanitize@url \@url }%
\providecommand \@url [1]{\endgroup\@href {#1}{\urlprefix }}%
\providecommand \urlprefix  [0]{URL }%
\providecommand \Eprint [0]{\href }%
\providecommand \doibase [0]{http://dx.doi.org/}%
\providecommand \selectlanguage [0]{\@gobble}%
\providecommand \bibinfo  [0]{\@secondoftwo}%
\providecommand \bibfield  [0]{\@secondoftwo}%
\providecommand \translation [1]{[#1]}%
\providecommand \BibitemOpen [0]{}%
\providecommand \bibitemStop [0]{}%
\providecommand \bibitemNoStop [0]{.\EOS\space}%
\providecommand \EOS [0]{\spacefactor3000\relax}%
\providecommand \BibitemShut  [1]{\csname bibitem#1\endcsname}%
\let\auto@bib@innerbib\@empty
\bibitem [{\citenamefont {Abanin}\ \emph {et~al.}(2019)\citenamefont {Abanin},
  \citenamefont {Altman}, \citenamefont {Bloch},\ and\ \citenamefont
  {Serbyn}}]{AbaninSerbyn2019}%
  \BibitemOpen
  \bibfield  {author} {\bibinfo {author} {\bibnamefont {Abanin}, \bibfnamefont
  {D.~A.}}, \bibinfo {author} {\bibfnamefont {E.}~\bibnamefont {Altman}},
  \bibinfo {author} {\bibfnamefont {I.}~\bibnamefont {Bloch}}, \ and\ \bibinfo
  {author} {\bibfnamefont {M.}~\bibnamefont {Serbyn}}} (\bibinfo {year}
  {2019}),\ \href {\doibase 10.1103/RevModPhys.91.021001} {\bibfield  {journal}
  {\bibinfo  {journal} {Rev. Mod. Phys.}\ }\textbf {\bibinfo {volume} {91}},\
  \bibinfo {pages} {021001}}\BibitemShut {NoStop}%
\bibitem [{\citenamefont {Abanin}\ and\ \citenamefont
  {Papi\'{c}}(2017)}]{AbaninPapic2017}%
  \BibitemOpen
  \bibfield  {author} {\bibinfo {author} {\bibnamefont {Abanin}, \bibfnamefont
  {D.~A.}}, \ and\ \bibinfo {author} {\bibfnamefont {Z.}~\bibnamefont
  {Papi\'{c}}}} (\bibinfo {year} {2017}),\ \href {\doibase
  10.1002/andp.201700169} {\bibfield  {journal} {\bibinfo  {journal} {Annals of
  Physics (Berlin)}\ }\textbf {\bibinfo {volume} {529}},\ \bibinfo {pages}
  {1700169}}\BibitemShut {NoStop}%
\bibitem [{\citenamefont {Aeberhard}(2011)}]{Aeberhard2011}%
  \BibitemOpen
  \bibfield  {author} {\bibinfo {author} {\bibnamefont {Aeberhard},
  \bibfnamefont {U.}}} (\bibinfo {year} {2011}),\ \href {\doibase
  10.1007/s10825-011-0375-6} {\bibfield  {journal} {\bibinfo  {journal} {J.
  Comput. Electron.}\ }\textbf {\bibinfo {volume} {10}}~(\bibinfo {number}
  {4}),\ \bibinfo {pages} {394}}\BibitemShut {NoStop}%
\bibitem [{\citenamefont {Agarwal}\ \emph {et~al.}(2017)\citenamefont
  {Agarwal}, \citenamefont {Altman}, \citenamefont {Demler}, \citenamefont
  {Gopalakrishnan}, ,\ and\ \citenamefont {Knap}}]{AgarwalKnap2017}%
  \BibitemOpen
  \bibfield  {author} {\bibinfo {author} {\bibnamefont {Agarwal}, \bibfnamefont
  {K.}}, \bibinfo {author} {\bibfnamefont {E.}~\bibnamefont {Altman}}, \bibinfo
  {author} {\bibfnamefont {E.}~\bibnamefont {Demler}}, \bibinfo {author}
  {\bibfnamefont {D.}~\bibnamefont {Gopalakrishnan}, \bibfnamefont
  {Sarang~Huse}}, , \ and\ \bibinfo {author} {\bibfnamefont {M.}~\bibnamefont
  {Knap}}} (\bibinfo {year} {2017}),\ \href {\doibase 10.1002/andp.201600326}
  {\bibfield  {journal} {\bibinfo  {journal} {Annalen der Physik}\ }\textbf
  {\bibinfo {volume} {529}},\ \bibinfo {pages} {1600326}}\BibitemShut {NoStop}%
\bibitem [{\citenamefont {Agarwal}\ \emph {et~al.}(2015)\citenamefont
  {Agarwal}, \citenamefont {Gopalakrishnan}, \citenamefont {Knap},
  \citenamefont {M{\"u}ller},\ and\ \citenamefont
  {Demler}}]{AgarwalDemler2015}%
  \BibitemOpen
  \bibfield  {author} {\bibinfo {author} {\bibnamefont {Agarwal}, \bibfnamefont
  {K.}}, \bibinfo {author} {\bibfnamefont {S.}~\bibnamefont {Gopalakrishnan}},
  \bibinfo {author} {\bibfnamefont {M.}~\bibnamefont {Knap}}, \bibinfo {author}
  {\bibfnamefont {M.}~\bibnamefont {M{\"u}ller}}, \ and\ \bibinfo {author}
  {\bibfnamefont {E.}~\bibnamefont {Demler}}} (\bibinfo {year} {2015}),\ \href
  {\doibase 10.1103/PhysRevLett.114.160401} {\bibfield  {journal} {\bibinfo
  {journal} {Phys. Rev. Lett.}\ }\textbf {\bibinfo {volume} {114}},\ \bibinfo
  {pages} {160401}}\BibitemShut {NoStop}%
\bibitem [{\citenamefont {Aharonov}\ and\ \citenamefont
  {Bohm}(1959)}]{AharonovBohm1959}%
  \BibitemOpen
  \bibfield  {author} {\bibinfo {author} {\bibnamefont {Aharonov},
  \bibfnamefont {Y.}}, \ and\ \bibinfo {author} {\bibfnamefont
  {D.}~\bibnamefont {Bohm}}} (\bibinfo {year} {1959}),\ \href {\doibase
  10.1103/PhysRev.115.485} {\bibfield  {journal} {\bibinfo  {journal} {Phys.
  Rev.}\ }\textbf {\bibinfo {volume} {115}},\ \bibinfo {pages}
  {485}}\BibitemShut {NoStop}%
\bibitem [{\citenamefont {Ajisaka}\ and\ \citenamefont
  {Barra}(2013)}]{AjisakaBarra2013}%
  \BibitemOpen
  \bibfield  {author} {\bibinfo {author} {\bibnamefont {Ajisaka}, \bibfnamefont
  {S.}}, \ and\ \bibinfo {author} {\bibfnamefont {F.}~\bibnamefont {Barra}}}
  (\bibinfo {year} {2013}),\ \href {\doibase 10.1103/PhysRevB.87.195114}
  {\bibfield  {journal} {\bibinfo  {journal} {Phys. Rev. B}\ }\textbf {\bibinfo
  {volume} {87}},\ \bibinfo {pages} {195114}}\BibitemShut {NoStop}%
\bibitem [{\citenamefont {Ajisaka}\ \emph {et~al.}(2012)\citenamefont
  {Ajisaka}, \citenamefont {Barra}, \citenamefont {Mej\'{\i}a-Monasterio},\
  and\ \citenamefont {Prosen}}]{AjisakaProsen2012}%
  \BibitemOpen
  \bibfield  {author} {\bibinfo {author} {\bibnamefont {Ajisaka}, \bibfnamefont
  {S.}}, \bibinfo {author} {\bibfnamefont {F.}~\bibnamefont {Barra}}, \bibinfo
  {author} {\bibfnamefont {C.}~\bibnamefont {Mej\'{\i}a-Monasterio}}, \ and\
  \bibinfo {author} {\bibfnamefont {T.}~\bibnamefont {Prosen}}} (\bibinfo
  {year} {2012}),\ \href {\doibase 10.1103/PhysRevB.86.125111} {\bibfield
  {journal} {\bibinfo  {journal} {Phys. Rev. B}\ }\textbf {\bibinfo {volume}
  {86}},\ \bibinfo {pages} {125111}}\BibitemShut {NoStop}%
\bibitem [{\citenamefont {Akemann}\ \emph {et~al.}(2019)\citenamefont
  {Akemann}, \citenamefont {Kieburg}, \citenamefont {Mielke},\ and\
  \citenamefont {Prosen}}]{AkemannProsen2019}%
  \BibitemOpen
  \bibfield  {author} {\bibinfo {author} {\bibnamefont {Akemann}, \bibfnamefont
  {G.}}, \bibinfo {author} {\bibfnamefont {M.}~\bibnamefont {Kieburg}},
  \bibinfo {author} {\bibfnamefont {A.}~\bibnamefont {Mielke}}, \ and\ \bibinfo
  {author} {\bibfnamefont {T.}~\bibnamefont {Prosen}}} (\bibinfo {year}
  {2019}),\ \href {\doibase 10.1103/PhysRevLett.123.254101} {\bibfield
  {journal} {\bibinfo  {journal} {Phys. Rev. Lett.}\ }\textbf {\bibinfo
  {volume} {123}},\ \bibinfo {pages} {254101}}\BibitemShut {NoStop}%
\bibitem [{\citenamefont {Akera}(1993)}]{Akera1993}%
  \BibitemOpen
  \bibfield  {author} {\bibinfo {author} {\bibnamefont {Akera}, \bibfnamefont
  {H.}}} (\bibinfo {year} {1993}),\ \href {\doibase 10.1103/PhysRevB.47.6835}
  {\bibfield  {journal} {\bibinfo  {journal} {Phys. Rev. B}\ }\textbf {\bibinfo
  {volume} {47}},\ \bibinfo {pages} {6835}}\BibitemShut {NoStop}%
\bibitem [{\citenamefont {Al-Assam}\ \emph {et~al.}(2017)\citenamefont
  {Al-Assam}, \citenamefont {Clark},\ and\ \citenamefont
  {Jaksch}}]{AlAssamJaksch2017}%
  \BibitemOpen
  \bibfield  {author} {\bibinfo {author} {\bibnamefont {Al-Assam},
  \bibfnamefont {S.}}, \bibinfo {author} {\bibfnamefont {S.~R.}\ \bibnamefont
  {Clark}}, \ and\ \bibinfo {author} {\bibfnamefont {D.}~\bibnamefont
  {Jaksch}}} (\bibinfo {year} {2017}),\ \href {\doibase
  10.1088/1742-5468/aa7df3} {\bibfield  {journal} {\bibinfo  {journal} {Journal
  of Statistical Mechanics: Theory and Experiment}\ }\textbf {\bibinfo {volume}
  {2017}}~(\bibinfo {number} {9}),\ \bibinfo {pages} {093102}}\BibitemShut
  {NoStop}%
\bibitem [{\citenamefont {Albert}\ and\ \citenamefont
  {Jiang}(2014)}]{albert2014a}%
  \BibitemOpen
  \bibfield  {author} {\bibinfo {author} {\bibnamefont {Albert}, \bibfnamefont
  {V.~V.}}, \ and\ \bibinfo {author} {\bibfnamefont {L.}~\bibnamefont {Jiang}}}
  (\bibinfo {year} {2014}),\ \href {\doibase 10.1103/PhysRevA.89.022118}
  {\bibfield  {journal} {\bibinfo  {journal} {Phys. Rev. A}\ }\textbf {\bibinfo
  {volume} {89}},\ \bibinfo {pages} {022118}}\BibitemShut {NoStop}%
\bibitem [{\citenamefont {Aleiner}\ \emph {et~al.}(2002)\citenamefont
  {Aleiner}, \citenamefont {Brouwer},\ and\ \citenamefont
  {Glazman}}]{AleinerGlazman2002}%
  \BibitemOpen
  \bibfield  {author} {\bibinfo {author} {\bibnamefont {Aleiner}, \bibfnamefont
  {I.}}, \bibinfo {author} {\bibfnamefont {P.}~\bibnamefont {Brouwer}}, \ and\
  \bibinfo {author} {\bibfnamefont {L.}~\bibnamefont {Glazman}}} (\bibinfo
  {year} {2002}),\ \href {\doibase
  https://doi.org/10.1016/S0370-1573(01)00063-1} {\bibfield  {journal}
  {\bibinfo  {journal} {Physics Reports}\ }\textbf {\bibinfo {volume}
  {358}}~(\bibinfo {number} {5}),\ \bibinfo {pages} {309}}\BibitemShut
  {NoStop}%
\bibitem [{\citenamefont {Alicki}(1979)}]{Alicki_1979}%
  \BibitemOpen
  \bibfield  {author} {\bibinfo {author} {\bibnamefont {Alicki}, \bibfnamefont
  {R.}}} (\bibinfo {year} {1979}),\ \href {\doibase 10.1088/0305-4470/12/5/007}
  {\bibfield  {journal} {\bibinfo  {journal} {Journal of Physics A:
  Mathematical and General}\ }\textbf {\bibinfo {volume} {12}}~(\bibinfo
  {number} {5}),\ \bibinfo {pages} {L103}}\BibitemShut {NoStop}%
\bibitem [{\citenamefont {Amato}\ \emph {et~al.}(2020)\citenamefont {Amato},
  \citenamefont {Breuer}, \citenamefont {Wimberger}, \citenamefont
  {Rodr\'{\i}guez},\ and\ \citenamefont {Buchleitner}}]{amato2020a}%
  \BibitemOpen
  \bibfield  {author} {\bibinfo {author} {\bibnamefont {Amato}, \bibfnamefont
  {G.}}, \bibinfo {author} {\bibfnamefont {H.-P.}\ \bibnamefont {Breuer}},
  \bibinfo {author} {\bibfnamefont {S.}~\bibnamefont {Wimberger}}, \bibinfo
  {author} {\bibfnamefont {A.}~\bibnamefont {Rodr\'{\i}guez}}, \ and\ \bibinfo
  {author} {\bibfnamefont {A.}~\bibnamefont {Buchleitner}}} (\bibinfo {year}
  {2020}),\ \href {\doibase 10.1103/PhysRevA.102.022207} {\bibfield  {journal}
  {\bibinfo  {journal} {Phys. Rev. A}\ }\textbf {\bibinfo {volume} {102}},\
  \bibinfo {pages} {022207}}\BibitemShut {NoStop}%
\bibitem [{\citenamefont {Amico}\ \emph {et~al.}(2017)\citenamefont {Amico},
  \citenamefont {Birkl}, \citenamefont {Boshier},\ and\ \citenamefont
  {Kwek}}]{AmicoKwek2017}%
  \BibitemOpen
  \bibfield  {author} {\bibinfo {author} {\bibnamefont {Amico}, \bibfnamefont
  {L.}}, \bibinfo {author} {\bibfnamefont {G.}~\bibnamefont {Birkl}}, \bibinfo
  {author} {\bibfnamefont {M.}~\bibnamefont {Boshier}}, \ and\ \bibinfo
  {author} {\bibfnamefont {L.-C.}\ \bibnamefont {Kwek}}} (\bibinfo {year}
  {2017}),\ \href {\doibase 10.1088/1367-2630/aa5a6d} {\bibfield  {journal}
  {\bibinfo  {journal} {New Journal of Physics}\ }\textbf {\bibinfo {volume}
  {19}}~(\bibinfo {number} {2}),\ \bibinfo {pages} {020201}}\BibitemShut
  {NoStop}%
\bibitem [{\citenamefont {Amico}\ \emph {et~al.}(2021)\citenamefont {Amico},
  \citenamefont {Boshier}, \citenamefont {Birkl}, \citenamefont {Minguzzi},
  \citenamefont {Miniatura}, \citenamefont {Kwek}, \citenamefont {Aghamalyan},
  \citenamefont {Ahufinger}, \citenamefont {Anderson}, \citenamefont {Andrei},
  \citenamefont {Arnold}, \citenamefont {Baker}, \citenamefont {Bell},
  \citenamefont {Bland}, \citenamefont {Brantut}, \citenamefont {Cassettari},
  \citenamefont {Chetcuti}, \citenamefont {Chevy}, \citenamefont {De~Palo},
  \citenamefont {Dumke}, \citenamefont {Edwards}, \citenamefont {Folman},
  \citenamefont {Fortagh}, \citenamefont {Gardiner}, \citenamefont {Garraway},
  \citenamefont {Gauthier}, \citenamefont {G{\"u}nther}, \citenamefont {Haug},
  \citenamefont {Hufnagel}, \citenamefont {Keil}, \citenamefont {von Klitzing},
  \citenamefont {Lebrat}, \citenamefont {Li}, \citenamefont {Longchambon},
  \citenamefont {Mompart}, \citenamefont {Morsch}, \citenamefont {Naldesi},
  \citenamefont {Neely}, \citenamefont {Olshanii}, \citenamefont {Orignac},
  \citenamefont {Pandey}, \citenamefont {Pérez-Obiol}, \citenamefont {Perrin},
  \citenamefont {Piroli}, \citenamefont {Polo}, \citenamefont {Pritchard},
  \citenamefont {Proukakis}, \citenamefont {Rylands}, \citenamefont
  {Rubinsztein-Dunlop}, \citenamefont {Scazza}, \citenamefont {Stringari},
  \citenamefont {Tosto}, \citenamefont {Trombettoni}, \citenamefont {Victorin},
  \citenamefont {Wilkowski}, \citenamefont {Xhani},\ and\ \citenamefont
  {Yakimenko}}]{AmicoYakimenko2020}%
  \BibitemOpen
  \bibfield  {author} {\bibinfo {author} {\bibnamefont {Amico}, \bibfnamefont
  {L.}}, \bibinfo {author} {\bibfnamefont {M.}~\bibnamefont {Boshier}},
  \bibinfo {author} {\bibfnamefont {G.}~\bibnamefont {Birkl}}, \bibinfo
  {author} {\bibfnamefont {A.}~\bibnamefont {Minguzzi}}, \bibinfo {author}
  {\bibfnamefont {C.}~\bibnamefont {Miniatura}}, \bibinfo {author}
  {\bibfnamefont {L.-C.}\ \bibnamefont {Kwek}}, \bibinfo {author}
  {\bibfnamefont {D.}~\bibnamefont {Aghamalyan}}, \bibinfo {author}
  {\bibfnamefont {V.}~\bibnamefont {Ahufinger}}, \bibinfo {author}
  {\bibfnamefont {D.}~\bibnamefont {Anderson}}, \bibinfo {author}
  {\bibfnamefont {N.}~\bibnamefont {Andrei}}, \bibinfo {author} {\bibfnamefont
  {A.~S.}\ \bibnamefont {Arnold}}, \bibinfo {author} {\bibfnamefont
  {M.}~\bibnamefont {Baker}}, \bibinfo {author} {\bibfnamefont
  {T.}~\bibnamefont {Bell}}, \bibinfo {author} {\bibfnamefont {T.}~\bibnamefont
  {Bland}}, \bibinfo {author} {\bibfnamefont {J.}~\bibnamefont {Brantut}},
  \bibinfo {author} {\bibfnamefont {D.}~\bibnamefont {Cassettari}}, \bibinfo
  {author} {\bibfnamefont {W.~J.}\ \bibnamefont {Chetcuti}}, \bibinfo {author}
  {\bibfnamefont {R.}~\bibnamefont {Chevy}, \bibfnamefont {F.~Citro}}, \bibinfo
  {author} {\bibfnamefont {S.}~\bibnamefont {De~Palo}}, \bibinfo {author}
  {\bibfnamefont {R.}~\bibnamefont {Dumke}}, \bibinfo {author} {\bibfnamefont
  {M.}~\bibnamefont {Edwards}}, \bibinfo {author} {\bibfnamefont
  {R.}~\bibnamefont {Folman}}, \bibinfo {author} {\bibfnamefont
  {J.}~\bibnamefont {Fortagh}}, \bibinfo {author} {\bibfnamefont {S.~A.}\
  \bibnamefont {Gardiner}}, \bibinfo {author} {\bibfnamefont {B.}~\bibnamefont
  {Garraway}}, \bibinfo {author} {\bibfnamefont {G.}~\bibnamefont {Gauthier}},
  \bibinfo {author} {\bibfnamefont {A.}~\bibnamefont {G{\"u}nther}}, \bibinfo
  {author} {\bibfnamefont {T.}~\bibnamefont {Haug}}, \bibinfo {author}
  {\bibfnamefont {C.}~\bibnamefont {Hufnagel}}, \bibinfo {author}
  {\bibfnamefont {M.}~\bibnamefont {Keil}}, \bibinfo {author} {\bibfnamefont
  {P.}~\bibnamefont {von Klitzing}, \bibfnamefont {W.~Ireland}}, \bibinfo
  {author} {\bibfnamefont {M.}~\bibnamefont {Lebrat}}, \bibinfo {author}
  {\bibfnamefont {W.}~\bibnamefont {Li}}, \bibinfo {author} {\bibfnamefont
  {L.}~\bibnamefont {Longchambon}}, \bibinfo {author} {\bibfnamefont
  {J.}~\bibnamefont {Mompart}}, \bibinfo {author} {\bibfnamefont
  {O.}~\bibnamefont {Morsch}}, \bibinfo {author} {\bibfnamefont
  {P.}~\bibnamefont {Naldesi}}, \bibinfo {author} {\bibfnamefont {T.~W.}\
  \bibnamefont {Neely}}, \bibinfo {author} {\bibfnamefont {M.}~\bibnamefont
  {Olshanii}}, \bibinfo {author} {\bibfnamefont {E.}~\bibnamefont {Orignac}},
  \bibinfo {author} {\bibfnamefont {S.}~\bibnamefont {Pandey}}, \bibinfo
  {author} {\bibfnamefont {A.}~\bibnamefont {Pérez-Obiol}}, \bibinfo {author}
  {\bibfnamefont {H.}~\bibnamefont {Perrin}}, \bibinfo {author} {\bibfnamefont
  {L.}~\bibnamefont {Piroli}}, \bibinfo {author} {\bibfnamefont
  {J.}~\bibnamefont {Polo}}, \bibinfo {author} {\bibfnamefont {A.}~\bibnamefont
  {Pritchard}}, \bibinfo {author} {\bibfnamefont {N.~P.}\ \bibnamefont
  {Proukakis}}, \bibinfo {author} {\bibfnamefont {C.}~\bibnamefont {Rylands}},
  \bibinfo {author} {\bibfnamefont {H.}~\bibnamefont {Rubinsztein-Dunlop}},
  \bibinfo {author} {\bibfnamefont {F.}~\bibnamefont {Scazza}}, \bibinfo
  {author} {\bibfnamefont {S.}~\bibnamefont {Stringari}}, \bibinfo {author}
  {\bibfnamefont {F.}~\bibnamefont {Tosto}}, \bibinfo {author} {\bibfnamefont
  {A.}~\bibnamefont {Trombettoni}}, \bibinfo {author} {\bibfnamefont
  {N.}~\bibnamefont {Victorin}}, \bibinfo {author} {\bibfnamefont
  {D.}~\bibnamefont {Wilkowski}}, \bibinfo {author} {\bibfnamefont
  {K.}~\bibnamefont {Xhani}}, \ and\ \bibinfo {author} {\bibfnamefont
  {A.}~\bibnamefont {Yakimenko}}} (\bibinfo {year} {2021}),\ \href {\doibase
  10.1116/5.0026178} {\bibfield  {journal} {\bibinfo  {journal} {AVS Quantum
  Sci.}\ }\textbf {\bibinfo {volume} {3}},\ \bibinfo {pages}
  {039201}}\BibitemShut {NoStop}%
\bibitem [{\citenamefont {Anantram}\ and\ \citenamefont
  {L\'eonard}(2006)}]{AnantramLeonard2006}%
  \BibitemOpen
  \bibfield  {author} {\bibinfo {author} {\bibnamefont {Anantram},
  \bibfnamefont {M.~P.}}, \ and\ \bibinfo {author} {\bibfnamefont
  {F.}~\bibnamefont {L\'eonard}}} (\bibinfo {year} {2006}),\ \href {\doibase
  10.1088/0034-4885/69/3/R01} {\bibfield  {journal} {\bibinfo  {journal}
  {Reports on Progress in Physics}\ }\textbf {\bibinfo {volume} {69}},\
  \bibinfo {pages} {507}}\BibitemShut {NoStop}%
\bibitem [{\citenamefont {Anderson}(1958)}]{Anderson1958}%
  \BibitemOpen
  \bibfield  {author} {\bibinfo {author} {\bibnamefont {Anderson},
  \bibfnamefont {P.~W.}}} (\bibinfo {year} {1958}),\ \href {\doibase
  10.1103/PhysRev.109.1492} {\bibfield  {journal} {\bibinfo  {journal} {Phys.
  Rev.}\ }\textbf {\bibinfo {volume} {109}},\ \bibinfo {pages}
  {1492}}\BibitemShut {NoStop}%
\bibitem [{\citenamefont {Anto-Sztrikacs}\ and\ \citenamefont
  {Segal}(2021)}]{anto_sztrikacs2021a}%
  \BibitemOpen
  \bibfield  {author} {\bibinfo {author} {\bibnamefont {Anto-Sztrikacs},
  \bibfnamefont {N.}}, \ and\ \bibinfo {author} {\bibfnamefont
  {D.}~\bibnamefont {Segal}}} (\bibinfo {year} {2021}),\ \href {\doibase
  10.1088/1367-2630/ac02df} {\bibfield  {journal} {\bibinfo  {journal} {New
  Journal of Physics}\ }\textbf {\bibinfo {volume} {23}}~(\bibinfo {number}
  {6}),\ \bibinfo {pages} {063036}}\BibitemShut {NoStop}%
\bibitem [{\citenamefont {Aoki}\ and\ \citenamefont
  {Kusnezov}(2001)}]{Aoki2001}%
  \BibitemOpen
  \bibfield  {author} {\bibinfo {author} {\bibnamefont {Aoki}, \bibfnamefont
  {K.}}, \ and\ \bibinfo {author} {\bibfnamefont {D.}~\bibnamefont {Kusnezov}}}
  (\bibinfo {year} {2001}),\ \href {\doibase 10.1103/PhysRevLett.86.4029}
  {\bibfield  {journal} {\bibinfo  {journal} {Phys. Rev. Lett.}\ }\textbf
  {\bibinfo {volume} {86}},\ \bibinfo {pages} {4029}}\BibitemShut {NoStop}%
\bibitem [{\citenamefont {Araki}\ and\ \citenamefont
  {Woods}(1963)}]{ArakiWoods1963}%
  \BibitemOpen
  \bibfield  {author} {\bibinfo {author} {\bibnamefont {Araki}, \bibfnamefont
  {H.}}, \ and\ \bibinfo {author} {\bibfnamefont {E.~J.}\ \bibnamefont
  {Woods}}} (\bibinfo {year} {1963}),\ \href {\doibase 10.1063/1.1704002}
  {\bibfield  {journal} {\bibinfo  {journal} {J. Math. Phys.}\ }\textbf
  {\bibinfo {volume} {4}},\ \bibinfo {pages} {637}}\BibitemShut {NoStop}%
\bibitem [{\citenamefont {Arrachea}\ \emph {et~al.}(2009)\citenamefont
  {Arrachea}, \citenamefont {Lozano},\ and\ \citenamefont
  {Aligia}}]{ArracheaAligia2009}%
  \BibitemOpen
  \bibfield  {author} {\bibinfo {author} {\bibnamefont {Arrachea},
  \bibfnamefont {L.}}, \bibinfo {author} {\bibfnamefont {G.~S.}\ \bibnamefont
  {Lozano}}, \ and\ \bibinfo {author} {\bibfnamefont {A.~A.}\ \bibnamefont
  {Aligia}}} (\bibinfo {year} {2009}),\ \href {\doibase
  10.1103/PhysRevB.80.014425} {\bibfield  {journal} {\bibinfo  {journal} {Phys.
  Rev. B}\ }\textbf {\bibinfo {volume} {80}},\ \bibinfo {pages}
  {014425}}\BibitemShut {NoStop}%
\bibitem [{\citenamefont {Asadian}\ \emph {et~al.}(2013)\citenamefont
  {Asadian}, \citenamefont {Manzano}, \citenamefont {Tiersch},\ and\
  \citenamefont {Briegel}}]{AsadianBriegel2013}%
  \BibitemOpen
  \bibfield  {author} {\bibinfo {author} {\bibnamefont {Asadian}, \bibfnamefont
  {A.}}, \bibinfo {author} {\bibfnamefont {D.}~\bibnamefont {Manzano}},
  \bibinfo {author} {\bibfnamefont {M.}~\bibnamefont {Tiersch}}, \ and\
  \bibinfo {author} {\bibfnamefont {H.~J.}\ \bibnamefont {Briegel}}} (\bibinfo
  {year} {2013}),\ \href {\doibase 10.1103/PhysRevE.87.012109} {\bibfield
  {journal} {\bibinfo  {journal} {Phys. Rev. E}\ }\textbf {\bibinfo {volume}
  {87}},\ \bibinfo {pages} {012109}}\BibitemShut {NoStop}%
\bibitem [{\citenamefont {Atala}\ \emph {et~al.}(2014)\citenamefont {Atala},
  \citenamefont {Aidelsburger}, \citenamefont {Lohse}, \citenamefont
  {Barreiro}, \citenamefont {Paredes},\ and\ \citenamefont
  {Bloch}}]{AtalaBloch2014}%
  \BibitemOpen
  \bibfield  {author} {\bibinfo {author} {\bibnamefont {Atala}, \bibfnamefont
  {M.}}, \bibinfo {author} {\bibfnamefont {M.}~\bibnamefont {Aidelsburger}},
  \bibinfo {author} {\bibfnamefont {M.}~\bibnamefont {Lohse}}, \bibinfo
  {author} {\bibfnamefont {J.~T.}\ \bibnamefont {Barreiro}}, \bibinfo {author}
  {\bibfnamefont {B.}~\bibnamefont {Paredes}}, \ and\ \bibinfo {author}
  {\bibfnamefont {I.}~\bibnamefont {Bloch}}} (\bibinfo {year} {2014}),\ \href
  {\doibase 10.1038/nphys2998} {\bibfield  {journal} {\bibinfo  {journal} {Nat.
  Phys.}\ }\textbf {\bibinfo {volume} {10}}~(\bibinfo {number} {8}),\ \bibinfo
  {pages} {588}}\BibitemShut {NoStop}%
\bibitem [{\citenamefont {Aubry}\ and\ \citenamefont
  {Andr\'e}(1980)}]{AubryAndre1980}%
  \BibitemOpen
  \bibfield  {author} {\bibinfo {author} {\bibnamefont {Aubry}, \bibfnamefont
  {S.}}, \ and\ \bibinfo {author} {\bibfnamefont {G.}~\bibnamefont {Andr\'e}}}
  (\bibinfo {year} {1980}),\ \href@noop {} {\bibfield  {journal} {\bibinfo
  {journal} {Ann. Isr. Phys. Soc.}\ }\textbf {\bibinfo {volume} {3}},\ \bibinfo
  {pages} {133}}\BibitemShut {NoStop}%
\bibitem [{\citenamefont {Audenaert}\ \emph {et~al.}(2002)\citenamefont
  {Audenaert}, \citenamefont {Eisert}, \citenamefont {Plenio},\ and\
  \citenamefont {Werner}}]{AudenaertWerner2002}%
  \BibitemOpen
  \bibfield  {author} {\bibinfo {author} {\bibnamefont {Audenaert},
  \bibfnamefont {K.}}, \bibinfo {author} {\bibfnamefont {J.}~\bibnamefont
  {Eisert}}, \bibinfo {author} {\bibfnamefont {M.~B.}\ \bibnamefont {Plenio}},
  \ and\ \bibinfo {author} {\bibfnamefont {R.~F.}\ \bibnamefont {Werner}}}
  (\bibinfo {year} {2002}),\ \href {\doibase 10.1103/PhysRevA.66.042327}
  {\bibfield  {journal} {\bibinfo  {journal} {Phys. Rev. A}\ }\textbf {\bibinfo
  {volume} {66}},\ \bibinfo {pages} {042327}}\BibitemShut {NoStop}%
\bibitem [{\citenamefont {Aurell}(2018)}]{aurell2018a}%
  \BibitemOpen
  \bibfield  {author} {\bibinfo {author} {\bibnamefont {Aurell}, \bibfnamefont
  {E.}}} (\bibinfo {year} {2018}),\ \href {\doibase 10.1103/PhysRevE.97.042112}
  {\bibfield  {journal} {\bibinfo  {journal} {Phys. Rev. E}\ }\textbf {\bibinfo
  {volume} {97}},\ \bibinfo {pages} {042112}}\BibitemShut {NoStop}%
\bibitem [{\citenamefont {Baer}\ and\ \citenamefont
  {Kosloff}(1997)}]{BaerKosloff1997}%
  \BibitemOpen
  \bibfield  {author} {\bibinfo {author} {\bibnamefont {Baer}, \bibfnamefont
  {R.}}, \ and\ \bibinfo {author} {\bibfnamefont {R.}~\bibnamefont {Kosloff}}}
  (\bibinfo {year} {1997}),\ \href {\doibase 10.1063/1.473950} {\bibfield
  {journal} {\bibinfo  {journal} {J. Chem. Phys.}\ }\textbf {\bibinfo {volume}
  {106}},\ \bibinfo {pages} {8862}}\BibitemShut {NoStop}%
\bibitem [{\citenamefont {Bagchi}(2015)}]{Bagchi2015}%
  \BibitemOpen
  \bibfield  {author} {\bibinfo {author} {\bibnamefont {Bagchi}, \bibfnamefont
  {D.}}} (\bibinfo {year} {2015}),\ \href {\doibase
  10.1088/1742-5468/2015/02/p02015} {\bibfield  {journal} {\bibinfo  {journal}
  {Journal of Statistical Mechanics: Theory and Experiment}\ }\textbf {\bibinfo
  {volume} {2015}}~(\bibinfo {number} {2}),\ \bibinfo {pages}
  {P02015}}\BibitemShut {NoStop}%
\bibitem [{\citenamefont {Bairey}\ \emph {et~al.}(2020)\citenamefont {Bairey},
  \citenamefont {Guo}, \citenamefont {Poletti}, \citenamefont {Lindner},\ and\
  \citenamefont {Arad}}]{BaireyArad2020}%
  \BibitemOpen
  \bibfield  {author} {\bibinfo {author} {\bibnamefont {Bairey}, \bibfnamefont
  {E.}}, \bibinfo {author} {\bibfnamefont {C.}~\bibnamefont {Guo}}, \bibinfo
  {author} {\bibfnamefont {D.}~\bibnamefont {Poletti}}, \bibinfo {author}
  {\bibfnamefont {N.~H.}\ \bibnamefont {Lindner}}, \ and\ \bibinfo {author}
  {\bibfnamefont {I.}~\bibnamefont {Arad}}} (\bibinfo {year} {2020}),\ \href
  {\doibase 10.1088/1367-2630/ab73cd} {\bibfield  {journal} {\bibinfo
  {journal} {New Journal of Physics}\ }\textbf {\bibinfo {volume}
  {22}}~(\bibinfo {number} {3}),\ \bibinfo {pages} {032001}}\BibitemShut
  {NoStop}%
\bibitem [{\citenamefont {Balachandran}\ \emph {et~al.}(2018)\citenamefont
  {Balachandran}, \citenamefont {Benenti}, \citenamefont {Pereira},
  \citenamefont {Casati},\ and\ \citenamefont {Poletti}}]{Balachandran2018}%
  \BibitemOpen
  \bibfield  {author} {\bibinfo {author} {\bibnamefont {Balachandran},
  \bibfnamefont {V.}}, \bibinfo {author} {\bibfnamefont {G.}~\bibnamefont
  {Benenti}}, \bibinfo {author} {\bibfnamefont {E.}~\bibnamefont {Pereira}},
  \bibinfo {author} {\bibfnamefont {G.}~\bibnamefont {Casati}}, \ and\ \bibinfo
  {author} {\bibfnamefont {D.}~\bibnamefont {Poletti}}} (\bibinfo {year}
  {2018}),\ \href {\doibase 10.1103/PhysRevLett.120.200603} {\bibfield
  {journal} {\bibinfo  {journal} {Physical Review Letters}\ }\textbf {\bibinfo
  {volume} {120}}~(\bibinfo {number} {20}),\ \bibinfo {pages}
  {200603}}\BibitemShut {NoStop}%
\bibitem [{\citenamefont {Balachandran}\ \emph
  {et~al.}(2019{\natexlab{a}})\citenamefont {Balachandran}, \citenamefont
  {Benenti}, \citenamefont {Pereira}, \citenamefont {Casati},\ and\
  \citenamefont {Poletti}}]{Balachandran2019a}%
  \BibitemOpen
  \bibfield  {author} {\bibinfo {author} {\bibnamefont {Balachandran},
  \bibfnamefont {V.}}, \bibinfo {author} {\bibfnamefont {G.}~\bibnamefont
  {Benenti}}, \bibinfo {author} {\bibfnamefont {E.}~\bibnamefont {Pereira}},
  \bibinfo {author} {\bibfnamefont {G.}~\bibnamefont {Casati}}, \ and\ \bibinfo
  {author} {\bibfnamefont {D.}~\bibnamefont {Poletti}}} (\bibinfo {year}
  {2019}{\natexlab{a}}),\ \href {\doibase 10.1103/PhysRevE.99.032136}
  {\bibfield  {journal} {\bibinfo  {journal} {Physical Review E}\ }\textbf
  {\bibinfo {volume} {99}},\ \bibinfo {pages} {032136}}\BibitemShut {NoStop}%
\bibitem [{\citenamefont {Balachandran}\ \emph
  {et~al.}(2019{\natexlab{b}})\citenamefont {Balachandran}, \citenamefont
  {Clark}, \citenamefont {Goold},\ and\ \citenamefont
  {Poletti}}]{Balachandran2019}%
  \BibitemOpen
  \bibfield  {author} {\bibinfo {author} {\bibnamefont {Balachandran},
  \bibfnamefont {V.}}, \bibinfo {author} {\bibfnamefont {S.~R.}\ \bibnamefont
  {Clark}}, \bibinfo {author} {\bibfnamefont {J.}~\bibnamefont {Goold}}, \ and\
  \bibinfo {author} {\bibfnamefont {D.}~\bibnamefont {Poletti}}} (\bibinfo
  {year} {2019}{\natexlab{b}}),\ \href {\doibase
  10.1103/PhysRevLett.123.020603} {\bibfield  {journal} {\bibinfo  {journal}
  {Physical Review Letters}\ }\textbf {\bibinfo {volume} {123}}~(\bibinfo
  {number} {2}),\ \bibinfo {pages} {020603}}\BibitemShut {NoStop}%
\bibitem [{\citenamefont {Banchi}\ \emph {et~al.}(2014)\citenamefont {Banchi},
  \citenamefont {Giorda},\ and\ \citenamefont {Zanardi}}]{BanchiZanardi2014}%
  \BibitemOpen
  \bibfield  {author} {\bibinfo {author} {\bibnamefont {Banchi}, \bibfnamefont
  {L.}}, \bibinfo {author} {\bibfnamefont {P.}~\bibnamefont {Giorda}}, \ and\
  \bibinfo {author} {\bibfnamefont {P.}~\bibnamefont {Zanardi}}} (\bibinfo
  {year} {2014}),\ \href {\doibase 10.1103/PhysRevE.89.022102} {\bibfield
  {journal} {\bibinfo  {journal} {Phys. Rev. E}\ }\textbf {\bibinfo {volume}
  {89}},\ \bibinfo {pages} {022102}}\BibitemShut {NoStop}%
\bibitem [{\citenamefont {Bar~Lev}\ \emph {et~al.}(2015)\citenamefont
  {Bar~Lev}, \citenamefont {Cohen},\ and\ \citenamefont
  {Reichman}}]{BarLevReichman2015}%
  \BibitemOpen
  \bibfield  {author} {\bibinfo {author} {\bibnamefont {Bar~Lev}, \bibfnamefont
  {Y.}}, \bibinfo {author} {\bibfnamefont {G.}~\bibnamefont {Cohen}}, \ and\
  \bibinfo {author} {\bibfnamefont {D.~R.}\ \bibnamefont {Reichman}}} (\bibinfo
  {year} {2015}),\ \href {\doibase 10.1103/PhysRevLett.114.100601} {\bibfield
  {journal} {\bibinfo  {journal} {Phys. Rev. Lett.}\ }\textbf {\bibinfo
  {volume} {114}},\ \bibinfo {pages} {100601}}\BibitemShut {NoStop}%
\bibitem [{\citenamefont {Barato}\ and\ \citenamefont
  {Seifert}(2015)}]{barato2015a}%
  \BibitemOpen
  \bibfield  {author} {\bibinfo {author} {\bibnamefont {Barato}, \bibfnamefont
  {A.~C.}}, \ and\ \bibinfo {author} {\bibfnamefont {U.}~\bibnamefont
  {Seifert}}} (\bibinfo {year} {2015}),\ \href {\doibase
  10.1103/PhysRevLett.114.158101} {\bibfield  {journal} {\bibinfo  {journal}
  {Physical Review Letters}\ }\textbf {\bibinfo {volume} {114}},\ \bibinfo
  {pages} {158101}}\BibitemShut {NoStop}%
\bibitem [{\citenamefont {Bargmann}(1961)}]{Bargmann1961}%
  \BibitemOpen
  \bibfield  {author} {\bibinfo {author} {\bibnamefont {Bargmann},
  \bibfnamefont {V.}}} (\bibinfo {year} {1961}),\ \href {\doibase
  10.1002/cpa.3160140303} {\bibfield  {journal} {\bibinfo  {journal} {Comm.
  Pure and App. Math.}\ }\textbf {\bibinfo {volume} {14}},\ \bibinfo {pages}
  {187}}\BibitemShut {NoStop}%
\bibitem [{\citenamefont {Barra}(2015)}]{Barra2015}%
  \BibitemOpen
  \bibfield  {author} {\bibinfo {author} {\bibnamefont {Barra}, \bibfnamefont
  {F.}}} (\bibinfo {year} {2015}),\ \href {\doibase 10.1038/srep14873}
  {\bibfield  {journal} {\bibinfo  {journal} {Scientific Reports}\ }\textbf
  {\bibinfo {volume} {5}},\ \bibinfo {pages} {14873}}\BibitemShut {NoStop}%
\bibitem [{\citenamefont {Bartels}\ and\ \citenamefont
  {Stewart}(1972)}]{BaterlsStewart1972}%
  \BibitemOpen
  \bibfield  {author} {\bibinfo {author} {\bibnamefont {Bartels}, \bibfnamefont
  {R.~H.}}, \ and\ \bibinfo {author} {\bibfnamefont {G.~W.}\ \bibnamefont
  {Stewart}}} (\bibinfo {year} {1972}),\ \href {\doibase 10.1145/361573.361582}
  {\bibfield  {journal} {\bibinfo  {journal} {Commun. ACM}\ }\textbf {\bibinfo
  {volume} {15}},\ \bibinfo {pages} {820}}\BibitemShut {NoStop}%
\bibitem [{\citenamefont {Barthel}\ and\ \citenamefont
  {Zhang}(2022)}]{Barthel2021}%
  \BibitemOpen
  \bibfield  {author} {\bibinfo {author} {\bibnamefont {Barthel}, \bibfnamefont
  {T.}}, \ and\ \bibinfo {author} {\bibfnamefont {Y.}~\bibnamefont {Zhang}}}
  (\bibinfo {year} {2022}),\ \href@noop {} {\bibfield  {journal} {\bibinfo
  {journal} {arXiv:2112.08344}\ }}\Eprint {http://arxiv.org/abs/2112.08344}
  {arXiv:2112.08344 [quant-ph]} \BibitemShut {NoStop}%
\bibitem [{\citenamefont {Basko}\ \emph {et~al.}(2006)\citenamefont {Basko},
  \citenamefont {Aleiner},\ and\ \citenamefont
  {Altshuler}}]{BaskoAltshuler2006}%
  \BibitemOpen
  \bibfield  {author} {\bibinfo {author} {\bibnamefont {Basko}, \bibfnamefont
  {D.~M.}}, \bibinfo {author} {\bibfnamefont {L.}~\bibnamefont {Aleiner}}, \
  and\ \bibinfo {author} {\bibfnamefont {B.~L.}\ \bibnamefont {Altshuler}}}
  (\bibinfo {year} {2006}),\ \href {\doibase 10.1016/j.aop.2005.11.014}
  {\bibfield  {journal} {\bibinfo  {journal} {Annals of Physics}\ }\textbf
  {\bibinfo {volume} {321}},\ \bibinfo {pages} {1126}}\BibitemShut {NoStop}%
\bibitem [{\citenamefont {Baumgartner}\ and\ \citenamefont
  {Narnhofer}(2008)}]{Baumgartner2008}%
  \BibitemOpen
  \bibfield  {author} {\bibinfo {author} {\bibnamefont {Baumgartner},
  \bibfnamefont {B.}}, \ and\ \bibinfo {author} {\bibfnamefont
  {H.}~\bibnamefont {Narnhofer}}} (\bibinfo {year} {2008}),\ \href {\doibase
  10.1088/1751-8113/41/39/395303} {\bibfield  {journal} {\bibinfo  {journal}
  {Journal of Physics A: Mathematical and Theoretical}\ }\textbf {\bibinfo
  {volume} {41}}~(\bibinfo {number} {39}),\ \bibinfo {pages}
  {395303}}\BibitemShut {NoStop}%
\bibitem [{\citenamefont {Benenti}\ \emph {et~al.}(2016)\citenamefont
  {Benenti}, \citenamefont {Casati}, \citenamefont {Mej{\'i}a-Monasterio},\
  and\ \citenamefont {Peyrard}}]{BenentiPeyrard2016}%
  \BibitemOpen
  \bibfield  {author} {\bibinfo {author} {\bibnamefont {Benenti}, \bibfnamefont
  {G.}}, \bibinfo {author} {\bibfnamefont {G.}~\bibnamefont {Casati}}, \bibinfo
  {author} {\bibfnamefont {C.}~\bibnamefont {Mej{\'i}a-Monasterio}}, \ and\
  \bibinfo {author} {\bibfnamefont {M.}~\bibnamefont {Peyrard}}} (\bibinfo
  {year} {2016}),\ \enquote {\bibinfo {title} {From thermal rectifiers to
  thermoelectric devices},}\ in\ \href {\doibase 10.1007/978-3-319-29261-8_10}
  {\emph {\bibinfo {booktitle} {Thermal Transport in Low Dimensions: From
  Statistical Physics to Nanoscale Heat Transfer}}},\ \bibinfo {editor} {edited
  by\ \bibinfo {editor} {\bibfnamefont {S.}~\bibnamefont {Lepri}}},\
  Chap.~\bibinfo {chapter} {10}\ (\bibinfo  {publisher} {Springer International
  Publishing},\ \bibinfo {address} {Cham})\ pp.\ \bibinfo {pages}
  {365--407}\BibitemShut {NoStop}%
\bibitem [{\citenamefont {Benenti}\ \emph
  {et~al.}(2009{\natexlab{a}})\citenamefont {Benenti}, \citenamefont {Casati},
  \citenamefont {Prosen},\ and\ \citenamefont {Rossini}}]{BenentiRossini2009}%
  \BibitemOpen
  \bibfield  {author} {\bibinfo {author} {\bibnamefont {Benenti}, \bibfnamefont
  {G.}}, \bibinfo {author} {\bibfnamefont {G.}~\bibnamefont {Casati}}, \bibinfo
  {author} {\bibfnamefont {T.}~\bibnamefont {Prosen}}, \ and\ \bibinfo {author}
  {\bibfnamefont {D.}~\bibnamefont {Rossini}}} (\bibinfo {year}
  {2009}{\natexlab{a}}),\ \href {\doibase 10.1209/0295-5075/85/37001}
  {\bibfield  {journal} {\bibinfo  {journal} {{EPL} (Europhysics Letters)}\
  }\textbf {\bibinfo {volume} {85}}~(\bibinfo {number} {3}),\ \bibinfo {pages}
  {37001}}\BibitemShut {NoStop}%
\bibitem [{\citenamefont {Benenti}\ \emph
  {et~al.}(2009{\natexlab{b}})\citenamefont {Benenti}, \citenamefont {Casati},
  \citenamefont {Prosen}, \citenamefont {Rossini},\ and\ \citenamefont
  {\v{Z}nidari\v{c}}}]{BenentiZnidaric2009}%
  \BibitemOpen
  \bibfield  {author} {\bibinfo {author} {\bibnamefont {Benenti}, \bibfnamefont
  {G.}}, \bibinfo {author} {\bibfnamefont {G.}~\bibnamefont {Casati}}, \bibinfo
  {author} {\bibfnamefont {T.}~\bibnamefont {Prosen}}, \bibinfo {author}
  {\bibfnamefont {D.}~\bibnamefont {Rossini}}, \ and\ \bibinfo {author}
  {\bibfnamefont {M.}~\bibnamefont {\v{Z}nidari\v{c}}}} (\bibinfo {year}
  {2009}{\natexlab{b}}),\ \href {\doibase 10.1103/PhysRevB.80.035110}
  {\bibfield  {journal} {\bibinfo  {journal} {Phys. Rev. B}\ }\textbf {\bibinfo
  {volume} {80}},\ \bibinfo {pages} {035110}}\BibitemShut {NoStop}%
\bibitem [{\citenamefont {Benenti}\ \emph {et~al.}(2017)\citenamefont
  {Benenti}, \citenamefont {Casati}, \citenamefont {Saito},\ and\ \citenamefont
  {Whitney}}]{BenentiWhitney2017}%
  \BibitemOpen
  \bibfield  {author} {\bibinfo {author} {\bibnamefont {Benenti}, \bibfnamefont
  {G.}}, \bibinfo {author} {\bibfnamefont {G.}~\bibnamefont {Casati}}, \bibinfo
  {author} {\bibfnamefont {K.}~\bibnamefont {Saito}}, \ and\ \bibinfo {author}
  {\bibfnamefont {R.~S.}\ \bibnamefont {Whitney}}} (\bibinfo {year} {2017}),\
  \href {\doibase 10.1016/j.physrep.2017.05.008} {\bibfield  {journal}
  {\bibinfo  {journal} {Physics Reports}\ }\textbf {\bibinfo {volume} {694}},\
  \bibinfo {pages} {1}}\BibitemShut {NoStop}%
\bibitem [{\citenamefont {Benito}\ \emph
  {et~al.}(2016{\natexlab{a}})\citenamefont {Benito}, \citenamefont {Niklas},\
  and\ \citenamefont {Kohler}}]{benito2016b}%
  \BibitemOpen
  \bibfield  {author} {\bibinfo {author} {\bibnamefont {Benito}, \bibfnamefont
  {M.}}, \bibinfo {author} {\bibfnamefont {M.}~\bibnamefont {Niklas}}, \ and\
  \bibinfo {author} {\bibfnamefont {S.}~\bibnamefont {Kohler}}} (\bibinfo
  {year} {2016}{\natexlab{a}}),\ \href {\doibase 10.1103/PhysRevB.94.195433}
  {\bibfield  {journal} {\bibinfo  {journal} {Phys. Rev. B}\ }\textbf {\bibinfo
  {volume} {94}},\ \bibinfo {pages} {195433}}\BibitemShut {NoStop}%
\bibitem [{\citenamefont {Benito}\ \emph
  {et~al.}(2016{\natexlab{b}})\citenamefont {Benito}, \citenamefont {Niklas},
  \citenamefont {Platero},\ and\ \citenamefont {Kohler}}]{benito2016a}%
  \BibitemOpen
  \bibfield  {author} {\bibinfo {author} {\bibnamefont {Benito}, \bibfnamefont
  {M.}}, \bibinfo {author} {\bibfnamefont {M.}~\bibnamefont {Niklas}}, \bibinfo
  {author} {\bibfnamefont {G.}~\bibnamefont {Platero}}, \ and\ \bibinfo
  {author} {\bibfnamefont {S.}~\bibnamefont {Kohler}}} (\bibinfo {year}
  {2016}{\natexlab{b}}),\ \href {\doibase 10.1103/PhysRevB.93.115432}
  {\bibfield  {journal} {\bibinfo  {journal} {Phys. Rev. B}\ }\textbf {\bibinfo
  {volume} {93}},\ \bibinfo {pages} {115432}}\BibitemShut {NoStop}%
\bibitem [{\citenamefont {Bera}\ \emph {et~al.}(2017)\citenamefont {Bera},
  \citenamefont {De~Tomasi}, \citenamefont {Weiner},\ and\ \citenamefont
  {Evers}}]{BeraEvers2017}%
  \BibitemOpen
  \bibfield  {author} {\bibinfo {author} {\bibnamefont {Bera}, \bibfnamefont
  {S.}}, \bibinfo {author} {\bibfnamefont {G.}~\bibnamefont {De~Tomasi}},
  \bibinfo {author} {\bibfnamefont {F.}~\bibnamefont {Weiner}}, \ and\ \bibinfo
  {author} {\bibfnamefont {F.}~\bibnamefont {Evers}}} (\bibinfo {year}
  {2017}),\ \href {\doibase 10.1103/PhysRevLett.118.196801} {\bibfield
  {journal} {\bibinfo  {journal} {Phys. Rev. Lett.}\ }\textbf {\bibinfo
  {volume} {118}},\ \bibinfo {pages} {196801}}\BibitemShut {NoStop}%
\bibitem [{\citenamefont {Bermudez}\ \emph {et~al.}(2013)\citenamefont
  {Bermudez}, \citenamefont {Bruderer},\ and\ \citenamefont
  {Plenio}}]{BermudezPlenio2013}%
  \BibitemOpen
  \bibfield  {author} {\bibinfo {author} {\bibnamefont {Bermudez},
  \bibfnamefont {A.}}, \bibinfo {author} {\bibfnamefont {M.}~\bibnamefont
  {Bruderer}}, \ and\ \bibinfo {author} {\bibfnamefont {M.~B.}\ \bibnamefont
  {Plenio}}} (\bibinfo {year} {2013}),\ \href {\doibase
  10.1103/PhysRevLett.111.040601} {\bibfield  {journal} {\bibinfo  {journal}
  {Phys. Rev. Lett.}\ }\textbf {\bibinfo {volume} {111}},\ \bibinfo {pages}
  {040601}}\BibitemShut {NoStop}%
\bibitem [{\citenamefont {Bertini}\ \emph {et~al.}(2021)\citenamefont
  {Bertini}, \citenamefont {Heidrich-Meisner}, \citenamefont {Karrasch},
  \citenamefont {Prosen}, \citenamefont {Steinigeweg},\ and\ \citenamefont
  {\ifmmode \check{Z}\else \v{Z}\fi{}nidari\ifmmode~\check{c}\else
  \v{c}\fi{}}}]{BertiniZnidaric2020}%
  \BibitemOpen
  \bibfield  {author} {\bibinfo {author} {\bibnamefont {Bertini}, \bibfnamefont
  {B.}}, \bibinfo {author} {\bibfnamefont {F.}~\bibnamefont
  {Heidrich-Meisner}}, \bibinfo {author} {\bibfnamefont {C.}~\bibnamefont
  {Karrasch}}, \bibinfo {author} {\bibfnamefont {T.}~\bibnamefont {Prosen}},
  \bibinfo {author} {\bibfnamefont {R.}~\bibnamefont {Steinigeweg}}, \ and\
  \bibinfo {author} {\bibfnamefont {M.}~\bibnamefont {\ifmmode \check{Z}\else
  \v{Z}\fi{}nidari\ifmmode~\check{c}\else \v{c}\fi{}}}} (\bibinfo {year}
  {2021}),\ \href {\doibase 10.1103/RevModPhys.93.025003} {\bibfield  {journal}
  {\bibinfo  {journal} {Rev. Mod. Phys.}\ }\textbf {\bibinfo {volume} {93}},\
  \bibinfo {pages} {025003}}\BibitemShut {NoStop}%
\bibitem [{\citenamefont {Bhandari}\ \emph {et~al.}(2021)\citenamefont
  {Bhandari}, \citenamefont {Fazio}, \citenamefont {Taddei},\ and\
  \citenamefont {Arrachea}}]{Bhandari2021}%
  \BibitemOpen
  \bibfield  {author} {\bibinfo {author} {\bibnamefont {Bhandari},
  \bibfnamefont {B.}}, \bibinfo {author} {\bibfnamefont {R.}~\bibnamefont
  {Fazio}}, \bibinfo {author} {\bibfnamefont {F.}~\bibnamefont {Taddei}}, \
  and\ \bibinfo {author} {\bibfnamefont {L.}~\bibnamefont {Arrachea}}}
  (\bibinfo {year} {2021}),\ \href {\doibase 10.1103/PhysRevB.104.035425}
  {\bibfield  {journal} {\bibinfo  {journal} {Phys. Rev. B}\ }\textbf {\bibinfo
  {volume} {104}},\ \bibinfo {pages} {035425}}\BibitemShut {NoStop}%
\bibitem [{\citenamefont {Bhaseen}\ \emph {et~al.}(2012)\citenamefont
  {Bhaseen}, \citenamefont {Mayoh}, \citenamefont {Simons},\ and\ \citenamefont
  {Keeling}}]{bhaseen2012a}%
  \BibitemOpen
  \bibfield  {author} {\bibinfo {author} {\bibnamefont {Bhaseen}, \bibfnamefont
  {M.~J.}}, \bibinfo {author} {\bibfnamefont {J.}~\bibnamefont {Mayoh}},
  \bibinfo {author} {\bibfnamefont {B.~D.}\ \bibnamefont {Simons}}, \ and\
  \bibinfo {author} {\bibfnamefont {J.}~\bibnamefont {Keeling}}} (\bibinfo
  {year} {2012}),\ \href {\doibase 10.1103/PhysRevA.85.013817} {\bibfield
  {journal} {\bibinfo  {journal} {Physical Review A}\ }\textbf {\bibinfo
  {volume} {85}},\ \bibinfo {pages} {013817}}\BibitemShut {NoStop}%
\bibitem [{\citenamefont {Biella}\ \emph {et~al.}(2019)\citenamefont {Biella},
  \citenamefont {Collura}, \citenamefont {Rossini}, \citenamefont {De~Luca},\
  and\ \citenamefont {Mazza}}]{BiellaMazza2019}%
  \BibitemOpen
  \bibfield  {author} {\bibinfo {author} {\bibnamefont {Biella}, \bibfnamefont
  {A.}}, \bibinfo {author} {\bibfnamefont {M.}~\bibnamefont {Collura}},
  \bibinfo {author} {\bibfnamefont {D.}~\bibnamefont {Rossini}}, \bibinfo
  {author} {\bibfnamefont {A.}~\bibnamefont {De~Luca}}, \ and\ \bibinfo
  {author} {\bibfnamefont {L.}~\bibnamefont {Mazza}}} (\bibinfo {year}
  {2019}),\ \href {\doibase 10.1038/s41467-019-12784-4} {\bibfield  {journal}
  {\bibinfo  {journal} {Nature Communications}\ }\textbf {\bibinfo {volume}
  {10}},\ \bibinfo {pages} {4820}}\BibitemShut {NoStop}%
\bibitem [{\citenamefont {Biella}\ \emph {et~al.}(2016)\citenamefont {Biella},
  \citenamefont {De~Luca}, \citenamefont {Viti}, \citenamefont {Rossini},
  \citenamefont {Mazza},\ and\ \citenamefont {Fazio}}]{BiellaFazio2016}%
  \BibitemOpen
  \bibfield  {author} {\bibinfo {author} {\bibnamefont {Biella}, \bibfnamefont
  {A.}}, \bibinfo {author} {\bibfnamefont {A.}~\bibnamefont {De~Luca}},
  \bibinfo {author} {\bibfnamefont {J.}~\bibnamefont {Viti}}, \bibinfo {author}
  {\bibfnamefont {D.}~\bibnamefont {Rossini}}, \bibinfo {author} {\bibfnamefont
  {L.}~\bibnamefont {Mazza}}, \ and\ \bibinfo {author} {\bibfnamefont
  {R.}~\bibnamefont {Fazio}}} (\bibinfo {year} {2016}),\ \href {\doibase
  10.1103/PhysRevB.93.205121} {\bibfield  {journal} {\bibinfo  {journal} {Phys.
  Rev. B}\ }\textbf {\bibinfo {volume} {93}},\ \bibinfo {pages}
  {205121}}\BibitemShut {NoStop}%
\bibitem [{\citenamefont {Biggerstaff}\ \emph {et~al.}(2016)\citenamefont
  {Biggerstaff}, \citenamefont {Heilmann}, \citenamefont {Zecevik},
  \citenamefont {Gr{\"{a}}fe}, \citenamefont {Broome}, \citenamefont
  {Fedrizzi}, \citenamefont {Nolte}, \citenamefont {Szameit}, \citenamefont
  {White},\ and\ \citenamefont {Kassal}}]{Biggerstaff2016}%
  \BibitemOpen
  \bibfield  {author} {\bibinfo {author} {\bibnamefont {Biggerstaff},
  \bibfnamefont {D.~N.}}, \bibinfo {author} {\bibfnamefont {R.}~\bibnamefont
  {Heilmann}}, \bibinfo {author} {\bibfnamefont {A.~A.}\ \bibnamefont
  {Zecevik}}, \bibinfo {author} {\bibfnamefont {M.}~\bibnamefont
  {Gr{\"{a}}fe}}, \bibinfo {author} {\bibfnamefont {M.~A.}\ \bibnamefont
  {Broome}}, \bibinfo {author} {\bibfnamefont {A.}~\bibnamefont {Fedrizzi}},
  \bibinfo {author} {\bibfnamefont {S.}~\bibnamefont {Nolte}}, \bibinfo
  {author} {\bibfnamefont {A.}~\bibnamefont {Szameit}}, \bibinfo {author}
  {\bibfnamefont {A.~G.}\ \bibnamefont {White}}, \ and\ \bibinfo {author}
  {\bibfnamefont {I.}~\bibnamefont {Kassal}}} (\bibinfo {year} {2016}),\ \href
  {\doibase 10.1038/ncomms11282} {\bibfield  {journal} {\bibinfo  {journal}
  {Nature Communications}\ }\textbf {\bibinfo {volume} {7}}~(\bibinfo {number}
  {1}),\ \bibinfo {pages} {11282}}\BibitemShut {NoStop}%
\bibitem [{\citenamefont {Blasone}\ \emph {et~al.}(2011)\citenamefont
  {Blasone}, \citenamefont {Jizba},\ and\ \citenamefont
  {Vitiello}}]{BlasoneVitiello2011}%
  \BibitemOpen
  \bibfield  {author} {\bibinfo {author} {\bibnamefont {Blasone}, \bibfnamefont
  {M.}}, \bibinfo {author} {\bibfnamefont {P.}~\bibnamefont {Jizba}}, \ and\
  \bibinfo {author} {\bibfnamefont {G.}~\bibnamefont {Vitiello}}} (\bibinfo
  {year} {2011}),\ \href {\doibase 10.1142/p592} {\emph {\bibinfo {title}
  {{Quantum Field Theory and Its Macroscopic Manifestations Boson Condensation,
  Ordered Patterns and Topological Defects}}}}\ (\bibinfo  {publisher} {World
  Scientific},\ \bibinfo {address} {Singapore})\BibitemShut {NoStop}%
\bibitem [{\citenamefont {Blatt}\ and\ \citenamefont
  {Roos}(2012)}]{BlattRoos2012}%
  \BibitemOpen
  \bibfield  {author} {\bibinfo {author} {\bibnamefont {Blatt}, \bibfnamefont
  {R.}}, \ and\ \bibinfo {author} {\bibfnamefont {C.~F.}\ \bibnamefont {Roos}}}
  (\bibinfo {year} {2012}),\ \href {\doibase 10.1038/nphys2252} {\bibfield
  {journal} {\bibinfo  {journal} {Nature Physics}\ }\textbf {\bibinfo {volume}
  {8}},\ \bibinfo {pages} {277}}\BibitemShut {NoStop}%
\bibitem [{\citenamefont {Bloch}\ \emph {et~al.}(2008)\citenamefont {Bloch},
  \citenamefont {Dalibard},\ and\ \citenamefont {Zwerger}}]{BlochZwerger2008}%
  \BibitemOpen
  \bibfield  {author} {\bibinfo {author} {\bibnamefont {Bloch}, \bibfnamefont
  {I.}}, \bibinfo {author} {\bibfnamefont {J.}~\bibnamefont {Dalibard}}, \ and\
  \bibinfo {author} {\bibfnamefont {W.}~\bibnamefont {Zwerger}}} (\bibinfo
  {year} {2008}),\ \href {\doibase 10.1103/RevModPhys.80.885} {\bibfield
  {journal} {\bibinfo  {journal} {Rev. Mod. Phys.}\ }\textbf {\bibinfo {volume}
  {80}},\ \bibinfo {pages} {885}}\BibitemShut {NoStop}%
\bibitem [{\citenamefont {Boers}\ \emph {et~al.}(2007)\citenamefont {Boers},
  \citenamefont {Goedeke}, \citenamefont {Hinrichs},\ and\ \citenamefont
  {Holthaus}}]{BoersHolthaus2007}%
  \BibitemOpen
  \bibfield  {author} {\bibinfo {author} {\bibnamefont {Boers}, \bibfnamefont
  {D.~J.}}, \bibinfo {author} {\bibfnamefont {B.}~\bibnamefont {Goedeke}},
  \bibinfo {author} {\bibfnamefont {D.}~\bibnamefont {Hinrichs}}, \ and\
  \bibinfo {author} {\bibfnamefont {M.}~\bibnamefont {Holthaus}}} (\bibinfo
  {year} {2007}),\ \href {\doibase 10.1103/PhysRevA.75.063404} {\bibfield
  {journal} {\bibinfo  {journal} {Phys. Rev. A}\ }\textbf {\bibinfo {volume}
  {75}},\ \bibinfo {pages} {063404}}\BibitemShut {NoStop}%
\bibitem [{\citenamefont {Boese}\ \emph {et~al.}(2002)\citenamefont {Boese},
  \citenamefont {Hofstetter},\ and\ \citenamefont
  {Schoeller}}]{BoeseSchoeller2002}%
  \BibitemOpen
  \bibfield  {author} {\bibinfo {author} {\bibnamefont {Boese}, \bibfnamefont
  {D.}}, \bibinfo {author} {\bibfnamefont {W.}~\bibnamefont {Hofstetter}}, \
  and\ \bibinfo {author} {\bibfnamefont {H.}~\bibnamefont {Schoeller}}}
  (\bibinfo {year} {2002}),\ \href {\doibase 10.1103/PhysRevB.66.125315}
  {\bibfield  {journal} {\bibinfo  {journal} {Phys. Rev. B}\ }\textbf {\bibinfo
  {volume} {66}},\ \bibinfo {pages} {125315}}\BibitemShut {NoStop}%
\bibitem [{\citenamefont {{Bohigas}}\ \emph {et~al.}(1984)\citenamefont
  {{Bohigas}}, \citenamefont {{Giannoni}},\ and\ \citenamefont
  {{Schmit}}}]{BGS1984}%
  \BibitemOpen
  \bibfield  {author} {\bibinfo {author} {\bibnamefont {{Bohigas}},
  \bibfnamefont {O.}}, \bibinfo {author} {\bibfnamefont {M.~J.}\ \bibnamefont
  {{Giannoni}}}, \ and\ \bibinfo {author} {\bibfnamefont {C.}~\bibnamefont
  {{Schmit}}}} (\bibinfo {year} {1984}),\ \href {\doibase
  10.1103/PhysRevLett.52.1} {\bibfield  {journal} {\bibinfo  {journal} {\prl}\
  }\textbf {\bibinfo {volume} {52}}~(\bibinfo {number} {1}),\ \bibinfo {pages}
  {1}}\BibitemShut {NoStop}%
\bibitem [{\citenamefont {Bolsterli}\ \emph {et~al.}(1970)\citenamefont
  {Bolsterli}, \citenamefont {Rich},\ and\ \citenamefont
  {Visscher}}]{Bolsterli1970}%
  \BibitemOpen
  \bibfield  {author} {\bibinfo {author} {\bibnamefont {Bolsterli},
  \bibfnamefont {M.}}, \bibinfo {author} {\bibfnamefont {M.}~\bibnamefont
  {Rich}}, \ and\ \bibinfo {author} {\bibfnamefont {W.~M.}\ \bibnamefont
  {Visscher}}} (\bibinfo {year} {1970}),\ \href {\doibase
  10.1103/PhysRevA.1.1086} {\bibfield  {journal} {\bibinfo  {journal} {Phys.
  Rev. A}\ }\textbf {\bibinfo {volume} {1}},\ \bibinfo {pages}
  {1086}}\BibitemShut {NoStop}%
\bibitem [{\citenamefont {Bombelli}\ \emph {et~al.}(1986)\citenamefont
  {Bombelli}, \citenamefont {Koul}, \citenamefont {Lee},\ and\ \citenamefont
  {Sorkin}}]{BombelliSorkin1986}%
  \BibitemOpen
  \bibfield  {author} {\bibinfo {author} {\bibnamefont {Bombelli},
  \bibfnamefont {L.}}, \bibinfo {author} {\bibfnamefont {R.~K.}\ \bibnamefont
  {Koul}}, \bibinfo {author} {\bibfnamefont {J.}~\bibnamefont {Lee}}, \ and\
  \bibinfo {author} {\bibfnamefont {R.~D.}\ \bibnamefont {Sorkin}}} (\bibinfo
  {year} {1986}),\ \href {\doibase 10.1103/PhysRevD.34.373} {\bibfield
  {journal} {\bibinfo  {journal} {Phys. Rev. D}\ }\textbf {\bibinfo {volume}
  {34}},\ \bibinfo {pages} {373}}\BibitemShut {NoStop}%
\bibitem [{\citenamefont {Bonnes}\ and\ \citenamefont
  {Lauchli}(2014)}]{BonnesLauchli2014}%
  \BibitemOpen
  \bibfield  {author} {\bibinfo {author} {\bibnamefont {Bonnes}, \bibfnamefont
  {L.}}, \ and\ \bibinfo {author} {\bibfnamefont {A.}~\bibnamefont {Lauchli}}}
  (\bibinfo {year} {2014}),\ \href {https://arxiv.org/abs/1411.4831} {\bibinfo
  {journal} {arXiv:1411.4831}\ }\BibitemShut {NoStop}%
\bibitem [{\citenamefont {Bordia}\ \emph {et~al.}(2017)\citenamefont {Bordia},
  \citenamefont {L{\"u}schen}, \citenamefont {Scherg}, \citenamefont
  {Gopalakrishnan}, \citenamefont {Knap}, \citenamefont {Schneider},\ and\
  \citenamefont {Bloch}}]{BordiaBloch2017}%
  \BibitemOpen
\bibfield  {journal} {  }\bibfield  {author} {\bibinfo {author} {\bibnamefont
  {Bordia}, \bibfnamefont {P.}}, \bibinfo {author} {\bibfnamefont
  {H.}~\bibnamefont {L{\"u}schen}}, \bibinfo {author} {\bibfnamefont
  {S.}~\bibnamefont {Scherg}}, \bibinfo {author} {\bibfnamefont
  {S.}~\bibnamefont {Gopalakrishnan}}, \bibinfo {author} {\bibfnamefont
  {M.}~\bibnamefont {Knap}}, \bibinfo {author} {\bibfnamefont {U.}~\bibnamefont
  {Schneider}}, \ and\ \bibinfo {author} {\bibfnamefont {I.}~\bibnamefont
  {Bloch}}} (\bibinfo {year} {2017}),\ \href {\doibase
  10.1103/PhysRevX.7.041047} {\bibfield  {journal} {\bibinfo  {journal} {Phys.
  Rev. X}\ }\textbf {\bibinfo {volume} {7}},\ \bibinfo {pages}
  {041047}}\BibitemShut {NoStop}%
\bibitem [{\citenamefont {Bours}\ \emph {et~al.}(2019)\citenamefont {Bours},
  \citenamefont {Sothmann}, \citenamefont {Carrega}, \citenamefont {Strambini},
  \citenamefont {Braggio}, \citenamefont {Hankiewicz}, \citenamefont
  {Molenkamp},\ and\ \citenamefont {Giazotto}}]{BoursGiazotto2019}%
  \BibitemOpen
  \bibfield  {author} {\bibinfo {author} {\bibnamefont {Bours}, \bibfnamefont
  {L.}}, \bibinfo {author} {\bibfnamefont {B.}~\bibnamefont {Sothmann}},
  \bibinfo {author} {\bibfnamefont {M.}~\bibnamefont {Carrega}}, \bibinfo
  {author} {\bibfnamefont {E.}~\bibnamefont {Strambini}}, \bibinfo {author}
  {\bibfnamefont {A.}~\bibnamefont {Braggio}}, \bibinfo {author} {\bibfnamefont
  {E.~M.}\ \bibnamefont {Hankiewicz}}, \bibinfo {author} {\bibfnamefont
  {L.~W.}\ \bibnamefont {Molenkamp}}, \ and\ \bibinfo {author} {\bibfnamefont
  {F.}~\bibnamefont {Giazotto}}} (\bibinfo {year} {2019}),\ \href {\doibase
  10.1103/PhysRevApplied.11.044073} {\bibfield  {journal} {\bibinfo  {journal}
  {Phys. Rev. Applied}\ }\textbf {\bibinfo {volume} {11}},\ \bibinfo {pages}
  {044073}}\BibitemShut {NoStop}%
\bibitem [{\citenamefont {Braggio}\ \emph {et~al.}(2009)\citenamefont
  {Braggio}, \citenamefont {Flindt},\ and\ \citenamefont
  {Novotn{\'{y}}}}]{braggio2009a}%
  \BibitemOpen
  \bibfield  {author} {\bibinfo {author} {\bibnamefont {Braggio}, \bibfnamefont
  {A.}}, \bibinfo {author} {\bibfnamefont {C.}~\bibnamefont {Flindt}}, \ and\
  \bibinfo {author} {\bibfnamefont {T.}~\bibnamefont {Novotn{\'{y}}}}}
  (\bibinfo {year} {2009}),\ \href {\doibase 10.1088/1742-5468/2009/01/p01048}
  {\bibfield  {journal} {\bibinfo  {journal} {Journal of Statistical Mechanics:
  Theory and Experiment}\ }\textbf {\bibinfo {volume} {2009}}~(\bibinfo
  {number} {01}),\ \bibinfo {pages} {P01048}}\BibitemShut {NoStop}%
\bibitem [{\citenamefont {Brand\~{a}o}\ \emph {et~al.}(2015)\citenamefont
  {Brand\~{a}o}, \citenamefont {Cubitt}, \citenamefont {Lucia}, \citenamefont
  {Michalakis},\ and\ \citenamefont {Perez-Garcia}}]{BrandaoPerezGarcia2015}%
  \BibitemOpen
  \bibfield  {author} {\bibinfo {author} {\bibnamefont {Brand\~{a}o},
  \bibfnamefont {F.~G. S.~L.}}, \bibinfo {author} {\bibfnamefont {T.~S.}\
  \bibnamefont {Cubitt}}, \bibinfo {author} {\bibfnamefont {A.}~\bibnamefont
  {Lucia}}, \bibinfo {author} {\bibfnamefont {S.}~\bibnamefont {Michalakis}}, \
  and\ \bibinfo {author} {\bibfnamefont {D.}~\bibnamefont {Perez-Garcia}}}
  (\bibinfo {year} {2015}),\ \href {\doibase 10.1063/1.4932612} {\bibfield
  {journal} {\bibinfo  {journal} {Journal of Mathematical Physics}\ }\textbf
  {\bibinfo {volume} {56}},\ \bibinfo {pages} {102202}}\BibitemShut {NoStop}%
\bibitem [{\citenamefont {Brandes}(2005)}]{brandes2005a}%
  \BibitemOpen
  \bibfield  {author} {\bibinfo {author} {\bibnamefont {Brandes}, \bibfnamefont
  {T.}}} (\bibinfo {year} {2005}),\ \href {\doibase
  https://doi.org/10.1016/j.physrep.2004.12.002} {\bibfield  {journal}
  {\bibinfo  {journal} {Physics Reports}\ }\textbf {\bibinfo {volume}
  {408}}~(\bibinfo {number} {5}),\ \bibinfo {pages} {315}}\BibitemShut
  {NoStop}%
\bibitem [{\citenamefont {Brandes}(2008)}]{brandes2008a}%
  \BibitemOpen
  \bibfield  {author} {\bibinfo {author} {\bibnamefont {Brandes}, \bibfnamefont
  {T.}}} (\bibinfo {year} {2008}),\ \href {\doibase 10.1002/andp.200810306}
  {\bibfield  {journal} {\bibinfo  {journal} {Annalen der Physik (Berlin)}\
  }\textbf {\bibinfo {volume} {17}},\ \bibinfo {pages} {477}}\BibitemShut
  {NoStop}%
\bibitem [{\citenamefont {Brandes}(2010)}]{brandes2010a}%
  \BibitemOpen
  \bibfield  {author} {\bibinfo {author} {\bibnamefont {Brandes}, \bibfnamefont
  {T.}}} (\bibinfo {year} {2010}),\ \href {\doibase
  10.1103/PhysRevLett.105.060602} {\bibfield  {journal} {\bibinfo  {journal}
  {Physical Review Letters}\ }\textbf {\bibinfo {volume} {105}}~(\bibinfo
  {number} {6}),\ \bibinfo {pages} {060602}}\BibitemShut {NoStop}%
\bibitem [{\citenamefont {Brantut}\ \emph {et~al.}(2013)\citenamefont
  {Brantut}, \citenamefont {Grenier}, \citenamefont {Meineke}, \citenamefont
  {Stadler}, \citenamefont {Krinner}, \citenamefont {Kollath}, \citenamefont
  {Esslinger},\ and\ \citenamefont {Georges}}]{BrantutGeorges2013}%
  \BibitemOpen
  \bibfield  {author} {\bibinfo {author} {\bibnamefont {Brantut}, \bibfnamefont
  {J.-P.}}, \bibinfo {author} {\bibfnamefont {C.}~\bibnamefont {Grenier}},
  \bibinfo {author} {\bibfnamefont {J.}~\bibnamefont {Meineke}}, \bibinfo
  {author} {\bibfnamefont {D.}~\bibnamefont {Stadler}}, \bibinfo {author}
  {\bibfnamefont {S.}~\bibnamefont {Krinner}}, \bibinfo {author} {\bibfnamefont
  {C.}~\bibnamefont {Kollath}}, \bibinfo {author} {\bibfnamefont
  {T.}~\bibnamefont {Esslinger}}, \ and\ \bibinfo {author} {\bibfnamefont
  {A.}~\bibnamefont {Georges}}} (\bibinfo {year} {2013}),\ \href {\doibase
  10.1126/science.1242308} {\bibfield  {journal} {\bibinfo  {journal}
  {Science}\ }\textbf {\bibinfo {volume} {342}}~(\bibinfo {number} {6159}),\
  \bibinfo {pages} {713}}\BibitemShut {NoStop}%
\bibitem [{\citenamefont {Brantut}\ \emph {et~al.}(2012)\citenamefont
  {Brantut}, \citenamefont {Meineke}, \citenamefont {Stadler}, \citenamefont
  {Krinner},\ and\ \citenamefont {Esslinger}}]{BrantutEsslinger2012}%
  \BibitemOpen
  \bibfield  {author} {\bibinfo {author} {\bibnamefont {Brantut}, \bibfnamefont
  {J.-P.}}, \bibinfo {author} {\bibfnamefont {J.}~\bibnamefont {Meineke}},
  \bibinfo {author} {\bibfnamefont {D.}~\bibnamefont {Stadler}}, \bibinfo
  {author} {\bibfnamefont {S.}~\bibnamefont {Krinner}}, \ and\ \bibinfo
  {author} {\bibfnamefont {T.}~\bibnamefont {Esslinger}}} (\bibinfo {year}
  {2012}),\ \href {\doibase 10.1126/science.1223175} {\bibfield  {journal}
  {\bibinfo  {journal} {Science}\ }\textbf {\bibinfo {volume} {337}}~(\bibinfo
  {number} {6098}),\ \bibinfo {pages} {1069}}\BibitemShut {NoStop}%
\bibitem [{\citenamefont {Brattain}(1951)}]{Brattain1951}%
  \BibitemOpen
  \bibfield  {author} {\bibinfo {author} {\bibnamefont {Brattain},
  \bibfnamefont {W.~H.}}} (\bibinfo {year} {1951}),\ \href {\doibase
  10.1103/RevModPhys.23.203} {\bibfield  {journal} {\bibinfo  {journal} {Rev.
  Mod. Phys.}\ }\textbf {\bibinfo {volume} {23}},\ \bibinfo {pages}
  {203}}\BibitemShut {NoStop}%
\bibitem [{\citenamefont {Brenes}\ \emph {et~al.}(2018)\citenamefont {Brenes},
  \citenamefont {Mascarenhas}, \citenamefont {Rigol},\ and\ \citenamefont
  {Goold}}]{BrenesGoold2018}%
  \BibitemOpen
  \bibfield  {author} {\bibinfo {author} {\bibnamefont {Brenes}, \bibfnamefont
  {M.}}, \bibinfo {author} {\bibfnamefont {E.}~\bibnamefont {Mascarenhas}},
  \bibinfo {author} {\bibfnamefont {M.}~\bibnamefont {Rigol}}, \ and\ \bibinfo
  {author} {\bibfnamefont {J.}~\bibnamefont {Goold}}} (\bibinfo {year}
  {2018}),\ \href {\doibase 10.1103/PhysRevB.98.235128} {\bibfield  {journal}
  {\bibinfo  {journal} {Phys. Rev. B}\ }\textbf {\bibinfo {volume} {98}},\
  \bibinfo {pages} {235128}}\BibitemShut {NoStop}%
\bibitem [{\citenamefont {Brenes}\ \emph {et~al.}(2020)\citenamefont {Brenes},
  \citenamefont {Mendoza-Arenas}, \citenamefont {Purkayastha}, \citenamefont
  {Mitchison}, \citenamefont {Clark},\ and\ \citenamefont
  {Goold}}]{BrenesGoold2020}%
  \BibitemOpen
  \bibfield  {author} {\bibinfo {author} {\bibnamefont {Brenes}, \bibfnamefont
  {M.}}, \bibinfo {author} {\bibfnamefont {J.~J.}\ \bibnamefont
  {Mendoza-Arenas}}, \bibinfo {author} {\bibfnamefont {A.}~\bibnamefont
  {Purkayastha}}, \bibinfo {author} {\bibfnamefont {M.~T.}\ \bibnamefont
  {Mitchison}}, \bibinfo {author} {\bibfnamefont {S.~R.}\ \bibnamefont
  {Clark}}, \ and\ \bibinfo {author} {\bibfnamefont {J.}~\bibnamefont {Goold}}}
  (\bibinfo {year} {2020}),\ \href {\doibase 10.1103/PhysRevX.10.031040}
  {\bibfield  {journal} {\bibinfo  {journal} {Phys. Rev. X}\ }\textbf {\bibinfo
  {volume} {10}},\ \bibinfo {pages} {031040}}\BibitemShut {NoStop}%
\bibitem [{\citenamefont {Breuer}(2012)}]{breuer2012a}%
  \BibitemOpen
  \bibfield  {author} {\bibinfo {author} {\bibnamefont {Breuer}, \bibfnamefont
  {H.-P.}}} (\bibinfo {year} {2012}),\ \href {\doibase
  10.1088/0953-4075/45/15/154001} {\bibfield  {journal} {\bibinfo  {journal}
  {Journal of Physics B: Atomic, Molecular and Optical Physics}\ }\textbf
  {\bibinfo {volume} {45}},\ \bibinfo {pages} {154001}}\BibitemShut {NoStop}%
\bibitem [{\citenamefont {Breuer}\ \emph {et~al.}(2016)\citenamefont {Breuer},
  \citenamefont {Laine}, \citenamefont {Piilo},\ and\ \citenamefont
  {Vacchini}}]{Breuer2015}%
  \BibitemOpen
  \bibfield  {author} {\bibinfo {author} {\bibnamefont {Breuer}, \bibfnamefont
  {H.-P.}}, \bibinfo {author} {\bibfnamefont {E.-M.}\ \bibnamefont {Laine}},
  \bibinfo {author} {\bibfnamefont {J.}~\bibnamefont {Piilo}}, \ and\ \bibinfo
  {author} {\bibfnamefont {B.}~\bibnamefont {Vacchini}}} (\bibinfo {year}
  {2016}),\ \href {\doibase 10.1103/RevModPhys.88.021002} {\bibfield  {journal}
  {\bibinfo  {journal} {Rev. Mod. Phys.}\ }\textbf {\bibinfo {volume} {88}},\
  \bibinfo {pages} {021002}}\BibitemShut {NoStop}%
\bibitem [{\citenamefont {Breuer}\ and\ \citenamefont
  {Petruccione}(2002)}]{Breuer2002}%
  \BibitemOpen
  \bibfield  {author} {\bibinfo {author} {\bibnamefont {Breuer}, \bibfnamefont
  {H.-P.}}, \ and\ \bibinfo {author} {\bibfnamefont {F.}~\bibnamefont
  {Petruccione}}} (\bibinfo {year} {2002}),\ \href {\doibase
  10.1093/acprof:oso/9780199213900.001.0001} {\emph {\bibinfo {title} {{The
  Theory of Open Quantum Systems}}}}\ (\bibinfo  {publisher} {Oxford University
  Press, USA})\BibitemShut {NoStop}%
\bibitem [{\citenamefont {Bruder}\ \emph {et~al.}(1996)\citenamefont {Bruder},
  \citenamefont {Fazio},\ and\ \citenamefont
  {Schoeller}}]{BruderSchoeller1996}%
  \BibitemOpen
  \bibfield  {author} {\bibinfo {author} {\bibnamefont {Bruder}, \bibfnamefont
  {C.}}, \bibinfo {author} {\bibfnamefont {R.}~\bibnamefont {Fazio}}, \ and\
  \bibinfo {author} {\bibfnamefont {H.}~\bibnamefont {Schoeller}}} (\bibinfo
  {year} {1996}),\ \href {\doibase 10.1103/PhysRevLett.76.114} {\bibfield
  {journal} {\bibinfo  {journal} {Phys. Rev. Lett.}\ }\textbf {\bibinfo
  {volume} {76}},\ \bibinfo {pages} {114}}\BibitemShut {NoStop}%
\bibitem [{\citenamefont {Bruderer}\ \emph {et~al.}(2014)\citenamefont
  {Bruderer}, \citenamefont {Contreras-Pulido}, \citenamefont {Thaller},
  \citenamefont {Sironi}, \citenamefont {Obreschkow},\ and\ \citenamefont
  {Plenio}}]{bruderer2014a}%
  \BibitemOpen
  \bibfield  {author} {\bibinfo {author} {\bibnamefont {Bruderer},
  \bibfnamefont {M.}}, \bibinfo {author} {\bibfnamefont {L.~D.}\ \bibnamefont
  {Contreras-Pulido}}, \bibinfo {author} {\bibfnamefont {M.}~\bibnamefont
  {Thaller}}, \bibinfo {author} {\bibfnamefont {L.}~\bibnamefont {Sironi}},
  \bibinfo {author} {\bibfnamefont {D.}~\bibnamefont {Obreschkow}}, \ and\
  \bibinfo {author} {\bibfnamefont {M.~B.}\ \bibnamefont {Plenio}}} (\bibinfo
  {year} {2014}),\ \href {\doibase 10.1088/1367-2630/16/3/033030} {\bibfield
  {journal} {\bibinfo  {journal} {New Journal of Physics}\ }\textbf {\bibinfo
  {volume} {16}}~(\bibinfo {number} {3}),\ \bibinfo {pages}
  {033030}}\BibitemShut {NoStop}%
\bibitem [{\citenamefont {Bu{\v{c}}a}\ \emph {et~al.}(2022)\citenamefont
  {Bu{\v{c}}a}, \citenamefont {Booker},\ and\ \citenamefont
  {Jaksch}}]{Buca2021}%
  \BibitemOpen
  \bibfield  {author} {\bibinfo {author} {\bibnamefont {Bu{\v{c}}a},
  \bibfnamefont {B.}}, \bibinfo {author} {\bibfnamefont {C.}~\bibnamefont
  {Booker}}, \ and\ \bibinfo {author} {\bibfnamefont {D.}~\bibnamefont
  {Jaksch}}} (\bibinfo {year} {2022}),\ \href {\doibase
  10.21468/SciPostPhys.12.3.097} {\bibfield  {journal} {\bibinfo  {journal}
  {SciPost Phys.}\ }\textbf {\bibinfo {volume} {12}},\ \bibinfo {pages}
  {97}}\BibitemShut {NoStop}%
\bibitem [{\citenamefont {Bu{\v{c}}a}\ \emph {et~al.}(2020)\citenamefont
  {Bu{\v{c}}a}, \citenamefont {Booker}, \citenamefont {Medenjak},\ and\
  \citenamefont {Jaksch}}]{Buca2020}%
  \BibitemOpen
  \bibfield  {author} {\bibinfo {author} {\bibnamefont {Bu{\v{c}}a},
  \bibfnamefont {B.}}, \bibinfo {author} {\bibfnamefont {C.}~\bibnamefont
  {Booker}}, \bibinfo {author} {\bibfnamefont {M.}~\bibnamefont {Medenjak}}, \
  and\ \bibinfo {author} {\bibfnamefont {D.}~\bibnamefont {Jaksch}}} (\bibinfo
  {year} {2020}),\ \href {\doibase 10.1088/1367-2630/abd124} {\bibfield
  {journal} {\bibinfo  {journal} {New Journal of Physics}\ }\textbf {\bibinfo
  {volume} {22}}~(\bibinfo {number} {12}),\ \bibinfo {pages}
  {123040}}\BibitemShut {NoStop}%
\bibitem [{\citenamefont {Bu{\v{c}}a}\ and\ \citenamefont
  {Prosen}(2012)}]{Buca2012}%
  \BibitemOpen
  \bibfield  {author} {\bibinfo {author} {\bibnamefont {Bu{\v{c}}a},
  \bibfnamefont {B.}}, \ and\ \bibinfo {author} {\bibfnamefont
  {T.}~\bibnamefont {Prosen}}} (\bibinfo {year} {2012}),\ \href {\doibase
  10.1088/1367-2630/14/7/073007} {\bibfield  {journal} {\bibinfo  {journal}
  {New Journal of Physics}\ }\textbf {\bibinfo {volume} {14}}~(\bibinfo
  {number} {7}),\ \bibinfo {pages} {073007}}\BibitemShut {NoStop}%
\bibitem [{\citenamefont {Bu{\v{c}}a}\ and\ \citenamefont
  {Prosen}(2014)}]{BucaProsen2014}%
  \BibitemOpen
  \bibfield  {author} {\bibinfo {author} {\bibnamefont {Bu{\v{c}}a},
  \bibfnamefont {B.}}, \ and\ \bibinfo {author} {\bibfnamefont
  {T.}~\bibnamefont {Prosen}}} (\bibinfo {year} {2014}),\ \href {\doibase
  10.1103/PhysRevLett.112.067201} {\bibfield  {journal} {\bibinfo  {journal}
  {Phys. Rev. Lett.}\ }\textbf {\bibinfo {volume} {112}},\ \bibinfo {pages}
  {067201}}\BibitemShut {NoStop}%
\bibitem [{\citenamefont {Bu{\v{c}}a}\ and\ \citenamefont
  {Prosen}(2016)}]{Buca2016}%
  \BibitemOpen
  \bibfield  {author} {\bibinfo {author} {\bibnamefont {Bu{\v{c}}a},
  \bibfnamefont {B.}}, \ and\ \bibinfo {author} {\bibfnamefont
  {T.}~\bibnamefont {Prosen}}} (\bibinfo {year} {2016}),\ \href {\doibase
  10.1088/1742-5468/2016/02/023102} {\bibfield  {journal} {\bibinfo  {journal}
  {Journal of Statistical Mechanics: Theory and Experiment}\ }\textbf {\bibinfo
  {volume} {2016}}~(\bibinfo {number} {2}),\ \bibinfo {pages}
  {023102}}\BibitemShut {NoStop}%
\bibitem [{\citenamefont {Bu{\v{c}}a}\ \emph {et~al.}(2019)\citenamefont
  {Bu{\v{c}}a}, \citenamefont {Tindall},\ and\ \citenamefont
  {Jaksch}}]{BucaJaksch2019}%
  \BibitemOpen
  \bibfield  {author} {\bibinfo {author} {\bibnamefont {Bu{\v{c}}a},
  \bibfnamefont {B.}}, \bibinfo {author} {\bibfnamefont {J.}~\bibnamefont
  {Tindall}}, \ and\ \bibinfo {author} {\bibfnamefont {D.}~\bibnamefont
  {Jaksch}}} (\bibinfo {year} {2019}),\ \href {\doibase
  10.1038/s41467-019-09757-y} {\bibfield  {journal} {\bibinfo  {journal}
  {Nature Comm.}\ }\textbf {\bibinfo {volume} {10}},\ \bibinfo {pages}
  {1730}}\BibitemShut {NoStop}%
\bibitem [{\citenamefont {Bulla}\ \emph {et~al.}(2008)\citenamefont {Bulla},
  \citenamefont {Costi},\ and\ \citenamefont {Pruschke}}]{BullaPruschke2008}%
  \BibitemOpen
  \bibfield  {author} {\bibinfo {author} {\bibnamefont {Bulla}, \bibfnamefont
  {R.}}, \bibinfo {author} {\bibfnamefont {T.~A.}\ \bibnamefont {Costi}}, \
  and\ \bibinfo {author} {\bibfnamefont {T.}~\bibnamefont {Pruschke}}}
  (\bibinfo {year} {2008}),\ \href {\doibase 10.1103/RevModPhys.80.395}
  {\bibfield  {journal} {\bibinfo  {journal} {Rev. Mod. Phys.}\ }\textbf
  {\bibinfo {volume} {80}},\ \bibinfo {pages} {395}}\BibitemShut {NoStop}%
\bibitem [{\citenamefont {Bulla}\ \emph {et~al.}(2005)\citenamefont {Bulla},
  \citenamefont {Lee}, \citenamefont {Tong},\ and\ \citenamefont
  {Vojta}}]{BullaVojta2005}%
  \BibitemOpen
  \bibfield  {author} {\bibinfo {author} {\bibnamefont {Bulla}, \bibfnamefont
  {R.}}, \bibinfo {author} {\bibfnamefont {H.-J.}\ \bibnamefont {Lee}},
  \bibinfo {author} {\bibfnamefont {N.-H.}\ \bibnamefont {Tong}}, \ and\
  \bibinfo {author} {\bibfnamefont {M.}~\bibnamefont {Vojta}}} (\bibinfo {year}
  {2005}),\ \href {\doibase 10.1103/PhysRevB.71.045122} {\bibfield  {journal}
  {\bibinfo  {journal} {Phys. Rev. B}\ }\textbf {\bibinfo {volume} {71}},\
  \bibinfo {pages} {045122}}\BibitemShut {NoStop}%
\bibitem [{\citenamefont {B{\"u}ttiker}(1986)}]{buettiker1986a}%
  \BibitemOpen
  \bibfield  {author} {\bibinfo {author} {\bibnamefont {B{\"u}ttiker},
  \bibfnamefont {M.}}} (\bibinfo {year} {1986}),\ \href {\doibase
  10.1103/PhysRevLett.57.1761} {\bibfield  {journal} {\bibinfo  {journal}
  {Physical Review Letters}\ }\textbf {\bibinfo {volume} {57}},\ \bibinfo
  {pages} {1761}}\BibitemShut {NoStop}%
\bibitem [{\citenamefont {Cabot}\ \emph {et~al.}(2017)\citenamefont {Cabot},
  \citenamefont {Galve},\ and\ \citenamefont {Zambrini}}]{cabot2017a}%
  \BibitemOpen
  \bibfield  {author} {\bibinfo {author} {\bibnamefont {Cabot}, \bibfnamefont
  {A.}}, \bibinfo {author} {\bibfnamefont {F.}~\bibnamefont {Galve}}, \ and\
  \bibinfo {author} {\bibfnamefont {R.}~\bibnamefont {Zambrini}}} (\bibinfo
  {year} {2017}),\ \href {\doibase 10.1088/1367-2630/aa8b9c} {\bibfield
  {journal} {\bibinfo  {journal} {New Journal of Physics}\ }\textbf {\bibinfo
  {volume} {19}}~(\bibinfo {number} {11}),\ \bibinfo {pages}
  {113007}}\BibitemShut {NoStop}%
\bibitem [{\citenamefont {Caffarel}\ and\ \citenamefont
  {Krauth}(1994)}]{CaffarelKrauth1994}%
  \BibitemOpen
  \bibfield  {author} {\bibinfo {author} {\bibnamefont {Caffarel},
  \bibfnamefont {M.}}, \ and\ \bibinfo {author} {\bibfnamefont
  {W.}~\bibnamefont {Krauth}}} (\bibinfo {year} {1994}),\ \href {\doibase
  10.1103/PhysRevLett.72.1545} {\bibfield  {journal} {\bibinfo  {journal}
  {Phys. Rev. Lett.}\ }\textbf {\bibinfo {volume} {72}},\ \bibinfo {pages}
  {1545}}\BibitemShut {NoStop}%
\bibitem [{\citenamefont {Cai}\ and\ \citenamefont
  {Barthel}(2013)}]{CaiBarthel2013}%
  \BibitemOpen
  \bibfield  {author} {\bibinfo {author} {\bibnamefont {Cai}, \bibfnamefont
  {Z.}}, \ and\ \bibinfo {author} {\bibfnamefont {T.}~\bibnamefont {Barthel}}}
  (\bibinfo {year} {2013}),\ \href {\doibase 10.1103/PhysRevLett.111.150403}
  {\bibfield  {journal} {\bibinfo  {journal} {Phys. Rev. Lett.}\ }\textbf
  {\bibinfo {volume} {111}},\ \bibinfo {pages} {150403}}\BibitemShut {NoStop}%
\bibitem [{\citenamefont {Campisi}\ \emph {et~al.}(2011)\citenamefont
  {Campisi}, \citenamefont {H{\"a}nggi},\ and\ \citenamefont
  {Talkner}}]{campisi2011a}%
  \BibitemOpen
  \bibfield  {author} {\bibinfo {author} {\bibnamefont {Campisi}, \bibfnamefont
  {M.}}, \bibinfo {author} {\bibfnamefont {P.}~\bibnamefont {H{\"a}nggi}}, \
  and\ \bibinfo {author} {\bibfnamefont {P.}~\bibnamefont {Talkner}}} (\bibinfo
  {year} {2011}),\ \href {\doibase 10.1103/RevModPhys.83.771} {\bibfield
  {journal} {\bibinfo  {journal} {Rev. Mod. Phys.}\ }\textbf {\bibinfo {volume}
  {83}}~(\bibinfo {number} {3}),\ \bibinfo {pages} {771}}\BibitemShut {NoStop}%
\bibitem [{\citenamefont {Can}(2019)}]{can2019a}%
  \BibitemOpen
  \bibfield  {author} {\bibinfo {author} {\bibnamefont {Can}, \bibfnamefont
  {T.}}} (\bibinfo {year} {2019}),\ \href {\doibase 10.1088/1751-8121/ab4d26}
  {\bibfield  {journal} {\bibinfo  {journal} {Journal of Physics A:
  Mathematical and Theoretical}\ }\textbf {\bibinfo {volume} {52}}~(\bibinfo
  {number} {48}),\ \bibinfo {pages} {485302}}\BibitemShut {NoStop}%
\bibitem [{\citenamefont {Can}\ \emph {et~al.}(2019)\citenamefont {Can},
  \citenamefont {Oganesyan}, \citenamefont {Orgad},\ and\ \citenamefont
  {Gopalakrishnan}}]{CanGopalakrishnan2019}%
  \BibitemOpen
  \bibfield  {author} {\bibinfo {author} {\bibnamefont {Can}, \bibfnamefont
  {T.}}, \bibinfo {author} {\bibfnamefont {V.}~\bibnamefont {Oganesyan}},
  \bibinfo {author} {\bibfnamefont {D.}~\bibnamefont {Orgad}}, \ and\ \bibinfo
  {author} {\bibfnamefont {S.}~\bibnamefont {Gopalakrishnan}}} (\bibinfo {year}
  {2019}),\ \href {\doibase 10.1103/PhysRevLett.123.234103} {\bibfield
  {journal} {\bibinfo  {journal} {Phys. Rev. Lett.}\ }\textbf {\bibinfo
  {volume} {123}},\ \bibinfo {pages} {234103}}\BibitemShut {NoStop}%
\bibitem [{\citenamefont {Carmichael}(1993)}]{Carmichael1993}%
  \BibitemOpen
  \bibfield  {author} {\bibinfo {author} {\bibnamefont {Carmichael},
  \bibfnamefont {H.~J.}}} (\bibinfo {year} {1993}),\ \href {\doibase
  10.1007/978-3-540-47620-7} {\emph {\bibinfo {title} {{An Open Systems
  Approach to Quantum Optics}}}}\ (\bibinfo  {publisher} {Springer,
  Berlin})\BibitemShut {NoStop}%
\bibitem [{\citenamefont {Caroli}\ \emph {et~al.}(1971)\citenamefont {Caroli},
  \citenamefont {Combescot}, \citenamefont {Nozieres},\ and\ \citenamefont
  {{Saint-James}}}]{CaroliSaint-James1971}%
  \BibitemOpen
  \bibfield  {author} {\bibinfo {author} {\bibnamefont {Caroli}, \bibfnamefont
  {C.}}, \bibinfo {author} {\bibfnamefont {R.}~\bibnamefont {Combescot}},
  \bibinfo {author} {\bibfnamefont {P.}~\bibnamefont {Nozieres}}, \ and\
  \bibinfo {author} {\bibfnamefont {D.}~\bibnamefont {{Saint-James}}}}
  (\bibinfo {year} {1971}),\ \href {\doibase 10.1088/0022-3719/4/8/018}
  {\bibfield  {journal} {\bibinfo  {journal} {J. Phys. C: Solid State Phys.}\
  }\textbf {\bibinfo {volume} {4}}~(\bibinfo {number} {8}),\ \bibinfo {pages}
  {916}}\BibitemShut {NoStop}%
\bibitem [{\citenamefont {Carollo}\ \emph {et~al.}(2018)\citenamefont
  {Carollo}, \citenamefont {Garrahan},\ and\ \citenamefont
  {Lesanovsky}}]{carollo2018a}%
  \BibitemOpen
  \bibfield  {author} {\bibinfo {author} {\bibnamefont {Carollo}, \bibfnamefont
  {F.}}, \bibinfo {author} {\bibfnamefont {J.~P.}\ \bibnamefont {Garrahan}}, \
  and\ \bibinfo {author} {\bibfnamefont {I.}~\bibnamefont {Lesanovsky}}}
  (\bibinfo {year} {2018}),\ \href {\doibase 10.1103/PhysRevB.98.094301}
  {\bibfield  {journal} {\bibinfo  {journal} {Phys. Rev. B}\ }\textbf {\bibinfo
  {volume} {98}},\ \bibinfo {pages} {094301}}\BibitemShut {NoStop}%
\bibitem [{\citenamefont {Carollo}\ \emph {et~al.}(2017)\citenamefont
  {Carollo}, \citenamefont {Garrahan}, \citenamefont {Lesanovsky},\ and\
  \citenamefont {P\'erez-Espigares}}]{carollo2017a}%
  \BibitemOpen
  \bibfield  {author} {\bibinfo {author} {\bibnamefont {Carollo}, \bibfnamefont
  {F.}}, \bibinfo {author} {\bibfnamefont {J.~P.}\ \bibnamefont {Garrahan}},
  \bibinfo {author} {\bibfnamefont {I.}~\bibnamefont {Lesanovsky}}, \ and\
  \bibinfo {author} {\bibfnamefont {C.}~\bibnamefont {P\'erez-Espigares}}}
  (\bibinfo {year} {2017}),\ \href {\doibase 10.1103/PhysRevE.96.052118}
  {\bibfield  {journal} {\bibinfo  {journal} {Phys. Rev. E}\ }\textbf {\bibinfo
  {volume} {96}},\ \bibinfo {pages} {052118}}\BibitemShut {NoStop}%
\bibitem [{\citenamefont {Caruso}\ \emph {et~al.}(2009)\citenamefont {Caruso},
  \citenamefont {Chin}, \citenamefont {Datta}, \citenamefont {Huelga},\ and\
  \citenamefont {Plenio}}]{Caruso2009}%
  \BibitemOpen
  \bibfield  {author} {\bibinfo {author} {\bibnamefont {Caruso}, \bibfnamefont
  {F.}}, \bibinfo {author} {\bibfnamefont {A.~W.}\ \bibnamefont {Chin}},
  \bibinfo {author} {\bibfnamefont {A.}~\bibnamefont {Datta}}, \bibinfo
  {author} {\bibfnamefont {S.~F.}\ \bibnamefont {Huelga}}, \ and\ \bibinfo
  {author} {\bibfnamefont {M.~B.}\ \bibnamefont {Plenio}}} (\bibinfo {year}
  {2009}),\ \href {\doibase 10.1063/1.3223548} {\bibfield  {journal} {\bibinfo
  {journal} {The Journal of Chemical Physics}\ }\textbf {\bibinfo {volume}
  {131}}~(\bibinfo {number} {10}),\ \bibinfo {pages} {105106}},\ \Eprint
  {http://arxiv.org/abs/0901.4454} {0901.4454} \BibitemShut {NoStop}%
\bibitem [{\citenamefont {Casagrande}\ \emph {et~al.}(2021)\citenamefont
  {Casagrande}, \citenamefont {Poletti},\ and\ \citenamefont
  {Landi}}]{PeresCasagrandeLandi2020}%
  \BibitemOpen
  \bibfield  {author} {\bibinfo {author} {\bibnamefont {Casagrande},
  \bibfnamefont {H.~P.}}, \bibinfo {author} {\bibfnamefont {D.}~\bibnamefont
  {Poletti}}, \ and\ \bibinfo {author} {\bibfnamefont {G.~T.}\ \bibnamefont
  {Landi}}} (\bibinfo {year} {2021}),\ \href {\doibase
  https://doi.org/10.1016/j.cpc.2021.108060} {\bibfield  {journal} {\bibinfo
  {journal} {Computer Physics Communications}\ }\textbf {\bibinfo {volume}
  {267}},\ \bibinfo {pages} {108060}}\BibitemShut {NoStop}%
\bibitem [{\citenamefont {Casati}\ \emph {et~al.}(1984)\citenamefont {Casati},
  \citenamefont {Ford}, \citenamefont {Vivaldi},\ and\ \citenamefont
  {Visscher}}]{Casati1984}%
  \BibitemOpen
  \bibfield  {author} {\bibinfo {author} {\bibnamefont {Casati}, \bibfnamefont
  {G.}}, \bibinfo {author} {\bibfnamefont {J.}~\bibnamefont {Ford}}, \bibinfo
  {author} {\bibfnamefont {F.}~\bibnamefont {Vivaldi}}, \ and\ \bibinfo
  {author} {\bibfnamefont {W.~M.}\ \bibnamefont {Visscher}}} (\bibinfo {year}
  {1984}),\ \href {\doibase 10.1103/PhysRevLett.52.1861} {\bibfield  {journal}
  {\bibinfo  {journal} {Physical Review Letters}\ }\textbf {\bibinfo {volume}
  {52}}~(\bibinfo {number} {21}),\ \bibinfo {pages} {1861}}\BibitemShut
  {NoStop}%
\bibitem [{\citenamefont {Casati}\ \emph {et~al.}(1980)\citenamefont {Casati},
  \citenamefont {Valz-Gris},\ and\ \citenamefont {Guarnieri}}]{Casati1980}%
  \BibitemOpen
  \bibfield  {author} {\bibinfo {author} {\bibnamefont {Casati}, \bibfnamefont
  {G.}}, \bibinfo {author} {\bibfnamefont {F.}~\bibnamefont {Valz-Gris}}, \
  and\ \bibinfo {author} {\bibfnamefont {I.}~\bibnamefont {Guarnieri}}}
  (\bibinfo {year} {1980}),\ \href {\doibase 10.1007/BF02798790} {\bibfield
  {journal} {\bibinfo  {journal} {Lett. Nuovo Cimento}\ }\textbf {\bibinfo
  {volume} {28}},\ \bibinfo {pages} {279}}\BibitemShut {NoStop}%
\bibitem [{\citenamefont {Cattaneo}\ \emph {et~al.}(2021)\citenamefont
  {Cattaneo}, \citenamefont {De~Chiara}, \citenamefont {Maniscalco},
  \citenamefont {Zambrini},\ and\ \citenamefont {Giorgi}}]{Cattaneo2021}%
  \BibitemOpen
  \bibfield  {author} {\bibinfo {author} {\bibnamefont {Cattaneo},
  \bibfnamefont {M.}}, \bibinfo {author} {\bibfnamefont {G.}~\bibnamefont
  {De~Chiara}}, \bibinfo {author} {\bibfnamefont {S.}~\bibnamefont
  {Maniscalco}}, \bibinfo {author} {\bibfnamefont {R.}~\bibnamefont
  {Zambrini}}, \ and\ \bibinfo {author} {\bibfnamefont {G.~L.}\ \bibnamefont
  {Giorgi}}} (\bibinfo {year} {2021}),\ \href {\doibase
  10.1103/PhysRevLett.126.130403} {\bibfield  {journal} {\bibinfo  {journal}
  {Phys. Rev. Lett.}\ }\textbf {\bibinfo {volume} {126}},\ \bibinfo {pages}
  {130403}}\BibitemShut {NoStop}%
\bibitem [{\citenamefont {Caux}\ and\ \citenamefont
  {Mossel}(2011)}]{CauxMossel2011}%
  \BibitemOpen
  \bibfield  {author} {\bibinfo {author} {\bibnamefont {Caux}, \bibfnamefont
  {J.-S.}}, \ and\ \bibinfo {author} {\bibfnamefont {J.}~\bibnamefont
  {Mossel}}} (\bibinfo {year} {2011}),\ \href {\doibase
  10.1088/1742-5468/2011/02/p02023} {\bibfield  {journal} {\bibinfo  {journal}
  {Journal of Statistical Mechanics: Theory and Experiment}\ }\textbf {\bibinfo
  {volume} {2011}}~(\bibinfo {number} {02}),\ \bibinfo {pages}
  {P02023}}\BibitemShut {NoStop}%
\bibitem [{\citenamefont {Cavaliere}\ \emph {et~al.}(2004)\citenamefont
  {Cavaliere}, \citenamefont {Braggio}, \citenamefont {Stockburger},
  \citenamefont {Sassetti},\ and\ \citenamefont
  {Kramer}}]{CavaliereKramer2004}%
  \BibitemOpen
  \bibfield  {author} {\bibinfo {author} {\bibnamefont {Cavaliere},
  \bibfnamefont {F.}}, \bibinfo {author} {\bibfnamefont {A.}~\bibnamefont
  {Braggio}}, \bibinfo {author} {\bibfnamefont {J.~T.}\ \bibnamefont
  {Stockburger}}, \bibinfo {author} {\bibfnamefont {M.}~\bibnamefont
  {Sassetti}}, \ and\ \bibinfo {author} {\bibfnamefont {B.}~\bibnamefont
  {Kramer}}} (\bibinfo {year} {2004}),\ \href {\doibase
  10.1103/PhysRevLett.93.036803} {\bibfield  {journal} {\bibinfo  {journal}
  {Phys. Rev. Lett.}\ }\textbf {\bibinfo {volume} {93}},\ \bibinfo {pages}
  {036803}}\BibitemShut {NoStop}%
\bibitem [{\citenamefont {Cejnar}\ \emph {et~al.}(2021)\citenamefont {Cejnar},
  \citenamefont {Str{\'{a}}nsk{\'{y}}}, \citenamefont {Macek},\ and\
  \citenamefont {Kloc}}]{cejnar2021a}%
  \BibitemOpen
  \bibfield  {author} {\bibinfo {author} {\bibnamefont {Cejnar}, \bibfnamefont
  {P.}}, \bibinfo {author} {\bibfnamefont {P.}~\bibnamefont
  {Str{\'{a}}nsk{\'{y}}}}, \bibinfo {author} {\bibfnamefont {M.}~\bibnamefont
  {Macek}}, \ and\ \bibinfo {author} {\bibfnamefont {M.}~\bibnamefont {Kloc}}}
  (\bibinfo {year} {2021}),\ \href {\doibase 10.1088/1751-8121/abdfe8}
  {\bibfield  {journal} {\bibinfo  {journal} {Journal of Physics A:
  Mathematical and Theoretical}\ }\textbf {\bibinfo {volume} {54}}~(\bibinfo
  {number} {13}),\ \bibinfo {pages} {133001}}\BibitemShut {NoStop}%
\bibitem [{\citenamefont {Chang}\ \emph {et~al.}(2006)\citenamefont {Chang},
  \citenamefont {Okawa}, \citenamefont {Majumdar},\ and\ \citenamefont
  {Zettl}}]{Chang2006}%
  \BibitemOpen
  \bibfield  {author} {\bibinfo {author} {\bibnamefont {Chang}, \bibfnamefont
  {C.~W.}}, \bibinfo {author} {\bibfnamefont {D.}~\bibnamefont {Okawa}},
  \bibinfo {author} {\bibfnamefont {A.}~\bibnamefont {Majumdar}}, \ and\
  \bibinfo {author} {\bibfnamefont {A.}~\bibnamefont {Zettl}}} (\bibinfo {year}
  {2006}),\ \href {\doibase 10.1126/science.1132898} {\bibfield  {journal}
  {\bibinfo  {journal} {Science}\ }\textbf {\bibinfo {volume} {314}},\ \bibinfo
  {pages} {1121}}\BibitemShut {NoStop}%
\bibitem [{\citenamefont {Chen}\ \emph {et~al.}(1999)\citenamefont {Chen},
  \citenamefont {Reed}, \citenamefont {Rawlett},\ and\ \citenamefont
  {Tour}}]{ChenTour1999}%
  \BibitemOpen
  \bibfield  {author} {\bibinfo {author} {\bibnamefont {Chen}, \bibfnamefont
  {J.}}, \bibinfo {author} {\bibfnamefont {M.~A.}\ \bibnamefont {Reed}},
  \bibinfo {author} {\bibfnamefont {A.~M.}\ \bibnamefont {Rawlett}}, \ and\
  \bibinfo {author} {\bibfnamefont {J.~M.}\ \bibnamefont {Tour}}} (\bibinfo
  {year} {1999}),\ \href {\doibase 10.1126/science.286.5444.1550} {\bibfield
  {journal} {\bibinfo  {journal} {Science}\ }\textbf {\bibinfo {volume}
  {286}},\ \bibinfo {pages} {1550}}\BibitemShut {NoStop}%
\bibitem [{\citenamefont {Chen}\ \emph {et~al.}(2020)\citenamefont {Chen},
  \citenamefont {Balachandran}, \citenamefont {Guo},\ and\ \citenamefont
  {Poletti}}]{ChenPoletti2020}%
  \BibitemOpen
  \bibfield  {author} {\bibinfo {author} {\bibnamefont {Chen}, \bibfnamefont
  {T.}}, \bibinfo {author} {\bibfnamefont {V.}~\bibnamefont {Balachandran}},
  \bibinfo {author} {\bibfnamefont {C.}~\bibnamefont {Guo}}, \ and\ \bibinfo
  {author} {\bibfnamefont {D.}~\bibnamefont {Poletti}}} (\bibinfo {year}
  {2020}),\ \href {\doibase 10.1103/PhysRevE.102.012155} {\bibfield  {journal}
  {\bibinfo  {journal} {Phys. Rev. E}\ }\textbf {\bibinfo {volume} {102}},\
  \bibinfo {pages} {012155}}\BibitemShut {NoStop}%
\bibitem [{\citenamefont {{Chen, Shunda}}\ \emph {et~al.}(2015)\citenamefont
  {{Chen, Shunda}}, \citenamefont {{Pereira, Emmanuel}},\ and\ \citenamefont
  {{Casati, Giulio}}}]{Chen2015}%
  \BibitemOpen
  \bibfield  {author} {\bibinfo {author} {\bibnamefont {{Chen, Shunda}},},
  \bibinfo {author} {\bibnamefont {{Pereira, Emmanuel}}}, \ and\ \bibinfo
  {author} {\bibnamefont {{Casati, Giulio}}}} (\bibinfo {year} {2015}),\ \href
  {\doibase 10.1209/0295-5075/111/30004} {\bibfield  {journal} {\bibinfo
  {journal} {EPL}\ }\textbf {\bibinfo {volume} {111}}~(\bibinfo {number} {3}),\
  \bibinfo {pages} {30004}}\BibitemShut {NoStop}%
\bibitem [{\citenamefont {Chiaracane}\ \emph {et~al.}(2021)\citenamefont
  {Chiaracane}, \citenamefont {Pietracaprina}, \citenamefont {Purkayastha},\
  and\ \citenamefont {Goold}}]{ChiaracaneGoold2021}%
  \BibitemOpen
  \bibfield  {author} {\bibinfo {author} {\bibnamefont {Chiaracane},
  \bibfnamefont {C.}}, \bibinfo {author} {\bibfnamefont {F.}~\bibnamefont
  {Pietracaprina}}, \bibinfo {author} {\bibfnamefont {A.}~\bibnamefont
  {Purkayastha}}, \ and\ \bibinfo {author} {\bibfnamefont {J.}~\bibnamefont
  {Goold}}} (\bibinfo {year} {2021}),\ \href {\doibase
  10.1103/PhysRevB.103.184205} {\bibfield  {journal} {\bibinfo  {journal}
  {Phys. Rev. B}\ }\textbf {\bibinfo {volume} {103}},\ \bibinfo {pages}
  {184205}}\BibitemShut {NoStop}%
\bibitem [{\citenamefont {Chiaracane}\ \emph {et~al.}(2022)\citenamefont
  {Chiaracane}, \citenamefont {Purkayastha}, \citenamefont {Mitchison},\ and\
  \citenamefont {Goold}}]{ChiaracaneGoold2021b}%
  \BibitemOpen
  \bibfield  {author} {\bibinfo {author} {\bibnamefont {Chiaracane},
  \bibfnamefont {C.}}, \bibinfo {author} {\bibfnamefont {A.}~\bibnamefont
  {Purkayastha}}, \bibinfo {author} {\bibfnamefont {M.~T.}\ \bibnamefont
  {Mitchison}}, \ and\ \bibinfo {author} {\bibfnamefont {J.}~\bibnamefont
  {Goold}}} (\bibinfo {year} {2022}),\ \href {\doibase
  10.1103/PhysRevB.105.134203} {\bibfield  {journal} {\bibinfo  {journal}
  {Phys. Rev. B}\ }\textbf {\bibinfo {volume} {105}},\ \bibinfo {pages}
  {134203}}\BibitemShut {NoStop}%
\bibitem [{\citenamefont {Chin}\ \emph
  {et~al.}(2010{\natexlab{a}})\citenamefont {Chin}, \citenamefont {Datta},
  \citenamefont {Caruso}, \citenamefont {Huelga},\ and\ \citenamefont
  {Plenio}}]{Chin2010}%
  \BibitemOpen
  \bibfield  {author} {\bibinfo {author} {\bibnamefont {Chin}, \bibfnamefont
  {A.~W.}}, \bibinfo {author} {\bibfnamefont {A.}~\bibnamefont {Datta}},
  \bibinfo {author} {\bibfnamefont {F.}~\bibnamefont {Caruso}}, \bibinfo
  {author} {\bibfnamefont {S.~F.}\ \bibnamefont {Huelga}}, \ and\ \bibinfo
  {author} {\bibfnamefont {M.~B.}\ \bibnamefont {Plenio}}} (\bibinfo {year}
  {2010}{\natexlab{a}}),\ \href {\doibase 10.1088/1367-2630/12/6/065002}
  {\bibfield  {journal} {\bibinfo  {journal} {New Journal of Physics}\ }\textbf
  {\bibinfo {volume} {12}}~(\bibinfo {number} {6}),\ \bibinfo {pages}
  {065002}}\BibitemShut {NoStop}%
\bibitem [{\citenamefont {Chin}\ \emph {et~al.}(2012)\citenamefont {Chin},
  \citenamefont {Huelga},\ and\ \citenamefont {Plenio}}]{Chin2012}%
  \BibitemOpen
  \bibfield  {author} {\bibinfo {author} {\bibnamefont {Chin}, \bibfnamefont
  {A.~W.}}, \bibinfo {author} {\bibfnamefont {S.~F.}\ \bibnamefont {Huelga}}, \
  and\ \bibinfo {author} {\bibfnamefont {M.~B.}\ \bibnamefont {Plenio}}}
  (\bibinfo {year} {2012}),\ \href {\doibase 10.1098/rsta.2011.0224} {\bibfield
   {journal} {\bibinfo  {journal} {Philosophical Transactions of the Royal
  Society A: Mathematical, Physical and Engineering Sciences}\ }\textbf
  {\bibinfo {volume} {370}}~(\bibinfo {number} {1972}),\ \bibinfo {pages}
  {3638}}\BibitemShut {NoStop}%
\bibitem [{\citenamefont {Chin}\ \emph
  {et~al.}(2010{\natexlab{b}})\citenamefont {Chin}, \citenamefont {Rivas},
  \citenamefont {Huelga},\ and\ \citenamefont {Plenio}}]{ChinPlenio2010}%
  \BibitemOpen
  \bibfield  {author} {\bibinfo {author} {\bibnamefont {Chin}, \bibfnamefont
  {A.~W.}}, \bibinfo {author} {\bibfnamefont {A.}~\bibnamefont {Rivas}},
  \bibinfo {author} {\bibfnamefont {S.~F.}\ \bibnamefont {Huelga}}, \ and\
  \bibinfo {author} {\bibfnamefont {M.~B.}\ \bibnamefont {Plenio}}} (\bibinfo
  {year} {2010}{\natexlab{b}}),\ \href {\doibase 10.1063/1.3490188} {\bibfield
  {journal} {\bibinfo  {journal} {Jour. Math. Phys.}\ }\textbf {\bibinfo
  {volume} {51}},\ \bibinfo {pages} {092109}}\BibitemShut {NoStop}%
\bibitem [{\citenamefont {Chioquetta}\ \emph {et~al.}(2021)\citenamefont
  {Chioquetta}, \citenamefont {Pereira}, \citenamefont {Landi},\ and\
  \citenamefont {Drumond}}]{ChioquettaDrumond2020}%
  \BibitemOpen
  \bibfield  {author} {\bibinfo {author} {\bibnamefont {Chioquetta},
  \bibfnamefont {A.}}, \bibinfo {author} {\bibfnamefont {E.}~\bibnamefont
  {Pereira}}, \bibinfo {author} {\bibfnamefont {G.~T.}\ \bibnamefont {Landi}},
  \ and\ \bibinfo {author} {\bibfnamefont {R.~C.}\ \bibnamefont {Drumond}}}
  (\bibinfo {year} {2021}),\ \href {\doibase 10.1103/PhysRevE.103.032108}
  {\bibfield  {journal} {\bibinfo  {journal} {Phys. Rev. E}\ }\textbf {\bibinfo
  {volume} {103}},\ \bibinfo {pages} {032108}}\BibitemShut {NoStop}%
\bibitem [{\citenamefont {Choi}(1975)}]{Choi}%
  \BibitemOpen
  \bibfield  {author} {\bibinfo {author} {\bibnamefont {Choi}, \bibfnamefont
  {M.-D.}}} (\bibinfo {year} {1975}),\ \href {\doibase
  https://doi.org/10.1016/0024-3795(75)90075-0} {\bibfield  {journal} {\bibinfo
   {journal} {Linear Algebra and its Applications}\ }\textbf {\bibinfo {volume}
  {10}}~(\bibinfo {number} {3}),\ \bibinfo {pages} {285 }}\BibitemShut
  {NoStop}%
\bibitem [{\citenamefont {Clark}\ \emph {et~al.}(2010)\citenamefont {Clark},
  \citenamefont {Prior}, \citenamefont {Hartmann}, \citenamefont {Jaksch},\
  and\ \citenamefont {B.}}]{ClarkPlenio2010}%
  \BibitemOpen
  \bibfield  {author} {\bibinfo {author} {\bibnamefont {Clark}, \bibfnamefont
  {S.~R.}}, \bibinfo {author} {\bibfnamefont {J.}~\bibnamefont {Prior}},
  \bibinfo {author} {\bibfnamefont {M.~J.}\ \bibnamefont {Hartmann}}, \bibinfo
  {author} {\bibfnamefont {D.}~\bibnamefont {Jaksch}}, \ and\ \bibinfo {author}
  {\bibfnamefont {P.~M.}\ \bibnamefont {B.}}} (\bibinfo {year} {2010}),\ \href
  {\doibase 10.1088/1367-2630/12/2/025005} {\bibfield  {journal} {\bibinfo
  {journal} {New Journal of Physics}\ }\textbf {\bibinfo {volume} {12}},\
  \bibinfo {pages} {025005}}\BibitemShut {NoStop}%
\bibitem [{\citenamefont {Coffey}\ \emph {et~al.}(2004)\citenamefont {Coffey},
  \citenamefont {Kalmykov},\ and\ \citenamefont {Waldron}}]{Coffey2004}%
  \BibitemOpen
  \bibfield  {author} {\bibinfo {author} {\bibnamefont {Coffey}, \bibfnamefont
  {W.~T.}}, \bibinfo {author} {\bibfnamefont {Y.~P.}\ \bibnamefont {Kalmykov}},
  \ and\ \bibinfo {author} {\bibfnamefont {J.~T.}\ \bibnamefont {Waldron}}}
  (\bibinfo {year} {2004}),\ \href {\doibase 10.1142/8195} {\emph {\bibinfo
  {title} {{The Langevin Equation. With Applications to Stochastic Problems in
  Physics, Chemistry and Electrical Engineering}}}},\ \bibinfo {edition} {2nd}\
  ed.\ (\bibinfo  {publisher} {World Scientific Publishing Co, Pte. Ltd.},\
  \bibinfo {address} {Singapore})\BibitemShut {NoStop}%
\bibitem [{\citenamefont {Cohen-Tannoudji}\ and\ \citenamefont
  {Dalibar}(1986)}]{cohen_tannoudji1986a}%
  \BibitemOpen
  \bibfield  {author} {\bibinfo {author} {\bibnamefont {Cohen-Tannoudji},
  \bibfnamefont {C.}}, \ and\ \bibinfo {author} {\bibfnamefont
  {J.}~\bibnamefont {Dalibar}}} (\bibinfo {year} {1986}),\ \href {\doibase
  10.1209/0295-5075/1/9/004} {\bibfield  {journal} {\bibinfo  {journal}
  {Europhysics Letters}\ }\textbf {\bibinfo {volume} {1}}~(\bibinfo {number}
  {9}),\ \bibinfo {pages} {441}}\BibitemShut {NoStop}%
\bibitem [{\citenamefont {Contreras-Pulido}\ \emph {et~al.}(2014)\citenamefont
  {Contreras-Pulido}, \citenamefont {Bruderer}, \citenamefont {Huelga},\ and\
  \citenamefont {Plenio}}]{Contreras-Pulido2014}%
  \BibitemOpen
  \bibfield  {author} {\bibinfo {author} {\bibnamefont {Contreras-Pulido},
  \bibfnamefont {L.~D.}}, \bibinfo {author} {\bibfnamefont {M.}~\bibnamefont
  {Bruderer}}, \bibinfo {author} {\bibfnamefont {S.~F.}\ \bibnamefont
  {Huelga}}, \ and\ \bibinfo {author} {\bibfnamefont {M.~B.}\ \bibnamefont
  {Plenio}}} (\bibinfo {year} {2014}),\ \href {\doibase
  10.1088/1367-2630/16/11/113061} {\bibfield  {journal} {\bibinfo  {journal}
  {New Journal of Physics}\ }\textbf {\bibinfo {volume} {16}}~(\bibinfo
  {number} {11}),\ \bibinfo {pages} {113061}}\BibitemShut {NoStop}%
\bibitem [{\citenamefont {Cookmeyer}\ \emph {et~al.}(2020)\citenamefont
  {Cookmeyer}, \citenamefont {Motruk},\ and\ \citenamefont
  {Moore}}]{CookmeyerMoore2020}%
  \BibitemOpen
  \bibfield  {author} {\bibinfo {author} {\bibnamefont {Cookmeyer},
  \bibfnamefont {T.}}, \bibinfo {author} {\bibfnamefont {J.}~\bibnamefont
  {Motruk}}, \ and\ \bibinfo {author} {\bibfnamefont {J.~E.}\ \bibnamefont
  {Moore}}} (\bibinfo {year} {2020}),\ \href {\doibase
  10.1103/PhysRevB.101.174203} {\bibfield  {journal} {\bibinfo  {journal}
  {Phys. Rev. B}\ }\textbf {\bibinfo {volume} {101}},\ \bibinfo {pages}
  {174203}}\BibitemShut {NoStop}%
\bibitem [{\citenamefont {Cormick}\ and\ \citenamefont
  {Schmiegelow}(2016)}]{Cormick2016}%
  \BibitemOpen
  \bibfield  {author} {\bibinfo {author} {\bibnamefont {Cormick}, \bibfnamefont
  {C.}}, \ and\ \bibinfo {author} {\bibfnamefont {C.~T.}\ \bibnamefont
  {Schmiegelow}}} (\bibinfo {year} {2016}),\ \href {\doibase
  10.1103/PhysRevA.94.053406} {\bibfield  {journal} {\bibinfo  {journal} {Phys.
  Rev. A}\ }\textbf {\bibinfo {volume} {94}},\ \bibinfo {pages}
  {053406}}\BibitemShut {NoStop}%
\bibitem [{\citenamefont {Correa}\ \emph {et~al.}(2019)\citenamefont {Correa},
  \citenamefont {Xu},\ and\ \citenamefont {Adesso}}]{correa2019a}%
  \BibitemOpen
  \bibfield  {author} {\bibinfo {author} {\bibnamefont {Correa}, \bibfnamefont
  {L.~A.}}, \bibinfo {author} {\bibfnamefont {B.}~\bibnamefont {Xu}}, \ and\
  \bibinfo {author} {\bibfnamefont {B.~M.~G.}\ \bibnamefont {Adesso}}}
  (\bibinfo {year} {2019}),\ \href {\doibase 10.1063/1.5114690} {\bibfield
  {journal} {\bibinfo  {journal} {Journal of Chemical Physics}\ }\textbf
  {\bibinfo {volume} {151}},\ \bibinfo {pages} {094107}}\BibitemShut {NoStop}%
\bibitem [{\citenamefont {Crooks}(1999)}]{crooks1999a}%
  \BibitemOpen
  \bibfield  {author} {\bibinfo {author} {\bibnamefont {Crooks}, \bibfnamefont
  {G.~E.}}} (\bibinfo {year} {1999}),\ \href {\doibase
  10.1103/PhysRevE.60.2721} {\bibfield  {journal} {\bibinfo  {journal}
  {Physical Review E}\ }\textbf {\bibinfo {volume} {60}},\ \bibinfo {pages}
  {2721}}\BibitemShut {NoStop}%
\bibitem [{\citenamefont {Cubitt}\ \emph {et~al.}(2015)\citenamefont {Cubitt},
  \citenamefont {Lucia}, \citenamefont {Michalakis},\ and\ \citenamefont
  {Perez-Garcia}}]{CubittPerezGarcia2015}%
  \BibitemOpen
  \bibfield  {author} {\bibinfo {author} {\bibnamefont {Cubitt}, \bibfnamefont
  {T.~S.}}, \bibinfo {author} {\bibfnamefont {A.}~\bibnamefont {Lucia}},
  \bibinfo {author} {\bibfnamefont {S.}~\bibnamefont {Michalakis}}, \ and\
  \bibinfo {author} {\bibfnamefont {D.}~\bibnamefont {Perez-Garcia}}} (\bibinfo
  {year} {2015}),\ \href {\doibase 10.1007/s00220-015-2355-3} {\bibfield
  {journal} {\bibinfo  {journal} {Commun. Math. Phys.}\ }\textbf {\bibinfo
  {volume} {337}},\ \bibinfo {pages} {1275}}\BibitemShut {NoStop}%
\bibitem [{\citenamefont {Cuetara}\ \emph {et~al.}(2016)\citenamefont
  {Cuetara}, \citenamefont {Esposito},\ and\ \citenamefont
  {Schaller}}]{bulnes_cuetara2016a}%
  \BibitemOpen
  \bibfield  {author} {\bibinfo {author} {\bibnamefont {Cuetara}, \bibfnamefont
  {G.~B.}}, \bibinfo {author} {\bibfnamefont {M.}~\bibnamefont {Esposito}}, \
  and\ \bibinfo {author} {\bibfnamefont {G.}~\bibnamefont {Schaller}}}
  (\bibinfo {year} {2016}),\ \href {\doibase 10.3390/e18120447} {\bibfield
  {journal} {\bibinfo  {journal} {Entropy}\ }\textbf {\bibinfo {volume}
  {18}}~(\bibinfo {number} {12}),\ \bibinfo {pages} {447}}\BibitemShut
  {NoStop}%
\bibitem [{\citenamefont {Cui}\ \emph {et~al.}(2015)\citenamefont {Cui},
  \citenamefont {Cirac},\ and\ \citenamefont {Ba\~nuls}}]{CiracBanuls2015}%
  \BibitemOpen
  \bibfield  {author} {\bibinfo {author} {\bibnamefont {Cui}, \bibfnamefont
  {J.}}, \bibinfo {author} {\bibfnamefont {J.~I.}\ \bibnamefont {Cirac}}, \
  and\ \bibinfo {author} {\bibfnamefont {M.~C.}\ \bibnamefont {Ba\~nuls}}}
  (\bibinfo {year} {2015}),\ \href {\doibase 10.1103/PhysRevLett.114.220601}
  {\bibfield  {journal} {\bibinfo  {journal} {Phys. Rev. Lett.}\ }\textbf
  {\bibinfo {volume} {114}},\ \bibinfo {pages} {220601}}\BibitemShut {NoStop}%
\bibitem [{\citenamefont {Da~Fonseca}\ and\ \citenamefont
  {Kowalenko}(2020)}]{FonsecaKowalenko2020}%
  \BibitemOpen
  \bibfield  {author} {\bibinfo {author} {\bibnamefont {Da~Fonseca},
  \bibfnamefont {C.}}, \ and\ \bibinfo {author} {\bibfnamefont
  {V.}~\bibnamefont {Kowalenko}}} (\bibinfo {year} {2020}),\ \href {\doibase
  10.1007/s10474-019-00970-1} {\bibfield  {journal} {\bibinfo  {journal} {Acta
  Math. Hungar.}\ }\textbf {\bibinfo {volume} {160}},\ \bibinfo {pages}
  {376}}\BibitemShut {NoStop}%
\bibitem [{\citenamefont {D'Abbruzzo}\ and\ \citenamefont
  {Rossini}(2021)}]{DAbbruzzo2021}%
  \BibitemOpen
  \bibfield  {author} {\bibinfo {author} {\bibnamefont {D'Abbruzzo},
  \bibfnamefont {A.}}, \ and\ \bibinfo {author} {\bibfnamefont
  {D.}~\bibnamefont {Rossini}}} (\bibinfo {year} {2021}),\ \href {\doibase
  10.1103/PhysRevA.103.052209} {\bibfield  {journal} {\bibinfo  {journal}
  {Phys. Rev. A}\ }\textbf {\bibinfo {volume} {103}},\ \bibinfo {pages}
  {052209}}\BibitemShut {NoStop}%
\bibitem [{\citenamefont {D'Alessio}\ \emph {et~al.}(2016)\citenamefont
  {D'Alessio}, \citenamefont {Kafri}, \citenamefont {Polkovnikov},\ and\
  \citenamefont {Rigol}}]{AlessioRigol2016}%
  \BibitemOpen
  \bibfield  {author} {\bibinfo {author} {\bibnamefont {D'Alessio},
  \bibfnamefont {L.}}, \bibinfo {author} {\bibfnamefont {Y.}~\bibnamefont
  {Kafri}}, \bibinfo {author} {\bibfnamefont {A.}~\bibnamefont {Polkovnikov}},
  \ and\ \bibinfo {author} {\bibfnamefont {M.}~\bibnamefont {Rigol}}} (\bibinfo
  {year} {2016}),\ \href {\doibase 10.1080/00018732.2016.1198134} {\bibfield
  {journal} {\bibinfo  {journal} {Advances in Physics}\ }\textbf {\bibinfo
  {volume} {65}}~(\bibinfo {number} {3}),\ \bibinfo {pages} {239}}\BibitemShut
  {NoStop}%
\bibitem [{\citenamefont {Daley}(2014)}]{Daley2014}%
  \BibitemOpen
  \bibfield  {author} {\bibinfo {author} {\bibnamefont {Daley}, \bibfnamefont
  {A.~J.}}} (\bibinfo {year} {2014}),\ \href {\doibase
  10.1080/00018732.2014.933502} {\bibfield  {journal} {\bibinfo  {journal}
  {Advances in Physics}\ }\textbf {\bibinfo {volume} {63}}~(\bibinfo {number}
  {2}),\ \bibinfo {pages} {77}}\BibitemShut {NoStop}%
\bibitem [{\citenamefont {Dalibard}\ \emph {et~al.}(1992)\citenamefont
  {Dalibard}, \citenamefont {Castin},\ and\ \citenamefont
  {M\o{}lmer}}]{DalibardMolmer1992}%
  \BibitemOpen
  \bibfield  {author} {\bibinfo {author} {\bibnamefont {Dalibard},
  \bibfnamefont {J.}}, \bibinfo {author} {\bibfnamefont {Y.}~\bibnamefont
  {Castin}}, \ and\ \bibinfo {author} {\bibfnamefont {K.}~\bibnamefont
  {M\o{}lmer}}} (\bibinfo {year} {1992}),\ \href {\doibase
  10.1103/PhysRevLett.68.580} {\bibfield  {journal} {\bibinfo  {journal} {Phys.
  Rev. Lett.}\ }\textbf {\bibinfo {volume} {68}},\ \bibinfo {pages}
  {580}}\BibitemShut {NoStop}%
\bibitem [{\citenamefont {Dalidovich}\ and\ \citenamefont
  {Kennett}(2009)}]{DalidovichKennett2009}%
  \BibitemOpen
  \bibfield  {author} {\bibinfo {author} {\bibnamefont {Dalidovich},
  \bibfnamefont {D.}}, \ and\ \bibinfo {author} {\bibfnamefont {M.~P.}\
  \bibnamefont {Kennett}}} (\bibinfo {year} {2009}),\ \href {\doibase
  10.1103/PhysRevA.79.053611} {\bibfield  {journal} {\bibinfo  {journal} {Phys.
  Rev. A}\ }\textbf {\bibinfo {volume} {79}},\ \bibinfo {pages}
  {053611}}\BibitemShut {NoStop}%
\bibitem [{\citenamefont {Damanet}\ \emph {et~al.}(2019)\citenamefont
  {Damanet}, \citenamefont {Mascarenhas}, \citenamefont {Pekker},\ and\
  \citenamefont {Daley}}]{DamanetDaley2019b}%
  \BibitemOpen
  \bibfield  {author} {\bibinfo {author} {\bibnamefont {Damanet}, \bibfnamefont
  {F.}}, \bibinfo {author} {\bibfnamefont {E.}~\bibnamefont {Mascarenhas}},
  \bibinfo {author} {\bibfnamefont {D.}~\bibnamefont {Pekker}}, \ and\ \bibinfo
  {author} {\bibfnamefont {A.~J.}\ \bibnamefont {Daley}}} (\bibinfo {year}
  {2019}),\ \href {\doibase 10.1088/1367-2630/ab4f5d} {\bibfield  {journal}
  {\bibinfo  {journal} {New Journal of Physics}\ }\textbf {\bibinfo {volume}
  {21}}~(\bibinfo {number} {11}),\ \bibinfo {pages} {115001}}\BibitemShut
  {NoStop}%
\bibitem [{\citenamefont {Das~Sarma}\ \emph {et~al.}(1988)\citenamefont
  {Das~Sarma}, \citenamefont {He},\ and\ \citenamefont
  {Xie}}]{DasSarmaXie1988}%
  \BibitemOpen
  \bibfield  {author} {\bibinfo {author} {\bibnamefont {Das~Sarma},
  \bibfnamefont {S.}}, \bibinfo {author} {\bibfnamefont {S.}~\bibnamefont
  {He}}, \ and\ \bibinfo {author} {\bibfnamefont {X.~C.}\ \bibnamefont {Xie}}}
  (\bibinfo {year} {1988}),\ \href {\doibase 10.1103/PhysRevLett.61.2144}
  {\bibfield  {journal} {\bibinfo  {journal} {Phys. Rev. Lett.}\ }\textbf
  {\bibinfo {volume} {61}},\ \bibinfo {pages} {2144}}\BibitemShut {NoStop}%
\bibitem [{\citenamefont {{De Chiara}}\ \emph {et~al.}(2018)\citenamefont {{De
  Chiara}}, \citenamefont {Landi}, \citenamefont {Hewgill}, \citenamefont
  {Reid}, \citenamefont {Ferraro}, \citenamefont {Roncaglia},\ and\
  \citenamefont {Antezza}}]{DeChiara2018}%
  \BibitemOpen
  \bibfield  {author} {\bibinfo {author} {\bibnamefont {{De Chiara}},
  \bibfnamefont {G.}}, \bibinfo {author} {\bibfnamefont {G.}~\bibnamefont
  {Landi}}, \bibinfo {author} {\bibfnamefont {A.}~\bibnamefont {Hewgill}},
  \bibinfo {author} {\bibfnamefont {B.}~\bibnamefont {Reid}}, \bibinfo {author}
  {\bibfnamefont {A.}~\bibnamefont {Ferraro}}, \bibinfo {author} {\bibfnamefont
  {A.~J.}\ \bibnamefont {Roncaglia}}, \ and\ \bibinfo {author} {\bibfnamefont
  {M.}~\bibnamefont {Antezza}}} (\bibinfo {year} {2018}),\ \href {\doibase
  https://doi.org/10.1088/1367-2630/aaecee} {\bibfield  {journal} {\bibinfo
  {journal} {New Journal of Physics}\ }\textbf {\bibinfo {volume} {20}},\
  \bibinfo {pages} {113024}}\BibitemShut {NoStop}%
\bibitem [{\citenamefont {De~Nardis}\ \emph {et~al.}(2019)\citenamefont
  {De~Nardis}, \citenamefont {Medenjak}, \citenamefont {Karrasch},\ and\
  \citenamefont {Ilievski}}]{DeNardisIlievski2019}%
  \BibitemOpen
  \bibfield  {author} {\bibinfo {author} {\bibnamefont {De~Nardis},
  \bibfnamefont {J.}}, \bibinfo {author} {\bibfnamefont {M.}~\bibnamefont
  {Medenjak}}, \bibinfo {author} {\bibfnamefont {C.}~\bibnamefont {Karrasch}},
  \ and\ \bibinfo {author} {\bibfnamefont {E.}~\bibnamefont {Ilievski}}}
  (\bibinfo {year} {2019}),\ \href {\doibase 10.1103/PhysRevLett.123.186601}
  {\bibfield  {journal} {\bibinfo  {journal} {Phys. Rev. Lett.}\ }\textbf
  {\bibinfo {volume} {123}},\ \bibinfo {pages} {186601}}\BibitemShut {NoStop}%
\bibitem [{\citenamefont {Denisov}\ \emph {et~al.}(2019)\citenamefont
  {Denisov}, \citenamefont {Laptyeva}, \citenamefont {Tarnowski}, \citenamefont
  {Chru\'{s}ci\'{n}ski},\ and\ \citenamefont {\.{Z}yczkowski}}]{denisov2019a}%
  \BibitemOpen
  \bibfield  {author} {\bibinfo {author} {\bibnamefont {Denisov}, \bibfnamefont
  {S.}}, \bibinfo {author} {\bibfnamefont {T.}~\bibnamefont {Laptyeva}},
  \bibinfo {author} {\bibfnamefont {W.}~\bibnamefont {Tarnowski}}, \bibinfo
  {author} {\bibfnamefont {D.}~\bibnamefont {Chru\'{s}ci\'{n}ski}}, \ and\
  \bibinfo {author} {\bibfnamefont {K.}~\bibnamefont {\.{Z}yczkowski}}}
  (\bibinfo {year} {2019}),\ \href {\doibase 10.1103/PhysRevLett.123.140403}
  {\bibfield  {journal} {\bibinfo  {journal} {Phys. Rev. Lett.}\ }\textbf
  {\bibinfo {volume} {123}},\ \bibinfo {pages} {140403}}\BibitemShut {NoStop}%
\bibitem [{\citenamefont {Denniston}\ and\ \citenamefont
  {Tang}(1995)}]{DennistonTang1995}%
  \BibitemOpen
  \bibfield  {author} {\bibinfo {author} {\bibnamefont {Denniston},
  \bibfnamefont {C.}}, \ and\ \bibinfo {author} {\bibfnamefont
  {C.}~\bibnamefont {Tang}}} (\bibinfo {year} {1995}),\ \href {\doibase
  10.1103/PhysRevLett.75.3930} {\bibfield  {journal} {\bibinfo  {journal}
  {Phys. Rev. Lett.}\ }\textbf {\bibinfo {volume} {75}}~(\bibinfo {number}
  {21}),\ \bibinfo {pages} {3930}}\BibitemShut {NoStop}%
\bibitem [{\citenamefont {Dhar}(2008)}]{Dhar2008}%
  \BibitemOpen
  \bibfield  {author} {\bibinfo {author} {\bibnamefont {Dhar}, \bibfnamefont
  {A.}}} (\bibinfo {year} {2008}),\ \href {\doibase 10.1080/00018730802538522}
  {\bibfield  {journal} {\bibinfo  {journal} {Advances in Physics}\ }\textbf
  {\bibinfo {volume} {57}}~(\bibinfo {number} {5}),\ \bibinfo {pages}
  {457}}\BibitemShut {NoStop}%
\bibitem [{\citenamefont {Dhar}\ \emph {et~al.}(2012)\citenamefont {Dhar},
  \citenamefont {Saito},\ and\ \citenamefont {H{\"a}nggi}}]{DharHanggi2012}%
  \BibitemOpen
  \bibfield  {author} {\bibinfo {author} {\bibnamefont {Dhar}, \bibfnamefont
  {A.}}, \bibinfo {author} {\bibfnamefont {K.}~\bibnamefont {Saito}}, \ and\
  \bibinfo {author} {\bibfnamefont {P.}~\bibnamefont {H{\"a}nggi}}} (\bibinfo
  {year} {2012}),\ \href {\doibase 10.1103/PhysRevE.85.011126} {\bibfield
  {journal} {\bibinfo  {journal} {Phys. Rev. E}\ }\textbf {\bibinfo {volume}
  {85}}~(\bibinfo {number} {1}),\ \bibinfo {pages} {011126}}\BibitemShut
  {NoStop}%
\bibitem [{\citenamefont {Dolfi}\ \emph {et~al.}(2014)\citenamefont {Dolfi},
  \citenamefont {Bauer}, \citenamefont {Keller}, \citenamefont {Kosenkov},
  \citenamefont {Ewart}, \citenamefont {Kantian}, \citenamefont {Giamarchi},\
  and\ \citenamefont {Troyer}}]{DolfiTroyer2014}%
  \BibitemOpen
  \bibfield  {author} {\bibinfo {author} {\bibnamefont {Dolfi}, \bibfnamefont
  {M.}}, \bibinfo {author} {\bibfnamefont {B.}~\bibnamefont {Bauer}}, \bibinfo
  {author} {\bibfnamefont {S.}~\bibnamefont {Keller}}, \bibinfo {author}
  {\bibfnamefont {A.}~\bibnamefont {Kosenkov}}, \bibinfo {author}
  {\bibfnamefont {T.}~\bibnamefont {Ewart}}, \bibinfo {author} {\bibfnamefont
  {A.}~\bibnamefont {Kantian}}, \bibinfo {author} {\bibfnamefont
  {T.}~\bibnamefont {Giamarchi}}, \ and\ \bibinfo {author} {\bibfnamefont
  {M.}~\bibnamefont {Troyer}}} (\bibinfo {year} {2014}),\ \href {\doibase
  https://doi.org/10.1016/j.cpc.2014.08.019} {\bibfield  {journal} {\bibinfo
  {journal} {Computer Physics Communications}\ }\textbf {\bibinfo {volume}
  {185}}~(\bibinfo {number} {12}),\ \bibinfo {pages} {3430}}\BibitemShut
  {NoStop}%
\bibitem [{\citenamefont {Donohue}\ and\ \citenamefont
  {Giamarchi}(2001)}]{DonohueGiamarchi2001}%
  \BibitemOpen
  \bibfield  {author} {\bibinfo {author} {\bibnamefont {Donohue}, \bibfnamefont
  {P.}}, \ and\ \bibinfo {author} {\bibfnamefont {T.}~\bibnamefont
  {Giamarchi}}} (\bibinfo {year} {2001}),\ \href {\doibase
  10.1103/PhysRevB.63.180508} {\bibfield  {journal} {\bibinfo  {journal} {Phys.
  Rev. B}\ }\textbf {\bibinfo {volume} {63}}~(\bibinfo {number} {18}),\
  \bibinfo {pages} {180508}}\BibitemShut {NoStop}%
\bibitem [{\citenamefont {Dorda}\ \emph {et~al.}(2014)\citenamefont {Dorda},
  \citenamefont {Nuss}, \citenamefont {von~der Linden},\ and\ \citenamefont
  {Arrigoni}}]{DordaArrigoni2014}%
  \BibitemOpen
  \bibfield  {author} {\bibinfo {author} {\bibnamefont {Dorda}, \bibfnamefont
  {A.}}, \bibinfo {author} {\bibfnamefont {M.}~\bibnamefont {Nuss}}, \bibinfo
  {author} {\bibfnamefont {W.}~\bibnamefont {von~der Linden}}, \ and\ \bibinfo
  {author} {\bibfnamefont {E.}~\bibnamefont {Arrigoni}}} (\bibinfo {year}
  {2014}),\ \href {\doibase 10.1103/PhysRevB.89.165105} {\bibfield  {journal}
  {\bibinfo  {journal} {Phys. Rev. B}\ }\textbf {\bibinfo {volume} {89}},\
  \bibinfo {pages} {165105}}\BibitemShut {NoStop}%
\bibitem [{\citenamefont {Droenner}\ and\ \citenamefont
  {Carmele}(2017)}]{DroennerCarmele2017}%
  \BibitemOpen
  \bibfield  {author} {\bibinfo {author} {\bibnamefont {Droenner},
  \bibfnamefont {L.}}, \ and\ \bibinfo {author} {\bibfnamefont
  {A.}~\bibnamefont {Carmele}}} (\bibinfo {year} {2017}),\ \href {\doibase
  10.1103/PhysRevB.96.184421} {\bibfield  {journal} {\bibinfo  {journal} {Phys.
  Rev. B}\ }\textbf {\bibinfo {volume} {96}},\ \bibinfo {pages}
  {184421}}\BibitemShut {NoStop}%
\bibitem [{\citenamefont {Dubi}\ and\ \citenamefont
  {Di~Ventra}(2011)}]{DubiDiVentra2011}%
  \BibitemOpen
  \bibfield  {author} {\bibinfo {author} {\bibnamefont {Dubi}, \bibfnamefont
  {Y.}}, \ and\ \bibinfo {author} {\bibfnamefont {M.}~\bibnamefont
  {Di~Ventra}}} (\bibinfo {year} {2011}),\ \href {\doibase
  10.1103/RevModPhys.83.131} {\bibfield  {journal} {\bibinfo  {journal} {Rev.
  Mod. Phys.}\ }\textbf {\bibinfo {volume} {83}},\ \bibinfo {pages}
  {131}}\BibitemShut {NoStop}%
\bibitem [{\citenamefont {Dum}\ \emph {et~al.}(1992{\natexlab{a}})\citenamefont
  {Dum}, \citenamefont {Parkins}, \citenamefont {Zoller},\ and\ \citenamefont
  {Gardiner}}]{DumGardiner1992}%
  \BibitemOpen
  \bibfield  {author} {\bibinfo {author} {\bibnamefont {Dum}, \bibfnamefont
  {R.}}, \bibinfo {author} {\bibfnamefont {A.~S.}\ \bibnamefont {Parkins}},
  \bibinfo {author} {\bibfnamefont {P.}~\bibnamefont {Zoller}}, \ and\ \bibinfo
  {author} {\bibfnamefont {C.~W.}\ \bibnamefont {Gardiner}}} (\bibinfo {year}
  {1992}{\natexlab{a}}),\ \href {\doibase 10.1103/PhysRevA.46.4382} {\bibfield
  {journal} {\bibinfo  {journal} {Phys. Rev. A}\ }\textbf {\bibinfo {volume}
  {46}},\ \bibinfo {pages} {4382}}\BibitemShut {NoStop}%
\bibitem [{\citenamefont {Dum}\ \emph {et~al.}(1992{\natexlab{b}})\citenamefont
  {Dum}, \citenamefont {Zoller},\ and\ \citenamefont {Ritsch}}]{DumRitsch1992}%
  \BibitemOpen
  \bibfield  {author} {\bibinfo {author} {\bibnamefont {Dum}, \bibfnamefont
  {R.}}, \bibinfo {author} {\bibfnamefont {P.}~\bibnamefont {Zoller}}, \ and\
  \bibinfo {author} {\bibfnamefont {H.}~\bibnamefont {Ritsch}}} (\bibinfo
  {year} {1992}{\natexlab{b}}),\ \href {\doibase 10.1103/PhysRevA.45.4879}
  {\bibfield  {journal} {\bibinfo  {journal} {Phys. Rev. A}\ }\textbf {\bibinfo
  {volume} {45}},\ \bibinfo {pages} {4879}}\BibitemShut {NoStop}%
\bibitem [{\citenamefont {D{\"u}mcke}\ and\ \citenamefont
  {Spohn}(1979)}]{duemcke1979a}%
  \BibitemOpen
  \bibfield  {author} {\bibinfo {author} {\bibnamefont {D{\"u}mcke},
  \bibfnamefont {R.}}, \ and\ \bibinfo {author} {\bibfnamefont
  {H.}~\bibnamefont {Spohn}}} (\bibinfo {year} {1979}),\ \href {\doibase
  https://doi.org/10.1007/BF01325208} {\bibfield  {journal} {\bibinfo
  {journal} {Zeitschrift f{\"u}r Physik B}\ }\textbf {\bibinfo {volume} {34}},\
  \bibinfo {pages} {419}}\BibitemShut {NoStop}%
\bibitem [{\citenamefont {Dwiputra}\ and\ \citenamefont
  {Zen}(2021)}]{Dwiputra2021}%
  \BibitemOpen
  \bibfield  {author} {\bibinfo {author} {\bibnamefont {Dwiputra},
  \bibfnamefont {D.}}, \ and\ \bibinfo {author} {\bibfnamefont {F.~P.}\
  \bibnamefont {Zen}}} (\bibinfo {year} {2021}),\ \href {\doibase
  10.1103/PhysRevA.104.022205} {\bibfield  {journal} {\bibinfo  {journal}
  {Phys. Rev. A}\ }\textbf {\bibinfo {volume} {104}},\ \bibinfo {pages}
  {022205}}\BibitemShut {NoStop}%
\bibitem [{\citenamefont {Dzhioev}\ and\ \citenamefont
  {Kosov}(2011)}]{Dzhioev2011}%
  \BibitemOpen
  \bibfield  {author} {\bibinfo {author} {\bibnamefont {Dzhioev}, \bibfnamefont
  {A.~A.}}, \ and\ \bibinfo {author} {\bibfnamefont {D.~S.}\ \bibnamefont
  {Kosov}}} (\bibinfo {year} {2011}),\ \href {\doibase 10.1063/1.3548065}
  {\bibfield  {journal} {\bibinfo  {journal} {Journal of Chemical Physics}\
  }\textbf {\bibinfo {volume} {134}}~(\bibinfo {number} {4}),\ \bibinfo {pages}
  {1}}\BibitemShut {NoStop}%
\bibitem [{\citenamefont {Economou}(2006)}]{economou2006}%
  \BibitemOpen
  \bibfield  {author} {\bibinfo {author} {\bibnamefont {Economou},
  \bibfnamefont {E.~N.}}} (\bibinfo {year} {2006}),\ \href {\doibase
  10.1007/3-540-28841-4} {\emph {\bibinfo {title} {Green's functions in quantum
  physics}}}\ (\bibinfo  {publisher} {Springer},\ \bibinfo {address} {Berlin
  Heidelberg})\BibitemShut {NoStop}%
\bibitem [{\citenamefont {Ehrlich}\ and\ \citenamefont
  {Schaller}(2021)}]{ehrlich2021a}%
  \BibitemOpen
  \bibfield  {author} {\bibinfo {author} {\bibnamefont {Ehrlich}, \bibfnamefont
  {T.}}, \ and\ \bibinfo {author} {\bibfnamefont {G.}~\bibnamefont {Schaller}}}
  (\bibinfo {year} {2021}),\ \href {\doibase 10.1103/PhysRevB.104.045424}
  {\bibfield  {journal} {\bibinfo  {journal} {Phys. Rev. B}\ }\textbf {\bibinfo
  {volume} {104}},\ \bibinfo {pages} {045424}}\BibitemShut {NoStop}%
\bibitem [{\citenamefont {Eisert}\ \emph {et~al.}(2010)\citenamefont {Eisert},
  \citenamefont {Cramer},\ and\ \citenamefont {Plenio}}]{EisertPlenio2010}%
  \BibitemOpen
  \bibfield  {author} {\bibinfo {author} {\bibnamefont {Eisert}, \bibfnamefont
  {J.}}, \bibinfo {author} {\bibfnamefont {M.}~\bibnamefont {Cramer}}, \ and\
  \bibinfo {author} {\bibfnamefont {M.~B.}\ \bibnamefont {Plenio}}} (\bibinfo
  {year} {2010}),\ \href {\doibase 10.1103/RevModPhys.82.277} {\bibfield
  {journal} {\bibinfo  {journal} {Rev. Mod. Phys.}\ }\textbf {\bibinfo {volume}
  {82}},\ \bibinfo {pages} {277}}\BibitemShut {NoStop}%
\bibitem [{\citenamefont {Englert}\ and\ \citenamefont
  {Morigi}(2002)}]{Englert2002}%
  \BibitemOpen
  \bibfield  {author} {\bibinfo {author} {\bibnamefont {Englert}, \bibfnamefont
  {B.-G.}}, \ and\ \bibinfo {author} {\bibfnamefont {G.}~\bibnamefont
  {Morigi}}} (\bibinfo {year} {2002}),\ in\ \href {\doibase
  10.1007/3-540-45855-7} {\emph {\bibinfo {booktitle} {Coherent Evolution in
  Noisy Environments - Lecture Notes in Physics}}},\ \bibinfo {editor} {edited
  by\ \bibinfo {editor} {\bibfnamefont {A.}~\bibnamefont {Buchleitner}}\ and\
  \bibinfo {editor} {\bibfnamefont {K.}~\bibnamefont {Hornberger}}}\ (\bibinfo
  {publisher} {Springer},\ \bibinfo {address} {Berlin, Heidelberg})\ p.\
  \bibinfo {pages} {611}\BibitemShut {NoStop}%
\bibitem [{\citenamefont {Esaki}(1958)}]{Esaki1958}%
  \BibitemOpen
  \bibfield  {author} {\bibinfo {author} {\bibnamefont {Esaki}, \bibfnamefont
  {L.}}} (\bibinfo {year} {1958}),\ \href {\doibase 10.1103/PhysRev.109.603}
  {\bibfield  {journal} {\bibinfo  {journal} {Phys. Rev.}\ }\textbf {\bibinfo
  {volume} {109}},\ \bibinfo {pages} {603}}\BibitemShut {NoStop}%
\bibitem [{\citenamefont {Esaki}\ and\ \citenamefont
  {Stiles}(1966)}]{EsakiStiles1966}%
  \BibitemOpen
  \bibfield  {author} {\bibinfo {author} {\bibnamefont {Esaki}, \bibfnamefont
  {L.}}, \ and\ \bibinfo {author} {\bibfnamefont {P.~J.}\ \bibnamefont
  {Stiles}}} (\bibinfo {year} {1966}),\ \href {\doibase
  10.1103/PhysRevLett.16.1108} {\bibfield  {journal} {\bibinfo  {journal}
  {Phys. Rev. Lett.}\ }\textbf {\bibinfo {volume} {16}},\ \bibinfo {pages}
  {1108}}\BibitemShut {NoStop}%
\bibitem [{\citenamefont {Esaki}\ and\ \citenamefont
  {Tsu}(1970)}]{EsakiTsu1970}%
  \BibitemOpen
  \bibfield  {author} {\bibinfo {author} {\bibnamefont {Esaki}, \bibfnamefont
  {L.}}, \ and\ \bibinfo {author} {\bibfnamefont {R.}~\bibnamefont {Tsu}}}
  (\bibinfo {year} {1970}),\ \href {\doibase
  https://doi.org/10.1147/rd.141.0061} {\bibfield  {journal} {\bibinfo
  {journal} {IBM J. Res. Develop.}\ }\textbf {\bibinfo {volume} {14}},\
  \bibinfo {pages} {61}}\BibitemShut {NoStop}%
\bibitem [{\citenamefont {Esposito}\ \emph {et~al.}(2007)\citenamefont
  {Esposito}, \citenamefont {Harbola},\ and\ \citenamefont
  {Mukamel}}]{esposito2007b}%
  \BibitemOpen
  \bibfield  {author} {\bibinfo {author} {\bibnamefont {Esposito},
  \bibfnamefont {M.}}, \bibinfo {author} {\bibfnamefont {U.}~\bibnamefont
  {Harbola}}, \ and\ \bibinfo {author} {\bibfnamefont {S.}~\bibnamefont
  {Mukamel}}} (\bibinfo {year} {2007}),\ \href {\doibase
  10.1103/PhysRevB.75.155316} {\bibfield  {journal} {\bibinfo  {journal}
  {Physical Review B}\ }\textbf {\bibinfo {volume} {75}}~(\bibinfo {number}
  {15}),\ \bibinfo {pages} {155316}}\BibitemShut {NoStop}%
\bibitem [{\citenamefont {Esposito}\ \emph {et~al.}(2009)\citenamefont
  {Esposito}, \citenamefont {Harbola},\ and\ \citenamefont
  {Mukamel}}]{Esposito2009}%
  \BibitemOpen
  \bibfield  {author} {\bibinfo {author} {\bibnamefont {Esposito},
  \bibfnamefont {M.}}, \bibinfo {author} {\bibfnamefont {U.}~\bibnamefont
  {Harbola}}, \ and\ \bibinfo {author} {\bibfnamefont {S.}~\bibnamefont
  {Mukamel}}} (\bibinfo {year} {2009}),\ \href {\doibase
  10.1103/RevModPhys.81.1665} {\bibfield  {journal} {\bibinfo  {journal}
  {Reviews of Modern Physics}\ }\textbf {\bibinfo {volume} {81}}~(\bibinfo
  {number} {4}),\ \bibinfo {pages} {1665}}\BibitemShut {NoStop}%
\bibitem [{\citenamefont {Esposito}\ \emph {et~al.}(2010)\citenamefont
  {Esposito}, \citenamefont {Lindenberg},\ and\ \citenamefont {den
  Broeck}}]{esposito2010b}%
  \BibitemOpen
  \bibfield  {author} {\bibinfo {author} {\bibnamefont {Esposito},
  \bibfnamefont {M.}}, \bibinfo {author} {\bibfnamefont {K.}~\bibnamefont
  {Lindenberg}}, \ and\ \bibinfo {author} {\bibfnamefont {C.~V.}\ \bibnamefont
  {den Broeck}}} (\bibinfo {year} {2010}),\ \href {\doibase
  10.1088/1367-2630/12/1/013013} {\bibfield  {journal} {\bibinfo  {journal}
  {New Journal of Physics}\ }\textbf {\bibinfo {volume} {12}},\ \bibinfo
  {pages} {013013}}\BibitemShut {NoStop}%
\bibitem [{\citenamefont {Esposito}\ and\ \citenamefont
  {Mukamel}(2006)}]{esposito2006a}%
  \BibitemOpen
  \bibfield  {author} {\bibinfo {author} {\bibnamefont {Esposito},
  \bibfnamefont {M.}}, \ and\ \bibinfo {author} {\bibfnamefont
  {S.}~\bibnamefont {Mukamel}}} (\bibinfo {year} {2006}),\ \href {\doibase
  10.1103/PhysRevE.73.046129} {\bibfield  {journal} {\bibinfo  {journal}
  {Physical Review E}\ }\textbf {\bibinfo {volume} {73}}~(\bibinfo {number}
  {4}),\ \bibinfo {pages} {046129}}\BibitemShut {NoStop}%
\bibitem [{\citenamefont {Essler}\ and\ \citenamefont
  {Piroli}(2020)}]{Essler2020}%
  \BibitemOpen
  \bibfield  {author} {\bibinfo {author} {\bibnamefont {Essler}, \bibfnamefont
  {F.~H.~L.}}, \ and\ \bibinfo {author} {\bibfnamefont {L.}~\bibnamefont
  {Piroli}}} (\bibinfo {year} {2020}),\ \href {\doibase
  10.1103/PhysRevE.102.062210} {\bibfield  {journal} {\bibinfo  {journal}
  {Phys. Rev. E}\ }\textbf {\bibinfo {volume} {102}},\ \bibinfo {pages}
  {062210}}\BibitemShut {NoStop}%
\bibitem [{\citenamefont {Evans}(1977)}]{Evans1977}%
  \BibitemOpen
  \bibfield  {author} {\bibinfo {author} {\bibnamefont {Evans}, \bibfnamefont
  {D.~E.}}} (\bibinfo {year} {1977}),\ \href {\doibase 10.1007/BF01614091}
  {\bibfield  {journal} {\bibinfo  {journal} {Comm. Math. Phys.}\ }\textbf
  {\bibinfo {volume} {54}},\ \bibinfo {pages} {293}}\BibitemShut {NoStop}%
\bibitem [{\citenamefont {Evans}\ and\ \citenamefont
  {Hanche-Olsen}(1979)}]{Evans1979}%
  \BibitemOpen
  \bibfield  {author} {\bibinfo {author} {\bibnamefont {Evans}, \bibfnamefont
  {D.~E.}}, \ and\ \bibinfo {author} {\bibfnamefont {H.}~\bibnamefont
  {Hanche-Olsen}}} (\bibinfo {year} {1979}),\ \href {\doibase
  10.1016/0022-1236(79)90054-5} {\bibfield  {journal} {\bibinfo  {journal}
  {Journal of Functional Analysis}\ }\textbf {\bibinfo {volume} {32}}~(\bibinfo
  {number} {2}),\ \bibinfo {pages} {207}}\BibitemShut {NoStop}%
\bibitem [{\citenamefont {Farhi}\ \emph {et~al.}(2001)\citenamefont {Farhi},
  \citenamefont {Goldstone}, \citenamefont {Gutmann}, \citenamefont {Lapan},
  \citenamefont {Lundgren},\ and\ \citenamefont {Preda}}]{farhi2001a}%
  \BibitemOpen
  \bibfield  {author} {\bibinfo {author} {\bibnamefont {Farhi}, \bibfnamefont
  {E.}}, \bibinfo {author} {\bibfnamefont {J.}~\bibnamefont {Goldstone}},
  \bibinfo {author} {\bibfnamefont {S.}~\bibnamefont {Gutmann}}, \bibinfo
  {author} {\bibfnamefont {J.}~\bibnamefont {Lapan}}, \bibinfo {author}
  {\bibfnamefont {A.}~\bibnamefont {Lundgren}}, \ and\ \bibinfo {author}
  {\bibfnamefont {D.}~\bibnamefont {Preda}}} (\bibinfo {year} {2001}),\ \href
  {https://science.sciencemag.org/content/292/5516/472.abstract} {\bibfield
  {journal} {\bibinfo  {journal} {Science}\ }\textbf {\bibinfo {volume}
  {292}},\ \bibinfo {pages} {472}}\BibitemShut {NoStop}%
\bibitem [{\citenamefont {Farina}\ \emph {et~al.}(2020)\citenamefont {Farina},
  \citenamefont {De~Filippis}, \citenamefont {Cataudella}, \citenamefont
  {Polini},\ and\ \citenamefont {Giovannetti}}]{farina2020a}%
  \BibitemOpen
  \bibfield  {author} {\bibinfo {author} {\bibnamefont {Farina}, \bibfnamefont
  {D.}}, \bibinfo {author} {\bibfnamefont {G.}~\bibnamefont {De~Filippis}},
  \bibinfo {author} {\bibfnamefont {V.}~\bibnamefont {Cataudella}}, \bibinfo
  {author} {\bibfnamefont {M.}~\bibnamefont {Polini}}, \ and\ \bibinfo {author}
  {\bibfnamefont {V.}~\bibnamefont {Giovannetti}}} (\bibinfo {year} {2020}),\
  \href {\doibase 10.1103/PhysRevA.102.052208} {\bibfield  {journal} {\bibinfo
  {journal} {Phys. Rev. A}\ }\textbf {\bibinfo {volume} {102}},\ \bibinfo
  {pages} {052208}}\BibitemShut {NoStop}%
\bibitem [{\citenamefont {Farina}\ and\ \citenamefont
  {Giovannetti}(2019)}]{farina2019a}%
  \BibitemOpen
  \bibfield  {author} {\bibinfo {author} {\bibnamefont {Farina}, \bibfnamefont
  {D.}}, \ and\ \bibinfo {author} {\bibfnamefont {V.}~\bibnamefont
  {Giovannetti}}} (\bibinfo {year} {2019}),\ \href {\doibase
  10.1103/PhysRevA.100.012107} {\bibfield  {journal} {\bibinfo  {journal}
  {Phys. Rev. A}\ }\textbf {\bibinfo {volume} {100}},\ \bibinfo {pages}
  {012107}}\BibitemShut {NoStop}%
\bibitem [{\citenamefont {Finazzi}\ \emph {et~al.}(2015)\citenamefont
  {Finazzi}, \citenamefont {Le~Boit\'e}, \citenamefont {Storme}, \citenamefont
  {Baksic},\ and\ \citenamefont {Ciuti}}]{FinazziCiuti2015}%
  \BibitemOpen
  \bibfield  {author} {\bibinfo {author} {\bibnamefont {Finazzi}, \bibfnamefont
  {S.}}, \bibinfo {author} {\bibfnamefont {A.}~\bibnamefont {Le~Boit\'e}},
  \bibinfo {author} {\bibfnamefont {F.}~\bibnamefont {Storme}}, \bibinfo
  {author} {\bibfnamefont {A.}~\bibnamefont {Baksic}}, \ and\ \bibinfo {author}
  {\bibfnamefont {C.}~\bibnamefont {Ciuti}}} (\bibinfo {year} {2015}),\ \href
  {\doibase 10.1103/PhysRevLett.115.080604} {\bibfield  {journal} {\bibinfo
  {journal} {Phys. Rev. Lett.}\ }\textbf {\bibinfo {volume} {115}},\ \bibinfo
  {pages} {080604}}\BibitemShut {NoStop}%
\bibitem [{\citenamefont {Fisher}\ \emph {et~al.}(1989)\citenamefont {Fisher},
  \citenamefont {Weichman}, \citenamefont {Grinstein},\ and\ \citenamefont
  {Fisher}}]{FisherFisher1989}%
  \BibitemOpen
  \bibfield  {author} {\bibinfo {author} {\bibnamefont {Fisher}, \bibfnamefont
  {M.~P.~A.}}, \bibinfo {author} {\bibfnamefont {P.~B.}\ \bibnamefont
  {Weichman}}, \bibinfo {author} {\bibfnamefont {G.}~\bibnamefont {Grinstein}},
  \ and\ \bibinfo {author} {\bibfnamefont {D.~S.}\ \bibnamefont {Fisher}}}
  (\bibinfo {year} {1989}),\ \href {\doibase 10.1103/PhysRevB.40.546}
  {\bibfield  {journal} {\bibinfo  {journal} {Phys. Rev. B}\ }\textbf {\bibinfo
  {volume} {40}},\ \bibinfo {pages} {546}}\BibitemShut {NoStop}%
\bibitem [{\citenamefont {Fishman}\ \emph {et~al.}(2020)\citenamefont
  {Fishman}, \citenamefont {White},\ and\ \citenamefont
  {Stoudenmire}}]{itensor}%
  \BibitemOpen
  \bibfield  {author} {\bibinfo {author} {\bibnamefont {Fishman}, \bibfnamefont
  {M.}}, \bibinfo {author} {\bibfnamefont {S.~R.}\ \bibnamefont {White}}, \
  and\ \bibinfo {author} {\bibfnamefont {E.~M.}\ \bibnamefont {Stoudenmire}}}
  (\bibinfo {year} {2020}),\ \href {https://arxiv.org/abs/2007.14822} {\bibinfo
   {journal} {arXiv:2007.14822}\ }\BibitemShut {NoStop}%
\bibitem [{\citenamefont {Fleming}\ and\ \citenamefont
  {Cummings}(2011)}]{fleming2011a}%
  \BibitemOpen
\bibfield  {journal} {  }\bibfield  {author} {\bibinfo {author} {\bibnamefont
  {Fleming}, \bibfnamefont {C.~H.}}, \ and\ \bibinfo {author} {\bibfnamefont
  {N.~I.}\ \bibnamefont {Cummings}}} (\bibinfo {year} {2011}),\ \href {\doibase
  10.1103/PhysRevE.83.031117} {\bibfield  {journal} {\bibinfo  {journal} {Phys.
  Rev. E}\ }\textbf {\bibinfo {volume} {83}},\ \bibinfo {pages}
  {031117}}\BibitemShut {NoStop}%
\bibitem [{\citenamefont {Flindt}\ \emph {et~al.}(2007)\citenamefont {Flindt},
  \citenamefont {Braggio},\ and\ \citenamefont {Novotn{\'{y}}}}]{flindt2007a}%
  \BibitemOpen
  \bibfield  {author} {\bibinfo {author} {\bibnamefont {Flindt}, \bibfnamefont
  {C.}}, \bibinfo {author} {\bibfnamefont {A.}~\bibnamefont {Braggio}}, \ and\
  \bibinfo {author} {\bibfnamefont {T.}~\bibnamefont {Novotn{\'{y}}}}}
  (\bibinfo {year} {2007}),\ \href {https://doi.org/10.1063/1.2759735}
  {\bibfield  {journal} {\bibinfo  {journal} {AIP Conf. Proc.}\ }\textbf
  {\bibinfo {volume} {922}},\ \bibinfo {pages} {531}}\BibitemShut {NoStop}%
\bibitem [{\citenamefont {Flindt}\ \emph {et~al.}(2009)\citenamefont {Flindt},
  \citenamefont {Fricke}, \citenamefont {Hohls}, \citenamefont {Novotn\'y},
  \citenamefont {Netocny}, \citenamefont {Brandes},\ and\ \citenamefont
  {Haug}}]{flindt2009a}%
  \BibitemOpen
  \bibfield  {author} {\bibinfo {author} {\bibnamefont {Flindt}, \bibfnamefont
  {C.}}, \bibinfo {author} {\bibfnamefont {C.}~\bibnamefont {Fricke}}, \bibinfo
  {author} {\bibfnamefont {F.}~\bibnamefont {Hohls}}, \bibinfo {author}
  {\bibfnamefont {T.}~\bibnamefont {Novotn\'y}}, \bibinfo {author}
  {\bibfnamefont {K.}~\bibnamefont {Netocny}}, \bibinfo {author} {\bibfnamefont
  {T.}~\bibnamefont {Brandes}}, \ and\ \bibinfo {author} {\bibfnamefont
  {R.~J.}\ \bibnamefont {Haug}}} (\bibinfo {year} {2009}),\ \href
  {https://doi.org/10.1073/pnas.0901002106} {\bibfield  {journal} {\bibinfo
  {journal} {PNAS}\ }\textbf {\bibinfo {volume} {106}},\ \bibinfo {pages}
  {10116}}\BibitemShut {NoStop}%
\bibitem [{\citenamefont {Flindt}\ \emph {et~al.}(2010)\citenamefont {Flindt},
  \citenamefont {Novotn\'y}, \citenamefont {Braggio},\ and\ \citenamefont
  {Jauho}}]{FlindtAnttiPekka2010}%
  \BibitemOpen
  \bibfield  {author} {\bibinfo {author} {\bibnamefont {Flindt}, \bibfnamefont
  {C.}}, \bibinfo {author} {\bibfnamefont {T.}~\bibnamefont {Novotn\'y}},
  \bibinfo {author} {\bibfnamefont {A.}~\bibnamefont {Braggio}}, \ and\
  \bibinfo {author} {\bibfnamefont {A.-P.}\ \bibnamefont {Jauho}}} (\bibinfo
  {year} {2010}),\ \href {\doibase 10.1103/PhysRevB.82.155407} {\bibfield
  {journal} {\bibinfo  {journal} {Phys. Rev. B}\ }\textbf {\bibinfo {volume}
  {82}},\ \bibinfo {pages} {155407}}\BibitemShut {NoStop}%
\bibitem [{\citenamefont {Flindt}\ \emph {et~al.}(2008)\citenamefont {Flindt},
  \citenamefont {Novotn\'y}, \citenamefont {Braggio}, \citenamefont
  {Sassetti},\ and\ \citenamefont {Jauho}}]{flindt2008a}%
  \BibitemOpen
  \bibfield  {author} {\bibinfo {author} {\bibnamefont {Flindt}, \bibfnamefont
  {C.}}, \bibinfo {author} {\bibfnamefont {T.}~\bibnamefont {Novotn\'y}},
  \bibinfo {author} {\bibfnamefont {A.}~\bibnamefont {Braggio}}, \bibinfo
  {author} {\bibfnamefont {M.}~\bibnamefont {Sassetti}}, \ and\ \bibinfo
  {author} {\bibfnamefont {A.-P.}\ \bibnamefont {Jauho}}} (\bibinfo {year}
  {2008}),\ \href {\doibase 10.1103/PhysRevLett.100.150601} {\bibfield
  {journal} {\bibinfo  {journal} {Phys. Rev. Lett.}\ }\textbf {\bibinfo
  {volume} {100}},\ \bibinfo {pages} {150601}}\BibitemShut {NoStop}%
\bibitem [{\citenamefont {Flindt}\ \emph {et~al.}(2004)\citenamefont {Flindt},
  \citenamefont {Novotn\'y},\ and\ \citenamefont {Jauho}}]{flindt2004a}%
  \BibitemOpen
  \bibfield  {author} {\bibinfo {author} {\bibnamefont {Flindt}, \bibfnamefont
  {C.}}, \bibinfo {author} {\bibfnamefont {T.}~\bibnamefont {Novotn\'y}}, \
  and\ \bibinfo {author} {\bibfnamefont {A.-P.}\ \bibnamefont {Jauho}}}
  (\bibinfo {year} {2004}),\ \href {\doibase 10.1103/PhysRevB.70.205334}
  {\bibfield  {journal} {\bibinfo  {journal} {Phys. Rev. B}\ }\textbf {\bibinfo
  {volume} {70}},\ \bibinfo {pages} {205334}}\BibitemShut {NoStop}%
\bibitem [{\citenamefont {Fornieri}\ \emph {et~al.}(2015)\citenamefont
  {Fornieri}, \citenamefont {Martínez-P\'{e}rez},\ and\ \citenamefont
  {Giazotto}}]{FornieriGiazotto2015}%
  \BibitemOpen
  \bibfield  {author} {\bibinfo {author} {\bibnamefont {Fornieri},
  \bibfnamefont {A.}}, \bibinfo {author} {\bibfnamefont {M.~J.}\ \bibnamefont
  {Martínez-P\'{e}rez}}, \ and\ \bibinfo {author} {\bibfnamefont
  {F.}~\bibnamefont {Giazotto}}} (\bibinfo {year} {2015}),\ \href {\doibase
  10.1063/1.4915899} {\bibfield  {journal} {\bibinfo  {journal} {AIP Advances}\
  }\textbf {\bibinfo {volume} {5}},\ \bibinfo {pages} {053301}}\BibitemShut
  {NoStop}%
\bibitem [{\citenamefont {Fourier}(1822)}]{Fourier1822}%
  \BibitemOpen
  \bibfield  {author} {\bibinfo {author} {\bibnamefont {Fourier}, \bibfnamefont
  {J.}}} (\bibinfo {year} {1822}),\ \href@noop {} {\emph {\bibinfo {title}
  {{Th\'eorie Analytique de la Chaleur}}}}\ (\bibinfo  {publisher} {Firmin
  Didot},\ \bibinfo {address} {Paris})\BibitemShut {NoStop}%
\bibitem [{\citenamefont {Friedman}\ \emph {et~al.}(2018)\citenamefont
  {Friedman}, \citenamefont {Agarwalla},\ and\ \citenamefont
  {Segal}}]{friedman2018a}%
  \BibitemOpen
  \bibfield  {author} {\bibinfo {author} {\bibnamefont {Friedman},
  \bibfnamefont {H.~M.}}, \bibinfo {author} {\bibfnamefont {B.~K.}\
  \bibnamefont {Agarwalla}}, \ and\ \bibinfo {author} {\bibfnamefont
  {D.}~\bibnamefont {Segal}}} (\bibinfo {year} {2018}),\ \href {\doibase
  10.1088/1367-2630/aad5fc} {\bibfield  {journal} {\bibinfo  {journal} {New
  Journal of Physics}\ }\textbf {\bibinfo {volume} {20}}~(\bibinfo {number}
  {8}),\ \bibinfo {pages} {083026}}\BibitemShut {NoStop}%
\bibitem [{\citenamefont {Friedman}\ and\ \citenamefont
  {Segal}(2019)}]{friedman2019a}%
  \BibitemOpen
  \bibfield  {author} {\bibinfo {author} {\bibnamefont {Friedman},
  \bibfnamefont {H.~M.}}, \ and\ \bibinfo {author} {\bibfnamefont
  {D.}~\bibnamefont {Segal}}} (\bibinfo {year} {2019}),\ \href {\doibase
  10.1103/PhysRevE.100.062112} {\bibfield  {journal} {\bibinfo  {journal}
  {Phys. Rev. E}\ }\textbf {\bibinfo {volume} {100}},\ \bibinfo {pages}
  {062112}}\BibitemShut {NoStop}%
\bibitem [{\citenamefont {Frigerio}(1978)}]{Frigerio1978}%
  \BibitemOpen
  \bibfield  {author} {\bibinfo {author} {\bibnamefont {Frigerio},
  \bibfnamefont {A.}}} (\bibinfo {year} {1978}),\ \href {\doibase
  10.1007/BF01196936} {\bibfield  {journal} {\bibinfo  {journal}
  {Communications in Mathematical Physics}\ }\textbf {\bibinfo {volume}
  {63}}~(\bibinfo {number} {3}),\ \bibinfo {pages} {269 }}\BibitemShut
  {NoStop}%
\bibitem [{\citenamefont {Fujisawa}\ \emph {et~al.}(2006)\citenamefont
  {Fujisawa}, \citenamefont {Hayashi}, \citenamefont {Tomita},\ and\
  \citenamefont {Hirayama}}]{fujisawa2006a}%
  \BibitemOpen
  \bibfield  {author} {\bibinfo {author} {\bibnamefont {Fujisawa},
  \bibfnamefont {T.}}, \bibinfo {author} {\bibfnamefont {T.}~\bibnamefont
  {Hayashi}}, \bibinfo {author} {\bibfnamefont {R.}~\bibnamefont {Tomita}}, \
  and\ \bibinfo {author} {\bibfnamefont {Y.}~\bibnamefont {Hirayama}}}
  (\bibinfo {year} {2006}),\ \href {\doibase 10.1126/science.1126788}
  {\bibfield  {journal} {\bibinfo  {journal} {Science}\ }\textbf {\bibinfo
  {volume} {312}},\ \bibinfo {pages} {1634}}\BibitemShut {NoStop}%
\bibitem [{\citenamefont {Galperin}\ \emph {et~al.}(2005)\citenamefont
  {Galperin}, \citenamefont {Ratner},\ and\ \citenamefont
  {Nitzan}}]{GalperinNitzan2005}%
  \BibitemOpen
  \bibfield  {author} {\bibinfo {author} {\bibnamefont {Galperin},
  \bibfnamefont {M.}}, \bibinfo {author} {\bibfnamefont {M.~A.}\ \bibnamefont
  {Ratner}}, \ and\ \bibinfo {author} {\bibfnamefont {A.}~\bibnamefont
  {Nitzan}}} (\bibinfo {year} {2005}),\ \href {\doibase 10.1021/nl048216c}
  {\bibfield  {journal} {\bibinfo  {journal} {Nano Letters}\ }\textbf {\bibinfo
  {volume} {5}},\ \bibinfo {pages} {125}}\BibitemShut {NoStop}%
\bibitem [{\citenamefont {Galperin}\ \emph {et~al.}(2007)\citenamefont
  {Galperin}, \citenamefont {Ratner},\ and\ \citenamefont
  {Nitzan}}]{galperin2007a}%
  \BibitemOpen
  \bibfield  {author} {\bibinfo {author} {\bibnamefont {Galperin},
  \bibfnamefont {M.}}, \bibinfo {author} {\bibfnamefont {M.~A.}\ \bibnamefont
  {Ratner}}, \ and\ \bibinfo {author} {\bibfnamefont {A.}~\bibnamefont
  {Nitzan}}} (\bibinfo {year} {2007}),\ \href {\doibase
  10.1088/0953-8984/19/10/103201} {\bibfield  {journal} {\bibinfo  {journal}
  {Journal of Physics: Condensed Matter}\ }\textbf {\bibinfo {volume}
  {19}}~(\bibinfo {number} {10}),\ \bibinfo {pages} {103201}}\BibitemShut
  {NoStop}%
\bibitem [{\citenamefont {Galperin}\ \emph {et~al.}(2008)\citenamefont
  {Galperin}, \citenamefont {Ratner}, \citenamefont {Nitzan},\ and\
  \citenamefont {Troisi}}]{galperin2008a}%
  \BibitemOpen
  \bibfield  {author} {\bibinfo {author} {\bibnamefont {Galperin},
  \bibfnamefont {M.}}, \bibinfo {author} {\bibfnamefont {M.~A.}\ \bibnamefont
  {Ratner}}, \bibinfo {author} {\bibfnamefont {A.}~\bibnamefont {Nitzan}}, \
  and\ \bibinfo {author} {\bibfnamefont {A.}~\bibnamefont {Troisi}}} (\bibinfo
  {year} {2008}),\ \href {\doibase 10.1126/science.1146556} {\bibfield
  {journal} {\bibinfo  {journal} {Science}\ }\textbf {\bibinfo {volume}
  {319}}~(\bibinfo {number} {5866}),\ \bibinfo {pages} {1056}}\BibitemShut
  {NoStop}%
\bibitem [{\citenamefont {Ganeshan}\ \emph {et~al.}(2015)\citenamefont
  {Ganeshan}, \citenamefont {Pixley},\ and\ \citenamefont
  {Das~Sarma}}]{GaneshanDasSarma2015}%
  \BibitemOpen
  \bibfield  {author} {\bibinfo {author} {\bibnamefont {Ganeshan},
  \bibfnamefont {S.}}, \bibinfo {author} {\bibfnamefont {J.~H.}\ \bibnamefont
  {Pixley}}, \ and\ \bibinfo {author} {\bibfnamefont {S.}~\bibnamefont
  {Das~Sarma}}} (\bibinfo {year} {2015}),\ \href {\doibase
  10.1103/PhysRevLett.114.146601} {\bibfield  {journal} {\bibinfo  {journal}
  {Phys. Rev. Lett.}\ }\textbf {\bibinfo {volume} {114}},\ \bibinfo {pages}
  {146601}}\BibitemShut {NoStop}%
\bibitem [{\citenamefont {Gangat}\ \emph {et~al.}(2017)\citenamefont {Gangat},
  \citenamefont {I},\ and\ \citenamefont {Kao}}]{Gangat2017}%
  \BibitemOpen
  \bibfield  {author} {\bibinfo {author} {\bibnamefont {Gangat}, \bibfnamefont
  {A.~A.}}, \bibinfo {author} {\bibfnamefont {T.}~\bibnamefont {I}}, \ and\
  \bibinfo {author} {\bibfnamefont {Y.-J.}\ \bibnamefont {Kao}}} (\bibinfo
  {year} {2017}),\ \href {\doibase 10.1103/PhysRevLett.119.010501} {\bibfield
  {journal} {\bibinfo  {journal} {Phys. Rev. Lett.}\ }\textbf {\bibinfo
  {volume} {119}},\ \bibinfo {pages} {010501}}\BibitemShut {NoStop}%
\bibitem [{\citenamefont {Garc{\'{\i}}a-Ripoll}(2006)}]{GarciaRipoll2006}%
  \BibitemOpen
  \bibfield  {author} {\bibinfo {author} {\bibnamefont {Garc{\'{\i}}a-Ripoll},
  \bibfnamefont {J.~J.}}} (\bibinfo {year} {2006}),\ \href {\doibase
  10.1088/1367-2630/8/12/305} {\bibfield  {journal} {\bibinfo  {journal} {New
  Journal of Physics}\ }\textbf {\bibinfo {volume} {8}}~(\bibinfo {number}
  {12}),\ \bibinfo {pages} {305}}\BibitemShut {NoStop}%
\bibitem [{\citenamefont {Garc{\'{\i}}a-Ripoll}\ \emph
  {et~al.}(2009)\citenamefont {Garc{\'{\i}}a-Ripoll}, \citenamefont {D{\"u}rr},
  \citenamefont {Syassen}, \citenamefont {Bauer}, \citenamefont {Lettner},
  \citenamefont {Rempe},\ and\ \citenamefont {Cirac}}]{GarciaRipollCirac2009}%
  \BibitemOpen
  \bibfield  {author} {\bibinfo {author} {\bibnamefont {Garc{\'{\i}}a-Ripoll},
  \bibfnamefont {J.~J.}}, \bibinfo {author} {\bibfnamefont {S.}~\bibnamefont
  {D{\"u}rr}}, \bibinfo {author} {\bibfnamefont {N.}~\bibnamefont {Syassen}},
  \bibinfo {author} {\bibfnamefont {D.~M.}\ \bibnamefont {Bauer}}, \bibinfo
  {author} {\bibfnamefont {M.}~\bibnamefont {Lettner}}, \bibinfo {author}
  {\bibfnamefont {G.}~\bibnamefont {Rempe}}, \ and\ \bibinfo {author}
  {\bibfnamefont {J.~I.}\ \bibnamefont {Cirac}}} (\bibinfo {year} {2009}),\
  \href {\doibase 10.1088/1367-2630/11/1/013053} {\bibfield  {journal}
  {\bibinfo  {journal} {New Journal of Physics}\ }\textbf {\bibinfo {volume}
  {11}}~(\bibinfo {number} {1}),\ \bibinfo {pages} {013053}}\BibitemShut
  {NoStop}%
\bibitem [{\citenamefont {Gardiner}\ and\ \citenamefont
  {Zoller}(2004)}]{Gardiner2004}%
  \BibitemOpen
  \bibfield  {author} {\bibinfo {author} {\bibnamefont {Gardiner},
  \bibfnamefont {C.}}, \ and\ \bibinfo {author} {\bibfnamefont
  {P.}~\bibnamefont {Zoller}}} (\bibinfo {year} {2004}),\ \href
  {https://link.springer.com/gp/book/9783540223016} {\emph {\bibinfo {title}
  {{Quantum noise}}}},\ \bibinfo {edition} {3rd}\ ed.\ (\bibinfo  {publisher}
  {Springer})\BibitemShut {NoStop}%
\bibitem [{\citenamefont {Garrahan}\ and\ \citenamefont
  {Lesanovsky}(2010)}]{garrahan2010a}%
  \BibitemOpen
  \bibfield  {author} {\bibinfo {author} {\bibnamefont {Garrahan},
  \bibfnamefont {J.~P.}}, \ and\ \bibinfo {author} {\bibfnamefont
  {I.}~\bibnamefont {Lesanovsky}}} (\bibinfo {year} {2010}),\ \href {\doibase
  10.1103/PhysRevLett.104.160601} {\bibfield  {journal} {\bibinfo  {journal}
  {Physical Review Letters}\ }\textbf {\bibinfo {volume} {104}}~(\bibinfo
  {number} {16}),\ \bibinfo {pages} {160601}}\BibitemShut {NoStop}%
\bibitem [{\citenamefont {Garraway}(1997{\natexlab{a}})}]{Garraway1997}%
  \BibitemOpen
  \bibfield  {author} {\bibinfo {author} {\bibnamefont {Garraway},
  \bibfnamefont {B.~M.}}} (\bibinfo {year} {1997}{\natexlab{a}}),\ \href
  {\doibase 10.1103/PhysRevA.55.4636} {\bibfield  {journal} {\bibinfo
  {journal} {Phys. Rev. A}\ }\textbf {\bibinfo {volume} {55}},\ \bibinfo
  {pages} {4636}}\BibitemShut {NoStop}%
\bibitem [{\citenamefont {Garraway}(1997{\natexlab{b}})}]{garraway1997a}%
  \BibitemOpen
  \bibfield  {author} {\bibinfo {author} {\bibnamefont {Garraway},
  \bibfnamefont {B.~M.}}} (\bibinfo {year} {1997}{\natexlab{b}}),\ \href
  {\doibase 10.1103/PhysRevA.55.2290} {\bibfield  {journal} {\bibinfo
  {journal} {Phys. Rev. A}\ }\textbf {\bibinfo {volume} {55}},\ \bibinfo
  {pages} {2290}}\BibitemShut {NoStop}%
\bibitem [{\citenamefont {Garraway}(2011)}]{Garraway2011}%
  \BibitemOpen
  \bibfield  {author} {\bibinfo {author} {\bibnamefont {Garraway},
  \bibfnamefont {B.~M.}}} (\bibinfo {year} {2011}),\ \href {\doibase
  10.1098/rsta.2010.0333} {\bibfield  {journal} {\bibinfo  {journal}
  {Philosophical Transactions of the Royal Society A: Mathematical, Physical
  and Engineering Sciences}\ }\textbf {\bibinfo {volume} {369}}~(\bibinfo
  {number} {1939}),\ \bibinfo {pages} {1137}}\BibitemShut {NoStop}%
\bibitem [{\citenamefont {Garst}\ and\ \citenamefont
  {Rosch}(2001)}]{GarstRosch2001}%
  \BibitemOpen
  \bibfield  {author} {\bibinfo {author} {\bibnamefont {Garst}, \bibfnamefont
  {M.}}, \ and\ \bibinfo {author} {\bibfnamefont {A.}~\bibnamefont {Rosch}}}
  (\bibinfo {year} {2001}),\ \href {\doibase 10.1209/epl/i2001-00382-3}
  {\bibfield  {journal} {\bibinfo  {journal} {Europhysics Letters ({EPL})}\
  }\textbf {\bibinfo {volume} {55}}~(\bibinfo {number} {1}),\ \bibinfo {pages}
  {66}}\BibitemShut {NoStop}%
\bibitem [{\citenamefont {Gaspard}\ and\ \citenamefont
  {Nagaoka}(1999)}]{gaspard1999a}%
  \BibitemOpen
  \bibfield  {author} {\bibinfo {author} {\bibnamefont {Gaspard}, \bibfnamefont
  {P.}}, \ and\ \bibinfo {author} {\bibfnamefont {M.}~\bibnamefont {Nagaoka}}}
  (\bibinfo {year} {1999}),\ \href {\doibase 10.1063/1.479867} {\bibfield
  {journal} {\bibinfo  {journal} {The Journal of Chemical Physics}\ }\textbf
  {\bibinfo {volume} {111}},\ \bibinfo {pages} {5668}}\BibitemShut {NoStop}%
\bibitem [{\citenamefont {Gaudioso}\ \emph {et~al.}(2000)\citenamefont
  {Gaudioso}, \citenamefont {Lauhon},\ and\ \citenamefont
  {Ho}}]{GaudiosoHo2000}%
  \BibitemOpen
  \bibfield  {author} {\bibinfo {author} {\bibnamefont {Gaudioso},
  \bibfnamefont {J.}}, \bibinfo {author} {\bibfnamefont {L.~J.}\ \bibnamefont
  {Lauhon}}, \ and\ \bibinfo {author} {\bibfnamefont {W.}~\bibnamefont {Ho}}}
  (\bibinfo {year} {2000}),\ \href {\doibase 10.1103/PhysRevLett.85.1918}
  {\bibfield  {journal} {\bibinfo  {journal} {Phys. Rev. Lett.}\ }\textbf
  {\bibinfo {volume} {85}},\ \bibinfo {pages} {1918}}\BibitemShut {NoStop}%
\bibitem [{\citenamefont {Gautschi}(2005)}]{Gautschi2005}%
  \BibitemOpen
  \bibfield  {author} {\bibinfo {author} {\bibnamefont {Gautschi},
  \bibfnamefont {W.}}} (\bibinfo {year} {2005}),\ \href {\doibase
  10.1016/j.cam.2004.03.029} {\bibfield  {journal} {\bibinfo  {journal} {Jour.
  Comp. App. Math.}\ }\textbf {\bibinfo {volume} {178}},\ \bibinfo {pages}
  {215}}\BibitemShut {NoStop}%
\bibitem [{\citenamefont {Gelbwaser-Klimovsky}\ and\ \citenamefont
  {Aspuru-Guzik}(2015)}]{gelbwaser_klimovsky2015a}%
  \BibitemOpen
  \bibfield  {author} {\bibinfo {author} {\bibnamefont {Gelbwaser-Klimovsky},
  \bibfnamefont {D.}}, \ and\ \bibinfo {author} {\bibfnamefont
  {A.}~\bibnamefont {Aspuru-Guzik}}} (\bibinfo {year} {2015}),\ \href {\doibase
  10.1021/acs.jpclett.5b01404} {\bibfield  {journal} {\bibinfo  {journal} {The
  Journal of Physical Chemistry Letters}\ }\textbf {\bibinfo {volume} {6}},\
  \bibinfo {pages} {3477}}\BibitemShut {NoStop}%
\bibitem [{\citenamefont {Gerland}\ \emph {et~al.}(2000)\citenamefont
  {Gerland}, \citenamefont {von Delft}, \citenamefont {Costi},\ and\
  \citenamefont {Oreg}}]{GerlandOreg2000}%
  \BibitemOpen
  \bibfield  {author} {\bibinfo {author} {\bibnamefont {Gerland}, \bibfnamefont
  {U.}}, \bibinfo {author} {\bibfnamefont {J.}~\bibnamefont {von Delft}},
  \bibinfo {author} {\bibfnamefont {T.~A.}\ \bibnamefont {Costi}}, \ and\
  \bibinfo {author} {\bibfnamefont {Y.}~\bibnamefont {Oreg}}} (\bibinfo {year}
  {2000}),\ \href {\doibase 10.1103/PhysRevLett.84.3710} {\bibfield  {journal}
  {\bibinfo  {journal} {Phys. Rev. Lett.}\ }\textbf {\bibinfo {volume} {84}},\
  \bibinfo {pages} {3710}}\BibitemShut {NoStop}%
\bibitem [{\citenamefont {Giamarchi}\ and\ \citenamefont
  {Schulz}(1987)}]{GiamarchiSchulz1987}%
  \BibitemOpen
  \bibfield  {author} {\bibinfo {author} {\bibnamefont {Giamarchi},
  \bibfnamefont {T.}}, \ and\ \bibinfo {author} {\bibfnamefont {H.~J.}\
  \bibnamefont {Schulz}}} (\bibinfo {year} {1987}),\ \href {\doibase
  10.1209/0295-5075/3/12/007} {\bibfield  {journal} {\bibinfo  {journal}
  {Europhysics Letters ({EPL})}\ }\textbf {\bibinfo {volume} {3}}~(\bibinfo
  {number} {12}),\ \bibinfo {pages} {1287}}\BibitemShut {NoStop}%
\bibitem [{\citenamefont {Giamarchi}\ and\ \citenamefont
  {Schulz}(1988)}]{GiamarchiSchulz1988}%
  \BibitemOpen
  \bibfield  {author} {\bibinfo {author} {\bibnamefont {Giamarchi},
  \bibfnamefont {T.}}, \ and\ \bibinfo {author} {\bibfnamefont {H.~J.}\
  \bibnamefont {Schulz}}} (\bibinfo {year} {1988}),\ \href {\doibase
  10.1103/PhysRevB.37.325} {\bibfield  {journal} {\bibinfo  {journal} {Phys.
  Rev. B}\ }\textbf {\bibinfo {volume} {37}},\ \bibinfo {pages}
  {325}}\BibitemShut {NoStop}%
\bibitem [{\citenamefont {Gingrich}\ \emph {et~al.}(2016)\citenamefont
  {Gingrich}, \citenamefont {Horowitz}, \citenamefont {Perunov},\ and\
  \citenamefont {England}}]{gingrich2016a}%
  \BibitemOpen
  \bibfield  {author} {\bibinfo {author} {\bibnamefont {Gingrich},
  \bibfnamefont {T.~R.}}, \bibinfo {author} {\bibfnamefont {J.~M.}\
  \bibnamefont {Horowitz}}, \bibinfo {author} {\bibfnamefont {N.}~\bibnamefont
  {Perunov}}, \ and\ \bibinfo {author} {\bibfnamefont {J.~L.}\ \bibnamefont
  {England}}} (\bibinfo {year} {2016}),\ \href {\doibase
  10.1103/PhysRevLett.116.120601} {\bibfield  {journal} {\bibinfo  {journal}
  {Phys. Rev. Lett.}\ }\textbf {\bibinfo {volume} {116}},\ \bibinfo {pages}
  {120601}}\BibitemShut {NoStop}%
\bibitem [{\citenamefont {Golub}\ \emph {et~al.}(1979)\citenamefont {Golub},
  \citenamefont {Nash},\ and\ \citenamefont {Van~Loan}}]{GolubVanLoan1979}%
  \BibitemOpen
  \bibfield  {author} {\bibinfo {author} {\bibnamefont {Golub}, \bibfnamefont
  {G.~H.}}, \bibinfo {author} {\bibfnamefont {S.}~\bibnamefont {Nash}}, \ and\
  \bibinfo {author} {\bibfnamefont {C.~F.}\ \bibnamefont {Van~Loan}}} (\bibinfo
  {year} {1979}),\ \href {\doibase 10.1109/TAC.1979.1102170} {\bibfield
  {journal} {\bibinfo  {journal} {IEEE Trans. Autom. Control}\ }\textbf
  {\bibinfo {volume} {24}},\ \bibinfo {pages} {909}}\BibitemShut {NoStop}%
\bibitem [{\citenamefont {Gonz{\'{a}}lez}\ \emph {et~al.}(2017)\citenamefont
  {Gonz{\'{a}}lez}, \citenamefont {Correa}, \citenamefont {Nocerino},
  \citenamefont {Palao}, \citenamefont {Alonso},\ and\ \citenamefont
  {Adesso}}]{Gonzalez2017}%
  \BibitemOpen
  \bibfield  {author} {\bibinfo {author} {\bibnamefont {Gonz{\'{a}}lez},
  \bibfnamefont {J.~O.}}, \bibinfo {author} {\bibfnamefont {L.~A.}\
  \bibnamefont {Correa}}, \bibinfo {author} {\bibfnamefont {G.}~\bibnamefont
  {Nocerino}}, \bibinfo {author} {\bibfnamefont {J.~P.}\ \bibnamefont {Palao}},
  \bibinfo {author} {\bibfnamefont {D.}~\bibnamefont {Alonso}}, \ and\ \bibinfo
  {author} {\bibfnamefont {G.}~\bibnamefont {Adesso}}} (\bibinfo {year}
  {2017}),\ \href {\doibase 10.1142/S1230161217400108} {\bibfield  {journal}
  {\bibinfo  {journal} {Open Systems {\&} Information Dynamics}\ }\textbf
  {\bibinfo {volume} {24}},\ \bibinfo {pages} {1740010}}\BibitemShut {NoStop}%
\bibitem [{\citenamefont {Goold}\ \emph {et~al.}(2016)\citenamefont {Goold},
  \citenamefont {Huber}, \citenamefont {Riera}, \citenamefont {del Rio},\ and\
  \citenamefont {Skrzypczyk}}]{GooldSkrzypczyk2016}%
  \BibitemOpen
  \bibfield  {author} {\bibinfo {author} {\bibnamefont {Goold}, \bibfnamefont
  {J.}}, \bibinfo {author} {\bibfnamefont {M.}~\bibnamefont {Huber}}, \bibinfo
  {author} {\bibfnamefont {A.}~\bibnamefont {Riera}}, \bibinfo {author}
  {\bibfnamefont {L.}~\bibnamefont {del Rio}}, \ and\ \bibinfo {author}
  {\bibfnamefont {P.}~\bibnamefont {Skrzypczyk}}} (\bibinfo {year} {2016}),\
  \href {\doibase 10.1088/1751-8113/49/14/143001} {\bibfield  {journal}
  {\bibinfo  {journal} {Journal of Physics A: Mathematical and Theoretical}\
  }\textbf {\bibinfo {volume} {49}}~(\bibinfo {number} {14}),\ \bibinfo {pages}
  {143001}}\BibitemShut {NoStop}%
\bibitem [{\citenamefont {Gopalakrishnan}\ \emph {et~al.}(2016)\citenamefont
  {Gopalakrishnan}, \citenamefont {Agarwal}, \citenamefont {Demler},
  \citenamefont {Huse},\ and\ \citenamefont {Knap}}]{GopalakrishnanKnap2016}%
  \BibitemOpen
  \bibfield  {author} {\bibinfo {author} {\bibnamefont {Gopalakrishnan},
  \bibfnamefont {S.}}, \bibinfo {author} {\bibfnamefont {K.}~\bibnamefont
  {Agarwal}}, \bibinfo {author} {\bibfnamefont {E.~A.}\ \bibnamefont {Demler}},
  \bibinfo {author} {\bibfnamefont {D.~A.}\ \bibnamefont {Huse}}, \ and\
  \bibinfo {author} {\bibfnamefont {M.}~\bibnamefont {Knap}}} (\bibinfo {year}
  {2016}),\ \href {\doibase 10.1103/PhysRevB.93.134206} {\bibfield  {journal}
  {\bibinfo  {journal} {Phys. Rev. B}\ }\textbf {\bibinfo {volume} {93}},\
  \bibinfo {pages} {134206}}\BibitemShut {NoStop}%
\bibitem [{\citenamefont {Gopalakrishnan}\ and\ \citenamefont
  {Vasseur}(2019)}]{GopalakrishnanVasseur2019}%
  \BibitemOpen
  \bibfield  {author} {\bibinfo {author} {\bibnamefont {Gopalakrishnan},
  \bibfnamefont {S.}}, \ and\ \bibinfo {author} {\bibfnamefont
  {R.}~\bibnamefont {Vasseur}}} (\bibinfo {year} {2019}),\ \href {\doibase
  10.1103/PhysRevLett.122.127202} {\bibfield  {journal} {\bibinfo  {journal}
  {Phys. Rev. Lett.}\ }\textbf {\bibinfo {volume} {122}},\ \bibinfo {pages}
  {127202}}\BibitemShut {NoStop}%
\bibitem [{\citenamefont {Gorelik}\ \emph {et~al.}(1998)\citenamefont
  {Gorelik}, \citenamefont {Isacsson}, \citenamefont {Voinova}, \citenamefont
  {Kasemo}, \citenamefont {Shekhter},\ and\ \citenamefont
  {Jonson}}]{gorelik1998a}%
  \BibitemOpen
  \bibfield  {author} {\bibinfo {author} {\bibnamefont {Gorelik}, \bibfnamefont
  {L.~Y.}}, \bibinfo {author} {\bibfnamefont {A.}~\bibnamefont {Isacsson}},
  \bibinfo {author} {\bibfnamefont {M.~V.}\ \bibnamefont {Voinova}}, \bibinfo
  {author} {\bibfnamefont {B.}~\bibnamefont {Kasemo}}, \bibinfo {author}
  {\bibfnamefont {R.~I.}\ \bibnamefont {Shekhter}}, \ and\ \bibinfo {author}
  {\bibfnamefont {M.}~\bibnamefont {Jonson}}} (\bibinfo {year} {1998}),\ \href
  {\doibase 10.1103/PhysRevLett.80.4526} {\bibfield  {journal} {\bibinfo
  {journal} {Phys. Rev. Lett.}\ }\textbf {\bibinfo {volume} {80}},\ \bibinfo
  {pages} {4526}}\BibitemShut {NoStop}%
\bibitem [{\citenamefont {Gorini}\ \emph {et~al.}(1976)\citenamefont {Gorini},
  \citenamefont {Kossakowski},\ and\ \citenamefont {Sudarshan}}]{Gorini1976}%
  \BibitemOpen
  \bibfield  {author} {\bibinfo {author} {\bibnamefont {Gorini}, \bibfnamefont
  {V.}}, \bibinfo {author} {\bibfnamefont {A.}~\bibnamefont {Kossakowski}}, \
  and\ \bibinfo {author} {\bibfnamefont {E.~C.~G.}\ \bibnamefont {Sudarshan}}}
  (\bibinfo {year} {1976}),\ \href {\doibase 10.1063/1.522979} {\bibfield
  {journal} {\bibinfo  {journal} {Journal of Mathematical Physics}\ }\textbf
  {\bibinfo {volume} {17}},\ \bibinfo {pages} {821}}\BibitemShut {NoStop}%
\bibitem [{\citenamefont {Gornyi}\ \emph {et~al.}(2005)\citenamefont {Gornyi},
  \citenamefont {Mirlin},\ and\ \citenamefont {Polyakov}}]{GornyiPolyakov2005}%
  \BibitemOpen
  \bibfield  {author} {\bibinfo {author} {\bibnamefont {Gornyi}, \bibfnamefont
  {I.~V.}}, \bibinfo {author} {\bibfnamefont {A.~D.}\ \bibnamefont {Mirlin}}, \
  and\ \bibinfo {author} {\bibfnamefont {D.~G.}\ \bibnamefont {Polyakov}}}
  (\bibinfo {year} {2005}),\ \href {\doibase 10.1103/PhysRevLett.95.206603}
  {\bibfield  {journal} {\bibinfo  {journal} {Phys. Rev. Lett.}\ }\textbf
  {\bibinfo {volume} {95}},\ \bibinfo {pages} {206603}}\BibitemShut {NoStop}%
\bibitem [{\citenamefont {Granato}(1990)}]{Granato1990}%
  \BibitemOpen
  \bibfield  {author} {\bibinfo {author} {\bibnamefont {Granato}, \bibfnamefont
  {E.}}} (\bibinfo {year} {1990}),\ \href {\doibase 10.1103/PhysRevB.42.4797}
  {\bibfield  {journal} {\bibinfo  {journal} {Phys. Rev. B}\ }\textbf {\bibinfo
  {volume} {42}},\ \bibinfo {pages} {4797}}\BibitemShut {NoStop}%
\bibitem [{\citenamefont {Green}(1952)}]{Green1952}%
  \BibitemOpen
  \bibfield  {author} {\bibinfo {author} {\bibnamefont {Green}, \bibfnamefont
  {M.~S.}}} (\bibinfo {year} {1952}),\ \href {\doibase 10.1063/1.1700722}
  {\bibfield  {journal} {\bibinfo  {journal} {J. Chem. Phys.}\ }\textbf
  {\bibinfo {volume} {20}},\ \bibinfo {pages} {1281}}\BibitemShut {NoStop}%
\bibitem [{\citenamefont {Green}(1954)}]{Green1954}%
  \BibitemOpen
  \bibfield  {author} {\bibinfo {author} {\bibnamefont {Green}, \bibfnamefont
  {M.~S.}}} (\bibinfo {year} {1954}),\ \href {\doibase 10.1063/1.1740082}
  {\bibfield  {journal} {\bibinfo  {journal} {J. Chem. Phys.}\ }\textbf
  {\bibinfo {volume} {22}},\ \bibinfo {pages} {398}}\BibitemShut {NoStop}%
\bibitem [{\citenamefont {Greiner}\ \emph {et~al.}(2002)\citenamefont
  {Greiner}, \citenamefont {Mandel}, \citenamefont {Esslinger}, \citenamefont
  {H\"ansch},\ and\ \citenamefont {Bloch}}]{GreinerBloch2002}%
  \BibitemOpen
  \bibfield  {author} {\bibinfo {author} {\bibnamefont {Greiner}, \bibfnamefont
  {M.}}, \bibinfo {author} {\bibfnamefont {O.}~\bibnamefont {Mandel}}, \bibinfo
  {author} {\bibfnamefont {T.}~\bibnamefont {Esslinger}}, \bibinfo {author}
  {\bibfnamefont {T.}~\bibnamefont {H\"ansch}}, \ and\ \bibinfo {author}
  {\bibfnamefont {I.}~\bibnamefont {Bloch}}} (\bibinfo {year} {2002}),\ \href
  {\doibase 10.1038/415039a} {\bibfield  {journal} {\bibinfo  {journal}
  {Nature}\ }\textbf {\bibinfo {volume} {415}},\ \bibinfo {pages}
  {39}}\BibitemShut {NoStop}%
\bibitem [{\citenamefont {Griffiths}(1969)}]{Griffiths1969}%
  \BibitemOpen
  \bibfield  {author} {\bibinfo {author} {\bibnamefont {Griffiths},
  \bibfnamefont {R.~B.}}} (\bibinfo {year} {1969}),\ \href {\doibase
  10.1103/PhysRevLett.23.17} {\bibfield  {journal} {\bibinfo  {journal} {Phys.
  Rev. Lett.}\ }\textbf {\bibinfo {volume} {23}},\ \bibinfo {pages}
  {17}}\BibitemShut {NoStop}%
\bibitem [{\citenamefont {Grifoni}\ and\ \citenamefont
  {H{\"a}nggi}(1998)}]{GrifoniHanggi1998}%
  \BibitemOpen
  \bibfield  {author} {\bibinfo {author} {\bibnamefont {Grifoni}, \bibfnamefont
  {M.}}, \ and\ \bibinfo {author} {\bibfnamefont {P.}~\bibnamefont
  {H{\"a}nggi}}} (\bibinfo {year} {1998}),\ \href {\doibase
  https://doi.org/10.1016/S0370-1573(98)00022-2} {\bibfield  {journal}
  {\bibinfo  {journal} {Physics Reports}\ }\textbf {\bibinfo {volume}
  {304}}~(\bibinfo {number} {5}),\ \bibinfo {pages} {229}}\BibitemShut
  {NoStop}%
\bibitem [{\citenamefont {Gross}\ and\ \citenamefont
  {Haroche}(1982)}]{gross1982a}%
  \BibitemOpen
  \bibfield  {author} {\bibinfo {author} {\bibnamefont {Gross}, \bibfnamefont
  {M.}}, \ and\ \bibinfo {author} {\bibfnamefont {S.}~\bibnamefont {Haroche}}}
  (\bibinfo {year} {1982}),\ \href {\doibase 10.1016/0370-1573(82)90102-8}
  {\bibfield  {journal} {\bibinfo  {journal} {Physics Reports}\ }\textbf
  {\bibinfo {volume} {93}},\ \bibinfo {pages} {301}}\BibitemShut {NoStop}%
\bibitem [{\citenamefont {Guimar{\~{a}}es}\ \emph {et~al.}(2016)\citenamefont
  {Guimar{\~{a}}es}, \citenamefont {de~Oliveira},\ and\ \citenamefont
  {Landi}}]{Guimar2016}%
  \BibitemOpen
  \bibfield  {author} {\bibinfo {author} {\bibnamefont {Guimar{\~{a}}es},
  \bibfnamefont {P.~H.}}, \bibinfo {author} {\bibfnamefont {M.~J.}\
  \bibnamefont {de~Oliveira}}, \ and\ \bibinfo {author} {\bibfnamefont {G.~T.}\
  \bibnamefont {Landi}}} (\bibinfo {year} {2016}),\ \href {\doibase
  10.1103/PhysRevE.94.032139} {\bibfield  {journal} {\bibinfo  {journal}
  {Physical Review E}\ }\textbf {\bibinfo {volume} {94}},\ \bibinfo {pages}
  {032139}}\BibitemShut {NoStop}%
\bibitem [{\citenamefont {Gullans}\ and\ \citenamefont
  {Huse}(2019)}]{GullansHuse2019b}%
  \BibitemOpen
  \bibfield  {author} {\bibinfo {author} {\bibnamefont {Gullans}, \bibfnamefont
  {M.~J.}}, \ and\ \bibinfo {author} {\bibfnamefont {D.~A.}\ \bibnamefont
  {Huse}}} (\bibinfo {year} {2019}),\ \href {\doibase
  10.1103/PhysRevX.9.021007} {\bibfield  {journal} {\bibinfo  {journal} {Phys.
  Rev. X}\ }\textbf {\bibinfo {volume} {9}},\ \bibinfo {pages}
  {021007}}\BibitemShut {NoStop}%
\bibitem [{\citenamefont {Guo}\ \emph {et~al.}(2015)\citenamefont {Guo},
  \citenamefont {Mukherjee},\ and\ \citenamefont {Poletti}}]{GuoPoletti2015}%
  \BibitemOpen
  \bibfield  {author} {\bibinfo {author} {\bibnamefont {Guo}, \bibfnamefont
  {C.}}, \bibinfo {author} {\bibfnamefont {M.}~\bibnamefont {Mukherjee}}, \
  and\ \bibinfo {author} {\bibfnamefont {D.}~\bibnamefont {Poletti}}} (\bibinfo
  {year} {2015}),\ \href {\doibase 10.1103/PhysRevA.92.023637} {\bibfield
  {journal} {\bibinfo  {journal} {Phys. Rev. A}\ }\textbf {\bibinfo {volume}
  {92}},\ \bibinfo {pages} {023637}}\BibitemShut {NoStop}%
\bibitem [{\citenamefont {Guo}\ and\ \citenamefont
  {Poletti}(2016)}]{GuoPoletti2016}%
  \BibitemOpen
  \bibfield  {author} {\bibinfo {author} {\bibnamefont {Guo}, \bibfnamefont
  {C.}}, \ and\ \bibinfo {author} {\bibfnamefont {D.}~\bibnamefont {Poletti}}}
  (\bibinfo {year} {2016}),\ \href {\doibase 10.1103/PhysRevA.94.033610}
  {\bibfield  {journal} {\bibinfo  {journal} {Phys. Rev. A}\ }\textbf {\bibinfo
  {volume} {94}},\ \bibinfo {pages} {033610}}\BibitemShut {NoStop}%
\bibitem [{\citenamefont {Guo}\ and\ \citenamefont
  {Poletti}(2017{\natexlab{a}})}]{GuoPoletti2017b}%
  \BibitemOpen
  \bibfield  {author} {\bibinfo {author} {\bibnamefont {Guo}, \bibfnamefont
  {C.}}, \ and\ \bibinfo {author} {\bibfnamefont {D.}~\bibnamefont {Poletti}}}
  (\bibinfo {year} {2017}{\natexlab{a}}),\ \href {\doibase
  10.1103/PhysRevB.96.165409} {\bibfield  {journal} {\bibinfo  {journal} {Phys.
  Rev. B}\ }\textbf {\bibinfo {volume} {96}},\ \bibinfo {pages}
  {165409}}\BibitemShut {NoStop}%
\bibitem [{\citenamefont {Guo}\ and\ \citenamefont
  {Poletti}(2017{\natexlab{b}})}]{GuoPoletti2017}%
  \BibitemOpen
  \bibfield  {author} {\bibinfo {author} {\bibnamefont {Guo}, \bibfnamefont
  {C.}}, \ and\ \bibinfo {author} {\bibfnamefont {D.}~\bibnamefont {Poletti}}}
  (\bibinfo {year} {2017}{\natexlab{b}}),\ \href {\doibase
  10.1103/PhysRevA.95.052107} {\bibfield  {journal} {\bibinfo  {journal} {Phys.
  Rev. A}\ }\textbf {\bibinfo {volume} {95}},\ \bibinfo {pages}
  {052107}}\BibitemShut {NoStop}%
\bibitem [{\citenamefont {Guo}\ and\ \citenamefont
  {Poletti}(2018)}]{GuoPoletti2018b}%
  \BibitemOpen
  \bibfield  {author} {\bibinfo {author} {\bibnamefont {Guo}, \bibfnamefont
  {C.}}, \ and\ \bibinfo {author} {\bibfnamefont {D.}~\bibnamefont {Poletti}}}
  (\bibinfo {year} {2018}),\ \href {\doibase 10.1103/PhysRevA.98.052126}
  {\bibfield  {journal} {\bibinfo  {journal} {Phys. Rev. A}\ }\textbf {\bibinfo
  {volume} {98}},\ \bibinfo {pages} {052126}}\BibitemShut {NoStop}%
\bibitem [{\citenamefont {Guo}\ and\ \citenamefont
  {Poletti}(2019)}]{GuoPoletti2019}%
  \BibitemOpen
  \bibfield  {author} {\bibinfo {author} {\bibnamefont {Guo}, \bibfnamefont
  {C.}}, \ and\ \bibinfo {author} {\bibfnamefont {D.}~\bibnamefont {Poletti}}}
  (\bibinfo {year} {2019}),\ \href {\doibase 10.1103/PhysRevB.100.134304}
  {\bibfield  {journal} {\bibinfo  {journal} {Phys. Rev. B}\ }\textbf {\bibinfo
  {volume} {100}},\ \bibinfo {pages} {134304}}\BibitemShut {NoStop}%
\bibitem [{\citenamefont {Guo}\ \emph {et~al.}(2018)\citenamefont {Guo},
  \citenamefont {de~Vega}, \citenamefont {Schollw{\"o}ck},\ and\ \citenamefont
  {Poletti}}]{GuoPoletti2018}%
  \BibitemOpen
  \bibfield  {author} {\bibinfo {author} {\bibnamefont {Guo}, \bibfnamefont
  {C.}}, \bibinfo {author} {\bibfnamefont {I.}~\bibnamefont {de~Vega}},
  \bibinfo {author} {\bibfnamefont {U.}~\bibnamefont {Schollw{\"o}ck}}, \ and\
  \bibinfo {author} {\bibfnamefont {D.}~\bibnamefont {Poletti}}} (\bibinfo
  {year} {2018}),\ \href {\doibase 10.1103/PhysRevA.97.053610} {\bibfield
  {journal} {\bibinfo  {journal} {Phys. Rev. A}\ }\textbf {\bibinfo {volume}
  {97}},\ \bibinfo {pages} {053610}}\BibitemShut {NoStop}%
\bibitem [{\citenamefont {Gustavsson}\ \emph {et~al.}(2006)\citenamefont
  {Gustavsson}, \citenamefont {Leturcq}, \citenamefont {Simovi\v{c}},
  \citenamefont {Schleser}, \citenamefont {Ihn}, \citenamefont {Studerus},
  \citenamefont {Ensslin}, \citenamefont {Driscoll},\ and\ \citenamefont
  {Gossard}}]{gustavsson2006a}%
  \BibitemOpen
  \bibfield  {author} {\bibinfo {author} {\bibnamefont {Gustavsson},
  \bibfnamefont {S.}}, \bibinfo {author} {\bibfnamefont {R.}~\bibnamefont
  {Leturcq}}, \bibinfo {author} {\bibfnamefont {B.}~\bibnamefont
  {Simovi\v{c}}}, \bibinfo {author} {\bibfnamefont {R.}~\bibnamefont
  {Schleser}}, \bibinfo {author} {\bibfnamefont {T.}~\bibnamefont {Ihn}},
  \bibinfo {author} {\bibfnamefont {P.}~\bibnamefont {Studerus}}, \bibinfo
  {author} {\bibfnamefont {K.}~\bibnamefont {Ensslin}}, \bibinfo {author}
  {\bibfnamefont {D.~C.}\ \bibnamefont {Driscoll}}, \ and\ \bibinfo {author}
  {\bibfnamefont {A.~C.}\ \bibnamefont {Gossard}}} (\bibinfo {year} {2006}),\
  \href {\doibase 10.1103/PhysRevLett.96.076605} {\bibfield  {journal}
  {\bibinfo  {journal} {Phys. Rev. Lett.}\ }\textbf {\bibinfo {volume} {96}},\
  \bibinfo {pages} {076605}}\BibitemShut {NoStop}%
\bibitem [{\citenamefont {Hackenbroich}\ and\ \citenamefont
  {Weidenm{\"u}ller}(1996)}]{HackenbroichWeidenmuller1996}%
  \BibitemOpen
  \bibfield  {author} {\bibinfo {author} {\bibnamefont {Hackenbroich},
  \bibfnamefont {G.}}, \ and\ \bibinfo {author} {\bibfnamefont {H.~A.}\
  \bibnamefont {Weidenm{\"u}ller}}} (\bibinfo {year} {1996}),\ \href {\doibase
  10.1103/PhysRevLett.76.110} {\bibfield  {journal} {\bibinfo  {journal} {Phys.
  Rev. Lett.}\ }\textbf {\bibinfo {volume} {76}},\ \bibinfo {pages}
  {110}}\BibitemShut {NoStop}%
\bibitem [{\citenamefont {Halbritter}\ \emph {et~al.}(2008)\citenamefont
  {Halbritter}, \citenamefont {Makk}, \citenamefont {Csonka},\ and\
  \citenamefont {Mih\'aly}}]{HalbritterWihaly2008}%
  \BibitemOpen
  \bibfield  {author} {\bibinfo {author} {\bibnamefont {Halbritter},
  \bibfnamefont {A.}}, \bibinfo {author} {\bibfnamefont {P.}~\bibnamefont
  {Makk}}, \bibinfo {author} {\bibfnamefont {S.}~\bibnamefont {Csonka}}, \ and\
  \bibinfo {author} {\bibfnamefont {G.}~\bibnamefont {Mih\'aly}}} (\bibinfo
  {year} {2008}),\ \href {\doibase 10.1103/PhysRevB.77.075402} {\bibfield
  {journal} {\bibinfo  {journal} {Phys. Rev. B}\ }\textbf {\bibinfo {volume}
  {77}},\ \bibinfo {pages} {075402}}\BibitemShut {NoStop}%
\bibitem [{\citenamefont {Hamazaki}\ \emph {et~al.}(2020)\citenamefont
  {Hamazaki}, \citenamefont {Kawabata}, \citenamefont {Kura},\ and\
  \citenamefont {Ueda}}]{HamazakiUeda2020}%
  \BibitemOpen
  \bibfield  {author} {\bibinfo {author} {\bibnamefont {Hamazaki},
  \bibfnamefont {R.}}, \bibinfo {author} {\bibfnamefont {K.}~\bibnamefont
  {Kawabata}}, \bibinfo {author} {\bibfnamefont {N.}~\bibnamefont {Kura}}, \
  and\ \bibinfo {author} {\bibfnamefont {M.}~\bibnamefont {Ueda}}} (\bibinfo
  {year} {2020}),\ \href {\doibase 10.1103/PhysRevResearch.2.023286} {\bibfield
   {journal} {\bibinfo  {journal} {Phys. Rev. Research}\ }\textbf {\bibinfo
  {volume} {2}},\ \bibinfo {pages} {023286}}\BibitemShut {NoStop}%
\bibitem [{\citenamefont {Harper}(1955)}]{Harper1955}%
  \BibitemOpen
  \bibfield  {author} {\bibinfo {author} {\bibnamefont {Harper}, \bibfnamefont
  {P.~G.}}} (\bibinfo {year} {1955}),\ \href {\doibase
  10.1088/0370-1298/68/10/305} {\bibfield  {journal} {\bibinfo  {journal}
  {Proc. Phys. Soc. London Sec. A}\ }\textbf {\bibinfo {volume} {68}},\
  \bibinfo {pages} {874}}\BibitemShut {NoStop}%
\bibitem [{\citenamefont {Hartmann}\ and\ \citenamefont
  {Carleo}(2019)}]{HartmannCarleo2019}%
  \BibitemOpen
  \bibfield  {author} {\bibinfo {author} {\bibnamefont {Hartmann},
  \bibfnamefont {M.~J.}}, \ and\ \bibinfo {author} {\bibfnamefont
  {G.}~\bibnamefont {Carleo}}} (\bibinfo {year} {2019}),\ \href {\doibase
  10.1103/PhysRevLett.122.250502} {\bibfield  {journal} {\bibinfo  {journal}
  {Phys. Rev. Lett.}\ }\textbf {\bibinfo {volume} {122}},\ \bibinfo {pages}
  {250502}}\BibitemShut {NoStop}%
\bibitem [{\citenamefont {Hartmann}\ \emph {et~al.}(2009)\citenamefont
  {Hartmann}, \citenamefont {Prior}, \citenamefont {Clark},\ and\ \citenamefont
  {Plenio}}]{HartmannPlenio2009}%
  \BibitemOpen
  \bibfield  {author} {\bibinfo {author} {\bibnamefont {Hartmann},
  \bibfnamefont {M.~J.}}, \bibinfo {author} {\bibfnamefont {J.}~\bibnamefont
  {Prior}}, \bibinfo {author} {\bibfnamefont {S.~R.}\ \bibnamefont {Clark}}, \
  and\ \bibinfo {author} {\bibfnamefont {M.~B.}\ \bibnamefont {Plenio}}}
  (\bibinfo {year} {2009}),\ \href {\doibase 10.1103/PhysRevLett.102.057202}
  {\bibfield  {journal} {\bibinfo  {journal} {Phys. Rev. Lett.}\ }\textbf
  {\bibinfo {volume} {102}},\ \bibinfo {pages} {057202}}\BibitemShut {NoStop}%
\bibitem [{\citenamefont {Hartmann}\ and\ \citenamefont
  {Strunz}(2020)}]{hartmann2020a}%
  \BibitemOpen
  \bibfield  {author} {\bibinfo {author} {\bibnamefont {Hartmann},
  \bibfnamefont {R.}}, \ and\ \bibinfo {author} {\bibfnamefont {W.~T.}\
  \bibnamefont {Strunz}}} (\bibinfo {year} {2020}),\ \href {\doibase
  10.1103/PhysRevA.101.012103} {\bibfield  {journal} {\bibinfo  {journal}
  {Phys. Rev. A}\ }\textbf {\bibinfo {volume} {101}},\ \bibinfo {pages}
  {012103}}\BibitemShut {NoStop}%
\bibitem [{\citenamefont {Hasan}\ and\ \citenamefont
  {Kane}(2010)}]{hasan2010a}%
  \BibitemOpen
  \bibfield  {author} {\bibinfo {author} {\bibnamefont {Hasan}, \bibfnamefont
  {M.~Z.}}, \ and\ \bibinfo {author} {\bibfnamefont {C.~L.}\ \bibnamefont
  {Kane}}} (\bibinfo {year} {2010}),\ \href {\doibase
  10.1103/RevModPhys.82.3045} {\bibfield  {journal} {\bibinfo  {journal} {Rev.
  Mod. Phys.}\ }\textbf {\bibinfo {volume} {82}},\ \bibinfo {pages}
  {3045}}\BibitemShut {NoStop}%
\bibitem [{\citenamefont {Hasegawa}\ and\ \citenamefont
  {Van~Vu}(2019)}]{hasegawa2019a}%
  \BibitemOpen
  \bibfield  {author} {\bibinfo {author} {\bibnamefont {Hasegawa},
  \bibfnamefont {Y.}}, \ and\ \bibinfo {author} {\bibfnamefont
  {T.}~\bibnamefont {Van~Vu}}} (\bibinfo {year} {2019}),\ \href {\doibase
  10.1103/PhysRevLett.123.110602} {\bibfield  {journal} {\bibinfo  {journal}
  {Phys. Rev. Lett.}\ }\textbf {\bibinfo {volume} {123}},\ \bibinfo {pages}
  {110602}}\BibitemShut {NoStop}%
\bibitem [{\citenamefont {Haug}\ and\ \citenamefont
  {Jauho}(2008)}]{HaugJauho2008}%
  \BibitemOpen
  \bibfield  {author} {\bibinfo {author} {\bibnamefont {Haug}, \bibfnamefont
  {H.}}, \ and\ \bibinfo {author} {\bibfnamefont {A.-P.}\ \bibnamefont
  {Jauho}}} (\bibinfo {year} {2008}),\ \href {\doibase
  10.1007/978-3-540-73564-9} {\emph {\bibinfo {title} {Quantum {{Kinetics}} in
  {{Transport}} and {{Optics}} of {{Semiconductors}}}}}\ (\bibinfo  {publisher}
  {{Springer-Verlag}},\ \bibinfo {address} {{Berlin Heidelberg}})\BibitemShut
  {NoStop}%
\bibitem [{\citenamefont {Heidrich-Meisner}\ \emph {et~al.}(2005)\citenamefont
  {Heidrich-Meisner}, \citenamefont {Honecker},\ and\ \citenamefont
  {Brenig}}]{HeidrichMeisnerBrenig2005}%
  \BibitemOpen
  \bibfield  {author} {\bibinfo {author} {\bibnamefont {Heidrich-Meisner},
  \bibfnamefont {F.}}, \bibinfo {author} {\bibfnamefont {A.}~\bibnamefont
  {Honecker}}, \ and\ \bibinfo {author} {\bibfnamefont {W.}~\bibnamefont
  {Brenig}}} (\bibinfo {year} {2005}),\ \href {\doibase
  10.1103/PhysRevB.71.184415} {\bibfield  {journal} {\bibinfo  {journal} {Phys.
  Rev. B}\ }\textbf {\bibinfo {volume} {71}},\ \bibinfo {pages}
  {184415}}\BibitemShut {NoStop}%
\bibitem [{\citenamefont {Heij}\ \emph {et~al.}(1999)\citenamefont {Heij},
  \citenamefont {Dixon}, \citenamefont {Hadley},\ and\ \citenamefont
  {Mooij}}]{HeijMooij1999}%
  \BibitemOpen
  \bibfield  {author} {\bibinfo {author} {\bibnamefont {Heij}, \bibfnamefont
  {C.~P.}}, \bibinfo {author} {\bibfnamefont {D.~C.}\ \bibnamefont {Dixon}},
  \bibinfo {author} {\bibfnamefont {P.}~\bibnamefont {Hadley}}, \ and\ \bibinfo
  {author} {\bibfnamefont {J.~E.}\ \bibnamefont {Mooij}}} (\bibinfo {year}
  {1999}),\ \href {\doibase 10.1063/1.123449} {\bibfield  {journal} {\bibinfo
  {journal} {Applied Physics Letters}\ }\textbf {\bibinfo {volume} {74}},\
  \bibinfo {pages} {1042}}\BibitemShut {NoStop}%
\bibitem [{\citenamefont {Hiltscher}\ \emph {et~al.}(2010)\citenamefont
  {Hiltscher}, \citenamefont {Governale},\ and\ \citenamefont
  {K{\"o}nig}}]{hiltscher2010a}%
  \BibitemOpen
  \bibfield  {author} {\bibinfo {author} {\bibnamefont {Hiltscher},
  \bibfnamefont {B.}}, \bibinfo {author} {\bibfnamefont {M.}~\bibnamefont
  {Governale}}, \ and\ \bibinfo {author} {\bibfnamefont {J.}~\bibnamefont
  {K{\"o}nig}}} (\bibinfo {year} {2010}),\ \href {\doibase
  10.1103/PhysRevB.82.165452} {\bibfield  {journal} {\bibinfo  {journal} {Phys.
  Rev. B}\ }\textbf {\bibinfo {volume} {82}},\ \bibinfo {pages}
  {165452}}\BibitemShut {NoStop}%
\bibitem [{\citenamefont {Hiramoto}\ and\ \citenamefont
  {Abe}(1988)}]{HiramotoAbe1988}%
  \BibitemOpen
  \bibfield  {author} {\bibinfo {author} {\bibnamefont {Hiramoto},
  \bibfnamefont {H.}}, \ and\ \bibinfo {author} {\bibfnamefont
  {S.}~\bibnamefont {Abe}}} (\bibinfo {year} {1988}),\ \href {\doibase
  10.1143/JPSJ.57.230} {\bibfield  {journal} {\bibinfo  {journal} {Journal of
  the Physical Society of Japan}\ }\textbf {\bibinfo {volume} {57}},\ \bibinfo
  {pages} {230}}\BibitemShut {NoStop}%
\bibitem [{\citenamefont {Hiramoto}\ and\ \citenamefont
  {Kohmoto}(1992)}]{HiramotoKohmoto1992}%
  \BibitemOpen
  \bibfield  {author} {\bibinfo {author} {\bibnamefont {Hiramoto},
  \bibfnamefont {H.}}, \ and\ \bibinfo {author} {\bibfnamefont
  {M.}~\bibnamefont {Kohmoto}}} (\bibinfo {year} {1992}),\ \href {\doibase
  10.1142/S0217979292000153} {\bibfield  {journal} {\bibinfo  {journal}
  {International Journal of Modern Physics B}\ }\textbf {\bibinfo {volume}
  {6}},\ \bibinfo {pages} {281}}\BibitemShut {NoStop}%
\bibitem [{\citenamefont {Hofer}\ \emph {et~al.}(2017)\citenamefont {Hofer},
  \citenamefont {Perarnau-Llobet}, \citenamefont {David}, \citenamefont
  {Miranda}, \citenamefont {Haack}, \citenamefont {Silva}, \citenamefont
  {Brask},\ and\ \citenamefont {Brunner}}]{hofer2017a}%
  \BibitemOpen
  \bibfield  {author} {\bibinfo {author} {\bibnamefont {Hofer}, \bibfnamefont
  {P.~P.}}, \bibinfo {author} {\bibfnamefont {M.}~\bibnamefont
  {Perarnau-Llobet}}, \bibinfo {author} {\bibfnamefont {L.}~\bibnamefont
  {David}}, \bibinfo {author} {\bibfnamefont {M.}~\bibnamefont {Miranda}},
  \bibinfo {author} {\bibfnamefont {G.}~\bibnamefont {Haack}}, \bibinfo
  {author} {\bibfnamefont {R.}~\bibnamefont {Silva}}, \bibinfo {author}
  {\bibfnamefont {J.~B.}\ \bibnamefont {Brask}}, \ and\ \bibinfo {author}
  {\bibfnamefont {N.}~\bibnamefont {Brunner}}} (\bibinfo {year} {2017}),\ \href
  {\doibase 10.1088/1367-2630/aa964f} {\bibfield  {journal} {\bibinfo
  {journal} {New Journal of Physics}\ }\textbf {\bibinfo {volume} {19}},\
  \bibinfo {pages} {123037}}\BibitemShut {NoStop}%
\bibitem [{\citenamefont {Hofstadter}(1976)}]{Hofstadter1976}%
  \BibitemOpen
  \bibfield  {author} {\bibinfo {author} {\bibnamefont {Hofstadter},
  \bibfnamefont {D.~R.}}} (\bibinfo {year} {1976}),\ \href {\doibase
  10.1103/PhysRevB.14.2239} {\bibfield  {journal} {\bibinfo  {journal} {Phys.
  Rev. B}\ }\textbf {\bibinfo {volume} {14}},\ \bibinfo {pages}
  {2239}}\BibitemShut {NoStop}%
\bibitem [{\citenamefont {Hofstetter}\ \emph {et~al.}(2001)\citenamefont
  {Hofstetter}, \citenamefont {K{\"o}nig},\ and\ \citenamefont
  {Schoeller}}]{HofstetterSchoeller2001}%
  \BibitemOpen
  \bibfield  {author} {\bibinfo {author} {\bibnamefont {Hofstetter},
  \bibfnamefont {W.}}, \bibinfo {author} {\bibfnamefont {J.}~\bibnamefont
  {K{\"o}nig}}, \ and\ \bibinfo {author} {\bibfnamefont {H.}~\bibnamefont
  {Schoeller}}} (\bibinfo {year} {2001}),\ \href {\doibase
  10.1103/PhysRevLett.87.156803} {\bibfield  {journal} {\bibinfo  {journal}
  {Phys. Rev. Lett.}\ }\textbf {\bibinfo {volume} {87}},\ \bibinfo {pages}
  {156803}}\BibitemShut {NoStop}%
\bibitem [{\citenamefont {Holleitner}\ \emph {et~al.}(2001)\citenamefont
  {Holleitner}, \citenamefont {Decker}, \citenamefont {Qin}, \citenamefont
  {Eberl},\ and\ \citenamefont {Blick}}]{HolleitnerBlick2001}%
  \BibitemOpen
  \bibfield  {author} {\bibinfo {author} {\bibnamefont {Holleitner},
  \bibfnamefont {A.~W.}}, \bibinfo {author} {\bibfnamefont {C.~R.}\
  \bibnamefont {Decker}}, \bibinfo {author} {\bibfnamefont {H.}~\bibnamefont
  {Qin}}, \bibinfo {author} {\bibfnamefont {K.}~\bibnamefont {Eberl}}, \ and\
  \bibinfo {author} {\bibfnamefont {R.~H.}\ \bibnamefont {Blick}}} (\bibinfo
  {year} {2001}),\ \href {\doibase 10.1103/PhysRevLett.87.256802} {\bibfield
  {journal} {\bibinfo  {journal} {Phys. Rev. Lett.}\ }\textbf {\bibinfo
  {volume} {87}},\ \bibinfo {pages} {256802}}\BibitemShut {NoStop}%
\bibitem [{\citenamefont {Hu}\ \emph {et~al.}(2009)\citenamefont {Hu},
  \citenamefont {Ruan},\ and\ \citenamefont {Chen}}]{Hu2009}%
  \BibitemOpen
  \bibfield  {author} {\bibinfo {author} {\bibnamefont {Hu}, \bibfnamefont
  {J.}}, \bibinfo {author} {\bibfnamefont {X.}~\bibnamefont {Ruan}}, \ and\
  \bibinfo {author} {\bibfnamefont {Y.~P.}\ \bibnamefont {Chen}}} (\bibinfo
  {year} {2009}),\ \href {http://pubs.acs.org/doi/abs/10.1021/nl901231s}
  {\bibfield  {journal} {\bibinfo  {journal} {Nano Letters}\ }\textbf {\bibinfo
  {volume} {9}}~(\bibinfo {number} {7}),\ \bibinfo {pages} {2730}}\BibitemShut
  {NoStop}%
\bibitem [{\citenamefont {Hubig}\ \emph {et~al.}(2017)\citenamefont {Hubig},
  \citenamefont {McCulloch},\ and\ \citenamefont
  {Schollw{\"o}ck}}]{HubigSchollwock2017}%
  \BibitemOpen
  \bibfield  {author} {\bibinfo {author} {\bibnamefont {Hubig}, \bibfnamefont
  {C.}}, \bibinfo {author} {\bibfnamefont {I.~P.}\ \bibnamefont {McCulloch}}, \
  and\ \bibinfo {author} {\bibfnamefont {U.}~\bibnamefont {Schollw{\"o}ck}}}
  (\bibinfo {year} {2017}),\ \href {\doibase 10.1103/PhysRevB.95.035129}
  {\bibfield  {journal} {\bibinfo  {journal} {Phys. Rev. B}\ }\textbf {\bibinfo
  {volume} {95}},\ \bibinfo {pages} {035129}}\BibitemShut {NoStop}%
\bibitem [{\citenamefont {Huh}\ \emph {et~al.}(2014)\citenamefont {Huh},
  \citenamefont {Mostame}, \citenamefont {Fujita}, \citenamefont {Yung},\ and\
  \citenamefont {Aspuru-Guzik}}]{huh2014a}%
  \BibitemOpen
  \bibfield  {author} {\bibinfo {author} {\bibnamefont {Huh}, \bibfnamefont
  {J.}}, \bibinfo {author} {\bibfnamefont {S.}~\bibnamefont {Mostame}},
  \bibinfo {author} {\bibfnamefont {T.}~\bibnamefont {Fujita}}, \bibinfo
  {author} {\bibfnamefont {M.-H.}\ \bibnamefont {Yung}}, \ and\ \bibinfo
  {author} {\bibfnamefont {A.}~\bibnamefont {Aspuru-Guzik}}} (\bibinfo {year}
  {2014}),\ \href {\doibase 10.1088/1367-2630/16/12/123008} {\bibfield
  {journal} {\bibinfo  {journal} {New Journal of Physics}\ }\textbf {\bibinfo
  {volume} {16}},\ \bibinfo {pages} {123008}}\BibitemShut {NoStop}%
\bibitem [{\citenamefont {Huse}\ \emph {et~al.}(2014)\citenamefont {Huse},
  \citenamefont {Nandkishore},\ and\ \citenamefont
  {Oganesyan}}]{HuseOganesyan2014}%
  \BibitemOpen
  \bibfield  {author} {\bibinfo {author} {\bibnamefont {Huse}, \bibfnamefont
  {D.~A.}}, \bibinfo {author} {\bibfnamefont {R.}~\bibnamefont {Nandkishore}},
  \ and\ \bibinfo {author} {\bibfnamefont {V.}~\bibnamefont {Oganesyan}}}
  (\bibinfo {year} {2014}),\ \href {\doibase 10.1103/PhysRevB.90.174202}
  {\bibfield  {journal} {\bibinfo  {journal} {Phys. Rev. B}\ }\textbf {\bibinfo
  {volume} {90}},\ \bibinfo {pages} {174202}}\BibitemShut {NoStop}%
\bibitem [{\citenamefont {Husmann}\ \emph {et~al.}(2018)\citenamefont
  {Husmann}, \citenamefont {Lebrat}, \citenamefont {H{\"a}usler}, \citenamefont
  {Brantut}, \citenamefont {Corman},\ and\ \citenamefont
  {Esslinger}}]{HusmannEsslinger2018}%
  \BibitemOpen
  \bibfield  {author} {\bibinfo {author} {\bibnamefont {Husmann}, \bibfnamefont
  {D.}}, \bibinfo {author} {\bibfnamefont {M.}~\bibnamefont {Lebrat}}, \bibinfo
  {author} {\bibfnamefont {S.}~\bibnamefont {H{\"a}usler}}, \bibinfo {author}
  {\bibfnamefont {J.-P.}\ \bibnamefont {Brantut}}, \bibinfo {author}
  {\bibfnamefont {L.}~\bibnamefont {Corman}}, \ and\ \bibinfo {author}
  {\bibfnamefont {T.}~\bibnamefont {Esslinger}}} (\bibinfo {year} {2018}),\
  \href {\doibase 10.1073/pnas.1803336115} {\bibfield  {journal} {\bibinfo
  {journal} {Proceedings of the National Academy of Sciences}\ }\textbf
  {\bibinfo {volume} {115}}~(\bibinfo {number} {34}),\ \bibinfo {pages}
  {8563}}\BibitemShut {NoStop}%
\bibitem [{\citenamefont {Husmann}\ \emph {et~al.}(2015)\citenamefont
  {Husmann}, \citenamefont {Uchino}, \citenamefont {Krinner}, \citenamefont
  {Lebrat}, \citenamefont {Giamarchi}, \citenamefont {Esslinger},\ and\
  \citenamefont {Brantut}}]{HusmannBrantut2015}%
  \BibitemOpen
  \bibfield  {author} {\bibinfo {author} {\bibnamefont {Husmann}, \bibfnamefont
  {D.}}, \bibinfo {author} {\bibfnamefont {S.}~\bibnamefont {Uchino}}, \bibinfo
  {author} {\bibfnamefont {S.}~\bibnamefont {Krinner}}, \bibinfo {author}
  {\bibfnamefont {M.}~\bibnamefont {Lebrat}}, \bibinfo {author} {\bibfnamefont
  {T.}~\bibnamefont {Giamarchi}}, \bibinfo {author} {\bibfnamefont
  {T.}~\bibnamefont {Esslinger}}, \ and\ \bibinfo {author} {\bibfnamefont
  {J.-P.}\ \bibnamefont {Brantut}}} (\bibinfo {year} {2015}),\ \href {\doibase
  10.1126/science.aac9584} {\bibfield  {journal} {\bibinfo  {journal}
  {Science}\ }\textbf {\bibinfo {volume} {350}}~(\bibinfo {number} {6267}),\
  \bibinfo {pages} {1498}}\BibitemShut {NoStop}%
\bibitem [{\citenamefont {Hussein}\ and\ \citenamefont
  {Kohler}(2014)}]{hussein2014a}%
  \BibitemOpen
  \bibfield  {author} {\bibinfo {author} {\bibnamefont {Hussein}, \bibfnamefont
  {R.}}, \ and\ \bibinfo {author} {\bibfnamefont {S.}~\bibnamefont {Kohler}}}
  (\bibinfo {year} {2014}),\ \href {\doibase 10.1103/PhysRevB.89.205424}
  {\bibfield  {journal} {\bibinfo  {journal} {Phys. Rev. B}\ }\textbf {\bibinfo
  {volume} {89}},\ \bibinfo {pages} {205424}}\BibitemShut {NoStop}%
\bibitem [{\citenamefont {H{\"u}tzen}\ \emph {et~al.}(2012)\citenamefont
  {H{\"u}tzen}, \citenamefont {Weiss}, \citenamefont {Thorwart},\ and\
  \citenamefont {Egger}}]{HutzenThorwart2012}%
  \BibitemOpen
  \bibfield  {author} {\bibinfo {author} {\bibnamefont {H{\"u}tzen},
  \bibfnamefont {R.}}, \bibinfo {author} {\bibfnamefont {S.}~\bibnamefont
  {Weiss}}, \bibinfo {author} {\bibfnamefont {M.}~\bibnamefont {Thorwart}}, \
  and\ \bibinfo {author} {\bibfnamefont {R.}~\bibnamefont {Egger}}} (\bibinfo
  {year} {2012}),\ \href {\doibase 10.1103/PhysRevB.85.121408} {\bibfield
  {journal} {\bibinfo  {journal} {Phys. Rev. B}\ }\textbf {\bibinfo {volume}
  {85}},\ \bibinfo {pages} {121408}}\BibitemShut {NoStop}%
\bibitem [{\citenamefont {Iadecola}\ and\ \citenamefont
  {\v{Z}nidari\v{c}}(2019)}]{IadecolaZnidaric2019}%
  \BibitemOpen
  \bibfield  {author} {\bibinfo {author} {\bibnamefont {Iadecola},
  \bibfnamefont {T.}}, \ and\ \bibinfo {author} {\bibfnamefont
  {M.}~\bibnamefont {\v{Z}nidari\v{c}}}} (\bibinfo {year} {2019}),\ \href
  {\doibase 10.1103/PhysRevLett.123.036403} {\bibfield  {journal} {\bibinfo
  {journal} {Phys. Rev. Lett.}\ }\textbf {\bibinfo {volume} {123}},\ \bibinfo
  {pages} {036403}}\BibitemShut {NoStop}%
\bibitem [{\citenamefont {Iles-Smith}\ \emph {et~al.}(2016)\citenamefont
  {Iles-Smith}, \citenamefont {Dijkstra}, \citenamefont {Lambert},\ and\
  \citenamefont {Nazir}}]{ilessmith2016a}%
  \BibitemOpen
  \bibfield  {author} {\bibinfo {author} {\bibnamefont {Iles-Smith},
  \bibfnamefont {J.}}, \bibinfo {author} {\bibfnamefont {A.~G.}\ \bibnamefont
  {Dijkstra}}, \bibinfo {author} {\bibfnamefont {N.}~\bibnamefont {Lambert}}, \
  and\ \bibinfo {author} {\bibfnamefont {A.}~\bibnamefont {Nazir}}} (\bibinfo
  {year} {2016}),\ \href {\doibase 10.1063/1.4940218} {\bibfield  {journal}
  {\bibinfo  {journal} {The Journal of Chemical Physics}\ }\textbf {\bibinfo
  {volume} {144}},\ \bibinfo {pages} {044110}}\BibitemShut {NoStop}%
\bibitem [{\citenamefont {Iles-Smith}\ \emph {et~al.}(2014)\citenamefont
  {Iles-Smith}, \citenamefont {Lambert},\ and\ \citenamefont
  {Nazir}}]{IlesSmithNazir2014}%
  \BibitemOpen
  \bibfield  {author} {\bibinfo {author} {\bibnamefont {Iles-Smith},
  \bibfnamefont {J.}}, \bibinfo {author} {\bibfnamefont {N.}~\bibnamefont
  {Lambert}}, \ and\ \bibinfo {author} {\bibfnamefont {A.}~\bibnamefont
  {Nazir}}} (\bibinfo {year} {2014}),\ \href {\doibase
  10.1103/PhysRevA.90.032114} {\bibfield  {journal} {\bibinfo  {journal} {Phys.
  Rev. A}\ }\textbf {\bibinfo {volume} {90}},\ \bibinfo {pages}
  {032114}}\BibitemShut {NoStop}%
\bibitem [{\citenamefont {Ilievski}\ and\ \citenamefont
  {De~Nardis}(2017)}]{IlievskiDeNardis2017}%
  \BibitemOpen
  \bibfield  {author} {\bibinfo {author} {\bibnamefont {Ilievski},
  \bibfnamefont {E.}}, \ and\ \bibinfo {author} {\bibfnamefont
  {J.}~\bibnamefont {De~Nardis}}} (\bibinfo {year} {2017}),\ \href {\doibase
  10.1103/PhysRevB.96.081118} {\bibfield  {journal} {\bibinfo  {journal} {Phys.
  Rev. B}\ }\textbf {\bibinfo {volume} {96}},\ \bibinfo {pages}
  {081118}}\BibitemShut {NoStop}%
\bibitem [{\citenamefont {Imamo\u{g}lu}(1994)}]{Imamoglu1996}%
  \BibitemOpen
  \bibfield  {author} {\bibinfo {author} {\bibnamefont {Imamo\u{g}lu},
  \bibfnamefont {A.}}} (\bibinfo {year} {1994}),\ \href {\doibase
  10.1103/PhysRevA.50.3650} {\bibfield  {journal} {\bibinfo  {journal} {Phys.
  Rev. A}\ }\textbf {\bibinfo {volume} {50}},\ \bibinfo {pages}
  {3650}}\BibitemShut {NoStop}%
\bibitem [{\citenamefont {Iorio}\ \emph {et~al.}(2021)\citenamefont {Iorio},
  \citenamefont {Strambini}, \citenamefont {Haack}, \citenamefont {Campisi},\
  and\ \citenamefont {Giazotto}}]{IorioGiazotto2021}%
  \BibitemOpen
  \bibfield  {author} {\bibinfo {author} {\bibnamefont {Iorio}, \bibfnamefont
  {A.}}, \bibinfo {author} {\bibfnamefont {E.}~\bibnamefont {Strambini}},
  \bibinfo {author} {\bibfnamefont {G.}~\bibnamefont {Haack}}, \bibinfo
  {author} {\bibfnamefont {M.}~\bibnamefont {Campisi}}, \ and\ \bibinfo
  {author} {\bibfnamefont {F.}~\bibnamefont {Giazotto}}} (\bibinfo {year}
  {2021}),\ \href {\doibase 10.1103/PhysRevApplied.15.054050} {\bibfield
  {journal} {\bibinfo  {journal} {Phys. Rev. Applied}\ }\textbf {\bibinfo
  {volume} {15}},\ \bibinfo {pages} {054050}}\BibitemShut {NoStop}%
\bibitem [{\citenamefont {Iyer}\ \emph {et~al.}(2013)\citenamefont {Iyer},
  \citenamefont {Oganesyan}, \citenamefont {Refael},\ and\ \citenamefont
  {Huse}}]{IyerHuse2013}%
  \BibitemOpen
  \bibfield  {author} {\bibinfo {author} {\bibnamefont {Iyer}, \bibfnamefont
  {S.}}, \bibinfo {author} {\bibfnamefont {V.}~\bibnamefont {Oganesyan}},
  \bibinfo {author} {\bibfnamefont {G.}~\bibnamefont {Refael}}, \ and\ \bibinfo
  {author} {\bibfnamefont {D.~A.}\ \bibnamefont {Huse}}} (\bibinfo {year}
  {2013}),\ \href {\doibase 10.1103/PhysRevB.87.134202} {\bibfield  {journal}
  {\bibinfo  {journal} {Phys. Rev. B}\ }\textbf {\bibinfo {volume} {87}},\
  \bibinfo {pages} {134202}}\BibitemShut {NoStop}%
\bibitem [{\citenamefont {Jacobs}\ and\ \citenamefont
  {Steck}(2006)}]{Jacobs2006}%
  \BibitemOpen
  \bibfield  {author} {\bibinfo {author} {\bibnamefont {Jacobs}, \bibfnamefont
  {K.}}, \ and\ \bibinfo {author} {\bibfnamefont {D.~A.}\ \bibnamefont
  {Steck}}} (\bibinfo {year} {2006}),\ \href {\doibase
  10.1080/00107510601101934} {\bibfield  {journal} {\bibinfo  {journal}
  {Contemporary Physics}\ }\textbf {\bibinfo {volume} {47}}~(\bibinfo {number}
  {5}),\ \bibinfo {pages} {279–303}}\BibitemShut {NoStop}%
\bibitem [{\citenamefont {Jaksch}\ \emph {et~al.}(1998)\citenamefont {Jaksch},
  \citenamefont {Bruder}, \citenamefont {Cirac}, \citenamefont {Gardiner},\
  and\ \citenamefont {Zoller}}]{JakschZoller1998}%
  \BibitemOpen
  \bibfield  {author} {\bibinfo {author} {\bibnamefont {Jaksch}, \bibfnamefont
  {D.}}, \bibinfo {author} {\bibfnamefont {C.}~\bibnamefont {Bruder}}, \bibinfo
  {author} {\bibfnamefont {J.~I.}\ \bibnamefont {Cirac}}, \bibinfo {author}
  {\bibfnamefont {C.~W.}\ \bibnamefont {Gardiner}}, \ and\ \bibinfo {author}
  {\bibfnamefont {P.}~\bibnamefont {Zoller}}} (\bibinfo {year} {1998}),\ \href
  {\doibase 10.1103/PhysRevLett.81.3108} {\bibfield  {journal} {\bibinfo
  {journal} {Phys. Rev. Lett.}\ }\textbf {\bibinfo {volume} {81}},\ \bibinfo
  {pages} {3108}}\BibitemShut {NoStop}%
\bibitem [{\citenamefont {Jamiolkowski}(1972)}]{Jamiolkowski}%
  \BibitemOpen
  \bibfield  {author} {\bibinfo {author} {\bibnamefont {Jamiolkowski},
  \bibfnamefont {A.}}} (\bibinfo {year} {1972}),\ \href {\doibase
  https://doi.org/10.1016/0034-4877(72)90011-0} {\bibfield  {journal} {\bibinfo
   {journal} {Reports on Mathematical Physics}\ }\textbf {\bibinfo {volume}
  {3}}~(\bibinfo {number} {4}),\ \bibinfo {pages} {275 }}\BibitemShut {NoStop}%
\bibitem [{\citenamefont {Jang}\ \emph {et~al.}(2002)\citenamefont {Jang},
  \citenamefont {Cao},\ and\ \citenamefont {Silbey}}]{jang2002a}%
  \BibitemOpen
  \bibfield  {author} {\bibinfo {author} {\bibnamefont {Jang}, \bibfnamefont
  {S.}}, \bibinfo {author} {\bibfnamefont {J.}~\bibnamefont {Cao}}, \ and\
  \bibinfo {author} {\bibfnamefont {R.~J.}\ \bibnamefont {Silbey}}} (\bibinfo
  {year} {2002}),\ \href {\doibase 10.1063/1.1445105} {\bibfield  {journal}
  {\bibinfo  {journal} {The Journal of Chemical Physics}\ }\textbf {\bibinfo
  {volume} {116}},\ \bibinfo {pages} {2705}}\BibitemShut {NoStop}%
\bibitem [{\citenamefont {Jaschke}\ \emph {et~al.}(2019)\citenamefont
  {Jaschke}, \citenamefont {Carr},\ and\ \citenamefont
  {de~Vega}}]{JaschkeDeVega2019}%
  \BibitemOpen
  \bibfield  {author} {\bibinfo {author} {\bibnamefont {Jaschke}, \bibfnamefont
  {D.}}, \bibinfo {author} {\bibfnamefont {L.~D.}\ \bibnamefont {Carr}}, \ and\
  \bibinfo {author} {\bibfnamefont {I.}~\bibnamefont {de~Vega}}} (\bibinfo
  {year} {2019}),\ \href {\doibase 10.1088/2058-9565/ab1a71} {\bibfield
  {journal} {\bibinfo  {journal} {Quantum Science and Technology}\ }\textbf
  {\bibinfo {volume} {4}}~(\bibinfo {number} {3}),\ \bibinfo {pages}
  {034002}}\BibitemShut {NoStop}%
\bibitem [{\citenamefont {Jaschke}\ \emph {et~al.}(2018)\citenamefont
  {Jaschke}, \citenamefont {Wall},\ and\ \citenamefont
  {Carr}}]{JaschkeCarr2018}%
  \BibitemOpen
  \bibfield  {author} {\bibinfo {author} {\bibnamefont {Jaschke}, \bibfnamefont
  {D.}}, \bibinfo {author} {\bibfnamefont {M.~L.}\ \bibnamefont {Wall}}, \ and\
  \bibinfo {author} {\bibfnamefont {L.~D.}\ \bibnamefont {Carr}}} (\bibinfo
  {year} {2018}),\ \href {\doibase https://doi.org/10.1016/j.cpc.2017.12.015}
  {\bibfield  {journal} {\bibinfo  {journal} {Computer Physics Communications}\
  }\textbf {\bibinfo {volume} {225}},\ \bibinfo {pages} {59}}\BibitemShut
  {NoStop}%
\bibitem [{\citenamefont {Jepsen}\ \emph {et~al.}(2020)\citenamefont {Jepsen},
  \citenamefont {Amato-Grill}, \citenamefont {Dimitrova}, \citenamefont {Ho},
  \citenamefont {Demler},\ and\ \citenamefont {Ketterle}}]{JepsenKetterle2020}%
  \BibitemOpen
  \bibfield  {author} {\bibinfo {author} {\bibnamefont {Jepsen}, \bibfnamefont
  {P.~N.}}, \bibinfo {author} {\bibfnamefont {J.}~\bibnamefont {Amato-Grill}},
  \bibinfo {author} {\bibfnamefont {I.}~\bibnamefont {Dimitrova}}, \bibinfo
  {author} {\bibfnamefont {W.~W.}\ \bibnamefont {Ho}}, \bibinfo {author}
  {\bibfnamefont {E.}~\bibnamefont {Demler}}, \ and\ \bibinfo {author}
  {\bibfnamefont {W.}~\bibnamefont {Ketterle}}} (\bibinfo {year} {2020}),\
  \href {\doibase 10.1038/s41586-020-3033-y} {\bibfield  {journal} {\bibinfo
  {journal} {Nature}\ }\textbf {\bibinfo {volume} {588}},\ \bibinfo {pages}
  {403}}\BibitemShut {NoStop}%
\bibitem [{\citenamefont {Jesenko}\ and\ \citenamefont
  {\v{Z}nidari\v{c}}(2011)}]{JesenkoZnidaric2011}%
  \BibitemOpen
  \bibfield  {author} {\bibinfo {author} {\bibnamefont {Jesenko}, \bibfnamefont
  {S.}}, \ and\ \bibinfo {author} {\bibfnamefont {M.}~\bibnamefont
  {\v{Z}nidari\v{c}}}} (\bibinfo {year} {2011}),\ \href {\doibase
  10.1103/PhysRevB.84.174438} {\bibfield  {journal} {\bibinfo  {journal} {Phys.
  Rev. B}\ }\textbf {\bibinfo {volume} {84}},\ \bibinfo {pages}
  {174438}}\BibitemShut {NoStop}%
\bibitem [{\citenamefont {Jin}\ \emph {et~al.}(2016)\citenamefont {Jin},
  \citenamefont {Biella}, \citenamefont {Viyuela}, \citenamefont {Mazza},
  \citenamefont {Keeling}, \citenamefont {Fazio},\ and\ \citenamefont
  {Rossini}}]{JinRossini2016}%
  \BibitemOpen
  \bibfield  {author} {\bibinfo {author} {\bibnamefont {Jin}, \bibfnamefont
  {J.}}, \bibinfo {author} {\bibfnamefont {A.}~\bibnamefont {Biella}}, \bibinfo
  {author} {\bibfnamefont {O.}~\bibnamefont {Viyuela}}, \bibinfo {author}
  {\bibfnamefont {L.}~\bibnamefont {Mazza}}, \bibinfo {author} {\bibfnamefont
  {J.}~\bibnamefont {Keeling}}, \bibinfo {author} {\bibfnamefont
  {R.}~\bibnamefont {Fazio}}, \ and\ \bibinfo {author} {\bibfnamefont
  {D.}~\bibnamefont {Rossini}}} (\bibinfo {year} {2016}),\ \href {\doibase
  10.1103/PhysRevX.6.031011} {\bibfield  {journal} {\bibinfo  {journal} {Phys.
  Rev. X}\ }\textbf {\bibinfo {volume} {6}},\ \bibinfo {pages}
  {031011}}\BibitemShut {NoStop}%
\bibitem [{\citenamefont {Jin}\ \emph {et~al.}(2010)\citenamefont {Jin},
  \citenamefont {Tu}, \citenamefont {Zhang},\ and\ \citenamefont
  {Yan}}]{jin2010a}%
  \BibitemOpen
  \bibfield  {author} {\bibinfo {author} {\bibnamefont {Jin}, \bibfnamefont
  {J.}}, \bibinfo {author} {\bibfnamefont {M.~W.-Y.}\ \bibnamefont {Tu}},
  \bibinfo {author} {\bibfnamefont {W.-M.}\ \bibnamefont {Zhang}}, \ and\
  \bibinfo {author} {\bibfnamefont {Y.}~\bibnamefont {Yan}}} (\bibinfo {year}
  {2010}),\ \href {\doibase 10.1088/1367-2630/12/8/083013} {\bibfield
  {journal} {\bibinfo  {journal} {New Journal of Physics}\ }\textbf {\bibinfo
  {volume} {12}},\ \bibinfo {pages} {083013}}\BibitemShut {NoStop}%
\bibitem [{\citenamefont {Jin}\ \emph {et~al.}(2020)\citenamefont {Jin},
  \citenamefont {Filippone},\ and\ \citenamefont {Giamarchi}}]{jin2020a}%
  \BibitemOpen
  \bibfield  {author} {\bibinfo {author} {\bibnamefont {Jin}, \bibfnamefont
  {T.}}, \bibinfo {author} {\bibfnamefont {M.}~\bibnamefont {Filippone}}, \
  and\ \bibinfo {author} {\bibfnamefont {T.}~\bibnamefont {Giamarchi}}}
  (\bibinfo {year} {2020}),\ \href {\doibase 10.1103/PhysRevB.102.205131}
  {\bibfield  {journal} {\bibinfo  {journal} {Phys. Rev. B}\ }\textbf {\bibinfo
  {volume} {102}},\ \bibinfo {pages} {205131}}\BibitemShut {NoStop}%
\bibitem [{\citenamefont {Joachim}\ \emph {et~al.}(2000)\citenamefont
  {Joachim}, \citenamefont {Gimzewski},\ and\ \citenamefont
  {Aviram}}]{joachim2000a}%
  \BibitemOpen
  \bibfield  {author} {\bibinfo {author} {\bibnamefont {Joachim}, \bibfnamefont
  {C.}}, \bibinfo {author} {\bibfnamefont {J.}~\bibnamefont {Gimzewski}}, \
  and\ \bibinfo {author} {\bibfnamefont {A.}~\bibnamefont {Aviram}}} (\bibinfo
  {year} {2000}),\ \href {\doibase 10.1038/35046000} {\bibfield  {journal}
  {\bibinfo  {journal} {Nature}\ }\textbf {\bibinfo {volume} {408}},\ \bibinfo
  {pages} {541–548}}\BibitemShut {NoStop}%
\bibitem [{\citenamefont {Jordan}\ and\ \citenamefont
  {Wigner}(1928)}]{JordanWigner1928}%
  \BibitemOpen
  \bibfield  {author} {\bibinfo {author} {\bibnamefont {Jordan}, \bibfnamefont
  {P.}}, \ and\ \bibinfo {author} {\bibfnamefont {E.}~\bibnamefont {Wigner}}}
  (\bibinfo {year} {1928}),\ \href {\doibase 10.1007/BF01331938} {\bibfield
  {journal} {\bibinfo  {journal} {Z. Phys.}\ }\textbf {\bibinfo {volume}
  {47}},\ \bibinfo {pages} {631}}\BibitemShut {NoStop}%
\bibitem [{\citenamefont {Jussiau}\ \emph {et~al.}(2019)\citenamefont
  {Jussiau}, \citenamefont {Hasegawa},\ and\ \citenamefont
  {Whitney}}]{jussiau2019a}%
  \BibitemOpen
  \bibfield  {author} {\bibinfo {author} {\bibnamefont {Jussiau}, \bibfnamefont
  {E.}}, \bibinfo {author} {\bibfnamefont {M.}~\bibnamefont {Hasegawa}}, \ and\
  \bibinfo {author} {\bibfnamefont {R.~S.}\ \bibnamefont {Whitney}}} (\bibinfo
  {year} {2019}),\ \href {\doibase 10.1103/PhysRevB.100.115411} {\bibfield
  {journal} {\bibinfo  {journal} {Phys. Rev. B}\ }\textbf {\bibinfo {volume}
  {100}},\ \bibinfo {pages} {115411}}\BibitemShut {NoStop}%
\bibitem [{\citenamefont {Kaestner}\ and\ \citenamefont
  {Kashcheyevs}(2015)}]{KaestnerKashcheyevs2015}%
  \BibitemOpen
  \bibfield  {author} {\bibinfo {author} {\bibnamefont {Kaestner},
  \bibfnamefont {B.}}, \ and\ \bibinfo {author} {\bibfnamefont
  {V.}~\bibnamefont {Kashcheyevs}}} (\bibinfo {year} {2015}),\ \href {\doibase
  10.1088/0034-4885/78/10/103901} {\bibfield  {journal} {\bibinfo  {journal}
  {Reports on Progress in Physics}\ }\textbf {\bibinfo {volume} {78}}~(\bibinfo
  {number} {10}),\ \bibinfo {pages} {103901}}\BibitemShut {NoStop}%
\bibitem [{\citenamefont {Kalantar}\ \emph {et~al.}(2021)\citenamefont
  {Kalantar}, \citenamefont {Agarwalla},\ and\ \citenamefont
  {Segal}}]{Bijay2021}%
  \BibitemOpen
  \bibfield  {author} {\bibinfo {author} {\bibnamefont {Kalantar},
  \bibfnamefont {N.}}, \bibinfo {author} {\bibfnamefont {B.~K.}\ \bibnamefont
  {Agarwalla}}, \ and\ \bibinfo {author} {\bibfnamefont {D.}~\bibnamefont
  {Segal}}} (\bibinfo {year} {2021}),\ \href {\doibase
  10.1103/PhysRevE.103.052130} {\bibfield  {journal} {\bibinfo  {journal}
  {Phys. Rev. E}\ }\textbf {\bibinfo {volume} {103}},\ \bibinfo {pages}
  {052130}}\BibitemShut {NoStop}%
\bibitem [{\citenamefont {Kalugin}\ \emph {et~al.}(1986)\citenamefont
  {Kalugin}, \citenamefont {Kitaev},\ and\ \citenamefont
  {Levitov}}]{KaluginLevitov1986}%
  \BibitemOpen
  \bibfield  {author} {\bibinfo {author} {\bibnamefont {Kalugin}, \bibfnamefont
  {P.~A.}}, \bibinfo {author} {\bibfnamefont {A.~Y.}\ \bibnamefont {Kitaev}}, \
  and\ \bibinfo {author} {\bibfnamefont {L.}~\bibnamefont {Levitov}}} (\bibinfo
  {year} {1986}),\ \href
  {http://www.jetp.ras.ru/cgi-bin/e/index/e/64/2/p410?a=list} {\bibfield
  {journal} {\bibinfo  {journal} {Sov. Phys. JETP}\ }\textbf {\bibinfo {volume}
  {64}},\ \bibinfo {pages} {410}}\BibitemShut {NoStop}%
\bibitem [{\citenamefont {Kamiya}(2015)}]{Kamiya2015}%
  \BibitemOpen
  \bibfield  {author} {\bibinfo {author} {\bibnamefont {Kamiya}, \bibfnamefont
  {N.}}} (\bibinfo {year} {2015}),\ \href {\doibase 10.1093/ptep/ptv034}
  {\bibfield  {journal} {\bibinfo  {journal} {Progress of Theoretical and
  Experimental Physics}\ }\textbf {\bibinfo {volume} {2015}}~(\bibinfo {number}
  {4}),\ \bibinfo {pages} {043A02}}\BibitemShut {NoStop}%
\bibitem [{\citenamefont {Kao}\ \emph {et~al.}(2015)\citenamefont {Kao},
  \citenamefont {Hsieh},\ and\ \citenamefont {Chen}}]{KaoChen2015}%
  \BibitemOpen
  \bibfield  {author} {\bibinfo {author} {\bibnamefont {Kao}, \bibfnamefont
  {Y.-J.}}, \bibinfo {author} {\bibfnamefont {Y.-D.}\ \bibnamefont {Hsieh}}, \
  and\ \bibinfo {author} {\bibfnamefont {P.}~\bibnamefont {Chen}}} (\bibinfo
  {year} {2015}),\ \href {\doibase 10.1088/1742-6596/640/1/012040} {\bibfield
  {journal} {\bibinfo  {journal} {Journal of Physics: Conference Series}\
  }\textbf {\bibinfo {volume} {640}},\ \bibinfo {pages} {012040}}\BibitemShut
  {NoStop}%
\bibitem [{\citenamefont {Kardar}(1986)}]{Kardar1986}%
  \BibitemOpen
  \bibfield  {author} {\bibinfo {author} {\bibnamefont {Kardar}, \bibfnamefont
  {M.}}} (\bibinfo {year} {1986}),\ \href {\doibase 10.1103/PhysRevB.33.3125}
  {\bibfield  {journal} {\bibinfo  {journal} {Phys. Rev. B}\ }\textbf {\bibinfo
  {volume} {33}},\ \bibinfo {pages} {3125}}\BibitemShut {NoStop}%
\bibitem [{\citenamefont {Karevski}\ and\ \citenamefont
  {Platini}(2009)}]{Karevski2009}%
  \BibitemOpen
  \bibfield  {author} {\bibinfo {author} {\bibnamefont {Karevski},
  \bibfnamefont {D.}}, \ and\ \bibinfo {author} {\bibfnamefont
  {T.}~\bibnamefont {Platini}}} (\bibinfo {year} {2009}),\ \href {\doibase
  10.1103/PhysRevLett.102.207207} {\bibfield  {journal} {\bibinfo  {journal}
  {Physical Review Letters}\ }\textbf {\bibinfo {volume} {102}}~(\bibinfo
  {number} {20}),\ \bibinfo {pages} {207207}}\BibitemShut {NoStop}%
\bibitem [{\citenamefont {Karevski}\ \emph {et~al.}(2013)\citenamefont
  {Karevski}, \citenamefont {Popkov},\ and\ \citenamefont
  {Sch{\"{u}}tz}}]{Karevski2013}%
  \BibitemOpen
  \bibfield  {author} {\bibinfo {author} {\bibnamefont {Karevski},
  \bibfnamefont {D.}}, \bibinfo {author} {\bibfnamefont {V.}~\bibnamefont
  {Popkov}}, \ and\ \bibinfo {author} {\bibfnamefont {G.~M.}\ \bibnamefont
  {Sch{\"{u}}tz}}} (\bibinfo {year} {2013}),\ \href {\doibase
  10.1103/PhysRevLett.110.047201} {\bibfield  {journal} {\bibinfo  {journal}
  {Physical Review Letters}\ }\textbf {\bibinfo {volume} {110}}~(\bibinfo
  {number} {4}),\ \bibinfo {pages} {047201}}\BibitemShut {NoStop}%
\bibitem [{\citenamefont {Karrasch}(2017)}]{Karrasch2017}%
  \BibitemOpen
  \bibfield  {author} {\bibinfo {author} {\bibnamefont {Karrasch},
  \bibfnamefont {C.}}} (\bibinfo {year} {2017}),\ \href {\doibase
  10.1088/1367-2630/aa631a} {\bibfield  {journal} {\bibinfo  {journal} {New
  Journal of Physics}\ }\textbf {\bibinfo {volume} {19}}~(\bibinfo {number}
  {3}),\ \bibinfo {pages} {033027}}\BibitemShut {NoStop}%
\bibitem [{\citenamefont {Karrasch}\ \emph {et~al.}(2012)\citenamefont
  {Karrasch}, \citenamefont {Bardarson},\ and\ \citenamefont
  {Moore}}]{KarraschMoore2012}%
  \BibitemOpen
  \bibfield  {author} {\bibinfo {author} {\bibnamefont {Karrasch},
  \bibfnamefont {C.}}, \bibinfo {author} {\bibfnamefont {J.~H.}\ \bibnamefont
  {Bardarson}}, \ and\ \bibinfo {author} {\bibfnamefont {J.~E.}\ \bibnamefont
  {Moore}}} (\bibinfo {year} {2012}),\ \href {\doibase
  10.1103/PhysRevLett.108.227206} {\bibfield  {journal} {\bibinfo  {journal}
  {Phys. Rev. Lett.}\ }\textbf {\bibinfo {volume} {108}},\ \bibinfo {pages}
  {227206}}\BibitemShut {NoStop}%
\bibitem [{\citenamefont {Karrasch}\ \emph {et~al.}(2013)\citenamefont
  {Karrasch}, \citenamefont {Ilan},\ and\ \citenamefont
  {Moore}}]{KarraschMoore2013}%
  \BibitemOpen
  \bibfield  {author} {\bibinfo {author} {\bibnamefont {Karrasch},
  \bibfnamefont {C.}}, \bibinfo {author} {\bibfnamefont {R.}~\bibnamefont
  {Ilan}}, \ and\ \bibinfo {author} {\bibfnamefont {J.~E.}\ \bibnamefont
  {Moore}}} (\bibinfo {year} {2013}),\ \href {\doibase
  10.1103/PhysRevB.88.195129} {\bibfield  {journal} {\bibinfo  {journal} {Phys.
  Rev. B}\ }\textbf {\bibinfo {volume} {88}},\ \bibinfo {pages}
  {195129}}\BibitemShut {NoStop}%
\bibitem [{\citenamefont {Karrasch}\ \emph {et~al.}(2016)\citenamefont
  {Karrasch}, \citenamefont {Kennes},\ and\ \citenamefont
  {Heidrich-Meisner}}]{KarraschHeidrichMeisner2016}%
  \BibitemOpen
  \bibfield  {author} {\bibinfo {author} {\bibnamefont {Karrasch},
  \bibfnamefont {C.}}, \bibinfo {author} {\bibfnamefont {D.~M.}\ \bibnamefont
  {Kennes}}, \ and\ \bibinfo {author} {\bibfnamefont {F.}~\bibnamefont
  {Heidrich-Meisner}}} (\bibinfo {year} {2016}),\ \href {\doibase
  10.1103/PhysRevLett.117.116401} {\bibfield  {journal} {\bibinfo  {journal}
  {Phys. Rev. Lett.}\ }\textbf {\bibinfo {volume} {117}},\ \bibinfo {pages}
  {116401}}\BibitemShut {NoStop}%
\bibitem [{\citenamefont {Karrlein}\ and\ \citenamefont
  {Grabert}(1997)}]{karrlein1997a}%
  \BibitemOpen
  \bibfield  {author} {\bibinfo {author} {\bibnamefont {Karrlein},
  \bibfnamefont {R.}}, \ and\ \bibinfo {author} {\bibfnamefont
  {H.}~\bibnamefont {Grabert}}} (\bibinfo {year} {1997}),\ \href {\doibase
  10.1103/PhysRevE.55.153} {\bibfield  {journal} {\bibinfo  {journal} {Phys.
  Rev. E}\ }\textbf {\bibinfo {volume} {55}},\ \bibinfo {pages}
  {153}}\BibitemShut {NoStop}%
\bibitem [{\citenamefont {Kast}\ and\ \citenamefont
  {Ankerhold}(2013)}]{kast2013a}%
  \BibitemOpen
  \bibfield  {author} {\bibinfo {author} {\bibnamefont {Kast}, \bibfnamefont
  {D.}}, \ and\ \bibinfo {author} {\bibfnamefont {J.}~\bibnamefont
  {Ankerhold}}} (\bibinfo {year} {2013}),\ \href {\doibase
  10.1103/PhysRevB.87.134301} {\bibfield  {journal} {\bibinfo  {journal} {Phys.
  Rev. B}\ }\textbf {\bibinfo {volume} {87}},\ \bibinfo {pages}
  {134301}}\BibitemShut {NoStop}%
\bibitem [{\citenamefont {Kastner}(1992)}]{Kastner1992}%
  \BibitemOpen
  \bibfield  {author} {\bibinfo {author} {\bibnamefont {Kastner}, \bibfnamefont
  {M.~A.}}} (\bibinfo {year} {1992}),\ \href {\doibase
  10.1103/RevModPhys.64.849} {\bibfield  {journal} {\bibinfo  {journal} {Rev.
  Mod. Phys.}\ }\textbf {\bibinfo {volume} {64}},\ \bibinfo {pages}
  {849}}\BibitemShut {NoStop}%
\bibitem [{\citenamefont {Katz}\ and\ \citenamefont
  {Kosloff}(2016)}]{GatzKosloff2016}%
  \BibitemOpen
  \bibfield  {author} {\bibinfo {author} {\bibnamefont {Katz}, \bibfnamefont
  {G.}}, \ and\ \bibinfo {author} {\bibfnamefont {R.}~\bibnamefont {Kosloff}}}
  (\bibinfo {year} {2016}),\ \href {\doibase 10.3390/e18050186} {\bibfield
  {journal} {\bibinfo  {journal} {Entropy}\ }\textbf {\bibinfo {volume} {18}},\
  \bibinfo {pages} {186}}\BibitemShut {NoStop}%
\bibitem [{\citenamefont {Kessler}\ \emph {et~al.}(2012)\citenamefont
  {Kessler}, \citenamefont {Giedke}, \citenamefont {Imamo\u{g}lu},
  \citenamefont {Yelin}, \citenamefont {Lukin},\ and\ \citenamefont
  {Cirac}}]{Kessler2012}%
  \BibitemOpen
  \bibfield  {author} {\bibinfo {author} {\bibnamefont {Kessler}, \bibfnamefont
  {E.~M.}}, \bibinfo {author} {\bibfnamefont {G.}~\bibnamefont {Giedke}},
  \bibinfo {author} {\bibfnamefont {A.}~\bibnamefont {Imamo\u{g}lu}}, \bibinfo
  {author} {\bibfnamefont {S.~F.}\ \bibnamefont {Yelin}}, \bibinfo {author}
  {\bibfnamefont {M.~D.}\ \bibnamefont {Lukin}}, \ and\ \bibinfo {author}
  {\bibfnamefont {J.~I.}\ \bibnamefont {Cirac}}} (\bibinfo {year} {2012}),\
  \href {\doibase 10.1103/PhysRevA.86.012116} {\bibfield  {journal} {\bibinfo
  {journal} {Phys. Rev. A}\ }\textbf {\bibinfo {volume} {86}},\ \bibinfo
  {pages} {012116}}\BibitemShut {NoStop}%
\bibitem [{\citenamefont {Khemani}\ \emph {et~al.}(2016)\citenamefont
  {Khemani}, \citenamefont {Pollmann},\ and\ \citenamefont
  {Sondhi}}]{KhemaniSondhi2016}%
  \BibitemOpen
  \bibfield  {author} {\bibinfo {author} {\bibnamefont {Khemani}, \bibfnamefont
  {V.}}, \bibinfo {author} {\bibfnamefont {F.}~\bibnamefont {Pollmann}}, \ and\
  \bibinfo {author} {\bibfnamefont {S.~L.}\ \bibnamefont {Sondhi}}} (\bibinfo
  {year} {2016}),\ \href {\doibase 10.1103/PhysRevLett.116.247204} {\bibfield
  {journal} {\bibinfo  {journal} {Phys. Rev. Lett.}\ }\textbf {\bibinfo
  {volume} {116}},\ \bibinfo {pages} {247204}}\BibitemShut {NoStop}%
\bibitem [{\citenamefont {Kilgour}\ \emph {et~al.}(2019)\citenamefont
  {Kilgour}, \citenamefont {Agarwalla},\ and\ \citenamefont
  {Segal}}]{kilgour2019a}%
  \BibitemOpen
  \bibfield  {author} {\bibinfo {author} {\bibnamefont {Kilgour}, \bibfnamefont
  {M.}}, \bibinfo {author} {\bibfnamefont {B.~K.}\ \bibnamefont {Agarwalla}}, \
  and\ \bibinfo {author} {\bibfnamefont {D.}~\bibnamefont {Segal}}} (\bibinfo
  {year} {2019}),\ \href {\doibase 10.1063/1.5084949} {\bibfield  {journal}
  {\bibinfo  {journal} {The Journal of Chemical Physics}\ }\textbf {\bibinfo
  {volume} {150}}~(\bibinfo {number} {8}),\ \bibinfo {pages}
  {084111}}\BibitemShut {NoStop}%
\bibitem [{\citenamefont {Kilgour}\ and\ \citenamefont
  {Segal}(2015)}]{Kilgour2015}%
  \BibitemOpen
  \bibfield  {author} {\bibinfo {author} {\bibnamefont {Kilgour}, \bibfnamefont
  {M.}}, \ and\ \bibinfo {author} {\bibfnamefont {D.}~\bibnamefont {Segal}}}
  (\bibinfo {year} {2015}),\ \href {\doibase 10.1063/1.4926395} {\bibfield
  {journal} {\bibinfo  {journal} {Journal of Chemical Physics}\ }\textbf
  {\bibinfo {volume} {143}},\ \bibinfo {pages} {024111}},\ \Eprint
  {http://arxiv.org/abs/1505.00645} {1505.00645} \BibitemShut {NoStop}%
\bibitem [{\citenamefont {Kilgour}\ and\ \citenamefont
  {Segal}(2016)}]{Kilgour2016}%
  \BibitemOpen
  \bibfield  {author} {\bibinfo {author} {\bibnamefont {Kilgour}, \bibfnamefont
  {M.}}, \ and\ \bibinfo {author} {\bibfnamefont {D.}~\bibnamefont {Segal}}}
  (\bibinfo {year} {2016}),\ \href {\doibase 10.1063/1.4944470} {\bibfield
  {journal} {\bibinfo  {journal} {The Journal of Chemical Physics}\ }\textbf
  {\bibinfo {volume} {144}}~(\bibinfo {number} {12}),\ \bibinfo {pages}
  {124107}}\BibitemShut {NoStop}%
\bibitem [{\citenamefont {Kim}\ \emph {et~al.}(2015)\citenamefont {Kim},
  \citenamefont {Prada}, \citenamefont {Qin}, \citenamefont {Kim},\ and\
  \citenamefont {Blick}}]{kim2015a}%
  \BibitemOpen
  \bibfield  {author} {\bibinfo {author} {\bibnamefont {Kim}, \bibfnamefont
  {C.}}, \bibinfo {author} {\bibfnamefont {M.}~\bibnamefont {Prada}}, \bibinfo
  {author} {\bibfnamefont {H.}~\bibnamefont {Qin}}, \bibinfo {author}
  {\bibfnamefont {H.-S.}\ \bibnamefont {Kim}}, \ and\ \bibinfo {author}
  {\bibfnamefont {R.~H.}\ \bibnamefont {Blick}}} (\bibinfo {year} {2015}),\
  \href {\doibase 10.1063/1.4908151} {\bibfield  {journal} {\bibinfo  {journal}
  {Applied Physics Letters}\ }\textbf {\bibinfo {volume} {106}}~(\bibinfo
  {number} {6}),\ \bibinfo {pages} {061909}}\BibitemShut {NoStop}%
\bibitem [{\citenamefont {Kim}\ and\ \citenamefont
  {Hershfield}(2002)}]{KimHershfield2002}%
  \BibitemOpen
  \bibfield  {author} {\bibinfo {author} {\bibnamefont {Kim}, \bibfnamefont
  {T.-S.}}, \ and\ \bibinfo {author} {\bibfnamefont {S.}~\bibnamefont
  {Hershfield}}} (\bibinfo {year} {2002}),\ \href {\doibase
  10.1103/PhysRevLett.88.136601} {\bibfield  {journal} {\bibinfo  {journal}
  {Phys. Rev. Lett.}\ }\textbf {\bibinfo {volume} {88}},\ \bibinfo {pages}
  {136601}}\BibitemShut {NoStop}%
\bibitem [{\citenamefont {Kimura}\ \emph {et~al.}(2017)\citenamefont {Kimura},
  \citenamefont {Ajisaka},\ and\ \citenamefont {Watanabe}}]{Kimura2017}%
  \BibitemOpen
  \bibfield  {author} {\bibinfo {author} {\bibnamefont {Kimura}, \bibfnamefont
  {G.}}, \bibinfo {author} {\bibfnamefont {S.}~\bibnamefont {Ajisaka}}, \ and\
  \bibinfo {author} {\bibfnamefont {K.}~\bibnamefont {Watanabe}}} (\bibinfo
  {year} {2017}),\ \href {\doibase 10.1142/S1230161217400091} {\bibfield
  {journal} {\bibinfo  {journal} {Open Systems \& Information Dynamics}\
  }\textbf {\bibinfo {volume} {24}}~(\bibinfo {number} {04}),\ \bibinfo {pages}
  {1740009}}\BibitemShut {NoStop}%
\bibitem [{\citenamefont {Kir\v{s}anskas}\ \emph {et~al.}(2018)\citenamefont
  {Kir\v{s}anskas}, \citenamefont {Francki\'e},\ and\ \citenamefont
  {Wacker}}]{kirsanskas2018a}%
  \BibitemOpen
  \bibfield  {author} {\bibinfo {author} {\bibnamefont {Kir\v{s}anskas},
  \bibfnamefont {G.}}, \bibinfo {author} {\bibfnamefont {M.}~\bibnamefont
  {Francki\'e}}, \ and\ \bibinfo {author} {\bibfnamefont {A.}~\bibnamefont
  {Wacker}}} (\bibinfo {year} {2018}),\ \href {\doibase
  10.1103/PhysRevB.97.035432} {\bibfield  {journal} {\bibinfo  {journal} {Phys.
  Rev. B}\ }\textbf {\bibinfo {volume} {97}},\ \bibinfo {pages}
  {035432}}\BibitemShut {NoStop}%
\bibitem [{\citenamefont {Kleinekath{\"o}fer}(2004)}]{kleinekathoefer2004a}%
  \BibitemOpen
  \bibfield  {author} {\bibinfo {author} {\bibnamefont {Kleinekath{\"o}fer},
  \bibfnamefont {U.}}} (\bibinfo {year} {2004}),\ \href {\doibase
  10.1063/1.1770619} {\bibfield  {journal} {\bibinfo  {journal} {The Journal of
  Chemical Physics}\ }\textbf {\bibinfo {volume} {121}},\ \bibinfo {pages}
  {2505}}\BibitemShut {NoStop}%
\bibitem [{\citenamefont {Kleinherbers}\ \emph {et~al.}(2020)\citenamefont
  {Kleinherbers}, \citenamefont {Szpak}, \citenamefont {K{\"o}nig},\ and\
  \citenamefont {Sch{\"u}tzhold}}]{kleinherbers2020a}%
  \BibitemOpen
  \bibfield  {author} {\bibinfo {author} {\bibnamefont {Kleinherbers},
  \bibfnamefont {E.}}, \bibinfo {author} {\bibfnamefont {N.}~\bibnamefont
  {Szpak}}, \bibinfo {author} {\bibfnamefont {J.}~\bibnamefont {K{\"o}nig}}, \
  and\ \bibinfo {author} {\bibfnamefont {R.}~\bibnamefont {Sch{\"u}tzhold}}}
  (\bibinfo {year} {2020}),\ \href {\doibase 10.1103/PhysRevB.101.125131}
  {\bibfield  {journal} {\bibinfo  {journal} {Phys. Rev. B}\ }\textbf {\bibinfo
  {volume} {101}},\ \bibinfo {pages} {125131}}\BibitemShut {NoStop}%
\bibitem [{\citenamefont {Klich}(2003)}]{klich2003a}%
  \BibitemOpen
  \bibfield  {author} {\bibinfo {author} {\bibnamefont {Klich}, \bibfnamefont
  {I.}}} (\bibinfo {year} {2003}),\ in\ \href {\doibase
  10.1007/978-94-010-0089-5_19} {\emph {\bibinfo {booktitle} {Quantum Noise in
  Mesoscopic Physics}}},\ \bibinfo {series} {NATO Science Series},
  Vol.~\bibinfo {volume} {97},\ \bibinfo {editor} {edited by\ \bibinfo {editor}
  {\bibfnamefont {Y.~V.}\ \bibnamefont {Nazarov}}}\ (\bibinfo  {publisher}
  {Springer},\ \bibinfo {address} {Dordrecht})\ p.\ \bibinfo {pages}
  {397}\BibitemShut {NoStop}%
\bibitem [{\citenamefont {Kobayashi}\ \emph {et~al.}(2003)\citenamefont
  {Kobayashi}, \citenamefont {Aikawa}, \citenamefont {Katsumoto},\ and\
  \citenamefont {Iye}}]{KobayashiIye2003}%
  \BibitemOpen
  \bibfield  {author} {\bibinfo {author} {\bibnamefont {Kobayashi},
  \bibfnamefont {K.}}, \bibinfo {author} {\bibfnamefont {H.}~\bibnamefont
  {Aikawa}}, \bibinfo {author} {\bibfnamefont {S.}~\bibnamefont {Katsumoto}}, \
  and\ \bibinfo {author} {\bibfnamefont {Y.}~\bibnamefont {Iye}}} (\bibinfo
  {year} {2003}),\ \href {\doibase 10.1103/PhysRevB.68.235304} {\bibfield
  {journal} {\bibinfo  {journal} {Phys. Rev. B}\ }\textbf {\bibinfo {volume}
  {68}},\ \bibinfo {pages} {235304}}\BibitemShut {NoStop}%
\bibitem [{\citenamefont {Kohler}\ \emph {et~al.}(2005)\citenamefont {Kohler},
  \citenamefont {Lehmann},\ and\ \citenamefont
  {H{\"a}nggi}}]{KohlerHanggi2005}%
  \BibitemOpen
  \bibfield  {author} {\bibinfo {author} {\bibnamefont {Kohler}, \bibfnamefont
  {S.}}, \bibinfo {author} {\bibfnamefont {J.}~\bibnamefont {Lehmann}}, \ and\
  \bibinfo {author} {\bibfnamefont {P.}~\bibnamefont {H{\"a}nggi}}} (\bibinfo
  {year} {2005}),\ \href {\doibase
  https://doi.org/10.1016/j.physrep.2004.11.002} {\bibfield  {journal}
  {\bibinfo  {journal} {Physics Reports}\ }\textbf {\bibinfo {volume}
  {406}}~(\bibinfo {number} {6}),\ \bibinfo {pages} {379}}\BibitemShut
  {NoStop}%
\bibitem [{\citenamefont {Kohmoto}\ \emph {et~al.}(1983)\citenamefont
  {Kohmoto}, \citenamefont {Kadanoff},\ and\ \citenamefont
  {Tang}}]{KohmotoTang1983}%
  \BibitemOpen
  \bibfield  {author} {\bibinfo {author} {\bibnamefont {Kohmoto}, \bibfnamefont
  {M.}}, \bibinfo {author} {\bibfnamefont {L.~P.}\ \bibnamefont {Kadanoff}}, \
  and\ \bibinfo {author} {\bibfnamefont {C.}~\bibnamefont {Tang}}} (\bibinfo
  {year} {1983}),\ \href {\doibase 10.1103/PhysRevLett.50.1870} {\bibfield
  {journal} {\bibinfo  {journal} {Phys. Rev. Lett.}\ }\textbf {\bibinfo
  {volume} {50}},\ \bibinfo {pages} {1870}}\BibitemShut {NoStop}%
\bibitem [{\citenamefont {Kohmoto}\ \emph {et~al.}(1987)\citenamefont
  {Kohmoto}, \citenamefont {Sutherland},\ and\ \citenamefont
  {Tang}}]{KohmotoTang1987}%
  \BibitemOpen
  \bibfield  {author} {\bibinfo {author} {\bibnamefont {Kohmoto}, \bibfnamefont
  {M.}}, \bibinfo {author} {\bibfnamefont {B.}~\bibnamefont {Sutherland}}, \
  and\ \bibinfo {author} {\bibfnamefont {C.}~\bibnamefont {Tang}}} (\bibinfo
  {year} {1987}),\ \href {\doibase 10.1103/PhysRevB.35.1020} {\bibfield
  {journal} {\bibinfo  {journal} {Phys. Rev. B}\ }\textbf {\bibinfo {volume}
  {35}},\ \bibinfo {pages} {1020}}\BibitemShut {NoStop}%
\bibitem [{\citenamefont {Kohn}(1964)}]{Kohn1964}%
  \BibitemOpen
  \bibfield  {author} {\bibinfo {author} {\bibnamefont {Kohn}, \bibfnamefont
  {W.}}} (\bibinfo {year} {1964}),\ \href {\doibase 10.1103/PhysRev.133.A171}
  {\bibfield  {journal} {\bibinfo  {journal} {Phys. Rev.}\ }\textbf {\bibinfo
  {volume} {133}},\ \bibinfo {pages} {A171}}\BibitemShut {NoStop}%
\bibitem [{\citenamefont {Kolovsky}(2020)}]{kolovsky2020a}%
  \BibitemOpen
  \bibfield  {author} {\bibinfo {author} {\bibnamefont {Kolovsky},
  \bibfnamefont {A.~R.}}} (\bibinfo {year} {2020}),\ \href {\doibase
  10.1103/PhysRevB.102.174310} {\bibfield  {journal} {\bibinfo  {journal}
  {Phys. Rev. B}\ }\textbf {\bibinfo {volume} {102}},\ \bibinfo {pages}
  {174310}}\BibitemShut {NoStop}%
\bibitem [{\citenamefont {K{\"{o}}nig}\ and\ \citenamefont
  {Weig}(2012)}]{koenig2012a}%
  \BibitemOpen
  \bibfield  {author} {\bibinfo {author} {\bibnamefont {K{\"{o}}nig},
  \bibfnamefont {D.~R.}}, \ and\ \bibinfo {author} {\bibfnamefont {E.~M.}\
  \bibnamefont {Weig}}} (\bibinfo {year} {2012}),\ \href {\doibase
  10.1063/1.4767359} {\bibfield  {journal} {\bibinfo  {journal} {Applied
  Physics Letters}\ }\textbf {\bibinfo {volume} {101}}~(\bibinfo {number}
  {21}),\ \bibinfo {pages} {213111}}\BibitemShut {NoStop}%
\bibitem [{\citenamefont {K{\"o}nig}\ and\ \citenamefont
  {Gefen}(2001)}]{KonigGefen2001}%
  \BibitemOpen
  \bibfield  {author} {\bibinfo {author} {\bibnamefont {K{\"o}nig},
  \bibfnamefont {J.}}, \ and\ \bibinfo {author} {\bibfnamefont
  {Y.}~\bibnamefont {Gefen}}} (\bibinfo {year} {2001}),\ \href {\doibase
  10.1103/PhysRevLett.86.3855} {\bibfield  {journal} {\bibinfo  {journal}
  {Phys. Rev. Lett.}\ }\textbf {\bibinfo {volume} {86}},\ \bibinfo {pages}
  {3855}}\BibitemShut {NoStop}%
\bibitem [{\citenamefont {K{\"o}nig}\ and\ \citenamefont
  {Gefen}(2002)}]{KonigGefen2002}%
  \BibitemOpen
  \bibfield  {author} {\bibinfo {author} {\bibnamefont {K{\"o}nig},
  \bibfnamefont {J.}}, \ and\ \bibinfo {author} {\bibfnamefont
  {Y.}~\bibnamefont {Gefen}}} (\bibinfo {year} {2002}),\ \href {\doibase
  10.1103/PhysRevB.65.045316} {\bibfield  {journal} {\bibinfo  {journal} {Phys.
  Rev. B}\ }\textbf {\bibinfo {volume} {65}},\ \bibinfo {pages}
  {045316}}\BibitemShut {NoStop}%
\bibitem [{\citenamefont {Kosloff}(2013)}]{Kosloff2013}%
  \BibitemOpen
  \bibfield  {author} {\bibinfo {author} {\bibnamefont {Kosloff}, \bibfnamefont
  {R.}}} (\bibinfo {year} {2013}),\ \href {\doibase 10.3390/e15062100}
  {\bibfield  {journal} {\bibinfo  {journal} {Entropy}\ }\textbf {\bibinfo
  {volume} {15}}~(\bibinfo {number} {6}),\ \bibinfo {pages} {2100}}\BibitemShut
  {NoStop}%
\bibitem [{\citenamefont {Kouachi}(2006)}]{Kouachi2006}%
  \BibitemOpen
  \bibfield  {author} {\bibinfo {author} {\bibnamefont {Kouachi}, \bibfnamefont
  {S.}}} (\bibinfo {year} {2006}),\ \href {\doibase 10.13001/1081-3810.1223}
  {\bibfield  {journal} {\bibinfo  {journal} {Electronic Journal of Linear
  Algebra}\ }\textbf {\bibinfo {volume} {15}},\ \bibinfo {pages}
  {115}}\BibitemShut {NoStop}%
\bibitem [{\citenamefont {Kraus}(1971)}]{kraus1971a}%
  \BibitemOpen
  \bibfield  {author} {\bibinfo {author} {\bibnamefont {Kraus}, \bibfnamefont
  {K.}}} (\bibinfo {year} {1971}),\ \href {\doibase
  https://doi.org/10.1016/0003-4916(71)90108-4} {\bibfield  {journal} {\bibinfo
   {journal} {Annals of Physics}\ }\textbf {\bibinfo {volume} {64}}~(\bibinfo
  {number} {2}),\ \bibinfo {pages} {311 }}\BibitemShut {NoStop}%
\bibitem [{\citenamefont {Krause}\ \emph {et~al.}(2015)\citenamefont {Krause},
  \citenamefont {Brandes}, \citenamefont {Esposito},\ and\ \citenamefont
  {Schaller}}]{krause2015a}%
  \BibitemOpen
  \bibfield  {author} {\bibinfo {author} {\bibnamefont {Krause}, \bibfnamefont
  {T.}}, \bibinfo {author} {\bibfnamefont {T.}~\bibnamefont {Brandes}},
  \bibinfo {author} {\bibfnamefont {M.}~\bibnamefont {Esposito}}, \ and\
  \bibinfo {author} {\bibfnamefont {G.}~\bibnamefont {Schaller}}} (\bibinfo
  {year} {2015}),\ \href {\doibase 10.1063/1.4916359} {\bibfield  {journal}
  {\bibinfo  {journal} {The Journal of Chemical Physics}\ }\textbf {\bibinfo
  {volume} {142}},\ \bibinfo {pages} {134106}}\BibitemShut {NoStop}%
\bibitem [{\citenamefont {Krinner}\ \emph {et~al.}(2016)\citenamefont
  {Krinner}, \citenamefont {Lebrat}, \citenamefont {Husmann}, \citenamefont
  {Grenier}, \citenamefont {Brantut},\ and\ \citenamefont
  {Esslinger}}]{KrinnerEsslinger2016}%
  \BibitemOpen
  \bibfield  {author} {\bibinfo {author} {\bibnamefont {Krinner}, \bibfnamefont
  {S.}}, \bibinfo {author} {\bibfnamefont {M.}~\bibnamefont {Lebrat}}, \bibinfo
  {author} {\bibfnamefont {D.}~\bibnamefont {Husmann}}, \bibinfo {author}
  {\bibfnamefont {C.}~\bibnamefont {Grenier}}, \bibinfo {author} {\bibfnamefont
  {J.-P.}\ \bibnamefont {Brantut}}, \ and\ \bibinfo {author} {\bibfnamefont
  {T.}~\bibnamefont {Esslinger}}} (\bibinfo {year} {2016}),\ \href {\doibase
  10.1073/pnas.1601812113} {\bibfield  {journal} {\bibinfo  {journal}
  {Proceedings of the National Academy of Sciences}\ }\textbf {\bibinfo
  {volume} {113}}~(\bibinfo {number} {29}),\ \bibinfo {pages}
  {8144}}\BibitemShut {NoStop}%
\bibitem [{\citenamefont {Krinner}\ \emph {et~al.}(2015)\citenamefont
  {Krinner}, \citenamefont {Stadler}, \citenamefont {Husmann}, \citenamefont
  {Brantut},\ and\ \citenamefont {Esslinger}}]{KrinnerEsslinger2015}%
  \BibitemOpen
  \bibfield  {author} {\bibinfo {author} {\bibnamefont {Krinner}, \bibfnamefont
  {S.}}, \bibinfo {author} {\bibfnamefont {D.}~\bibnamefont {Stadler}},
  \bibinfo {author} {\bibfnamefont {D.}~\bibnamefont {Husmann}}, \bibinfo
  {author} {\bibfnamefont {J.-P.}\ \bibnamefont {Brantut}}, \ and\ \bibinfo
  {author} {\bibfnamefont {T.}~\bibnamefont {Esslinger}}} (\bibinfo {year}
  {2015}),\ \href {\doibase 10.1038/nature14049} {\bibfield  {journal}
  {\bibinfo  {journal} {Nature}\ }\textbf {\bibinfo {volume} {517}},\ \bibinfo
  {pages} {64}}\BibitemShut {NoStop}%
\bibitem [{\citenamefont {Kubala}\ and\ \citenamefont
  {K{\"o}nig}(2002)}]{KubalaKonig2002}%
  \BibitemOpen
  \bibfield  {author} {\bibinfo {author} {\bibnamefont {Kubala}, \bibfnamefont
  {B.}}, \ and\ \bibinfo {author} {\bibfnamefont {J.}~\bibnamefont
  {K{\"o}nig}}} (\bibinfo {year} {2002}),\ \href {\doibase
  10.1103/PhysRevB.65.245301} {\bibfield  {journal} {\bibinfo  {journal} {Phys.
  Rev. B}\ }\textbf {\bibinfo {volume} {65}},\ \bibinfo {pages}
  {245301}}\BibitemShut {NoStop}%
\bibitem [{\citenamefont {Kubo}(1957)}]{Kubo1957}%
  \BibitemOpen
  \bibfield  {author} {\bibinfo {author} {\bibnamefont {Kubo}, \bibfnamefont
  {R.}}} (\bibinfo {year} {1957}),\ \href {\doibase 10.1143/JPSJ.12.570}
  {\bibfield  {journal} {\bibinfo  {journal} {J. Phys. Soc. Jpn.}\ }\textbf
  {\bibinfo {volume} {12}},\ \bibinfo {pages} {570}}\BibitemShut {NoStop}%
\bibitem [{\citenamefont {Kulshreshtha}\ \emph {et~al.}(2019)\citenamefont
  {Kulshreshtha}, \citenamefont {Pal}, \citenamefont {Wahl},\ and\
  \citenamefont {Simon}}]{KulshreshthaSimon2019}%
  \BibitemOpen
  \bibfield  {author} {\bibinfo {author} {\bibnamefont {Kulshreshtha},
  \bibfnamefont {A.~K.}}, \bibinfo {author} {\bibfnamefont {A.}~\bibnamefont
  {Pal}}, \bibinfo {author} {\bibfnamefont {T.~B.}\ \bibnamefont {Wahl}}, \
  and\ \bibinfo {author} {\bibfnamefont {S.~H.}\ \bibnamefont {Simon}}}
  (\bibinfo {year} {2019}),\ \href {\doibase 10.1103/PhysRevB.99.104201}
  {\bibfield  {journal} {\bibinfo  {journal} {Phys. Rev. B}\ }\textbf {\bibinfo
  {volume} {99}},\ \bibinfo {pages} {104201}}\BibitemShut {NoStop}%
\bibitem [{\citenamefont {Kuo}\ and\ \citenamefont
  {Chang}(2010)}]{KuoChang2010}%
  \BibitemOpen
  \bibfield  {author} {\bibinfo {author} {\bibnamefont {Kuo}, \bibfnamefont
  {D.~M.-T.}}, \ and\ \bibinfo {author} {\bibfnamefont {Y.-c.}\ \bibnamefont
  {Chang}}} (\bibinfo {year} {2010}),\ \href {\doibase
  10.1103/PhysRevB.81.205321} {\bibfield  {journal} {\bibinfo  {journal} {Phys.
  Rev. B}\ }\textbf {\bibinfo {volume} {81}},\ \bibinfo {pages}
  {205321}}\BibitemShut {NoStop}%
\bibitem [{\citenamefont {Kurzmann}\ \emph {et~al.}(2019)\citenamefont
  {Kurzmann}, \citenamefont {Stegmann}, \citenamefont {Kerski}, \citenamefont
  {Schott}, \citenamefont {Ludwig}, \citenamefont {Wieck}, \citenamefont
  {K{\"o}nig}, \citenamefont {Lorke},\ and\ \citenamefont
  {Geller}}]{Kurzmann2019}%
  \BibitemOpen
  \bibfield  {author} {\bibinfo {author} {\bibnamefont {Kurzmann},
  \bibfnamefont {A.}}, \bibinfo {author} {\bibfnamefont {P.}~\bibnamefont
  {Stegmann}}, \bibinfo {author} {\bibfnamefont {J.}~\bibnamefont {Kerski}},
  \bibinfo {author} {\bibfnamefont {R.}~\bibnamefont {Schott}}, \bibinfo
  {author} {\bibfnamefont {A.}~\bibnamefont {Ludwig}}, \bibinfo {author}
  {\bibfnamefont {A.~D.}\ \bibnamefont {Wieck}}, \bibinfo {author}
  {\bibfnamefont {J.}~\bibnamefont {K{\"o}nig}}, \bibinfo {author}
  {\bibfnamefont {A.}~\bibnamefont {Lorke}}, \ and\ \bibinfo {author}
  {\bibfnamefont {M.}~\bibnamefont {Geller}}} (\bibinfo {year} {2019}),\ \href
  {\doibase 10.1103/PhysRevLett.122.247403} {\bibfield  {journal} {\bibinfo
  {journal} {Phys. Rev. Lett.}\ }\textbf {\bibinfo {volume} {122}},\ \bibinfo
  {pages} {247403}}\BibitemShut {NoStop}%
\bibitem [{\citenamefont {Lacerda}\ \emph {et~al.}(2021)\citenamefont
  {Lacerda}, \citenamefont {Goold},\ and\ \citenamefont {Landi}}]{Lacerda2021}%
  \BibitemOpen
  \bibfield  {author} {\bibinfo {author} {\bibnamefont {Lacerda}, \bibfnamefont
  {A.~M.}}, \bibinfo {author} {\bibfnamefont {J.}~\bibnamefont {Goold}}, \ and\
  \bibinfo {author} {\bibfnamefont {G.~T.}\ \bibnamefont {Landi}}} (\bibinfo
  {year} {2021}),\ \href {\doibase 10.1103/PhysRevB.104.174203} {\bibfield
  {journal} {\bibinfo  {journal} {Phys. Rev. B}\ }\textbf {\bibinfo {volume}
  {104}},\ \bibinfo {pages} {174203}}\BibitemShut {NoStop}%
\bibitem [{\citenamefont {Lacerda}\ \emph {et~al.}(2022)\citenamefont
  {Lacerda}, \citenamefont {Purkayastha}, \citenamefont {Kewming},
  \citenamefont {Landi},\ and\ \citenamefont {Goold}}]{Lacerda2022}%
  \BibitemOpen
  \bibfield  {author} {\bibinfo {author} {\bibnamefont {Lacerda}, \bibfnamefont
  {A.~M.}}, \bibinfo {author} {\bibfnamefont {A.}~\bibnamefont {Purkayastha}},
  \bibinfo {author} {\bibfnamefont {M.}~\bibnamefont {Kewming}}, \bibinfo
  {author} {\bibfnamefont {G.~T.}\ \bibnamefont {Landi}}, \ and\ \bibinfo
  {author} {\bibfnamefont {J.}~\bibnamefont {Goold}}} (\bibinfo {year}
  {2022}),\ \href@noop {} {\bibfield  {journal} {\bibinfo  {journal}
  {arXiv:2206.01090}\ }}\Eprint {http://arxiv.org/abs/2206.01090v2}
  {arXiv:2206.01090v2 [quant-ph]} \BibitemShut {NoStop}%
\bibitem [{\citenamefont {Laird}(1991)}]{laird1991a}%
  \BibitemOpen
  \bibfield  {author} {\bibinfo {author} {\bibnamefont {Laird}, \bibfnamefont
  {B.~B.}}} (\bibinfo {year} {1991}),\ \href {\doibase 10.1063/1.460626}
  {\bibfield  {journal} {\bibinfo  {journal} {The Journal of Chemical Physics}\
  }\textbf {\bibinfo {volume} {94}},\ \bibinfo {pages} {4391}}\BibitemShut
  {NoStop}%
\bibitem [{\citenamefont {Landauer}(1957)}]{landauer1957a}%
  \BibitemOpen
  \bibfield  {author} {\bibinfo {author} {\bibnamefont {Landauer},
  \bibfnamefont {R.}}} (\bibinfo {year} {1957}),\ \href {\doibase
  10.1147/rd.13.0223} {\bibfield  {journal} {\bibinfo  {journal} {IBM Journal
  of Research and Development}\ }\textbf {\bibinfo {volume} {1}},\ \bibinfo
  {pages} {223}}\BibitemShut {NoStop}%
\bibitem [{\citenamefont {Landi}\ and\ \citenamefont
  {Karevski}(2015)}]{Landi2015a}%
  \BibitemOpen
  \bibfield  {author} {\bibinfo {author} {\bibnamefont {Landi}, \bibfnamefont
  {G.~T.}}, \ and\ \bibinfo {author} {\bibfnamefont {D.}~\bibnamefont
  {Karevski}}} (\bibinfo {year} {2015}),\ \href {\doibase
  10.1103/PhysRevB.91.174422} {\bibfield  {journal} {\bibinfo  {journal}
  {Physical Review B}\ }\textbf {\bibinfo {volume} {91}},\ \bibinfo {pages}
  {174422}}\BibitemShut {NoStop}%
\bibitem [{\citenamefont {Landi}\ \emph {et~al.}(2014)\citenamefont {Landi},
  \citenamefont {Novais}, \citenamefont {de~Oliveira},\ and\ \citenamefont
  {Karevski}}]{LandiKarevski2014}%
  \BibitemOpen
  \bibfield  {author} {\bibinfo {author} {\bibnamefont {Landi}, \bibfnamefont
  {G.~T.}}, \bibinfo {author} {\bibfnamefont {E.}~\bibnamefont {Novais}},
  \bibinfo {author} {\bibfnamefont {M.~J.}\ \bibnamefont {de~Oliveira}}, \ and\
  \bibinfo {author} {\bibfnamefont {D.}~\bibnamefont {Karevski}}} (\bibinfo
  {year} {2014}),\ \href {\doibase 10.1103/PhysRevE.90.042142} {\bibfield
  {journal} {\bibinfo  {journal} {Phys. Rev. E}\ }\textbf {\bibinfo {volume}
  {90}},\ \bibinfo {pages} {042142}}\BibitemShut {NoStop}%
\bibitem [{\citenamefont {Landi}\ and\ \citenamefont
  {Paternostro}(2021)}]{LandiPaternostro2021}%
  \BibitemOpen
  \bibfield  {author} {\bibinfo {author} {\bibnamefont {Landi}, \bibfnamefont
  {G.~T.}}, \ and\ \bibinfo {author} {\bibfnamefont {M.}~\bibnamefont
  {Paternostro}}} (\bibinfo {year} {2021}),\ \href {\doibase
  10.1103/RevModPhys.93.035008} {\bibfield  {journal} {\bibinfo  {journal}
  {Rev. Mod. Phys.}\ }\textbf {\bibinfo {volume} {93}},\ \bibinfo {pages}
  {035008}}\BibitemShut {NoStop}%
\bibitem [{\citenamefont {Lang}\ and\ \citenamefont
  {Firsov}(1963)}]{lang1963a}%
  \BibitemOpen
  \bibfield  {author} {\bibinfo {author} {\bibnamefont {Lang}, \bibfnamefont
  {I.~G.}}, \ and\ \bibinfo {author} {\bibfnamefont {Y.~A.}\ \bibnamefont
  {Firsov}}} (\bibinfo {year} {1963}),\ \href
  {http://www.jetp.ras.ru/cgi-bin/e/index/e/16/5/p1301?a=list} {\bibfield
  {journal} {\bibinfo  {journal} {Soviet Physics JETP}\ }\textbf {\bibinfo
  {volume} {16}}~(\bibinfo {number} {5}),\ \bibinfo {pages} {1301}}\BibitemShut
  {NoStop}%
\bibitem [{\citenamefont {Lange}\ and\ \citenamefont
  {Timm}(2021)}]{lange2021a}%
  \BibitemOpen
  \bibfield  {author} {\bibinfo {author} {\bibnamefont {Lange}, \bibfnamefont
  {S.}}, \ and\ \bibinfo {author} {\bibfnamefont {C.}~\bibnamefont {Timm}}}
  (\bibinfo {year} {2021}),\ \href {\doibase 10.1063/5.0033486} {\bibfield
  {journal} {\bibinfo  {journal} {Chaos}\ }\textbf {\bibinfo {volume}
  {31}}~(\bibinfo {number} {2}),\ \bibinfo {pages} {023101}}\BibitemShut
  {NoStop}%
\bibitem [{\citenamefont {Latorre}\ \emph {et~al.}(2004)\citenamefont
  {Latorre}, \citenamefont {Rico},\ and\ \citenamefont
  {Vidal}}]{LaTorreVidal2004}%
  \BibitemOpen
  \bibfield  {author} {\bibinfo {author} {\bibnamefont {Latorre}, \bibfnamefont
  {J.~I.}}, \bibinfo {author} {\bibfnamefont {E.}~\bibnamefont {Rico}}, \ and\
  \bibinfo {author} {\bibfnamefont {G.}~\bibnamefont {Vidal}}} (\bibinfo {year}
  {2004}),\ \href {\doibase 10.26421/QIC4.1-4} {\bibfield  {journal} {\bibinfo
  {journal} {Quant.Inf.Comput.}\ }\textbf {\bibinfo {volume} {4}},\ \bibinfo
  {pages} {48}}\BibitemShut {NoStop}%
\bibitem [{\citenamefont {Lebowitz}\ and\ \citenamefont
  {Scaramazza}(2018)}]{LebowitzScaramazza2018}%
  \BibitemOpen
  \bibfield  {author} {\bibinfo {author} {\bibnamefont {Lebowitz},
  \bibfnamefont {J.~L.}}, \ and\ \bibinfo {author} {\bibfnamefont {J.~A.}\
  \bibnamefont {Scaramazza}}} (\bibinfo {year} {2018}),\ \href
  {https://arxiv.org/abs/1801.07153} {\bibinfo  {journal} {arXiv:1801.07153}\
  }\BibitemShut {NoStop}%
\bibitem [{\citenamefont {Lebrat}\ \emph {et~al.}(2018)\citenamefont {Lebrat},
  \citenamefont {Gri\v{s}ins}, \citenamefont {Husmann}, \citenamefont
  {H{\"a}usler}, \citenamefont {Corman}, \citenamefont {Giamarchi},
  \citenamefont {Brantut},\ and\ \citenamefont
  {Esslinger}}]{LebratEsslinger2018}%
  \BibitemOpen
\bibfield  {journal} {  }\bibfield  {author} {\bibinfo {author} {\bibnamefont
  {Lebrat}, \bibfnamefont {M.}}, \bibinfo {author} {\bibfnamefont
  {P.}~\bibnamefont {Gri\v{s}ins}}, \bibinfo {author} {\bibfnamefont
  {D.}~\bibnamefont {Husmann}}, \bibinfo {author} {\bibfnamefont
  {S.}~\bibnamefont {H{\"a}usler}}, \bibinfo {author} {\bibfnamefont
  {L.}~\bibnamefont {Corman}}, \bibinfo {author} {\bibfnamefont
  {T.}~\bibnamefont {Giamarchi}}, \bibinfo {author} {\bibfnamefont {J.-P.}\
  \bibnamefont {Brantut}}, \ and\ \bibinfo {author} {\bibfnamefont
  {T.}~\bibnamefont {Esslinger}}} (\bibinfo {year} {2018}),\ \href {\doibase
  10.1103/PhysRevX.8.011053} {\bibfield  {journal} {\bibinfo  {journal} {Phys.
  Rev. X}\ }\textbf {\bibinfo {volume} {8}},\ \bibinfo {pages}
  {011053}}\BibitemShut {NoStop}%
\bibitem [{\citenamefont {Lee}\ \emph {et~al.}(2020)\citenamefont {Lee},
  \citenamefont {Balachandran}, \citenamefont {Guo},\ and\ \citenamefont
  {Poletti}}]{LeePoletti2019}%
  \BibitemOpen
  \bibfield  {author} {\bibinfo {author} {\bibnamefont {Lee}, \bibfnamefont
  {K.}}, \bibinfo {author} {\bibfnamefont {V.}~\bibnamefont {Balachandran}},
  \bibinfo {author} {\bibfnamefont {C.}~\bibnamefont {Guo}}, \ and\ \bibinfo
  {author} {\bibfnamefont {D.}~\bibnamefont {Poletti}}} (\bibinfo {year}
  {2020}),\ \href {\doibase 10.3390/e22111311} {\bibfield  {journal} {\bibinfo
  {journal} {Entropy}\ }\textbf {\bibinfo {volume} {22}},\ \bibinfo {pages}
  {1311}}\BibitemShut {NoStop}%
\bibitem [{\citenamefont {Lee}\ \emph {et~al.}(2022)\citenamefont {Lee},
  \citenamefont {Balachandran}, \citenamefont {Guo},\ and\ \citenamefont
  {Poletti}}]{LeePoletti2021b}%
  \BibitemOpen
  \bibfield  {author} {\bibinfo {author} {\bibnamefont {Lee}, \bibfnamefont
  {K.~H.}}, \bibinfo {author} {\bibfnamefont {V.}~\bibnamefont {Balachandran}},
  \bibinfo {author} {\bibfnamefont {C.}~\bibnamefont {Guo}}, \ and\ \bibinfo
  {author} {\bibfnamefont {D.}~\bibnamefont {Poletti}}} (\bibinfo {year}
  {2022}),\ \href {\doibase 10.1103/PhysRevE.105.024120} {\bibfield  {journal}
  {\bibinfo  {journal} {Phys. Rev. E}\ }\textbf {\bibinfo {volume} {105}},\
  \bibinfo {pages} {024120}}\BibitemShut {NoStop}%
\bibitem [{\citenamefont {Lee}\ \emph {et~al.}(2021)\citenamefont {Lee},
  \citenamefont {Balachandran},\ and\ \citenamefont
  {Poletti}}]{LeePoletti2021}%
  \BibitemOpen
  \bibfield  {author} {\bibinfo {author} {\bibnamefont {Lee}, \bibfnamefont
  {K.~H.}}, \bibinfo {author} {\bibfnamefont {V.}~\bibnamefont {Balachandran}},
  \ and\ \bibinfo {author} {\bibfnamefont {D.}~\bibnamefont {Poletti}}}
  (\bibinfo {year} {2021}),\ \href {\doibase 10.1103/PhysRevE.103.052143}
  {\bibfield  {journal} {\bibinfo  {journal} {Phys. Rev. E}\ }\textbf {\bibinfo
  {volume} {103}},\ \bibinfo {pages} {052143}}\BibitemShut {NoStop}%
\bibitem [{\citenamefont {Leggett}\ \emph {et~al.}(1987)\citenamefont
  {Leggett}, \citenamefont {Chakravarty}, \citenamefont {Dorsey}, \citenamefont
  {Fisher}, \citenamefont {Garg},\ and\ \citenamefont
  {Zwerger}}]{LeggettZwerger1987}%
  \BibitemOpen
  \bibfield  {author} {\bibinfo {author} {\bibnamefont {Leggett}, \bibfnamefont
  {A.~J.}}, \bibinfo {author} {\bibfnamefont {S.}~\bibnamefont {Chakravarty}},
  \bibinfo {author} {\bibfnamefont {A.~T.}\ \bibnamefont {Dorsey}}, \bibinfo
  {author} {\bibfnamefont {M.~P.~A.}\ \bibnamefont {Fisher}}, \bibinfo {author}
  {\bibfnamefont {A.}~\bibnamefont {Garg}}, \ and\ \bibinfo {author}
  {\bibfnamefont {W.}~\bibnamefont {Zwerger}}} (\bibinfo {year} {1987}),\ \href
  {\doibase 10.1103/RevModPhys.59.1} {\bibfield  {journal} {\bibinfo  {journal}
  {Rev. Mod. Phys.}\ }\textbf {\bibinfo {volume} {59}},\ \bibinfo {pages}
  {1}}\BibitemShut {NoStop}%
\bibitem [{\citenamefont {{Lenar\v{c}i\v{c}}}\ and\ \citenamefont
  {Prosen}(2015)}]{LenarcicProsen2015}%
  \BibitemOpen
  \bibfield  {author} {\bibinfo {author} {\bibnamefont {{Lenar\v{c}i\v{c}}},
  \bibfnamefont {Z.}}, \ and\ \bibinfo {author} {\bibfnamefont
  {T.}~\bibnamefont {Prosen}}} (\bibinfo {year} {2015}),\ \href {\doibase
  10.1103/PhysRevE.91.030103} {\bibfield  {journal} {\bibinfo  {journal} {Phys.
  Rev. E}\ }\textbf {\bibinfo {volume} {91}},\ \bibinfo {pages}
  {030103}}\BibitemShut {NoStop}%
\bibitem [{\citenamefont {Lepri}\ and\ \citenamefont
  {Casati}(2011)}]{LepriCasati2011}%
  \BibitemOpen
  \bibfield  {author} {\bibinfo {author} {\bibnamefont {Lepri}, \bibfnamefont
  {S.}}, \ and\ \bibinfo {author} {\bibfnamefont {G.}~\bibnamefont {Casati}}}
  (\bibinfo {year} {2011}),\ \href {\doibase 10.1103/PhysRevLett.106.164101}
  {\bibfield  {journal} {\bibinfo  {journal} {Phys. Rev. Lett.}\ }\textbf
  {\bibinfo {volume} {106}},\ \bibinfo {pages} {164101}}\BibitemShut {NoStop}%
\bibitem [{\citenamefont {Levitov}\ \emph {et~al.}(1996)\citenamefont
  {Levitov}, \citenamefont {Lee},\ and\ \citenamefont
  {Lesovik}}]{levitov1996a}%
  \BibitemOpen
  \bibfield  {author} {\bibinfo {author} {\bibnamefont {Levitov}, \bibfnamefont
  {L.~S.}}, \bibinfo {author} {\bibfnamefont {H.}~\bibnamefont {Lee}}, \ and\
  \bibinfo {author} {\bibfnamefont {G.~B.}\ \bibnamefont {Lesovik}}} (\bibinfo
  {year} {1996}),\ \href {\doibase 10.1063/1.531672} {\bibfield  {journal}
  {\bibinfo  {journal} {Journal of Mathematical Physics}\ }\textbf {\bibinfo
  {volume} {37}},\ \bibinfo {pages} {4845}}\BibitemShut {NoStop}%
\bibitem [{\citenamefont {Levitov}\ and\ \citenamefont
  {Lesovik}(1993)}]{levitov1993a}%
  \BibitemOpen
  \bibfield  {author} {\bibinfo {author} {\bibnamefont {Levitov}, \bibfnamefont
  {L.~S.}}, \ and\ \bibinfo {author} {\bibfnamefont {G.~B.}\ \bibnamefont
  {Lesovik}}} (\bibinfo {year} {1993}),\ \href
  {http://jetpletters.ru/ps/1186/article_17907.shtml} {\bibfield  {journal}
  {\bibinfo  {journal} {JETP Letters}\ }\textbf {\bibinfo {volume} {58}},\
  \bibinfo {pages} {230}}\BibitemShut {NoStop}%
\bibitem [{\citenamefont {Levy}\ and\ \citenamefont
  {Kosloff}(2014)}]{Levy2014}%
  \BibitemOpen
  \bibfield  {author} {\bibinfo {author} {\bibnamefont {Levy}, \bibfnamefont
  {A.}}, \ and\ \bibinfo {author} {\bibfnamefont {R.}~\bibnamefont {Kosloff}}}
  (\bibinfo {year} {2014}),\ \href {\doibase 10.1209/0295-5075/107/20004}
  {\bibfield  {journal} {\bibinfo  {journal} {European Physics Letters}\
  }\textbf {\bibinfo {volume} {107}},\ \bibinfo {pages} {20004}}\BibitemShut
  {NoStop}%
\bibitem [{\citenamefont {Li}\ \emph {et~al.}(2004)\citenamefont {Li},
  \citenamefont {Wang},\ and\ \citenamefont {Casati}}]{Li2004a}%
  \BibitemOpen
  \bibfield  {author} {\bibinfo {author} {\bibnamefont {Li}, \bibfnamefont
  {B.}}, \bibinfo {author} {\bibfnamefont {L.}~\bibnamefont {Wang}}, \ and\
  \bibinfo {author} {\bibfnamefont {G.}~\bibnamefont {Casati}}} (\bibinfo
  {year} {2004}),\ \href {\doibase 10.1103/PhysRevLett.93.184301} {\bibfield
  {journal} {\bibinfo  {journal} {Physical Review Letters}\ }\textbf {\bibinfo
  {volume} {93}}~(\bibinfo {number} {18}),\ \bibinfo {pages}
  {184301}}\BibitemShut {NoStop}%
\bibitem [{\citenamefont {Li}\ \emph {et~al.}(2006)\citenamefont {Li},
  \citenamefont {Wang},\ and\ \citenamefont {Casati}}]{LiCasati2006}%
  \BibitemOpen
  \bibfield  {author} {\bibinfo {author} {\bibnamefont {Li}, \bibfnamefont
  {B.}}, \bibinfo {author} {\bibfnamefont {L.}~\bibnamefont {Wang}}, \ and\
  \bibinfo {author} {\bibfnamefont {G.}~\bibnamefont {Casati}}} (\bibinfo
  {year} {2006}),\ \href {\doibase 10.1063/1.2191730} {\bibfield  {journal}
  {\bibinfo  {journal} {Applied Physics Letters}\ }\textbf {\bibinfo {volume}
  {88}},\ \bibinfo {pages} {143501}}\BibitemShut {NoStop}%
\bibitem [{\citenamefont {Li}\ \emph {et~al.}(2021)\citenamefont {Li},
  \citenamefont {Prosen},\ and\ \citenamefont {Chan}}]{LiChan2021}%
  \BibitemOpen
  \bibfield  {author} {\bibinfo {author} {\bibnamefont {Li}, \bibfnamefont
  {J.}}, \bibinfo {author} {\bibfnamefont {T.~c.~v.}\ \bibnamefont {Prosen}}, \
  and\ \bibinfo {author} {\bibfnamefont {A.}~\bibnamefont {Chan}}} (\bibinfo
  {year} {2021}),\ \href {\doibase 10.1103/PhysRevLett.127.170602} {\bibfield
  {journal} {\bibinfo  {journal} {Phys. Rev. Lett.}\ }\textbf {\bibinfo
  {volume} {127}},\ \bibinfo {pages} {170602}}\BibitemShut {NoStop}%
\bibitem [{\citenamefont {Li}\ \emph {et~al.}(2012)\citenamefont {Li},
  \citenamefont {Ren}, \citenamefont {Wang}, \citenamefont {Zhang},
  \citenamefont {H{\"a}nggi},\ and\ \citenamefont {Li}}]{LiLi2012}%
  \BibitemOpen
  \bibfield  {author} {\bibinfo {author} {\bibnamefont {Li}, \bibfnamefont
  {N.}}, \bibinfo {author} {\bibfnamefont {J.}~\bibnamefont {Ren}}, \bibinfo
  {author} {\bibfnamefont {L.}~\bibnamefont {Wang}}, \bibinfo {author}
  {\bibfnamefont {G.}~\bibnamefont {Zhang}}, \bibinfo {author} {\bibfnamefont
  {P.}~\bibnamefont {H{\"a}nggi}}, \ and\ \bibinfo {author} {\bibfnamefont
  {B.}~\bibnamefont {Li}}} (\bibinfo {year} {2012}),\ \href {\doibase
  10.1103/RevModPhys.84.1045} {\bibfield  {journal} {\bibinfo  {journal} {Rev.
  Mod. Phys.}\ }\textbf {\bibinfo {volume} {84}},\ \bibinfo {pages}
  {1045}}\BibitemShut {NoStop}%
\bibitem [{\citenamefont {Li}\ \emph {et~al.}(2015)\citenamefont {Li},
  \citenamefont {Ganeshan}, \citenamefont {Pixley},\ and\ \citenamefont
  {Das~Sarma}}]{LiDasSarma2015}%
  \BibitemOpen
  \bibfield  {author} {\bibinfo {author} {\bibnamefont {Li}, \bibfnamefont
  {X.}}, \bibinfo {author} {\bibfnamefont {S.}~\bibnamefont {Ganeshan}},
  \bibinfo {author} {\bibfnamefont {J.~H.}\ \bibnamefont {Pixley}}, \ and\
  \bibinfo {author} {\bibfnamefont {S.}~\bibnamefont {Das~Sarma}}} (\bibinfo
  {year} {2015}),\ \href {\doibase 10.1103/PhysRevLett.115.186601} {\bibfield
  {journal} {\bibinfo  {journal} {Phys. Rev. Lett.}\ }\textbf {\bibinfo
  {volume} {115}},\ \bibinfo {pages} {186601}}\BibitemShut {NoStop}%
\bibitem [{\citenamefont {Li}\ \emph {et~al.}(2017)\citenamefont {Li},
  \citenamefont {Li},\ and\ \citenamefont {Das~Sarma}}]{LiDasSarma2017}%
  \BibitemOpen
  \bibfield  {author} {\bibinfo {author} {\bibnamefont {Li}, \bibfnamefont
  {X.}}, \bibinfo {author} {\bibfnamefont {X.}~\bibnamefont {Li}}, \ and\
  \bibinfo {author} {\bibfnamefont {S.}~\bibnamefont {Das~Sarma}}} (\bibinfo
  {year} {2017}),\ \href {\doibase 10.1103/PhysRevB.96.085119} {\bibfield
  {journal} {\bibinfo  {journal} {Phys. Rev. B}\ }\textbf {\bibinfo {volume}
  {96}},\ \bibinfo {pages} {085119}}\BibitemShut {NoStop}%
\bibitem [{\citenamefont {Lidar}\ \emph {et~al.}(2001)\citenamefont {Lidar},
  \citenamefont {Bihary},\ and\ \citenamefont {Whaley}}]{lidar2001a}%
  \BibitemOpen
  \bibfield  {author} {\bibinfo {author} {\bibnamefont {Lidar}, \bibfnamefont
  {D.~A.}}, \bibinfo {author} {\bibfnamefont {Z.}~\bibnamefont {Bihary}}, \
  and\ \bibinfo {author} {\bibfnamefont {K.~B.}\ \bibnamefont {Whaley}}}
  (\bibinfo {year} {2001}),\ \href {\doibase 10.1016/S0301-0104(01)00330-5}
  {\bibfield  {journal} {\bibinfo  {journal} {Chemical Physics}\ }\textbf
  {\bibinfo {volume} {268}},\ \bibinfo {pages} {35}}\BibitemShut {NoStop}%
\bibitem [{\citenamefont {Lieb}\ \emph {et~al.}(1961)\citenamefont {Lieb},
  \citenamefont {Schultz},\ and\ \citenamefont {Mattis}}]{Lieb1961}%
  \BibitemOpen
  \bibfield  {author} {\bibinfo {author} {\bibnamefont {Lieb}, \bibfnamefont
  {E.~H.}}, \bibinfo {author} {\bibfnamefont {T.}~\bibnamefont {Schultz}}, \
  and\ \bibinfo {author} {\bibfnamefont {D.}~\bibnamefont {Mattis}}} (\bibinfo
  {year} {1961}),\ \href
  {http://www.sciencedirect.com/science/article/pii/0003491661901154}
  {\bibfield  {journal} {\bibinfo  {journal} {Annals of Physics}\ }\textbf
  {\bibinfo {volume} {16}},\ \bibinfo {pages} {407}}\BibitemShut {NoStop}%
\bibitem [{\citenamefont {Lindblad}(1976)}]{Lindblad1976}%
  \BibitemOpen
  \bibfield  {author} {\bibinfo {author} {\bibnamefont {Lindblad},
  \bibfnamefont {G.}}} (\bibinfo {year} {1976}),\ \href
  {http://link.springer.com/article/10.1007/BF01608499} {\bibfield  {journal}
  {\bibinfo  {journal} {Communications in Mathematical Physics}\ }\textbf
  {\bibinfo {volume} {48}},\ \bibinfo {pages} {119}}\BibitemShut {NoStop}%
\bibitem [{\citenamefont {Linden}\ \emph {et~al.}(2010)\citenamefont {Linden},
  \citenamefont {Popescu},\ and\ \citenamefont {Skrzypczyk}}]{linden2010a}%
  \BibitemOpen
  \bibfield  {author} {\bibinfo {author} {\bibnamefont {Linden}, \bibfnamefont
  {N.}}, \bibinfo {author} {\bibfnamefont {S.}~\bibnamefont {Popescu}}, \ and\
  \bibinfo {author} {\bibfnamefont {P.}~\bibnamefont {Skrzypczyk}}} (\bibinfo
  {year} {2010}),\ \href {\doibase 10.1103/PhysRevLett.105.130401} {\bibfield
  {journal} {\bibinfo  {journal} {Physical Review Letters}\ }\textbf {\bibinfo
  {volume} {105}}~(\bibinfo {number} {13}),\ \bibinfo {pages}
  {130401}}\BibitemShut {NoStop}%
\bibitem [{\citenamefont {Liu}\ and\ \citenamefont {Segal}(2021)}]{liu2021a}%
  \BibitemOpen
  \bibfield  {author} {\bibinfo {author} {\bibnamefont {Liu}, \bibfnamefont
  {J.}}, \ and\ \bibinfo {author} {\bibfnamefont {D.}~\bibnamefont {Segal}}}
  (\bibinfo {year} {2021}),\ \href {\doibase 10.1103/PhysRevE.103.032138}
  {\bibfield  {journal} {\bibinfo  {journal} {Phys. Rev. E}\ }\textbf {\bibinfo
  {volume} {103}},\ \bibinfo {pages} {032138}}\BibitemShut {NoStop}%
\bibitem [{\citenamefont {Ljubotina}\ \emph {et~al.}(2017)\citenamefont
  {Ljubotina}, \citenamefont {Znidaric},\ and\ \citenamefont
  {Prosen}}]{LjubotinaProsen2017}%
  \BibitemOpen
  \bibfield  {author} {\bibinfo {author} {\bibnamefont {Ljubotina},
  \bibfnamefont {M.}}, \bibinfo {author} {\bibfnamefont {M.}~\bibnamefont
  {Znidaric}}, \ and\ \bibinfo {author} {\bibfnamefont {T.}~\bibnamefont
  {Prosen}}} (\bibinfo {year} {2017}),\ \href {\doibase 10.1038/ncomms16117}
  {\bibfield  {journal} {\bibinfo  {journal} {Nature Comm.}\ }\textbf {\bibinfo
  {volume} {8}},\ \bibinfo {pages} {16117}}\BibitemShut {NoStop}%
\bibitem [{\citenamefont {Longhi}(2007)}]{longhi2007a}%
  \BibitemOpen
  \bibfield  {author} {\bibinfo {author} {\bibnamefont {Longhi}, \bibfnamefont
  {S.}}} (\bibinfo {year} {2007}),\ \href {\doibase 10.1140/epjb/e2007-00143-2}
  {\bibfield  {journal} {\bibinfo  {journal} {The European Physical Journal B}\
  }\textbf {\bibinfo {volume} {57}},\ \bibinfo {pages} {45}}\BibitemShut
  {NoStop}%
\bibitem [{\citenamefont {Loss}\ and\ \citenamefont
  {Sukhorukov}(2000)}]{LossSukhorukov2000}%
  \BibitemOpen
  \bibfield  {author} {\bibinfo {author} {\bibnamefont {Loss}, \bibfnamefont
  {D.}}, \ and\ \bibinfo {author} {\bibfnamefont {E.~V.}\ \bibnamefont
  {Sukhorukov}}} (\bibinfo {year} {2000}),\ \href {\doibase
  10.1103/PhysRevLett.84.1035} {\bibfield  {journal} {\bibinfo  {journal}
  {Phys. Rev. Lett.}\ }\textbf {\bibinfo {volume} {84}},\ \bibinfo {pages}
  {1035}}\BibitemShut {NoStop}%
\bibitem [{\citenamefont {Lu}\ \emph {et~al.}(2019)\citenamefont {Lu},
  \citenamefont {Cao}, \citenamefont {Hansmann},\ and\ \citenamefont
  {Haverkort}}]{LuHaverkort2019}%
  \BibitemOpen
  \bibfield  {author} {\bibinfo {author} {\bibnamefont {Lu}, \bibfnamefont
  {Y.}}, \bibinfo {author} {\bibfnamefont {X.}~\bibnamefont {Cao}}, \bibinfo
  {author} {\bibfnamefont {P.}~\bibnamefont {Hansmann}}, \ and\ \bibinfo
  {author} {\bibfnamefont {M.~W.}\ \bibnamefont {Haverkort}}} (\bibinfo {year}
  {2019}),\ \href {\doibase 10.1103/PhysRevB.100.115134} {\bibfield  {journal}
  {\bibinfo  {journal} {Phys. Rev. B}\ }\textbf {\bibinfo {volume} {100}},\
  \bibinfo {pages} {115134}}\BibitemShut {NoStop}%
\bibitem [{\citenamefont {Luitz}\ and\ \citenamefont
  {Lev}(2017)}]{LuitzLev2017}%
  \BibitemOpen
  \bibfield  {author} {\bibinfo {author} {\bibnamefont {Luitz}, \bibfnamefont
  {D.~J.}}, \ and\ \bibinfo {author} {\bibfnamefont {Y.~B.}\ \bibnamefont
  {Lev}}} (\bibinfo {year} {2017}),\ \href {\doibase 10.1002/andp.201600350}
  {\bibfield  {journal} {\bibinfo  {journal} {Annals of Physics (Berlin)}\
  }\textbf {\bibinfo {volume} {529}},\ \bibinfo {pages} {1600350}}\BibitemShut
  {NoStop}%
\bibitem [{\citenamefont {Luo}\ \emph {et~al.}(2020)\citenamefont {Luo},
  \citenamefont {Chen}, \citenamefont {Carrasquilla},\ and\ \citenamefont
  {Clark}}]{LuoClark2020}%
  \BibitemOpen
  \bibfield  {author} {\bibinfo {author} {\bibnamefont {Luo}, \bibfnamefont
  {D.}}, \bibinfo {author} {\bibfnamefont {Z.}~\bibnamefont {Chen}}, \bibinfo
  {author} {\bibfnamefont {J.}~\bibnamefont {Carrasquilla}}, \ and\ \bibinfo
  {author} {\bibfnamefont {B.~K.}\ \bibnamefont {Clark}}} (\bibinfo {year}
  {2020}),\ \href {https://arxiv.org/abs/2009.05580} {\bibinfo  {journal}
  {arXiv:2009.05580}\ }\BibitemShut {NoStop}%
\bibitem [{\citenamefont {L{\"u}schen}\ \emph {et~al.}(2017)\citenamefont
  {L{\"u}schen}, \citenamefont {Bordia}, \citenamefont {Scherg}, \citenamefont
  {Alet}, \citenamefont {Altman}, \citenamefont {Schneider},\ and\
  \citenamefont {Bloch}}]{LuschenBloch2017}%
  \BibitemOpen
\bibfield  {journal} {  }\bibfield  {author} {\bibinfo {author} {\bibnamefont
  {L{\"u}schen}, \bibfnamefont {H.~P.}}, \bibinfo {author} {\bibfnamefont
  {P.}~\bibnamefont {Bordia}}, \bibinfo {author} {\bibfnamefont
  {S.}~\bibnamefont {Scherg}}, \bibinfo {author} {\bibfnamefont
  {F.}~\bibnamefont {Alet}}, \bibinfo {author} {\bibfnamefont {E.}~\bibnamefont
  {Altman}}, \bibinfo {author} {\bibfnamefont {U.}~\bibnamefont {Schneider}}, \
  and\ \bibinfo {author} {\bibfnamefont {I.}~\bibnamefont {Bloch}}} (\bibinfo
  {year} {2017}),\ \href {\doibase 10.1103/PhysRevLett.119.260401} {\bibfield
  {journal} {\bibinfo  {journal} {Phys. Rev. Lett.}\ }\textbf {\bibinfo
  {volume} {119}},\ \bibinfo {pages} {260401}}\BibitemShut {NoStop}%
\bibitem [{\citenamefont {L{\"u}schen}\ \emph {et~al.}(2018)\citenamefont
  {L{\"u}schen}, \citenamefont {Scherg}, \citenamefont {Kohlert}, \citenamefont
  {Schreiber}, \citenamefont {Bordia}, \citenamefont {Li}, \citenamefont
  {Das~Sarma},\ and\ \citenamefont {Bloch}}]{LuschenBloch2018}%
  \BibitemOpen
  \bibfield  {author} {\bibinfo {author} {\bibnamefont {L{\"u}schen},
  \bibfnamefont {H.~P.}}, \bibinfo {author} {\bibfnamefont {S.}~\bibnamefont
  {Scherg}}, \bibinfo {author} {\bibfnamefont {T.}~\bibnamefont {Kohlert}},
  \bibinfo {author} {\bibfnamefont {M.}~\bibnamefont {Schreiber}}, \bibinfo
  {author} {\bibfnamefont {P.}~\bibnamefont {Bordia}}, \bibinfo {author}
  {\bibfnamefont {X.}~\bibnamefont {Li}}, \bibinfo {author} {\bibfnamefont
  {S.}~\bibnamefont {Das~Sarma}}, \ and\ \bibinfo {author} {\bibfnamefont
  {I.}~\bibnamefont {Bloch}}} (\bibinfo {year} {2018}),\ \href {\doibase
  10.1103/PhysRevLett.120.160404} {\bibfield  {journal} {\bibinfo  {journal}
  {Phys. Rev. Lett.}\ }\textbf {\bibinfo {volume} {120}},\ \bibinfo {pages}
  {160404}}\BibitemShut {NoStop}%
\bibitem [{\citenamefont {Mac\'e}\ \emph {et~al.}(2016)\citenamefont {Mac\'e},
  \citenamefont {Jagannathan},\ and\ \citenamefont
  {Pi\'echon}}]{MacePiechon2016}%
  \BibitemOpen
  \bibfield  {author} {\bibinfo {author} {\bibnamefont {Mac\'e}, \bibfnamefont
  {N.}}, \bibinfo {author} {\bibfnamefont {A.}~\bibnamefont {Jagannathan}}, \
  and\ \bibinfo {author} {\bibfnamefont {F.}~\bibnamefont {Pi\'echon}}}
  (\bibinfo {year} {2016}),\ \href {\doibase 10.1103/PhysRevB.93.205153}
  {\bibfield  {journal} {\bibinfo  {journal} {Phys. Rev. B}\ }\textbf {\bibinfo
  {volume} {93}},\ \bibinfo {pages} {205153}}\BibitemShut {NoStop}%
\bibitem [{\citenamefont {Mac\'{e}}\ \emph {et~al.}(2019)\citenamefont
  {Mac\'{e}}, \citenamefont {Laflorencie},\ and\ \citenamefont
  {Alet}}]{MaceAlet2019}%
  \BibitemOpen
  \bibfield  {author} {\bibinfo {author} {\bibnamefont {Mac\'{e}},
  \bibfnamefont {N.}}, \bibinfo {author} {\bibfnamefont {N.}~\bibnamefont
  {Laflorencie}}, \ and\ \bibinfo {author} {\bibfnamefont {F.}~\bibnamefont
  {Alet}}} (\bibinfo {year} {2019}),\ \href {\doibase
  10.21468/SciPostPhys.6.4.050} {\bibfield  {journal} {\bibinfo  {journal}
  {SciPost Phys.}\ }\textbf {\bibinfo {volume} {6}},\ \bibinfo {pages}
  {50}}\BibitemShut {NoStop}%
\bibitem [{\citenamefont {Maier}\ \emph {et~al.}(2019)\citenamefont {Maier},
  \citenamefont {Brydges}, \citenamefont {Jurcevic}, \citenamefont {Trautmann},
  \citenamefont {Hempel}, \citenamefont {Lanyon}, \citenamefont {Hauke},
  \citenamefont {Blatt},\ and\ \citenamefont {Roos}}]{Maier2019}%
  \BibitemOpen
  \bibfield  {author} {\bibinfo {author} {\bibnamefont {Maier}, \bibfnamefont
  {C.}}, \bibinfo {author} {\bibfnamefont {T.}~\bibnamefont {Brydges}},
  \bibinfo {author} {\bibfnamefont {P.}~\bibnamefont {Jurcevic}}, \bibinfo
  {author} {\bibfnamefont {N.}~\bibnamefont {Trautmann}}, \bibinfo {author}
  {\bibfnamefont {C.}~\bibnamefont {Hempel}}, \bibinfo {author} {\bibfnamefont
  {B.~P.}\ \bibnamefont {Lanyon}}, \bibinfo {author} {\bibfnamefont
  {P.}~\bibnamefont {Hauke}}, \bibinfo {author} {\bibfnamefont
  {R.}~\bibnamefont {Blatt}}, \ and\ \bibinfo {author} {\bibfnamefont {C.~F.}\
  \bibnamefont {Roos}}} (\bibinfo {year} {2019}),\ \href {\doibase
  10.1103/PhysRevLett.122.050501} {\bibfield  {journal} {\bibinfo  {journal}
  {Phys. Rev. Lett.}\ }\textbf {\bibinfo {volume} {122}},\ \bibinfo {pages}
  {050501}}\BibitemShut {NoStop}%
\bibitem [{\citenamefont {M\"akel\"a}\ and\ \citenamefont
  {M\"ott\"onen}(2013)}]{maekelae2013a}%
  \BibitemOpen
  \bibfield  {author} {\bibinfo {author} {\bibnamefont {M\"akel\"a},
  \bibfnamefont {H.}}, \ and\ \bibinfo {author} {\bibfnamefont
  {M.}~\bibnamefont {M\"ott\"onen}}} (\bibinfo {year} {2013}),\ \href {\doibase
  10.1103/PhysRevA.88.052111} {\bibfield  {journal} {\bibinfo  {journal} {Phys.
  Rev. A}\ }\textbf {\bibinfo {volume} {88}},\ \bibinfo {pages}
  {052111}}\BibitemShut {NoStop}%
\bibitem [{\citenamefont {Malouf}\ \emph {et~al.}(2020)\citenamefont {Malouf},
  \citenamefont {Goold}, \citenamefont {Adesso},\ and\ \citenamefont
  {Landi}}]{Malouf2018}%
  \BibitemOpen
  \bibfield  {author} {\bibinfo {author} {\bibnamefont {Malouf}, \bibfnamefont
  {W.~T.~B.}}, \bibinfo {author} {\bibfnamefont {J.}~\bibnamefont {Goold}},
  \bibinfo {author} {\bibfnamefont {G.}~\bibnamefont {Adesso}}, \ and\ \bibinfo
  {author} {\bibfnamefont {G.}~\bibnamefont {Landi}}} (\bibinfo {year}
  {2020}),\ \href {\doibase 10.1088/1751-8121/ab93fd} {\bibfield  {journal}
  {\bibinfo  {journal} {Journal of Physics A: Mathematical and Theoretical}\
  }\textbf {\bibinfo {volume} {53}},\ \bibinfo {pages} {305302}}\BibitemShut
  {NoStop}%
\bibitem [{\citenamefont {Mandel}\ and\ \citenamefont
  {Wolf}(1995)}]{mandel1995}%
  \BibitemOpen
  \bibfield  {author} {\bibinfo {author} {\bibnamefont {Mandel}, \bibfnamefont
  {L.}}, \ and\ \bibinfo {author} {\bibfnamefont {E.}~\bibnamefont {Wolf}}}
  (\bibinfo {year} {1995}),\ \href@noop {} {\emph {\bibinfo {title} {Optical
  coherence and quantum optics}}}\ (\bibinfo  {publisher} {Cambridge University
  Press})\BibitemShut {NoStop}%
\bibitem [{\citenamefont {Maniscalco}(2007)}]{maniscalco2007a}%
  \BibitemOpen
  \bibfield  {author} {\bibinfo {author} {\bibnamefont {Maniscalco},
  \bibfnamefont {S.}}} (\bibinfo {year} {2007}),\ \href {\doibase
  10.1103/PhysRevA.75.062103} {\bibfield  {journal} {\bibinfo  {journal} {Phys.
  Rev. A}\ }\textbf {\bibinfo {volume} {75}},\ \bibinfo {pages}
  {062103}}\BibitemShut {NoStop}%
\bibitem [{\citenamefont {Manzano}(2013)}]{DanielManzano2013}%
  \BibitemOpen
  \bibfield  {author} {\bibinfo {author} {\bibnamefont {Manzano}, \bibfnamefont
  {D.}}} (\bibinfo {year} {2013}),\ \href {\doibase
  10.1371/journal.pone.0057041} {\bibfield  {journal} {\bibinfo  {journal}
  {PLoS ONE}\ }\textbf {\bibinfo {volume} {8}}~(\bibinfo {number} {2}),\
  \bibinfo {pages} {e57041}}\BibitemShut {NoStop}%
\bibitem [{\citenamefont {Manzano}\ \emph {et~al.}(2016)\citenamefont
  {Manzano}, \citenamefont {Chuang},\ and\ \citenamefont
  {Cao}}]{ManzanoCao2016}%
  \BibitemOpen
  \bibfield  {author} {\bibinfo {author} {\bibnamefont {Manzano}, \bibfnamefont
  {D.}}, \bibinfo {author} {\bibfnamefont {C.}~\bibnamefont {Chuang}}, \ and\
  \bibinfo {author} {\bibfnamefont {J.}~\bibnamefont {Cao}}} (\bibinfo {year}
  {2016}),\ \href {\doibase 10.1088/1367-2630/18/4/043044} {\bibfield
  {journal} {\bibinfo  {journal} {New Journal of Physics}\ }\textbf {\bibinfo
  {volume} {18}}~(\bibinfo {number} {4}),\ \bibinfo {pages}
  {043044}}\BibitemShut {NoStop}%
\bibitem [{\citenamefont {Manzano}\ and\ \citenamefont
  {Hurtado}(2014)}]{ManzanoHurtado2014}%
  \BibitemOpen
  \bibfield  {author} {\bibinfo {author} {\bibnamefont {Manzano}, \bibfnamefont
  {D.}}, \ and\ \bibinfo {author} {\bibfnamefont {P.~I.}\ \bibnamefont
  {Hurtado}}} (\bibinfo {year} {2014}),\ \href {\doibase
  10.1103/PhysRevB.90.125138} {\bibfield  {journal} {\bibinfo  {journal} {Phys.
  Rev. B}\ }\textbf {\bibinfo {volume} {90}},\ \bibinfo {pages}
  {125138}}\BibitemShut {NoStop}%
\bibitem [{\citenamefont {Manzano}\ and\ \citenamefont
  {Hurtado}(2018)}]{ManzanoHurtado2018}%
  \BibitemOpen
  \bibfield  {author} {\bibinfo {author} {\bibnamefont {Manzano}, \bibfnamefont
  {D.}}, \ and\ \bibinfo {author} {\bibfnamefont {P.~I.}\ \bibnamefont
  {Hurtado}}} (\bibinfo {year} {2018}),\ \href {\doibase
  10.1080/00018732.2018.1519981} {\bibfield  {journal} {\bibinfo  {journal}
  {Advances in Physics}\ }\textbf {\bibinfo {volume} {67}}~(\bibinfo {number}
  {1}),\ \bibinfo {pages} {1}}\BibitemShut {NoStop}%
\bibitem [{\citenamefont {Marcos-Vicioso}\ \emph {et~al.}(2018)\citenamefont
  {Marcos-Vicioso}, \citenamefont {L\'opez-Jurado}, \citenamefont
  {Ruiz-Garcia},\ and\ \citenamefont {S\'anchez}}]{MarcosViciosoSanchez2018}%
  \BibitemOpen
  \bibfield  {author} {\bibinfo {author} {\bibnamefont {Marcos-Vicioso},
  \bibfnamefont {A.}}, \bibinfo {author} {\bibfnamefont {C.}~\bibnamefont
  {L\'opez-Jurado}}, \bibinfo {author} {\bibfnamefont {M.}~\bibnamefont
  {Ruiz-Garcia}}, \ and\ \bibinfo {author} {\bibfnamefont {R.}~\bibnamefont
  {S\'anchez}}} (\bibinfo {year} {2018}),\ \href {\doibase
  10.1103/PhysRevB.98.035414} {\bibfield  {journal} {\bibinfo  {journal} {Phys.
  Rev. B}\ }\textbf {\bibinfo {volume} {98}},\ \bibinfo {pages}
  {035414}}\BibitemShut {NoStop}%
\bibitem [{\citenamefont {Martensen}\ and\ \citenamefont
  {Schaller}(2019)}]{martensen2019a}%
  \BibitemOpen
  \bibfield  {author} {\bibinfo {author} {\bibnamefont {Martensen},
  \bibfnamefont {N.}}, \ and\ \bibinfo {author} {\bibfnamefont
  {G.}~\bibnamefont {Schaller}}} (\bibinfo {year} {2019}),\ \href {\doibase
  10.1140/epjb/e2019-90585-0} {\bibfield  {journal} {\bibinfo  {journal}
  {European Physical Journal B}\ }\textbf {\bibinfo {volume} {92}},\ \bibinfo
  {pages} {30}}\BibitemShut {NoStop}%
\bibitem [{\citenamefont {Martin}\ and\ \citenamefont
  {Schwinger}(1959)}]{martin1959a}%
  \BibitemOpen
  \bibfield  {author} {\bibinfo {author} {\bibnamefont {Martin}, \bibfnamefont
  {P.~C.}}, \ and\ \bibinfo {author} {\bibfnamefont {J.}~\bibnamefont
  {Schwinger}}} (\bibinfo {year} {1959}),\ \href {\doibase
  10.1103/PhysRev.115.1342} {\bibfield  {journal} {\bibinfo  {journal}
  {Physical Review}\ }\textbf {\bibinfo {volume} {115}}~(\bibinfo {number}
  {6}),\ \bibinfo {pages} {1342}}\BibitemShut {NoStop}%
\bibitem [{\citenamefont {Martinazzo}\ \emph {et~al.}(2011)\citenamefont
  {Martinazzo}, \citenamefont {Vacchini}, \citenamefont {Hughes},\ and\
  \citenamefont {Burghardt}}]{MartinazzoBurghardt2011}%
  \BibitemOpen
  \bibfield  {author} {\bibinfo {author} {\bibnamefont {Martinazzo},
  \bibfnamefont {R.}}, \bibinfo {author} {\bibfnamefont {B.}~\bibnamefont
  {Vacchini}}, \bibinfo {author} {\bibfnamefont {K.~H.}\ \bibnamefont
  {Hughes}}, \ and\ \bibinfo {author} {\bibfnamefont {I.}~\bibnamefont
  {Burghardt}}} (\bibinfo {year} {2011}),\ \href {\doibase 10.1063/1.3532408}
  {\bibfield  {journal} {\bibinfo  {journal} {J. Chem. Phys.}\ }\textbf
  {\bibinfo {volume} {134}},\ \bibinfo {pages} {011101}}\BibitemShut {NoStop}%
\bibitem [{\citenamefont {Martínez-P\'{e}rez}\ \emph
  {et~al.}(2015)\citenamefont {Martínez-P\'{e}rez}, \citenamefont {Fornieri},\
  and\ \citenamefont {Giazotto}}]{MartinezPerezGiazotto2015}%
  \BibitemOpen
  \bibfield  {author} {\bibinfo {author} {\bibnamefont {Martínez-P\'{e}rez},
  \bibfnamefont {M.~J.}}, \bibinfo {author} {\bibfnamefont {A.}~\bibnamefont
  {Fornieri}}, \ and\ \bibinfo {author} {\bibfnamefont {F.}~\bibnamefont
  {Giazotto}}} (\bibinfo {year} {2015}),\ \href {\doibase
  10.1038/nnano.2015.11} {\bibfield  {journal} {\bibinfo  {journal} {Nature
  Nanotech}\ }\textbf {\bibinfo {volume} {10}},\ \bibinfo {pages}
  {303}}\BibitemShut {NoStop}%
\bibitem [{\citenamefont {Mascarenhas}\ \emph {et~al.}(2019)\citenamefont
  {Mascarenhas}, \citenamefont {Damanet}, \citenamefont {Flannigan},
  \citenamefont {Tagliacozzo}, \citenamefont {Daley}, \citenamefont {Goold},\
  and\ \citenamefont {de~Vega}}]{MascharenasDeVega2019}%
  \BibitemOpen
  \bibfield  {author} {\bibinfo {author} {\bibnamefont {Mascarenhas},
  \bibfnamefont {E.}}, \bibinfo {author} {\bibfnamefont {F.}~\bibnamefont
  {Damanet}}, \bibinfo {author} {\bibfnamefont {S.}~\bibnamefont {Flannigan}},
  \bibinfo {author} {\bibfnamefont {L.}~\bibnamefont {Tagliacozzo}}, \bibinfo
  {author} {\bibfnamefont {A.~J.}\ \bibnamefont {Daley}}, \bibinfo {author}
  {\bibfnamefont {J.}~\bibnamefont {Goold}}, \ and\ \bibinfo {author}
  {\bibfnamefont {I.}~\bibnamefont {de~Vega}}} (\bibinfo {year} {2019}),\ \href
  {\doibase 10.1103/PhysRevB.99.245134} {\bibfield  {journal} {\bibinfo
  {journal} {Phys. Rev. B}\ }\textbf {\bibinfo {volume} {99}},\ \bibinfo
  {pages} {245134}}\BibitemShut {NoStop}%
\bibitem [{\citenamefont {Mascarenhas}\ \emph {et~al.}(2015)\citenamefont
  {Mascarenhas}, \citenamefont {Flayac},\ and\ \citenamefont
  {Savona}}]{MascarenhasSavona2015}%
  \BibitemOpen
  \bibfield  {author} {\bibinfo {author} {\bibnamefont {Mascarenhas},
  \bibfnamefont {E.}}, \bibinfo {author} {\bibfnamefont {H.}~\bibnamefont
  {Flayac}}, \ and\ \bibinfo {author} {\bibfnamefont {V.}~\bibnamefont
  {Savona}}} (\bibinfo {year} {2015}),\ \href {\doibase
  10.1103/PhysRevA.92.022116} {\bibfield  {journal} {\bibinfo  {journal} {Phys.
  Rev. A}\ }\textbf {\bibinfo {volume} {92}},\ \bibinfo {pages}
  {022116}}\BibitemShut {NoStop}%
\bibitem [{\citenamefont {Mascarenhas}\ \emph {et~al.}(2017)\citenamefont
  {Mascarenhas}, \citenamefont {Giudice},\ and\ \citenamefont
  {Savona}}]{MascarenhasSavona2017}%
  \BibitemOpen
  \bibfield  {author} {\bibinfo {author} {\bibnamefont {Mascarenhas},
  \bibfnamefont {E.}}, \bibinfo {author} {\bibfnamefont {G.}~\bibnamefont
  {Giudice}}, \ and\ \bibinfo {author} {\bibfnamefont {V.}~\bibnamefont
  {Savona}}} (\bibinfo {year} {2017}),\ \href {\doibase
  10.22331/q-2017-12-20-40} {\bibfield  {journal} {\bibinfo  {journal}
  {{Quantum}}\ }\textbf {\bibinfo {volume} {1}},\ \bibinfo {pages}
  {40}}\BibitemShut {NoStop}%
\bibitem [{\citenamefont {Mascarenhas}\ \emph {et~al.}(2016)\citenamefont
  {Mascarenhas}, \citenamefont {Santos}, \citenamefont {Auff\`eves},\ and\
  \citenamefont {Gerace}}]{MascharenasGerace2016}%
  \BibitemOpen
  \bibfield  {author} {\bibinfo {author} {\bibnamefont {Mascarenhas},
  \bibfnamefont {E.}}, \bibinfo {author} {\bibfnamefont {M.~F.}\ \bibnamefont
  {Santos}}, \bibinfo {author} {\bibfnamefont {A.}~\bibnamefont {Auff\`eves}},
  \ and\ \bibinfo {author} {\bibfnamefont {D.}~\bibnamefont {Gerace}}}
  (\bibinfo {year} {2016}),\ \href {\doibase 10.1103/PhysRevA.93.043821}
  {\bibfield  {journal} {\bibinfo  {journal} {Phys. Rev. A}\ }\textbf {\bibinfo
  {volume} {93}},\ \bibinfo {pages} {043821}}\BibitemShut {NoStop}%
\bibitem [{\citenamefont {Mascherpa}\ \emph {et~al.}(2020)\citenamefont
  {Mascherpa}, \citenamefont {Smirne}, \citenamefont {Somoza}, \citenamefont
  {Fern\'andez-Acebal}, \citenamefont {Donadi}, \citenamefont {Tamascelli},
  \citenamefont {Huelga},\ and\ \citenamefont {Plenio}}]{mascherpa2020a}%
  \BibitemOpen
  \bibfield  {author} {\bibinfo {author} {\bibnamefont {Mascherpa},
  \bibfnamefont {F.}}, \bibinfo {author} {\bibfnamefont {A.}~\bibnamefont
  {Smirne}}, \bibinfo {author} {\bibfnamefont {A.~D.}\ \bibnamefont {Somoza}},
  \bibinfo {author} {\bibfnamefont {P.}~\bibnamefont {Fern\'andez-Acebal}},
  \bibinfo {author} {\bibfnamefont {S.}~\bibnamefont {Donadi}}, \bibinfo
  {author} {\bibfnamefont {D.}~\bibnamefont {Tamascelli}}, \bibinfo {author}
  {\bibfnamefont {S.~F.}\ \bibnamefont {Huelga}}, \ and\ \bibinfo {author}
  {\bibfnamefont {M.~B.}\ \bibnamefont {Plenio}}} (\bibinfo {year} {2020}),\
  \href {\doibase 10.1103/PhysRevA.101.052108} {\bibfield  {journal} {\bibinfo
  {journal} {Phys. Rev. A}\ }\textbf {\bibinfo {volume} {101}},\ \bibinfo
  {pages} {052108}}\BibitemShut {NoStop}%
\bibitem [{\citenamefont {Mastropietro}(2013)}]{Mastropietro2013}%
  \BibitemOpen
  \bibfield  {author} {\bibinfo {author} {\bibnamefont {Mastropietro},
  \bibfnamefont {V.}}} (\bibinfo {year} {2013}),\ \href {\doibase
  10.1103/PhysRevE.87.042121} {\bibfield  {journal} {\bibinfo  {journal} {Phys.
  Rev. E}\ }\textbf {\bibinfo {volume} {87}},\ \bibinfo {pages}
  {042121}}\BibitemShut {NoStop}%
\bibitem [{\citenamefont {Mazur}(1969)}]{Mazur1969}%
  \BibitemOpen
  \bibfield  {author} {\bibinfo {author} {\bibnamefont {Mazur}, \bibfnamefont
  {P.}}} (\bibinfo {year} {1969}),\ \href {\doibase
  10.1016/0031-8914(69)90185-2} {\bibfield  {journal} {\bibinfo  {journal}
  {Physica}\ }\textbf {\bibinfo {volume} {43}},\ \bibinfo {pages}
  {533}}\BibitemShut {NoStop}%
\bibitem [{\citenamefont {McCauley}\ \emph {et~al.}(2020)\citenamefont
  {McCauley}, \citenamefont {Cruikshank}, \citenamefont {Bondar},\ and\
  \citenamefont {Jacobs}}]{mccauley2020a}%
  \BibitemOpen
  \bibfield  {author} {\bibinfo {author} {\bibnamefont {McCauley},
  \bibfnamefont {G.}}, \bibinfo {author} {\bibfnamefont {B.}~\bibnamefont
  {Cruikshank}}, \bibinfo {author} {\bibfnamefont {D.~I.}\ \bibnamefont
  {Bondar}}, \ and\ \bibinfo {author} {\bibfnamefont {K.}~\bibnamefont
  {Jacobs}}} (\bibinfo {year} {2020}),\ \href {\doibase
  10.1038/s41534-020-00299-6} {\bibfield  {journal} {\bibinfo  {journal} {npj
  Quantum Information}\ }\textbf {\bibinfo {volume} {6}},\ \bibinfo {pages}
  {74}}\BibitemShut {NoStop}%
\bibitem [{\citenamefont {McCulloch}(2007)}]{McCulloch2007}%
  \BibitemOpen
  \bibfield  {author} {\bibinfo {author} {\bibnamefont {McCulloch},
  \bibfnamefont {I.~P.}}} (\bibinfo {year} {2007}),\ \href {\doibase
  10.1088/1742-5468/2007/10/p10014} {\bibfield  {journal} {\bibinfo  {journal}
  {Journal of Statistical Mechanics: Theory and Experiment}\ }\textbf {\bibinfo
  {volume} {2007}}~(\bibinfo {number} {10}),\ \bibinfo {pages}
  {P10014}}\BibitemShut {NoStop}%
\bibitem [{\citenamefont {McCutcheon}\ \emph {et~al.}(2011)\citenamefont
  {McCutcheon}, \citenamefont {Dattani}, \citenamefont {Gauger}, \citenamefont
  {Lovett},\ and\ \citenamefont {Nazir}}]{mccutcheon2011a}%
  \BibitemOpen
  \bibfield  {author} {\bibinfo {author} {\bibnamefont {McCutcheon},
  \bibfnamefont {D.~P.~S.}}, \bibinfo {author} {\bibfnamefont {N.~S.}\
  \bibnamefont {Dattani}}, \bibinfo {author} {\bibfnamefont {E.~M.}\
  \bibnamefont {Gauger}}, \bibinfo {author} {\bibfnamefont {B.~W.}\
  \bibnamefont {Lovett}}, \ and\ \bibinfo {author} {\bibfnamefont
  {A.}~\bibnamefont {Nazir}}} (\bibinfo {year} {2011}),\ \href {\doibase
  10.1103/PhysRevB.84.081305} {\bibfield  {journal} {\bibinfo  {journal} {Phys.
  Rev. B}\ }\textbf {\bibinfo {volume} {84}},\ \bibinfo {pages}
  {081305}}\BibitemShut {NoStop}%
\bibitem [{\citenamefont {Medvedyeva}\ \emph
  {et~al.}(2016{\natexlab{a}})\citenamefont {Medvedyeva}, \citenamefont
  {Essler},\ and\ \citenamefont {Prosen}}]{MedvedyevaProsen2016}%
  \BibitemOpen
  \bibfield  {author} {\bibinfo {author} {\bibnamefont {Medvedyeva},
  \bibfnamefont {M.~V.}}, \bibinfo {author} {\bibfnamefont {F.~H.~L.}\
  \bibnamefont {Essler}}, \ and\ \bibinfo {author} {\bibfnamefont
  {T.}~\bibnamefont {Prosen}}} (\bibinfo {year} {2016}{\natexlab{a}}),\ \href
  {\doibase 10.1103/PhysRevLett.117.137202} {\bibfield  {journal} {\bibinfo
  {journal} {Phys. Rev. Lett.}\ }\textbf {\bibinfo {volume} {117}},\ \bibinfo
  {pages} {137202}}\BibitemShut {NoStop}%
\bibitem [{\citenamefont {Medvedyeva}\ \emph
  {et~al.}(2016{\natexlab{b}})\citenamefont {Medvedyeva}, \citenamefont
  {Prosen},\ and\ \citenamefont {\v{Z}nidari\v{c}}}]{MedvedyevaZnidaric2016}%
  \BibitemOpen
  \bibfield  {author} {\bibinfo {author} {\bibnamefont {Medvedyeva},
  \bibfnamefont {M.~V.}}, \bibinfo {author} {\bibfnamefont {T.}~\bibnamefont
  {Prosen}}, \ and\ \bibinfo {author} {\bibfnamefont {M.}~\bibnamefont
  {\v{Z}nidari\v{c}}}} (\bibinfo {year} {2016}{\natexlab{b}}),\ \href {\doibase
  10.1103/PhysRevB.93.094205} {\bibfield  {journal} {\bibinfo  {journal} {Phys.
  Rev. B}\ }\textbf {\bibinfo {volume} {93}},\ \bibinfo {pages}
  {094205}}\BibitemShut {NoStop}%
\bibitem [{\citenamefont {Meir}\ and\ \citenamefont
  {Wingreen}(1992)}]{MeirWingreen1992}%
  \BibitemOpen
  \bibfield  {author} {\bibinfo {author} {\bibnamefont {Meir}, \bibfnamefont
  {Y.}}, \ and\ \bibinfo {author} {\bibfnamefont {N.~S.}\ \bibnamefont
  {Wingreen}}} (\bibinfo {year} {1992}),\ \href {\doibase
  10.1103/PhysRevLett.68.2512} {\bibfield  {journal} {\bibinfo  {journal}
  {Phys. Rev. Lett.}\ }\textbf {\bibinfo {volume} {68}}~(\bibinfo {number}
  {16}),\ \bibinfo {pages} {2512}}\BibitemShut {NoStop}%
\bibitem [{\citenamefont {Meir}\ \emph {et~al.}(1993)\citenamefont {Meir},
  \citenamefont {Wingreen},\ and\ \citenamefont {Lee}}]{meir1993a}%
  \BibitemOpen
  \bibfield  {author} {\bibinfo {author} {\bibnamefont {Meir}, \bibfnamefont
  {Y.}}, \bibinfo {author} {\bibfnamefont {N.~S.}\ \bibnamefont {Wingreen}}, \
  and\ \bibinfo {author} {\bibfnamefont {P.~A.}\ \bibnamefont {Lee}}} (\bibinfo
  {year} {1993}),\ \href {\doibase 10.1103/PhysRevLett.70.2601} {\bibfield
  {journal} {\bibinfo  {journal} {Phys. Rev. Lett.}\ }\textbf {\bibinfo
  {volume} {70}},\ \bibinfo {pages} {2601}}\BibitemShut {NoStop}%
\bibitem [{\citenamefont {Mendoza-Arenas}\ \emph
  {et~al.}(2013{\natexlab{a}})\citenamefont {Mendoza-Arenas}, \citenamefont
  {Al-Assam}, \citenamefont {Clark},\ and\ \citenamefont
  {Jaksch}}]{Mendoza-Arenas2013}%
  \BibitemOpen
  \bibfield  {author} {\bibinfo {author} {\bibnamefont {Mendoza-Arenas},
  \bibfnamefont {J.~J.}}, \bibinfo {author} {\bibfnamefont {S.}~\bibnamefont
  {Al-Assam}}, \bibinfo {author} {\bibfnamefont {S.~R.}\ \bibnamefont {Clark}},
  \ and\ \bibinfo {author} {\bibfnamefont {D.}~\bibnamefont {Jaksch}}}
  (\bibinfo {year} {2013}{\natexlab{a}}),\ \href {\doibase
  10.1088/1742-5468/2013/07/P07007} {\bibfield  {journal} {\bibinfo  {journal}
  {Journal of Statistical Mechanics: Theory and Experiment}\ }\textbf {\bibinfo
  {volume} {2013}}~(\bibinfo {number} {07}),\ \bibinfo {pages}
  {P07007}}\BibitemShut {NoStop}%
\bibitem [{\citenamefont {Mendoza-Arenas}\ \emph {et~al.}(2015)\citenamefont
  {Mendoza-Arenas}, \citenamefont {Clark},\ and\ \citenamefont
  {Jaksch}}]{Mendoza-Arenas2014a}%
  \BibitemOpen
  \bibfield  {author} {\bibinfo {author} {\bibnamefont {Mendoza-Arenas},
  \bibfnamefont {J.~J.}}, \bibinfo {author} {\bibfnamefont {S.~R.}\
  \bibnamefont {Clark}}, \ and\ \bibinfo {author} {\bibfnamefont
  {D.}~\bibnamefont {Jaksch}}} (\bibinfo {year} {2015}),\ \href {\doibase
  10.1103/PhysRevE.91.042129} {\bibfield  {journal} {\bibinfo  {journal} {Phys.
  Rev. E}\ }\textbf {\bibinfo {volume} {91}},\ \bibinfo {pages}
  {042129}}\BibitemShut {NoStop}%
\bibitem [{\citenamefont {Mendoza-Arenas}\ \emph
  {et~al.}(2013{\natexlab{b}})\citenamefont {Mendoza-Arenas}, \citenamefont
  {Grujic}, \citenamefont {Jaksch},\ and\ \citenamefont
  {Clark}}]{MendozaArenaClark2014}%
  \BibitemOpen
  \bibfield  {author} {\bibinfo {author} {\bibnamefont {Mendoza-Arenas},
  \bibfnamefont {J.~J.}}, \bibinfo {author} {\bibfnamefont {T.}~\bibnamefont
  {Grujic}}, \bibinfo {author} {\bibfnamefont {D.}~\bibnamefont {Jaksch}}, \
  and\ \bibinfo {author} {\bibfnamefont {S.~R.}\ \bibnamefont {Clark}}}
  (\bibinfo {year} {2013}{\natexlab{b}}),\ \href {\doibase
  10.1103/PhysRevB.87.235130} {\bibfield  {journal} {\bibinfo  {journal} {Phys.
  Rev. B}\ }\textbf {\bibinfo {volume} {87}},\ \bibinfo {pages}
  {235130}}\BibitemShut {NoStop}%
\bibitem [{\citenamefont {Mendoza-Arenas}\ \emph {et~al.}(2014)\citenamefont
  {Mendoza-Arenas}, \citenamefont {Mitchison}, \citenamefont {Clark},
  \citenamefont {Prior}, \citenamefont {Jaksch},\ and\ \citenamefont
  {Plenio}}]{Mendoza_ArenasPlenio2014}%
  \BibitemOpen
  \bibfield  {author} {\bibinfo {author} {\bibnamefont {Mendoza-Arenas},
  \bibfnamefont {J.~J.}}, \bibinfo {author} {\bibfnamefont {M.~T.}\
  \bibnamefont {Mitchison}}, \bibinfo {author} {\bibfnamefont {S.~R.}\
  \bibnamefont {Clark}}, \bibinfo {author} {\bibfnamefont {J.}~\bibnamefont
  {Prior}}, \bibinfo {author} {\bibfnamefont {D.}~\bibnamefont {Jaksch}}, \
  and\ \bibinfo {author} {\bibfnamefont {M.~B.}\ \bibnamefont {Plenio}}}
  (\bibinfo {year} {2014}),\ \href {\doibase 10.1088/1367-2630/16/5/053016}
  {\bibfield  {journal} {\bibinfo  {journal} {New Journal of Physics}\ }\textbf
  {\bibinfo {volume} {16}}~(\bibinfo {number} {5}),\ \bibinfo {pages}
  {053016}}\BibitemShut {NoStop}%
\bibitem [{\citenamefont {Mendoza-Arenas}\ \emph {et~al.}(2019)\citenamefont
  {Mendoza-Arenas}, \citenamefont {\v{Z}nidari\v{c}}, \citenamefont {Varma},
  \citenamefont {Goold}, \citenamefont {Clark},\ and\ \citenamefont
  {Scardicchio}}]{MendozaArenasScardicchio2019}%
  \BibitemOpen
  \bibfield  {author} {\bibinfo {author} {\bibnamefont {Mendoza-Arenas},
  \bibfnamefont {J.~J.}}, \bibinfo {author} {\bibfnamefont {M.}~\bibnamefont
  {\v{Z}nidari\v{c}}}, \bibinfo {author} {\bibfnamefont {V.~K.}\ \bibnamefont
  {Varma}}, \bibinfo {author} {\bibfnamefont {J.}~\bibnamefont {Goold}},
  \bibinfo {author} {\bibfnamefont {S.~R.}\ \bibnamefont {Clark}}, \ and\
  \bibinfo {author} {\bibfnamefont {A.}~\bibnamefont {Scardicchio}}} (\bibinfo
  {year} {2019}),\ \href {\doibase 10.1103/PhysRevB.99.094435} {\bibfield
  {journal} {\bibinfo  {journal} {Phys. Rev. B}\ }\textbf {\bibinfo {volume}
  {99}},\ \bibinfo {pages} {094435}}\BibitemShut {NoStop}%
\bibitem [{\citenamefont {Michaelis}\ \emph {et~al.}(2006)\citenamefont
  {Michaelis}, \citenamefont {Emary},\ and\ \citenamefont
  {Beenakker}}]{MichaelisBeenakker2006}%
  \BibitemOpen
  \bibfield  {author} {\bibinfo {author} {\bibnamefont {Michaelis},
  \bibfnamefont {B.}}, \bibinfo {author} {\bibfnamefont {C.}~\bibnamefont
  {Emary}}, \ and\ \bibinfo {author} {\bibfnamefont {C.~W.~J.}\ \bibnamefont
  {Beenakker}}} (\bibinfo {year} {2006}),\ \href {\doibase
  10.1209/epl/i2005-10458-6} {\bibfield  {journal} {\bibinfo  {journal}
  {Europhysics Letters ({EPL})}\ }\textbf {\bibinfo {volume} {73}}~(\bibinfo
  {number} {5}),\ \bibinfo {pages} {677}}\BibitemShut {NoStop}%
\bibitem [{\citenamefont {Michel}\ \emph {et~al.}(2004)\citenamefont {Michel},
  \citenamefont {Gemmer},\ and\ \citenamefont {Mahler}}]{MichelMahler2004}%
  \BibitemOpen
  \bibfield  {author} {\bibinfo {author} {\bibnamefont {Michel}, \bibfnamefont
  {M.}}, \bibinfo {author} {\bibfnamefont {J.}~\bibnamefont {Gemmer}}, \ and\
  \bibinfo {author} {\bibfnamefont {G.}~\bibnamefont {Mahler}}} (\bibinfo
  {year} {2004}),\ \href {\doibase 10.1140/epjb/e2005-00014-x} {\bibfield
  {journal} {\bibinfo  {journal} {Eur. Phys. J. B}\ }\textbf {\bibinfo {volume}
  {42}},\ \bibinfo {pages} {555}}\BibitemShut {NoStop}%
\bibitem [{\citenamefont {Mierzejewski}\ \emph {et~al.}(2019)\citenamefont
  {Mierzejewski}, \citenamefont {Prelov\v{s}ek},\ and\ \citenamefont
  {Bon\v{c}a}}]{MierzejewskiBonca2019}%
  \BibitemOpen
  \bibfield  {author} {\bibinfo {author} {\bibnamefont {Mierzejewski},
  \bibfnamefont {M.}}, \bibinfo {author} {\bibfnamefont {P.}~\bibnamefont
  {Prelov\v{s}ek}}, \ and\ \bibinfo {author} {\bibfnamefont {J.}~\bibnamefont
  {Bon\v{c}a}}} (\bibinfo {year} {2019}),\ \href {\doibase
  10.1103/PhysRevLett.122.206601} {\bibfield  {journal} {\bibinfo  {journal}
  {Phys. Rev. Lett.}\ }\textbf {\bibinfo {volume} {122}},\ \bibinfo {pages}
  {206601}}\BibitemShut {NoStop}%
\bibitem [{\citenamefont {Mierzejewski}\ \emph {et~al.}(2020)\citenamefont
  {Mierzejewski}, \citenamefont {\ifmmode~\acute{S}\else \'{S}\fi{}roda},
  \citenamefont {Herbrych},\ and\ \citenamefont {Prelov\ifmmode~\check{s}\else
  \v{s}\fi{}ek}}]{MierzejewskiPrelovsek2020}%
  \BibitemOpen
  \bibfield  {author} {\bibinfo {author} {\bibnamefont {Mierzejewski},
  \bibfnamefont {M.}}, \bibinfo {author} {\bibfnamefont {M.}~\bibnamefont
  {\ifmmode~\acute{S}\else \'{S}\fi{}roda}}, \bibinfo {author} {\bibfnamefont
  {J.}~\bibnamefont {Herbrych}}, \ and\ \bibinfo {author} {\bibfnamefont
  {P.}~\bibnamefont {Prelov\ifmmode~\check{s}\else \v{s}\fi{}ek}}} (\bibinfo
  {year} {2020}),\ \href {\doibase 10.1103/PhysRevB.102.161111} {\bibfield
  {journal} {\bibinfo  {journal} {Phys. Rev. B}\ }\textbf {\bibinfo {volume}
  {102}},\ \bibinfo {pages} {161111}}\BibitemShut {NoStop}%
\bibitem [{\citenamefont {Millen}\ and\ \citenamefont
  {Xuereb}(2016)}]{MillenXuereb2016}%
  \BibitemOpen
  \bibfield  {author} {\bibinfo {author} {\bibnamefont {Millen}, \bibfnamefont
  {J.}}, \ and\ \bibinfo {author} {\bibfnamefont {A.}~\bibnamefont {Xuereb}}}
  (\bibinfo {year} {2016}),\ \href {\doibase 10.1088/1367-2630/18/1/011002}
  {\bibfield  {journal} {\bibinfo  {journal} {New Journal of Physics}\ }\textbf
  {\bibinfo {volume} {18}}~(\bibinfo {number} {1}),\ \bibinfo {pages}
  {011002}}\BibitemShut {NoStop}%
\bibitem [{\citenamefont {Minganti}\ \emph {et~al.}(2018)\citenamefont
  {Minganti}, \citenamefont {Biella}, \citenamefont {Bartolo},\ and\
  \citenamefont {Ciuti}}]{minganti2018a}%
  \BibitemOpen
  \bibfield  {author} {\bibinfo {author} {\bibnamefont {Minganti},
  \bibfnamefont {F.}}, \bibinfo {author} {\bibfnamefont {A.}~\bibnamefont
  {Biella}}, \bibinfo {author} {\bibfnamefont {N.}~\bibnamefont {Bartolo}}, \
  and\ \bibinfo {author} {\bibfnamefont {C.}~\bibnamefont {Ciuti}}} (\bibinfo
  {year} {2018}),\ \href {\doibase 10.1103/PhysRevA.98.042118} {\bibfield
  {journal} {\bibinfo  {journal} {Phys. Rev. A}\ }\textbf {\bibinfo {volume}
  {98}},\ \bibinfo {pages} {042118}}\BibitemShut {NoStop}%
\bibitem [{\citenamefont {Mitchison}\ and\ \citenamefont
  {Plenio}(2018)}]{Mitchison2018}%
  \BibitemOpen
  \bibfield  {author} {\bibinfo {author} {\bibnamefont {Mitchison},
  \bibfnamefont {M.~T.}}, \ and\ \bibinfo {author} {\bibfnamefont {M.~B.}\
  \bibnamefont {Plenio}}} (\bibinfo {year} {2018}),\ \href {\doibase
  10.1088/1367-2630/aa9f70} {\bibfield  {journal} {\bibinfo  {journal} {New
  Journal of Physics}\ }\textbf {\bibinfo {volume} {20}},\ \bibinfo {pages}
  {033005}}\BibitemShut {NoStop}%
\bibitem [{\citenamefont {Moca}\ \emph {et~al.}(2022)\citenamefont {Moca},
  \citenamefont {Werner}, \citenamefont {Legeza}, \citenamefont {Prosen},
  \citenamefont {Kormos},\ and\ \citenamefont {Zar\'and}}]{Moca2022}%
  \BibitemOpen
  \bibfield  {author} {\bibinfo {author} {\bibnamefont {Moca}, \bibfnamefont
  {C.~P.}}, \bibinfo {author} {\bibfnamefont {M.~A.}\ \bibnamefont {Werner}},
  \bibinfo {author} {\bibfnamefont {O.}~\bibnamefont {Legeza}}, \bibinfo
  {author} {\bibfnamefont {T.}~\bibnamefont {Prosen}}, \bibinfo {author}
  {\bibfnamefont {M.}~\bibnamefont {Kormos}}, \ and\ \bibinfo {author}
  {\bibfnamefont {G.}~\bibnamefont {Zar\'and}}} (\bibinfo {year} {2022}),\
  \href {\doibase 10.1103/PhysRevB.105.195144} {\bibfield  {journal} {\bibinfo
  {journal} {Phys. Rev. B}\ }\textbf {\bibinfo {volume} {105}},\ \bibinfo
  {pages} {195144}}\BibitemShut {NoStop}%
\bibitem [{\citenamefont {Modak}\ \emph {et~al.}(2018)\citenamefont {Modak},
  \citenamefont {Ghosh},\ and\ \citenamefont {Mukerjee}}]{ModakMukerjee2018}%
  \BibitemOpen
  \bibfield  {author} {\bibinfo {author} {\bibnamefont {Modak}, \bibfnamefont
  {R.}}, \bibinfo {author} {\bibfnamefont {S.}~\bibnamefont {Ghosh}}, \ and\
  \bibinfo {author} {\bibfnamefont {S.}~\bibnamefont {Mukerjee}}} (\bibinfo
  {year} {2018}),\ \href {\doibase 10.1103/PhysRevB.97.104204} {\bibfield
  {journal} {\bibinfo  {journal} {Phys. Rev. B}\ }\textbf {\bibinfo {volume}
  {97}},\ \bibinfo {pages} {104204}}\BibitemShut {NoStop}%
\bibitem [{\citenamefont {Modak}\ and\ \citenamefont
  {Mukerjee}(2015)}]{ModakMukerjee2015}%
  \BibitemOpen
  \bibfield  {author} {\bibinfo {author} {\bibnamefont {Modak}, \bibfnamefont
  {R.}}, \ and\ \bibinfo {author} {\bibfnamefont {S.}~\bibnamefont {Mukerjee}}}
  (\bibinfo {year} {2015}),\ \href {\doibase 10.1103/PhysRevLett.115.230401}
  {\bibfield  {journal} {\bibinfo  {journal} {Phys. Rev. Lett.}\ }\textbf
  {\bibinfo {volume} {115}},\ \bibinfo {pages} {230401}}\BibitemShut {NoStop}%
\bibitem [{\citenamefont {M\o{}lmer}\ \emph {et~al.}(1993)\citenamefont
  {M\o{}lmer}, \citenamefont {Castin},\ and\ \citenamefont
  {Dalibard}}]{MolmerDalibard1993}%
  \BibitemOpen
  \bibfield  {author} {\bibinfo {author} {\bibnamefont {M\o{}lmer},
  \bibfnamefont {K.}}, \bibinfo {author} {\bibfnamefont {Y.}~\bibnamefont
  {Castin}}, \ and\ \bibinfo {author} {\bibfnamefont {J.}~\bibnamefont
  {Dalibard}}} (\bibinfo {year} {1993}),\ \href {\doibase
  10.1364/JOSAB.10.000524} {\bibfield  {journal} {\bibinfo  {journal} {Journal
  of the Optical Society of America B}\ }\textbf {\bibinfo {volume} {10}},\
  \bibinfo {pages} {524}}\BibitemShut {NoStop}%
\bibitem [{\citenamefont {Monthus}(2017)}]{Monthus2017}%
  \BibitemOpen
  \bibfield  {author} {\bibinfo {author} {\bibnamefont {Monthus}, \bibfnamefont
  {C.}}} (\bibinfo {year} {2017}),\ \href {\doibase 10.1088/1742-5468/aa6a2f}
  {\bibfield  {journal} {\bibinfo  {journal} {Journal of Statistical Mechanics:
  Theory and Experiment}\ }\textbf {\bibinfo {volume} {2017}}~(\bibinfo
  {number} {4}),\ \bibinfo {pages} {043303}}\BibitemShut {NoStop}%
\bibitem [{\citenamefont {Mori}\ and\ \citenamefont
  {Miyashita}(2008)}]{mori2008a}%
  \BibitemOpen
  \bibfield  {author} {\bibinfo {author} {\bibnamefont {Mori}, \bibfnamefont
  {T.}}, \ and\ \bibinfo {author} {\bibfnamefont {S.}~\bibnamefont
  {Miyashita}}} (\bibinfo {year} {2008}),\ \href {\doibase
  10.1143/JPSJ.77.124005} {\bibfield  {journal} {\bibinfo  {journal} {Journal
  of the Physical Society of Japan}\ }\textbf {\bibinfo {volume}
  {77}}~(\bibinfo {number} {12}),\ \bibinfo {pages} {124005}}\BibitemShut
  {NoStop}%
\bibitem [{\citenamefont {Mori}\ and\ \citenamefont
  {Shirai}(2020)}]{MoriShirai2020}%
  \BibitemOpen
  \bibfield  {author} {\bibinfo {author} {\bibnamefont {Mori}, \bibfnamefont
  {T.}}, \ and\ \bibinfo {author} {\bibfnamefont {T.}~\bibnamefont {Shirai}}}
  (\bibinfo {year} {2020}),\ \href {\doibase 10.1103/PhysRevLett.125.230604}
  {\bibfield  {journal} {\bibinfo  {journal} {Phys. Rev. Lett.}\ }\textbf
  {\bibinfo {volume} {125}},\ \bibinfo {pages} {230604}}\BibitemShut {NoStop}%
\bibitem [{\citenamefont {Morrison}\ and\ \citenamefont
  {Parkins}(2008)}]{morrison2008a}%
  \BibitemOpen
  \bibfield  {author} {\bibinfo {author} {\bibnamefont {Morrison},
  \bibfnamefont {S.}}, \ and\ \bibinfo {author} {\bibfnamefont {A.~S.}\
  \bibnamefont {Parkins}}} (\bibinfo {year} {2008}),\ \href {\doibase
  10.1103/PhysRevLett.100.040403} {\bibfield  {journal} {\bibinfo  {journal}
  {Physical Review Letters}\ }\textbf {\bibinfo {volume} {100}},\ \bibinfo
  {pages} {040403}}\BibitemShut {NoStop}%
\bibitem [{\citenamefont {Moskalenko}\ \emph {et~al.}(2009)\citenamefont
  {Moskalenko}, \citenamefont {Gordeev}, \citenamefont {Koentjoro},
  \citenamefont {Raithby}, \citenamefont {French}, \citenamefont {Marken},\
  and\ \citenamefont {Savel'ev}}]{moskalenko2009a}%
  \BibitemOpen
  \bibfield  {author} {\bibinfo {author} {\bibnamefont {Moskalenko},
  \bibfnamefont {A.~V.}}, \bibinfo {author} {\bibfnamefont {S.~N.}\
  \bibnamefont {Gordeev}}, \bibinfo {author} {\bibfnamefont {O.~F.}\
  \bibnamefont {Koentjoro}}, \bibinfo {author} {\bibfnamefont {P.~R.}\
  \bibnamefont {Raithby}}, \bibinfo {author} {\bibfnamefont {R.~W.}\
  \bibnamefont {French}}, \bibinfo {author} {\bibfnamefont {F.}~\bibnamefont
  {Marken}}, \ and\ \bibinfo {author} {\bibfnamefont {S.~E.}\ \bibnamefont
  {Savel'ev}}} (\bibinfo {year} {2009}),\ \href {\doibase
  10.1103/PhysRevB.79.241403} {\bibfield  {journal} {\bibinfo  {journal} {Phys.
  Rev. B}\ }\textbf {\bibinfo {volume} {79}},\ \bibinfo {pages}
  {241403}}\BibitemShut {NoStop}%
\bibitem [{\citenamefont {Mott}(1969)}]{Mott1969}%
  \BibitemOpen
  \bibfield  {author} {\bibinfo {author} {\bibnamefont {Mott}, \bibfnamefont
  {N.~F.}}} (\bibinfo {year} {1969}),\ \href {\doibase
  10.1080/14786436908216338} {\bibfield  {journal} {\bibinfo  {journal}
  {Philosophical Magazine}\ }\textbf {\bibinfo {volume} {19}},\ \bibinfo
  {pages} {835}}\BibitemShut {NoStop}%
\bibitem [{\citenamefont {Mourokh}\ \emph {et~al.}(2002)\citenamefont
  {Mourokh}, \citenamefont {Horing},\ and\ \citenamefont
  {Smirnov}}]{MourokhSmirnov2002}%
  \BibitemOpen
  \bibfield  {author} {\bibinfo {author} {\bibnamefont {Mourokh}, \bibfnamefont
  {L.~G.}}, \bibinfo {author} {\bibfnamefont {N.~J.~M.}\ \bibnamefont
  {Horing}}, \ and\ \bibinfo {author} {\bibfnamefont {A.~Y.}\ \bibnamefont
  {Smirnov}}} (\bibinfo {year} {2002}),\ \href {\doibase
  10.1103/PhysRevB.66.085332} {\bibfield  {journal} {\bibinfo  {journal} {Phys.
  Rev. B}\ }\textbf {\bibinfo {volume} {66}},\ \bibinfo {pages}
  {085332}}\BibitemShut {NoStop}%
\bibitem [{\citenamefont {Mu}\ \emph {et~al.}(2017)\citenamefont {Mu},
  \citenamefont {Agarwalla}, \citenamefont {Schaller},\ and\ \citenamefont
  {Segal}}]{mu2017a}%
  \BibitemOpen
  \bibfield  {author} {\bibinfo {author} {\bibnamefont {Mu}, \bibfnamefont
  {A.}}, \bibinfo {author} {\bibfnamefont {B.~K.}\ \bibnamefont {Agarwalla}},
  \bibinfo {author} {\bibfnamefont {G.}~\bibnamefont {Schaller}}, \ and\
  \bibinfo {author} {\bibfnamefont {D.}~\bibnamefont {Segal}}} (\bibinfo {year}
  {2017}),\ \href {\doibase 10.1088/1367-2630/aa9b75} {\bibfield  {journal}
  {\bibinfo  {journal} {New Journal of Physics}\ }\textbf {\bibinfo {volume}
  {19}},\ \bibinfo {pages} {123034}}\BibitemShut {NoStop}%
\bibitem [{\citenamefont {M{\"u}hlbacher}\ and\ \citenamefont
  {Rabani}(2008)}]{MuhlbacherRabani2008}%
  \BibitemOpen
  \bibfield  {author} {\bibinfo {author} {\bibnamefont {M{\"u}hlbacher},
  \bibfnamefont {L.}}, \ and\ \bibinfo {author} {\bibfnamefont
  {E.}~\bibnamefont {Rabani}}} (\bibinfo {year} {2008}),\ \href {\doibase
  10.1103/PhysRevLett.100.176403} {\bibfield  {journal} {\bibinfo  {journal}
  {Phys. Rev. Lett.}\ }\textbf {\bibinfo {volume} {100}},\ \bibinfo {pages}
  {176403}}\BibitemShut {NoStop}%
\bibitem [{\citenamefont {Mujica}\ \emph {et~al.}(2003)\citenamefont {Mujica},
  \citenamefont {Nitzan}, \citenamefont {Datta}, \citenamefont {Ratner},\ and\
  \citenamefont {Kubiak}}]{MujicaKubiak2003}%
  \BibitemOpen
  \bibfield  {author} {\bibinfo {author} {\bibnamefont {Mujica}, \bibfnamefont
  {V.}}, \bibinfo {author} {\bibfnamefont {A.}~\bibnamefont {Nitzan}}, \bibinfo
  {author} {\bibfnamefont {S.}~\bibnamefont {Datta}}, \bibinfo {author}
  {\bibfnamefont {M.~A.}\ \bibnamefont {Ratner}}, \ and\ \bibinfo {author}
  {\bibfnamefont {C.~P.}\ \bibnamefont {Kubiak}}} (\bibinfo {year} {2003}),\
  \href {\doibase 10.1021/jp0216427} {\bibfield  {journal} {\bibinfo  {journal}
  {J. Phys. Chem. B}\ }\textbf {\bibinfo {volume} {107}},\ \bibinfo {pages}
  {91}}\BibitemShut {NoStop}%
\bibitem [{\citenamefont {Nagy}\ and\ \citenamefont
  {Savona}(2018)}]{NagySavona2018}%
  \BibitemOpen
  \bibfield  {author} {\bibinfo {author} {\bibnamefont {Nagy}, \bibfnamefont
  {A.}}, \ and\ \bibinfo {author} {\bibfnamefont {V.}~\bibnamefont {Savona}}}
  (\bibinfo {year} {2018}),\ \href {\doibase 10.1103/PhysRevA.97.052129}
  {\bibfield  {journal} {\bibinfo  {journal} {Phys. Rev. A}\ }\textbf {\bibinfo
  {volume} {97}},\ \bibinfo {pages} {052129}}\BibitemShut {NoStop}%
\bibitem [{\citenamefont {Nagy}\ and\ \citenamefont
  {Savona}(2019)}]{NagySavona2019}%
  \BibitemOpen
  \bibfield  {author} {\bibinfo {author} {\bibnamefont {Nagy}, \bibfnamefont
  {A.}}, \ and\ \bibinfo {author} {\bibfnamefont {V.}~\bibnamefont {Savona}}}
  (\bibinfo {year} {2019}),\ \href {\doibase 10.1103/PhysRevLett.122.250501}
  {\bibfield  {journal} {\bibinfo  {journal} {Phys. Rev. Lett.}\ }\textbf
  {\bibinfo {volume} {122}},\ \bibinfo {pages} {250501}}\BibitemShut {NoStop}%
\bibitem [{\citenamefont {Nakagawa}\ \emph {et~al.}(2021)\citenamefont
  {Nakagawa}, \citenamefont {Kawakami},\ and\ \citenamefont
  {Ueda}}]{Nakagawa2021}%
  \BibitemOpen
  \bibfield  {author} {\bibinfo {author} {\bibnamefont {Nakagawa},
  \bibfnamefont {M.}}, \bibinfo {author} {\bibfnamefont {N.}~\bibnamefont
  {Kawakami}}, \ and\ \bibinfo {author} {\bibfnamefont {M.}~\bibnamefont
  {Ueda}}} (\bibinfo {year} {2021}),\ \href {\doibase
  10.1103/PhysRevLett.126.110404} {\bibfield  {journal} {\bibinfo  {journal}
  {Phys. Rev. Lett.}\ }\textbf {\bibinfo {volume} {126}},\ \bibinfo {pages}
  {110404}}\BibitemShut {NoStop}%
\bibitem [{\citenamefont {Nakajima}(1958)}]{nakajima1958a}%
  \BibitemOpen
  \bibfield  {author} {\bibinfo {author} {\bibnamefont {Nakajima},
  \bibfnamefont {S.}}} (\bibinfo {year} {1958}),\ \href {\doibase
  10.1143/PTP.20.948} {\bibfield  {journal} {\bibinfo  {journal} {Progress of
  Theoretical Physics}\ }\textbf {\bibinfo {volume} {20}}~(\bibinfo {number}
  {6}),\ \bibinfo {pages} {948}}\BibitemShut {NoStop}%
\bibitem [{\citenamefont {Nandkishore}\ and\ \citenamefont
  {Huse}(2015)}]{NandkishoreHuse2015}%
  \BibitemOpen
  \bibfield  {author} {\bibinfo {author} {\bibnamefont {Nandkishore},
  \bibfnamefont {R.}}, \ and\ \bibinfo {author} {\bibfnamefont
  {D.}~\bibnamefont {Huse}}} (\bibinfo {year} {2015}),\ \href {\doibase
  10.1146/annurev-conmatphys-031214-014726} {\bibfield  {journal} {\bibinfo
  {journal} {Annual Review of Condensed Matter Physics}\ }\textbf {\bibinfo
  {volume} {6}},\ \bibinfo {pages} {15}}\BibitemShut {NoStop}%
\bibitem [{\citenamefont {Navez}\ and\ \citenamefont
  {Sch{\"u}tzhold}(2010)}]{navez2010a}%
  \BibitemOpen
  \bibfield  {author} {\bibinfo {author} {\bibnamefont {Navez}, \bibfnamefont
  {P.}}, \ and\ \bibinfo {author} {\bibfnamefont {R.}~\bibnamefont
  {Sch{\"u}tzhold}}} (\bibinfo {year} {2010}),\ \href {\doibase
  10.1103/PhysRevA.82.063603} {\bibfield  {journal} {\bibinfo  {journal} {Phys.
  Rev. A}\ }\textbf {\bibinfo {volume} {82}},\ \bibinfo {pages}
  {063603}}\BibitemShut {NoStop}%
\bibitem [{\citenamefont {Nazarov}\ and\ \citenamefont
  {Blanter}(2009)}]{nazarov2009}%
  \BibitemOpen
  \bibfield  {author} {\bibinfo {author} {\bibnamefont {Nazarov}, \bibfnamefont
  {Y.~V.}}, \ and\ \bibinfo {author} {\bibfnamefont {Y.~M.}\ \bibnamefont
  {Blanter}}} (\bibinfo {year} {2009}),\ \href {\doibase
  10.1017/CBO9780511626906} {\emph {\bibinfo {title} {Quantum Transport:
  Introduction to Nanoscience}}}\ (\bibinfo  {publisher} {Cambridge University
  Press},\ \bibinfo {address} {Cambridge})\BibitemShut {NoStop}%
\bibitem [{\citenamefont {Nazir}\ and\ \citenamefont
  {Schaller}(2019)}]{nazir2019a}%
  \BibitemOpen
  \bibfield  {author} {\bibinfo {author} {\bibnamefont {Nazir}, \bibfnamefont
  {A.}}, \ and\ \bibinfo {author} {\bibfnamefont {G.}~\bibnamefont {Schaller}}}
  (\bibinfo {year} {2019}),\ in\ \href {\doibase 10.1007/978-3-319-99046-0}
  {\emph {\bibinfo {booktitle} {Thermodynamics in the quantum regime -- Recent
  progress and outlook}}},\ \bibinfo {series and number} {Fundamental Theories
  of Physics},\ \bibinfo {editor} {edited by\ \bibinfo {editor} {\bibfnamefont
  {F.}~\bibnamefont {Binder}}, \bibinfo {editor} {\bibfnamefont {L.~A.}\
  \bibnamefont {Correa}}, \bibinfo {editor} {\bibfnamefont {C.}~\bibnamefont
  {Gogolin}}, \bibinfo {editor} {\bibfnamefont {J.}~\bibnamefont {Anders}}, \
  and\ \bibinfo {editor} {\bibfnamefont {G.}~\bibnamefont {Adesso}}}\ (\bibinfo
   {publisher} {Springer},\ \bibinfo {address} {Cham})\ p.\ \bibinfo {pages}
  {551}\BibitemShut {NoStop}%
\bibitem [{\citenamefont {Newns}(1969)}]{Newns1969}%
  \BibitemOpen
  \bibfield  {author} {\bibinfo {author} {\bibnamefont {Newns}, \bibfnamefont
  {D.~M.}}} (\bibinfo {year} {1969}),\ \href {\doibase
  10.1103/PhysRev.178.1123} {\bibfield  {journal} {\bibinfo  {journal} {Phys.
  Rev.}\ }\textbf {\bibinfo {volume} {178}},\ \bibinfo {pages}
  {1123}}\BibitemShut {NoStop}%
\bibitem [{\citenamefont {Nielsen}\ and\ \citenamefont
  {Chuang}(2000)}]{NielsenChuang2000}%
  \BibitemOpen
  \bibfield  {author} {\bibinfo {author} {\bibnamefont {Nielsen}, \bibfnamefont
  {M.~A.}}, \ and\ \bibinfo {author} {\bibfnamefont {I.~L.}\ \bibnamefont
  {Chuang}}} (\bibinfo {year} {2000}),\ \href {\doibase
  10.1017/CBO9780511976667} {\emph {\bibinfo {title} {Quantum Computation and
  Quantum Information}}}\ (\bibinfo  {publisher} {Cambridge University
  Press})\BibitemShut {NoStop}%
\bibitem [{\citenamefont {Nigro}(2019)}]{nigro2019a}%
  \BibitemOpen
  \bibfield  {author} {\bibinfo {author} {\bibnamefont {Nigro}, \bibfnamefont
  {D.}}} (\bibinfo {year} {2019}),\ \href {\doibase 10.1088/1742-5468/ab0c1c}
  {\bibfield  {journal} {\bibinfo  {journal} {Journal of Statistical Mechanics:
  Theory and Experiment}\ }\textbf {\bibinfo {volume} {2019}},\ \bibinfo
  {pages} {043202}}\BibitemShut {NoStop}%
\bibitem [{\citenamefont {Nikoli{\'c}}\ \emph {et~al.}(2012)\citenamefont
  {Nikoli{\'c}}, \citenamefont {Saha}, \citenamefont {Markussen},\ and\
  \citenamefont {Thygesen}}]{NikolicThygesen2012}%
  \BibitemOpen
  \bibfield  {author} {\bibinfo {author} {\bibnamefont {Nikoli{\'c}},
  \bibfnamefont {B.~K.}}, \bibinfo {author} {\bibfnamefont {K.~K.}\
  \bibnamefont {Saha}}, \bibinfo {author} {\bibfnamefont {T.}~\bibnamefont
  {Markussen}}, \ and\ \bibinfo {author} {\bibfnamefont {K.~S.}\ \bibnamefont
  {Thygesen}}} (\bibinfo {year} {2012}),\ \href {\doibase
  10.1007/s10825-012-0386-y} {\bibfield  {journal} {\bibinfo  {journal} {J.
  Comput. Electron.}\ }\textbf {\bibinfo {volume} {11}}~(\bibinfo {number}
  {1}),\ \bibinfo {pages} {78}}\BibitemShut {NoStop}%
\bibitem [{\citenamefont {Nishiyama}(2000)}]{Nishiyama2000}%
  \BibitemOpen
  \bibfield  {author} {\bibinfo {author} {\bibnamefont {Nishiyama},
  \bibfnamefont {Y.}}} (\bibinfo {year} {2000}),\ \href {\doibase
  10.1007/s100510070144} {\bibfield  {journal} {\bibinfo  {journal} {Eur. Phys.
  J. B}\ }\textbf {\bibinfo {volume} {17}}~(\bibinfo {number} {2}),\ \bibinfo
  {pages} {295}}\BibitemShut {NoStop}%
\bibitem [{\citenamefont {Nitzan}\ and\ \citenamefont
  {Ratner}(2003)}]{NitzanRatner2003}%
  \BibitemOpen
  \bibfield  {author} {\bibinfo {author} {\bibnamefont {Nitzan}, \bibfnamefont
  {A.}}, \ and\ \bibinfo {author} {\bibfnamefont {M.~A.}\ \bibnamefont
  {Ratner}}} (\bibinfo {year} {2003}),\ \href {\doibase
  10.1126/science.1081572} {\bibfield  {journal} {\bibinfo  {journal}
  {Science}\ }\textbf {\bibinfo {volume} {300}},\ \bibinfo {pages}
  {1384}}\BibitemShut {NoStop}%
\bibitem [{\citenamefont {Nordsieck}\ \emph {et~al.}(1940)\citenamefont
  {Nordsieck}, \citenamefont {Lamb},\ and\ \citenamefont
  {Uhlenbeck}}]{Nordsieck1940}%
  \BibitemOpen
  \bibfield  {author} {\bibinfo {author} {\bibnamefont {Nordsieck},
  \bibfnamefont {A.}}, \bibinfo {author} {\bibfnamefont {W.~E.}\ \bibnamefont
  {Lamb}}, \ and\ \bibinfo {author} {\bibfnamefont {G.~E.}\ \bibnamefont
  {Uhlenbeck}}} (\bibinfo {year} {1940}),\ \href {\doibase
  10.1016/S0031-8914(40)90102-1} {\bibfield  {journal} {\bibinfo  {journal}
  {Physica}\ }\textbf {\bibinfo {volume} {7}}~(\bibinfo {number} {4}),\
  \bibinfo {pages} {344}}\BibitemShut {NoStop}%
\bibitem [{\citenamefont {Novotn\'y}\ \emph {et~al.}(2003)\citenamefont
  {Novotn\'y}, \citenamefont {Donarini},\ and\ \citenamefont
  {Jauho}}]{novotny2003a}%
  \BibitemOpen
  \bibfield  {author} {\bibinfo {author} {\bibnamefont {Novotn\'y},
  \bibfnamefont {T.}}, \bibinfo {author} {\bibfnamefont {A.}~\bibnamefont
  {Donarini}}, \ and\ \bibinfo {author} {\bibfnamefont {A.-P.}\ \bibnamefont
  {Jauho}}} (\bibinfo {year} {2003}),\ \href {\doibase
  10.1103/PhysRevLett.90.256801} {\bibfield  {journal} {\bibinfo  {journal}
  {Phys. Rev. Lett.}\ }\textbf {\bibinfo {volume} {90}},\ \bibinfo {pages}
  {256801}}\BibitemShut {NoStop}%
\bibitem [{\citenamefont {N{\"u}\ss{}eler}\ \emph {et~al.}(2020)\citenamefont
  {N{\"u}\ss{}eler}, \citenamefont {Dhand}, \citenamefont {Huelga},\ and\
  \citenamefont {Plenio}}]{NuselerPlenio2019}%
  \BibitemOpen
  \bibfield  {author} {\bibinfo {author} {\bibnamefont {N{\"u}\ss{}eler},
  \bibfnamefont {A.}}, \bibinfo {author} {\bibfnamefont {I.}~\bibnamefont
  {Dhand}}, \bibinfo {author} {\bibfnamefont {S.~F.}\ \bibnamefont {Huelga}}, \
  and\ \bibinfo {author} {\bibfnamefont {M.~B.}\ \bibnamefont {Plenio}}}
  (\bibinfo {year} {2020}),\ \href {\doibase 10.1103/PhysRevB.101.155134}
  {\bibfield  {journal} {\bibinfo  {journal} {Phys. Rev. B}\ }\textbf {\bibinfo
  {volume} {101}},\ \bibinfo {pages} {155134}}\BibitemShut {NoStop}%
\bibitem [{\citenamefont {Olaya-Castro}\ \emph {et~al.}(2008)\citenamefont
  {Olaya-Castro}, \citenamefont {Lee}, \citenamefont {Olsen},\ and\
  \citenamefont {Johnson}}]{OlayaCastro2008}%
  \BibitemOpen
  \bibfield  {author} {\bibinfo {author} {\bibnamefont {Olaya-Castro},
  \bibfnamefont {A.}}, \bibinfo {author} {\bibfnamefont {C.~F.}\ \bibnamefont
  {Lee}}, \bibinfo {author} {\bibfnamefont {F.~F.}\ \bibnamefont {Olsen}}, \
  and\ \bibinfo {author} {\bibfnamefont {N.~F.}\ \bibnamefont {Johnson}}}
  (\bibinfo {year} {2008}),\ \href {\doibase 10.1103/PhysRevB.78.085115}
  {\bibfield  {journal} {\bibinfo  {journal} {Phys. Rev. B}\ }\textbf {\bibinfo
  {volume} {78}},\ \bibinfo {pages} {085115}}\BibitemShut {NoStop}%
\bibitem [{\citenamefont {Orignac}\ and\ \citenamefont
  {Giamarchi}(2001)}]{OrignacGiamarchi2001}%
  \BibitemOpen
  \bibfield  {author} {\bibinfo {author} {\bibnamefont {Orignac}, \bibfnamefont
  {E.}}, \ and\ \bibinfo {author} {\bibfnamefont {T.}~\bibnamefont
  {Giamarchi}}} (\bibinfo {year} {2001}),\ \href {\doibase
  10.1103/PhysRevB.64.144515} {\bibfield  {journal} {\bibinfo  {journal} {Phys.
  Rev. B}\ }\textbf {\bibinfo {volume} {64}}~(\bibinfo {number} {14}),\
  \bibinfo {pages} {144515}}\BibitemShut {NoStop}%
\bibitem [{\citenamefont {Or\'{u}s}(2014)}]{Orus2014}%
  \BibitemOpen
  \bibfield  {author} {\bibinfo {author} {\bibnamefont {Or\'{u}s},
  \bibfnamefont {R.}}} (\bibinfo {year} {2014}),\ \href {\doibase
  10.1016/j.aop.2014.06.013} {\bibfield  {journal} {\bibinfo  {journal} {Annals
  of Phys.}\ }\textbf {\bibinfo {volume} {349}},\ \bibinfo {pages}
  {117}}\BibitemShut {NoStop}%
\bibitem [{\citenamefont {Ostlund}\ \emph {et~al.}(1983)\citenamefont
  {Ostlund}, \citenamefont {Pandit}, \citenamefont {Rand}, \citenamefont
  {Schellnhuber},\ and\ \citenamefont {Siggia}}]{Ostlund1983}%
  \BibitemOpen
  \bibfield  {author} {\bibinfo {author} {\bibnamefont {Ostlund}, \bibfnamefont
  {S.}}, \bibinfo {author} {\bibfnamefont {R.}~\bibnamefont {Pandit}}, \bibinfo
  {author} {\bibfnamefont {D.}~\bibnamefont {Rand}}, \bibinfo {author}
  {\bibfnamefont {H.~J.}\ \bibnamefont {Schellnhuber}}, \ and\ \bibinfo
  {author} {\bibfnamefont {E.~D.}\ \bibnamefont {Siggia}}} (\bibinfo {year}
  {1983}),\ \href {\doibase 10.1103/PhysRevLett.50.1873} {\bibfield  {journal}
  {\bibinfo  {journal} {Phys. Rev. Lett.}\ }\textbf {\bibinfo {volume} {50}},\
  \bibinfo {pages} {1873}}\BibitemShut {NoStop}%
\bibitem [{\citenamefont {Paeckel}\ \emph {et~al.}(2019)\citenamefont
  {Paeckel}, \citenamefont {K{\"o}hler}, \citenamefont {Swoboda}, \citenamefont
  {Manmana}, \citenamefont {Schollw{\"o}ck},\ and\ \citenamefont
  {Hubig}}]{PaeckelHubig2019}%
  \BibitemOpen
  \bibfield  {author} {\bibinfo {author} {\bibnamefont {Paeckel}, \bibfnamefont
  {S.}}, \bibinfo {author} {\bibfnamefont {T.}~\bibnamefont {K{\"o}hler}},
  \bibinfo {author} {\bibfnamefont {A.}~\bibnamefont {Swoboda}}, \bibinfo
  {author} {\bibfnamefont {S.~R.}\ \bibnamefont {Manmana}}, \bibinfo {author}
  {\bibfnamefont {U.}~\bibnamefont {Schollw{\"o}ck}}, \ and\ \bibinfo {author}
  {\bibfnamefont {C.}~\bibnamefont {Hubig}}} (\bibinfo {year} {2019}),\ \href
  {\doibase https://doi.org/10.1016/j.aop.2019.167998} {\bibfield  {journal}
  {\bibinfo  {journal} {Annals of Physics}\ }\textbf {\bibinfo {volume}
  {411}},\ \bibinfo {pages} {167998}}\BibitemShut {NoStop}%
\bibitem [{\citenamefont {Palmer}(1977)}]{palmer1977a}%
  \BibitemOpen
  \bibfield  {author} {\bibinfo {author} {\bibnamefont {Palmer}, \bibfnamefont
  {P.~F.}}} (\bibinfo {year} {1977}),\ \href {\doibase 10.1063/1.523296}
  {\bibfield  {journal} {\bibinfo  {journal} {Journal of Mathematical Physics}\
  }\textbf {\bibinfo {volume} {18}}~(\bibinfo {number} {3}),\ \bibinfo {pages}
  {527}}\BibitemShut {NoStop}%
\bibitem [{\citenamefont {Palmero}\ \emph {et~al.}(2019)\citenamefont
  {Palmero}, \citenamefont {Xu}, \citenamefont {Guo},\ and\ \citenamefont
  {Poletti}}]{PalmeroPoletti2019}%
  \BibitemOpen
  \bibfield  {author} {\bibinfo {author} {\bibnamefont {Palmero}, \bibfnamefont
  {M.}}, \bibinfo {author} {\bibfnamefont {X.}~\bibnamefont {Xu}}, \bibinfo
  {author} {\bibfnamefont {C.}~\bibnamefont {Guo}}, \ and\ \bibinfo {author}
  {\bibfnamefont {D.}~\bibnamefont {Poletti}}} (\bibinfo {year} {2019}),\ \href
  {\doibase 10.1103/PhysRevE.100.022111} {\bibfield  {journal} {\bibinfo
  {journal} {Phys. Rev. E}\ }\textbf {\bibinfo {volume} {100}},\ \bibinfo
  {pages} {022111}}\BibitemShut {NoStop}%
\bibitem [{\citenamefont {Parameswaran}\ and\ \citenamefont
  {Vasseur}(2018)}]{ParameswaranVasseur2018}%
  \BibitemOpen
  \bibfield  {author} {\bibinfo {author} {\bibnamefont {Parameswaran},
  \bibfnamefont {S.~A.}}, \ and\ \bibinfo {author} {\bibfnamefont
  {R.}~\bibnamefont {Vasseur}}} (\bibinfo {year} {2018}),\ \href {\doibase
  10.1088/1361-6633/aac9ed} {\bibfield  {journal} {\bibinfo  {journal} {Reports
  on Progress in Physics}\ }\textbf {\bibinfo {volume} {81}}~(\bibinfo {number}
  {8}),\ \bibinfo {pages} {082501}}\BibitemShut {NoStop}%
\bibitem [{\citenamefont {Park}\ \emph {et~al.}(2000)\citenamefont {Park},
  \citenamefont {Park}, \citenamefont {Lim}, \citenamefont {Anderson},
  \citenamefont {Alivisatos},\ and\ \citenamefont {McEuen}}]{park2000a}%
  \BibitemOpen
  \bibfield  {author} {\bibinfo {author} {\bibnamefont {Park}, \bibfnamefont
  {H.}}, \bibinfo {author} {\bibfnamefont {J.}~\bibnamefont {Park}}, \bibinfo
  {author} {\bibfnamefont {A.~K.~L.}\ \bibnamefont {Lim}}, \bibinfo {author}
  {\bibfnamefont {E.~H.}\ \bibnamefont {Anderson}}, \bibinfo {author}
  {\bibfnamefont {A.~P.}\ \bibnamefont {Alivisatos}}, \ and\ \bibinfo {author}
  {\bibfnamefont {P.~L.}\ \bibnamefont {McEuen}}} (\bibinfo {year} {2000}),\
  \href {\doibase 10.1038/35024031} {\bibfield  {journal} {\bibinfo  {journal}
  {Nature}\ }\textbf {\bibinfo {volume} {407}},\ \bibinfo {pages}
  {57}}\BibitemShut {NoStop}%
\bibitem [{\citenamefont {Pekola}\ \emph {et~al.}(2013)\citenamefont {Pekola},
  \citenamefont {Saira}, \citenamefont {Maisi}, \citenamefont {Kemppinen},
  \citenamefont {M{\"o}tt{\"o}nen}, \citenamefont {Pashkin},\ and\
  \citenamefont {Averin}}]{PekolaAverin2013}%
  \BibitemOpen
  \bibfield  {author} {\bibinfo {author} {\bibnamefont {Pekola}, \bibfnamefont
  {J.~P.}}, \bibinfo {author} {\bibfnamefont {O.-P.}\ \bibnamefont {Saira}},
  \bibinfo {author} {\bibfnamefont {V.~F.}\ \bibnamefont {Maisi}}, \bibinfo
  {author} {\bibfnamefont {A.}~\bibnamefont {Kemppinen}}, \bibinfo {author}
  {\bibfnamefont {M.}~\bibnamefont {M{\"o}tt{\"o}nen}}, \bibinfo {author}
  {\bibfnamefont {Y.~A.}\ \bibnamefont {Pashkin}}, \ and\ \bibinfo {author}
  {\bibfnamefont {D.~V.}\ \bibnamefont {Averin}}} (\bibinfo {year} {2013}),\
  \href {\doibase 10.1103/RevModPhys.85.1421} {\bibfield  {journal} {\bibinfo
  {journal} {Rev. Mod. Phys.}\ }\textbf {\bibinfo {volume} {85}},\ \bibinfo
  {pages} {1421}}\BibitemShut {NoStop}%
\bibitem [{\citenamefont {Perarnau-Llobet}\ \emph {et~al.}(2018)\citenamefont
  {Perarnau-Llobet}, \citenamefont {Wilming}, \citenamefont {Riera},
  \citenamefont {Gallego},\ and\ \citenamefont
  {Eisert}}]{perarnau_llobet2018a}%
  \BibitemOpen
  \bibfield  {author} {\bibinfo {author} {\bibnamefont {Perarnau-Llobet},
  \bibfnamefont {M.}}, \bibinfo {author} {\bibfnamefont {H.}~\bibnamefont
  {Wilming}}, \bibinfo {author} {\bibfnamefont {A.}~\bibnamefont {Riera}},
  \bibinfo {author} {\bibfnamefont {R.}~\bibnamefont {Gallego}}, \ and\
  \bibinfo {author} {\bibfnamefont {J.}~\bibnamefont {Eisert}}} (\bibinfo
  {year} {2018}),\ \href {\doibase 10.1103/PhysRevLett.120.120602} {\bibfield
  {journal} {\bibinfo  {journal} {Phys. Rev. Lett.}\ }\textbf {\bibinfo
  {volume} {120}},\ \bibinfo {pages} {120602}}\BibitemShut {NoStop}%
\bibitem [{\citenamefont {Pereira}(2017{\natexlab{a}})}]{Pereira2017}%
  \BibitemOpen
  \bibfield  {author} {\bibinfo {author} {\bibnamefont {Pereira}, \bibfnamefont
  {E.}}} (\bibinfo {year} {2017}{\natexlab{a}}),\ \href {\doibase
  10.1103/PhysRevE.95.030104} {\bibfield  {journal} {\bibinfo  {journal} {Phys.
  Rev. E}\ }\textbf {\bibinfo {volume} {95}},\ \bibinfo {pages}
  {030104}}\BibitemShut {NoStop}%
\bibitem [{\citenamefont {Pereira}(2017{\natexlab{b}})}]{Pereira2017b}%
  \BibitemOpen
  \bibfield  {author} {\bibinfo {author} {\bibnamefont {Pereira}, \bibfnamefont
  {E.}}} (\bibinfo {year} {2017}{\natexlab{b}}),\ \href {\doibase
  10.1103/PhysRevE.96.012114} {\bibfield  {journal} {\bibinfo  {journal} {Phys.
  Rev. E}\ }\textbf {\bibinfo {volume} {96}},\ \bibinfo {pages}
  {012114}}\BibitemShut {NoStop}%
\bibitem [{\citenamefont {Pereira}(2018)}]{Pereira2018}%
  \BibitemOpen
  \bibfield  {author} {\bibinfo {author} {\bibnamefont {Pereira}, \bibfnamefont
  {E.}}} (\bibinfo {year} {2018}),\ \href {\doibase 10.1103/PhysRevE.97.022115}
  {\bibfield  {journal} {\bibinfo  {journal} {Physical Review E}\ }\textbf
  {\bibinfo {volume} {97}}~(\bibinfo {number} {2}),\ \bibinfo {pages}
  {022115}}\BibitemShut {NoStop}%
\bibitem [{\citenamefont {Pereira}(2019)}]{Pereira2019}%
  \BibitemOpen
  \bibfield  {author} {\bibinfo {author} {\bibnamefont {Pereira}, \bibfnamefont
  {E.}}} (\bibinfo {year} {2019}),\ \href {\doibase 10.1103/PhysRevE.99.032116}
  {\bibfield  {journal} {\bibinfo  {journal} {Phys. Rev. E}\ }\textbf {\bibinfo
  {volume} {99}},\ \bibinfo {pages} {032116}}\BibitemShut {NoStop}%
\bibitem [{\citenamefont {Pereira}\ and\ \citenamefont
  {\'Avila}(2013)}]{Pereira2013}%
  \BibitemOpen
  \bibfield  {author} {\bibinfo {author} {\bibnamefont {Pereira}, \bibfnamefont
  {E.}}, \ and\ \bibinfo {author} {\bibfnamefont {R.~R.}\ \bibnamefont
  {\'Avila}}} (\bibinfo {year} {2013}),\ \href {\doibase
  10.1103/PhysRevE.88.032139} {\bibfield  {journal} {\bibinfo  {journal} {Phys.
  Rev. E}\ }\textbf {\bibinfo {volume} {88}},\ \bibinfo {pages}
  {032139}}\BibitemShut {NoStop}%
\bibitem [{\citenamefont {Pi\v{z}orn}\ and\ \citenamefont
  {Prosen}(2009)}]{ProsenPizorn2009}%
  \BibitemOpen
  \bibfield  {author} {\bibinfo {author} {\bibnamefont {Pi\v{z}orn},
  \bibfnamefont {I.}}, \ and\ \bibinfo {author} {\bibfnamefont
  {T.}~\bibnamefont {Prosen}}} (\bibinfo {year} {2009}),\ \href {\doibase
  10.1103/PhysRevB.79.184416} {\bibfield  {journal} {\bibinfo  {journal} {Phys.
  Rev. B}\ }\textbf {\bibinfo {volume} {79}},\ \bibinfo {pages}
  {184416}}\BibitemShut {NoStop}%
\bibitem [{\citenamefont {Pleasance}\ \emph {et~al.}(2020)\citenamefont
  {Pleasance}, \citenamefont {Garraway},\ and\ \citenamefont
  {Petruccione}}]{pleasance2020a}%
  \BibitemOpen
  \bibfield  {author} {\bibinfo {author} {\bibnamefont {Pleasance},
  \bibfnamefont {G.}}, \bibinfo {author} {\bibfnamefont {B.~M.}\ \bibnamefont
  {Garraway}}, \ and\ \bibinfo {author} {\bibfnamefont {F.}~\bibnamefont
  {Petruccione}}} (\bibinfo {year} {2020}),\ \href {\doibase
  10.1103/PhysRevResearch.2.043058} {\bibfield  {journal} {\bibinfo  {journal}
  {Phys. Rev. Research}\ }\textbf {\bibinfo {volume} {2}},\ \bibinfo {pages}
  {043058}}\BibitemShut {NoStop}%
\bibitem [{\citenamefont {Plenio}\ \emph {et~al.}(2005)\citenamefont {Plenio},
  \citenamefont {Eisert}, \citenamefont {Drei\ss{}ig},\ and\ \citenamefont
  {Cramer}}]{PlenioCramer2005}%
  \BibitemOpen
  \bibfield  {author} {\bibinfo {author} {\bibnamefont {Plenio}, \bibfnamefont
  {M.~B.}}, \bibinfo {author} {\bibfnamefont {J.}~\bibnamefont {Eisert}},
  \bibinfo {author} {\bibfnamefont {J.}~\bibnamefont {Drei\ss{}ig}}, \ and\
  \bibinfo {author} {\bibfnamefont {M.}~\bibnamefont {Cramer}}} (\bibinfo
  {year} {2005}),\ \href {\doibase 10.1103/PhysRevLett.94.060503} {\bibfield
  {journal} {\bibinfo  {journal} {Phys. Rev. Lett.}\ }\textbf {\bibinfo
  {volume} {94}},\ \bibinfo {pages} {060503}}\BibitemShut {NoStop}%
\bibitem [{\citenamefont {Plenio}\ and\ \citenamefont
  {Huelga}(2008)}]{Plenio2008}%
  \BibitemOpen
  \bibfield  {author} {\bibinfo {author} {\bibnamefont {Plenio}, \bibfnamefont
  {M.~B.}}, \ and\ \bibinfo {author} {\bibfnamefont {S.~F.}\ \bibnamefont
  {Huelga}}} (\bibinfo {year} {2008}),\ \href {\doibase
  10.1088/1367-2630/10/11/113019} {\bibfield  {journal} {\bibinfo  {journal}
  {New Journal of Physics}\ }\textbf {\bibinfo {volume} {10}}~(\bibinfo
  {number} {11}),\ \bibinfo {pages} {113019}}\BibitemShut {NoStop}%
\bibitem [{\citenamefont {Plenio}\ and\ \citenamefont
  {Knight}(1998)}]{PlenioKnight1998}%
  \BibitemOpen
  \bibfield  {author} {\bibinfo {author} {\bibnamefont {Plenio}, \bibfnamefont
  {M.~B.}}, \ and\ \bibinfo {author} {\bibfnamefont {P.~L.}\ \bibnamefont
  {Knight}}} (\bibinfo {year} {1998}),\ \href {\doibase
  10.1103/RevModPhys.70.101} {\bibfield  {journal} {\bibinfo  {journal} {Rev.
  Mod. Phys.}\ }\textbf {\bibinfo {volume} {70}},\ \bibinfo {pages}
  {101}}\BibitemShut {NoStop}%
\bibitem [{\citenamefont {Poletti}\ \emph {et~al.}(2013)\citenamefont
  {Poletti}, \citenamefont {Barmettler}, \citenamefont {Georges},\ and\
  \citenamefont {Kollath}}]{PolettiKollath2013}%
  \BibitemOpen
  \bibfield  {author} {\bibinfo {author} {\bibnamefont {Poletti}, \bibfnamefont
  {D.}}, \bibinfo {author} {\bibfnamefont {P.}~\bibnamefont {Barmettler}},
  \bibinfo {author} {\bibfnamefont {A.}~\bibnamefont {Georges}}, \ and\
  \bibinfo {author} {\bibfnamefont {C.}~\bibnamefont {Kollath}}} (\bibinfo
  {year} {2013}),\ \href {\doibase 10.1103/PhysRevLett.111.195301} {\bibfield
  {journal} {\bibinfo  {journal} {Phys. Rev. Lett.}\ }\textbf {\bibinfo
  {volume} {111}},\ \bibinfo {pages} {195301}}\BibitemShut {NoStop}%
\bibitem [{\citenamefont {Poletti}\ \emph {et~al.}(2012)\citenamefont
  {Poletti}, \citenamefont {Bernier}, \citenamefont {Georges},\ and\
  \citenamefont {Kollath}}]{PolettiKollath2012}%
  \BibitemOpen
  \bibfield  {author} {\bibinfo {author} {\bibnamefont {Poletti}, \bibfnamefont
  {D.}}, \bibinfo {author} {\bibfnamefont {J.-S.}\ \bibnamefont {Bernier}},
  \bibinfo {author} {\bibfnamefont {A.}~\bibnamefont {Georges}}, \ and\
  \bibinfo {author} {\bibfnamefont {C.}~\bibnamefont {Kollath}}} (\bibinfo
  {year} {2012}),\ \href {\doibase 10.1103/PhysRevLett.109.045302} {\bibfield
  {journal} {\bibinfo  {journal} {Phys. Rev. Lett.}\ }\textbf {\bibinfo
  {volume} {109}},\ \bibinfo {pages} {045302}}\BibitemShut {NoStop}%
\bibitem [{\citenamefont {Polkovnikov}(2011)}]{polkovnikov2011a}%
  \BibitemOpen
  \bibfield  {author} {\bibinfo {author} {\bibnamefont {Polkovnikov},
  \bibfnamefont {A.}}} (\bibinfo {year} {2011}),\ \href@noop {} {\bibfield
  {journal} {\bibinfo  {journal} {Annals of Physics}\ }\textbf {\bibinfo
  {volume} {326}},\ \bibinfo {pages} {486}}\BibitemShut {NoStop}%
\bibitem [{\citenamefont {P{\"o}ltl}\ \emph {et~al.}(2009)\citenamefont
  {P{\"o}ltl}, \citenamefont {Emary},\ and\ \citenamefont
  {Brandes}}]{PoltlBrandes2009}%
  \BibitemOpen
  \bibfield  {author} {\bibinfo {author} {\bibnamefont {P{\"o}ltl},
  \bibfnamefont {C.}}, \bibinfo {author} {\bibfnamefont {C.}~\bibnamefont
  {Emary}}, \ and\ \bibinfo {author} {\bibfnamefont {T.}~\bibnamefont
  {Brandes}}} (\bibinfo {year} {2009}),\ \href {\doibase
  10.1103/PhysRevB.80.115313} {\bibfield  {journal} {\bibinfo  {journal} {Phys.
  Rev. B}\ }\textbf {\bibinfo {volume} {80}},\ \bibinfo {pages}
  {115313}}\BibitemShut {NoStop}%
\bibitem [{\citenamefont {Ponomarev}\ \emph {et~al.}(2011)\citenamefont
  {Ponomarev}, \citenamefont {Denisov},\ and\ \citenamefont
  {H{\"a}nggi}}]{PonomarevHanggi2011}%
  \BibitemOpen
  \bibfield  {author} {\bibinfo {author} {\bibnamefont {Ponomarev},
  \bibfnamefont {A.~V.}}, \bibinfo {author} {\bibfnamefont {S.}~\bibnamefont
  {Denisov}}, \ and\ \bibinfo {author} {\bibfnamefont {P.}~\bibnamefont
  {H{\"a}nggi}}} (\bibinfo {year} {2011}),\ \href {\doibase
  10.1103/PhysRevLett.106.010405} {\bibfield  {journal} {\bibinfo  {journal}
  {Phys. Rev. Lett.}\ }\textbf {\bibinfo {volume} {106}},\ \bibinfo {pages}
  {010405}}\BibitemShut {NoStop}%
\bibitem [{\citenamefont {Pop}\ \emph {et~al.}(2005)\citenamefont {Pop},
  \citenamefont {Mann}, \citenamefont {Cao}, \citenamefont {Wang},
  \citenamefont {Goodson},\ and\ \citenamefont {Dai}}]{PopDai2005}%
  \BibitemOpen
  \bibfield  {author} {\bibinfo {author} {\bibnamefont {Pop}, \bibfnamefont
  {E.}}, \bibinfo {author} {\bibfnamefont {D.}~\bibnamefont {Mann}}, \bibinfo
  {author} {\bibfnamefont {J.}~\bibnamefont {Cao}}, \bibinfo {author}
  {\bibfnamefont {Q.}~\bibnamefont {Wang}}, \bibinfo {author} {\bibfnamefont
  {K.}~\bibnamefont {Goodson}}, \ and\ \bibinfo {author} {\bibfnamefont
  {H.}~\bibnamefont {Dai}}} (\bibinfo {year} {2005}),\ \href {\doibase
  10.1103/PhysRevLett.95.155505} {\bibfield  {journal} {\bibinfo  {journal}
  {Phys. Rev. Lett.}\ }\textbf {\bibinfo {volume} {95}},\ \bibinfo {pages}
  {155505}}\BibitemShut {NoStop}%
\bibitem [{\citenamefont {Popkov}\ \emph {et~al.}(2013)\citenamefont {Popkov},
  \citenamefont {Karevski},\ and\ \citenamefont {Sch{\"{u}}tz}}]{Popkov2013a}%
  \BibitemOpen
  \bibfield  {author} {\bibinfo {author} {\bibnamefont {Popkov}, \bibfnamefont
  {V.}}, \bibinfo {author} {\bibfnamefont {D.}~\bibnamefont {Karevski}}, \ and\
  \bibinfo {author} {\bibfnamefont {G.~M.}\ \bibnamefont {Sch{\"{u}}tz}}}
  (\bibinfo {year} {2013}),\ \href {\doibase 10.1103/PhysRevE.88.062118}
  {\bibfield  {journal} {\bibinfo  {journal} {Physical Review E}\ }\textbf
  {\bibinfo {volume} {88}}~(\bibinfo {number} {6}),\ \bibinfo {pages}
  {062118}}\BibitemShut {NoStop}%
\bibitem [{\citenamefont {Popkov}\ and\ \citenamefont
  {Presilla}(2016)}]{Popkov2016}%
  \BibitemOpen
  \bibfield  {author} {\bibinfo {author} {\bibnamefont {Popkov}, \bibfnamefont
  {V.}}, \ and\ \bibinfo {author} {\bibfnamefont {C.}~\bibnamefont {Presilla}}}
  (\bibinfo {year} {2016}),\ \href {\doibase 10.1103/PhysRevA.93.022111}
  {\bibfield  {journal} {\bibinfo  {journal} {Phys. Rev. A}\ }\textbf {\bibinfo
  {volume} {93}},\ \bibinfo {pages} {022111}}\BibitemShut {NoStop}%
\bibitem [{\citenamefont {Popkov}\ and\ \citenamefont
  {Prosen}(2015)}]{Popkov2015}%
  \BibitemOpen
  \bibfield  {author} {\bibinfo {author} {\bibnamefont {Popkov}, \bibfnamefont
  {V.}}, \ and\ \bibinfo {author} {\bibfnamefont {T.}~\bibnamefont {Prosen}}}
  (\bibinfo {year} {2015}),\ \href {\doibase 10.1103/PhysRevLett.114.127201}
  {\bibfield  {journal} {\bibinfo  {journal} {Phys. Rev. Lett.}\ }\textbf
  {\bibinfo {volume} {114}},\ \bibinfo {pages} {127201}}\BibitemShut {NoStop}%
\bibitem [{\citenamefont {Popkov}\ \emph
  {et~al.}(2020{\natexlab{a}})\citenamefont {Popkov}, \citenamefont {Prosen},\
  and\ \citenamefont {Zadnik}}]{PopkovZadnik2020b}%
  \BibitemOpen
  \bibfield  {author} {\bibinfo {author} {\bibnamefont {Popkov}, \bibfnamefont
  {V.}}, \bibinfo {author} {\bibfnamefont {T.}~\bibnamefont {Prosen}}, \ and\
  \bibinfo {author} {\bibfnamefont {L.}~\bibnamefont {Zadnik}}} (\bibinfo
  {year} {2020}{\natexlab{a}}),\ \href {\doibase
  10.1103/PhysRevLett.124.160403} {\bibfield  {journal} {\bibinfo  {journal}
  {Phys. Rev. Lett.}\ }\textbf {\bibinfo {volume} {124}},\ \bibinfo {pages}
  {160403}}\BibitemShut {NoStop}%
\bibitem [{\citenamefont {Popkov}\ \emph
  {et~al.}(2020{\natexlab{b}})\citenamefont {Popkov}, \citenamefont {Prosen},\
  and\ \citenamefont {Zadnik}}]{PopkovZadnik2020}%
  \BibitemOpen
  \bibfield  {author} {\bibinfo {author} {\bibnamefont {Popkov}, \bibfnamefont
  {V.}}, \bibinfo {author} {\bibfnamefont {T.}~\bibnamefont {Prosen}}, \ and\
  \bibinfo {author} {\bibfnamefont {L.}~\bibnamefont {Zadnik}}} (\bibinfo
  {year} {2020}{\natexlab{b}}),\ \href {\doibase 10.1103/PhysRevE.101.042122}
  {\bibfield  {journal} {\bibinfo  {journal} {Phys. Rev. E}\ }\textbf {\bibinfo
  {volume} {101}},\ \bibinfo {pages} {042122}}\BibitemShut {NoStop}%
\bibitem [{\citenamefont {Potter}\ \emph {et~al.}(2015)\citenamefont {Potter},
  \citenamefont {Vasseur},\ and\ \citenamefont
  {Parameswaran}}]{PotterParameswaran2015}%
  \BibitemOpen
  \bibfield  {author} {\bibinfo {author} {\bibnamefont {Potter}, \bibfnamefont
  {A.~C.}}, \bibinfo {author} {\bibfnamefont {R.}~\bibnamefont {Vasseur}}, \
  and\ \bibinfo {author} {\bibfnamefont {S.~A.}\ \bibnamefont {Parameswaran}}}
  (\bibinfo {year} {2015}),\ \href {\doibase 10.1103/PhysRevX.5.031033}
  {\bibfield  {journal} {\bibinfo  {journal} {Phys. Rev. X}\ }\textbf {\bibinfo
  {volume} {5}},\ \bibinfo {pages} {031033}}\BibitemShut {NoStop}%
\bibitem [{\citenamefont {Potts}\ \emph {et~al.}(2021)\citenamefont {Potts},
  \citenamefont {Kalaee},\ and\ \citenamefont {Wacker}}]{Potts2021}%
  \BibitemOpen
  \bibfield  {author} {\bibinfo {author} {\bibnamefont {Potts}, \bibfnamefont
  {P.~P.}}, \bibinfo {author} {\bibfnamefont {A.~A.~S.}\ \bibnamefont
  {Kalaee}}, \ and\ \bibinfo {author} {\bibfnamefont {A.}~\bibnamefont
  {Wacker}}} (\bibinfo {year} {2021}),\ \href {\doibase
  10.1088/1367-2630/ac3b2f} {\bibfield  {journal} {\bibinfo  {journal} {New
  Journal of Physics}\ }\textbf {\bibinfo {volume} {23}}~(\bibinfo {number}
  {12}),\ \bibinfo {pages} {123013}}\BibitemShut {NoStop}%
\bibitem [{\citenamefont {Poulsen}\ \emph {et~al.}(2022)\citenamefont
  {Poulsen}, \citenamefont {Santos}, \citenamefont {Kristensen},\ and\
  \citenamefont {Zinner}}]{PoulsenZinner2021}%
  \BibitemOpen
  \bibfield  {author} {\bibinfo {author} {\bibnamefont {Poulsen}, \bibfnamefont
  {K.}}, \bibinfo {author} {\bibfnamefont {A.~C.}\ \bibnamefont {Santos}},
  \bibinfo {author} {\bibfnamefont {L.~B.}\ \bibnamefont {Kristensen}}, \ and\
  \bibinfo {author} {\bibfnamefont {N.~T.}\ \bibnamefont {Zinner}}} (\bibinfo
  {year} {2022}),\ \href {\doibase 10.1103/PhysRevA.105.052605} {\bibfield
  {journal} {\bibinfo  {journal} {Phys. Rev. A}\ }\textbf {\bibinfo {volume}
  {105}},\ \bibinfo {pages} {052605}}\BibitemShut {NoStop}%
\bibitem [{\citenamefont {Prelov\v{s}ek}\ \emph {et~al.}(2016)\citenamefont
  {Prelov\v{s}ek}, \citenamefont {Bari\v{s}i\'c},\ and\ \citenamefont
  {\v{Z}nidari\v{c}}}]{PrelovsekZnidaric2016}%
  \BibitemOpen
  \bibfield  {author} {\bibinfo {author} {\bibnamefont {Prelov\v{s}ek},
  \bibfnamefont {P.}}, \bibinfo {author} {\bibfnamefont {O.~S.}\ \bibnamefont
  {Bari\v{s}i\'c}}, \ and\ \bibinfo {author} {\bibfnamefont {M.}~\bibnamefont
  {\v{Z}nidari\v{c}}}} (\bibinfo {year} {2016}),\ \href {\doibase
  10.1103/PhysRevB.94.241104} {\bibfield  {journal} {\bibinfo  {journal} {Phys.
  Rev. B}\ }\textbf {\bibinfo {volume} {94}},\ \bibinfo {pages}
  {241104}}\BibitemShut {NoStop}%
\bibitem [{\citenamefont {Prelov\v{s}ek}\ \emph {et~al.}(2018)\citenamefont
  {Prelov\v{s}ek}, \citenamefont {Bon\v{c}a},\ and\ \citenamefont
  {Mierzejewski}}]{PrelovsekMierzejewski2018}%
  \BibitemOpen
  \bibfield  {author} {\bibinfo {author} {\bibnamefont {Prelov\v{s}ek},
  \bibfnamefont {P.}}, \bibinfo {author} {\bibfnamefont {J.}~\bibnamefont
  {Bon\v{c}a}}, \ and\ \bibinfo {author} {\bibfnamefont {M.}~\bibnamefont
  {Mierzejewski}}} (\bibinfo {year} {2018}),\ \href {\doibase
  10.1103/PhysRevB.98.125119} {\bibfield  {journal} {\bibinfo  {journal} {Phys.
  Rev. B}\ }\textbf {\bibinfo {volume} {98}},\ \bibinfo {pages}
  {125119}}\BibitemShut {NoStop}%
\bibitem [{\citenamefont {Prior}\ \emph {et~al.}(2010)\citenamefont {Prior},
  \citenamefont {Chin}, \citenamefont {Huelga},\ and\ \citenamefont
  {Plenio}}]{PriorPlenio2010}%
  \BibitemOpen
  \bibfield  {author} {\bibinfo {author} {\bibnamefont {Prior}, \bibfnamefont
  {J.}}, \bibinfo {author} {\bibfnamefont {A.~W.}\ \bibnamefont {Chin}},
  \bibinfo {author} {\bibfnamefont {S.~F.}\ \bibnamefont {Huelga}}, \ and\
  \bibinfo {author} {\bibfnamefont {M.~B.}\ \bibnamefont {Plenio}}} (\bibinfo
  {year} {2010}),\ \href {\doibase 10.1103/PhysRevLett.105.050404} {\bibfield
  {journal} {\bibinfo  {journal} {Phys. Rev. Lett.}\ }\textbf {\bibinfo
  {volume} {105}},\ \bibinfo {pages} {050404}}\BibitemShut {NoStop}%
\bibitem [{\citenamefont {Prociuk}\ \emph {et~al.}(2010)\citenamefont
  {Prociuk}, \citenamefont {Phillips},\ and\ \citenamefont
  {Dunietz}}]{ProciukDunietz2010}%
  \BibitemOpen
  \bibfield  {author} {\bibinfo {author} {\bibnamefont {Prociuk}, \bibfnamefont
  {A.}}, \bibinfo {author} {\bibfnamefont {H.}~\bibnamefont {Phillips}}, \ and\
  \bibinfo {author} {\bibfnamefont {B.~D.}\ \bibnamefont {Dunietz}}} (\bibinfo
  {year} {2010}),\ \href {\doibase 10.1088/1742-6596/220/1/012008} {\bibfield
  {journal} {\bibinfo  {journal} {J. Phys.: Conf. Ser.}\ }\textbf {\bibinfo
  {volume} {220}},\ \bibinfo {pages} {012008}}\BibitemShut {NoStop}%
\bibitem [{\citenamefont {Prosen}(2008)}]{Prosen2008}%
  \BibitemOpen
  \bibfield  {author} {\bibinfo {author} {\bibnamefont {Prosen}, \bibfnamefont
  {T.}}} (\bibinfo {year} {2008}),\ \href {\doibase
  10.1088/1367-2630/10/4/043026} {\bibfield  {journal} {\bibinfo  {journal}
  {New Journal of Physics}\ }\textbf {\bibinfo {volume} {10}}~(\bibinfo
  {number} {4}),\ \bibinfo {pages} {043026}}\BibitemShut {NoStop}%
\bibitem [{\citenamefont {Prosen}(2010)}]{Prosen2010b}%
  \BibitemOpen
  \bibfield  {author} {\bibinfo {author} {\bibnamefont {Prosen}, \bibfnamefont
  {T.}}} (\bibinfo {year} {2010}),\ \href {\doibase
  10.1088/1742-5468/2010/07/p07020} {\bibfield  {journal} {\bibinfo  {journal}
  {Journal of Statistical Mechanics: Theory and Experiment}\ }\textbf {\bibinfo
  {volume} {2010}}~(\bibinfo {number} {07}),\ \bibinfo {pages}
  {P07020}}\BibitemShut {NoStop}%
\bibitem [{\citenamefont {Prosen}(2011{\natexlab{a}})}]{Prosen2011b}%
  \BibitemOpen
  \bibfield  {author} {\bibinfo {author} {\bibnamefont {Prosen}, \bibfnamefont
  {T.}}} (\bibinfo {year} {2011}{\natexlab{a}}),\ \href {\doibase
  10.1103/PhysRevLett.107.137201} {\bibfield  {journal} {\bibinfo  {journal}
  {Phys. Rev. Lett.}\ }\textbf {\bibinfo {volume} {107}},\ \bibinfo {pages}
  {137201}}\BibitemShut {NoStop}%
\bibitem [{\citenamefont {Prosen}(2011{\natexlab{b}})}]{Prosen2011}%
  \BibitemOpen
  \bibfield  {author} {\bibinfo {author} {\bibnamefont {Prosen}, \bibfnamefont
  {T.}}} (\bibinfo {year} {2011}{\natexlab{b}}),\ \href {\doibase
  10.1103/PhysRevLett.106.217206} {\bibfield  {journal} {\bibinfo  {journal}
  {Phys. Rev. Lett.}\ }\textbf {\bibinfo {volume} {106}},\ \bibinfo {pages}
  {217206}}\BibitemShut {NoStop}%
\bibitem [{\citenamefont {Prosen}(2014)}]{Prosen2014}%
  \BibitemOpen
  \bibfield  {author} {\bibinfo {author} {\bibnamefont {Prosen}, \bibfnamefont
  {T.}}} (\bibinfo {year} {2014}),\ \href {\doibase
  10.1103/PhysRevLett.112.030603} {\bibfield  {journal} {\bibinfo  {journal}
  {Phys. Rev. Lett.}\ }\textbf {\bibinfo {volume} {112}},\ \bibinfo {pages}
  {030603}}\BibitemShut {NoStop}%
\bibitem [{\citenamefont {Prosen}\ and\ \citenamefont
  {Ilievski}(2011)}]{ProsenIlievski2011}%
  \BibitemOpen
  \bibfield  {author} {\bibinfo {author} {\bibnamefont {Prosen}, \bibfnamefont
  {T.}}, \ and\ \bibinfo {author} {\bibfnamefont {E.}~\bibnamefont {Ilievski}}}
  (\bibinfo {year} {2011}),\ \href {\doibase 10.1103/PhysRevLett.107.060403}
  {\bibfield  {journal} {\bibinfo  {journal} {Phys. Rev. Lett.}\ }\textbf
  {\bibinfo {volume} {107}},\ \bibinfo {pages} {060403}}\BibitemShut {NoStop}%
\bibitem [{\citenamefont {Prosen}\ and\ \citenamefont
  {Ilievski}(2013)}]{ProsenIlievski2013}%
  \BibitemOpen
  \bibfield  {author} {\bibinfo {author} {\bibnamefont {Prosen}, \bibfnamefont
  {T.}}, \ and\ \bibinfo {author} {\bibfnamefont {E.}~\bibnamefont {Ilievski}}}
  (\bibinfo {year} {2013}),\ \href {\doibase 10.1103/PhysRevLett.111.057203}
  {\bibfield  {journal} {\bibinfo  {journal} {Phys. Rev. Lett.}\ }\textbf
  {\bibinfo {volume} {111}},\ \bibinfo {pages} {057203}}\BibitemShut {NoStop}%
\bibitem [{\citenamefont {Prosen}\ and\ \citenamefont
  {Pi\v{z}orn}(2007)}]{ProsenPizorn2007}%
  \BibitemOpen
  \bibfield  {author} {\bibinfo {author} {\bibnamefont {Prosen}, \bibfnamefont
  {T.}}, \ and\ \bibinfo {author} {\bibfnamefont {I.}~\bibnamefont
  {Pi\v{z}orn}}} (\bibinfo {year} {2007}),\ \href {\doibase
  10.1103/PhysRevA.76.032316} {\bibfield  {journal} {\bibinfo  {journal} {Phys.
  Rev. A}\ }\textbf {\bibinfo {volume} {76}},\ \bibinfo {pages}
  {032316}}\BibitemShut {NoStop}%
\bibitem [{\citenamefont {Prosen}\ and\ \citenamefont
  {Pi\v{z}orn}(2008)}]{ProsenPizorn2008}%
  \BibitemOpen
  \bibfield  {author} {\bibinfo {author} {\bibnamefont {Prosen}, \bibfnamefont
  {T.}}, \ and\ \bibinfo {author} {\bibfnamefont {I.}~\bibnamefont
  {Pi\v{z}orn}}} (\bibinfo {year} {2008}),\ \href {\doibase
  10.1103/PhysRevLett.101.105701} {\bibfield  {journal} {\bibinfo  {journal}
  {Phys. Rev. Lett.}\ }\textbf {\bibinfo {volume} {101}},\ \bibinfo {pages}
  {105701}}\BibitemShut {NoStop}%
\bibitem [{\citenamefont {Prosen}\ and\ \citenamefont
  {Seligman}(2010)}]{ProsenSeligman2010}%
  \BibitemOpen
  \bibfield  {author} {\bibinfo {author} {\bibnamefont {Prosen}, \bibfnamefont
  {T.}}, \ and\ \bibinfo {author} {\bibfnamefont {T.~H.}\ \bibnamefont
  {Seligman}}} (\bibinfo {year} {2010}),\ \href {\doibase
  10.1088/1751-8113/43/39/392004} {\bibfield  {journal} {\bibinfo  {journal}
  {Journal of Physics A: Mathematical and Theoretical}\ }\textbf {\bibinfo
  {volume} {43}}~(\bibinfo {number} {39}),\ \bibinfo {pages}
  {392004}}\BibitemShut {NoStop}%
\bibitem [{\citenamefont {Prosen}\ and\ \citenamefont
  {\v{Z}nidari\v{c}}(2009)}]{Prosen2009}%
  \BibitemOpen
  \bibfield  {author} {\bibinfo {author} {\bibnamefont {Prosen}, \bibfnamefont
  {T.}}, \ and\ \bibinfo {author} {\bibfnamefont {M.}~\bibnamefont
  {\v{Z}nidari\v{c}}}} (\bibinfo {year} {2009}),\ \href {\doibase
  10.1088/1742-5468/2009/02/P02035} {\bibfield  {journal} {\bibinfo  {journal}
  {Journal of Statistical Mechanics: Theory and Experiment}\ }\textbf {\bibinfo
  {volume} {2009}}~(\bibinfo {number} {02}),\ \bibinfo {pages}
  {P02035}}\BibitemShut {NoStop}%
\bibitem [{\citenamefont {Prosen}\ and\ \citenamefont
  {\v{Z}nidari\v{c}}(2012)}]{ProsenZnidaric2012}%
  \BibitemOpen
  \bibfield  {author} {\bibinfo {author} {\bibnamefont {Prosen}, \bibfnamefont
  {T.}}, \ and\ \bibinfo {author} {\bibfnamefont {M.}~\bibnamefont
  {\v{Z}nidari\v{c}}}} (\bibinfo {year} {2012}),\ \href {\doibase
  10.1103/PhysRevB.86.125118} {\bibfield  {journal} {\bibinfo  {journal} {Phys.
  Rev. B}\ }\textbf {\bibinfo {volume} {86}},\ \bibinfo {pages}
  {125118}}\BibitemShut {NoStop}%
\bibitem [{\citenamefont {Prosen}\ and\ \citenamefont
  {\v{Z}nidari\v{c}}(2013)}]{ProsenZnidaric2013}%
  \BibitemOpen
  \bibfield  {author} {\bibinfo {author} {\bibnamefont {Prosen}, \bibfnamefont
  {T.}}, \ and\ \bibinfo {author} {\bibfnamefont {M.}~\bibnamefont
  {\v{Z}nidari\v{c}}}} (\bibinfo {year} {2013}),\ \href {\doibase
  10.1103/PhysRevLett.111.124101} {\bibfield  {journal} {\bibinfo  {journal}
  {Phys. Rev. Lett.}\ }\textbf {\bibinfo {volume} {111}},\ \bibinfo {pages}
  {124101}}\BibitemShut {NoStop}%
\bibitem [{\citenamefont {Prosen}\ and\ \citenamefont
  {\v{Z}unkovi\v{c}}(2010)}]{ProsenZunkovic2010}%
  \BibitemOpen
  \bibfield  {author} {\bibinfo {author} {\bibnamefont {Prosen}, \bibfnamefont
  {T.}}, \ and\ \bibinfo {author} {\bibfnamefont {B.}~\bibnamefont
  {\v{Z}unkovi\v{c}}}} (\bibinfo {year} {2010}),\ \href {\doibase
  10.1088/1367-2630/12/2/025016} {\bibfield  {journal} {\bibinfo  {journal}
  {New Journal of Physics}\ }\textbf {\bibinfo {volume} {12}},\ \bibinfo
  {pages} {025016}}\BibitemShut {NoStop}%
\bibitem [{\citenamefont {Ptaszy\'{n}ski}\ and\ \citenamefont
  {Esposito}(2019)}]{ptaszynski2019a}%
  \BibitemOpen
  \bibfield  {author} {\bibinfo {author} {\bibnamefont {Ptaszy\'{n}ski},
  \bibfnamefont {K.}}, \ and\ \bibinfo {author} {\bibfnamefont
  {M.}~\bibnamefont {Esposito}}} (\bibinfo {year} {2019}),\ \href {\doibase
  10.1103/PhysRevLett.122.150603} {\bibfield  {journal} {\bibinfo  {journal}
  {Phys. Rev. Lett.}\ }\textbf {\bibinfo {volume} {122}},\ \bibinfo {pages}
  {150603}}\BibitemShut {NoStop}%
\bibitem [{\citenamefont {Purkayastha}(2019)}]{Purkayastha2019}%
  \BibitemOpen
  \bibfield  {author} {\bibinfo {author} {\bibnamefont {Purkayastha},
  \bibfnamefont {A.}}} (\bibinfo {year} {2019}),\ \href {\doibase
  10.1088/1742-5468/ab02f4} {\bibfield  {journal} {\bibinfo  {journal} {Journal
  of Statistical Mechanics: Theory and Experiment}\ }\textbf {\bibinfo {volume}
  {2019}}~(\bibinfo {number} {4}),\ \bibinfo {pages} {043101}}\BibitemShut
  {NoStop}%
\bibitem [{\citenamefont {Purkayastha}(2022)}]{Purkayastha2022}%
  \BibitemOpen
  \bibfield  {author} {\bibinfo {author} {\bibnamefont {Purkayastha},
  \bibfnamefont {A.}}} (\bibinfo {year} {2022}),\ \href@noop {} {\bibfield
  {journal} {\bibinfo  {journal} {arXiv:2201.00677}\ }}\Eprint
  {http://arxiv.org/abs/2201.00677} {arXiv:2201.00677 [quant-ph]} \BibitemShut
  {NoStop}%
\bibitem [{\citenamefont {Purkayastha}\ \emph
  {et~al.}(2016{\natexlab{a}})\citenamefont {Purkayastha}, \citenamefont
  {Dhar},\ and\ \citenamefont {Kulkarni}}]{PurkayasthaKulkarni2016}%
  \BibitemOpen
  \bibfield  {author} {\bibinfo {author} {\bibnamefont {Purkayastha},
  \bibfnamefont {A.}}, \bibinfo {author} {\bibfnamefont {A.}~\bibnamefont
  {Dhar}}, \ and\ \bibinfo {author} {\bibfnamefont {M.}~\bibnamefont
  {Kulkarni}}} (\bibinfo {year} {2016}{\natexlab{a}}),\ \href {\doibase
  10.1103/PhysRevA.94.052134} {\bibfield  {journal} {\bibinfo  {journal} {Phys.
  Rev. A}\ }\textbf {\bibinfo {volume} {94}},\ \bibinfo {pages}
  {052134}}\BibitemShut {NoStop}%
\bibitem [{\citenamefont {Purkayastha}\ \emph
  {et~al.}(2016{\natexlab{b}})\citenamefont {Purkayastha}, \citenamefont
  {Dhar},\ and\ \citenamefont {Kulkarni}}]{Purkayastha2016}%
  \BibitemOpen
  \bibfield  {author} {\bibinfo {author} {\bibnamefont {Purkayastha},
  \bibfnamefont {A.}}, \bibinfo {author} {\bibfnamefont {A.}~\bibnamefont
  {Dhar}}, \ and\ \bibinfo {author} {\bibfnamefont {M.}~\bibnamefont
  {Kulkarni}}} (\bibinfo {year} {2016}{\natexlab{b}}),\ \href {\doibase
  10.1103/PhysRevA.93.062114} {\bibfield  {journal} {\bibinfo  {journal}
  {Physical Review A}\ }\textbf {\bibinfo {volume} {93}}~(\bibinfo {number}
  {6}),\ \bibinfo {pages} {062114}}\BibitemShut {NoStop}%
\bibitem [{\citenamefont {Purkayastha}\ \emph {et~al.}(2017)\citenamefont
  {Purkayastha}, \citenamefont {Dhar},\ and\ \citenamefont
  {Kulkarni}}]{PurkayasthaKulkarni2017b}%
  \BibitemOpen
  \bibfield  {author} {\bibinfo {author} {\bibnamefont {Purkayastha},
  \bibfnamefont {A.}}, \bibinfo {author} {\bibfnamefont {A.}~\bibnamefont
  {Dhar}}, \ and\ \bibinfo {author} {\bibfnamefont {M.}~\bibnamefont
  {Kulkarni}}} (\bibinfo {year} {2017}),\ \href {\doibase
  10.1103/PhysRevB.96.180204} {\bibfield  {journal} {\bibinfo  {journal} {Phys.
  Rev. B}\ }\textbf {\bibinfo {volume} {96}},\ \bibinfo {pages}
  {180204}}\BibitemShut {NoStop}%
\bibitem [{\citenamefont {Purkayastha}\ and\ \citenamefont
  {Dubi}(2017)}]{PurkayasthaDubi2017}%
  \BibitemOpen
  \bibfield  {author} {\bibinfo {author} {\bibnamefont {Purkayastha},
  \bibfnamefont {A.}}, \ and\ \bibinfo {author} {\bibfnamefont
  {Y.}~\bibnamefont {Dubi}}} (\bibinfo {year} {2017}),\ \href {\doibase
  10.1103/PhysRevB.96.085425} {\bibfield  {journal} {\bibinfo  {journal} {Phys.
  Rev. B}\ }\textbf {\bibinfo {volume} {96}},\ \bibinfo {pages}
  {085425}}\BibitemShut {NoStop}%
\bibitem [{\citenamefont {Purkayastha}\ \emph
  {et~al.}(2021{\natexlab{a}})\citenamefont {Purkayastha}, \citenamefont
  {Guarnieri}, \citenamefont {Campbell}, \citenamefont {Prior},\ and\
  \citenamefont {Goold}}]{PurkayasthaGoold2020}%
  \BibitemOpen
  \bibfield  {author} {\bibinfo {author} {\bibnamefont {Purkayastha},
  \bibfnamefont {A.}}, \bibinfo {author} {\bibfnamefont {G.}~\bibnamefont
  {Guarnieri}}, \bibinfo {author} {\bibfnamefont {S.}~\bibnamefont {Campbell}},
  \bibinfo {author} {\bibfnamefont {J.}~\bibnamefont {Prior}}, \ and\ \bibinfo
  {author} {\bibfnamefont {J.}~\bibnamefont {Goold}}} (\bibinfo {year}
  {2021}{\natexlab{a}}),\ \href {\doibase 10.1103/PhysRevB.104.045417}
  {\bibfield  {journal} {\bibinfo  {journal} {Phys. Rev. B}\ }\textbf {\bibinfo
  {volume} {104}},\ \bibinfo {pages} {045417}}\BibitemShut {NoStop}%
\bibitem [{\citenamefont {Purkayastha}\ \emph
  {et~al.}(2021{\natexlab{b}})\citenamefont {Purkayastha}, \citenamefont
  {Saha},\ and\ \citenamefont {Agarwalla}}]{PurkayasthaAgarwalla2021}%
  \BibitemOpen
  \bibfield  {author} {\bibinfo {author} {\bibnamefont {Purkayastha},
  \bibfnamefont {A.}}, \bibinfo {author} {\bibfnamefont {M.}~\bibnamefont
  {Saha}}, \ and\ \bibinfo {author} {\bibfnamefont {B.~K.}\ \bibnamefont
  {Agarwalla}}} (\bibinfo {year} {2021}{\natexlab{b}}),\ \href {\doibase
  10.1103/PhysRevLett.127.240601} {\bibfield  {journal} {\bibinfo  {journal}
  {Phys. Rev. Lett.}\ }\textbf {\bibinfo {volume} {127}},\ \bibinfo {pages}
  {240601}}\BibitemShut {NoStop}%
\bibitem [{\citenamefont {Purkayastha}\ \emph {et~al.}(2018)\citenamefont
  {Purkayastha}, \citenamefont {Sanyal}, \citenamefont {Dhar},\ and\
  \citenamefont {Kulkarni}}]{PurkayasthaKulkarni2018}%
  \BibitemOpen
  \bibfield  {author} {\bibinfo {author} {\bibnamefont {Purkayastha},
  \bibfnamefont {A.}}, \bibinfo {author} {\bibfnamefont {S.}~\bibnamefont
  {Sanyal}}, \bibinfo {author} {\bibfnamefont {A.}~\bibnamefont {Dhar}}, \ and\
  \bibinfo {author} {\bibfnamefont {M.}~\bibnamefont {Kulkarni}}} (\bibinfo
  {year} {2018}),\ \href {\doibase 10.1103/PhysRevB.97.174206} {\bibfield
  {journal} {\bibinfo  {journal} {Phys. Rev. B}\ }\textbf {\bibinfo {volume}
  {97}},\ \bibinfo {pages} {174206}}\BibitemShut {NoStop}%
\bibitem [{\citenamefont {Queisser}\ \emph {et~al.}(2014)\citenamefont
  {Queisser}, \citenamefont {Krutitsky}, \citenamefont {Navez},\ and\
  \citenamefont {Sch{\"u}tzhold}}]{queisser2014a}%
  \BibitemOpen
  \bibfield  {author} {\bibinfo {author} {\bibnamefont {Queisser},
  \bibfnamefont {F.}}, \bibinfo {author} {\bibfnamefont {K.~V.}\ \bibnamefont
  {Krutitsky}}, \bibinfo {author} {\bibfnamefont {P.}~\bibnamefont {Navez}}, \
  and\ \bibinfo {author} {\bibfnamefont {R.}~\bibnamefont {Sch{\"u}tzhold}}}
  (\bibinfo {year} {2014}),\ \href {\doibase 10.1103/PhysRevA.89.033616}
  {\bibfield  {journal} {\bibinfo  {journal} {Phys. Rev. A}\ }\textbf {\bibinfo
  {volume} {89}},\ \bibinfo {pages} {033616}}\BibitemShut {NoStop}%
\bibitem [{\citenamefont {Rams}\ and\ \citenamefont
  {Zwolak}(2020)}]{RamsZwolak2020}%
  \BibitemOpen
  \bibfield  {author} {\bibinfo {author} {\bibnamefont {Rams}, \bibfnamefont
  {M.~M.}}, \ and\ \bibinfo {author} {\bibfnamefont {M.}~\bibnamefont
  {Zwolak}}} (\bibinfo {year} {2020}),\ \href {\doibase
  10.1103/PhysRevLett.124.137701} {\bibfield  {journal} {\bibinfo  {journal}
  {Phys. Rev. Lett.}\ }\textbf {\bibinfo {volume} {124}},\ \bibinfo {pages}
  {137701}}\BibitemShut {NoStop}%
\bibitem [{\citenamefont {Rau}(1963)}]{Rau1963}%
  \BibitemOpen
  \bibfield  {author} {\bibinfo {author} {\bibnamefont {Rau}, \bibfnamefont
  {J.}}} (\bibinfo {year} {1963}),\ \href {\doibase 10.1103/PhysRev.129.1880}
  {\bibfield  {journal} {\bibinfo  {journal} {Physical Review}\ }\textbf
  {\bibinfo {volume} {129}}~(\bibinfo {number} {4}),\ \bibinfo {pages}
  {1880}}\BibitemShut {NoStop}%
\bibitem [{\citenamefont {Rebentrost}\ \emph {et~al.}(2009)\citenamefont
  {Rebentrost}, \citenamefont {Mohseni}, \citenamefont {Kassal}, \citenamefont
  {Lloyd},\ and\ \citenamefont {Aspuru-Guzik}}]{RebentrostAspuruGuzik2009}%
  \BibitemOpen
  \bibfield  {author} {\bibinfo {author} {\bibnamefont {Rebentrost},
  \bibfnamefont {P.}}, \bibinfo {author} {\bibfnamefont {M.}~\bibnamefont
  {Mohseni}}, \bibinfo {author} {\bibfnamefont {I.}~\bibnamefont {Kassal}},
  \bibinfo {author} {\bibfnamefont {S.}~\bibnamefont {Lloyd}}, \ and\ \bibinfo
  {author} {\bibfnamefont {A.}~\bibnamefont {Aspuru-Guzik}}} (\bibinfo {year}
  {2009}),\ \href {\doibase 10.1088/1367-2630/11/3/033003} {\bibfield
  {journal} {\bibinfo  {journal} {New Journal of Physics}\ }\textbf {\bibinfo
  {volume} {11}}~(\bibinfo {number} {3}),\ \bibinfo {pages}
  {033003}}\BibitemShut {NoStop}%
\bibitem [{\citenamefont {Redfield}(1965)}]{Redfield1965}%
  \BibitemOpen
  \bibfield  {author} {\bibinfo {author} {\bibnamefont {Redfield},
  \bibfnamefont {A.~G.}}} (\bibinfo {year} {1965}),\ \href {\doibase
  10.1016/B978-1-4832-3114-3.50007-6} {\emph {\bibinfo {title} {Advances in
  Magnetic and Optical Resonance}}},\ Vol.~\bibinfo {volume} {1}\ (\bibinfo
  {publisher} {Academic Press Inc.})\BibitemShut {NoStop}%
\bibitem [{\citenamefont {Restrepo}\ \emph {et~al.}(2019)\citenamefont
  {Restrepo}, \citenamefont {B{\"o}hling}, \citenamefont {Cerrillo},\ and\
  \citenamefont {Schaller}}]{restrepo2019a}%
  \BibitemOpen
  \bibfield  {author} {\bibinfo {author} {\bibnamefont {Restrepo},
  \bibfnamefont {S.}}, \bibinfo {author} {\bibfnamefont {S.}~\bibnamefont
  {B{\"o}hling}}, \bibinfo {author} {\bibfnamefont {J.}~\bibnamefont
  {Cerrillo}}, \ and\ \bibinfo {author} {\bibfnamefont {G.}~\bibnamefont
  {Schaller}}} (\bibinfo {year} {2019}),\ \href {\doibase
  10.1103/PhysRevB.100.035109} {\bibfield  {journal} {\bibinfo  {journal}
  {Phys. Rev. B}\ }\textbf {\bibinfo {volume} {100}},\ \bibinfo {pages}
  {035109}}\BibitemShut {NoStop}%
\bibitem [{\citenamefont {del Rey}\ \emph {et~al.}(2013)\citenamefont {del
  Rey}, \citenamefont {Chin}, \citenamefont {Huelga},\ and\ \citenamefont
  {Plenio}}]{peaks1}%
  \BibitemOpen
  \bibfield  {author} {\bibinfo {author} {\bibnamefont {del Rey}, \bibfnamefont
  {M.}}, \bibinfo {author} {\bibfnamefont {A.~W.}\ \bibnamefont {Chin}},
  \bibinfo {author} {\bibfnamefont {S.~F.}\ \bibnamefont {Huelga}}, \ and\
  \bibinfo {author} {\bibfnamefont {M.~B.}\ \bibnamefont {Plenio}}} (\bibinfo
  {year} {2013}),\ \href {\doibase 10.1021/jz400058a} {\bibfield  {journal}
  {\bibinfo  {journal} {J. Phys. Chem. Lett.}\ }\textbf {\bibinfo {volume}
  {4}}~(\bibinfo {number} {6}),\ \bibinfo {pages} {903}}\BibitemShut {NoStop}%
\bibitem [{\citenamefont {Ribeiro}\ and\ \citenamefont
  {Vieira}(2015)}]{Ribeiro2015}%
  \BibitemOpen
  \bibfield  {author} {\bibinfo {author} {\bibnamefont {Ribeiro}, \bibfnamefont
  {P.}}, \ and\ \bibinfo {author} {\bibfnamefont {V.~R.}\ \bibnamefont
  {Vieira}}} (\bibinfo {year} {2015}),\ \href {\doibase
  10.1103/PhysRevB.92.100302} {\bibfield  {journal} {\bibinfo  {journal}
  {Physical Review B}\ }\textbf {\bibinfo {volume} {92}},\ \bibinfo {pages}
  {100302(R)}}\BibitemShut {NoStop}%
\bibitem [{\citenamefont {Rieder}\ \emph {et~al.}(1967)\citenamefont {Rieder},
  \citenamefont {Lebowitz},\ and\ \citenamefont {Lieb}}]{Rieder1967}%
  \BibitemOpen
  \bibfield  {author} {\bibinfo {author} {\bibnamefont {Rieder}, \bibfnamefont
  {Z.}}, \bibinfo {author} {\bibfnamefont {J.~L.}\ \bibnamefont {Lebowitz}}, \
  and\ \bibinfo {author} {\bibfnamefont {E.}~\bibnamefont {Lieb}}} (\bibinfo
  {year} {1967}),\ \href {\doibase 10.1063/1.1705319} {\bibfield  {journal}
  {\bibinfo  {journal} {Journal of Mathematical Physics}\ }\textbf {\bibinfo
  {volume} {8}}~(\bibinfo {number} {5}),\ \bibinfo {pages} {1073}}\BibitemShut
  {NoStop}%
\bibitem [{\citenamefont {Riera-Campeny}\ \emph {et~al.}(2022)\citenamefont
  {Riera-Campeny}, \citenamefont {Sanpera},\ and\ \citenamefont
  {Strasberg}}]{RieraStrasberg2021}%
  \BibitemOpen
  \bibfield  {author} {\bibinfo {author} {\bibnamefont {Riera-Campeny},
  \bibfnamefont {A.}}, \bibinfo {author} {\bibfnamefont {A.}~\bibnamefont
  {Sanpera}}, \ and\ \bibinfo {author} {\bibfnamefont {P.}~\bibnamefont
  {Strasberg}}} (\bibinfo {year} {2022}),\ \href {\doibase
  10.1103/PhysRevE.105.054119} {\bibfield  {journal} {\bibinfo  {journal}
  {Phys. Rev. E}\ }\textbf {\bibinfo {volume} {105}},\ \bibinfo {pages}
  {054119}}\BibitemShut {NoStop}%
\bibitem [{\citenamefont {Rivas}\ and\ \citenamefont
  {Huelga}(2012)}]{RivasHuelgaBook}%
  \BibitemOpen
  \bibfield  {author} {\bibinfo {author} {\bibnamefont {Rivas}, \bibfnamefont
  {A.}}, \ and\ \bibinfo {author} {\bibfnamefont {S.~F.}\ \bibnamefont
  {Huelga}}} (\bibinfo {year} {2012}),\ \href {\doibase
  10.1007/978-3-642-23354-8} {\emph {\bibinfo {title} {Open quantum systems: an
  introduction}}}\ (\bibinfo  {publisher} {Springer Heidelberg Dordrecht London
  New York})\BibitemShut {NoStop}%
\bibitem [{\citenamefont {Rivas}\ \emph {et~al.}(2014)\citenamefont {Rivas},
  \citenamefont {Huelga},\ and\ \citenamefont {Plenio}}]{Rivas2014}%
  \BibitemOpen
  \bibfield  {author} {\bibinfo {author} {\bibnamefont {Rivas}, \bibfnamefont
  {{\'{A}}.}}, \bibinfo {author} {\bibfnamefont {S.~F.}\ \bibnamefont
  {Huelga}}, \ and\ \bibinfo {author} {\bibfnamefont {M.~B.}\ \bibnamefont
  {Plenio}}} (\bibinfo {year} {2014}),\ \href {\doibase
  10.1088/0034-4885/77/9/094001} {\bibfield  {journal} {\bibinfo  {journal}
  {Reports on Progress in Physics}\ }\textbf {\bibinfo {volume} {77}}~(\bibinfo
  {number} {9}),\ \bibinfo {pages} {094001}}\BibitemShut {NoStop}%
\bibitem [{\citenamefont {Rivas}\ and\ \citenamefont
  {Martin-Delgado}(2017)}]{RivasMartinDelgado2017}%
  \BibitemOpen
  \bibfield  {author} {\bibinfo {author} {\bibnamefont {Rivas}, \bibfnamefont
  {{\'{A}}.}}, \ and\ \bibinfo {author} {\bibfnamefont {M.~A.}\ \bibnamefont
  {Martin-Delgado}}} (\bibinfo {year} {2017}),\ \href {\doibase
  10.1038/s41598-017-06722-x} {\bibfield  {journal} {\bibinfo  {journal}
  {Scien. Rep.}\ }\textbf {\bibinfo {volume} {7}},\ \bibinfo {pages}
  {6350}}\BibitemShut {NoStop}%
\bibitem [{\citenamefont {Rivas}\ \emph {et~al.}(2010)\citenamefont {Rivas},
  \citenamefont {Plato}, \citenamefont {Huelga},\ and\ \citenamefont
  {Plenio}}]{Rivas2010b}%
  \BibitemOpen
  \bibfield  {author} {\bibinfo {author} {\bibnamefont {Rivas}, \bibfnamefont
  {{\'{A}}.}}, \bibinfo {author} {\bibfnamefont {A.~D.~K.}\ \bibnamefont
  {Plato}}, \bibinfo {author} {\bibfnamefont {S.~F.}\ \bibnamefont {Huelga}}, \
  and\ \bibinfo {author} {\bibfnamefont {M.~B.}\ \bibnamefont {Plenio}}}
  (\bibinfo {year} {2010}),\ \href {\doibase 10.1088/1367-2630/12/11/113032}
  {\bibfield  {journal} {\bibinfo  {journal} {New Journal of Physics}\ }\textbf
  {\bibinfo {volume} {12}}~(\bibinfo {number} {11}),\ \bibinfo {pages}
  {113032}}\BibitemShut {NoStop}%
\bibitem [{\citenamefont {Roberts}\ and\ \citenamefont
  {Walker}(2011)}]{Roberts2011}%
  \BibitemOpen
  \bibfield  {author} {\bibinfo {author} {\bibnamefont {Roberts}, \bibfnamefont
  {N.~A.}}, \ and\ \bibinfo {author} {\bibfnamefont {D.~G.}\ \bibnamefont
  {Walker}}} (\bibinfo {year} {2011}),\ \href {\doibase
  10.1016/j.ijthermalsci.2010.12.004} {\bibfield  {journal} {\bibinfo
  {journal} {International Journal of Thermal Sciences}\ }\textbf {\bibinfo
  {volume} {50}}~(\bibinfo {number} {5}),\ \bibinfo {pages} {648}}\BibitemShut
  {NoStop}%
\bibitem [{\citenamefont {Rolf-Pissarczyk}\ \emph {et~al.}(2017)\citenamefont
  {Rolf-Pissarczyk}, \citenamefont {Yan}, \citenamefont {Malavolti},
  \citenamefont {Burgess}, \citenamefont {McMurtrie},\ and\ \citenamefont
  {Loth}}]{RolfPissarczykLoth2017}%
  \BibitemOpen
  \bibfield  {author} {\bibinfo {author} {\bibnamefont {Rolf-Pissarczyk},
  \bibfnamefont {S.}}, \bibinfo {author} {\bibfnamefont {S.}~\bibnamefont
  {Yan}}, \bibinfo {author} {\bibfnamefont {L.}~\bibnamefont {Malavolti}},
  \bibinfo {author} {\bibfnamefont {J.~A.~J.}\ \bibnamefont {Burgess}},
  \bibinfo {author} {\bibfnamefont {G.}~\bibnamefont {McMurtrie}}, \ and\
  \bibinfo {author} {\bibfnamefont {S.}~\bibnamefont {Loth}}} (\bibinfo {year}
  {2017}),\ \href {\doibase 10.1103/PhysRevLett.119.217201} {\bibfield
  {journal} {\bibinfo  {journal} {Phys. Rev. Lett.}\ }\textbf {\bibinfo
  {volume} {119}},\ \bibinfo {pages} {217201}}\BibitemShut {NoStop}%
\bibitem [{\citenamefont {Ros}\ \emph {et~al.}(2015)\citenamefont {Ros},
  \citenamefont {M{\"u}ller},\ and\ \citenamefont
  {Scardicchio}}]{RosScardicchio2015}%
  \BibitemOpen
  \bibfield  {author} {\bibinfo {author} {\bibnamefont {Ros}, \bibfnamefont
  {V.}}, \bibinfo {author} {\bibfnamefont {M.}~\bibnamefont {M{\"u}ller}}, \
  and\ \bibinfo {author} {\bibfnamefont {A.}~\bibnamefont {Scardicchio}}}
  (\bibinfo {year} {2015}),\ \href {\doibase 10.1016/j.nuclphysb.2014.12.014}
  {\bibfield  {journal} {\bibinfo  {journal} {Nuclear Physics B}\ }\textbf
  {\bibinfo {volume} {891}},\ \bibinfo {pages} {420}}\BibitemShut {NoStop}%
\bibitem [{\citenamefont {Roy}\ and\ \citenamefont
  {Sharma}(2019)}]{RoySharma2019}%
  \BibitemOpen
  \bibfield  {author} {\bibinfo {author} {\bibnamefont {Roy}, \bibfnamefont
  {N.}}, \ and\ \bibinfo {author} {\bibfnamefont {A.}~\bibnamefont {Sharma}}}
  (\bibinfo {year} {2019}),\ \href {\doibase 10.1103/PhysRevB.100.195143}
  {\bibfield  {journal} {\bibinfo  {journal} {Phys. Rev. B}\ }\textbf {\bibinfo
  {volume} {100}},\ \bibinfo {pages} {195143}}\BibitemShut {NoStop}%
\bibitem [{\citenamefont {Álvaro Rubio-García}\ \emph
  {et~al.}(2022)\citenamefont {Álvaro Rubio-García}, \citenamefont {Molina},\
  and\ \citenamefont {Dukelsky}}]{RubioGarcia2021}%
  \BibitemOpen
  \bibfield  {author} {\bibinfo {author} {\bibnamefont {Álvaro
  Rubio-García},}, \bibinfo {author} {\bibfnamefont {R.~A.}\ \bibnamefont
  {Molina}}, \ and\ \bibinfo {author} {\bibfnamefont {J.}~\bibnamefont
  {Dukelsky}}} (\bibinfo {year} {2022}),\ \href {\doibase
  10.21468/SciPostPhysCore.5.2.026} {\bibfield  {journal} {\bibinfo  {journal}
  {SciPost Phys. Core}\ }\textbf {\bibinfo {volume} {5}},\ \bibinfo {pages}
  {26}}\BibitemShut {NoStop}%
\bibitem [{\citenamefont {Ryndyk}(2016)}]{Ryndyk2016}%
  \BibitemOpen
  \bibfield  {author} {\bibinfo {author} {\bibnamefont {Ryndyk}, \bibfnamefont
  {D.}}} (\bibinfo {year} {2016}),\ \href {\doibase 10.1007/978-3-319-24088-6}
  {\emph {\bibinfo {title} {Theory of {{Quantum Transport}} at
  {{Nanoscale}}}}},\ Vol.\ \bibinfo {volume} {184}\ (\bibinfo  {publisher}
  {{Springer International Publishing}},\ \bibinfo {address}
  {{Cham}})\BibitemShut {NoStop}%
\bibitem [{\citenamefont {S\'a}\ \emph
  {et~al.}(2020{\natexlab{a}})\citenamefont {S\'a}, \citenamefont {Ribeiro},
  \citenamefont {Can},\ and\ \citenamefont {Prosen}}]{SaProsen2020b}%
  \BibitemOpen
  \bibfield  {author} {\bibinfo {author} {\bibnamefont {S\'a}, \bibfnamefont
  {L.}}, \bibinfo {author} {\bibfnamefont {P.}~\bibnamefont {Ribeiro}},
  \bibinfo {author} {\bibfnamefont {T.}~\bibnamefont {Can}}, \ and\ \bibinfo
  {author} {\bibfnamefont {T.}~\bibnamefont {Prosen}}} (\bibinfo {year}
  {2020}{\natexlab{a}}),\ \href {\doibase 10.1103/PhysRevB.102.134310}
  {\bibfield  {journal} {\bibinfo  {journal} {Phys. Rev. B}\ }\textbf {\bibinfo
  {volume} {102}},\ \bibinfo {pages} {134310}}\BibitemShut {NoStop}%
\bibitem [{\citenamefont {S\'a}\ \emph
  {et~al.}(2020{\natexlab{b}})\citenamefont {S\'a}, \citenamefont {Ribeiro},\
  and\ \citenamefont {Prosen}}]{SaProsen2020}%
  \BibitemOpen
  \bibfield  {author} {\bibinfo {author} {\bibnamefont {S\'a}, \bibfnamefont
  {L.}}, \bibinfo {author} {\bibfnamefont {P.}~\bibnamefont {Ribeiro}}, \ and\
  \bibinfo {author} {\bibfnamefont {T.}~\bibnamefont {Prosen}}} (\bibinfo
  {year} {2020}{\natexlab{b}}),\ \href {\doibase 10.1103/PhysRevX.10.021019}
  {\bibfield  {journal} {\bibinfo  {journal} {Phys. Rev. X}\ }\textbf {\bibinfo
  {volume} {10}},\ \bibinfo {pages} {021019}}\BibitemShut {NoStop}%
\bibitem [{\citenamefont {S{\'{a}}}\ \emph {et~al.}(2020)\citenamefont
  {S{\'{a}}}, \citenamefont {Ribeiro},\ and\ \citenamefont
  {Prosen}}]{SaProsen2020c}%
  \BibitemOpen
  \bibfield  {author} {\bibinfo {author} {\bibnamefont {S{\'{a}}},
  \bibfnamefont {L.}}, \bibinfo {author} {\bibfnamefont {P.}~\bibnamefont
  {Ribeiro}}, \ and\ \bibinfo {author} {\bibfnamefont {T.}~\bibnamefont
  {Prosen}}} (\bibinfo {year} {2020}),\ \href {\doibase
  10.1088/1751-8121/ab9337} {\bibfield  {journal} {\bibinfo  {journal} {Journal
  of Physics A: Mathematical and Theoretical}\ }\textbf {\bibinfo {volume}
  {53}}~(\bibinfo {number} {30}),\ \bibinfo {pages} {305303}}\BibitemShut
  {NoStop}%
\bibitem [{\citenamefont {Sachdev}(2011)}]{Sachdev2011}%
  \BibitemOpen
  \bibfield  {author} {\bibinfo {author} {\bibnamefont {Sachdev}, \bibfnamefont
  {S.}}} (\bibinfo {year} {2011}),\ \href {\doibase 10.1017/CBO9780511973765}
  {\emph {\bibinfo {title} {{Quantum Phase Transitions}}}}\ (\bibinfo
  {publisher} {Cambridge University Press},\ \bibinfo {address} {Cambridge
  UK})\BibitemShut {NoStop}%
\bibitem [{\citenamefont {Saha}\ \emph {et~al.}(2019)\citenamefont {Saha},
  \citenamefont {Maiti},\ and\ \citenamefont
  {Purkayastha}}]{SahaPurkayastha2019}%
  \BibitemOpen
  \bibfield  {author} {\bibinfo {author} {\bibnamefont {Saha}, \bibfnamefont
  {M.}}, \bibinfo {author} {\bibfnamefont {S.~K.}\ \bibnamefont {Maiti}}, \
  and\ \bibinfo {author} {\bibfnamefont {A.}~\bibnamefont {Purkayastha}}}
  (\bibinfo {year} {2019}),\ \href {\doibase 10.1103/PhysRevB.100.174201}
  {\bibfield  {journal} {\bibinfo  {journal} {Phys. Rev. B}\ }\textbf {\bibinfo
  {volume} {100}},\ \bibinfo {pages} {174201}}\BibitemShut {NoStop}%
\bibitem [{\citenamefont {Saha}\ \emph {et~al.}(2022)\citenamefont {Saha},
  \citenamefont {Venkatesh},\ and\ \citenamefont
  {Agarwalla}}]{SahaAgarwalla2022}%
  \BibitemOpen
  \bibfield  {author} {\bibinfo {author} {\bibnamefont {Saha}, \bibfnamefont
  {M.}}, \bibinfo {author} {\bibfnamefont {B.~P.}\ \bibnamefont {Venkatesh}}, \
  and\ \bibinfo {author} {\bibfnamefont {B.~K.}\ \bibnamefont {Agarwalla}}}
  (\bibinfo {year} {2022}),\ \href {\doibase 10.48550/arXiv.2202.14033}
  {\bibfield  {journal} {\bibinfo  {journal} {arxiv:2202.14033}\
  }10.48550/arXiv.2202.14033}\BibitemShut {NoStop}%
\bibitem [{\citenamefont {Saito}\ and\ \citenamefont
  {Utsumi}(2008)}]{saito2008a}%
  \BibitemOpen
  \bibfield  {author} {\bibinfo {author} {\bibnamefont {Saito}, \bibfnamefont
  {K.}}, \ and\ \bibinfo {author} {\bibfnamefont {Y.}~\bibnamefont {Utsumi}}}
  (\bibinfo {year} {2008}),\ \href {\doibase 10.1103/PhysRevB.78.115429}
  {\bibfield  {journal} {\bibinfo  {journal} {Phys. Rev. B}\ }\textbf {\bibinfo
  {volume} {78}},\ \bibinfo {pages} {115429}}\BibitemShut {NoStop}%
\bibitem [{\citenamefont {Scalapino}\ \emph {et~al.}(1993)\citenamefont
  {Scalapino}, \citenamefont {White},\ and\ \citenamefont
  {Zhang}}]{ScalapinoZhang1993}%
  \BibitemOpen
  \bibfield  {author} {\bibinfo {author} {\bibnamefont {Scalapino},
  \bibfnamefont {D.~J.}}, \bibinfo {author} {\bibfnamefont {S.~R.}\
  \bibnamefont {White}}, \ and\ \bibinfo {author} {\bibfnamefont
  {S.}~\bibnamefont {Zhang}}} (\bibinfo {year} {1993}),\ \href {\doibase
  10.1103/PhysRevB.47.7995} {\bibfield  {journal} {\bibinfo  {journal} {Phys.
  Rev. B}\ }\textbf {\bibinfo {volume} {47}},\ \bibinfo {pages}
  {7995}}\BibitemShut {NoStop}%
\bibitem [{\citenamefont {Scali}\ \emph {et~al.}(2021)\citenamefont {Scali},
  \citenamefont {Anders},\ and\ \citenamefont {Correa}}]{Scali2021}%
  \BibitemOpen
  \bibfield  {author} {\bibinfo {author} {\bibnamefont {Scali}, \bibfnamefont
  {S.}}, \bibinfo {author} {\bibfnamefont {J.}~\bibnamefont {Anders}}, \ and\
  \bibinfo {author} {\bibfnamefont {L.~A.}\ \bibnamefont {Correa}}} (\bibinfo
  {year} {2021}),\ \href {\doibase 10.22331/q-2021-05-01-451} {\bibfield
  {journal} {\bibinfo  {journal} {{Quantum}}\ }\textbf {\bibinfo {volume}
  {5}},\ \bibinfo {pages} {451}}\BibitemShut {NoStop}%
\bibitem [{\citenamefont {Scarani}\ \emph {et~al.}(2002)\citenamefont
  {Scarani}, \citenamefont {Ziman}, \citenamefont {{\v{S}}telmachovi{\v{c}}},
  \citenamefont {Gisin}, \citenamefont {Bu{\v{z}}ek},\ and\ \citenamefont
  {Bu{\v{z}}ek}}]{Scarani2002}%
  \BibitemOpen
  \bibfield  {author} {\bibinfo {author} {\bibnamefont {Scarani}, \bibfnamefont
  {V.}}, \bibinfo {author} {\bibfnamefont {M.}~\bibnamefont {Ziman}}, \bibinfo
  {author} {\bibfnamefont {P.}~\bibnamefont {{\v{S}}telmachovi{\v{c}}}},
  \bibinfo {author} {\bibfnamefont {N.}~\bibnamefont {Gisin}}, \bibinfo
  {author} {\bibfnamefont {V.}~\bibnamefont {Bu{\v{z}}ek}}, \ and\ \bibinfo
  {author} {\bibfnamefont {V.}~\bibnamefont {Bu{\v{z}}ek}}} (\bibinfo {year}
  {2002}),\ \href {\doibase 10.1103/PhysRevLett.88.097905} {\bibfield
  {journal} {\bibinfo  {journal} {Physical Review Letters}\ }\textbf {\bibinfo
  {volume} {88}}~(\bibinfo {number} {9}),\ \bibinfo {pages}
  {097905}}\BibitemShut {NoStop}%
\bibitem [{\citenamefont {Schaller}(2014)}]{schaller2014}%
  \BibitemOpen
  \bibfield  {author} {\bibinfo {author} {\bibnamefont {Schaller},
  \bibfnamefont {G.}}} (\bibinfo {year} {2014}),\ \href {\doibase
  10.1007/978-3-319-03877-3} {\emph {\bibinfo {title} {Open Quantum Systems Far
  from Equilibrium}}},\ \bibinfo {series} {Lecture Notes in Physics}, Vol.\
  \bibinfo {volume} {881}\ (\bibinfo  {publisher} {Springer},\ \bibinfo
  {address} {Cham})\BibitemShut {NoStop}%
\bibitem [{\citenamefont {Schaller}\ \emph {et~al.}(2018)\citenamefont
  {Schaller}, \citenamefont {Cerrillo}, \citenamefont {Engelhardt},\ and\
  \citenamefont {Strasberg}}]{SchallerStrasberg2018}%
  \BibitemOpen
  \bibfield  {author} {\bibinfo {author} {\bibnamefont {Schaller},
  \bibfnamefont {G.}}, \bibinfo {author} {\bibfnamefont {J.}~\bibnamefont
  {Cerrillo}}, \bibinfo {author} {\bibfnamefont {G.}~\bibnamefont
  {Engelhardt}}, \ and\ \bibinfo {author} {\bibfnamefont {P.}~\bibnamefont
  {Strasberg}}} (\bibinfo {year} {2018}),\ \href {\doibase
  10.1103/PhysRevB.97.195104} {\bibfield  {journal} {\bibinfo  {journal} {Phys.
  Rev. B}\ }\textbf {\bibinfo {volume} {97}},\ \bibinfo {pages}
  {195104}}\BibitemShut {NoStop}%
\bibitem [{\citenamefont {Schaller}\ \emph {et~al.}(2016)\citenamefont
  {Schaller}, \citenamefont {Giusteri},\ and\ \citenamefont
  {Celardo}}]{SchallerCelardo2016}%
  \BibitemOpen
  \bibfield  {author} {\bibinfo {author} {\bibnamefont {Schaller},
  \bibfnamefont {G.}}, \bibinfo {author} {\bibfnamefont {G.~G.}\ \bibnamefont
  {Giusteri}}, \ and\ \bibinfo {author} {\bibfnamefont {G.~L.}\ \bibnamefont
  {Celardo}}} (\bibinfo {year} {2016}),\ \href {\doibase
  10.1103/PhysRevE.94.032135} {\bibfield  {journal} {\bibinfo  {journal} {Phys.
  Rev. E}\ }\textbf {\bibinfo {volume} {94}},\ \bibinfo {pages}
  {032135}}\BibitemShut {NoStop}%
\bibitem [{\citenamefont {Schaller}\ \emph {et~al.}(2010)\citenamefont
  {Schaller}, \citenamefont {Kie\ss{}lich},\ and\ \citenamefont
  {Brandes}}]{schaller2010b}%
  \BibitemOpen
  \bibfield  {author} {\bibinfo {author} {\bibnamefont {Schaller},
  \bibfnamefont {G.}}, \bibinfo {author} {\bibfnamefont {G.}~\bibnamefont
  {Kie\ss{}lich}}, \ and\ \bibinfo {author} {\bibfnamefont {T.}~\bibnamefont
  {Brandes}}} (\bibinfo {year} {2010}),\ \href {\doibase
  10.1103/PhysRevB.81.205305} {\bibfield  {journal} {\bibinfo  {journal}
  {Physical Review B}\ }\textbf {\bibinfo {volume} {81}}~(\bibinfo {number}
  {20}),\ \bibinfo {pages} {205305}}\BibitemShut {NoStop}%
\bibitem [{\citenamefont {Schaller}\ \emph {et~al.}(2013)\citenamefont
  {Schaller}, \citenamefont {Krause}, \citenamefont {Brandes},\ and\
  \citenamefont {Esposito}}]{schaller2013a}%
  \BibitemOpen
  \bibfield  {author} {\bibinfo {author} {\bibnamefont {Schaller},
  \bibfnamefont {G.}}, \bibinfo {author} {\bibfnamefont {T.}~\bibnamefont
  {Krause}}, \bibinfo {author} {\bibfnamefont {T.}~\bibnamefont {Brandes}}, \
  and\ \bibinfo {author} {\bibfnamefont {M.}~\bibnamefont {Esposito}}}
  (\bibinfo {year} {2013}),\ \href {\doibase 10.1088/1367-2630/15/3/033032}
  {\bibfield  {journal} {\bibinfo  {journal} {New Journal of Physics}\ }\textbf
  {\bibinfo {volume} {15}},\ \bibinfo {pages} {033032}}\BibitemShut {NoStop}%
\bibitem [{\citenamefont {Schaller}\ \emph
  {et~al.}(2014{\natexlab{a}})\citenamefont {Schaller}, \citenamefont
  {Nietner},\ and\ \citenamefont {Brandes}}]{schaller2014b}%
  \BibitemOpen
  \bibfield  {author} {\bibinfo {author} {\bibnamefont {Schaller},
  \bibfnamefont {G.}}, \bibinfo {author} {\bibfnamefont {C.}~\bibnamefont
  {Nietner}}, \ and\ \bibinfo {author} {\bibfnamefont {T.}~\bibnamefont
  {Brandes}}} (\bibinfo {year} {2014}{\natexlab{a}}),\ \href {\doibase
  10.1088/1367-2630/16/12/125011} {\bibfield  {journal} {\bibinfo  {journal}
  {New Journal of Physics}\ }\textbf {\bibinfo {volume} {16}},\ \bibinfo
  {pages} {125011}}\BibitemShut {NoStop}%
\bibitem [{\citenamefont {Schaller}\ \emph {et~al.}(2022)\citenamefont
  {Schaller}, \citenamefont {Queisser}, \citenamefont {Szpak}, \citenamefont
  {K\"onig},\ and\ \citenamefont {Sch\"utzhold}}]{schaller2022a}%
  \BibitemOpen
  \bibfield  {author} {\bibinfo {author} {\bibnamefont {Schaller},
  \bibfnamefont {G.}}, \bibinfo {author} {\bibfnamefont {F.}~\bibnamefont
  {Queisser}}, \bibinfo {author} {\bibfnamefont {N.}~\bibnamefont {Szpak}},
  \bibinfo {author} {\bibfnamefont {J.}~\bibnamefont {K\"onig}}, \ and\
  \bibinfo {author} {\bibfnamefont {R.}~\bibnamefont {Sch\"utzhold}}} (\bibinfo
  {year} {2022}),\ \href {\doibase 10.1103/PhysRevB.105.115139} {\bibfield
  {journal} {\bibinfo  {journal} {Phys. Rev. B}\ }\textbf {\bibinfo {volume}
  {105}},\ \bibinfo {pages} {115139}}\BibitemShut {NoStop}%
\bibitem [{\citenamefont {Schaller}\ \emph
  {et~al.}(2014{\natexlab{b}})\citenamefont {Schaller}, \citenamefont {Vogl},\
  and\ \citenamefont {Brandes}}]{SchallerBrandes2014}%
  \BibitemOpen
  \bibfield  {author} {\bibinfo {author} {\bibnamefont {Schaller},
  \bibfnamefont {G.}}, \bibinfo {author} {\bibfnamefont {M.}~\bibnamefont
  {Vogl}}, \ and\ \bibinfo {author} {\bibfnamefont {T.}~\bibnamefont
  {Brandes}}} (\bibinfo {year} {2014}{\natexlab{b}}),\ \href {\doibase
  10.1088/0953-8984/26/26/265001} {\bibfield  {journal} {\bibinfo  {journal}
  {Journal of Physics: Condensed Matter}\ }\textbf {\bibinfo {volume}
  {26}}~(\bibinfo {number} {26}),\ \bibinfo {pages} {265001}}\BibitemShut
  {NoStop}%
\bibitem [{\citenamefont {Schaller}\ \emph {et~al.}(2009)\citenamefont
  {Schaller}, \citenamefont {Zedler},\ and\ \citenamefont
  {Brandes}}]{schaller2009a}%
  \BibitemOpen
  \bibfield  {author} {\bibinfo {author} {\bibnamefont {Schaller},
  \bibfnamefont {G.}}, \bibinfo {author} {\bibfnamefont {P.}~\bibnamefont
  {Zedler}}, \ and\ \bibinfo {author} {\bibfnamefont {T.}~\bibnamefont
  {Brandes}}} (\bibinfo {year} {2009}),\ \href {\doibase
  10.1103/PhysRevA.79.032110} {\bibfield  {journal} {\bibinfo  {journal} {Phys.
  Rev. A}\ }\textbf {\bibinfo {volume} {79}},\ \bibinfo {pages}
  {032110}}\BibitemShut {NoStop}%
\bibitem [{\citenamefont {Scheible}\ and\ \citenamefont
  {Blick}(2004)}]{scheible2004a}%
  \BibitemOpen
  \bibfield  {author} {\bibinfo {author} {\bibnamefont {Scheible},
  \bibfnamefont {D.~V.}}, \ and\ \bibinfo {author} {\bibfnamefont {R.~H.}\
  \bibnamefont {Blick}}} (\bibinfo {year} {2004}),\ \href {\doibase
  10.1063/1.1759371} {\bibfield  {journal} {\bibinfo  {journal} {Applied
  Physics Letters}\ }\textbf {\bibinfo {volume} {84}}~(\bibinfo {number}
  {23}),\ \bibinfo {pages} {4632}}\BibitemShut {NoStop}%
\bibitem [{\citenamefont {Scheibner}\ \emph {et~al.}(2008)\citenamefont
  {Scheibner}, \citenamefont {Konig}, \citenamefont {Reuter}, \citenamefont
  {Weick}, \citenamefont {Gould}, \citenamefont {Buhmann},\ and\ \citenamefont
  {Molenkamp}}]{ScheibnerMolenkamp2008}%
  \BibitemOpen
  \bibfield  {author} {\bibinfo {author} {\bibnamefont {Scheibner},
  \bibfnamefont {R.}}, \bibinfo {author} {\bibfnamefont {M.}~\bibnamefont
  {Konig}}, \bibinfo {author} {\bibfnamefont {D.}~\bibnamefont {Reuter}},
  \bibinfo {author} {\bibfnamefont {A.}~\bibnamefont {Weick}}, \bibinfo
  {author} {\bibfnamefont {C.}~\bibnamefont {Gould}}, \bibinfo {author}
  {\bibfnamefont {H.}~\bibnamefont {Buhmann}}, \ and\ \bibinfo {author}
  {\bibfnamefont {L.}~\bibnamefont {Molenkamp}}} (\bibinfo {year} {2008}),\
  \href {\doibase 10.1088/1367-2630/10/8/083016} {\bibfield  {journal}
  {\bibinfo  {journal} {New Journal of Physics}\ }\textbf {\bibinfo {volume}
  {10}},\ \bibinfo {pages} {083016}}\BibitemShut {NoStop}%
\bibitem [{\citenamefont {Scheie}\ \emph {et~al.}(2021)\citenamefont {Scheie},
  \citenamefont {Sherman}, \citenamefont {Dupont}, \citenamefont {Nagler},
  \citenamefont {Stone}, \citenamefont {Granroth}, \citenamefont {Moore},\ and\
  \citenamefont {Tennant}}]{ScheieTennat2021}%
  \BibitemOpen
  \bibfield  {author} {\bibinfo {author} {\bibnamefont {Scheie}, \bibfnamefont
  {A.}}, \bibinfo {author} {\bibfnamefont {N.~E.}\ \bibnamefont {Sherman}},
  \bibinfo {author} {\bibfnamefont {M.}~\bibnamefont {Dupont}}, \bibinfo
  {author} {\bibfnamefont {S.~E.}\ \bibnamefont {Nagler}}, \bibinfo {author}
  {\bibfnamefont {M.~B.}\ \bibnamefont {Stone}}, \bibinfo {author}
  {\bibfnamefont {G.~E.}\ \bibnamefont {Granroth}}, \bibinfo {author}
  {\bibfnamefont {J.~E.}\ \bibnamefont {Moore}}, \ and\ \bibinfo {author}
  {\bibfnamefont {D.~A.}\ \bibnamefont {Tennant}}} (\bibinfo {year} {2021}),\
  \href {\doibase 10.1038/s41567-021-01191-6} {\bibfield  {journal} {\bibinfo
  {journal} {Nature Physics}\ }\textbf {\bibinfo {volume} {17}},\ \bibinfo
  {pages} {726}}\BibitemShut {NoStop}%
\bibitem [{\citenamefont {Schlosshauer}(2007)}]{schlosshauer2007}%
  \BibitemOpen
  \bibfield  {author} {\bibinfo {author} {\bibnamefont {Schlosshauer},
  \bibfnamefont {M.}}} (\bibinfo {year} {2007}),\ \href {\doibase
  10.1007/978-3-540-35775-9} {\emph {\bibinfo {title} {Decoherence and the
  Quantum-To-Classical Transition}}}\ (\bibinfo  {publisher} {Springer-Verlag
  Berlin Heidelberg})\BibitemShut {NoStop}%
\bibitem [{\citenamefont {Schollw{\"o}ck}(2005)}]{Schollwock2005}%
  \BibitemOpen
  \bibfield  {author} {\bibinfo {author} {\bibnamefont {Schollw{\"o}ck},
  \bibfnamefont {U.}}} (\bibinfo {year} {2005}),\ \href {\doibase
  10.1103/RevModPhys.77.259} {\bibfield  {journal} {\bibinfo  {journal} {Rev.
  Mod. Phys.}\ }\textbf {\bibinfo {volume} {77}},\ \bibinfo {pages}
  {259}}\BibitemShut {NoStop}%
\bibitem [{\citenamefont {Schollw\"{o}ck}(2011)}]{Schollwock2011}%
  \BibitemOpen
  \bibfield  {author} {\bibinfo {author} {\bibnamefont {Schollw\"{o}ck},
  \bibfnamefont {U.}}} (\bibinfo {year} {2011}),\ \href {\doibase
  10.1016/j.aop.2010.09.012} {\bibfield  {journal} {\bibinfo  {journal} {Annals
  of Phys.}\ }\textbf {\bibinfo {volume} {326}},\ \bibinfo {pages}
  {96}}\BibitemShut {NoStop}%
\bibitem [{\citenamefont {Sch{\"o}nhammer}(2007)}]{schoenhammer2007a}%
  \BibitemOpen
  \bibfield  {author} {\bibinfo {author} {\bibnamefont {Sch{\"o}nhammer},
  \bibfnamefont {K.}}} (\bibinfo {year} {2007}),\ \href {\doibase
  10.1103/PhysRevB.75.205329} {\bibfield  {journal} {\bibinfo  {journal} {Phys.
  Rev. B}\ }\textbf {\bibinfo {volume} {75}},\ \bibinfo {pages}
  {205329}}\BibitemShut {NoStop}%
\bibitem [{\citenamefont {Schreiber}\ \emph {et~al.}(2015)\citenamefont
  {Schreiber}, \citenamefont {Hodgman}, \citenamefont {Bordia}, \citenamefont
  {L{\"u}schen}, \citenamefont {Fischer}, \citenamefont {Vosk}, \citenamefont
  {Altman}, \citenamefont {Schneider},\ and\ \citenamefont
  {Bloch}}]{SchreiberBloch2015}%
  \BibitemOpen
  \bibfield  {author} {\bibinfo {author} {\bibnamefont {Schreiber},
  \bibfnamefont {M.}}, \bibinfo {author} {\bibfnamefont {S.~S.}\ \bibnamefont
  {Hodgman}}, \bibinfo {author} {\bibfnamefont {P.}~\bibnamefont {Bordia}},
  \bibinfo {author} {\bibfnamefont {H.~P.}\ \bibnamefont {L{\"u}schen}},
  \bibinfo {author} {\bibfnamefont {M.~H.}\ \bibnamefont {Fischer}}, \bibinfo
  {author} {\bibfnamefont {R.}~\bibnamefont {Vosk}}, \bibinfo {author}
  {\bibfnamefont {E.}~\bibnamefont {Altman}}, \bibinfo {author} {\bibfnamefont
  {U.}~\bibnamefont {Schneider}}, \ and\ \bibinfo {author} {\bibfnamefont
  {I.}~\bibnamefont {Bloch}}} (\bibinfo {year} {2015}),\ \href {\doibase
  10.1126/science.aaa7432} {\bibfield  {journal} {\bibinfo  {journal}
  {Science}\ }\textbf {\bibinfo {volume} {349}},\ \bibinfo {pages}
  {842}}\BibitemShut {NoStop}%
\bibitem [{\citenamefont {Schr{\"o}der}\ \emph {et~al.}(2007)\citenamefont
  {Schr{\"o}der}, \citenamefont {Schreiber},\ and\ \citenamefont
  {Kleinekath{\"o}fer}}]{schroeder2007a}%
  \BibitemOpen
  \bibfield  {author} {\bibinfo {author} {\bibnamefont {Schr{\"o}der},
  \bibfnamefont {M.}}, \bibinfo {author} {\bibfnamefont {M.}~\bibnamefont
  {Schreiber}}, \ and\ \bibinfo {author} {\bibfnamefont {U.}~\bibnamefont
  {Kleinekath{\"o}fer}}} (\bibinfo {year} {2007}),\ \href {\doibase
  10.1063/1.2538754} {\bibfield  {journal} {\bibinfo  {journal} {The Journal of
  Chemical Physics}\ }\textbf {\bibinfo {volume} {126}},\ \bibinfo {pages}
  {114102}}\BibitemShut {NoStop}%
\bibitem [{\citenamefont {Schuab}\ \emph {et~al.}(2016)\citenamefont {Schuab},
  \citenamefont {Pereira},\ and\ \citenamefont {Landi}}]{Schuab2016a}%
  \BibitemOpen
  \bibfield  {author} {\bibinfo {author} {\bibnamefont {Schuab}, \bibfnamefont
  {L.}}, \bibinfo {author} {\bibfnamefont {E.}~\bibnamefont {Pereira}}, \ and\
  \bibinfo {author} {\bibfnamefont {G.~T.}\ \bibnamefont {Landi}}} (\bibinfo
  {year} {2016}),\ \href {\doibase 10.1103/PhysRevE.94.042122} {\bibfield
  {journal} {\bibinfo  {journal} {Physical Review E}\ }\textbf {\bibinfo
  {volume} {94}}~(\bibinfo {number} {4}),\ \bibinfo {pages}
  {042122}}\BibitemShut {NoStop}%
\bibitem [{\citenamefont {Schultz}\ and\ \citenamefont {von
  Oppen}(2009)}]{schultz2009a}%
  \BibitemOpen
  \bibfield  {author} {\bibinfo {author} {\bibnamefont {Schultz}, \bibfnamefont
  {M.~G.}}, \ and\ \bibinfo {author} {\bibfnamefont {F.}~\bibnamefont {von
  Oppen}}} (\bibinfo {year} {2009}),\ \href {\doibase
  10.1103/PhysRevB.80.033302} {\bibfield  {journal} {\bibinfo  {journal} {Phys.
  Rev. B}\ }\textbf {\bibinfo {volume} {80}},\ \bibinfo {pages}
  {033302}}\BibitemShut {NoStop}%
\bibitem [{\citenamefont {Schulz}\ \emph {et~al.}(2018)\citenamefont {Schulz},
  \citenamefont {Taylor}, \citenamefont {Hooley},\ and\ \citenamefont
  {Scardicchio}}]{SchulzScardicchio2018}%
  \BibitemOpen
  \bibfield  {author} {\bibinfo {author} {\bibnamefont {Schulz}, \bibfnamefont
  {M.}}, \bibinfo {author} {\bibfnamefont {S.~R.}\ \bibnamefont {Taylor}},
  \bibinfo {author} {\bibfnamefont {C.~A.}\ \bibnamefont {Hooley}}, \ and\
  \bibinfo {author} {\bibfnamefont {A.}~\bibnamefont {Scardicchio}}} (\bibinfo
  {year} {2018}),\ \href {\doibase 10.1103/PhysRevB.98.180201} {\bibfield
  {journal} {\bibinfo  {journal} {Phys. Rev. B}\ }\textbf {\bibinfo {volume}
  {98}},\ \bibinfo {pages} {180201}}\BibitemShut {NoStop}%
\bibitem [{\citenamefont {Schulz}\ \emph {et~al.}(2020)\citenamefont {Schulz},
  \citenamefont {Taylor}, \citenamefont {Scardicchio},\ and\ \citenamefont
  {{\v{Z}}nidari{\v{c}}}}]{SchulzZnidaric2020}%
  \BibitemOpen
  \bibfield  {author} {\bibinfo {author} {\bibnamefont {Schulz}, \bibfnamefont
  {M.}}, \bibinfo {author} {\bibfnamefont {S.~R.}\ \bibnamefont {Taylor}},
  \bibinfo {author} {\bibfnamefont {A.}~\bibnamefont {Scardicchio}}, \ and\
  \bibinfo {author} {\bibfnamefont {M.}~\bibnamefont {{\v{Z}}nidari{\v{c}}}}}
  (\bibinfo {year} {2020}),\ \href {\doibase 10.1088/1742-5468/ab6de0}
  {\bibfield  {journal} {\bibinfo  {journal} {Journal of Statistical Mechanics:
  Theory and Experiment}\ }\textbf {\bibinfo {volume} {2020}}~(\bibinfo
  {number} {2}),\ \bibinfo {pages} {023107}}\BibitemShut {NoStop}%
\bibitem [{\citenamefont {Schuster}\ \emph {et~al.}(1997)\citenamefont
  {Schuster}, \citenamefont {Buks}, \citenamefont {Heiblum}, \citenamefont
  {Mahalu}, \citenamefont {Umansky},\ and\ \citenamefont
  {Shtrikman}}]{SchusterShtrikman1997}%
  \BibitemOpen
  \bibfield  {author} {\bibinfo {author} {\bibnamefont {Schuster},
  \bibfnamefont {B.}}, \bibinfo {author} {\bibfnamefont {E.}~\bibnamefont
  {Buks}}, \bibinfo {author} {\bibfnamefont {M.}~\bibnamefont {Heiblum}},
  \bibinfo {author} {\bibfnamefont {D.}~\bibnamefont {Mahalu}}, \bibinfo
  {author} {\bibfnamefont {V.}~\bibnamefont {Umansky}}, \ and\ \bibinfo
  {author} {\bibfnamefont {H.}~\bibnamefont {Shtrikman}}} (\bibinfo {year}
  {1997}),\ \href {\doibase 10.1038/385417a0} {\bibfield  {journal} {\bibinfo
  {journal} {Nature}\ }\textbf {\bibinfo {volume} {385}},\ \bibinfo {pages}
  {417}}\BibitemShut {NoStop}%
\bibitem [{\citenamefont {Schwarz}\ \emph {et~al.}(2018)\citenamefont
  {Schwarz}, \citenamefont {Weymann}, \citenamefont {von Delft},\ and\
  \citenamefont {Weichselbaum}}]{SchwarzWeischselbaum2018}%
  \BibitemOpen
  \bibfield  {author} {\bibinfo {author} {\bibnamefont {Schwarz}, \bibfnamefont
  {F.}}, \bibinfo {author} {\bibfnamefont {I.}~\bibnamefont {Weymann}},
  \bibinfo {author} {\bibfnamefont {J.}~\bibnamefont {von Delft}}, \ and\
  \bibinfo {author} {\bibfnamefont {A.}~\bibnamefont {Weichselbaum}}} (\bibinfo
  {year} {2018}),\ \href {\doibase 10.1103/PhysRevLett.121.137702} {\bibfield
  {journal} {\bibinfo  {journal} {Phys. Rev. Lett.}\ }\textbf {\bibinfo
  {volume} {121}},\ \bibinfo {pages} {137702}}\BibitemShut {NoStop}%
\bibitem [{\citenamefont {Sciolla}\ \emph {et~al.}(2015)\citenamefont
  {Sciolla}, \citenamefont {Poletti},\ and\ \citenamefont
  {Kollath}}]{SciollaKollath2015}%
  \BibitemOpen
  \bibfield  {author} {\bibinfo {author} {\bibnamefont {Sciolla}, \bibfnamefont
  {B.}}, \bibinfo {author} {\bibfnamefont {D.}~\bibnamefont {Poletti}}, \ and\
  \bibinfo {author} {\bibfnamefont {C.}~\bibnamefont {Kollath}}} (\bibinfo
  {year} {2015}),\ \href {\doibase 10.1103/PhysRevLett.114.170401} {\bibfield
  {journal} {\bibinfo  {journal} {Phys. Rev. Lett.}\ }\textbf {\bibinfo
  {volume} {114}},\ \bibinfo {pages} {170401}}\BibitemShut {NoStop}%
\bibitem [{\citenamefont {Seah}\ \emph {et~al.}(2018)\citenamefont {Seah},
  \citenamefont {Nimmrichter},\ and\ \citenamefont {Scarani}}]{Seah2018}%
  \BibitemOpen
  \bibfield  {author} {\bibinfo {author} {\bibnamefont {Seah}, \bibfnamefont
  {S.}}, \bibinfo {author} {\bibfnamefont {S.}~\bibnamefont {Nimmrichter}}, \
  and\ \bibinfo {author} {\bibfnamefont {V.}~\bibnamefont {Scarani}}} (\bibinfo
  {year} {2018}),\ \href {\doibase 10.1103/PhysRevE.98.012131} {\bibfield
  {journal} {\bibinfo  {journal} {Phys. Rev. E}\ }\textbf {\bibinfo {volume}
  {98}},\ \bibinfo {pages} {012131}}\BibitemShut {NoStop}%
\bibitem [{\citenamefont {Seaman}\ \emph {et~al.}(2007)\citenamefont {Seaman},
  \citenamefont {Kr{\"a}mer}, \citenamefont {Anderson},\ and\ \citenamefont
  {Holland}}]{SeamanHolland2007}%
  \BibitemOpen
  \bibfield  {author} {\bibinfo {author} {\bibnamefont {Seaman}, \bibfnamefont
  {B.~T.}}, \bibinfo {author} {\bibfnamefont {M.}~\bibnamefont {Kr{\"a}mer}},
  \bibinfo {author} {\bibfnamefont {D.~Z.}\ \bibnamefont {Anderson}}, \ and\
  \bibinfo {author} {\bibfnamefont {M.~J.}\ \bibnamefont {Holland}}} (\bibinfo
  {year} {2007}),\ \href {\doibase 10.1103/PhysRevA.75.023615} {\bibfield
  {journal} {\bibinfo  {journal} {Phys. Rev. A}\ }\textbf {\bibinfo {volume}
  {75}},\ \bibinfo {pages} {023615}}\BibitemShut {NoStop}%
\bibitem [{\citenamefont {Secular}\ \emph {et~al.}(2020)\citenamefont
  {Secular}, \citenamefont {Gourianov}, \citenamefont {Lubasch}, \citenamefont
  {Dolgov}, \citenamefont {Clark},\ and\ \citenamefont
  {Jaksch}}]{SecularJaksch2020}%
  \BibitemOpen
  \bibfield  {author} {\bibinfo {author} {\bibnamefont {Secular}, \bibfnamefont
  {P.}}, \bibinfo {author} {\bibfnamefont {N.}~\bibnamefont {Gourianov}},
  \bibinfo {author} {\bibfnamefont {M.}~\bibnamefont {Lubasch}}, \bibinfo
  {author} {\bibfnamefont {S.}~\bibnamefont {Dolgov}}, \bibinfo {author}
  {\bibfnamefont {S.~R.}\ \bibnamefont {Clark}}, \ and\ \bibinfo {author}
  {\bibfnamefont {D.}~\bibnamefont {Jaksch}}} (\bibinfo {year} {2020}),\ \href
  {\doibase 10.1103/PhysRevB.101.235123} {\bibfield  {journal} {\bibinfo
  {journal} {Phys. Rev. B}\ }\textbf {\bibinfo {volume} {101}},\ \bibinfo
  {pages} {235123}}\BibitemShut {NoStop}%
\bibitem [{\citenamefont {Segal}(2008)}]{Segal2008}%
  \BibitemOpen
  \bibfield  {author} {\bibinfo {author} {\bibnamefont {Segal}, \bibfnamefont
  {D.}}} (\bibinfo {year} {2008}),\ \href {\doibase
  10.1103/PhysRevLett.100.105901} {\bibfield  {journal} {\bibinfo  {journal}
  {Phys. Rev. Lett.}\ }\textbf {\bibinfo {volume} {100}},\ \bibinfo {pages}
  {105901}}\BibitemShut {NoStop}%
\bibitem [{\citenamefont {Segal}\ and\ \citenamefont
  {Agarwalla}(2016)}]{SegalAgarwalla2016}%
  \BibitemOpen
  \bibfield  {author} {\bibinfo {author} {\bibnamefont {Segal}, \bibfnamefont
  {D.}}, \ and\ \bibinfo {author} {\bibfnamefont {B.~K.}\ \bibnamefont
  {Agarwalla}}} (\bibinfo {year} {2016}),\ \href {\doibase
  10.1146/annurev-physchem-040215-112103} {\bibfield  {journal} {\bibinfo
  {journal} {Annual Review of Physical Chemistry}\ }\textbf {\bibinfo {volume}
  {67}}~(\bibinfo {number} {1}),\ \bibinfo {pages} {185}}\BibitemShut {NoStop}%
\bibitem [{\citenamefont {Segal}\ \emph {et~al.}(2010)\citenamefont {Segal},
  \citenamefont {Millis},\ and\ \citenamefont {Reichman}}]{SegalReichman2010}%
  \BibitemOpen
  \bibfield  {author} {\bibinfo {author} {\bibnamefont {Segal}, \bibfnamefont
  {D.}}, \bibinfo {author} {\bibfnamefont {A.~J.}\ \bibnamefont {Millis}}, \
  and\ \bibinfo {author} {\bibfnamefont {D.~R.}\ \bibnamefont {Reichman}}}
  (\bibinfo {year} {2010}),\ \href {\doibase 10.1103/PhysRevB.82.205323}
  {\bibfield  {journal} {\bibinfo  {journal} {Phys. Rev. B}\ }\textbf {\bibinfo
  {volume} {82}},\ \bibinfo {pages} {205323}}\BibitemShut {NoStop}%
\bibitem [{\citenamefont {Serafini}(2017)}]{SerafiniBook}%
  \BibitemOpen
  \bibfield  {author} {\bibinfo {author} {\bibnamefont {Serafini},
  \bibfnamefont {A.}}} (\bibinfo {year} {2017}),\ \href {\doibase
  10.1201/9781315118727} {\emph {\bibinfo {title} {Quantum continuous variable:
  A primer of theoretical methods}}}\ (\bibinfo  {publisher} {CRC Press},\
  \bibinfo {address} {Boca Raton, Fl.})\BibitemShut {NoStop}%
\bibitem [{\citenamefont {Setiawan}\ \emph {et~al.}(2017)\citenamefont
  {Setiawan}, \citenamefont {Deng},\ and\ \citenamefont
  {Pixley}}]{SetiawanPixley2017}%
  \BibitemOpen
  \bibfield  {author} {\bibinfo {author} {\bibnamefont {Setiawan},
  \bibfnamefont {F.}}, \bibinfo {author} {\bibfnamefont {D.-L.}\ \bibnamefont
  {Deng}}, \ and\ \bibinfo {author} {\bibfnamefont {J.~H.}\ \bibnamefont
  {Pixley}}} (\bibinfo {year} {2017}),\ \href {\doibase
  10.1103/PhysRevB.96.104205} {\bibfield  {journal} {\bibinfo  {journal} {Phys.
  Rev. B}\ }\textbf {\bibinfo {volume} {96}},\ \bibinfo {pages}
  {104205}}\BibitemShut {NoStop}%
\bibitem [{\citenamefont {Shahbazyan}\ and\ \citenamefont
  {Raikh}(1994)}]{ShahbazyanRaikh1994}%
  \BibitemOpen
  \bibfield  {author} {\bibinfo {author} {\bibnamefont {Shahbazyan},
  \bibfnamefont {T.~V.}}, \ and\ \bibinfo {author} {\bibfnamefont {M.~E.}\
  \bibnamefont {Raikh}}} (\bibinfo {year} {1994}),\ \href {\doibase
  10.1103/PhysRevB.49.17123} {\bibfield  {journal} {\bibinfo  {journal} {Phys.
  Rev. B}\ }\textbf {\bibinfo {volume} {49}},\ \bibinfo {pages}
  {17123}}\BibitemShut {NoStop}%
\bibitem [{\citenamefont {Shastry}\ and\ \citenamefont
  {Sutherland}(1990)}]{ShastrySutherland1990}%
  \BibitemOpen
  \bibfield  {author} {\bibinfo {author} {\bibnamefont {Shastry}, \bibfnamefont
  {B.~S.}}, \ and\ \bibinfo {author} {\bibfnamefont {B.}~\bibnamefont
  {Sutherland}}} (\bibinfo {year} {1990}),\ \href {\doibase
  10.1103/PhysRevLett.65.243} {\bibfield  {journal} {\bibinfo  {journal} {Phys.
  Rev. Lett.}\ }\textbf {\bibinfo {volume} {65}},\ \bibinfo {pages}
  {243}}\BibitemShut {NoStop}%
\bibitem [{\citenamefont {Shekhter}\ \emph {et~al.}(2003)\citenamefont
  {Shekhter}, \citenamefont {Galperin}, \citenamefont {Gorelik}, \citenamefont
  {Isacsson},\ and\ \citenamefont {Jonson}}]{shekhter2003a}%
  \BibitemOpen
  \bibfield  {author} {\bibinfo {author} {\bibnamefont {Shekhter},
  \bibfnamefont {R.~I.}}, \bibinfo {author} {\bibfnamefont {Y.}~\bibnamefont
  {Galperin}}, \bibinfo {author} {\bibfnamefont {L.~Y.}\ \bibnamefont
  {Gorelik}}, \bibinfo {author} {\bibfnamefont {A.}~\bibnamefont {Isacsson}}, \
  and\ \bibinfo {author} {\bibfnamefont {M.}~\bibnamefont {Jonson}}} (\bibinfo
  {year} {2003}),\ \href {\doibase 10.1088/0953-8984/15/12/201} {\bibfield
  {journal} {\bibinfo  {journal} {Journal of Physics: Condensed Matter}\
  }\textbf {\bibinfo {volume} {15}}~(\bibinfo {number} {12}),\ \bibinfo {pages}
  {R441}}\BibitemShut {NoStop}%
\bibitem [{\citenamefont {Sierra}\ and\ \citenamefont
  {S\'anchez}(2014)}]{SierraSanchez2014}%
  \BibitemOpen
  \bibfield  {author} {\bibinfo {author} {\bibnamefont {Sierra}, \bibfnamefont
  {M.~A.}}, \ and\ \bibinfo {author} {\bibfnamefont {D.}~\bibnamefont
  {S\'anchez}}} (\bibinfo {year} {2014}),\ \href {\doibase
  10.1103/PhysRevB.90.115313} {\bibfield  {journal} {\bibinfo  {journal} {Phys.
  Rev. B}\ }\textbf {\bibinfo {volume} {90}},\ \bibinfo {pages}
  {115313}}\BibitemShut {NoStop}%
\bibitem [{\citenamefont {Sigrist}\ \emph {et~al.}(2004)\citenamefont
  {Sigrist}, \citenamefont {Fuhrer}, \citenamefont {Ihn}, \citenamefont
  {Ensslin}, \citenamefont {Ulloa}, \citenamefont {Wegscheider},\ and\
  \citenamefont {Bichler}}]{SigristBichler2004}%
  \BibitemOpen
  \bibfield  {author} {\bibinfo {author} {\bibnamefont {Sigrist}, \bibfnamefont
  {M.}}, \bibinfo {author} {\bibfnamefont {A.}~\bibnamefont {Fuhrer}}, \bibinfo
  {author} {\bibfnamefont {T.}~\bibnamefont {Ihn}}, \bibinfo {author}
  {\bibfnamefont {K.}~\bibnamefont {Ensslin}}, \bibinfo {author} {\bibfnamefont
  {S.~E.}\ \bibnamefont {Ulloa}}, \bibinfo {author} {\bibfnamefont
  {W.}~\bibnamefont {Wegscheider}}, \ and\ \bibinfo {author} {\bibfnamefont
  {M.}~\bibnamefont {Bichler}}} (\bibinfo {year} {2004}),\ \href {\doibase
  10.1103/PhysRevLett.93.066802} {\bibfield  {journal} {\bibinfo  {journal}
  {Phys. Rev. Lett.}\ }\textbf {\bibinfo {volume} {93}},\ \bibinfo {pages}
  {066802}}\BibitemShut {NoStop}%
\bibitem [{\citenamefont {Silva}\ \emph {et~al.}(2002)\citenamefont {Silva},
  \citenamefont {Oreg},\ and\ \citenamefont {Gefen}}]{SilvaGefen2002}%
  \BibitemOpen
  \bibfield  {author} {\bibinfo {author} {\bibnamefont {Silva}, \bibfnamefont
  {A.}}, \bibinfo {author} {\bibfnamefont {Y.}~\bibnamefont {Oreg}}, \ and\
  \bibinfo {author} {\bibfnamefont {Y.}~\bibnamefont {Gefen}}} (\bibinfo {year}
  {2002}),\ \href {\doibase 10.1103/PhysRevB.66.195316} {\bibfield  {journal}
  {\bibinfo  {journal} {Phys. Rev. B}\ }\textbf {\bibinfo {volume} {66}},\
  \bibinfo {pages} {195316}}\BibitemShut {NoStop}%
\bibitem [{\citenamefont {Silva}\ \emph {et~al.}(2020)\citenamefont {Silva},
  \citenamefont {Landi}, \citenamefont {Drumond},\ and\ \citenamefont
  {Pereira}}]{SauloPereira2020}%
  \BibitemOpen
  \bibfield  {author} {\bibinfo {author} {\bibnamefont {Silva}, \bibfnamefont
  {S.~H.~S.}}, \bibinfo {author} {\bibfnamefont {G.~T.}\ \bibnamefont {Landi}},
  \bibinfo {author} {\bibfnamefont {R.~C.}\ \bibnamefont {Drumond}}, \ and\
  \bibinfo {author} {\bibfnamefont {E.}~\bibnamefont {Pereira}}} (\bibinfo
  {year} {2020}),\ \href {\doibase 10.1103/PhysRevE.102.062146} {\bibfield
  {journal} {\bibinfo  {journal} {Phys. Rev. E}\ }\textbf {\bibinfo {volume}
  {102}},\ \bibinfo {pages} {062146}}\BibitemShut {NoStop}%
\bibitem [{\citenamefont {Silvi}\ \emph {et~al.}(2019)\citenamefont {Silvi},
  \citenamefont {Tschirsich}, \citenamefont {Gerster}, \citenamefont
  {Jünemann}, \citenamefont {Jaschke}, \citenamefont {Rizzi},\ and\
  \citenamefont {Montangero}}]{SilviMontangero2019}%
  \BibitemOpen
  \bibfield  {author} {\bibinfo {author} {\bibnamefont {Silvi}, \bibfnamefont
  {P.}}, \bibinfo {author} {\bibfnamefont {F.}~\bibnamefont {Tschirsich}},
  \bibinfo {author} {\bibfnamefont {M.}~\bibnamefont {Gerster}}, \bibinfo
  {author} {\bibfnamefont {J.}~\bibnamefont {Jünemann}}, \bibinfo {author}
  {\bibfnamefont {D.}~\bibnamefont {Jaschke}}, \bibinfo {author} {\bibfnamefont
  {M.}~\bibnamefont {Rizzi}}, \ and\ \bibinfo {author} {\bibfnamefont
  {S.}~\bibnamefont {Montangero}}} (\bibinfo {year} {2019}),\ \href {\doibase
  10.21468/SciPostPhysLectNotes.8} {\bibfield  {journal} {\bibinfo  {journal}
  {SciPost Phys. Lect. Notes}\ }\textbf {\bibinfo {volume} {8}},\ \bibinfo
  {pages} {1}}\BibitemShut {NoStop}%
\bibitem [{\citenamefont {Simine}\ and\ \citenamefont
  {Segal}(2012)}]{simine2012a}%
  \BibitemOpen
  \bibfield  {author} {\bibinfo {author} {\bibnamefont {Simine}, \bibfnamefont
  {L.}}, \ and\ \bibinfo {author} {\bibfnamefont {D.}~\bibnamefont {Segal}}}
  (\bibinfo {year} {2012}),\ \href {\doibase 10.1039/C2CP40851A} {\bibfield
  {journal} {\bibinfo  {journal} {Physical Chemistry Chemical Physics}\
  }\textbf {\bibinfo {volume} {14}},\ \bibinfo {pages} {13820}}\BibitemShut
  {NoStop}%
\bibitem [{\citenamefont {Spohn}(1978)}]{spohn1978b}%
  \BibitemOpen
  \bibfield  {author} {\bibinfo {author} {\bibnamefont {Spohn}, \bibfnamefont
  {H.}}} (\bibinfo {year} {1978}),\ \href {\doibase 10.1063/1.523789}
  {\bibfield  {journal} {\bibinfo  {journal} {Journal of Mathematical Phyics}\
  }\textbf {\bibinfo {volume} {19}},\ \bibinfo {pages} {1227}}\BibitemShut
  {NoStop}%
\bibitem [{\citenamefont {Srednicki}(1993)}]{Srednicki1993}%
  \BibitemOpen
  \bibfield  {author} {\bibinfo {author} {\bibnamefont {Srednicki},
  \bibfnamefont {M.}}} (\bibinfo {year} {1993}),\ \href {\doibase
  10.1103/PhysRevLett.71.666} {\bibfield  {journal} {\bibinfo  {journal} {Phys.
  Rev. Lett.}\ }\textbf {\bibinfo {volume} {71}},\ \bibinfo {pages}
  {666}}\BibitemShut {NoStop}%
\bibitem [{\citenamefont {Srednicki}(1994)}]{Srednicki1994}%
  \BibitemOpen
  \bibfield  {author} {\bibinfo {author} {\bibnamefont {Srednicki},
  \bibfnamefont {M.}}} (\bibinfo {year} {1994}),\ \href {\doibase
  10.1103/PhysRevE.50.888} {\bibfield  {journal} {\bibinfo  {journal} {Phys.
  Rev. E}\ }\textbf {\bibinfo {volume} {50}},\ \bibinfo {pages}
  {888}}\BibitemShut {NoStop}%
\bibitem [{\citenamefont {\'{S}roda}\ \emph {et~al.}(2019)\citenamefont
  {\'{S}roda}, \citenamefont {Prelov\v{s}ek},\ and\ \citenamefont
  {Mierzejewski}}]{SrodaMierzejewski2019}%
  \BibitemOpen
  \bibfield  {author} {\bibinfo {author} {\bibnamefont {\'{S}roda},
  \bibfnamefont {M.}}, \bibinfo {author} {\bibfnamefont {P.}~\bibnamefont
  {Prelov\v{s}ek}}, \ and\ \bibinfo {author} {\bibfnamefont {M.}~\bibnamefont
  {Mierzejewski}}} (\bibinfo {year} {2019}),\ \href {\doibase
  10.1103/PhysRevB.99.121110} {\bibfield  {journal} {\bibinfo  {journal} {Phys.
  Rev. B}\ }\textbf {\bibinfo {volume} {99}},\ \bibinfo {pages}
  {121110}}\BibitemShut {NoStop}%
\bibitem [{\citenamefont {Stadler}\ \emph {et~al.}(2012)\citenamefont
  {Stadler}, \citenamefont {Krinner}, \citenamefont {Meineke}, \citenamefont
  {Brantut},\ and\ \citenamefont {Esslinger}}]{StadlerEsslinger2012}%
  \BibitemOpen
  \bibfield  {author} {\bibinfo {author} {\bibnamefont {Stadler}, \bibfnamefont
  {D.}}, \bibinfo {author} {\bibfnamefont {S.}~\bibnamefont {Krinner}},
  \bibinfo {author} {\bibfnamefont {J.}~\bibnamefont {Meineke}}, \bibinfo
  {author} {\bibfnamefont {J.-P.}\ \bibnamefont {Brantut}}, \ and\ \bibinfo
  {author} {\bibfnamefont {T.}~\bibnamefont {Esslinger}}} (\bibinfo {year}
  {2012}),\ \href {\doibase 10.1038/nature11613} {\bibfield  {journal}
  {\bibinfo  {journal} {Nature}\ }\textbf {\bibinfo {volume} {491}},\ \bibinfo
  {pages} {736}}\BibitemShut {NoStop}%
\bibitem [{\citenamefont {Starr}(1936)}]{Starr1936}%
  \BibitemOpen
  \bibfield  {author} {\bibinfo {author} {\bibnamefont {Starr}, \bibfnamefont
  {C.}}} (\bibinfo {year} {1936}),\ \href {\doibase 10.1063/1.1745338}
  {\bibfield  {journal} {\bibinfo  {journal} {Physics}\ }\textbf {\bibinfo
  {volume} {7}},\ \bibinfo {pages} {15}}\BibitemShut {NoStop}%
\bibitem [{\citenamefont {Stegmann}\ \emph {et~al.}(2020)\citenamefont
  {Stegmann}, \citenamefont {K{\"o}nig},\ and\ \citenamefont
  {Sothmann}}]{Stegmann2020}%
  \BibitemOpen
  \bibfield  {author} {\bibinfo {author} {\bibnamefont {Stegmann},
  \bibfnamefont {P.}}, \bibinfo {author} {\bibfnamefont {J.}~\bibnamefont
  {K{\"o}nig}}, \ and\ \bibinfo {author} {\bibfnamefont {B.}~\bibnamefont
  {Sothmann}}} (\bibinfo {year} {2020}),\ \href {\doibase
  10.1103/PhysRevB.101.075411} {\bibfield  {journal} {\bibinfo  {journal}
  {Phys. Rev. B}\ }\textbf {\bibinfo {volume} {101}},\ \bibinfo {pages}
  {075411}}\BibitemShut {NoStop}%
\bibitem [{\citenamefont {Stegmann}\ \emph {et~al.}(2015)\citenamefont
  {Stegmann}, \citenamefont {Sothmann}, \citenamefont {Hucht},\ and\
  \citenamefont {K{\"o}nig}}]{Stegmann2015}%
  \BibitemOpen
  \bibfield  {author} {\bibinfo {author} {\bibnamefont {Stegmann},
  \bibfnamefont {P.}}, \bibinfo {author} {\bibfnamefont {B.}~\bibnamefont
  {Sothmann}}, \bibinfo {author} {\bibfnamefont {A.}~\bibnamefont {Hucht}}, \
  and\ \bibinfo {author} {\bibfnamefont {J.}~\bibnamefont {K{\"o}nig}}}
  (\bibinfo {year} {2015}),\ \href {\doibase 10.1103/PhysRevB.92.155413}
  {\bibfield  {journal} {\bibinfo  {journal} {Phys. Rev. B}\ }\textbf {\bibinfo
  {volume} {92}},\ \bibinfo {pages} {155413}}\BibitemShut {NoStop}%
\bibitem [{\citenamefont {Stinespring}(1955)}]{Stinespring1955}%
  \BibitemOpen
  \bibfield  {author} {\bibinfo {author} {\bibnamefont {Stinespring},
  \bibfnamefont {W.~F.}}} (\bibinfo {year} {1955}),\ \href {\doibase
  10.2307/2032342} {\bibfield  {journal} {\bibinfo  {journal} {Proceedings of
  the American Mathematical Society}\ }\textbf {\bibinfo {volume} {6}},\
  \bibinfo {pages} {211}}\BibitemShut {NoStop}%
\bibitem [{\citenamefont {Stopa}(2002)}]{Stopa2002}%
  \BibitemOpen
  \bibfield  {author} {\bibinfo {author} {\bibnamefont {Stopa}, \bibfnamefont
  {M.}}} (\bibinfo {year} {2002}),\ \href {\doibase
  10.1103/PhysRevLett.88.146802} {\bibfield  {journal} {\bibinfo  {journal}
  {Phys. Rev. Lett.}\ }\textbf {\bibinfo {volume} {88}},\ \bibinfo {pages}
  {146802}}\BibitemShut {NoStop}%
\bibitem [{\citenamefont {Stoudenmire}\ and\ \citenamefont
  {White}(2013)}]{StoudenmireWhite2013}%
  \BibitemOpen
  \bibfield  {author} {\bibinfo {author} {\bibnamefont {Stoudenmire},
  \bibfnamefont {E.~M.}}, \ and\ \bibinfo {author} {\bibfnamefont {S.~R.}\
  \bibnamefont {White}}} (\bibinfo {year} {2013}),\ \href {\doibase
  10.1103/PhysRevB.87.155137} {\bibfield  {journal} {\bibinfo  {journal} {Phys.
  Rev. B}\ }\textbf {\bibinfo {volume} {87}},\ \bibinfo {pages}
  {155137}}\BibitemShut {NoStop}%
\bibitem [{\citenamefont {Strasberg}(2019)}]{strasberg2019b}%
  \BibitemOpen
  \bibfield  {author} {\bibinfo {author} {\bibnamefont {Strasberg},
  \bibfnamefont {P.}}} (\bibinfo {year} {2019}),\ \href {\doibase
  10.1103/PhysRevLett.123.180604} {\bibfield  {journal} {\bibinfo  {journal}
  {Phys. Rev. Lett.}\ }\textbf {\bibinfo {volume} {123}},\ \bibinfo {pages}
  {180604}}\BibitemShut {NoStop}%
\bibitem [{\citenamefont {Strasberg}(2022)}]{strasberg2022}%
  \BibitemOpen
  \bibfield  {author} {\bibinfo {author} {\bibnamefont {Strasberg},
  \bibfnamefont {P.}}} (\bibinfo {year} {2022}),\ \href {\doibase
  10.1093/oso/9789192895585.001.0001} {\emph {\bibinfo {title} {Quantum
  Stochastic Thermodynamics}}}\ (\bibinfo  {publisher} {Oxford University
  Press},\ \bibinfo {address} {Oxford})\BibitemShut {NoStop}%
\bibitem [{\citenamefont {Strasberg}\ \emph {et~al.}(2017)\citenamefont
  {Strasberg}, \citenamefont {Schaller}, \citenamefont {Brandes},\ and\
  \citenamefont {Esposito}}]{strasberg2017a}%
  \BibitemOpen
  \bibfield  {author} {\bibinfo {author} {\bibnamefont {Strasberg},
  \bibfnamefont {P.}}, \bibinfo {author} {\bibfnamefont {G.}~\bibnamefont
  {Schaller}}, \bibinfo {author} {\bibfnamefont {T.}~\bibnamefont {Brandes}}, \
  and\ \bibinfo {author} {\bibfnamefont {M.}~\bibnamefont {Esposito}}}
  (\bibinfo {year} {2017}),\ \href {\doibase 10.1103/PhysRevX.7.021003}
  {\bibfield  {journal} {\bibinfo  {journal} {Physical Review X}\ }\textbf
  {\bibinfo {volume} {7}},\ \bibinfo {pages} {021003}}\BibitemShut {NoStop}%
\bibitem [{\citenamefont {Strasberg}\ \emph {et~al.}(2016)\citenamefont
  {Strasberg}, \citenamefont {Schaller}, \citenamefont {Lambert},\ and\
  \citenamefont {Brandes}}]{StrasbergBrandes2016}%
  \BibitemOpen
  \bibfield  {author} {\bibinfo {author} {\bibnamefont {Strasberg},
  \bibfnamefont {P.}}, \bibinfo {author} {\bibfnamefont {G.}~\bibnamefont
  {Schaller}}, \bibinfo {author} {\bibfnamefont {N.}~\bibnamefont {Lambert}}, \
  and\ \bibinfo {author} {\bibfnamefont {T.}~\bibnamefont {Brandes}}} (\bibinfo
  {year} {2016}),\ \href {\doibase 10.1088/1367-2630/18/7/073007} {\bibfield
  {journal} {\bibinfo  {journal} {New Journal of Physics}\ }\textbf {\bibinfo
  {volume} {18}}~(\bibinfo {number} {7}),\ \bibinfo {pages}
  {073007}}\BibitemShut {NoStop}%
\bibitem [{\citenamefont {Strasberg}\ \emph {et~al.}(2018)\citenamefont
  {Strasberg}, \citenamefont {Schaller}, \citenamefont {Schmidt},\ and\
  \citenamefont {Esposito}}]{StrasbergEsposito2018}%
  \BibitemOpen
  \bibfield  {author} {\bibinfo {author} {\bibnamefont {Strasberg},
  \bibfnamefont {P.}}, \bibinfo {author} {\bibfnamefont {G.}~\bibnamefont
  {Schaller}}, \bibinfo {author} {\bibfnamefont {T.~L.}\ \bibnamefont
  {Schmidt}}, \ and\ \bibinfo {author} {\bibfnamefont {M.}~\bibnamefont
  {Esposito}}} (\bibinfo {year} {2018}),\ \href {\doibase
  10.1103/PhysRevB.97.205405} {\bibfield  {journal} {\bibinfo  {journal} {Phys.
  Rev. B}\ }\textbf {\bibinfo {volume} {97}},\ \bibinfo {pages}
  {205405}}\BibitemShut {NoStop}%
\bibitem [{\citenamefont {Strasberg}\ \emph {et~al.}(2021)\citenamefont
  {Strasberg}, \citenamefont {W{\"a}chtler},\ and\ \citenamefont
  {Schaller}}]{strasberg2021a}%
  \BibitemOpen
  \bibfield  {author} {\bibinfo {author} {\bibnamefont {Strasberg},
  \bibfnamefont {P.}}, \bibinfo {author} {\bibfnamefont {C.~W.}\ \bibnamefont
  {W{\"a}chtler}}, \ and\ \bibinfo {author} {\bibfnamefont {G.}~\bibnamefont
  {Schaller}}} (\bibinfo {year} {2021}),\ \href {\doibase
  10.1103/PhysRevLett.126.180605} {\bibfield  {journal} {\bibinfo  {journal}
  {Phys. Rev. Lett.}\ }\textbf {\bibinfo {volume} {126}},\ \bibinfo {pages}
  {180605}}\BibitemShut {NoStop}%
\bibitem [{\citenamefont {Sutherland}\ and\ \citenamefont
  {Kohmoto}(1987)}]{SutherlandKohmoto1987}%
  \BibitemOpen
  \bibfield  {author} {\bibinfo {author} {\bibnamefont {Sutherland},
  \bibfnamefont {B.}}, \ and\ \bibinfo {author} {\bibfnamefont
  {M.}~\bibnamefont {Kohmoto}}} (\bibinfo {year} {1987}),\ \href {\doibase
  10.1103/PhysRevB.36.5877} {\bibfield  {journal} {\bibinfo  {journal} {Phys.
  Rev. B}\ }\textbf {\bibinfo {volume} {36}},\ \bibinfo {pages}
  {5877}}\BibitemShut {NoStop}%
\bibitem [{\citenamefont {Sutradhar}\ \emph {et~al.}(2019)\citenamefont
  {Sutradhar}, \citenamefont {Mukerjee}, \citenamefont {Pandit},\ and\
  \citenamefont {Banerjee}}]{SutradharBanerjee2019}%
  \BibitemOpen
  \bibfield  {author} {\bibinfo {author} {\bibnamefont {Sutradhar},
  \bibfnamefont {J.}}, \bibinfo {author} {\bibfnamefont {S.}~\bibnamefont
  {Mukerjee}}, \bibinfo {author} {\bibfnamefont {R.}~\bibnamefont {Pandit}}, \
  and\ \bibinfo {author} {\bibfnamefont {S.}~\bibnamefont {Banerjee}}}
  (\bibinfo {year} {2019}),\ \href {\doibase 10.1103/PhysRevB.99.224204}
  {\bibfield  {journal} {\bibinfo  {journal} {Phys. Rev. B}\ }\textbf {\bibinfo
  {volume} {99}},\ \bibinfo {pages} {224204}}\BibitemShut {NoStop}%
\bibitem [{\citenamefont {Suzuki}(1971)}]{Suzuki1971}%
  \BibitemOpen
  \bibfield  {author} {\bibinfo {author} {\bibnamefont {Suzuki}, \bibfnamefont
  {M.}}} (\bibinfo {year} {1971}),\ \href {\doibase
  10.1016/0031-8914(71)90226-6} {\bibfield  {journal} {\bibinfo  {journal}
  {Physica}\ }\textbf {\bibinfo {volume} {51}},\ \bibinfo {pages}
  {277}}\BibitemShut {NoStop}%
\bibitem [{\citenamefont {Svensson}\ \emph {et~al.}(2013)\citenamefont
  {Svensson}, \citenamefont {Hoffmann}, \citenamefont {Nakpathomkun},
  \citenamefont {Wu}, \citenamefont {Xu}, \citenamefont {Nilsson},
  \citenamefont {S{\'{a}}nchez}, \citenamefont {Kashcheyevs},\ and\
  \citenamefont {Linke}}]{SvenssonLinke2013}%
  \BibitemOpen
  \bibfield  {author} {\bibinfo {author} {\bibnamefont {Svensson},
  \bibfnamefont {S.~F.}}, \bibinfo {author} {\bibfnamefont {E.~A.}\
  \bibnamefont {Hoffmann}}, \bibinfo {author} {\bibfnamefont {N.}~\bibnamefont
  {Nakpathomkun}}, \bibinfo {author} {\bibfnamefont {P.~M.}\ \bibnamefont
  {Wu}}, \bibinfo {author} {\bibfnamefont {H.~Q.}\ \bibnamefont {Xu}}, \bibinfo
  {author} {\bibfnamefont {H.~A.}\ \bibnamefont {Nilsson}}, \bibinfo {author}
  {\bibfnamefont {D.}~\bibnamefont {S{\'{a}}nchez}}, \bibinfo {author}
  {\bibfnamefont {V.}~\bibnamefont {Kashcheyevs}}, \ and\ \bibinfo {author}
  {\bibfnamefont {H.}~\bibnamefont {Linke}}} (\bibinfo {year} {2013}),\ \href
  {\doibase 10.1088/1367-2630/15/10/105011} {\bibfield  {journal} {\bibinfo
  {journal} {New Journal of Physics}\ }\textbf {\bibinfo {volume}
  {15}}~(\bibinfo {number} {10}),\ \bibinfo {pages} {105011}}\BibitemShut
  {NoStop}%
\bibitem [{\citenamefont {Switkes}\ \emph {et~al.}(1999)\citenamefont
  {Switkes}, \citenamefont {Marcus}, \citenamefont {Campman},\ and\
  \citenamefont {Gossard}}]{SwitkesGossard1999}%
  \BibitemOpen
  \bibfield  {author} {\bibinfo {author} {\bibnamefont {Switkes}, \bibfnamefont
  {M.}}, \bibinfo {author} {\bibfnamefont {C.~M.}\ \bibnamefont {Marcus}},
  \bibinfo {author} {\bibfnamefont {K.}~\bibnamefont {Campman}}, \ and\
  \bibinfo {author} {\bibfnamefont {A.~C.}\ \bibnamefont {Gossard}}} (\bibinfo
  {year} {1999}),\ \href {\doibase 10.1126/science.283.5409.1905} {\bibfield
  {journal} {\bibinfo  {journal} {Science}\ }\textbf {\bibinfo {volume}
  {283}}~(\bibinfo {number} {5409}),\ \bibinfo {pages} {1905}}\BibitemShut
  {NoStop}%
\bibitem [{\citenamefont {Szeg{\H{o}}}(1939)}]{Szego1939}%
  \BibitemOpen
  \bibfield  {author} {\bibinfo {author} {\bibnamefont {Szeg{\H{o}}},
  \bibfnamefont {G.}}} (\bibinfo {year} {1939}),\ \href {\doibase
  10.1090/coll/023} {\emph {\bibinfo {title} {Orthogonal Polynomials}}}\
  (\bibinfo  {publisher} {American Mathematical Society})\BibitemShut {NoStop}%
\bibitem [{\citenamefont {Takahashi}\ and\ \citenamefont
  {Umezawa}(1996)}]{TakahashiUmezawa1996}%
  \BibitemOpen
  \bibfield  {author} {\bibinfo {author} {\bibnamefont {Takahashi},
  \bibfnamefont {Y.}}, \ and\ \bibinfo {author} {\bibfnamefont
  {H.}~\bibnamefont {Umezawa}}} (\bibinfo {year} {1996}),\ \href {\doibase
  10.1142/S0217979296000817} {\bibfield  {journal} {\bibinfo  {journal} {Int.
  Jour. Mod. Phys. B}\ }\textbf {\bibinfo {volume} {10}},\ \bibinfo {pages}
  {1755}}\BibitemShut {NoStop}%
\bibitem [{\citenamefont {Talarico}\ \emph {et~al.}(2020)\citenamefont
  {Talarico}, \citenamefont {Maniscalco},\ and\ \citenamefont
  {Gullo}}]{talarico2020a}%
  \BibitemOpen
  \bibfield  {author} {\bibinfo {author} {\bibnamefont {Talarico},
  \bibfnamefont {N.~W.}}, \bibinfo {author} {\bibfnamefont {S.}~\bibnamefont
  {Maniscalco}}, \ and\ \bibinfo {author} {\bibfnamefont {N.~L.}\ \bibnamefont
  {Gullo}}} (\bibinfo {year} {2020}),\ \href {\doibase
  10.1103/PhysRevB.101.045103} {\bibfield  {journal} {\bibinfo  {journal}
  {Phys. Rev. B}\ }\textbf {\bibinfo {volume} {101}},\ \bibinfo {pages}
  {045103}}\BibitemShut {NoStop}%
\bibitem [{\citenamefont {Talkner}\ and\ \citenamefont
  {H{\"a}nggi}(2020)}]{TalknerHanggi2020}%
  \BibitemOpen
  \bibfield  {author} {\bibinfo {author} {\bibnamefont {Talkner}, \bibfnamefont
  {P.}}, \ and\ \bibinfo {author} {\bibfnamefont {P.}~\bibnamefont
  {H{\"a}nggi}}} (\bibinfo {year} {2020}),\ \href {\doibase
  10.1103/RevModPhys.92.041002} {\bibfield  {journal} {\bibinfo  {journal}
  {Rev. Mod. Phys.}\ }\textbf {\bibinfo {volume} {92}},\ \bibinfo {pages}
  {041002}}\BibitemShut {NoStop}%
\bibitem [{\citenamefont {Tamascelli}\ \emph {et~al.}(2018)\citenamefont
  {Tamascelli}, \citenamefont {Smirne}, \citenamefont {Huelga},\ and\
  \citenamefont {Plenio}}]{TamascelliPlenio2018}%
  \BibitemOpen
  \bibfield  {author} {\bibinfo {author} {\bibnamefont {Tamascelli},
  \bibfnamefont {D.}}, \bibinfo {author} {\bibfnamefont {A.}~\bibnamefont
  {Smirne}}, \bibinfo {author} {\bibfnamefont {S.~F.}\ \bibnamefont {Huelga}},
  \ and\ \bibinfo {author} {\bibfnamefont {M.~B.}\ \bibnamefont {Plenio}}}
  (\bibinfo {year} {2018}),\ \href {\doibase 10.1103/PhysRevLett.120.030402}
  {\bibfield  {journal} {\bibinfo  {journal} {Phys. Rev. Lett.}\ }\textbf
  {\bibinfo {volume} {120}},\ \bibinfo {pages} {030402}}\BibitemShut {NoStop}%
\bibitem [{\citenamefont {Tamascelli}\ \emph {et~al.}(2019)\citenamefont
  {Tamascelli}, \citenamefont {Smirne}, \citenamefont {Lim}, \citenamefont
  {Huelga},\ and\ \citenamefont {Plenio}}]{TamascelliPlenio2019}%
  \BibitemOpen
  \bibfield  {author} {\bibinfo {author} {\bibnamefont {Tamascelli},
  \bibfnamefont {D.}}, \bibinfo {author} {\bibfnamefont {A.}~\bibnamefont
  {Smirne}}, \bibinfo {author} {\bibfnamefont {J.}~\bibnamefont {Lim}},
  \bibinfo {author} {\bibfnamefont {S.~F.}\ \bibnamefont {Huelga}}, \ and\
  \bibinfo {author} {\bibfnamefont {M.~B.}\ \bibnamefont {Plenio}}} (\bibinfo
  {year} {2019}),\ \href {\doibase 10.1103/PhysRevLett.123.090402} {\bibfield
  {journal} {\bibinfo  {journal} {Phys. Rev. Lett.}\ }\textbf {\bibinfo
  {volume} {123}},\ \bibinfo {pages} {090402}}\BibitemShut {NoStop}%
\bibitem [{\citenamefont {Tang}\ \emph {et~al.}(2018)\citenamefont {Tang},
  \citenamefont {Zhang},\ and\ \citenamefont {Wang}}]{TangWang2018}%
  \BibitemOpen
  \bibfield  {author} {\bibinfo {author} {\bibnamefont {Tang}, \bibfnamefont
  {G.}}, \bibinfo {author} {\bibfnamefont {L.}~\bibnamefont {Zhang}}, \ and\
  \bibinfo {author} {\bibfnamefont {J.}~\bibnamefont {Wang}}} (\bibinfo {year}
  {2018}),\ \href {\doibase 10.1103/PhysRevB.97.224311} {\bibfield  {journal}
  {\bibinfo  {journal} {Phys. Rev. B}\ }\textbf {\bibinfo {volume} {97}},\
  \bibinfo {pages} {224311}}\BibitemShut {NoStop}%
\bibitem [{\citenamefont {Tanimura}(1990)}]{tanimura1990a}%
  \BibitemOpen
  \bibfield  {author} {\bibinfo {author} {\bibnamefont {Tanimura},
  \bibfnamefont {Y.}}} (\bibinfo {year} {1990}),\ \href {\doibase
  10.1103/PhysRevA.41.6676} {\bibfield  {journal} {\bibinfo  {journal} {Phys.
  Rev. A}\ }\textbf {\bibinfo {volume} {41}},\ \bibinfo {pages}
  {6676}}\BibitemShut {NoStop}%
\bibitem [{\citenamefont {Tanimura}\ and\ \citenamefont
  {Kubo}(1989)}]{tanimura1989a}%
  \BibitemOpen
  \bibfield  {author} {\bibinfo {author} {\bibnamefont {Tanimura},
  \bibfnamefont {Y.}}, \ and\ \bibinfo {author} {\bibfnamefont
  {R.}~\bibnamefont {Kubo}}} (\bibinfo {year} {1989}),\ \href {\doibase
  10.1143/JPSJ.58.101} {\bibfield  {journal} {\bibinfo  {journal} {Journal of
  the Physical Society of Japan}\ }\textbf {\bibinfo {volume} {58}}~(\bibinfo
  {number} {1}),\ \bibinfo {pages} {101}}\BibitemShut {NoStop}%
\bibitem [{\citenamefont {Tao}(2009)}]{Tao2009}%
  \BibitemOpen
  \bibfield  {author} {\bibinfo {author} {\bibnamefont {Tao}, \bibfnamefont
  {N.~J.}}} (\bibinfo {year} {2009}),\ \href {\doibase
  10.1142/9789814287005_0019} {\emph {\bibinfo {title} {Electron transport in
  molecular junctions}}}\ (\bibinfo  {publisher} {World
  Scientific})\BibitemShut {NoStop}%
\bibitem [{\citenamefont {Taylor}\ and\ \citenamefont
  {Scardicchio}(2021)}]{TaylorScardicchio2020}%
  \BibitemOpen
  \bibfield  {author} {\bibinfo {author} {\bibnamefont {Taylor}, \bibfnamefont
  {S.~R.}}, \ and\ \bibinfo {author} {\bibfnamefont {A.}~\bibnamefont
  {Scardicchio}}} (\bibinfo {year} {2021}),\ \href {\doibase
  10.1103/PhysRevB.103.184202} {\bibfield  {journal} {\bibinfo  {journal}
  {Phys. Rev. B}\ }\textbf {\bibinfo {volume} {103}},\ \bibinfo {pages}
  {184202}}\BibitemShut {NoStop}%
\bibitem [{\citenamefont {Terraneo}\ \emph {et~al.}(2002)\citenamefont
  {Terraneo}, \citenamefont {Peyrard},\ and\ \citenamefont
  {Casati}}]{Terraneo2002}%
  \BibitemOpen
  \bibfield  {author} {\bibinfo {author} {\bibnamefont {Terraneo},
  \bibfnamefont {M.}}, \bibinfo {author} {\bibfnamefont {M.}~\bibnamefont
  {Peyrard}}, \ and\ \bibinfo {author} {\bibfnamefont {G.}~\bibnamefont
  {Casati}}} (\bibinfo {year} {2002}),\ \href {\doibase
  10.1103/PhysRevLett.88.094302} {\bibfield  {journal} {\bibinfo  {journal}
  {Physical Review Letters}\ }\textbf {\bibinfo {volume} {88}}~(\bibinfo
  {number} {9}),\ \bibinfo {pages} {094302}}\BibitemShut {NoStop}%
\bibitem [{\citenamefont {Thingna}\ and\ \citenamefont
  {Manzano}(2021)}]{Thingna2021}%
  \BibitemOpen
  \bibfield  {author} {\bibinfo {author} {\bibnamefont {Thingna}, \bibfnamefont
  {J.}}, \ and\ \bibinfo {author} {\bibfnamefont {D.}~\bibnamefont {Manzano}}}
  (\bibinfo {year} {2021}),\ \href {\doibase 10.1063/5.0045308} {\bibfield
  {journal} {\bibinfo  {journal} {Chaos: An Interdisciplinary Journal of
  Nonlinear Science}\ }\textbf {\bibinfo {volume} {31}}~(\bibinfo {number}
  {7}),\ \bibinfo {pages} {073114}}\BibitemShut {NoStop}%
\bibitem [{\citenamefont {Thingna}\ \emph {et~al.}(2016)\citenamefont
  {Thingna}, \citenamefont {Manzano},\ and\ \citenamefont
  {Cao}}]{ThingnaCao2016}%
  \BibitemOpen
  \bibfield  {author} {\bibinfo {author} {\bibnamefont {Thingna}, \bibfnamefont
  {J.}}, \bibinfo {author} {\bibfnamefont {D.}~\bibnamefont {Manzano}}, \ and\
  \bibinfo {author} {\bibfnamefont {J.}~\bibnamefont {Cao}}} (\bibinfo {year}
  {2016}),\ \href {\doibase 10.1038/srep28027} {\bibfield  {journal} {\bibinfo
  {journal} {Scientific Reports}\ }\textbf {\bibinfo {volume} {6}},\ \bibinfo
  {pages} {28027}}\BibitemShut {NoStop}%
\bibitem [{\citenamefont {Thingna}\ \emph {et~al.}(2020)\citenamefont
  {Thingna}, \citenamefont {Manzano},\ and\ \citenamefont
  {Cao}}]{ThingnaCao2020}%
  \BibitemOpen
  \bibfield  {author} {\bibinfo {author} {\bibnamefont {Thingna}, \bibfnamefont
  {J.}}, \bibinfo {author} {\bibfnamefont {D.}~\bibnamefont {Manzano}}, \ and\
  \bibinfo {author} {\bibfnamefont {J.}~\bibnamefont {Cao}}} (\bibinfo {year}
  {2020}),\ \href {\doibase 10.1088/1367-2630/aba0e4} {\bibfield  {journal}
  {\bibinfo  {journal} {New Journal of Physics}\ }\textbf {\bibinfo {volume}
  {22}}~(\bibinfo {number} {8}),\ \bibinfo {pages} {083026}}\BibitemShut
  {NoStop}%
\bibitem [{\citenamefont {Thingna}\ \emph {et~al.}(2012)\citenamefont
  {Thingna}, \citenamefont {Wang},\ and\ \citenamefont
  {H{\"a}nggi}}]{thingna2012a}%
  \BibitemOpen
  \bibfield  {author} {\bibinfo {author} {\bibnamefont {Thingna}, \bibfnamefont
  {J.}}, \bibinfo {author} {\bibfnamefont {J.-S.}\ \bibnamefont {Wang}}, \ and\
  \bibinfo {author} {\bibfnamefont {P.}~\bibnamefont {H{\"a}nggi}}} (\bibinfo
  {year} {2012}),\ \href {\doibase 10.1063/1.4718706} {\bibfield  {journal}
  {\bibinfo  {journal} {Journal of Chemical Physics}\ }\textbf {\bibinfo
  {volume} {136}},\ \bibinfo {pages} {194110}}\BibitemShut {NoStop}%
\bibitem [{\citenamefont {Thingna}\ \emph {et~al.}(2014)\citenamefont
  {Thingna}, \citenamefont {Zhou},\ and\ \citenamefont
  {Wang}}]{ThingnaWang2014}%
  \BibitemOpen
  \bibfield  {author} {\bibinfo {author} {\bibnamefont {Thingna}, \bibfnamefont
  {J.}}, \bibinfo {author} {\bibfnamefont {H.}~\bibnamefont {Zhou}}, \ and\
  \bibinfo {author} {\bibfnamefont {J.-S.}\ \bibnamefont {Wang}}} (\bibinfo
  {year} {2014}),\ \href {\doibase 10.1063/1.4901274} {\bibfield  {journal}
  {\bibinfo  {journal} {J. Chem. Phys}\ }\textbf {\bibinfo {volume} {141}},\
  \bibinfo {pages} {194101}}\BibitemShut {NoStop}%
\bibitem [{\citenamefont {Thouless}(1983)}]{Thouless1983}%
  \BibitemOpen
  \bibfield  {author} {\bibinfo {author} {\bibnamefont {Thouless},
  \bibfnamefont {D.~J.}}} (\bibinfo {year} {1983}),\ \href {\doibase
  10.1103/PhysRevB.27.6083} {\bibfield  {journal} {\bibinfo  {journal} {Phys.
  Rev. B}\ }\textbf {\bibinfo {volume} {27}},\ \bibinfo {pages}
  {6083}}\BibitemShut {NoStop}%
\bibitem [{\citenamefont {Timm}(2009)}]{timm2009a}%
  \BibitemOpen
  \bibfield  {author} {\bibinfo {author} {\bibnamefont {Timm}, \bibfnamefont
  {C.}}} (\bibinfo {year} {2009}),\ \href {\doibase 10.1103/PhysRevE.80.021140}
  {\bibfield  {journal} {\bibinfo  {journal} {Phys. Rev. E}\ }\textbf {\bibinfo
  {volume} {80}},\ \bibinfo {pages} {021140}}\BibitemShut {NoStop}%
\bibitem [{\citenamefont {Timpanaro}\ \emph {et~al.}(2019)\citenamefont
  {Timpanaro}, \citenamefont {Guarnieri}, \citenamefont {Goold},\ and\
  \citenamefont {Landi}}]{timpanaro2019a}%
  \BibitemOpen
  \bibfield  {author} {\bibinfo {author} {\bibnamefont {Timpanaro},
  \bibfnamefont {A.~M.}}, \bibinfo {author} {\bibfnamefont {G.}~\bibnamefont
  {Guarnieri}}, \bibinfo {author} {\bibfnamefont {J.}~\bibnamefont {Goold}}, \
  and\ \bibinfo {author} {\bibfnamefont {G.~T.}\ \bibnamefont {Landi}}}
  (\bibinfo {year} {2019}),\ \href {\doibase 10.1103/PhysRevLett.123.090604}
  {\bibfield  {journal} {\bibinfo  {journal} {Phys. Rev. Lett.}\ }\textbf
  {\bibinfo {volume} {123}},\ \bibinfo {pages} {090604}}\BibitemShut {NoStop}%
\bibitem [{\citenamefont {Topp}\ \emph {et~al.}(2015)\citenamefont {Topp},
  \citenamefont {Brandes},\ and\ \citenamefont {Schaller}}]{topp2015a}%
  \BibitemOpen
  \bibfield  {author} {\bibinfo {author} {\bibnamefont {Topp}, \bibfnamefont
  {G.~E.}}, \bibinfo {author} {\bibfnamefont {T.}~\bibnamefont {Brandes}}, \
  and\ \bibinfo {author} {\bibfnamefont {G.}~\bibnamefont {Schaller}}}
  (\bibinfo {year} {2015}),\ \href {\doibase 10.1209/0295-5075/110/67003}
  {\bibfield  {journal} {\bibinfo  {journal} {Europhysics Letters}\ }\textbf
  {\bibinfo {volume} {110}},\ \bibinfo {pages} {67003}}\BibitemShut {NoStop}%
\bibitem [{\citenamefont {Torres}(2014)}]{Torres2014}%
  \BibitemOpen
  \bibfield  {author} {\bibinfo {author} {\bibnamefont {Torres}, \bibfnamefont
  {J.~M.}}} (\bibinfo {year} {2014}),\ \href {\doibase
  10.1103/PhysRevA.89.052133} {\bibfield  {journal} {\bibinfo  {journal} {Phys.
  Rev. A}\ }\textbf {\bibinfo {volume} {89}},\ \bibinfo {pages}
  {052133}}\BibitemShut {NoStop}%
\bibitem [{\citenamefont {Torrontegui}\ and\ \citenamefont
  {Kosloff}(2016)}]{TorronteguiKosloff2016}%
  \BibitemOpen
  \bibfield  {author} {\bibinfo {author} {\bibnamefont {Torrontegui},
  \bibfnamefont {E.}}, \ and\ \bibinfo {author} {\bibfnamefont
  {R.}~\bibnamefont {Kosloff}}} (\bibinfo {year} {2016}),\ \href {\doibase
  10.1088/1367-2630/18/9/093001} {\bibfield  {journal} {\bibinfo  {journal}
  {New Journal of Physics}\ }\textbf {\bibinfo {volume} {18}}~(\bibinfo
  {number} {9}),\ \bibinfo {pages} {093001}}\BibitemShut {NoStop}%
\bibitem [{\citenamefont {Touchette}(2009)}]{touchette2009a}%
  \BibitemOpen
  \bibfield  {author} {\bibinfo {author} {\bibnamefont {Touchette},
  \bibfnamefont {H.}}} (\bibinfo {year} {2009}),\ \href {\doibase
  https://doi.org/10.1016/j.physrep.2009.05.002} {\bibfield  {journal}
  {\bibinfo  {journal} {Physics Reports}\ }\textbf {\bibinfo {volume}
  {478}}~(\bibinfo {number} {1}),\ \bibinfo {pages} {1}}\BibitemShut {NoStop}%
\bibitem [{\citenamefont {Tritt}(2004)}]{Tritt2004}%
  \BibitemOpen
  \bibfield  {author} {\bibinfo {author} {\bibnamefont {Tritt}, \bibfnamefont
  {T.~M.}}} (\bibinfo {year} {2004}),\ \href {\doibase 10.1007/b136496} {\emph
  {\bibinfo {title} {{Thermal conductivity: theory, properties and
  applications}}}}\ (\bibinfo  {publisher} {Kluwer Academic})\BibitemShut
  {NoStop}%
\bibitem [{\citenamefont {Trushechkin}(2021)}]{trushechkin2021a}%
  \BibitemOpen
  \bibfield  {author} {\bibinfo {author} {\bibnamefont {Trushechkin},
  \bibfnamefont {A.}}} (\bibinfo {year} {2021}),\ \href {\doibase
  10.1103/PhysRevA.103.062226} {\bibfield  {journal} {\bibinfo  {journal}
  {Phys. Rev. A}\ }\textbf {\bibinfo {volume} {103}},\ \bibinfo {pages}
  {062226}}\BibitemShut {NoStop}%
\bibitem [{\citenamefont {Turkington}(2013)}]{Vectorization}%
  \BibitemOpen
  \bibfield  {author} {\bibinfo {author} {\bibnamefont {Turkington},
  \bibfnamefont {D.~A.}}} (\bibinfo {year} {2013}),\ \href {\doibase
  10.1017/CBO9781139424400} {\emph {\bibinfo {title} {{Generalized
  Vectorization, Cross-Products, and Matrix Calculus}}}}\ (\bibinfo
  {publisher} {Cambridge University Press},\ \bibinfo {address}
  {Cambridge})\BibitemShut {NoStop}%
\bibitem [{\citenamefont {Ueda}\ \emph {et~al.}(2003)\citenamefont {Ueda},
  \citenamefont {Baba}, \citenamefont {Suzuki},\ and\ \citenamefont
  {Eto}}]{UedaEto2003}%
  \BibitemOpen
  \bibfield  {author} {\bibinfo {author} {\bibnamefont {Ueda}, \bibfnamefont
  {A.}}, \bibinfo {author} {\bibfnamefont {I.}~\bibnamefont {Baba}}, \bibinfo
  {author} {\bibfnamefont {K.}~\bibnamefont {Suzuki}}, \ and\ \bibinfo {author}
  {\bibfnamefont {M.}~\bibnamefont {Eto}}} (\bibinfo {year} {2003}),\ \href
  {\doibase 10.1143/JPSJS.72SA.157} {\bibfield  {journal} {\bibinfo  {journal}
  {J. Phys. Soc. Jpn}\ }\textbf {\bibinfo {volume} {72}},\ \bibinfo {pages}
  {Suppl. A 157}}\BibitemShut {NoStop}%
\bibitem [{\citenamefont {Urban}\ and\ \citenamefont
  {K{\"o}nig}(2009)}]{urban2009a}%
  \BibitemOpen
  \bibfield  {author} {\bibinfo {author} {\bibnamefont {Urban}, \bibfnamefont
  {D.}}, \ and\ \bibinfo {author} {\bibfnamefont {J.}~\bibnamefont
  {K{\"o}nig}}} (\bibinfo {year} {2009}),\ \href {\doibase
  10.1103/PhysRevB.79.165319} {\bibfield  {journal} {\bibinfo  {journal} {Phys.
  Rev. B}\ }\textbf {\bibinfo {volume} {79}},\ \bibinfo {pages}
  {165319}}\BibitemShut {NoStop}%
\bibitem [{\citenamefont {Urban}\ \emph {et~al.}(2008)\citenamefont {Urban},
  \citenamefont {K{\"o}nig},\ and\ \citenamefont {Fazio}}]{UrbanFazio2008}%
  \BibitemOpen
  \bibfield  {author} {\bibinfo {author} {\bibnamefont {Urban}, \bibfnamefont
  {D.}}, \bibinfo {author} {\bibfnamefont {J.}~\bibnamefont {K{\"o}nig}}, \
  and\ \bibinfo {author} {\bibfnamefont {R.}~\bibnamefont {Fazio}}} (\bibinfo
  {year} {2008}),\ \href {\doibase 10.1103/PhysRevB.78.075318} {\bibfield
  {journal} {\bibinfo  {journal} {Phys. Rev. B}\ }\textbf {\bibinfo {volume}
  {78}},\ \bibinfo {pages} {075318}}\BibitemShut {NoStop}%
\bibitem [{\citenamefont {Utsumi}\ \emph {et~al.}(2010)\citenamefont {Utsumi},
  \citenamefont {Golubev}, \citenamefont {Marthaler}, \citenamefont {Saito},
  \citenamefont {Fujisawa},\ and\ \citenamefont {Sch{\"o}n}}]{utsumi2010a}%
  \BibitemOpen
  \bibfield  {author} {\bibinfo {author} {\bibnamefont {Utsumi}, \bibfnamefont
  {Y.}}, \bibinfo {author} {\bibfnamefont {D.~S.}\ \bibnamefont {Golubev}},
  \bibinfo {author} {\bibfnamefont {M.}~\bibnamefont {Marthaler}}, \bibinfo
  {author} {\bibfnamefont {K.}~\bibnamefont {Saito}}, \bibinfo {author}
  {\bibfnamefont {T.}~\bibnamefont {Fujisawa}}, \ and\ \bibinfo {author}
  {\bibfnamefont {G.}~\bibnamefont {Sch{\"o}n}}} (\bibinfo {year} {2010}),\
  \href {\doibase 10.1103/PhysRevB.81.125331} {\bibfield  {journal} {\bibinfo
  {journal} {Physical Review B}\ }\textbf {\bibinfo {volume} {81}}~(\bibinfo
  {number} {12}),\ \bibinfo {pages} {125331}}\BibitemShut {NoStop}%
\bibitem [{\citenamefont {{Van den Broeck}}\ and\ \citenamefont
  {Esposito}(2015)}]{VANDENBROECK2015}%
  \BibitemOpen
  \bibfield  {author} {\bibinfo {author} {\bibnamefont {{Van den Broeck}},
  \bibfnamefont {C.}}, \ and\ \bibinfo {author} {\bibfnamefont
  {M.}~\bibnamefont {Esposito}}} (\bibinfo {year} {2015}),\ \href {\doibase
  https://doi.org/10.1016/j.physa.2014.04.035} {\bibfield  {journal} {\bibinfo
  {journal} {Physica A: Statistical Mechanics and its Applications}\ }\textbf
  {\bibinfo {volume} {418}},\ \bibinfo {pages} {6}},\ \bibinfo {note}
  {proceedings of the 13th International Summer School on Fundamental Problems
  in Statistical Physics}\BibitemShut {NoStop}%
\bibitem [{\citenamefont {{van Oudenaarden}}\ and\ \citenamefont
  {Mooij}(1996)}]{vanOudenaardenMooij1996}%
  \BibitemOpen
  \bibfield  {author} {\bibinfo {author} {\bibnamefont {{van Oudenaarden}},
  \bibfnamefont {A.}}, \ and\ \bibinfo {author} {\bibfnamefont {J.~E.}\
  \bibnamefont {Mooij}}} (\bibinfo {year} {1996}),\ \href {\doibase
  10.1103/PhysRevLett.76.4947} {\bibfield  {journal} {\bibinfo  {journal}
  {Phys. Rev. Lett.}\ }\textbf {\bibinfo {volume} {76}}~(\bibinfo {number}
  {26}),\ \bibinfo {pages} {4947}}\BibitemShut {NoStop}%
\bibitem [{\citenamefont {Varma}\ \emph
  {et~al.}(2017{\natexlab{a}})\citenamefont {Varma}, \citenamefont {{De
  Mulatier}},\ and\ \citenamefont {{\v{Z}}nidari{\v{c}}}}]{Varma2017}%
  \BibitemOpen
  \bibfield  {author} {\bibinfo {author} {\bibnamefont {Varma}, \bibfnamefont
  {V.~K.}}, \bibinfo {author} {\bibfnamefont {C.}~\bibnamefont {{De
  Mulatier}}}, \ and\ \bibinfo {author} {\bibfnamefont {M.}~\bibnamefont
  {{\v{Z}}nidari{\v{c}}}}} (\bibinfo {year} {2017}{\natexlab{a}}),\ \href
  {\doibase 10.1103/PhysRevE.96.032130} {\bibfield  {journal} {\bibinfo
  {journal} {Physical Review E}\ }\textbf {\bibinfo {volume} {96}}~(\bibinfo
  {number} {3}),\ \bibinfo {pages} {1}}\BibitemShut {NoStop}%
\bibitem [{\citenamefont {Varma}\ \emph
  {et~al.}(2017{\natexlab{b}})\citenamefont {Varma}, \citenamefont {Lerose},
  \citenamefont {Pietracaprina}, \citenamefont {Goold},\ and\ \citenamefont
  {Scardicchio}}]{VarmaScardicchio2017}%
  \BibitemOpen
  \bibfield  {author} {\bibinfo {author} {\bibnamefont {Varma}, \bibfnamefont
  {V.~K.}}, \bibinfo {author} {\bibfnamefont {A.}~\bibnamefont {Lerose}},
  \bibinfo {author} {\bibfnamefont {F.}~\bibnamefont {Pietracaprina}}, \bibinfo
  {author} {\bibfnamefont {J.}~\bibnamefont {Goold}}, \ and\ \bibinfo {author}
  {\bibfnamefont {A.}~\bibnamefont {Scardicchio}}} (\bibinfo {year}
  {2017}{\natexlab{b}}),\ \href {\doibase 10.1088/1742-5468/aa668b} {\bibfield
  {journal} {\bibinfo  {journal} {Journal of Statistical Mechanics: Theory and
  Experiment}\ }\textbf {\bibinfo {volume} {2017}}~(\bibinfo {number} {5}),\
  \bibinfo {pages} {053101}}\BibitemShut {NoStop}%
\bibitem [{\citenamefont {Varma}\ \emph
  {et~al.}(2017{\natexlab{c}})\citenamefont {Varma}, \citenamefont
  {de~Mulatier},\ and\ \citenamefont {\v{Z}nidari\v{c}}}]{VarmaZnidaric2017}%
  \BibitemOpen
  \bibfield  {author} {\bibinfo {author} {\bibnamefont {Varma}, \bibfnamefont
  {V.~K.}}, \bibinfo {author} {\bibfnamefont {C.}~\bibnamefont {de~Mulatier}},
  \ and\ \bibinfo {author} {\bibfnamefont {M.}~\bibnamefont
  {\v{Z}nidari\v{c}}}} (\bibinfo {year} {2017}{\natexlab{c}}),\ \href {\doibase
  10.1103/PhysRevE.96.032130} {\bibfield  {journal} {\bibinfo  {journal} {Phys.
  Rev. E}\ }\textbf {\bibinfo {volume} {96}},\ \bibinfo {pages}
  {032130}}\BibitemShut {NoStop}%
\bibitem [{\citenamefont {Varma}\ and\ \citenamefont
  {\v{Z}nidari\v{c}}(2019)}]{VarmaZnidaric2019}%
  \BibitemOpen
  \bibfield  {author} {\bibinfo {author} {\bibnamefont {Varma}, \bibfnamefont
  {V.~K.}}, \ and\ \bibinfo {author} {\bibfnamefont {M.}~\bibnamefont
  {\v{Z}nidari\v{c}}}} (\bibinfo {year} {2019}),\ \href {\doibase
  10.1103/PhysRevB.100.085105} {\bibfield  {journal} {\bibinfo  {journal}
  {Phys. Rev. B}\ }\textbf {\bibinfo {volume} {100}},\ \bibinfo {pages}
  {085105}}\BibitemShut {NoStop}%
\bibitem [{\citenamefont {\ifmmode \check{Z}\else
  \v{Z}\fi{}nidari\ifmmode~\check{c}\else \v{c}\fi{}}(2021)}]{Znidaric2020b}%
  \BibitemOpen
  \bibfield  {author} {\bibinfo {author} {\bibnamefont {\ifmmode \check{Z}\else
  \v{Z}\fi{}nidari\ifmmode~\check{c}\else \v{c}\fi{}}, \bibfnamefont {M.}}}
  (\bibinfo {year} {2021}),\ \href {\doibase 10.1103/PhysRevB.103.237101}
  {\bibfield  {journal} {\bibinfo  {journal} {Phys. Rev. B}\ }\textbf {\bibinfo
  {volume} {103}},\ \bibinfo {pages} {237101}}\BibitemShut {NoStop}%
\bibitem [{\citenamefont {de~Vega}\ and\ \citenamefont
  {Alonso}(2017)}]{devega2017a}%
  \BibitemOpen
  \bibfield  {author} {\bibinfo {author} {\bibnamefont {de~Vega}, \bibfnamefont
  {I.}}, \ and\ \bibinfo {author} {\bibfnamefont {D.}~\bibnamefont {Alonso}}}
  (\bibinfo {year} {2017}),\ \href {\doibase 10.1103/RevModPhys.89.015001}
  {\bibfield  {journal} {\bibinfo  {journal} {Rev. Mod. Phys.}\ }\textbf
  {\bibinfo {volume} {89}},\ \bibinfo {pages} {015001}}\BibitemShut {NoStop}%
\bibitem [{\citenamefont {de~Vega}\ and\ \citenamefont
  {Ba\~nuls}(2015)}]{DeVegaBanuls2015}%
  \BibitemOpen
  \bibfield  {author} {\bibinfo {author} {\bibnamefont {de~Vega}, \bibfnamefont
  {I.}}, \ and\ \bibinfo {author} {\bibfnamefont {M.-C.}\ \bibnamefont
  {Ba\~nuls}}} (\bibinfo {year} {2015}),\ \href {\doibase
  10.1103/PhysRevA.92.052116} {\bibfield  {journal} {\bibinfo  {journal} {Phys.
  Rev. A}\ }\textbf {\bibinfo {volume} {92}},\ \bibinfo {pages}
  {052116}}\BibitemShut {NoStop}%
\bibitem [{\citenamefont {de~Vega}\ \emph {et~al.}(2015)\citenamefont
  {de~Vega}, \citenamefont {Schollw{\"o}ck},\ and\ \citenamefont
  {Wolf}}]{DeVegaWolf2015}%
  \BibitemOpen
  \bibfield  {author} {\bibinfo {author} {\bibnamefont {de~Vega}, \bibfnamefont
  {I.}}, \bibinfo {author} {\bibfnamefont {U.}~\bibnamefont {Schollw{\"o}ck}},
  \ and\ \bibinfo {author} {\bibfnamefont {F.~A.}\ \bibnamefont {Wolf}}}
  (\bibinfo {year} {2015}),\ \href {\doibase 10.1103/PhysRevB.92.155126}
  {\bibfield  {journal} {\bibinfo  {journal} {Phys. Rev. B}\ }\textbf {\bibinfo
  {volume} {92}},\ \bibinfo {pages} {155126}}\BibitemShut {NoStop}%
\bibitem [{\citenamefont {Verstraete}\ \emph {et~al.}(2008)\citenamefont
  {Verstraete}, \citenamefont {Murg},\ and\ \citenamefont
  {Cirac}}]{VerstraeteCirac2008}%
  \BibitemOpen
  \bibfield  {author} {\bibinfo {author} {\bibnamefont {Verstraete},
  \bibfnamefont {F.}}, \bibinfo {author} {\bibfnamefont {V.}~\bibnamefont
  {Murg}}, \ and\ \bibinfo {author} {\bibfnamefont {J.~I.}\ \bibnamefont
  {Cirac}}} (\bibinfo {year} {2008}),\ \href {\doibase
  10.1080/14789940801912366} {\bibfield  {journal} {\bibinfo  {journal} {Adv.
  Phys.}\ }\textbf {\bibinfo {volume} {57}},\ \bibinfo {pages}
  {143}}\BibitemShut {NoStop}%
\bibitem [{\citenamefont {Vicentini}\ \emph {et~al.}(2019)\citenamefont
  {Vicentini}, \citenamefont {Biella}, \citenamefont {Regnault},\ and\
  \citenamefont {Ciuti}}]{VicentiniCiuti2019}%
  \BibitemOpen
  \bibfield  {author} {\bibinfo {author} {\bibnamefont {Vicentini},
  \bibfnamefont {F.}}, \bibinfo {author} {\bibfnamefont {A.}~\bibnamefont
  {Biella}}, \bibinfo {author} {\bibfnamefont {N.}~\bibnamefont {Regnault}}, \
  and\ \bibinfo {author} {\bibfnamefont {C.}~\bibnamefont {Ciuti}}} (\bibinfo
  {year} {2019}),\ \href {\doibase 10.1103/PhysRevLett.122.250503} {\bibfield
  {journal} {\bibinfo  {journal} {Phys. Rev. Lett.}\ }\textbf {\bibinfo
  {volume} {122}},\ \bibinfo {pages} {250503}}\BibitemShut {NoStop}%
\bibitem [{\citenamefont {Viciani}\ \emph {et~al.}(2015)\citenamefont
  {Viciani}, \citenamefont {Lima}, \citenamefont {Bellini},\ and\ \citenamefont
  {Caruso}}]{Viciani2015}%
  \BibitemOpen
  \bibfield  {author} {\bibinfo {author} {\bibnamefont {Viciani}, \bibfnamefont
  {S.}}, \bibinfo {author} {\bibfnamefont {M.}~\bibnamefont {Lima}}, \bibinfo
  {author} {\bibfnamefont {M.}~\bibnamefont {Bellini}}, \ and\ \bibinfo
  {author} {\bibfnamefont {F.}~\bibnamefont {Caruso}}} (\bibinfo {year}
  {2015}),\ \href {\doibase 10.1103/PhysRevLett.115.083601} {\bibfield
  {journal} {\bibinfo  {journal} {Phys. Rev. Lett.}\ }\textbf {\bibinfo
  {volume} {115}},\ \bibinfo {pages} {083601}}\BibitemShut {NoStop}%
\bibitem [{\citenamefont {Vidal}\ \emph {et~al.}(2003)\citenamefont {Vidal},
  \citenamefont {Latorre}, \citenamefont {Rico},\ and\ \citenamefont
  {Kitaev}}]{VidalKitaev2003}%
  \BibitemOpen
  \bibfield  {author} {\bibinfo {author} {\bibnamefont {Vidal}, \bibfnamefont
  {G.}}, \bibinfo {author} {\bibfnamefont {J.~I.}\ \bibnamefont {Latorre}},
  \bibinfo {author} {\bibfnamefont {E.}~\bibnamefont {Rico}}, \ and\ \bibinfo
  {author} {\bibfnamefont {A.}~\bibnamefont {Kitaev}}} (\bibinfo {year}
  {2003}),\ \href {\doibase 10.1103/PhysRevLett.90.227902} {\bibfield
  {journal} {\bibinfo  {journal} {Phys. Rev. Lett.}\ }\textbf {\bibinfo
  {volume} {90}},\ \bibinfo {pages} {227902}}\BibitemShut {NoStop}%
\bibitem [{\citenamefont {Vidal}\ \emph {et~al.}(1999)\citenamefont {Vidal},
  \citenamefont {Mouhanna},\ and\ \citenamefont
  {Giamarchi}}]{VidalGiamarchi1999}%
  \BibitemOpen
  \bibfield  {author} {\bibinfo {author} {\bibnamefont {Vidal}, \bibfnamefont
  {J.}}, \bibinfo {author} {\bibfnamefont {D.}~\bibnamefont {Mouhanna}}, \ and\
  \bibinfo {author} {\bibfnamefont {T.}~\bibnamefont {Giamarchi}}} (\bibinfo
  {year} {1999}),\ \href {\doibase 10.1103/PhysRevLett.83.3908} {\bibfield
  {journal} {\bibinfo  {journal} {Phys. Rev. Lett.}\ }\textbf {\bibinfo
  {volume} {83}},\ \bibinfo {pages} {3908}}\BibitemShut {NoStop}%
\bibitem [{\citenamefont {Vidal}\ \emph {et~al.}(2001)\citenamefont {Vidal},
  \citenamefont {Mouhanna},\ and\ \citenamefont
  {Giamarchi}}]{VidalGiamarchi2001}%
  \BibitemOpen
  \bibfield  {author} {\bibinfo {author} {\bibnamefont {Vidal}, \bibfnamefont
  {J.}}, \bibinfo {author} {\bibfnamefont {D.}~\bibnamefont {Mouhanna}}, \ and\
  \bibinfo {author} {\bibfnamefont {T.}~\bibnamefont {Giamarchi}}} (\bibinfo
  {year} {2001}),\ \href {\doibase 10.1103/PhysRevB.65.014201} {\bibfield
  {journal} {\bibinfo  {journal} {Phys. Rev. B}\ }\textbf {\bibinfo {volume}
  {65}},\ \bibinfo {pages} {014201}}\BibitemShut {NoStop}%
\bibitem [{\citenamefont {Vinjanampathy}\ and\ \citenamefont
  {Anders}(2016)}]{VinjanampathyAnders2016}%
  \BibitemOpen
  \bibfield  {author} {\bibinfo {author} {\bibnamefont {Vinjanampathy},
  \bibfnamefont {S.}}, \ and\ \bibinfo {author} {\bibfnamefont
  {J.}~\bibnamefont {Anders}}} (\bibinfo {year} {2016}),\ \href {\doibase
  10.1080/00107514.2016.1201896} {\bibfield  {journal} {\bibinfo  {journal}
  {Contemporary Physics}\ }\textbf {\bibinfo {volume} {57}}~(\bibinfo {number}
  {4}),\ \bibinfo {pages} {545}}\BibitemShut {NoStop}%
\bibitem [{\citenamefont {Vogl}\ \emph {et~al.}(2011)\citenamefont {Vogl},
  \citenamefont {Schaller},\ and\ \citenamefont {Brandes}}]{VoglBrandes2011}%
  \BibitemOpen
  \bibfield  {author} {\bibinfo {author} {\bibnamefont {Vogl}, \bibfnamefont
  {M.}}, \bibinfo {author} {\bibfnamefont {G.}~\bibnamefont {Schaller}}, \ and\
  \bibinfo {author} {\bibfnamefont {T.}~\bibnamefont {Brandes}}} (\bibinfo
  {year} {2011}),\ \href {\doibase 10.1016/j.aop.2011.07.008} {\bibfield
  {journal} {\bibinfo  {journal} {Annals of Physics}\ }\textbf {\bibinfo
  {volume} {326}},\ \bibinfo {pages} {2827}}\BibitemShut {NoStop}%
\bibitem [{\citenamefont {Vorberg}\ \emph {et~al.}(2013)\citenamefont
  {Vorberg}, \citenamefont {Wustmann}, \citenamefont {Ketzmerick},\ and\
  \citenamefont {Eckardt}}]{VorbergEckardt2013}%
  \BibitemOpen
  \bibfield  {author} {\bibinfo {author} {\bibnamefont {Vorberg}, \bibfnamefont
  {D.}}, \bibinfo {author} {\bibfnamefont {W.}~\bibnamefont {Wustmann}},
  \bibinfo {author} {\bibfnamefont {R.}~\bibnamefont {Ketzmerick}}, \ and\
  \bibinfo {author} {\bibfnamefont {A.}~\bibnamefont {Eckardt}}} (\bibinfo
  {year} {2013}),\ \href {\doibase 10.1103/PhysRevLett.111.240405} {\bibfield
  {journal} {\bibinfo  {journal} {Phys. Rev. Lett.}\ }\textbf {\bibinfo
  {volume} {111}},\ \bibinfo {pages} {240405}}\BibitemShut {NoStop}%
\bibitem [{\citenamefont {Vosk}\ \emph {et~al.}(2015)\citenamefont {Vosk},
  \citenamefont {Huse},\ and\ \citenamefont {Altman}}]{VoskAltman2015}%
  \BibitemOpen
  \bibfield  {author} {\bibinfo {author} {\bibnamefont {Vosk}, \bibfnamefont
  {R.}}, \bibinfo {author} {\bibfnamefont {D.~A.}\ \bibnamefont {Huse}}, \ and\
  \bibinfo {author} {\bibfnamefont {E.}~\bibnamefont {Altman}}} (\bibinfo
  {year} {2015}),\ \href {\doibase 10.1103/PhysRevX.5.031032} {\bibfield
  {journal} {\bibinfo  {journal} {Phys. Rev. X}\ }\textbf {\bibinfo {volume}
  {5}},\ \bibinfo {pages} {031032}}\BibitemShut {NoStop}%
\bibitem [{\citenamefont {W{\"a}chtler}\ and\ \citenamefont
  {Schaller}(2020)}]{waechtler2020a}%
  \BibitemOpen
  \bibfield  {author} {\bibinfo {author} {\bibnamefont {W{\"a}chtler},
  \bibfnamefont {C.~W.}}, \ and\ \bibinfo {author} {\bibfnamefont
  {G.}~\bibnamefont {Schaller}}} (\bibinfo {year} {2020}),\ \href {\doibase
  10.1103/PhysRevResearch.2.023178} {\bibfield  {journal} {\bibinfo  {journal}
  {Phys. Rev. Research}\ }\textbf {\bibinfo {volume} {2}},\ \bibinfo {pages}
  {023178}}\BibitemShut {NoStop}%
\bibitem [{\citenamefont {W{\"a}chtler}\ \emph {et~al.}(2019)\citenamefont
  {W{\"a}chtler}, \citenamefont {Strasberg}, \citenamefont {Klapp},
  \citenamefont {Schaller},\ and\ \citenamefont {Jarzynski}}]{waechtler2019a}%
  \BibitemOpen
  \bibfield  {author} {\bibinfo {author} {\bibnamefont {W{\"a}chtler},
  \bibfnamefont {C.~W.}}, \bibinfo {author} {\bibfnamefont {P.}~\bibnamefont
  {Strasberg}}, \bibinfo {author} {\bibfnamefont {S.~H.~L.}\ \bibnamefont
  {Klapp}}, \bibinfo {author} {\bibfnamefont {G.}~\bibnamefont {Schaller}}, \
  and\ \bibinfo {author} {\bibfnamefont {C.}~\bibnamefont {Jarzynski}}}
  (\bibinfo {year} {2019}),\ \href {\doibase 10.1088/1367-2630/ab2727}
  {\bibfield  {journal} {\bibinfo  {journal} {New Journal of Physics}\ }\textbf
  {\bibinfo {volume} {21}},\ \bibinfo {pages} {073009}}\BibitemShut {NoStop}%
\bibitem [{\citenamefont {Wagner}\ \emph {et~al.}(2017)\citenamefont {Wagner},
  \citenamefont {Strasberg}, \citenamefont {Bayer}, \citenamefont
  {Rugeramigabo}, \citenamefont {Brandes},\ and\ \citenamefont
  {Haug}}]{wagner2017a}%
  \BibitemOpen
  \bibfield  {author} {\bibinfo {author} {\bibnamefont {Wagner}, \bibfnamefont
  {T.}}, \bibinfo {author} {\bibfnamefont {P.}~\bibnamefont {Strasberg}},
  \bibinfo {author} {\bibfnamefont {J.~C.}\ \bibnamefont {Bayer}}, \bibinfo
  {author} {\bibfnamefont {E.~P.}\ \bibnamefont {Rugeramigabo}}, \bibinfo
  {author} {\bibfnamefont {T.}~\bibnamefont {Brandes}}, \ and\ \bibinfo
  {author} {\bibfnamefont {R.~J.}\ \bibnamefont {Haug}}} (\bibinfo {year}
  {2017}),\ \href {\doibase 10.1038/nnano.2016.225} {\bibfield  {journal}
  {\bibinfo  {journal} {Nature Nanotechnology}\ }\textbf {\bibinfo {volume}
  {12}},\ \bibinfo {pages} {218}}\BibitemShut {NoStop}%
\bibitem [{\citenamefont {Wahl}\ \emph {et~al.}(2017)\citenamefont {Wahl},
  \citenamefont {Pal},\ and\ \citenamefont {Simon}}]{WahlSimon2017}%
  \BibitemOpen
  \bibfield  {author} {\bibinfo {author} {\bibnamefont {Wahl}, \bibfnamefont
  {T.~B.}}, \bibinfo {author} {\bibfnamefont {A.}~\bibnamefont {Pal}}, \ and\
  \bibinfo {author} {\bibfnamefont {S.~H.}\ \bibnamefont {Simon}}} (\bibinfo
  {year} {2017}),\ \href {\doibase 10.1103/PhysRevX.7.021018} {\bibfield
  {journal} {\bibinfo  {journal} {Phys. Rev. X}\ }\textbf {\bibinfo {volume}
  {7}},\ \bibinfo {pages} {021018}}\BibitemShut {NoStop}%
\bibitem [{\citenamefont {Wang}\ \emph {et~al.}(2015)\citenamefont {Wang},
  \citenamefont {Ren},\ and\ \citenamefont {Cao}}]{wang2015a}%
  \BibitemOpen
  \bibfield  {author} {\bibinfo {author} {\bibnamefont {Wang}, \bibfnamefont
  {C.}}, \bibinfo {author} {\bibfnamefont {J.}~\bibnamefont {Ren}}, \ and\
  \bibinfo {author} {\bibfnamefont {J.}~\bibnamefont {Cao}}} (\bibinfo {year}
  {2015}),\ \href {\doibase 10.1038/srep11787} {\bibfield  {journal} {\bibinfo
  {journal} {Scientific Reports}\ }\textbf {\bibinfo {volume} {5}},\ \bibinfo
  {pages} {11787}}\BibitemShut {NoStop}%
\bibitem [{\citenamefont {Wang}\ and\ \citenamefont {Sun}(2015)}]{wang2015b}%
  \BibitemOpen
  \bibfield  {author} {\bibinfo {author} {\bibnamefont {Wang}, \bibfnamefont
  {C.}}, \ and\ \bibinfo {author} {\bibfnamefont {K.-W.}\ \bibnamefont {Sun}}}
  (\bibinfo {year} {2015}),\ \href {\doibase 10.1016/j.aop.2015.09.005}
  {\bibfield  {journal} {\bibinfo  {journal} {Annals of Physics}\ }\textbf
  {\bibinfo {volume} {362}},\ \bibinfo {pages} {703}}\BibitemShut {NoStop}%
\bibitem [{\citenamefont {Wang}\ \emph {et~al.}(2019)\citenamefont {Wang},
  \citenamefont {Yang}, \citenamefont {Chen}, \citenamefont {Li},\ and\
  \citenamefont {Zhang}}]{WangLifa2019}%
  \BibitemOpen
  \bibfield  {author} {\bibinfo {author} {\bibnamefont {Wang}, \bibfnamefont
  {H.}}, \bibinfo {author} {\bibfnamefont {Y.}~\bibnamefont {Yang}}, \bibinfo
  {author} {\bibfnamefont {H.}~\bibnamefont {Chen}}, \bibinfo {author}
  {\bibfnamefont {N.}~\bibnamefont {Li}}, \ and\ \bibinfo {author}
  {\bibfnamefont {L.}~\bibnamefont {Zhang}}} (\bibinfo {year} {2019}),\ \href
  {\doibase 10.1103/PhysRevE.99.062111} {\bibfield  {journal} {\bibinfo
  {journal} {Phys. Rev. E}\ }\textbf {\bibinfo {volume} {99}},\ \bibinfo
  {pages} {062111}}\BibitemShut {NoStop}%
\bibitem [{\citenamefont {Wang}\ \emph {et~al.}(2014)\citenamefont {Wang},
  \citenamefont {Agarwalla},\ and\ \citenamefont {Thingna}}]{wang2014a}%
  \BibitemOpen
  \bibfield  {author} {\bibinfo {author} {\bibnamefont {Wang}, \bibfnamefont
  {J.-S.}}, \bibinfo {author} {\bibfnamefont {B.~K.}\ \bibnamefont
  {Agarwalla}}, \ and\ \bibinfo {author} {\bibfnamefont {H.~L.}\ \bibnamefont
  {Thingna}}} (\bibinfo {year} {2014}),\ \href {\doibase
  10.1007/s11467-013-0340-x} {\bibfield  {journal} {\bibinfo  {journal}
  {Frontiers of Physics}\ }\textbf {\bibinfo {volume} {9}},\ \bibinfo {pages}
  {673}}\BibitemShut {NoStop}%
\bibitem [{\citenamefont {Wang}\ \emph {et~al.}(2020)\citenamefont {Wang},
  \citenamefont {Piazza},\ and\ \citenamefont {Luitz}}]{wang2020a}%
  \BibitemOpen
  \bibfield  {author} {\bibinfo {author} {\bibnamefont {Wang}, \bibfnamefont
  {K.}}, \bibinfo {author} {\bibfnamefont {F.}~\bibnamefont {Piazza}}, \ and\
  \bibinfo {author} {\bibfnamefont {D.~J.}\ \bibnamefont {Luitz}}} (\bibinfo
  {year} {2020}),\ \href {\doibase 10.1103/PhysRevLett.124.100604} {\bibfield
  {journal} {\bibinfo  {journal} {Phys. Rev. Lett.}\ }\textbf {\bibinfo
  {volume} {124}},\ \bibinfo {pages} {100604}}\BibitemShut {NoStop}%
\bibitem [{\citenamefont {Weidenm{\"u}ller}(2003)}]{Weidenmuller2003}%
  \BibitemOpen
  \bibfield  {author} {\bibinfo {author} {\bibnamefont {Weidenm{\"u}ller},
  \bibfnamefont {H.~A.}}} (\bibinfo {year} {2003}),\ \href {\doibase
  10.1103/PhysRevB.68.125326} {\bibfield  {journal} {\bibinfo  {journal} {Phys.
  Rev. B}\ }\textbf {\bibinfo {volume} {68}},\ \bibinfo {pages}
  {125326}}\BibitemShut {NoStop}%
\bibitem [{\citenamefont {Weimer}(2015)}]{Weimer2015}%
  \BibitemOpen
  \bibfield  {author} {\bibinfo {author} {\bibnamefont {Weimer}, \bibfnamefont
  {H.}}} (\bibinfo {year} {2015}),\ \href {\doibase
  10.1103/PhysRevLett.114.040402} {\bibfield  {journal} {\bibinfo  {journal}
  {Phys. Rev. Lett.}\ }\textbf {\bibinfo {volume} {114}},\ \bibinfo {pages}
  {040402}}\BibitemShut {NoStop}%
\bibitem [{\citenamefont {Weimer}\ \emph {et~al.}(2021)\citenamefont {Weimer},
  \citenamefont {Kshetrimayum},\ and\ \citenamefont {Or\'us}}]{WeimerOrus2019}%
  \BibitemOpen
  \bibfield  {author} {\bibinfo {author} {\bibnamefont {Weimer}, \bibfnamefont
  {H.}}, \bibinfo {author} {\bibfnamefont {A.}~\bibnamefont {Kshetrimayum}}, \
  and\ \bibinfo {author} {\bibfnamefont {R.}~\bibnamefont {Or\'us}}} (\bibinfo
  {year} {2021}),\ \href {\doibase 10.1103/RevModPhys.93.015008} {\bibfield
  {journal} {\bibinfo  {journal} {Rev. Mod. Phys.}\ }\textbf {\bibinfo {volume}
  {93}},\ \bibinfo {pages} {015008}}\BibitemShut {NoStop}%
\bibitem [{\citenamefont {Weiner}\ \emph {et~al.}(2019)\citenamefont {Weiner},
  \citenamefont {Evers},\ and\ \citenamefont {Bera}}]{WeinerBera2019}%
  \BibitemOpen
  \bibfield  {author} {\bibinfo {author} {\bibnamefont {Weiner}, \bibfnamefont
  {F.}}, \bibinfo {author} {\bibfnamefont {F.}~\bibnamefont {Evers}}, \ and\
  \bibinfo {author} {\bibfnamefont {S.}~\bibnamefont {Bera}}} (\bibinfo {year}
  {2019}),\ \href {\doibase 10.1103/PhysRevB.100.104204} {\bibfield  {journal}
  {\bibinfo  {journal} {Phys. Rev. B}\ }\textbf {\bibinfo {volume} {100}},\
  \bibinfo {pages} {104204}}\BibitemShut {NoStop}%
\bibitem [{\citenamefont {Weis}\ \emph {et~al.}(1993)\citenamefont {Weis},
  \citenamefont {Haug}, \citenamefont {Klitzing},\ and\ \citenamefont
  {Ploog}}]{WeisPloog1993}%
  \BibitemOpen
  \bibfield  {author} {\bibinfo {author} {\bibnamefont {Weis}, \bibfnamefont
  {J.}}, \bibinfo {author} {\bibfnamefont {R.~J.}\ \bibnamefont {Haug}},
  \bibinfo {author} {\bibfnamefont {K.~v.}\ \bibnamefont {Klitzing}}, \ and\
  \bibinfo {author} {\bibfnamefont {K.}~\bibnamefont {Ploog}}} (\bibinfo {year}
  {1993}),\ \href {\doibase 10.1103/PhysRevLett.71.4019} {\bibfield  {journal}
  {\bibinfo  {journal} {Phys. Rev. Lett.}\ }\textbf {\bibinfo {volume} {71}},\
  \bibinfo {pages} {4019}}\BibitemShut {NoStop}%
\bibitem [{\citenamefont {Weiss}(1993)}]{weiss1993}%
  \BibitemOpen
  \bibfield  {author} {\bibinfo {author} {\bibnamefont {Weiss}, \bibfnamefont
  {U.}}} (\bibinfo {year} {1993}),\ \href {\doibase 10.1142/8334} {\emph
  {\bibinfo {title} {Quantum Dissipative Systems}}},\ \bibinfo {series} {Series
  of Modern Condensed Matter Physics}, Vol.~\bibinfo {volume} {2}\ (\bibinfo
  {publisher} {World Scientific},\ \bibinfo {address} {Singapore})\BibitemShut
  {NoStop}%
\bibitem [{\citenamefont {Werlang}\ \emph {et~al.}(2014)\citenamefont
  {Werlang}, \citenamefont {Marchiori}, \citenamefont {Cornelio},\ and\
  \citenamefont {Valente}}]{WerlangValente2014}%
  \BibitemOpen
  \bibfield  {author} {\bibinfo {author} {\bibnamefont {Werlang}, \bibfnamefont
  {T.}}, \bibinfo {author} {\bibfnamefont {M.~A.}\ \bibnamefont {Marchiori}},
  \bibinfo {author} {\bibfnamefont {M.~F.}\ \bibnamefont {Cornelio}}, \ and\
  \bibinfo {author} {\bibfnamefont {D.}~\bibnamefont {Valente}}} (\bibinfo
  {year} {2014}),\ \href {\doibase 10.1103/PhysRevE.89.062109} {\bibfield
  {journal} {\bibinfo  {journal} {Phys. Rev. E}\ }\textbf {\bibinfo {volume}
  {89}},\ \bibinfo {pages} {062109}}\BibitemShut {NoStop}%
\bibitem [{\citenamefont {Werlang}\ and\ \citenamefont
  {Valente}(2015)}]{Werlang2015}%
  \BibitemOpen
  \bibfield  {author} {\bibinfo {author} {\bibnamefont {Werlang}, \bibfnamefont
  {T.}}, \ and\ \bibinfo {author} {\bibfnamefont {D.}~\bibnamefont {Valente}}}
  (\bibinfo {year} {2015}),\ \href {\doibase 10.1103/PhysRevE.91.012143}
  {\bibfield  {journal} {\bibinfo  {journal} {Phys. Rev. E}\ }\textbf {\bibinfo
  {volume} {91}},\ \bibinfo {pages} {012143}}\BibitemShut {NoStop}%
\bibitem [{\citenamefont {Werner}\ \emph {et~al.}(2016)\citenamefont {Werner},
  \citenamefont {Jaschke}, \citenamefont {Silvi}, \citenamefont {Kliesch},
  \citenamefont {Calarco}, \citenamefont {Eisert},\ and\ \citenamefont
  {Montangero}}]{WernerMontangero2016}%
  \BibitemOpen
  \bibfield  {author} {\bibinfo {author} {\bibnamefont {Werner}, \bibfnamefont
  {A.~H.}}, \bibinfo {author} {\bibfnamefont {D.}~\bibnamefont {Jaschke}},
  \bibinfo {author} {\bibfnamefont {P.}~\bibnamefont {Silvi}}, \bibinfo
  {author} {\bibfnamefont {M.}~\bibnamefont {Kliesch}}, \bibinfo {author}
  {\bibfnamefont {T.}~\bibnamefont {Calarco}}, \bibinfo {author} {\bibfnamefont
  {J.}~\bibnamefont {Eisert}}, \ and\ \bibinfo {author} {\bibfnamefont
  {S.}~\bibnamefont {Montangero}}} (\bibinfo {year} {2016}),\ \href {\doibase
  10.1103/PhysRevLett.116.237201} {\bibfield  {journal} {\bibinfo  {journal}
  {Phys. Rev. Lett.}\ }\textbf {\bibinfo {volume} {116}},\ \bibinfo {pages}
  {237201}}\BibitemShut {NoStop}%
\bibitem [{\citenamefont {Wichterich}\ \emph {et~al.}(2007)\citenamefont
  {Wichterich}, \citenamefont {Henrich}, \citenamefont {Breuer}, \citenamefont
  {Gemmer},\ and\ \citenamefont {Michel}}]{Wichterich2007}%
  \BibitemOpen
  \bibfield  {author} {\bibinfo {author} {\bibnamefont {Wichterich},
  \bibfnamefont {H.}}, \bibinfo {author} {\bibfnamefont {M.~J.}\ \bibnamefont
  {Henrich}}, \bibinfo {author} {\bibfnamefont {H.~P.}\ \bibnamefont {Breuer}},
  \bibinfo {author} {\bibfnamefont {J.}~\bibnamefont {Gemmer}}, \ and\ \bibinfo
  {author} {\bibfnamefont {M.}~\bibnamefont {Michel}}} (\bibinfo {year}
  {2007}),\ \href {\doibase 10.1103/PhysRevE.76.031115} {\bibfield  {journal}
  {\bibinfo  {journal} {Physical Review E}\ }\textbf {\bibinfo {volume}
  {76}}~(\bibinfo {number} {3}),\ \bibinfo {pages} {031115}}\BibitemShut
  {NoStop}%
\bibitem [{\citenamefont {Willms}(2008)}]{Willms2008}%
  \BibitemOpen
  \bibfield  {author} {\bibinfo {author} {\bibnamefont {Willms}, \bibfnamefont
  {A.~R.}}} (\bibinfo {year} {2008}),\ \href {\doibase 10.1137/070695411}
  {\bibfield  {journal} {\bibinfo  {journal} {SIAM J. Matrix Anal. Appl.}\
  }\textbf {\bibinfo {volume} {30}},\ \bibinfo {pages} {639}}\BibitemShut
  {NoStop}%
\bibitem [{\citenamefont {Wilson}(1975)}]{Wilson1975}%
  \BibitemOpen
  \bibfield  {author} {\bibinfo {author} {\bibnamefont {Wilson}, \bibfnamefont
  {K.~G.}}} (\bibinfo {year} {1975}),\ \href {\doibase
  10.1103/RevModPhys.47.773} {\bibfield  {journal} {\bibinfo  {journal} {Rev.
  Mod. Phys.}\ }\textbf {\bibinfo {volume} {47}},\ \bibinfo {pages}
  {773}}\BibitemShut {NoStop}%
\bibitem [{\citenamefont {Wiseman}\ and\ \citenamefont
  {Milburn}(2010)}]{wiseman2010}%
  \BibitemOpen
  \bibfield  {author} {\bibinfo {author} {\bibnamefont {Wiseman}, \bibfnamefont
  {H.~M.}}, \ and\ \bibinfo {author} {\bibfnamefont {G.~J.}\ \bibnamefont
  {Milburn}}} (\bibinfo {year} {2010}),\ \href {\doibase
  https://doi.org/10.1017/CBO9780511813948} {\emph {\bibinfo {title} {Quantum
  Measurement and Control}}}\ (\bibinfo  {publisher} {Cambridge University
  Press},\ \bibinfo {address} {Cambridge})\BibitemShut {NoStop}%
\bibitem [{\citenamefont {W\'ojtowicz}\ \emph {et~al.}(2020)\citenamefont
  {W\'ojtowicz}, \citenamefont {Elenewski}, \citenamefont {Rams},\ and\
  \citenamefont {Zwolak}}]{WojtowiczZwolak2020}%
  \BibitemOpen
  \bibfield  {author} {\bibinfo {author} {\bibnamefont {W\'ojtowicz},
  \bibfnamefont {G.}}, \bibinfo {author} {\bibfnamefont {J.~E.}\ \bibnamefont
  {Elenewski}}, \bibinfo {author} {\bibfnamefont {M.~M.}\ \bibnamefont {Rams}},
  \ and\ \bibinfo {author} {\bibfnamefont {M.}~\bibnamefont {Zwolak}}}
  (\bibinfo {year} {2020}),\ \href {\doibase 10.1103/PhysRevA.101.050301}
  {\bibfield  {journal} {\bibinfo  {journal} {Phys. Rev. A}\ }\textbf {\bibinfo
  {volume} {101}},\ \bibinfo {pages} {050301}}\BibitemShut {NoStop}%
\bibitem [{\citenamefont {Wolf}\ \emph {et~al.}(2014)\citenamefont {Wolf},
  \citenamefont {McCulloch},\ and\ \citenamefont
  {Schollw{\"o}ck}}]{WolfSchollwock2014}%
  \BibitemOpen
  \bibfield  {author} {\bibinfo {author} {\bibnamefont {Wolf}, \bibfnamefont
  {F.~A.}}, \bibinfo {author} {\bibfnamefont {I.~P.}\ \bibnamefont
  {McCulloch}}, \ and\ \bibinfo {author} {\bibfnamefont {U.}~\bibnamefont
  {Schollw{\"o}ck}}} (\bibinfo {year} {2014}),\ \href {\doibase
  10.1103/PhysRevB.90.235131} {\bibfield  {journal} {\bibinfo  {journal} {Phys.
  Rev. B}\ }\textbf {\bibinfo {volume} {90}},\ \bibinfo {pages}
  {235131}}\BibitemShut {NoStop}%
\bibitem [{\citenamefont {Wolf}(2006)}]{Wolf2006}%
  \BibitemOpen
  \bibfield  {author} {\bibinfo {author} {\bibnamefont {Wolf}, \bibfnamefont
  {M.~M.}}} (\bibinfo {year} {2006}),\ \href {\doibase
  10.1103/PhysRevLett.96.010404} {\bibfield  {journal} {\bibinfo  {journal}
  {Phys. Rev. Lett.}\ }\textbf {\bibinfo {volume} {96}},\ \bibinfo {pages}
  {010404}}\BibitemShut {NoStop}%
\bibitem [{\citenamefont {Wolff}\ \emph {et~al.}(2020)\citenamefont {Wolff},
  \citenamefont {Sheikhan},\ and\ \citenamefont {Kollath}}]{WolffKollath2020}%
  \BibitemOpen
  \bibfield  {author} {\bibinfo {author} {\bibnamefont {Wolff}, \bibfnamefont
  {S.}}, \bibinfo {author} {\bibfnamefont {A.}~\bibnamefont {Sheikhan}}, \ and\
  \bibinfo {author} {\bibfnamefont {C.}~\bibnamefont {Kollath}}} (\bibinfo
  {year} {2020}),\ \href {\doibase 10.21468/SciPostPhysCore.3.2.010} {\bibfield
   {journal} {\bibinfo  {journal} {SciPost Phys. Core}\ }\textbf {\bibinfo
  {volume} {3}},\ \bibinfo {pages} {10}}\BibitemShut {NoStop}%
\bibitem [{\citenamefont {Woods}\ \emph {et~al.}(2014)\citenamefont {Woods},
  \citenamefont {Groux}, \citenamefont {Chin}, \citenamefont {Huelga},\ and\
  \citenamefont {Plenio}}]{WoodsPlenio2014}%
  \BibitemOpen
  \bibfield  {author} {\bibinfo {author} {\bibnamefont {Woods}, \bibfnamefont
  {M.~P.}}, \bibinfo {author} {\bibfnamefont {R.}~\bibnamefont {Groux}},
  \bibinfo {author} {\bibfnamefont {A.~W.}\ \bibnamefont {Chin}}, \bibinfo
  {author} {\bibfnamefont {S.~F.}\ \bibnamefont {Huelga}}, \ and\ \bibinfo
  {author} {\bibfnamefont {M.~B.}\ \bibnamefont {Plenio}}} (\bibinfo {year}
  {2014}),\ \href {\doibase 10.1063/1.4866769} {\bibfield  {journal} {\bibinfo
  {journal} {Jour. Math. Phys.}\ }\textbf {\bibinfo {volume} {55}},\ \bibinfo
  {pages} {032101}}\BibitemShut {NoStop}%
\bibitem [{\citenamefont {Wu}\ and\ \citenamefont
  {Segal}(2009{\natexlab{a}})}]{wu2009a}%
  \BibitemOpen
  \bibfield  {author} {\bibinfo {author} {\bibnamefont {Wu}, \bibfnamefont
  {L.-A.}}, \ and\ \bibinfo {author} {\bibfnamefont {D.}~\bibnamefont {Segal}}}
  (\bibinfo {year} {2009}{\natexlab{a}}),\ \href {\doibase
  10.1088/1751-8113/42/2/025302} {\bibfield  {journal} {\bibinfo  {journal}
  {Journal of Physics A: Mathematical and Theoretical}\ }\textbf {\bibinfo
  {volume} {42}},\ \bibinfo {pages} {025302}}\BibitemShut {NoStop}%
\bibitem [{\citenamefont {Wu}\ and\ \citenamefont
  {Segal}(2009{\natexlab{b}})}]{WuSegal2009}%
  \BibitemOpen
  \bibfield  {author} {\bibinfo {author} {\bibnamefont {Wu}, \bibfnamefont
  {L.-A.}}, \ and\ \bibinfo {author} {\bibfnamefont {D.}~\bibnamefont {Segal}}}
  (\bibinfo {year} {2009}{\natexlab{b}}),\ \href {\doibase
  10.1103/PhysRevLett.102.095503} {\bibfield  {journal} {\bibinfo  {journal}
  {Phys. Rev. Lett.}\ }\textbf {\bibinfo {volume} {102}},\ \bibinfo {pages}
  {095503}}\BibitemShut {NoStop}%
\bibitem [{\citenamefont {Wu}\ \emph {et~al.}(2009)\citenamefont {Wu},
  \citenamefont {Yu},\ and\ \citenamefont {Segal}}]{WuSegal2009b}%
  \BibitemOpen
  \bibfield  {author} {\bibinfo {author} {\bibnamefont {Wu}, \bibfnamefont
  {L.-A.}}, \bibinfo {author} {\bibfnamefont {C.~X.}\ \bibnamefont {Yu}}, \
  and\ \bibinfo {author} {\bibfnamefont {D.}~\bibnamefont {Segal}}} (\bibinfo
  {year} {2009}),\ \href {\doibase 10.1103/PhysRevE.80.041103} {\bibfield
  {journal} {\bibinfo  {journal} {Phys. Rev. E}\ }\textbf {\bibinfo {volume}
  {80}},\ \bibinfo {pages} {041103}}\BibitemShut {NoStop}%
\bibitem [{\citenamefont {Wu}\ and\ \citenamefont {Eckardt}(2019)}]{wu2019a}%
  \BibitemOpen
  \bibfield  {author} {\bibinfo {author} {\bibnamefont {Wu}, \bibfnamefont
  {L.-N.}}, \ and\ \bibinfo {author} {\bibfnamefont {A.}~\bibnamefont
  {Eckardt}}} (\bibinfo {year} {2019}),\ \href {\doibase
  10.1103/PhysRevLett.123.030602} {\bibfield  {journal} {\bibinfo  {journal}
  {Phys. Rev. Lett.}\ }\textbf {\bibinfo {volume} {123}},\ \bibinfo {pages}
  {030602}}\BibitemShut {NoStop}%
\bibitem [{\citenamefont {Wu}\ and\ \citenamefont {Eckardt}(2020)}]{wu2020a}%
  \BibitemOpen
  \bibfield  {author} {\bibinfo {author} {\bibnamefont {Wu}, \bibfnamefont
  {L.-N.}}, \ and\ \bibinfo {author} {\bibfnamefont {A.}~\bibnamefont
  {Eckardt}}} (\bibinfo {year} {2020}),\ \href {\doibase
  10.1103/PhysRevB.101.220302} {\bibfield  {journal} {\bibinfo  {journal}
  {Phys. Rev. B}\ }\textbf {\bibinfo {volume} {101}},\ \bibinfo {pages}
  {220302}}\BibitemShut {NoStop}%
\bibitem [{\citenamefont {Wu}\ \emph {et~al.}(2019)\citenamefont {Wu},
  \citenamefont {Schnell}, \citenamefont {Tomasi}, \citenamefont {Heyl},\ and\
  \citenamefont {Eckardt}}]{WuEckardt2019}%
  \BibitemOpen
  \bibfield  {author} {\bibinfo {author} {\bibnamefont {Wu}, \bibfnamefont
  {L.-N.}}, \bibinfo {author} {\bibfnamefont {A.}~\bibnamefont {Schnell}},
  \bibinfo {author} {\bibfnamefont {G.~D.}\ \bibnamefont {Tomasi}}, \bibinfo
  {author} {\bibfnamefont {M.}~\bibnamefont {Heyl}}, \ and\ \bibinfo {author}
  {\bibfnamefont {A.}~\bibnamefont {Eckardt}}} (\bibinfo {year} {2019}),\ \href
  {\doibase 10.1088/1367-2630/ab25a4} {\bibfield  {journal} {\bibinfo
  {journal} {New Journal of Physics}\ }\textbf {\bibinfo {volume}
  {21}}~(\bibinfo {number} {6}),\ \bibinfo {pages} {063026}}\BibitemShut
  {NoStop}%
\bibitem [{\citenamefont {Xiao}\ \emph {et~al.}(2010)\citenamefont {Xiao},
  \citenamefont {Chang},\ and\ \citenamefont {Niu}}]{XiaoNiu2010}%
  \BibitemOpen
  \bibfield  {author} {\bibinfo {author} {\bibnamefont {Xiao}, \bibfnamefont
  {D.}}, \bibinfo {author} {\bibfnamefont {M.-C.}\ \bibnamefont {Chang}}, \
  and\ \bibinfo {author} {\bibfnamefont {Q.}~\bibnamefont {Niu}}} (\bibinfo
  {year} {2010}),\ \href {\doibase 10.1103/RevModPhys.82.1959} {\bibfield
  {journal} {\bibinfo  {journal} {Rev. Mod. Phys.}\ }\textbf {\bibinfo {volume}
  {82}},\ \bibinfo {pages} {1959}}\BibitemShut {NoStop}%
\bibitem [{\citenamefont {Xing}\ \emph {et~al.}(2020)\citenamefont {Xing},
  \citenamefont {Xu}, \citenamefont {Balachandran},\ and\ \citenamefont
  {Poletti}}]{XingPoletti2020}%
  \BibitemOpen
  \bibfield  {author} {\bibinfo {author} {\bibnamefont {Xing}, \bibfnamefont
  {B.}}, \bibinfo {author} {\bibfnamefont {X.}~\bibnamefont {Xu}}, \bibinfo
  {author} {\bibfnamefont {V.}~\bibnamefont {Balachandran}}, \ and\ \bibinfo
  {author} {\bibfnamefont {D.}~\bibnamefont {Poletti}}} (\bibinfo {year}
  {2020}),\ \href {\doibase 10.1103/PhysRevB.102.245433} {\bibfield  {journal}
  {\bibinfo  {journal} {Phys. Rev. B}\ }\textbf {\bibinfo {volume} {102}},\
  \bibinfo {pages} {245433}}\BibitemShut {NoStop}%
\bibitem [{\citenamefont {Xu}\ and\ \citenamefont {Dubi}(2015)}]{XuDubi2015}%
  \BibitemOpen
  \bibfield  {author} {\bibinfo {author} {\bibnamefont {Xu}, \bibfnamefont
  {B.}}, \ and\ \bibinfo {author} {\bibfnamefont {Y.}~\bibnamefont {Dubi}}}
  (\bibinfo {year} {2015}),\ \href {\doibase 10.1088/0953-8984/27/26/263202}
  {\bibfield  {journal} {\bibinfo  {journal} {Journal of Physics: Condensed
  Matter}\ }\textbf {\bibinfo {volume} {27}},\ \bibinfo {pages}
  {263202}}\BibitemShut {NoStop}%
\bibitem [{\citenamefont {Xu}\ \emph {et~al.}(2019{\natexlab{a}})\citenamefont
  {Xu}, \citenamefont {Shen}, \citenamefont {Yi},\ and\ \citenamefont
  {Wang}}]{xu2019a}%
  \BibitemOpen
  \bibfield  {author} {\bibinfo {author} {\bibnamefont {Xu}, \bibfnamefont
  {S.}}, \bibinfo {author} {\bibfnamefont {H.~Z.}\ \bibnamefont {Shen}},
  \bibinfo {author} {\bibfnamefont {X.~X.}\ \bibnamefont {Yi}}, \ and\ \bibinfo
  {author} {\bibfnamefont {W.}~\bibnamefont {Wang}}} (\bibinfo {year}
  {2019}{\natexlab{a}}),\ \href {\doibase 10.1103/PhysRevA.100.032108}
  {\bibfield  {journal} {\bibinfo  {journal} {Phys. Rev. A}\ }\textbf {\bibinfo
  {volume} {100}},\ \bibinfo {pages} {032108}}\BibitemShut {NoStop}%
\bibitem [{\citenamefont {Xu}\ \emph {et~al.}(2019{\natexlab{b}})\citenamefont
  {Xu}, \citenamefont {Choo}, \citenamefont {Balachandran},\ and\ \citenamefont
  {Poletti}}]{XuPoletti2019b}%
  \BibitemOpen
  \bibfield  {author} {\bibinfo {author} {\bibnamefont {Xu}, \bibfnamefont
  {X.}}, \bibinfo {author} {\bibfnamefont {K.}~\bibnamefont {Choo}}, \bibinfo
  {author} {\bibfnamefont {V.}~\bibnamefont {Balachandran}}, \ and\ \bibinfo
  {author} {\bibfnamefont {D.}~\bibnamefont {Poletti}}} (\bibinfo {year}
  {2019}{\natexlab{b}}),\ \href {\doibase 10.3390/e21030228} {\bibfield
  {journal} {\bibinfo  {journal} {Entropy}\ }\textbf {\bibinfo {volume}
  {228}},\ \bibinfo {pages} {21}}\BibitemShut {NoStop}%
\bibitem [{\citenamefont {Xu}\ \emph {et~al.}(2022)\citenamefont {Xu},
  \citenamefont {Guo},\ and\ \citenamefont {Poletti}}]{XuPoletti2021}%
  \BibitemOpen
  \bibfield  {author} {\bibinfo {author} {\bibnamefont {Xu}, \bibfnamefont
  {X.}}, \bibinfo {author} {\bibfnamefont {C.}~\bibnamefont {Guo}}, \ and\
  \bibinfo {author} {\bibfnamefont {D.}~\bibnamefont {Poletti}}} (\bibinfo
  {year} {2022}),\ \href {\doibase 10.1103/PhysRevA.105.L040203} {\bibfield
  {journal} {\bibinfo  {journal} {Phys. Rev. A}\ }\textbf {\bibinfo {volume}
  {105}},\ \bibinfo {pages} {L040203}}\BibitemShut {NoStop}%
\bibitem [{\citenamefont {Xu}\ \emph {et~al.}(2019{\natexlab{c}})\citenamefont
  {Xu}, \citenamefont {Thingna}, \citenamefont {Guo},\ and\ \citenamefont
  {Poletti}}]{XuPoletti2019a}%
  \BibitemOpen
  \bibfield  {author} {\bibinfo {author} {\bibnamefont {Xu}, \bibfnamefont
  {X.}}, \bibinfo {author} {\bibfnamefont {J.}~\bibnamefont {Thingna}},
  \bibinfo {author} {\bibfnamefont {C.}~\bibnamefont {Guo}}, \ and\ \bibinfo
  {author} {\bibfnamefont {D.}~\bibnamefont {Poletti}}} (\bibinfo {year}
  {2019}{\natexlab{c}}),\ \href {\doibase 10.1103/PhysRevA.99.012106}
  {\bibfield  {journal} {\bibinfo  {journal} {Phys. Rev. A}\ }\textbf {\bibinfo
  {volume} {99}},\ \bibinfo {pages} {012106}}\BibitemShut {NoStop}%
\bibitem [{\citenamefont {Xu}\ \emph {et~al.}(2017)\citenamefont {Xu},
  \citenamefont {Thingna},\ and\ \citenamefont {Wang}}]{XuWang2017}%
  \BibitemOpen
  \bibfield  {author} {\bibinfo {author} {\bibnamefont {Xu}, \bibfnamefont
  {X.}}, \bibinfo {author} {\bibfnamefont {J.}~\bibnamefont {Thingna}}, \ and\
  \bibinfo {author} {\bibfnamefont {J.-S.}\ \bibnamefont {Wang}}} (\bibinfo
  {year} {2017}),\ \href {\doibase 10.1103/PhysRevB.95.035428} {\bibfield
  {journal} {\bibinfo  {journal} {Phys. Rev. B}\ }\textbf {\bibinfo {volume}
  {95}},\ \bibinfo {pages} {035428}}\BibitemShut {NoStop}%
\bibitem [{\citenamefont {Xue-Ou}\ \emph {et~al.}(2008)\citenamefont {Xue-Ou},
  \citenamefont {Bing},\ and\ \citenamefont {Xiao-Lin}}]{ChenLei2008}%
  \BibitemOpen
  \bibfield  {author} {\bibinfo {author} {\bibnamefont {Xue-Ou}, \bibfnamefont
  {C.}}, \bibinfo {author} {\bibfnamefont {D.}~\bibnamefont {Bing}}, \ and\
  \bibinfo {author} {\bibfnamefont {L.}~\bibnamefont {Xiao-Lin}}} (\bibinfo
  {year} {2008}),\ \href {\doibase 10.1088/0256-307x/25/8/080} {\bibfield
  {journal} {\bibinfo  {journal} {Chinese Physics Letters}\ }\textbf {\bibinfo
  {volume} {25}}~(\bibinfo {number} {8}),\ \bibinfo {pages} {3032}}\BibitemShut
  {NoStop}%
\bibitem [{\citenamefont {Yacoby}\ \emph {et~al.}(1995)\citenamefont {Yacoby},
  \citenamefont {Heiblum}, \citenamefont {Mahalu},\ and\ \citenamefont
  {Shtrikman}}]{YacobyShtrikman1995}%
  \BibitemOpen
  \bibfield  {author} {\bibinfo {author} {\bibnamefont {Yacoby}, \bibfnamefont
  {A.}}, \bibinfo {author} {\bibfnamefont {M.}~\bibnamefont {Heiblum}},
  \bibinfo {author} {\bibfnamefont {D.}~\bibnamefont {Mahalu}}, \ and\ \bibinfo
  {author} {\bibfnamefont {H.}~\bibnamefont {Shtrikman}}} (\bibinfo {year}
  {1995}),\ \href {\doibase 10.1103/PhysRevLett.74.4047} {\bibfield  {journal}
  {\bibinfo  {journal} {Phys. Rev. Lett.}\ }\textbf {\bibinfo {volume} {74}},\
  \bibinfo {pages} {4047}}\BibitemShut {NoStop}%
\bibitem [{\citenamefont {Yamanaka}\ and\ \citenamefont
  {Sasamoto}(2021)}]{YamanakaSasamoto2021}%
  \BibitemOpen
  \bibfield  {author} {\bibinfo {author} {\bibnamefont {Yamanaka},
  \bibfnamefont {K.}}, \ and\ \bibinfo {author} {\bibfnamefont
  {T.}~\bibnamefont {Sasamoto}}} (\bibinfo {year} {2021}),\ \href
  {https://arxiv.org/abs/2104.11479} {\bibinfo  {journal} {arXiv:2104.11479}\
  }\BibitemShut {NoStop}%
\bibitem [{\citenamefont {Yan}\ \emph {et~al.}(2009)\citenamefont {Yan},
  \citenamefont {Wu},\ and\ \citenamefont {Li}}]{YanLi2009}%
  \BibitemOpen
\bibfield  {journal} {  }\bibfield  {author} {\bibinfo {author} {\bibnamefont
  {Yan}, \bibfnamefont {Y.}}, \bibinfo {author} {\bibfnamefont {C.-Q.}\
  \bibnamefont {Wu}}, \ and\ \bibinfo {author} {\bibfnamefont {B.}~\bibnamefont
  {Li}}} (\bibinfo {year} {2009}),\ \href {\doibase 10.1103/PhysRevB.79.014207}
  {\bibfield  {journal} {\bibinfo  {journal} {Phys. Rev. B}\ }\textbf {\bibinfo
  {volume} {79}},\ \bibinfo {pages} {014207}}\BibitemShut {NoStop}%
\bibitem [{\citenamefont {Yang}\ \emph {et~al.}(2020)\citenamefont {Yang},
  \citenamefont {Hsiang}, \citenamefont {Jordan},\ and\ \citenamefont
  {Hu}}]{yang2020a}%
  \BibitemOpen
  \bibfield  {author} {\bibinfo {author} {\bibnamefont {Yang}, \bibfnamefont
  {J.}}, \bibinfo {author} {\bibfnamefont {J.-T.}\ \bibnamefont {Hsiang}},
  \bibinfo {author} {\bibfnamefont {A.~N.}\ \bibnamefont {Jordan}}, \ and\
  \bibinfo {author} {\bibfnamefont {B.}~\bibnamefont {Hu}}} (\bibinfo {year}
  {2020}),\ \href {\doibase https://doi.org/10.1016/j.aop.2020.168289}
  {\bibfield  {journal} {\bibinfo  {journal} {Annals of Physics}\ }\textbf
  {\bibinfo {volume} {421}},\ \bibinfo {pages} {168289}}\BibitemShut {NoStop}%
\bibitem [{\citenamefont {Yang}\ \emph {et~al.}(2018)\citenamefont {Yang},
  \citenamefont {Chen}, \citenamefont {Wang}, \citenamefont {Li},\ and\
  \citenamefont {Zhang}}]{YangZhang2018}%
  \BibitemOpen
  \bibfield  {author} {\bibinfo {author} {\bibnamefont {Yang}, \bibfnamefont
  {Y.}}, \bibinfo {author} {\bibfnamefont {H.}~\bibnamefont {Chen}}, \bibinfo
  {author} {\bibfnamefont {H.}~\bibnamefont {Wang}}, \bibinfo {author}
  {\bibfnamefont {N.}~\bibnamefont {Li}}, \ and\ \bibinfo {author}
  {\bibfnamefont {L.}~\bibnamefont {Zhang}}} (\bibinfo {year} {2018}),\ \href
  {\doibase 10.1103/PhysRevE.98.042131} {\bibfield  {journal} {\bibinfo
  {journal} {Phys. Rev. E}\ }\textbf {\bibinfo {volume} {98}},\ \bibinfo
  {pages} {042131}}\BibitemShut {NoStop}%
\bibitem [{\citenamefont {Yeyati}\ and\ \citenamefont
  {B{\"u}ttiker}(1995)}]{YeyatiButtiker1995}%
  \BibitemOpen
  \bibfield  {author} {\bibinfo {author} {\bibnamefont {Yeyati}, \bibfnamefont
  {A.~L.}}, \ and\ \bibinfo {author} {\bibfnamefont {M.}~\bibnamefont
  {B{\"u}ttiker}}} (\bibinfo {year} {1995}),\ \href {\doibase
  10.1103/PhysRevB.52.R14360} {\bibfield  {journal} {\bibinfo  {journal} {Phys.
  Rev. B}\ }\textbf {\bibinfo {volume} {52}},\ \bibinfo {pages}
  {R14360}}\BibitemShut {NoStop}%
\bibitem [{\citenamefont {Yoshioka}\ and\ \citenamefont
  {Hamazaki}(2019)}]{YoshiokaHamazaki2019}%
  \BibitemOpen
  \bibfield  {author} {\bibinfo {author} {\bibnamefont {Yoshioka},
  \bibfnamefont {N.}}, \ and\ \bibinfo {author} {\bibfnamefont
  {R.}~\bibnamefont {Hamazaki}}} (\bibinfo {year} {2019}),\ \href {\doibase
  10.1103/PhysRevB.99.214306} {\bibfield  {journal} {\bibinfo  {journal} {Phys.
  Rev. B}\ }\textbf {\bibinfo {volume} {99}},\ \bibinfo {pages}
  {214306}}\BibitemShut {NoStop}%
\bibitem [{\citenamefont {Yueh}(2005)}]{Yueh2005}%
  \BibitemOpen
  \bibfield  {author} {\bibinfo {author} {\bibnamefont {Yueh}, \bibfnamefont
  {W.~C.}}} (\bibinfo {year} {2005}),\ \href
  {https://www.emis.de/journals/AMEN/2005/040903-7.pdf} {\bibfield  {journal}
  {\bibinfo  {journal} {Appl. Math. E-Notes}\ }\textbf {\bibinfo {volume}
  {5}},\ \bibinfo {pages} {66}}\BibitemShut {NoStop}%
\bibitem [{\citenamefont {Zanoci}\ and\ \citenamefont
  {Swingle}(2021)}]{ZanociSwingle2021}%
  \BibitemOpen
  \bibfield  {author} {\bibinfo {author} {\bibnamefont {Zanoci}, \bibfnamefont
  {C.}}, \ and\ \bibinfo {author} {\bibfnamefont {B.}~\bibnamefont {Swingle}}}
  (\bibinfo {year} {2021}),\ \href {\doibase 10.1103/PhysRevB.103.115148}
  {\bibfield  {journal} {\bibinfo  {journal} {Phys. Rev. B}\ }\textbf {\bibinfo
  {volume} {103}},\ \bibinfo {pages} {115148}}\BibitemShut {NoStop}%
\bibitem [{\citenamefont {Zedler}\ \emph {et~al.}(2009)\citenamefont {Zedler},
  \citenamefont {Schaller}, \citenamefont {Kie{\ss}lich}, \citenamefont
  {Emary},\ and\ \citenamefont {Brandes}}]{zedler2009a}%
  \BibitemOpen
  \bibfield  {author} {\bibinfo {author} {\bibnamefont {Zedler}, \bibfnamefont
  {P.}}, \bibinfo {author} {\bibfnamefont {G.}~\bibnamefont {Schaller}},
  \bibinfo {author} {\bibfnamefont {G.}~\bibnamefont {Kie{\ss}lich}}, \bibinfo
  {author} {\bibfnamefont {C.}~\bibnamefont {Emary}}, \ and\ \bibinfo {author}
  {\bibfnamefont {T.}~\bibnamefont {Brandes}}} (\bibinfo {year} {2009}),\ \href
  {\doibase 10.1103/PhysRevB.80.045309} {\bibfield  {journal} {\bibinfo
  {journal} {Physical Review B}\ }\textbf {\bibinfo {volume} {80}},\ \bibinfo
  {pages} {045309}}\BibitemShut {NoStop}%
\bibitem [{\citenamefont {Zhang}\ \emph {et~al.}(2009)\citenamefont {Zhang},
  \citenamefont {Yan}, \citenamefont {Wu}, \citenamefont {Wang},\ and\
  \citenamefont {Li}}]{LifaLi2009}%
  \BibitemOpen
  \bibfield  {author} {\bibinfo {author} {\bibnamefont {Zhang}, \bibfnamefont
  {L.}}, \bibinfo {author} {\bibfnamefont {Y.}~\bibnamefont {Yan}}, \bibinfo
  {author} {\bibfnamefont {C.-Q.}\ \bibnamefont {Wu}}, \bibinfo {author}
  {\bibfnamefont {J.-S.}\ \bibnamefont {Wang}}, \ and\ \bibinfo {author}
  {\bibfnamefont {B.}~\bibnamefont {Li}}} (\bibinfo {year} {2009}),\ \href
  {\doibase 10.1103/PhysRevB.80.172301} {\bibfield  {journal} {\bibinfo
  {journal} {Phys. Rev. B}\ }\textbf {\bibinfo {volume} {80}},\ \bibinfo
  {pages} {172301}}\BibitemShut {NoStop}%
\bibitem [{\citenamefont {Zhou}\ \emph {et~al.}(2000)\citenamefont {Zhou},
  \citenamefont {Kong}, \citenamefont {Yenilmez},\ and\ \citenamefont
  {Dai}}]{ZhouDai2000}%
  \BibitemOpen
  \bibfield  {author} {\bibinfo {author} {\bibnamefont {Zhou}, \bibfnamefont
  {C.}}, \bibinfo {author} {\bibfnamefont {J.}~\bibnamefont {Kong}}, \bibinfo
  {author} {\bibfnamefont {E.}~\bibnamefont {Yenilmez}}, \ and\ \bibinfo
  {author} {\bibfnamefont {H.}~\bibnamefont {Dai}}} (\bibinfo {year} {2000}),\
  \href {\doibase 10.1126/science.290.5496.1552} {\bibfield  {journal}
  {\bibinfo  {journal} {Science}\ }\textbf {\bibinfo {volume} {290}}~(\bibinfo
  {number} {5496}),\ \bibinfo {pages} {1552}}\BibitemShut {NoStop}%
\bibitem [{\citenamefont {Zimbovskaya}(2020)}]{Zimbovskaya2020}%
  \BibitemOpen
  \bibfield  {author} {\bibinfo {author} {\bibnamefont {Zimbovskaya},
  \bibfnamefont {N.~A.}}} (\bibinfo {year} {2020}),\ \href {\doibase
  10.1088/1361-648x/ab83e9} {\bibfield  {journal} {\bibinfo  {journal} {Journal
  of Physics: Condensed Matter}\ }\textbf {\bibinfo {volume} {32}}~(\bibinfo
  {number} {32}),\ \bibinfo {pages} {325302}}\BibitemShut {NoStop}%
\bibitem [{\citenamefont {Zimbovskaya}\ and\ \citenamefont
  {Pederson}(2011)}]{ZimbovskayaPederson2011}%
  \BibitemOpen
  \bibfield  {author} {\bibinfo {author} {\bibnamefont {Zimbovskaya},
  \bibfnamefont {N.~A.}}, \ and\ \bibinfo {author} {\bibfnamefont {M.~R.}\
  \bibnamefont {Pederson}}} (\bibinfo {year} {2011}),\ \href {\doibase
  10.1016/j.physrep.2011.08.002} {\bibfield  {journal} {\bibinfo  {journal}
  {Phys. Rep.}\ }\textbf {\bibinfo {volume} {509}}~(\bibinfo {number} {1}),\
  \bibinfo {pages} {1}}\BibitemShut {NoStop}%
\bibitem [{\citenamefont {Ziolkowska}\ and\ \citenamefont
  {Essler}(2020)}]{Ziolkowska2020}%
  \BibitemOpen
  \bibfield  {author} {\bibinfo {author} {\bibnamefont {Ziolkowska},
  \bibfnamefont {A.~A.}}, \ and\ \bibinfo {author} {\bibfnamefont {F.~H.}\
  \bibnamefont {Essler}}} (\bibinfo {year} {2020}),\ \href {\doibase
  10.21468/SciPostPhys.8.3.044} {\bibfield  {journal} {\bibinfo  {journal}
  {SciPost Phys.}\ }\textbf {\bibinfo {volume} {8}},\ \bibinfo {pages}
  {44}}\BibitemShut {NoStop}%
\bibitem [{\citenamefont
  {{\v{Z}nidari\v{c}}}(2010{\natexlab{a}})}]{Znidaric2010}%
  \BibitemOpen
  \bibfield  {author} {\bibinfo {author} {\bibnamefont {{\v{Z}nidari\v{c}}},
  \bibfnamefont {M.}}} (\bibinfo {year} {2010}{\natexlab{a}}),\ \href {\doibase
  10.1088/1742-5468/2010/05/l05002} {\bibfield  {journal} {\bibinfo  {journal}
  {Journal of Statistical Mechanics: Theory and Experiment}\ }\textbf {\bibinfo
  {volume} {2010}}~(\bibinfo {number} {05}),\ \bibinfo {pages}
  {L05002}}\BibitemShut {NoStop}%
\bibitem [{\citenamefont
  {{\v{Z}nidari\v{c}}}(2010{\natexlab{b}})}]{Znidaric2010b}%
  \BibitemOpen
  \bibfield  {author} {\bibinfo {author} {\bibnamefont {{\v{Z}nidari\v{c}}},
  \bibfnamefont {M.}}} (\bibinfo {year} {2010}{\natexlab{b}}),\ \href {\doibase
  10.1088/1751-8113/43/41/415004} {\bibfield  {journal} {\bibinfo  {journal}
  {Journal of Physics A: Mathematical and Theoretical}\ }\textbf {\bibinfo
  {volume} {43}},\ \bibinfo {pages} {415004}}\BibitemShut {NoStop}%
\bibitem [{\citenamefont {{\v{Z}nidari\v{c}}}(2011)}]{Znidaric2011c}%
  \BibitemOpen
  \bibfield  {author} {\bibinfo {author} {\bibnamefont {{\v{Z}nidari\v{c}}},
  \bibfnamefont {M.}}} (\bibinfo {year} {2011}),\ \href {\doibase
  10.1103/PhysRevE.83.011108} {\bibfield  {journal} {\bibinfo  {journal} {Phys.
  Rev. E}\ }\textbf {\bibinfo {volume} {83}},\ \bibinfo {pages}
  {011108}}\BibitemShut {NoStop}%
\bibitem [{\citenamefont
  {{\v{Z}}nidari{\v{c}}}(2011{\natexlab{a}})}]{Znidaric2011}%
  \BibitemOpen
  \bibfield  {author} {\bibinfo {author} {\bibnamefont {{\v{Z}}nidari{\v{c}}},
  \bibfnamefont {M.}}} (\bibinfo {year} {2011}{\natexlab{a}}),\ \href {\doibase
  10.1103/PhysRevLett.106.220601} {\bibfield  {journal} {\bibinfo  {journal}
  {Physical Review Letters}\ }\textbf {\bibinfo {volume} {106}}~(\bibinfo
  {number} {22}),\ \bibinfo {pages} {220601}}\BibitemShut {NoStop}%
\bibitem [{\citenamefont
  {{\v{Z}}nidari{\v{c}}}(2011{\natexlab{b}})}]{Znidaric2011b}%
  \BibitemOpen
  \bibfield  {author} {\bibinfo {author} {\bibnamefont {{\v{Z}}nidari{\v{c}}},
  \bibfnamefont {M.}}} (\bibinfo {year} {2011}{\natexlab{b}}),\ \href {\doibase
  10.1088/1742-5468/2011/12/P12008} {\bibinfo  {journal} {Journal of
  Statistical Mechanics}\ ,\ \bibinfo {pages} {P12008}}\BibitemShut {NoStop}%
\bibitem [{\citenamefont
  {{\v{Z}nidari\v{c}}}(2013{\natexlab{a}})}]{Znidaric2013}%
  \BibitemOpen
\bibfield  {journal} {  }\bibfield  {author} {\bibinfo {author} {\bibnamefont
  {{\v{Z}nidari\v{c}}}, \bibfnamefont {M.}}} (\bibinfo {year}
  {2013}{\natexlab{a}}),\ \href {\doibase 10.1103/PhysRevLett.110.070602}
  {\bibfield  {journal} {\bibinfo  {journal} {Phys. Rev. Lett.}\ }\textbf
  {\bibinfo {volume} {110}},\ \bibinfo {pages} {070602}}\BibitemShut {NoStop}%
\bibitem [{\citenamefont
  {{\v{Z}nidari\v{c}}}(2013{\natexlab{b}})}]{Znidaric2013b}%
  \BibitemOpen
  \bibfield  {author} {\bibinfo {author} {\bibnamefont {{\v{Z}nidari\v{c}}},
  \bibfnamefont {M.}}} (\bibinfo {year} {2013}{\natexlab{b}}),\ \href {\doibase
  10.1103/PhysRevB.88.205135} {\bibfield  {journal} {\bibinfo  {journal} {Phys.
  Rev. B}\ }\textbf {\bibinfo {volume} {88}},\ \bibinfo {pages}
  {205135}}\BibitemShut {NoStop}%
\bibitem [{\citenamefont
  {{\v{Z}nidari\v{c}}}(2014{\natexlab{a}})}]{znidaric2014b}%
  \BibitemOpen
  \bibfield  {author} {\bibinfo {author} {\bibnamefont {{\v{Z}nidari\v{c}}},
  \bibfnamefont {M.}}} (\bibinfo {year} {2014}{\natexlab{a}}),\ \href {\doibase
  10.1103/PhysRevLett.112.040602} {\bibfield  {journal} {\bibinfo  {journal}
  {Phys. Rev. Lett.}\ }\textbf {\bibinfo {volume} {112}},\ \bibinfo {pages}
  {040602}}\BibitemShut {NoStop}%
\bibitem [{\citenamefont
  {{\v{Z}nidari\v{c}}}(2014{\natexlab{b}})}]{znidaric2014c}%
  \BibitemOpen
  \bibfield  {author} {\bibinfo {author} {\bibnamefont {{\v{Z}nidari\v{c}}},
  \bibfnamefont {M.}}} (\bibinfo {year} {2014}{\natexlab{b}}),\ \href {\doibase
  10.1103/PhysRevE.89.042140} {\bibfield  {journal} {\bibinfo  {journal} {Phys.
  Rev. E}\ }\textbf {\bibinfo {volume} {89}},\ \bibinfo {pages}
  {042140}}\BibitemShut {NoStop}%
\bibitem [{\citenamefont {{\v{Z}nidari\v{c}}}(2015)}]{Znidaric2015}%
  \BibitemOpen
  \bibfield  {author} {\bibinfo {author} {\bibnamefont {{\v{Z}nidari\v{c}}},
  \bibfnamefont {M.}}} (\bibinfo {year} {2015}),\ \href {\doibase
  10.1103/PhysRevE.92.042143} {\bibfield  {journal} {\bibinfo  {journal} {Phys.
  Rev. E}\ }\textbf {\bibinfo {volume} {92}},\ \bibinfo {pages}
  {042143}}\BibitemShut {NoStop}%
\bibitem [{\citenamefont {{\v{Z}nidari\v{c}}}(2019)}]{Znidaric2019}%
  \BibitemOpen
  \bibfield  {author} {\bibinfo {author} {\bibnamefont {{\v{Z}nidari\v{c}}},
  \bibfnamefont {M.}}} (\bibinfo {year} {2019}),\ \href {\doibase
  10.1103/PhysRevB.99.035143} {\bibfield  {journal} {\bibinfo  {journal} {Phys.
  Rev. B}\ }\textbf {\bibinfo {volume} {99}},\ \bibinfo {pages}
  {035143}}\BibitemShut {NoStop}%
\bibitem [{\citenamefont {{\v{Z}nidari\v{c}}}(2020)}]{Znidaric2020}%
  \BibitemOpen
  \bibfield  {author} {\bibinfo {author} {\bibnamefont {{\v{Z}nidari\v{c}}},
  \bibfnamefont {M.}}} (\bibinfo {year} {2020}),\ \href {\doibase
  10.1103/PhysRevLett.125.180605} {\bibfield  {journal} {\bibinfo  {journal}
  {Phys. Rev. Lett.}\ }\textbf {\bibinfo {volume} {125}},\ \bibinfo {pages}
  {180605}}\BibitemShut {NoStop}%
\bibitem [{\citenamefont {{\v{Z}nidari\v{c}}}(2022)}]{Znidaric2022}%
  \BibitemOpen
  \bibfield  {author} {\bibinfo {author} {\bibnamefont {{\v{Z}nidari\v{c}}},
  \bibfnamefont {M.}}} (\bibinfo {year} {2022}),\ \href {\doibase
  10.1103/PhysRevB.105.045140} {\bibfield  {journal} {\bibinfo  {journal}
  {Phys. Rev. B}\ }\textbf {\bibinfo {volume} {105}},\ \bibinfo {pages}
  {045140}}\BibitemShut {NoStop}%
\bibitem [{\citenamefont {{\v{Z}nidari\v{c}}}\ and\ \citenamefont
  {Horvat}(2013)}]{ZnidaricHorvat2013}%
  \BibitemOpen
  \bibfield  {author} {\bibinfo {author} {\bibnamefont {{\v{Z}nidari\v{c}}},
  \bibfnamefont {M.}}, \ and\ \bibinfo {author} {\bibfnamefont
  {M.}~\bibnamefont {Horvat}}} (\bibinfo {year} {2013}),\ \href {\doibase
  10.1140/epjb/e2012-30730-9} {\bibfield  {journal} {\bibinfo  {journal} {Eur.
  Phys. Jour. B}\ }\textbf {\bibinfo {volume} {86}},\ \bibinfo {pages}
  {67}}\BibitemShut {NoStop}%
\bibitem [{\citenamefont {{\v{Z}nidari\v{c}}}\ and\ \citenamefont
  {Ljubotina}(2018)}]{ZnidaricLjubotina2018}%
  \BibitemOpen
  \bibfield  {author} {\bibinfo {author} {\bibnamefont {{\v{Z}nidari\v{c}}},
  \bibfnamefont {M.}}, \ and\ \bibinfo {author} {\bibfnamefont
  {M.}~\bibnamefont {Ljubotina}}} (\bibinfo {year} {2018}),\ \href {\doibase
  10.1073/pnas.1800589115} {\bibfield  {journal} {\bibinfo  {journal} {Proc.
  Nat. Acad. Sc.}\ }\textbf {\bibinfo {volume} {115}},\ \bibinfo {pages}
  {4595}}\BibitemShut {NoStop}%
\bibitem [{\citenamefont {{\v{Z}nidari\v{c}}}\ \emph
  {et~al.}(2017)\citenamefont {{\v{Z}nidari\v{c}}}, \citenamefont
  {Mendoza-Arenas}, \citenamefont {Clark},\ and\ \citenamefont
  {Goold}}]{ZnidaricGoold2017}%
  \BibitemOpen
  \bibfield  {author} {\bibinfo {author} {\bibnamefont {{\v{Z}nidari\v{c}}},
  \bibfnamefont {M.}}, \bibinfo {author} {\bibfnamefont {J.~J.}\ \bibnamefont
  {Mendoza-Arenas}}, \bibinfo {author} {\bibfnamefont {S.~R.}\ \bibnamefont
  {Clark}}, \ and\ \bibinfo {author} {\bibfnamefont {J.}~\bibnamefont {Goold}}}
  (\bibinfo {year} {2017}),\ \href {\doibase 10.1002/andp.201600298} {\bibfield
   {journal} {\bibinfo  {journal} {Annalen der Physik}\ }\textbf {\bibinfo
  {volume} {529}},\ \bibinfo {pages} {1600298}}\BibitemShut {NoStop}%
\bibitem [{\citenamefont {{\v{Z}nidari\v{c}}}\ \emph
  {et~al.}(2010)\citenamefont {{\v{Z}nidari\v{c}}}, \citenamefont {Prosen},
  \citenamefont {Benenti}, \citenamefont {Casati},\ and\ \citenamefont
  {Rossini}}]{ZnidaricRossini2010}%
  \BibitemOpen
  \bibfield  {author} {\bibinfo {author} {\bibnamefont {{\v{Z}nidari\v{c}}},
  \bibfnamefont {M.}}, \bibinfo {author} {\bibfnamefont {T.}~\bibnamefont
  {Prosen}}, \bibinfo {author} {\bibfnamefont {G.}~\bibnamefont {Benenti}},
  \bibinfo {author} {\bibfnamefont {G.}~\bibnamefont {Casati}}, \ and\ \bibinfo
  {author} {\bibfnamefont {D.}~\bibnamefont {Rossini}}} (\bibinfo {year}
  {2010}),\ \href {\doibase 10.1103/PhysRevE.81.051135} {\bibfield  {journal}
  {\bibinfo  {journal} {Phys. Rev. E}\ }\textbf {\bibinfo {volume} {81}},\
  \bibinfo {pages} {051135}}\BibitemShut {NoStop}%
\bibitem [{\citenamefont {{\v{Z}nidari\v{c}}}\ \emph
  {et~al.}(2016)\citenamefont {{\v{Z}nidari\v{c}}}, \citenamefont
  {Scardicchio},\ and\ \citenamefont {Varma}}]{ZnidaricVarma2016}%
  \BibitemOpen
  \bibfield  {author} {\bibinfo {author} {\bibnamefont {{\v{Z}nidari\v{c}}},
  \bibfnamefont {M.}}, \bibinfo {author} {\bibfnamefont {A.}~\bibnamefont
  {Scardicchio}}, \ and\ \bibinfo {author} {\bibfnamefont {V.~K.}\ \bibnamefont
  {Varma}}} (\bibinfo {year} {2016}),\ \href {\doibase
  10.1103/PhysRevLett.117.040601} {\bibfield  {journal} {\bibinfo  {journal}
  {Phys. Rev. Lett.}\ }\textbf {\bibinfo {volume} {117}},\ \bibinfo {pages}
  {040601}}\BibitemShut {NoStop}%
\bibitem [{\citenamefont {{\v{Z}nidari\v{c}}}\ \emph
  {et~al.}(2011)\citenamefont {{\v{Z}nidari\v{c}}}, \citenamefont
  {\v{Z}unkovi\v{c}},\ and\ \citenamefont
  {Prosen}}]{ZnidaricZunkovicProsen2011}%
  \BibitemOpen
  \bibfield  {author} {\bibinfo {author} {\bibnamefont {{\v{Z}nidari\v{c}}},
  \bibfnamefont {M.}}, \bibinfo {author} {\bibfnamefont {B.}~\bibnamefont
  {\v{Z}unkovi\v{c}}}, \ and\ \bibinfo {author} {\bibfnamefont
  {T.}~\bibnamefont {Prosen}}} (\bibinfo {year} {2011}),\ \href {\doibase
  10.1103/PhysRevE.84.051115} {\bibfield  {journal} {\bibinfo  {journal} {Phys.
  Rev. E}\ }\textbf {\bibinfo {volume} {84}},\ \bibinfo {pages}
  {051115}}\BibitemShut {NoStop}%
\bibitem [{\citenamefont {Zotos}\ \emph {et~al.}(1997)\citenamefont {Zotos},
  \citenamefont {Naef},\ and\ \citenamefont {Prelovsek}}]{ZotosPrelovsek1997}%
  \BibitemOpen
  \bibfield  {author} {\bibinfo {author} {\bibnamefont {Zotos}, \bibfnamefont
  {X.}}, \bibinfo {author} {\bibfnamefont {F.}~\bibnamefont {Naef}}, \ and\
  \bibinfo {author} {\bibfnamefont {P.}~\bibnamefont {Prelovsek}}} (\bibinfo
  {year} {1997}),\ \href {\doibase 10.1103/PhysRevB.55.11029} {\bibfield
  {journal} {\bibinfo  {journal} {Phys. Rev. B}\ }\textbf {\bibinfo {volume}
  {55}},\ \bibinfo {pages} {11029}}\BibitemShut {NoStop}%
\bibitem [{\citenamefont {Zwanzig}(1960)}]{zwanzig1960a}%
  \BibitemOpen
  \bibfield  {author} {\bibinfo {author} {\bibnamefont {Zwanzig}, \bibfnamefont
  {R.}}} (\bibinfo {year} {1960}),\ \href {\doibase 10.1063/1.1731409}
  {\bibfield  {journal} {\bibinfo  {journal} {The Journal of Chemical Physics}\
  }\textbf {\bibinfo {volume} {33}}~(\bibinfo {number} {5}),\ \bibinfo {pages}
  {1338}}\BibitemShut {NoStop}%
\end{thebibliography}%


\end{document}